\documentclass[a4paper,12pt,oneside,openright,nohints]{report}
\pdfoutput=1 %

\usepackage[T1]{fontenc}

\usepackage[UKenglish]{babel}

\usepackage{graphicx}
\usepackage[dvipsnames]{xcolor}

\usepackage{silence}

\usepackage[lmargin=3.0cm,rmargin=2.3cm]{geometry}

\usepackage{multicol}

\usepackage{setspace}

\usepackage{fancyhdr}

\usepackage{footmisc}

\usepackage{titlesec,titletoc}

\usepackage{epigraph}

\usepackage{pdfpages}

\usepackage{pdflscape}

\usepackage{afterpage}

\usepackage{sectsty}

\usepackage[unicode,
			breaklinks=true,  %
			colorlinks=true,  %
			linkcolor=GOTOlightblue,  %
			urlcolor=GOTOlightblue,    %
			citecolor=GOTOlightblue,   %
			pdfusetitle,      %
			pdfsubject={PhD Thesis},
			pdfkeywords={astronomy}
			]{hyperref}

\usepackage{bookmark}

\usepackage{nameref}

\usepackage{csquotes}

\usepackage[UKenglish]{isodate}

\usepackage[acronym,shortcuts]{glossaries}

\usepackage[backend=bibtex,      %
			bibencoding=ascii,   %
			style=authoryear,    %
			natbib=true,         %
			dashed=false,        %
			maxnames=2,          %
			url=false,           %
			doi=false,           %
			isbn=false,          %
			eprint=false,        %
			uniquename=init,
			giveninits=true,
			sorting=nyt,
			]{biblatex}

\usepackage[titletoc]{appendix}

\usepackage{float}

\usepackage{array}

\usepackage[style=base, labelfont={bf, color=GOTOdarkblue}]{caption}

\usepackage{captcont}

\usepackage[section]{placeins}

\usepackage{rotating}

\usepackage{rotfloat}

\usepackage{tablefootnote}

\usepackage{wrapfig}

\usepackage{booktabs}

\usepackage{longtable}

\usepackage{multirow}

\usepackage{makecell}

\usepackage{tocloft}

\usepackage{minitoc}

\usepackage{amsmath,amssymb,amsthm}

\usepackage{aas_macros}

\usepackage{siunitx}
\usepackage{nicefrac}

\usepackage{listings}

\usepackage{pifont}

\usepackage{stmaryrd}

\usepackage{marvosym}

\usepackage{xfrac}

\usepackage{verbatim}

\usepackage{datetime2}

\usepackage{lipsum}

\usepackage[textsize=footnotesize]{todonotes}
\usepackage{xifthen}

\usepackage{xpatch}

\newcommand{\uptriangle}{\ding{115}}

\newcommand{\elec}{\ensuremath{\textup{e}^-}}

\makeatletter
\renewcommand{\@makefntext}[1]{\leftskip=2em\hskip-2em\@makefnmark#1} %
\makeatother

\newcommand{\pkg}{\texttt}
\newcommand{\code}{\texttt}

\newcommand{\acro}{\glsfirst}
\newcommand{\acroadd}{\glsadd}

\newcommand{\textcolorbf}[2]{\textcolor{#1}{\textbf{#2}}}
\newenvironment{colsection}{}{}

\let\stdsection\section{}
\renewcommand\section{\newpage\stdsection}

\newcommand\makequote[3]{
    \begin{center}
        {\Huge \textbf{\textcolor{GOTOlightblue}{``~~}}}\\
            {\LARGE #1}\\
        \vspace{0.3cm}
        {\Huge \textbf{\textcolor{GOTOlightblue}{~~''}}}
    \end{center}
    \vspace{-0.7cm} \hspace{8cm}
    {\large --- #2\ifthenelse{\isempty{#3}}{}{, \textit{#3}}}
    }

\newcommand{\tls}{\hspace{-3.2pt}$\bullet$ \hspace{5pt}}

\makeatletter
\renewcommand{\@pnumwidth}{3em}
\renewcommand{\@tocrmarg}{4em}
\makeatother

\makeatletter
\patchcmd{\@makechapterhead}{\vskip 20}{\vskip 0}{}{} %
\makeatother

\mtcsettitle{minitoc}{}
\mtcsetrules{*}{off}
\mtcsetfont{minitoc}{section}{\normalsize\bfseries}
\mtcsetfont{minitoc}{subsection}{\normalsize}
\mtcsetoffset{minitoc}{-0.7em}
\let\oldappendices\appendices\def\appendices{\oldappendices\adjustmtc}
\newcommand\chaptoc{
    \begin{singlespacing}
    \begin{small}
    \vspace{-1.4cm}
    \noindent{\color{GOTOlightblue}\rule{\linewidth}{1.5pt}}
    \vspace{-1cm}
    {\hypersetup{linkcolor=black}
    \minitoc{}
    }
    \vspace{-0.5cm}
    \noindent{\color{GOTOlightblue}\rule{\linewidth}{1.5pt}}
    \begin{center}
        \hyperref[contents]{\textcolor{white}{\uptriangle}}
    \end{center}
    \end{small}
    \end{singlespacing}
    \newpage
    }

\AtBeginDocument{%
}
\newcommand\aref[1]{\autoref{#1}}
\newcommand\nref[1]{\autoref{#1} (\nameref{#1})}

\DeclareFieldFormat{citehyperref}{%
  \DeclareFieldAlias{bibhyperref}{noformat}%
  \bibhyperref{#1}}
\DeclareFieldFormat{textcitehyperref}{%
  \DeclareFieldAlias{bibhyperref}{noformat}%
  \bibhyperref{%
    #1%
    \ifbool{cbx:parens}
      {\bibcloseparen\global\boolfalse{cbx:parens}}
      {}}}
\savebibmacro{cite}
\savebibmacro{textcite}
\renewbibmacro*{cite}{%
  \printtext[citehyperref]{%
    \restorebibmacro{cite}%
    \usebibmacro{cite}}}
\renewbibmacro*{textcite}{%
  \ifboolexpr{
    (not test {\iffieldundef{prenote}} and
      test {\ifnumequal{\value{citecount}}{1}})
    or
    (not test {\iffieldundef{postnote}} and
      test {\ifnumequal{\value{citecount}}{\value{citetotal}}})
  }
    {\DeclareFieldAlias{textcitehyperref}{noformat}}
    {}%
  \printtext[textcitehyperref]{%
    \restorebibmacro{textcite}%
    \usebibmacro{textcite}}}

\renewbibmacro{in:}{}
\AtEveryBibitem{%
\clearfield{booktitle}%
\clearfield{editor}%
\clearfield{title}%
\clearfield{number}%
\clearfield{eid}%
}
\renewcommand{\newunitpunct}{\addcomma\space}
\renewcommand{\labelnamepunct}{\addcomma\space}
\renewcommand{\finentrypunct}{}
\DeclareFieldFormat{journaltitle}{#1,}
\DeclareFieldFormat{pages}{#1}
\xpatchbibmacro{date+extradate}{\printtext[parens]}{\setunit{\addcomma\space}\printtext}{}{}

\chapterfont{\color{GOTOdarkblue}}
\sectionfont{\color{GOTOdarkblue}}
\subsectionfont{\color{GOTOdarkblue}}
\subsubsectionfont{\color{GOTOdarkblue}}

\definecolor{GOTOlightblue}{RGB}{0, 108, 171}
\definecolor{GOTOdarkblue}{RGB}{0, 48, 80}
\definecolor{linkblue}{RGB}{0, 124, 207}

\definecolor{tab_blue}{HTML}{4E79A7}
\definecolor{tab_orange}{HTML}{F28E2B}
\definecolor{tab_red}{HTML}{E15759}
\definecolor{tab_cyan}{HTML}{76B7B2}
\definecolor{tab_green}{HTML}{59A14F}
\definecolor{tab_yellow}{HTML}{EDC948}
\definecolor{tab_purple}{HTML}{B07AA1}
\definecolor{tab_pink}{HTML}{FF9DA7}
\definecolor{tab_brown}{HTML}{9C755F}
\definecolor{tab_grey}{HTML}{BAB0AC}

\DeclareSIUnit\parsec{pc}
\DeclareSIUnit\lightyear{ly}
\DeclareSIUnit\arcmin{arcmin}
\DeclareSIUnit\arcsec{arcsec}
\DeclareSIUnit\pixel{pixel}
\DeclareSIUnit\photon{photons}
\DeclareSIUnit\mag{mag}
\DeclareSIUnit\erg{erg}
\DeclareSIUnit\jansky{Jy}
\DeclareSIUnit\solarmass{\ensuremath{\textup{M}_\odot}}
\DeclareSIUnit\solarlum{\ensuremath{\textup{L}_\odot}}
\DeclareSIUnit\solarrad{\ensuremath{\textup{R}_\odot}}
\DeclareSIUnit\electron{\ensuremath{\textup{e}^-}}

\makeglossaries{}
\glsdisablehyper{} %

\newacronym{goto}{GOTO}{Gravitational-wave Optical Transient Observer}
\newacronym{orm_lapalma}{ORM}{Observatorio del Roque de los Muchachos}
\newacronym{decam}{DECam}{Dark Energy Camera}
\newacronym{dlt40}{DLT40}{Distance Less Than 40Mpc survey}
\newacronym{lco}{LCO}{Las Cumbres Observatory}
\newacronym{eso}{ESO}{European Southern Observatory}
\newacronym{vista}{VISTA}{Visible and Infrared Survey Telescope for Astronomy}
\newacronym{ztf}{ZTF}{Zwicky Transient Facility}
\newacronym{pirate}{PIRATE}{Physics Innovations Astronomical Telescope Explorer}
\newacronym{wasp}{WASP}{Wide Angle Search for Planets}
\newacronym{ngts}{NGTS}{Next Generation Transit Survey}
\newacronym{w1m}{W1m}{Warwick 1-metre telescope}
\newacronym{wht}{WHT}{William Herschel Telescope}
\newacronym{pt5m}{pt5m}{point-five-metre telescope}
\newacronym{slodar}{SLODAR}{SLOpe Detection And Ranging}
\newacronym{gbm}{GBM}{Gamma-ray Burst Monitor}
\newacronym{bat}{BAT}{Burst Alert Telescope}
\newacronym{asassn}{ASAS-SN}{All-Sky Automated Survey for Supernovae}
\newacronym{ptf}{PTF}{Palomar Transient Factory}
\newacronym{lsst}{LSST}{Large Synoptic Survey Telescope}
\newacronym{sdss}{SDSS}{Sloan Digital Sky Survey}

\newacronym{tcs}{TCS}{Telescope Control System}
\newacronym{gtecs}{G\babelhyphen{nobreak}TeCS}{GOTO Telescope Control System}
\newacronym[sort=rts2]{rts2}{RTS2}{Remote Telescope System v2}

\newacronym{fov}{FoV}{Field of View}
\newacronym{too}{ToO}{Target of Opportunity}
\newacronym{ut}{UT}{Unit Telescope}
\newacronym{ota}{OTA}{Optical Tube Assembly}
\newacronym{far}{FAR}{False Alarm Rate}
\newacronym{lst}{LST}{Local Sidereal Time}
\newacronym{ra}{RA}{Right Ascension}
\newacronym{ncp}{NCP}{Northern Celestial Pole}
\newacronym{scp}{SCP}{Southern Celestial Pole}
\newacronym{icrs}{ICRS}{International Celestial Reference System}

\newacronym{cbc}{CBC}{Compact Binary Coalescence}
\newacronym{bns}{BNS}{Binary Neutron Star}
\newacronym{nsbh}{NSBH}{Neutron Star-Black Hole}
\newacronym{bbh}{BBH}{Binary Black Hole}
\newacronym{gwtc}{GWTC}{Gravitational-Wave Transient Catalog}

\newacronym{gw}{GW}{Gravitational Wave}
\newacronym{em}{EM}{Electromagnetic}
\newacronym{grb}{GRB}{Gamma-Ray Burst}
\newacronym{ligo}{LIGO}{Laser Interferometer Gravitational-Wave Observatory}
\newacronym{ego}{EGO}{European Gravitational Observatory}
\newacronym{kagra}{KAGRA}{Kamioka Gravitational Wave Detector}
\newacronym{lvc}{LVC}{LIGO-Virgo Collaboration}
\newacronym{lisa}{LISA}{Laser Interferometer Space Antenna}
\newacronym{o1}{O1}{LIGO First Observing Run}
\newacronym{o2}{O2}{LIGO Second Observing Run}
\newacronym{o3}{O3}{LIGO-Virgo Third Observing Run}
\newacronym{gwgc}{GWGC}{Gravitational Wave Galaxy Catalogue}
\newacronym{glade}{GLADE}{Galaxy List for the Advanced Detector Era}
\newacronym{ivoa}{IVOA}{International Virtual Observatory Alliance}

\newacronym{healpix}{HEALPix}{Hierarchical Equal Area isoLatitude Pixelisation}
\newacronym{gcn}{GCN}{Gamma-ray burst Coordinates Network}
\newacronym{ivorn}{IVORN}{International Virtual Observatory Resource Name}

\newacronym{sitech}{SiTech}{Sidereal Technology}
\newacronym{fli}{FLI}{Fingerlake Instruments}
\newacronym{omi}{OMI}{Optical Mechanics Inc}
\newacronym{nuc}{NUC}{Next Unit of Computing}
\newacronym[longplural=Uninterruptible Power Supplies]{ups}{UPS}{Uninterruptible Power Supply}
\newacronym{pdu}{PDU}{Power Distribution Unit}
\newacronym{mosfet}{MOSFET}{Metal-Oxide-Semiconductor Field-Effect Transistor}

\newacronym{ssh}{SSH}{Secure Shell}
\newacronym{api}{API}{Application Programming Interface}
\newacronym{plc}{PLC}{Programmable Logic Controller}
\newacronym{com}{COM}{Component Object Model}
\newacronym{tcpip}{TCP/IP}{Transmission Control Protocol/Internet Protocol}
\newacronym{snmp}{SNMP}{Simple Network Management Protocol}
\newacronym{sdk}{SDK}{Software Development Kit}
\newacronym{fits}{FITS}{Flexible Image Transport System}
\newacronym{sql}{SQL}{Structured Query Language}
\newacronym{orm}{ORM}{Object-Relational Mapping}
\newacronym{xml}{XML}{Extensible Markup Language}
\newacronym{gil}{GIL}{Global Interpreter Lock}

\newacronym{adc}{ADC}{Analogue-to-Digital Converter}
\newacronym{adu}{ADU}{Analogue-to-Digital Units}
\newacronym{ccd}{CCD}{Charge-Coupled Device}
\newacronym{ptc}{PTC}{Photon Transfer Curve}
\newacronym{qe}{QE}{Quantum Efficiency}
\newacronym{snr}{SNR}{Signal-to-Noise Ratio}
\newacronym{fwhm}{FWHM}{Full-width at Half Maximum}
\newacronym{zp}{ZP}{Zeropoint}
\newacronym{lm}{LM}{Limiting Magnitude}
\newacronym{hfd}{HFD}{Half Flux Diameter}
\newacronym{nfv}{NFV}{Near-Focus Value}

\begin{document}

\title{A telescope control and scheduling system for the Gravitational-wave Optical Transient Observer}
\author{Martin J Dyer}

\singlespacing{}
\pagenumbering{roman}
\pagestyle{empty}
\begin{center}
    \includegraphics[width=0.45\linewidth]{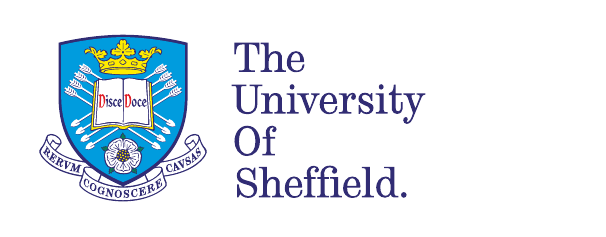}
    \includegraphics[width=0.45\linewidth]{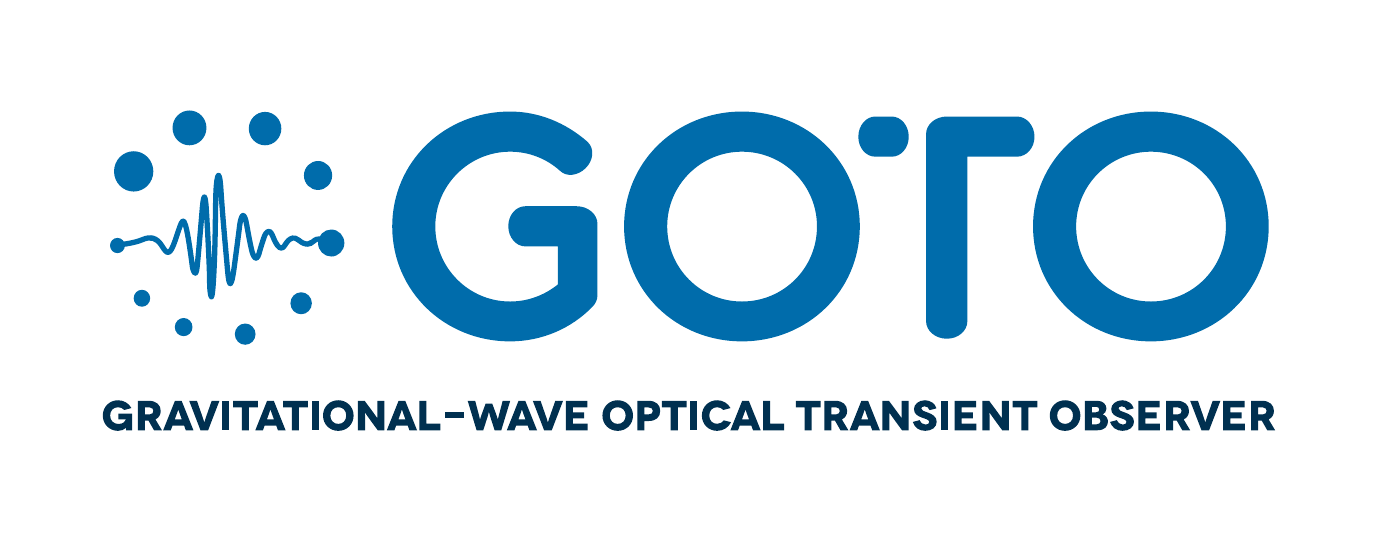}

    \vspace*{2.5cm}

    \begin{Huge}
        \textbf{A telescope control and
               scheduling system for the
               Gravitational-wave Optical Transient Observer
               }
    \end{Huge}

    \vspace*{2.5cm}

    \begin{LARGE}
        \text{Martin J. Dyer}
    \end{LARGE}

    \vspace*{1cm}

    \begin{Large}
        Department of Physics and Astronomy \\
        \smallskip
        The University of Sheffield
    \end{Large}

    \vspace*{2.5cm}

    \begin{large}
        \textit{A dissertation submitted in candidature for the degree of} \\
        \textit{Doctor of Philosophy at the University of Sheffield}
    \end{large}

    \vspace*{1cm}

    \begin{large}
        \text{26th September 2019}
    \end{large}

    \vfill
\end{center}

\cleardoublepage{}

\vspace*{2cm}

\begin{center}
    \begin{Huge}
        \textbf{\textcolor{GOTOlightblue}{``~~~~~~~~~~~~~~~~~~~~~~~~~~~~~~~~~~~~~~~~~~~~~}} \\
    \end{Huge}
    \begin{LARGE}
        \vspace{-\baselineskip}
        Astronomy compels the soul to look upwards \\
        and leads us from this world to another. \\
        \vspace{-\baselineskip}
    \end{LARGE}
    \begin{Huge}
        \textbf{\textcolor{GOTOlightblue}{~~~~~~~~~~~~~~~~~~~~~~~~~~~~~~~~~~~~~~~~~''}} \\
    \end{Huge}
\end{center}

\begin{small}
    \hspace{4cm} --- Plato, The \textit{Republic}, Book VII, c.\@ 380\@ \textsc{bce}

    \vspace{0.2cm}
    \hspace{4cm} --- Translation by Benjamin Jowett, {\hypersetup{urlcolor=black}\href{http://www.gutenberg.org/files/55201/55201-h/55201-h.htm}{\textit{The Republic of Plato}}}, 1888

    \vspace{0.2cm}
    \hspace{4cm} --- Quoted by Leonard Nimoy, \textit{Sid Meier's Civilization IV}, 2005
\end{small}

\vspace{3cm}

\begin{center}
    \begin{Huge}
        \textbf{\textcolor{GOTOlightblue}{``~~~~~~~~~~~~~~~~~~~~~~~~~~~~~}} \\
    \end{Huge}
    \begin{LARGE}
        \vspace{-\baselineskip}
        Astronomy's much more fun \\
        when you're not an astronomer. \\
        \vspace{-\baselineskip}
    \end{LARGE}
    \begin{Huge}
        \textbf{\textcolor{GOTOlightblue}{~~~~~~~~~~~~~~~~~~~~~~~~~~~~~~~~~''}} \\
    \end{Huge}
\end{center}

\begin{small}
    \hspace{4cm} --- Brian May, {\hypersetup{urlcolor=black}\href{https://brianmay.com/brian/magsandpress/voxmar91/voxmar91.html}{Interview with \textit{Vox} magazine}}, 1991

    \vspace{0.2cm}
    \hspace{4cm} --- Quoted by Sean Bean, \textit{Sid Meier's Civilization VI}, 2016
\end{small}

\doublespacing{}
\pagestyle{plain}
\chapter*{Declaration}

\begin{onehalfspace}

\noindent I declare that, unless otherwise stated, the work presented in this thesis is my own.

\bigskip

\noindent No part of this thesis has been accepted or is currently being submitted for any other qualification at the University of Sheffield or elsewhere.

\bigskip

\noindent Some of my work on the GOTO Telescope Control System has previously been published as \citet{Dyer}. This paper forms the basis of Chapters 3, 4 and 5 of this thesis. The other chapters have not been published previously, although the work within them will contribute to future publications.

\end{onehalfspace}

\chapter*{Acknowledgements}

\begin{onehalfspace}

This is the very last thing I'm writing before sending this to print, because I'm awful at writing these sort of things and I kept putting it off. Not that it really matters, of the very few people who will ever read this thesis I'm guessing most will skip over this bit. I always do.

\medskip

Anyway, I'll start by thanking my PhD supervisor Vik Dhillon, for being a fantastic supervisor and mentor over the past four years, as well as, when needed, a 5-star tour guide of La Palma. Thanks also to my second supervisor Ed Daw, whose Physical Computing class I enjoyed both teaching and learning from, as well as Stu, Dave, Steven, Liam, James, Lydia, Pablo and others in the group in Sheffield. I'd also like to thank members of the GOTO collaboration: Danny, Krzysztof, Paul, Joe, James and Ryan at Warwick; Duncan, Kendall, Evert, Travis and Alex at Monash; Gav, Mark and everyone else from the ever-increasing list of member institutions (especially the poor students who had to babysit GOTO for so long). Also my masters project supervisors at Durham, Richard and Tim, who first put me in touch with Vik --- without them I literally would not be where I am today.

\medskip

I'd like to thank all the various friends I've made over the years from Maidenhead, Borlase, Durham and Sheffield; including, but not at all limited to, Alex, Anna, Becky, Ben, Emma, Gemma, Harriet, Harry, Héloïse, James, Katie, Katherine, Laura, Mac, Oliver, Richard, Robin, Simon, Sophie and Varun (I wanted to include as many people as I could, apologies to anyone I forgot). A particular massive thanks to the other PhD students in my year at Sheffield: Becky, Héloïse and Katie. You were the best office-mates and compatriots I could have asked for, and while you all finished before me and went off to better things, it did mean I my own office in the final few weeks --- thanks!

\medskip

Of course, I must also thank my Mum and Dad, and my brother Phil, for their unwavering support and encouragement, as well as my Grandad Norman, my cousin Nye, and the rest of my family.

\newpage

\medskip

Finally, I would like to acknowledge the people who did the most to encourage my love of physics and astronomy that eventually culminated in this thesis: my secondary-school physics teacher Malcolm Brownsell, the members of the Maidenhead Astronomical Society, and my grandfather Eric Snelling, aka Bubba. May you all have clear skies.

\smallskip

\begin{flushright}
--- Martin Dyer, Sheffield, 25th September 2019
\end{flushright}

\medskip

PS Although the above is dated the 25th, and I did try to print out and submit on that day, I failed to do so --- the office didn't have any binding combs large enough for this epic tome. So I'm going to get it done today, the 26th, at the Rymans in town, and I'll leave this postscript as a permanent record. Let's hope today is the day!

\medskip

PPS As my final correction (I'm writing this on the 20th of December) I'd also like to say thanks to my viva examiners, Iain and James. The viva was far more enjoyable that I expected, and their feedback was very useful. And now, at last, it is finished.
\end{onehalfspace}

\pdfbookmark[section]{Summary}{summary}
\chapter*{Summary}

\begin{onehalfspace}

The detection of the first electromagnetic counterpart to a gravitational-wave signal in August 2017 marked the start of a new era of multi-messenger astrophysics. An unprecedented number of telescopes around the world were involved in hunting for the source of the signal, and although more gravitational-wave signals have been since detected, no further electromagnetic counterparts have been found.

\medskip

In this thesis, I present my work to help build a telescope dedicated to the hunt for these elusive sources: the Gravitational-wave Optical Transient Observer (GOTO). I detail the creation of the GOTO Telescope Control System, G-TeCS, which includes the software required to control multiple wide-field telescopes on a single robotic mount. G-TeCS also includes software that enables the telescope to complete a sky survey and transient alert follow-up observations completely autonomously, whilst monitoring the weather conditions and automatically fixing any hardware issues that arise. I go on to describe the routines used to determine target priorities, as well as how the all-sky survey grid is defined, how gravitational-wave and other transient alerts are received and processed, and how the optimum follow-up strategies for these events were determined.

\medskip

The first GOTO telescope, situated on La Palma in the Canary Islands, saw first light in June 2017. I detail the work I carried out on the site to help commission the prototype, and how the control software was developed during the commissioning phase. I also analyse the GOTO CCD cameras and optics, building a complete theoretical model of the system to confirm the performance of the prototype. Finally, I describe the results of simulations I carried out predicting the future of the GOTO project, with multiple robotic telescopes on La Palma and in Australia, and how the G-TeCS software might be modified to operate these telescopes as a single, global observatory.

\end{onehalfspace}

\chapter*{Contents}
\renewcommand{\contentsname}{}
\vspace{-3.5cm}

\dominitoc{}

{\hypersetup{linkcolor=black}
\pdfbookmark[section]{Contents}{contents}
\tableofcontents
}
\label{contents}
\clearpage

\chapter*{Figures}
\renewcommand{\listfigurename}{}
\vspace{-3.5cm}

{\hypersetup{linkcolor=black}
\pdfbookmark[section]{Figures}{figures}
\listoffigures
}
\clearpage

\chapter*{Tables}
\renewcommand{\listtablename}{}
\vspace{-3.5cm}

{\hypersetup{linkcolor=black}
\pdfbookmark[section]{Tables}{tables}
\listoftables
}
\clearpage

\chapter*{Acronyms and Abbreviations}
\vspace{-3.5cm}

\setlength{\glsdescwidth}{0.8\linewidth}

\pdfbookmark[section]{Acronyms and Abbreviations}{aaa}

\begingroup
\let\clearpage\relax
\vspace{-12pt}
\begin{singlespacing}
\printglossary[title={},
               type=\acronymtype,
               style=super,
               nopostdot,
               nonumberlist,
               ]
\end{singlespacing}
\endgroup

\cleardoublepage{}
\pagenumbering{arabic}
\pagestyle{fancy}

\chapter{Introduction}
\label{chap:intro}

\chaptoc{}

\section{Gravitational Waves}
\label{sec:gw}

\begin{colsection}

Einstein's theory of General Relativity describes gravity as the curvature of spacetime \citep{Einstein1914}, and he went on to describe the propagation of distortions within the spacetime `fabric' \citep{Einstein1916}. These \emph{gravitational waves} (GWs) \acroadd{gw} are produced by the acceleration of matter within the field of spacetime and propagate at the speed of light \citep{GW170817_gravity}, analogous to electromagnetic (EM) \acroadd{em} waves being produced by an accelerating charge. The existence of gravitational waves is a consequence of the finite propagation time of gravity in general relativity; there is no analogue to gravitational waves in Newtonian gravity as Newton described a force propagating instantaneously.

The result of Einstein's theory is the quadrupole formula \citep{Einstein1916}, which describes gravity propagating as a transverse wave, which alternately stretches and compresses spacetime in two orthogonal axes \citep{BIGcardiff}. A single object will never `observe' a gravitational wave, as it is embedded in the fabric, and the only way to detect the passing of gravitational waves is to look for changes in the relative positions of two or more objects as the wave passes through. A thought experiment considering the effects of gravitational waves on free-floating masses is shown in \aref{fig:wave}, for the two wave polarisation states. These perturbations are quantified by the strain, the fractional change in distance, which even for astronomical-scale events will be incredibly small --- the first direct detection of gravitational waves involved measuring strains of the order of $10^{-21}$ (see \aref{fig:chirp}). It is the goal of gravitational-wave detectors to observe these minute spatial perturbations as the wave passes through.

A detailed discussion of general relativity and gravitational-wave science is beyond the scope of this thesis, so this section gives only a brief introduction to the topic in order to explain the core purpose of the GOTO project.

\begin{figure}[t]
    \begin{center}
        \includegraphics[width=0.8\linewidth]{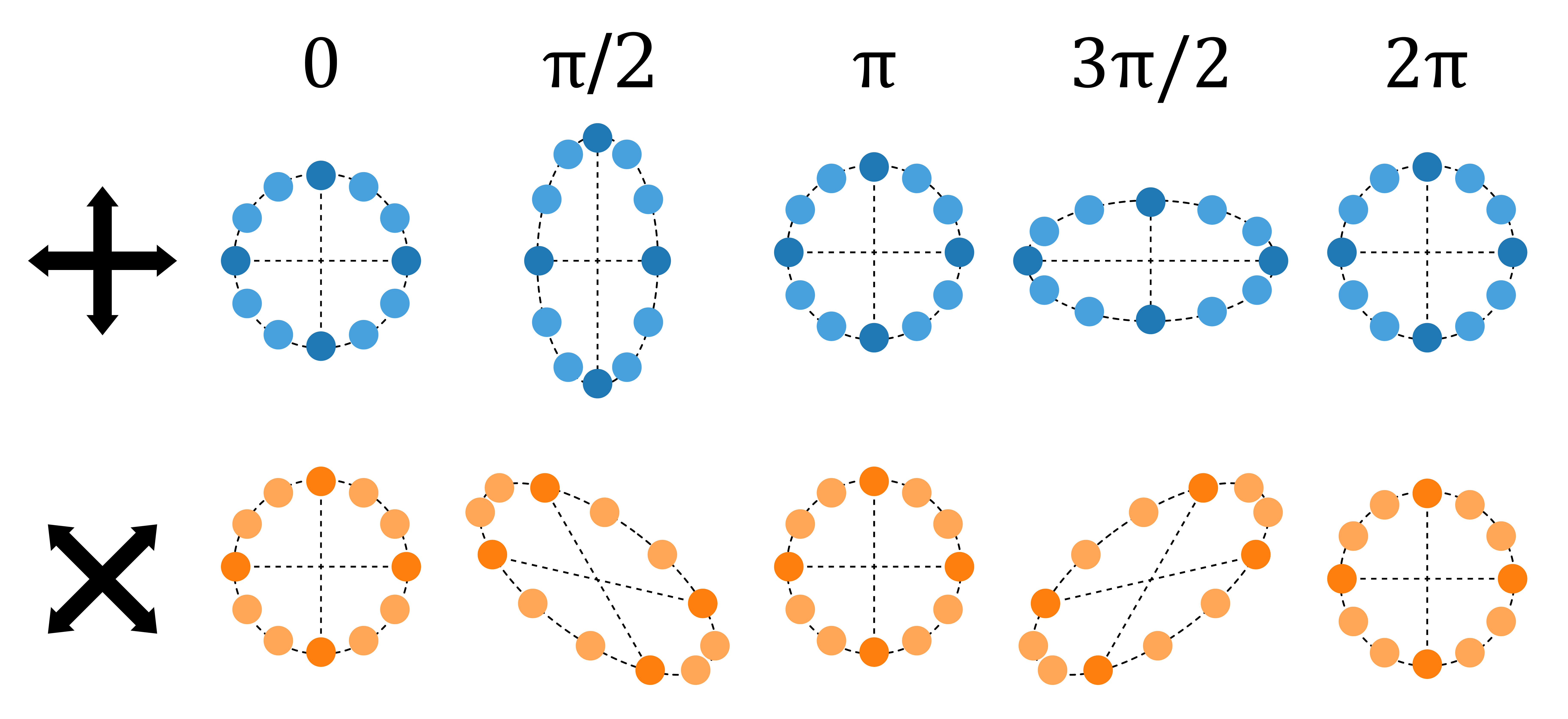}
    \end{center}
    \caption[Gravitational-wave polarisations]{
        Consider a 2-dimensional ring of free-floating particles in the $x$-$y$-plane. A gravitational-wave which passes through the ring in the $z$-direction (out of the page) will alternately stretch and compress the ring in two orthogonal axes out of phase. The upper image shows the effect of a plus-polarised wave, the lower image shows the effect of a cross-polarised wave. Adapted from \citet{BIGaustralia}.
        }\label{fig:wave}
\end{figure}

\newpage

\end{colsection}

\subsection{Detecting gravitational waves}
\label{sec:gw_detecting}
\begin{colsection}

As described above, gravitational waves manifest as alternately stretching and compressing spacetime along perpendicular axes. Several methods of directly detecting gravitational waves have been proposed, but the most successful design uses a Michelson interferometer to observe how two test masses move relative to each other as a wave passes through \citep{BIGbirmingham}. As shown in \aref{fig:detector}, an input laser is split into two by a beam splitter, and each beam is sent into one of two long perpendicular arms. In order to detect the tiny strains from gravitational waves these arms needs to be kilometres in length, and each arm acts as a laser cavity, reflecting the beam multiple times between two mirrored test masses, to further increase the effective distance. When they exit the arms, the beams are recombined to form a single output. Should the lengths of the arms change relative to each other, e.g.\ due to a gravitational wave passing through, the distance the beams travel will be different, which will produce a change in the resulting interference pattern produced when they recombine. The test masses are suspended by a complex vibration isolation system in order to reduce any outside interference, such as from man-made vibrations or seismic events.

\begin{figure}[t]
    \begin{center}
        \includegraphics[width=0.75\linewidth]{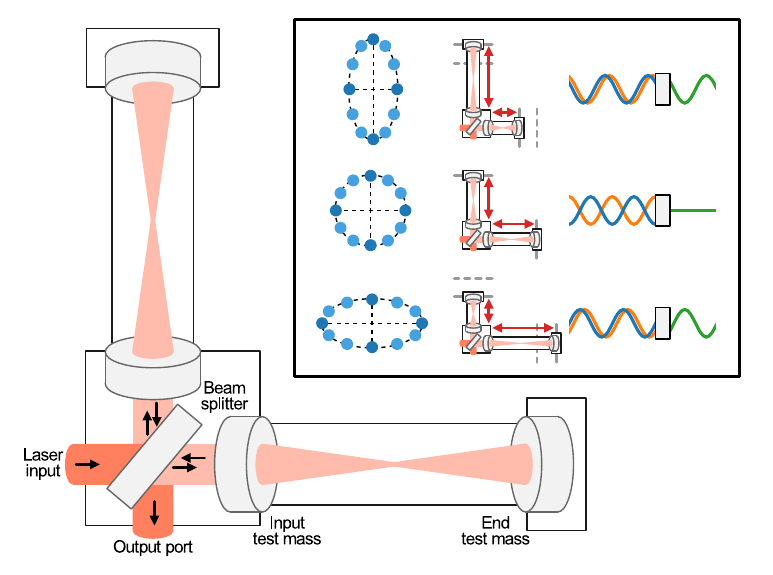}
    \end{center}
    \caption[A Michelson interferometer used as a gravitational-wave detector]{
        A Michelson interferometer used as a gravitational-wave detector. As a wave passes through, the relative lengths of the arms will change, as shown (highly exaggerated) in the inset. This will reduce or increase the distance the laser light travels through each arm, and therefore alter the output interference signal. Adapted from \citet{GW150914_detectors}.
        }\label{fig:detector}
\end{figure}

\begin{figure}[t]
    \begin{center}
        \includegraphics[width=0.95\linewidth]{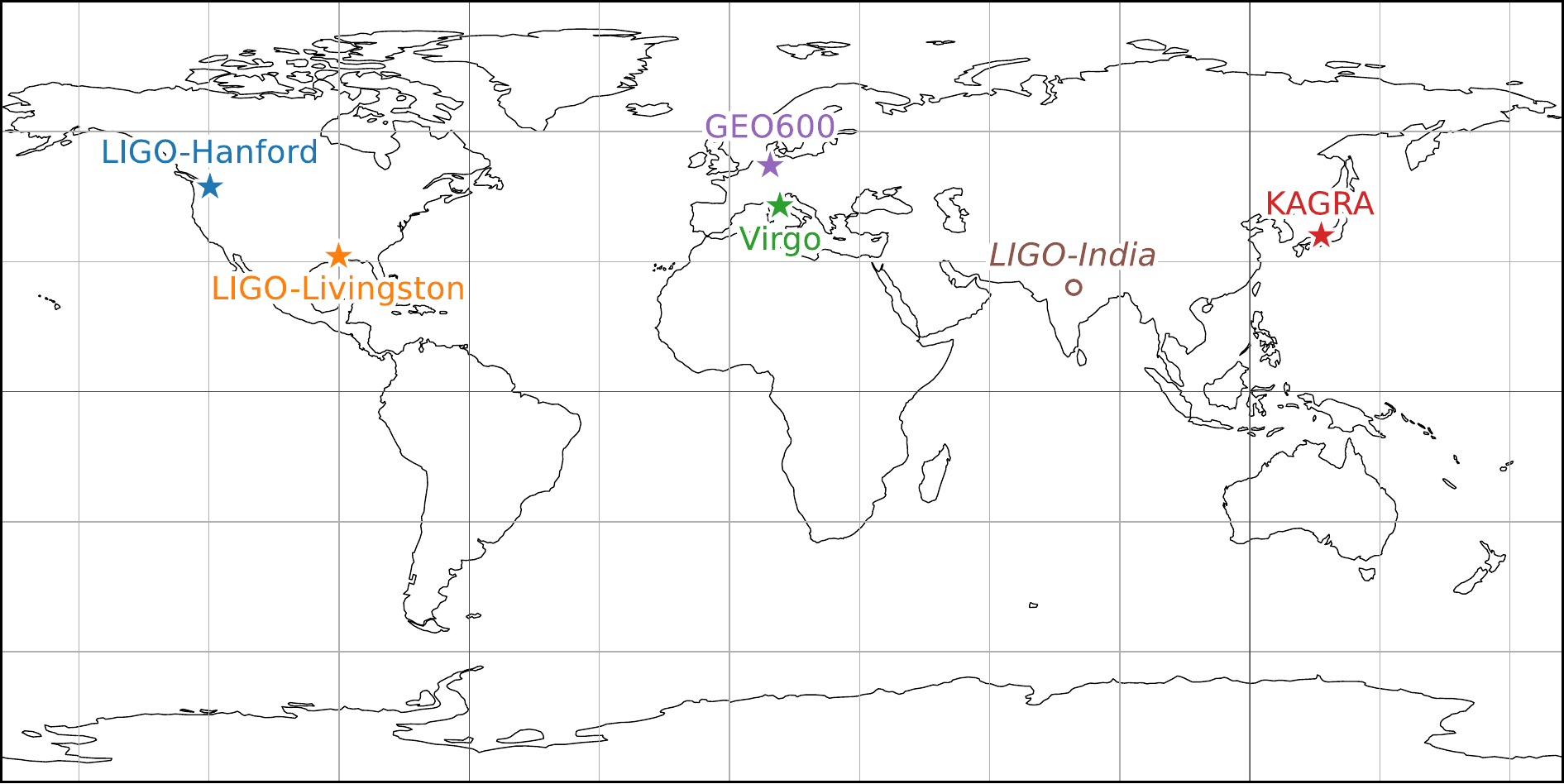}
    \end{center}
    \caption[Locations of gravitational-wave detectors]{
        Locations of current and proposed gravitational-wave detectors.
        }\label{fig:global}
\end{figure}

Several of these gravitational-wave detectors have been built around the world, as shown in \aref{fig:global}. Having multiple detectors acting together provides redundancy, and allows the source of the signal to be localised (see \aref{sec:gw_localisation}). There are currently three active detectors: the two Advanced \acro{ligo} detectors in the United States, at Hanford, Washington and Livingston, Louisiana \citep{LIGO}, and the \acro{ego} Advanced Virgo detector near Pisa, Italy \citep{Virgo}. These three detectors form a global network known as the LIGO-Virgo Collaboration \acroadd{lvc} \citep[LVC,][]{LIGO-Virgo}. In addition, the older German-British GEO600 detector in Germany is still active, primarily as a technology test system \citep{GEO600}. The \acro{kagra} is currently under construction in Japan \citep{KAGRA}, and is expected to join the global network before the end of 2019 \citep{LIGO-Virgo-KAGRA}. In the next decade, work should also begin on building a third LIGO detector, LIGO-India \citep{LIGO_India}, relocating what was previously a second interferometer at Hanford.

In the longer term, the next generation of larger and more sensitive gravitational-wave detectors is already being designed, including the Einstein Telescope \citep{EinsteinTelescope} and the Cosmic Explorer \citep{CosmicExplorer}. Space-based gravitational-wave detectors are also being planned, such as the Laser Interferometer Space Antenna \acroadd{lisa} \citep[LISA,][]{LISA}. Detectors in space would be free from the seismic noise that limits ground-based detectors at low frequencies, and could therefore detect lower-frequency gravitational waves. This could potentially include signals from supermassive black hole mergers, lower-mass white dwarf binaries within our own galaxy, and early detections of neutron star or black hole mergers that are subsequently observed by larger detectors on Earth.

\newpage

\end{colsection}

\subsection{Sources of gravitational waves}
\label{sec:gw_sources}
\begin{colsection}

Any accelerating mass will generate gravitational waves as it moves through spacetime, as long as the motion is not spherically symmetric \citep[such as a rotating disk or a uniformly expanding sphere;][]{BIGcardiff,BIGparis}. However, in practice it is only possible to detect gravitational waves from astronomical sources, as only they will produce large enough strains to be picked up by the detectors.

A continuous source of gravitational waves will be generated by two massive objects orbiting one another \citep{GW_sources}, and the loss of energy from the system in the form of gravitational waves will slowly cause the orbiting distance of the two objects to shrink. The first binary pulsar was discovered in 1974 \citep{HulseTaylor}, and after repeated observations it was apparent that the orbital period of the two stars was decreasing in perfect agreement with the predictions given by general relativity. This was the first real evidence, albeit indirect, of the existence of gravitational waves, and the discovery of the Hulse-Taylor pulsar was deemed so significant that its discoverers were awarded the Nobel Prize in 1993 \citep{HulseTaylor2}.

The loss of energy in the form of gravitational radiation will cause binary orbits to slowly decay \citep[unless counteracted by another process, such as mass transfer;][]{binary_masstransfer}. As the orbital distance decreases so will the period, resulting in the objects orbiting faster and the system emitting gravitational waves at higher frequencies. This will produce a characteristic `chirp' signal until the two objects collide \citep{GW_sources, BIGparis}. At the point of coalescence the system will release a huge burst of gravitational energy; at its peak the GW150914 signal reached a luminosity of \SI{3.6e49}{\watt} \citep[greater than the combined luminosity of all stars in the observable universe;][]{GW150914}. After the inspiral and merger gravitational waves are still detectable in the ``ring-down'' phase, as the resulting object gradually settles to form a stable sphere \citep{GW_ringdown}. More massive objects produce stronger signals, and so the ideal binary systems for gravitational-wave detections are from compact binary coalescence (CBC) \acroadd{cbc} events, which include binary neutron stars \acroadd{bns} (BNS), binary black holes \acroadd{bbh} (BBH) and neutron star-black hole (NSBH) \acroadd{nsbh} binaries.

Coalescing binaries are not the only predicted sources of gravitational-wave signals. Sources of gravitational waves are typically classed into three categories: bursts, continuous emission and the stochastic background. Along with coalescing binaries, core-collapse supernovae are predicted to produce bursts of gravitational radiation \citep{GW_supernovae}, as long as the explosions are not entirely symmetrical. Asymmetric, rapidly-spinning neutron stars (pulsars) should produce continuous, periodic emission of gravitational waves \citep{GW_pulsars}. A stochastic background of gravitational radiation from events throughout the history of the universe is also predicted, which, if measured, could provide insights into the physics of the early universe \citep{GW_background, GW_background2}. In this thesis, I will focus on gravitational-wave signals from compact binaries, as at the time of writing they are the only confirmed detections by the LIGO-Virgo interferometers.

\end{colsection}

\subsection{Gravitational-wave detection history}
\label{sec:gw_detections}
\begin{colsection}

The LIGO detectors became operational in 2002 and observed on-and-off until 2010 without detecting any gravitational-wave signals, after which they were taken offline in order to upgrade into the second-generation Advanced LIGO detectors \citep{LIGO_initial, LIGO_advanced}. The upgrade to the optics and lasers increased the strain sensitivity by a factor of 10, which corresponded to an increase in the search volume of space from which a signal could be detected by a factor of 1000 \citep{LIGO}.

The detectors were recommissioned in 2015, and first direct detection of gravitational waves occurred on the 14th of September 2015, while the two LIGO detectors were still in engineering mode. The signal, GW150914\footnote{Confirmed gravitational-wave detections are named in the form GW\textit{YYMMDD}.}, was produced by the merger of a binary black hole system approximately \SI{440}{\mega\parsec} away with component masses of \SI{35}{\solarmass} and \SI{30}{\solarmass} \citep{GW150914}. The `chirp' signals recorded in the LIGO detectors are shown in \aref{fig:chirp}. Note the strain on the $y$-axis is of the order $10^{-21}$, meaning that, as the wave passed through, the \SI{4}{\kilo\metre}-long LIGO detector arms changed in length by approximately \SI{4e-18}{\metre}, a fraction of the size of a proton ($\approx$ \SI{e-15}{\metre}).

\begin{figure}[t]
    \begin{center}
        \includegraphics[width=\linewidth]{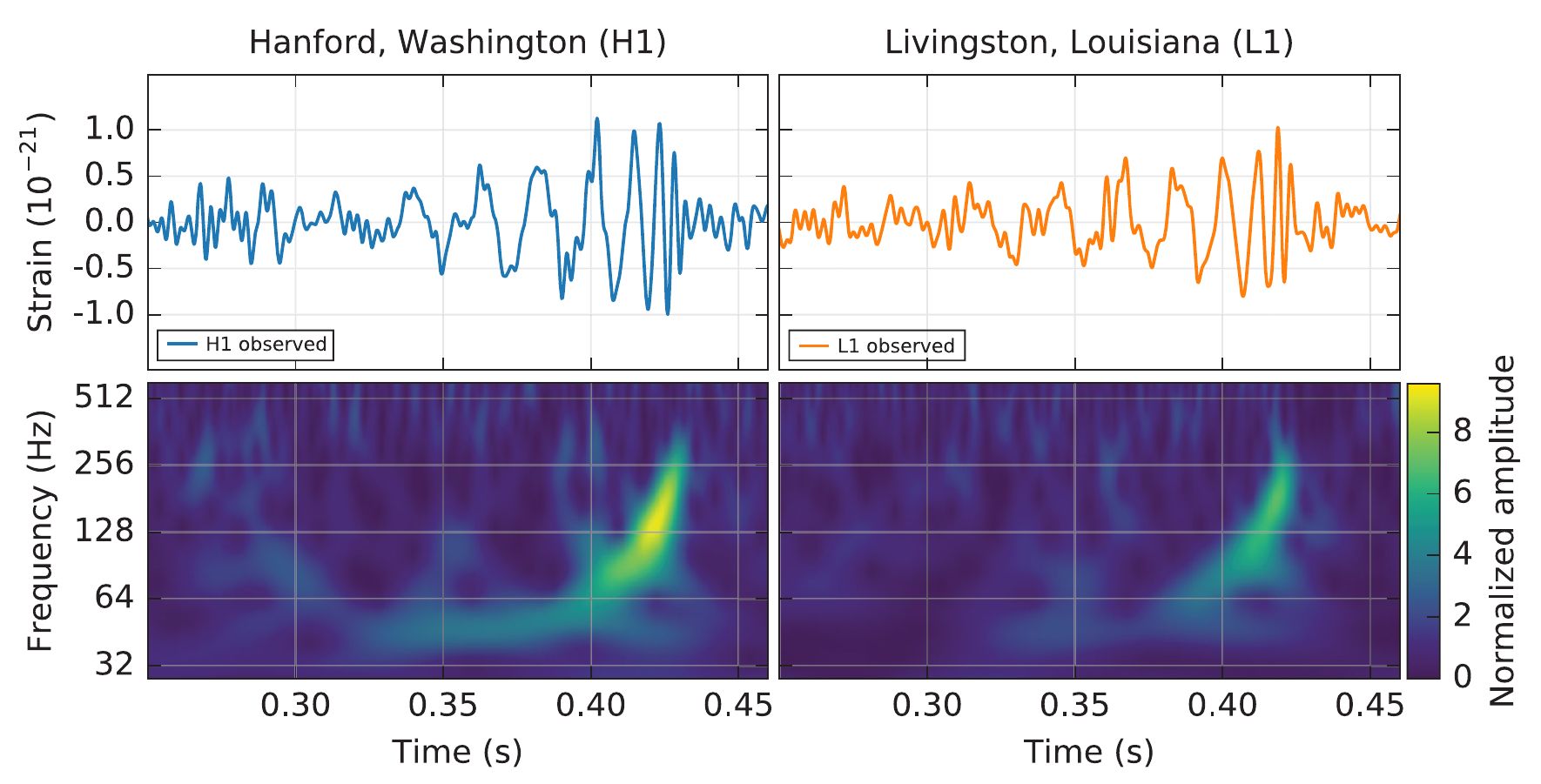}
    \end{center}
    \caption[The first detection of gravitational waves]{
        The first detection of gravitational waves, recorded in the LIGO Hanford detector (left) and then approximately \SI{7}{\milli\second} later in the LIGO Livingston detector (right). Adapted from \citet{GW150914}.
        }\label{fig:chirp}
\end{figure}

The first LIGO observing run (O1) \acroadd{o1} continued from September 2015 to January 2016, and during that time two further gravitational-wave signals were detected \citep{LIGO_O1}. All three detections were identified as being produced by coalescing black hole binaries, and although at the time one (LVT151012) was below the $5\sigma$ significance level, it has since been upgraded to a significant detection and reclassified as GW151012 \citep{GW_catalog}.

The second observing run (O2) \acroadd{o2} took place from November 2016 to August 2017. This run saw the first observation of gravitational waves from a binary neutron star, GW170817 \citep{GW170817}, as well as the addition of the Virgo detector to the network in the final month. In total eleven gravitational-wave events were detected during O1 and O2, ten from binary black holes and one (GW170817) from a binary neutron star. Together these eleven events form the first Gravitational-Wave Transient Catalogue \acroadd{gwtc} \citep[GWTC-1;][]{GW_catalog}. The source and remnant masses for each event are shown in \aref{fig:gw_masses}, compared to previous direct detections of neutron stars and stellar-mass black holes.

\begin{figure}[t]
    \begin{center}
        \includegraphics[width=\linewidth]{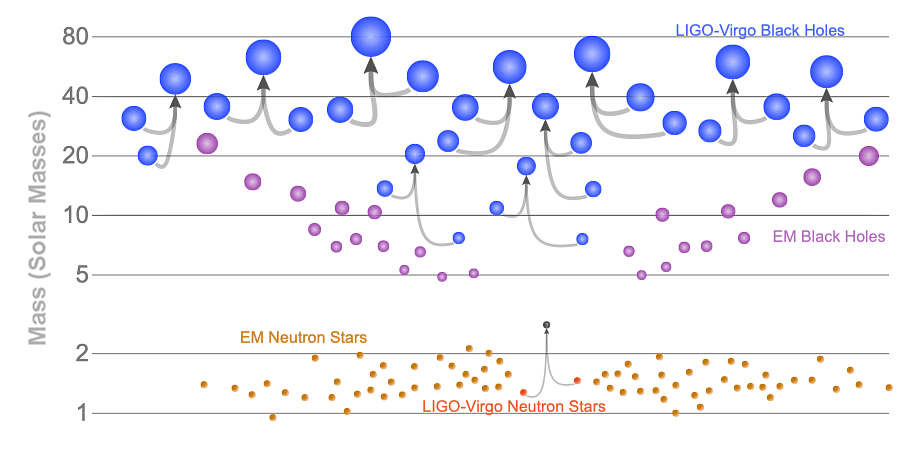}
    \end{center}
    \caption[Sources of gravitational waves detected during O1 and O2]{
        Masses of the sources and remnant objects from the 11 signals detected during O1 and O2, compared to previous electromagnetic detections. Note the ``mass gap'' between 2--\SI{5}{\solarmass}. Image credit: LIGO/Virgo/Northwestern/Frank Elavsky.
        }\label{fig:gw_masses}
\end{figure}

After a few short engineering runs the third observing run (O3) \acroadd{o3} began on 1 April 2019. At the time of writing, O3 is still ongoing, and after a short break in October 2019 is expected to run until May 2020. This is the first run to include the three LIGO-Virgo detectors from the beginning, and KAGRA is expected to join before the end of 2019 \citep{LIGO-Virgo-KAGRA}. O3 also marked the start of public alert releases\footnote{Public alerts are available at \url{https://gracedb.ligo.org/superevents/public/O3}.}; during O1 and O2 immediate alerts were only released to groups who had signed memoranda of understanding with the LVC (the GOTO Collaboration was one of these groups).

In the first 5 months of O3, from the start of April to the end of August 2019, the LVC released 32 alerts; 7 were ultimately retracted as false alarms, leaving 25 due to real astronomical signals. As O3 is currently ongoing the LVC has not yet published final values or mass estimates for any of these events. As such they are still treated as candidates, with provisional signal designations and preliminary classification probabilities. Of the 25 non-retracted events, 20 are currently classified as originating from binary black hole systems ($P_\text{BBH}>90\%$). Only one is likely from a binary neutron star \citep[S190425z;][]{S190425z}, one is classed as a likely neutron star-black hole binary \citep[S190814bv;][]{S190814bv}, and one \citep[S190426c;][]{S190426c} has an uncertain classification: a 49\% probability as coming from a binary neutron star, 24\% coming from a binary including a `MassGap' object \citep[a theorised object with a mass between a neutron star and a black hole;][]{GW_MassGap}, and 13\% from a neutron star-black hole binary (the remaining 14\% is the chance the signal is from a non-astrophysical source, i.e.\ detector noise). The remaining two events both have over 50\% non-astrophysical probability but have not been formally retracted by the LVC.\@

\end{colsection}

\subsection{On-sky localisation}
\label{sec:gw_localisation}
\begin{colsection}

One limitation with using interferometers to detect gravitational-wave signals is that alone they are very poor at localising the direction a signal originates from. It is possible to estimate a rough direction from the polarisation of the signal, and the distance to the origin can be estimated from the signal strength, but multiple detectors are needed to obtain more accurate sky localisations \citep{GW_localisation, GW_localisation2}. With two detectors, the difference between the arrival time of a signal at each allows the direction to the source to be narrowed down, based on the distance between the two detectors and knowing that gravitational waves propagate at the speed of light. However, this will only constrain the source to within an annulus on the celestial sphere, perpendicular to the line between the detectors. As shown in \aref{fig:triangulate}, at least three detectors are needed to triangulate the source location, and even then only to two positions on opposite sides of the sky (in practice the polarisation will suggest which of these two points is the more likely origin). The localisation skymaps for the GW170817 event, showing regions on the sky that the source is predicted to be within, are shown in \aref{fig:170817_skymaps}. They show how the contributions from multiple detectors dramatically improved the on-sky localisation.

Even with the three current detectors, sources that are detected by all three can typically only be localised to areas of tens to hundreds of square degrees. This also requires all three detectors to be observing at the same time, with no redundancy for down time due to maintenance or hardware problems. This is why the full network is anticipated to include five detectors across the globe, as described in \aref{sec:gw_detecting}.

\begin{figure}[t]
    \begin{center}
        \includegraphics[width=0.4\linewidth]{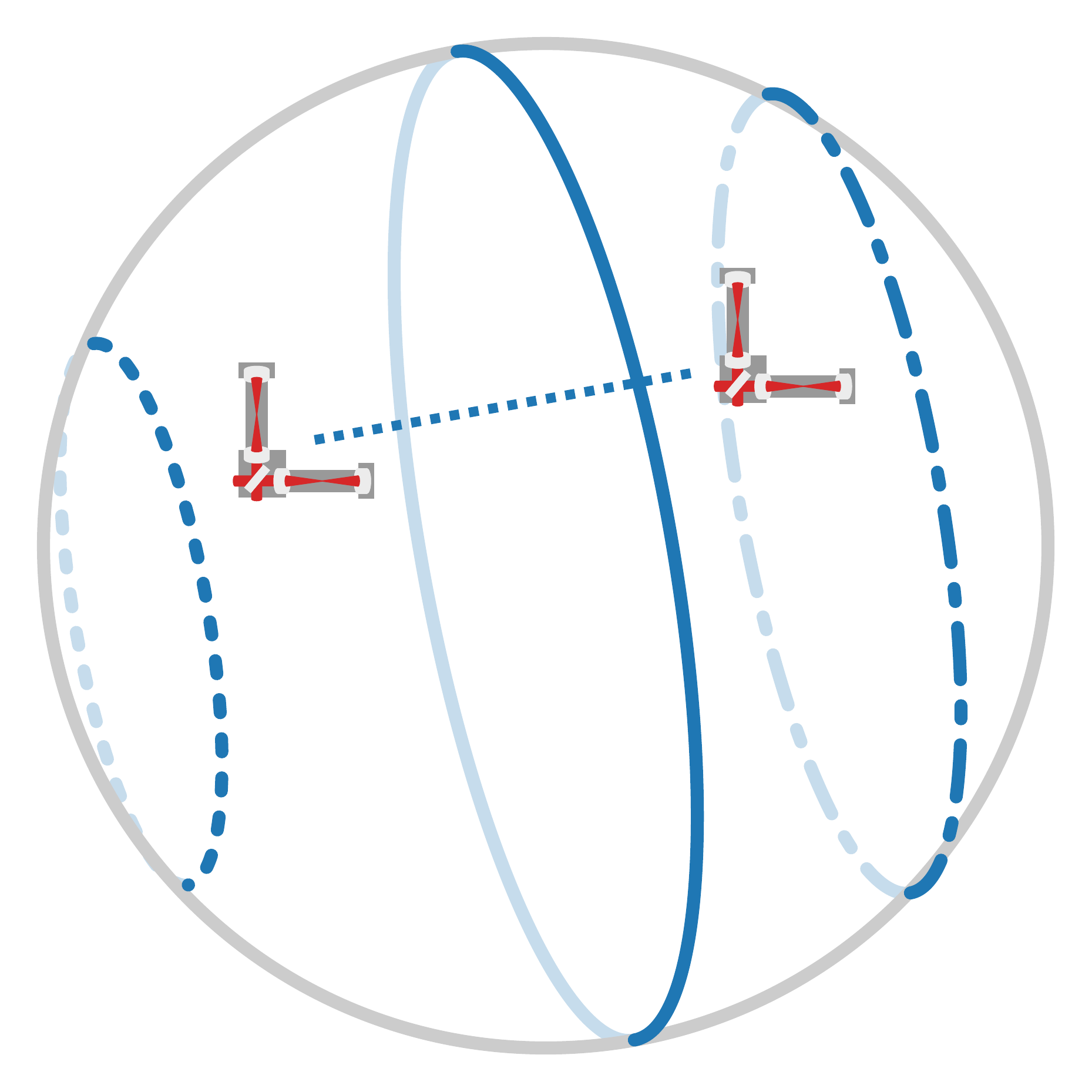}
        \includegraphics[width=0.4\linewidth]{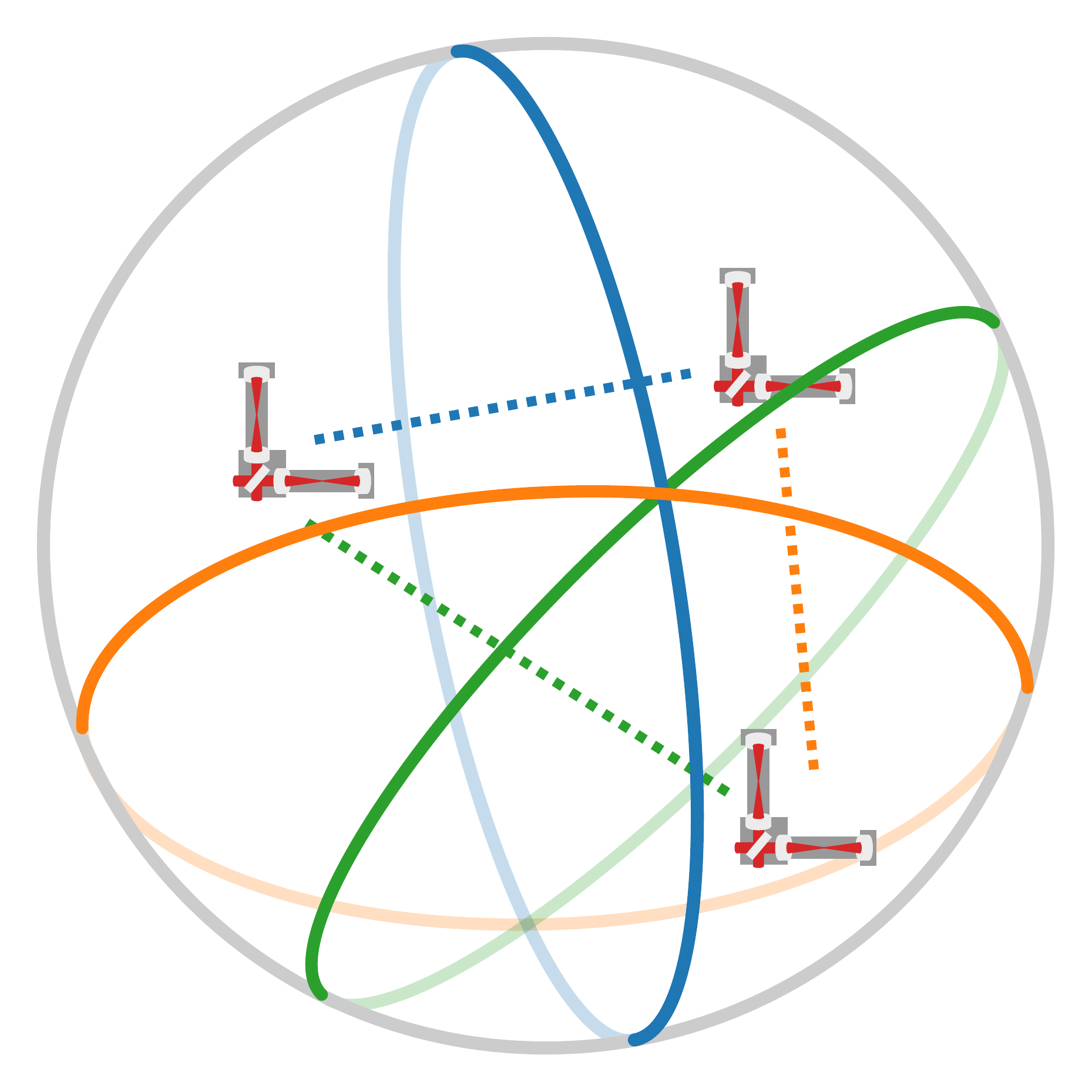}
    \end{center}
    \caption[Localising signals using gravitational-wave detectors]{
        Localising signals using gravitational-wave detectors. With just two detectors sources can only be localised to a ring on the sky (shown on the left, for three different sources). The addition of a third detector means sources can be triangulated to where the rings intercept (shown on the right).
        }\label{fig:triangulate}
\end{figure}

\newpage

\begin{figure}[p]
    \begin{center}
        \includegraphics[width=\linewidth]{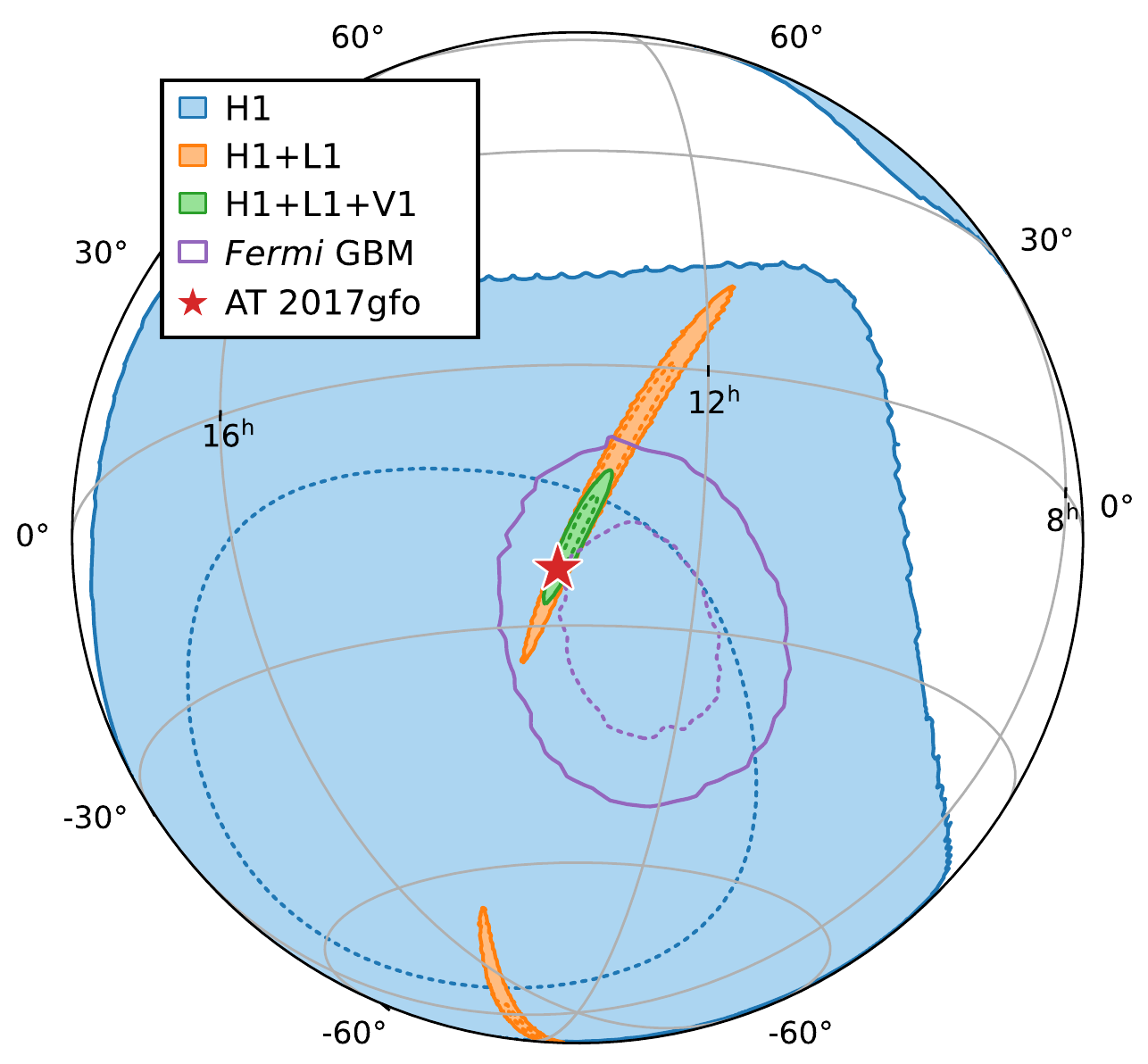}
    \end{center}
    \caption[Skymaps for GW170817]{
        Skymaps for GW170817, produced from the single Hanford detector (in \textcolorbf{NavyBlue}{blue}), both Hanford and Livingston (\textcolorbf{Orange}{orange}) and all three detectors (\textcolorbf{Green}{green}). The final \textit{Fermi} GBM skymap is also shown in \textcolorbf{Purple}{purple}, and the location of the counterpart source is marked by a \textcolorbf{Red}{red} star. Solid lines show 50\% confidence regions, dashed lines 90\% regions.
        }\label{fig:170817_skymaps}
\end{figure}

\newpage

\end{colsection}

\section{Multi-Messenger Astronomy}
\label{sec:multi}

\begin{colsection}

\emph{Multi-messenger astronomy} refers to detecting multiple signals from the same astrophysical source using two or more different `messengers'. Such messengers can include electromagnetic waves/photons, gravitational waves, neutrinos or cosmic rays. An example of a multi-messenger event is supernova SN 1987A, which was detected by neutrino detectors several hours before becoming visible in the electromagnetic spectrum \citep{SN1987A}. This thesis concentrates on the search for electromagnetic counterparts to gravitational-wave detections, of which at the time of writing only one has been found \citep[GW170817;][]{GW170817}.

Prior to the GW170817 detection, it had long been theorised that some gravitational-wave detections might have electromagnetic counterparts. Binary mergers involving neutrons stars (either neutron star binaries or neutron star-black hole mergers) were suggested as possible sources of short-duration gamma-ray bursts \citep{SGRBs}. Such events were expected to produce kilonovae, transient bursts of electromagnetic radiation that could also be visible in the optical \citep[these events were named ``kilo''-novae as they were predicted to reach luminosities approximately 1000 times that of a classical nova;][]{GW_kilonova}. Electromagnetic counterparts to binary black hole mergers were less expected; binary stellar-mass black holes may not be surrounded by much orbiting matter with which to interact, although certain systems with an orbiting disk might produce enough material for accretion and subsequent emission \citep{BBH_EM}.

\end{colsection}

\subsection{The benefits of multi-messenger observations}
\label{sec:mma_benefits}
\begin{colsection}

As explained in \aref{sec:gw_localisation}, gravitational-wave detectors have only a limited ability to localise the source of each detection. Wide-field electromagnetic monitors, such as the \textit{Fermi} \acro{gbm}, can provide independent localisation skymaps to help reduce the search area (\cite{GW170817_GRB}; note also the \textit{Fermi} skymap included in \aref{fig:170817_skymaps}). Ideally, the direct detection of a kilonova would allow precise localisation to a host galaxy, as well as a measure of redshift and therefore the distance to the source.

Electromagnetic observations of a counterpart to a gravitational-wave detection can give additional insights into the nature of the source and its environment, as well as further scientific breakthroughs. For example, observations of the kilonova associated with GW170817 \citep{GW170817, GW170817_followup} allowed analysis of the equation of state of neutron star material \citep{GW170817_NSscience}, an insight into the origin of heavy metals in the universe \citep{GW170818_heavy}, constraints on the nature of gravity \citep{GW170817_gravity}, and a new, independent measurement of the Hubble constant \citep{GW170817_hubble}.

At the time of writing, GW170817 remains the only gravitational-wave signal with an confirmed electromagnetic counterpart. As observations continue and new detectors come online it is only a matter of time until similar objects are found, which will assuredly lead to further astrophysical and cosmological breakthroughs. However, future counterparts may not be as easy to find.

\end{colsection}

\subsection{Finding optical counterparts to GW detections}
\label{sec:followup}
\begin{colsection}

GW170817 was a remarkably lucky event in several ways. The gravitational-wave signal was observed by both LIGO detectors, and although it was not observed by Virgo it was active at the time so the non-detection still helped narrow down the localisation area. This produced a fairly small skymap, covering just $31~\text{deg}^2$ \citep[see \aref{fig:170817_skymaps}]{GW170817}, which was well positioned for telescopes in the southern hemisphere to observe (although close to the sun, the area was visible for a few hours after sunset). The gravitational-wave detection also produced a very low luminosity distance of $40\pm\SI{8}{\mega\parsec}$, close enough to make an electromagnetic observation of the counterpart feasible.

\newpage

\begin{figure}[t]
    \begin{center}
        \includegraphics[width=0.8\linewidth]{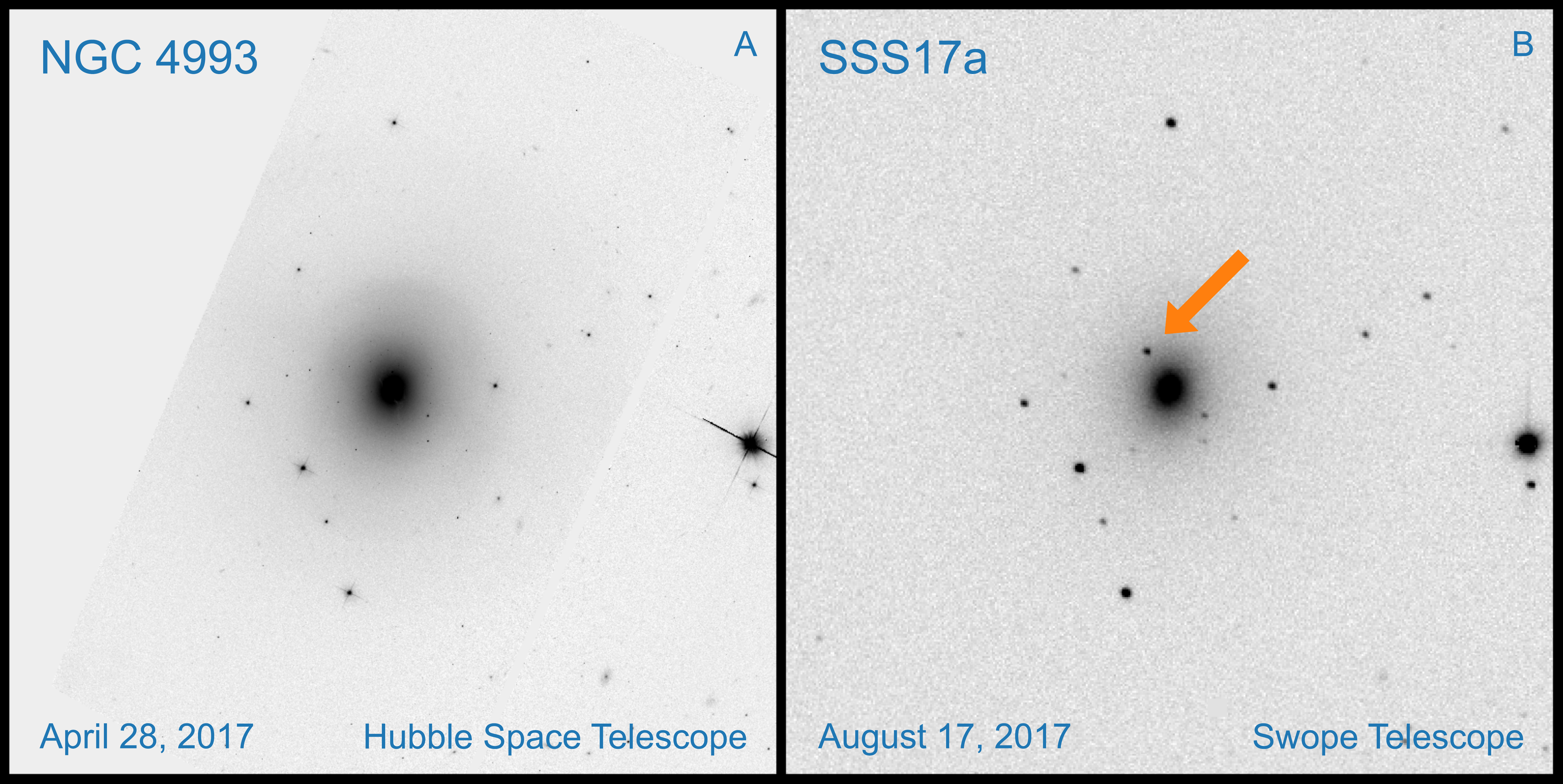}
    \end{center}
    \caption[Detection of the counterpart to GW170817]{
        Detection of the counterpart to GW170817 with the Swope telescope. On the left is an archival image of NGC 4993 from HST, and on the right the position of the kilonova is marked in the Swope discovery image. Adapted from \citet{GW170817_Swope}.
        }\label{fig:sss17a}
\end{figure}

The first of many observations of the counterpart transient, designated AT~2017gfo, was taken by the One-Meter, Two-Hemisphere collaboration using the 1-metre Swope telescope at Las Campanas Observatory in Chile. The discovery image is shown in \aref{fig:sss17a}, and the Swope team designated the transient SSS17a \citep{GW170817_Swope}. The Swope telescope only has a field of view of $\SI{30}{\arcmin}\times\SI{30}{\arcmin}$, but as the event was localised to a small, near-by region a list of potential host galaxies could be selected from the Gravitational Wave Galaxy Catalogue \acroadd{gwgc} \citep[GWGC;][]{GWGC}. The Swope observations were targeted to fields including these galaxies, as shown in \aref{fig:swope_decam}. The host galaxy, NGC 4993, was observed in the ninth pointing approximately 10 hours after the gravitational-wave detection, and as shown in \aref{fig:sss17a} the kilonova was clearly visible at the outer edge of the galaxy.

\begin{figure}[p]
    \begin{center}
        \includegraphics[width=\linewidth]{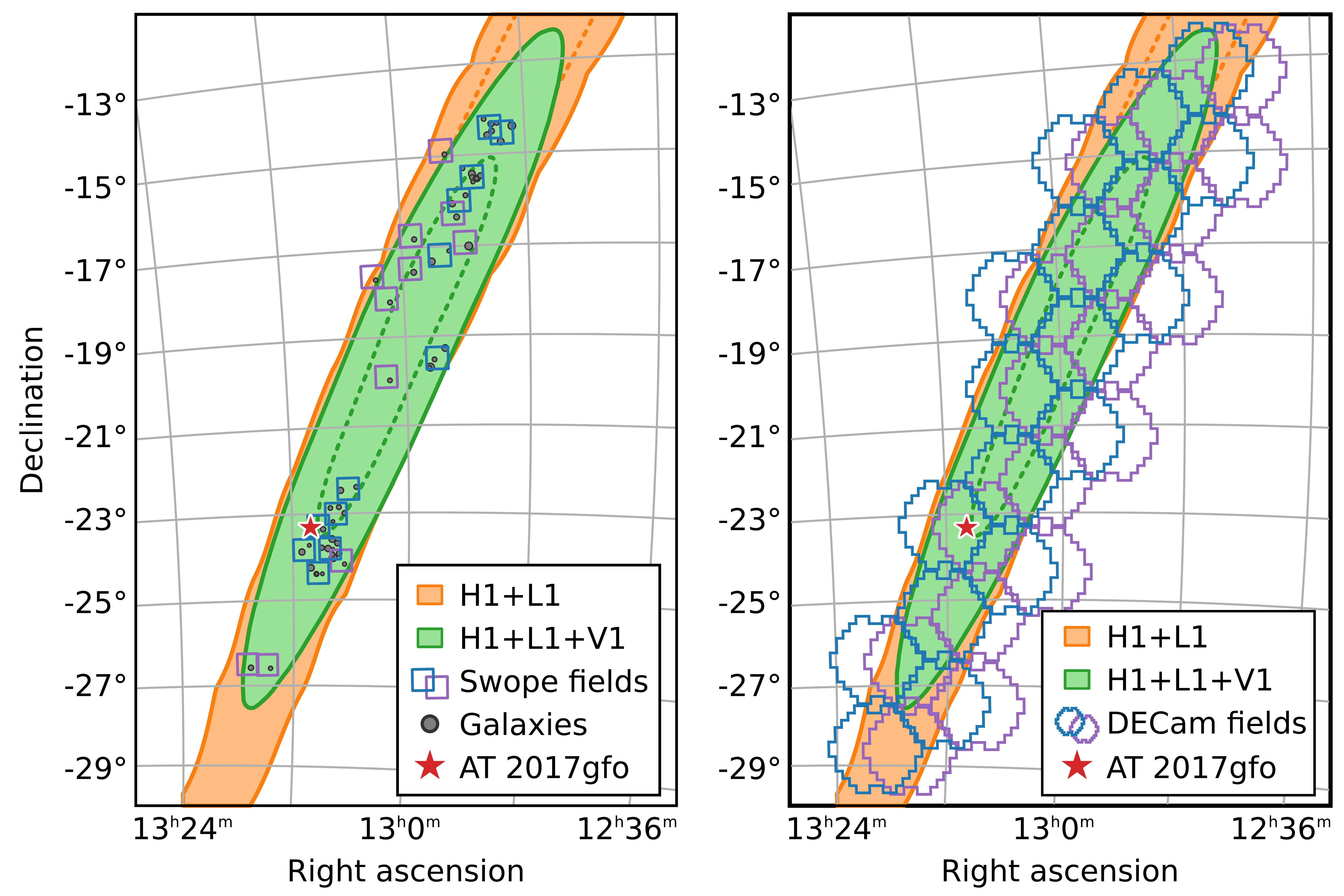}
    \end{center}
    \caption[Follow-up observations of GW170817 with Swope and DECam]{
        Follow-up observations of GW170817 with the Swope telescope \citep[left, adapted from][]{GW170817_Swope} and the Dark Energy Camera \citep[right, adapted from][]{GW170817_DECam}. The skymap contours are the same as in \aref{fig:170817_skymaps}; the Hanford-Livingston skymap is in \textcolorbf{Orange}{orange}, the final Hanford-Livingston-Virgo skymap is in \textcolorbf{Green}{green}, and the location of the counterpart is marked by the \textcolorbf{Red}{red} star. The Swope telescope has a small, $0.25~\text{deg}^2$ field of view, and so targeted its follow-up observations on concentrations of GWGC galaxies (\textcolorbf{darkgray}{grey} circles). The squares on the left show the fields observed by Swope, containing multiple galaxies (in \textcolorbf{NavyBlue}{blue}) or single galaxies (\textcolorbf{Purple}{purple}). Instead of targeting galaxies, DECam observed using a pre-defined grid of 18 pointings (\textcolorbf{NavyBlue}{blue} hexagons on the right), followed by a second set of offset observations (\textcolorbf{Purple}{purple}).
        }\label{fig:swope_decam}
\end{figure}

Five other groups independently observed the same transient within an hour of the Swope observation: the Dark Energy Camera \acroadd{decam}\citep[DECam,][]{GW170817_DECam}, the Distance Less Than \SI{40}{\mega\parsec} survey \acroadd{dlt40}\citep[DLT40,][]{GW170817_DLT40}, Las Cumbres Observatory \acroadd{lco}\citep[LCO,][]{GW170817_LCO}, the MASTER Global Robotic Net \citep{GW170817_MASTER} and the \acro{eso} Visible and Infrared Survey Telescope for Astronomy \acroadd{vista}\citep[VISTA,][]{GW170817_VISTA}. The observing strategy differed between groups depending on the field of view of the instruments. DLT40 was an existing supernova survey so targeted already-known galaxies, and the LCO and VISTA surveys both targeted their observations at possible host galaxies, just like Swope. On the other hand, DECam and MASTER had larger fields of view, and so could therefore cover the entire localisation region using a regular tiling pattern. The DECam tile pointings are shown on the right-hand plot of \aref{fig:swope_decam}.

The relatively small GW170817 skymap allowed telescopes with small fields of view, such as Swope, to efficiently cover the search area and locate the counterpart source. However, there is no guarantee that this will always be the case, and indeed subsequent events have not been as well localised. As previously mentioned in \aref{sec:gw_sources}, the second binary neutron star gravitational-wave detection, S190425z, occurred in April 2019, a few weeks into the O3 run \citep{S190425z}. Unlike GW170817, this was only a single-detector detection, observed only by LIGO-Livingston (again Virgo was also observing at the same time, but LIGO-Hanford was shut down for maintenance). The initial skymap for this event covered an area of approximately 10,000 square degrees, and the final skymap only reduced this to 7,500~sq~deg; still 250 times the size of the final GW170817 skymap (see \aref{fig:ztf}). For searching such large sky areas, containing thousands of galaxies, the method used by smaller telescopes such as Swope is impractical, and so dedicated wide-field survey telescopes are needed.

The \acro{ztf} is one such telescope, with a field of view of 47~sq~deg \citep{ZTF}. Over the first two nights following the S190425z detection ZTF covered approximately 8,000~sq~deg of the initial skymap \citep{S190425z_ZTF}, shown in \aref{fig:ztf}. This still only corresponded to 46\% of the localisation probability, reduced to 21\% in the final skymap, as unfortunately a large fraction of the skymap was located too close to the Sun to observe.

\newpage

\begin{figure}[t]
    \begin{center}
        \includegraphics[width=0.9\linewidth]{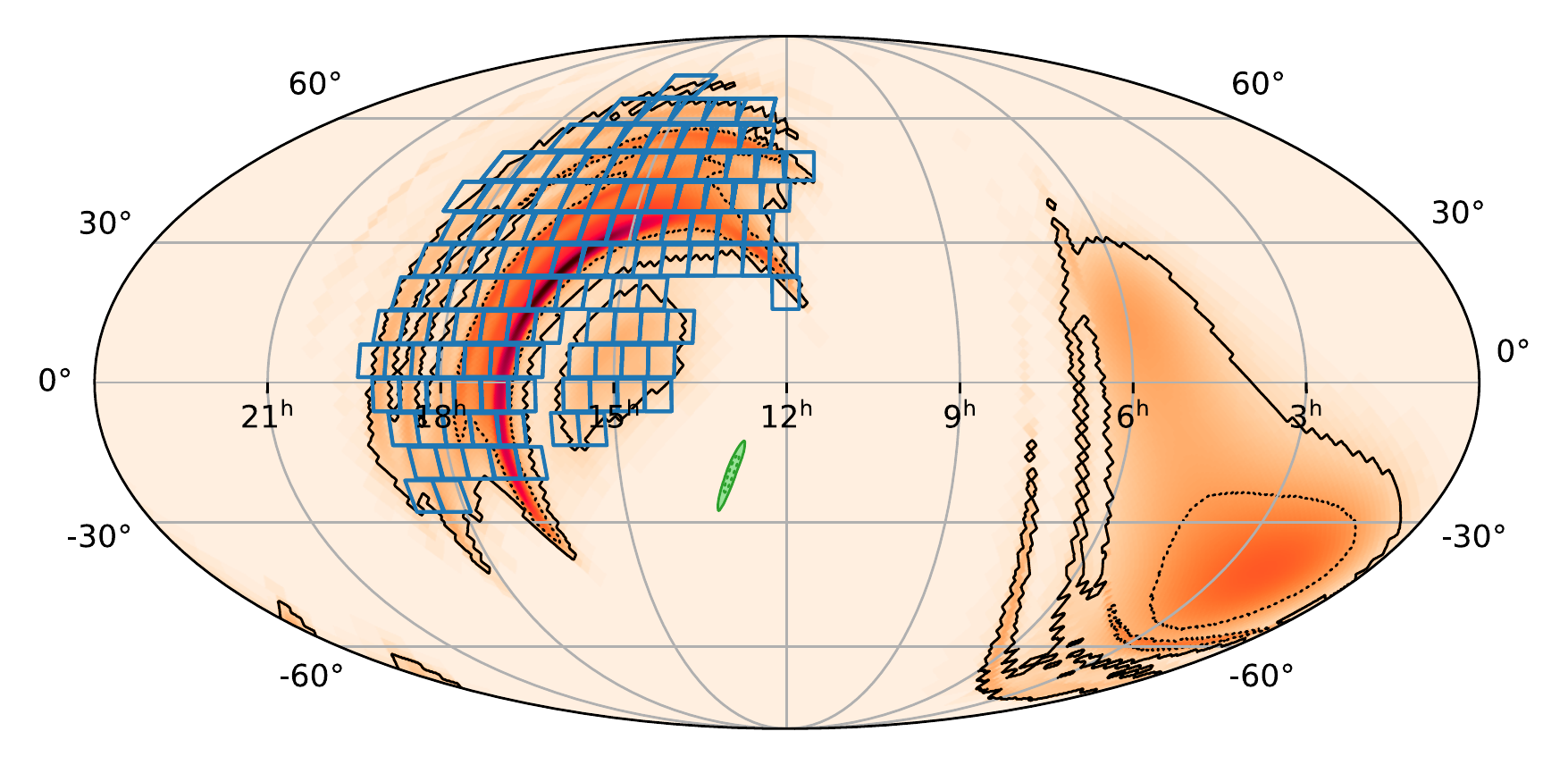}
    \end{center}
    \caption[Follow-up observations of S190425z with ZTF]{
        Follow-up observations of S190425z with the Zwicky Transient Facility. The ZTF tiled observations are shown in \textcolorbf{NavyBlue}{blue} over the initial LVC skymap in \textcolorbf{Orange}{orange}. Adapted from \citet{S190425z_ZTF}. For a size comparison to \aref{fig:swope_decam} the position of the final GW170817 skymap is also shown in \textcolorbf{Green}{green}.
        }\label{fig:ztf}
\end{figure}

If covering the large gravitational-wave skymap was not enough of a challenge, once electromagnetic observations have been taken a robust analysis pipeline is also needed in order to distinguish potential counterpart candidates from the large number of coincident transient and variable-star detections. ZTF found 340,000 transients when following-up S190425z, which were narrowed down to just 12 potential candidates \citep{S190425z_ZTF}. Ultimately, each was shown to be unconnected with the gravitational-wave signal, and in the end no counterpart was identified for this event by ZTF or any other project.

As a single-detector event S190425z was something of an extreme example, and as more gravitational-wave detectors come online the typical skymap size should decrease. The time and effort required to follow up S190425z stands in contrast to the relative ease with which the GW170817 counterpart was found. Small telescopes like Swope can contribute with galaxy-focused observations when the skymap is small enough, but for events like S190425z, where large searches are required, it is clear that dedicated, wide-field survey telescopes are required to have the best chance of finding any counterpart.

\newpage

\end{colsection}

\hfuzz=6pt %
\section[The Gravitational-wave Optical Transient Observer]{%
    \protect\scalebox{0.93}[1.0]{\mbox{The Gravitational-wave Optical Transient Observer}}
}
\hfuzz=.5pt %
\label{sec:goto}

\begin{colsection}

The \acro{goto}\footnote{\url{https://goto-observatory.org}} is a project dedicated to detecting optical counterparts of gravitational-wave sources. The GOTO collaboration was founded in 2014 and, as of 2019, contains 10 institutions from the UK, Australia, Thailand, Spain and Finland\footnote{The GOTO collaboration includes the University of Warwick, Monash University, Armagh Observatory and Planetarium, the University of Leicester, the University of Sheffield, the National Astronomical Research Institute of Thailand, the Instituto de Astrofísica de Canarias, the University of Manchester, the University of Turku and the University of Portsmouth.}. The first prototype GOTO telescope was inaugurated at the \acro{orm_lapalma} on La Palma, Canary Islands in July 2017, and is shown in \aref{fig:goto_photo}.

\begin{figure}[p]
    \begin{center}
        \includegraphics[width=0.9\linewidth]{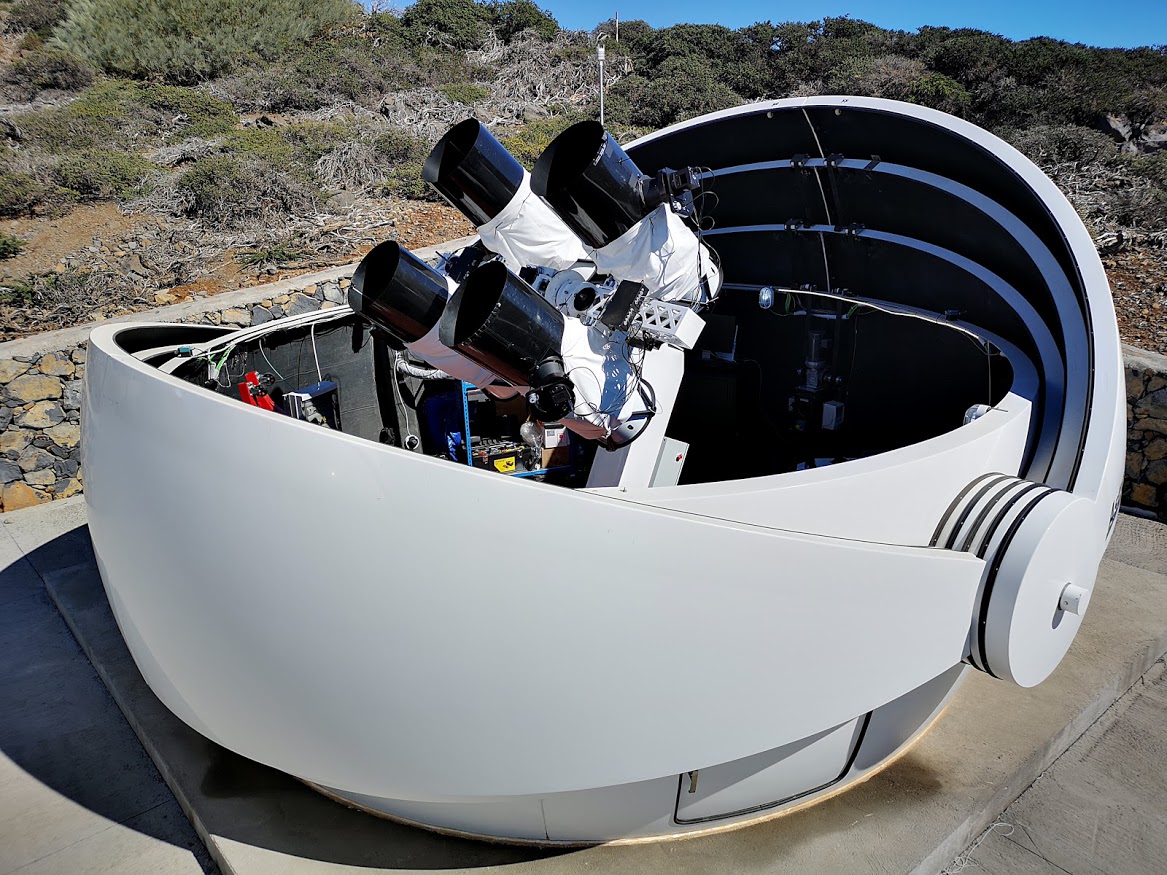}
    \end{center}
    \caption[The GOTO prototype on La Palma]{
        The GOTO prototype on La Palma, with four unit telescopes.
    }\label{fig:goto_photo}
\end{figure}

\begin{figure}[p]
    \begin{center}
        \includegraphics[width=0.9\linewidth]{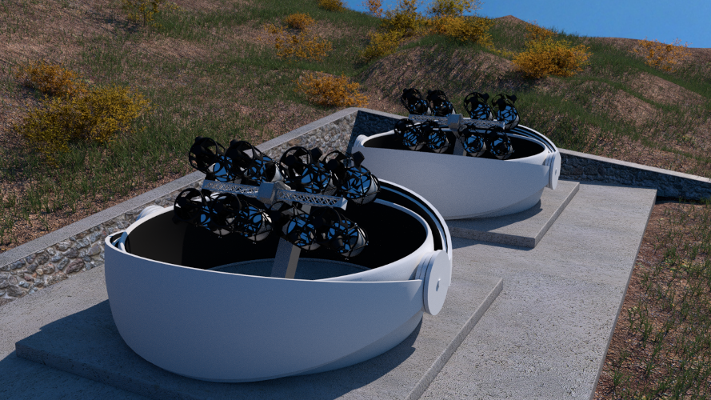}
    \end{center}
    \caption[A rendering of a complete GOTO node]{
        A rendering of a complete GOTO node with two independent mounts, each with eight unit telescopes.
    }\label{fig:goto_render}
\end{figure}

\end{colsection}

\subsection{Motivation}
\label{sec:goto_motivation}
\begin{colsection}

Even before the first detection of gravitational waves in 2015 it was recognised that, due to the issues described in the previous section, the best chance of reliably detecting electromagnetic counterparts quickly was with a network of dedicated, robotic, wide-field telescopes \citep{Darren}. On most nights the telescopes would carry out an all-sky survey on a fixed grid, but when a gravitational-wave alert was received they could quickly change to covering the skymap. As robotic telescopes, they would be quicker to respond than human-operated telescopes, meaning follow-up observations could begin automatically just minutes after an alert was issued.

Rapidly covering large areas of the sky maximises the chance of finding any possible counterpart before it fades from view. However, any such observations are going to be contaminated by a huge number of unrelated transient and variable objects (recall the hundreds of thousands of detections by ZTF when searching for the S190425z counterpart as described in \aref{sec:followup}). Therefore, the difficulty in finding any counterpart is not just in covering the large search area, but also in being able to distinguish the needle from the haystack of other astronomical transients and variables. The best way to reduce the number of candidate detections is temporally: any sources with detections prior to the gravitational-wave event time could not be the transient associated with it. In order to perform this temporal filtering, the project needs an as-recent-as-possible image of the same patch of sky, which necessitates the telescope carrying out an all-sky survey with as low a cadence (the time between observing each point of the sky) as possible.

\citet{Darren} outlined the original GOTO proposal, including the requirements for each telescope. An instantaneous field of view of 50--100 square degrees and a limiting magnitude of $R\simeq21$ in 5 minutes was suggested to be able to have the best chance of observing electromagnetic counterparts, with the field of view ideally split between multiple independent mounts to allow for covering the irregularly-shaped gravitational-wave skymaps. These telescopes would need to respond quickly to alerts, covering the skymaps within a single night before any possible counterpart faded from view, necessitating the use of fast-slewing, robotic mounts. Based on the sensitivity regions of the LIGO-Virgo network, the best sites for these telescopes would be in the North Atlantic (e.g.\@ the Canary Islands) and Australia, at least until the detectors in Japan and India come online (as described in \aref{sec:gw_detecting}).

GOTO is, of course, not the only project searching for gravitational-wave counterparts, and the optimal follow-up strategy has been an area of much analysis in recent years \citep[see, for example,][]{BlackGEM_strategy, ZTF_strategy, GW_strategy}. Other contemporary projects include the Zwicky Transient Facility at the Palomar Observatory in California, previously mentioned in \aref{sec:followup}, which captures 47~square~degrees to a depth of $r=20.5$ \citep{ZTF}, and the BlackGEM project, currently under construction at at La Silla Observatory in Chile, which aims to go deeper ($q=23$ in five minutes), albeit initially with a smaller footprint of $\sim$8~square~degrees. \aref{tab:rivals} shows a comparison of the planned GOTO network to other projects.

\newpage

\begin{table}[t]
    \begin{center}
        \begin{tabular}{r|ccccccl} %
                 & First & $D$ &    FoV    & Limiting  & Cost  &     Etendue    & \\
            Name & light & (m) & (deg$^2$) & magnitude & (M\$) & (m$^2$deg$^2$) & \\
            \midrule
            ATLAS &
            2015 &
            0.5$\times$2 &
            60 &
            $g=19.3$ (\SI{30}{\second}) &
            2 &
            12 &
            \tablefootnote{~~\citet{ATLAS}}
            \\
            Pan-STARRS1 &
            2008 &
            1.8 &
            7 &
            $g=22.0$ (\SI{43}{\second}) &
            25 &
            18 &
            \tablefootnote{~~\citet{Pan-STARRS}}
            \\
            ZTF &
            2017 &
            1.2 &
            47 &
            $g=20.8$ (\SI{30}{\second}) &
            24 &
            54 &
            \tablefootnote{~~\citet{ZTF}}
            \\
            \textit{BlackGEM} &
            \textit{2019} &
            0.65$\times$3 &
            8.1 &
            $q=23$ (\SI{300}{\second}) &
            3 &
            2.7 &
            \tablefootnote{~~\citet{BlackGEM}}
            \\
            \textit{LSST} &
            \textit{2020} &
            8.4 &
            9.6 &
            $g=25.6$ (\SI{15}{\second}) &
            500 &
            532 &
            \tablefootnote{~~\citet{LSST}}
            \\
            \\
            GOTO-4 &
            2017 &
            (0.4$\times$4) &
            18 &
            $g=19.5$ (\SI{60}{\second}) &
            1.0 &
            2.3
            \\
            \textit{GOTO-8} &
            \textit{2019} &
            (0.4$\times$8) &
            40 &
            $g=19.5$ (\SI{60}{\second}) &
            1.5 &
            5
            \\
            \textit{2$\times$GOTO-8} &
            \textit{2020} &
            (0.4$\times$8$)\times$2 &
            80 &
            $g=19.5$ (\SI{60}{\second}) &
            2.5 &
            10
            \\
            \textit{4$\times$GOTO-8} &
            \textit{2021} &
            (0.4$\times$8$)\times$4 &
            160 &
            $g=19.5$ (\SI{60}{\second}) &
            4.0 &
            20
            \\
        \end{tabular}
    \end{center}
    \caption[Comparison of projects involved in gravitational-wave follow-up]{
        A comparison of selected projects involved in following-up gravitational-wave detections. Given for each is the year it saw first light, the primary mirror diameter(s) (note GOTO has 4 or 8 small telescopes per mount, and ATLAS, BlackGEM and later GOTO stages include multiple mounts), the total instantaneous field of view, the limiting magnitude for a single exposure, estimated total cost, and etendue (the product of the primary mirror area(s) and the field of view). The projects in \textit{italics} are yet to be commissioned and so use predicted values. %
    }\label{tab:rivals}
\end{table}

\end{colsection}

\subsection{Hardware design}
\label{sec:goto_design}
\begin{colsection}

The GOTO prototype, as shown in \aref{fig:goto_photo}, uses an array of four \SI{40}{\cm} Unit Telescopes (UTs)\acroadd{ut}, attached to a boom-arm on a single robotic mount with a slew speed of \SI{4}{\degree} per second. Using multiple smaller instruments on one mount is a design already used by several survey and wide-field telescopes, such as the All-Sky Automated Survey for Supernovae \acroadd{asassn} \citep[ASAS-SN,][]{ASAS-SN} and SuperWASP \acroadd{wasp} \citep[Wide Angle Search for Planets,][]{SuperWASP}. The array design provides a cost-effective way of reaching the desired wide field of view with multiple small telescopes instead of one large one, and the modular nature also allows more unit telescopes to be added to a mount as more funding becomes available.

\newpage

The prototype unit telescopes and mount were constructed by APM Telescopes\footnote{\url{https://www.professional-telescopes.com}}. Each UT is a fast Wynne-Newtonian astrograph (see \aref{sec:optics}) with a focal ratio of f/2.5, and each uses off-the-shelf camera hardware from \acro{fli}\footnote{\url{https://www.flicamera.com}}. The FLI MicroLine cameras each use a 50 megapixel sensor (see \aref{sec:chip_layout}), which gives each UT a field of view of approximately 6 square degrees with a plate scale of \SI[per-mode=symbol]{1.24}{\arcsec\per\pixel}. Three \SI{60}{\second} GOTO images can reach a limiting magnitude of $g \approx 20$ \citep[see \aref{sec:onsky_comparison};][]{S190425z_GOTO}. A full GOTO telescope will have eight UTs on an GE-300 German equatorial (parallactic) mount, giving an overall field of view of \SI{40}{\square\deg} (accounting for some overlap between cameras). A sample frame taken with one UT is shown in \aref{fig:fov}, which also gives a comparison of the GOTO field of view to some of the other projects from \aref{tab:rivals}.

\begin{figure}[p]
    \begin{center}
        \includegraphics[width=\linewidth]{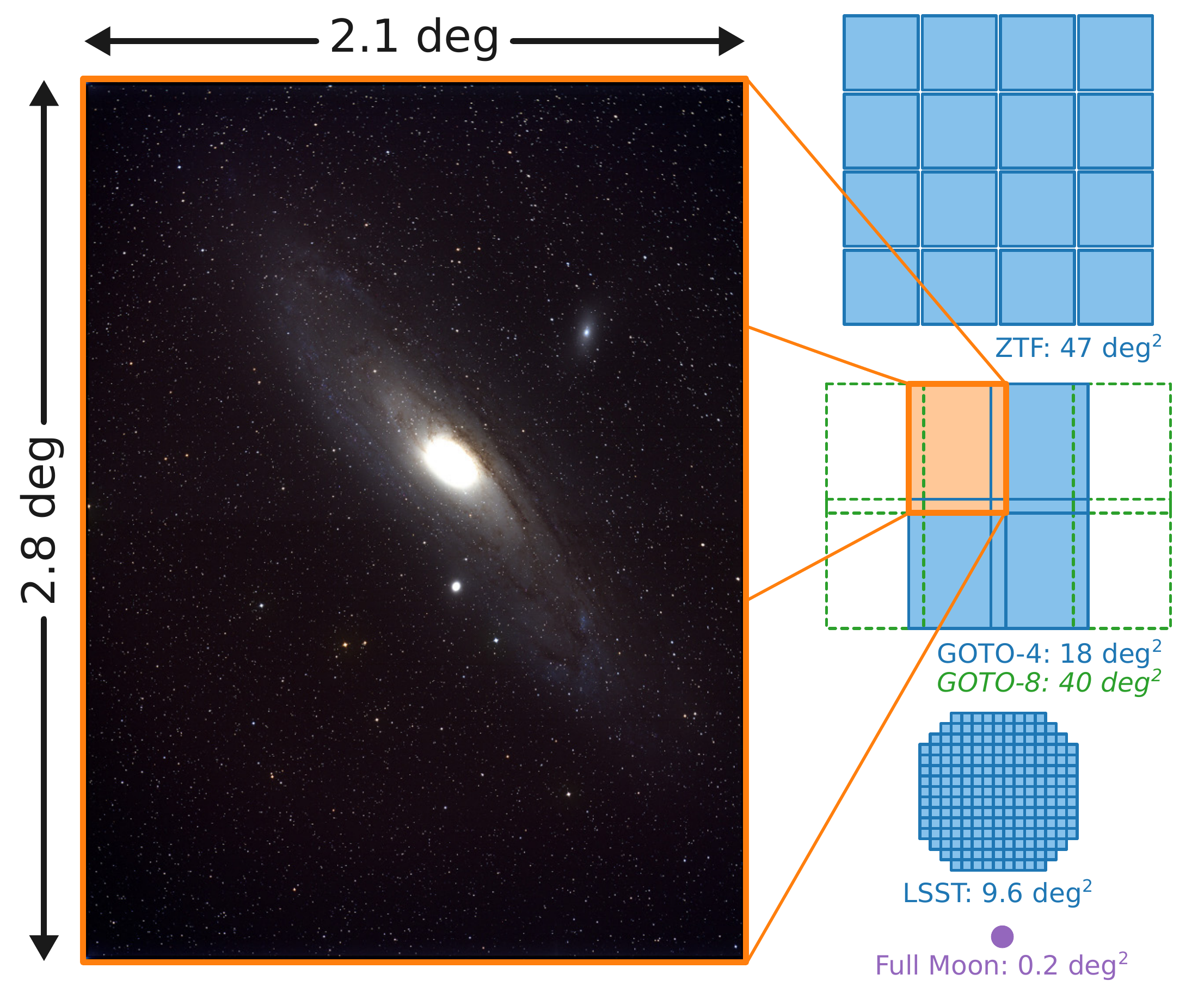}
    \end{center}
    \caption[GOTO's field of view compared to other projects]{
        GOTO's field of view compared to other projects. On the left is a commissioning image of M31 taken with one of GOTO's cameras, showing the wide field of view of a single unit telescope. Four of these UTs form the initial 18 square degree survey tile, to be increased to 40 square degrees in the full 8-UT system. On the right, the GOTO FoV is compared to two similar projects: the Zwicky Transient Facility \acroadd{ztf} \citep[ZTF,][]{ZTF} and the Large Synoptic Survey Telescope \acroadd{lsst} \citep[LSST,][]{LSST}.
    }\label{fig:fov}
\end{figure}

Each unit telescope is fitted with a filter wheel with several wide-band coloured filters (see \aref{sec:filters}), which can be used for additional source identification. The telescopes are housed in an Astrohaven clamshell dome\footnote{\url{https://www.astrohaven.com}}, which when fully open allows an unrestricted view of the sky. This means that the telescope does not need to waste time waiting for the dome to move when slewing to a new position, and can instead quickly move from observing one portion of the sky to another.

Each GOTO site is anticipated to host two domes, as shown in \aref{fig:goto_render}, with a total of 16 unit telescopes giving an instantaneous field of view of approximately \SI{80}{\square\deg}. Having two independent mounts will allow the sky to be surveyed at a higher cadence (every 2--3 days, see \aref{sec:survey_sims}), and also gives more options for survey and transient follow-up strategies. For example, the two mounts could observe different patches of the sky in order to cover the skymap as fast as possible, or they could combine to observe the same field to a greater depth. Alternately, each mount could observe using a different filter, to get immediate multi-colour information on any detected sources.

\newpage

\end{colsection}

\subsection{Image processing and candidate detection}
\label{sec:gotophoto}
\begin{colsection}

The GOTO project produces a huge amount of data to be handled and processed. Each image taken by the  50~megapixel cameras is approximately 100~MB;\@ GOTO typically takes three \SI{60}{\second} exposures per pointing and on average observes $\sim$130 targets each night. Just the prototype 4-UT system hence produces approximately 150~GB of data each night, and a full multi-site GOTO system would produce close to 400~TB per year. For real-time transient detection, each set of images needs to be processed in the approximately three minutes between each observation, and due to the wide field of view each image will contain many thousands of sources. Processing the images is therefore not an easy task. In order to do this, a real-time data flow system has been developed called GOTOflow, which is used to run the GOTO pipeline GOTOphoto. The key components of the pipeline are shown in \aref{fig:gotoflow}.

\begin{figure}[t]
    \begin{center}
        \includegraphics[width=\linewidth]{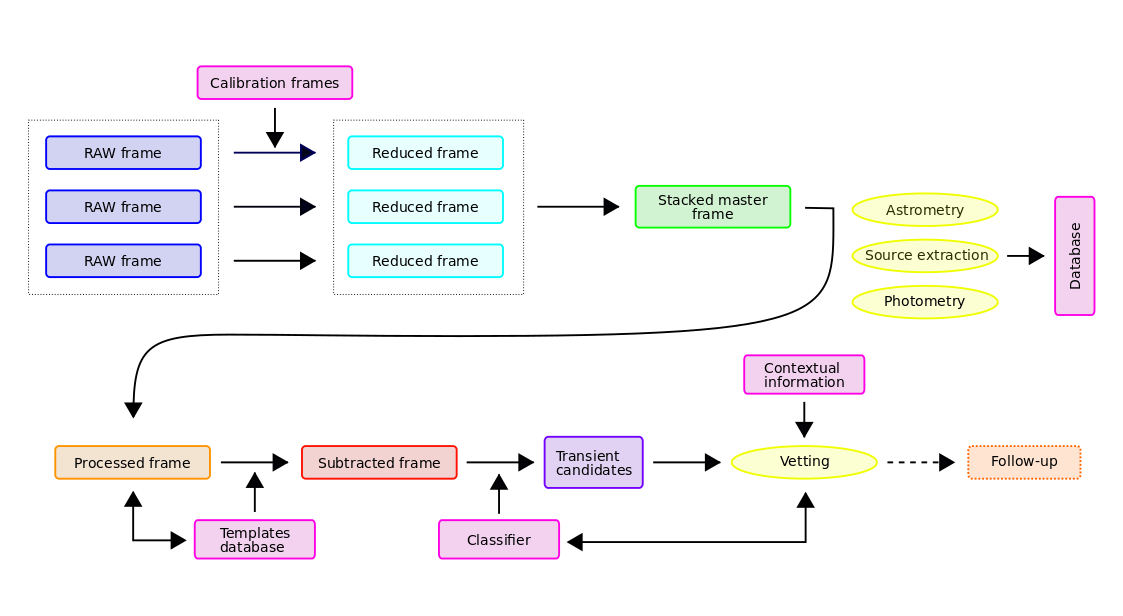}
    \end{center}
    \caption[The GOTO dataflow]{
        A flowchart showing the key components of the GOTO dataflow.
    }\label{fig:gotoflow}
\end{figure}

\newpage

\begin{figure}[t]
    \begin{center}
        \includegraphics[width=\linewidth]{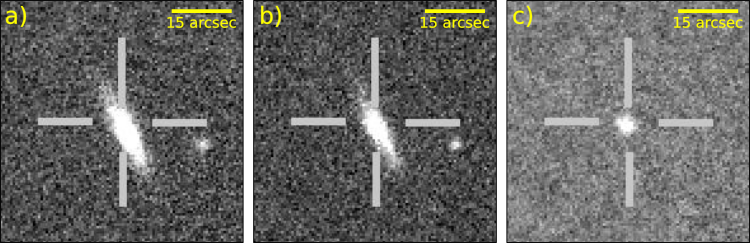}
    \end{center}
    \caption[The detection of SN 2019bpc through difference imaging]{
        The detection of supernova SN 2019bpc by the GOTOphoto difference imaging pipeline. The new exposure on the left (a) has the reference image (b) subtracted to give the difference image on the right (c), where the new source is clearly visible.
    }\label{fig:diffimg}
\end{figure}

GOTOphoto calibrates each image and then combines each set of three to increase the effective depth. New or changed sources are detected using difference imaging, as shown in \aref{fig:diffimg}. This requires a master reference image at each position in the sky, which is continuously built up from the all-sky survey. Any apparently new sources are then added to a detection database, and are also checked against historic observations to discount any sources which had been observed previously (for example, variable stars on the edge of the detection depth might fade in and out of visibility). Candidate sources are also checked against other catalogues such as Pan-STARRS \citep{Pan-STARRS}, as well as against lists of known minor planets and other transients.

New candidates are presented for human vetting through a web interface called the GOTO Marshal. Collaboration members can check each candidate and flag them either as potential astrophysical sources or junk detections. Work is ongoing across the collaboration on machine-learning projects for automatic transient detection and identification, which would use the human responses from the Marshal to train an automatic classifier.

\newpage

\end{colsection}

\subsection{Deployment and future expansion}
\label{sec:goto_expansion}
\begin{colsection}

The 4-UT GOTO prototype shown in \aref{fig:goto_photo} was inaugurated in July 2017. The telescope is located at the \acro{orm_lapalma} on La Palma in the Canary Islands, at the site shown in \aref{fig:orm}. After some hardware issues (see \aref{sec:hardware_commissioning}), the telescope is now fully operational: since February 2019 GOTO has been carrying out an all-sky survey, and since the start of the third LIGO-Virgo observing run (O3, see \aref{sec:gw_detections}) in April 2019 it has been following-up gravitational-wave events when they occur.

The next stage in the GOTO project will be the addition of the second set of four unit telescopes to the existing mount, due in late 2019. Funding for a second mount with another set of four unit telescopes has already been secured, and a second dome has already been constructed on La Palma. This second mount is expected to be commissioned in 2020.

La Palma is one of the best observing sites in the northern hemisphere, and is already home to several telescopes operated by GOTO collaboration members. It was therefore an obvious choice for the location of the first GOTO node. Ultimately, a second, complimentary node is planned to be built in the southern hemisphere, most likely in Australia. Having a site in both hemispheres allows the entire sky to be surveyed, and, as Australia is on the opposite side of the Earth from La Palma, the two sites would provide almost 24-hour coverage of gravitational-wave alerts. This second node would also host two GOTO telescopes with 8 UTs each, and the two sites combined would be able to survey the entire visible sky every 1--2 days (see \aref{sec:survey_sims}). As GOTO grows, it is anticipated that the telescopes at both sites will be operated as a single observatory, meaning observation scheduling will be optimised for each site and the output data will be unified into a single detection database.

\begin{figure}[p]
    \begin{center}
        \includegraphics[width=\linewidth]{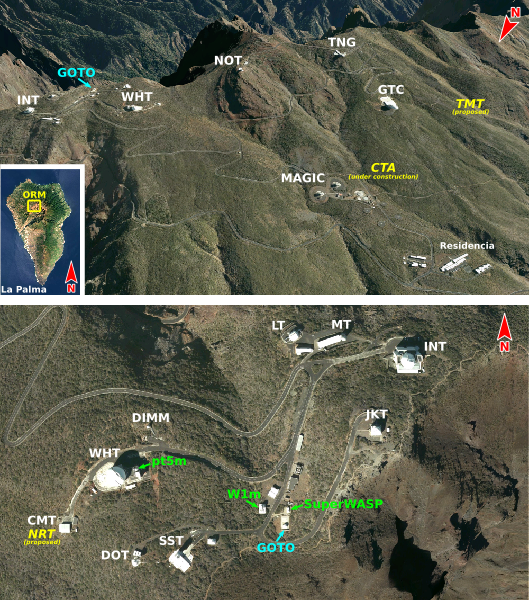}
    \end{center}
    \caption[The location of GOTO and other telescopes on La Palma]{
        The location of GOTO and other telescopes at the Observatorio del Roque de los Muchachos on La Palma, including the sites of proposed and under-construction projects in \textcolorbf{YellowOrange}{yellow}. GOTO, marked in \textcolorbf{BlueGreen}{blue}, is located on the east side of the observatory, close to the edge of the caldera. The lower image shows the eastern part of the ORM and the telescopes surrounding GOTO, including the other Warwick-operated telescopes (W1m and SuperWASP) and the Sheffield/Durham-operated pt5m (on the roof of the WHT building) in \textcolorbf{ForestGreen}{green}. Satellite images taken from Google Maps.
    }\label{fig:orm}
\end{figure}

\newpage

\end{colsection}

\section{Thesis outline}
\label{sec:outline}

\begin{colsection}

This thesis details my work on the GOTO project carried out between 2015 and 2019.

My primary role within the collaboration was to develop the software needed for GOTO to operate as an autonomous telescope. The hardware design of GOTO --- with multiple unit telescopes attached to each mount --- required custom software which could operate all of the cameras synchronously. Each unit telescope is also equipped with a filter wheel and a focuser, which also need to be controlled in parallel. Additional control software was required to move the mount, open and close the dome, and control any other pieces of on-site hardware.

The next stage of the control system development was writing the software that would operate the telescope without human involvement. The routine nightly operations (move to a target, take an exposure, repeat until sunrise) would need to be supported, and in addition standard observational tasks, such as taking calibration frames and focusing the telescopes, needed to be automated. Crucially, the control system required robust and reliable systems for monitoring the weather and the hardware status, and in the case of an emergency it had to be able to close the dome or recover from any problems without immediate human intervention.

In order for a telescope to observe autonomously it needs to be able to know what targets to observe and when. Therefore GOTO needed an automatic scheduling system, which had to be able to handle two distinct operational modes:\@ carrying out the all-sky survey for the majority of the time, but quickly switching to follow up any gravitational-wave alerts (or other transient events) when triggered. For the all-sky survey a series of tiled pointings had to be to be defined, and alert observations are mapped onto the same grid to enable the pipeline to carry out difference imaging. Reacting quickly to any alerts is vital, so GOTO needed a robust alert monitoring system that could determine the optimal follow-up strategy for each event.

Finally, as described in \aref{sec:goto_expansion}, GOTO is envisioned as a modular project with an increasing number of telescopes and sites, eventually operating as a multi-site observatory. This future expansion had to be taken into account in the design of both the hardware control and scheduling systems.

This thesis is arranged as follows:
\begin{itemize}
    \item In \nref{chap:hardware} I describe my work characterising the GOTO hardware and optical systems prior to its deployment on La Palma.
    \item In \nref{chap:gtecs} I introduce the software I developed to control the GOTO hardware.
    \item In \nref{chap:autonomous} I describe the additional autonomous level of software I wrote to allow GOTO to operate as a robotic telescope.
    \item In \nref{chap:scheduling} I examine in detail the functions used to prioritise and schedule GOTO observations.
    \item In \nref{chap:tiling} I describe how GOTO observations are mapped onto an all-sky grid, and how it is used to observe gravitational-wave skymaps.
    \item In \nref{chap:alerts} I describe the software systems used to receive and process astronomical alerts, including gravitational-wave detections.
    \item In \nref{chap:commissioning} I detail my work during the deployment of the GOTO prototype on La Palma and subsequent control system development.
    \item In \nref{chap:multiscope} I examine the future expansion plans of GOTO, and detail simulations I carried out to quantify the benefits of multiple telescopes and sites.
    \item Finally, in \nref{chap:conclusion} I present concluding remarks and some suggestions for future project development.
\end{itemize}

\end{colsection}

\chapter{Hardware Characterisation}
\label{chap:hardware}

\chaptoc{}

\section{Introduction}
\label{sec:hardware_intro}

\begin{colsection}

In this chapter I detail my work characterising and modelling the GOTO hardware. This work was carried out predominantly in the first year and a half of my PhD, prior to GOTO's commissioning in 2017.
\begin{itemize}
    \item In \nref{sec:detectors} I describe and give the results of the in-lab detector characterisation tests I ran on the GOTO CCD cameras.
    \item In \nref{sec:throughput} I detail the throughput model of the GOTO optical system that I created.
    \item In \nref{sec:photometry} I apply the results of the previous two sections to predict the photometric properties of GOTO images, before comparing them to real observations taken once GOTO was fully operational.
\end{itemize}
All work described in this chapter is my own unless otherwise indicated, and has not been published elsewhere.

\newpage

\end{colsection}

\section{Detector properties}
\label{sec:detectors}

\begin{colsection}

CCD cameras have a variety of characteristic parameters advertised by the manufacturers, including the amount of detector noise. As described in \aref{sec:goto_design}, GOTO uses MicroLine ML50100 CCD cameras manufactured by \acro{fli}, which contain KAF-50100 CCD sensors manufactured by ON Semiconductor. Both manufactures produce specification sheets advertising expected parameters\footnote{ML50100 available at \url{http://www.flicamera.com/spec_sheets/ML50100.pdf}.}\textsuperscript{,}\footnote{KAF-50100 available at \href{http://www.onsemi.com/pub/Collateral/KAF-50100-D.PDF}{\texttt{http://www.onsemi.com/pub/Collateral/KAF-50100-D.pdf}}.}. Confirming these under laboratory conditions is important before the detectors are used to take scientific images on the telescope. FLI carried out a limited series of tests on the cameras before selling them, but carrying out our own tests ensures that our cameras meet the specifications, and also allows independent measurements of the key parameters.

\end{colsection}

\subsection{Sources of CCD noise}
\label{sec:noise}
\begin{colsection}

There are many sources of noise in images taken with CCD cameras. The most important noise sources for astronomical images are \citep{CCDs}:
\begin{itemize}
    \item \emph{Shot noise} derived from counting photo-electrons from the source and background.
    \item \emph{Dark current noise} from thermally generated electrons within the sensor.
    \item \emph{Read-out noise} from the detector and CCD controller electronics.
    \item \emph{Fixed-pattern noise} from different sensitivities between pixels.
    \item \emph{Bias}, an offset in counts added to each pixel which can vary with time and position on the detector.
\end{itemize}

The shot noise ($\sigma_\text{N}$) arises as photons from the target object arrive at the sensor at irregular intervals. The photon arrival time is a Poisson distribution, and, if the number of electrons counted is $N$, for large numbers it tends towards a Gaussian distribution with mean $N$ and standard deviation $\sigma_\text{N} = \sqrt{N}$. When taking on-sky astronomical observations there are two sources of shot noise, from the target object ($\sigma_\text{obj}$) and the background sky ($\sigma_\text{sky}$).

Dark current noise ($\sigma_\text{DC}$) is due to electrons produced by thermal excitations, which are indistinguishable from photo-electrons and increase with exposure time. This is also a photon counting measurement, so the noise $\sigma_\text{DC} = \sqrt{D}$ where $D$ is the dark current per pixel. The dark current depends on temperature: cooling the cameras reduces the thermal excitations and therefore reduces the dark current.

Read-out noise ($\sigma_\text{RO}$) depends on the quality of the CCD outputs and readout electronics, and on the speed data is read out from the CCD.\ The FLI MicroLine cameras read out at a fixed frequency of \SI{8}{\mega\hertz} per pixel, but other astronomical cameras have variable read-out speeds. Since read-out noise is a property of the output electronics, it is independent of signal or the exposure time used, and therefore it can be represented by a constant value for each frame, $\sigma_\text{RO} = R$, measured in electrons per pixel. The MicroLine cameras have two channels with independent readouts, so each will have an independent read-out noise (see \aref{sec:chip_layout}).

Fixed-pattern noise ($\sigma_\text{FP}$, also called flat-field noise) is due to the small differences in size and response between pixels. It increases linearly with the electron count, including source ($N$), background ($N_\text{sky}$) and dark ($D$) electrons (the fixed-pattern noise can be further broken down into the photo response non-uniformity and dark signal non-uniformity, but we will consider it as a single noise source). It can be parametrised as $\sigma_\text{FP} = k_\text{FP}(N+N_\text{sky}+D)$, where $k_\text{FP}$ is a dimensionless constant describing the fixed-pattern noise as a fraction of the full-well capacity. Scientific CCD cameras typically have very small non-uniformities between pixels, so $k_\text{FP}$ is usually $<1\%$, but this noise source can dominate when the signal count is high. However, as it is linearly related to the number of counts recorded fixed-pattern noise can easily be removed by flat fielding.

Finally, the bias level is an offset in counts applied to each pixel independent of the input signal. A large bias level is applied to each pixel by CCD manufacturers to prevent negative counts from being recorded due to fluctuations in the read-out noise. Across a frame the bias level will sometimes show structure, but it is simple to remove by subtracting a master bias frame. The bias level can change by a few counts during a night due to changes in the temperature, and it should be measured regularly, as any large changes might indicate a problem with the detector. The MicroLine cameras also include an overscan region (see \aref{sec:chip_layout}), which gives an independent measurement of the typical bias level for every image.

The noise sources described above (aside from the bias) are all independent Gaussian random variables, and therefore are added in quadrature to get the total noise per pixel

\begin{equation}
    \begin{split}
        \sigma_\text{Total}^2 & = \sigma_\text{obj}^2 +
                                  \sigma_\text{sky}^2 +
                                  \sigma_\text{DC}^2 +
                                  \sigma_\text{RO}^2 +
                                  \sigma_\text{FP}^2 \\
                              & = N + N_\text{sky} + D + R^2 + k_\text{FP}^2{(N+N_\text{sky}+D)}^2.
    \end{split}
    \label{eq:noise}
\end{equation}

\end{colsection}

\subsection{In-lab tests}
\label{sec:camera_tests}
\begin{colsection}

The initial deployment of GOTO was delayed for several months, due to delays on-site and manufacturing the unit telescopes (see \aref{sec:hardware_commissioning}). The first set of four cameras, however, had already been purchased from FLI, and the delay gave time to test them in the lab in Sheffield in 2016. The second set of cameras were also purchased before the second four unit telescopes; these were also brought to Sheffield in 2018 so the same tests could be repeated. A list of the nine FLI cameras bought for GOTO is given in \aref{tab:cameras}. Each camera is given a name (Camera 1, Camera 2 etc.) based on the order of their serial numbers. These names are used throughout this section but do not necessarily match which GOTO unit telescope they were assigned to, and the cameras on La Palma are sometimes swapped around to allow for repairs.

\begin{table}[t]
    \begin{center}
        \begin{tabular}{c|ccc} %
            Name     & Serial number & Set & Tested \\
            \midrule
            Camera 1 & ML0010316     &   1 & May---June 2016     \\
            Camera 2 & ML0330316     &   1 & March---May 2016    \\
            Camera 3 & ML0420516     &   1 & May---June 2016     \\
            Camera 4 & ML0430516     &   1 & May---June 2016     \\
            Camera 5 & ML5644917     &   2 & May---June 2018     \\
            Camera 6 & ML6054917     &   2 & May---June 2018     \\
            Camera 7 & ML6094917     &   2 & May---June 2018     \\
            Camera 8 & ML6304917     &   2 & May---June 2018     \\
            Camera 9 & ML6314917     &   2 & \textit{not tested} \\
        \end{tabular}
    \end{center}
    \caption[List of GOTO cameras]{
        A list of the 9 GOTO cameras, with assigned names, serial numbers and dates when the tests were carried out. The ninth camera (bought as a spare) was retained by Warwick for their own use, and was not tested in Sheffield.
    }\label{tab:cameras}
\end{table}

The characterisation tests consisted of taking a series of calibration frames with each camera. Three types of images were needed:

\begin{itemize}
    \item Zero-second dark exposures, to construct bias frames (see \aref{sec:bias}).
    \item Long (30 minute) dark exposures at different temperatures, to measure the dark current (see \aref{sec:dc}).
    \item Flat illuminated frames at different exposure times, to construct photon transfer curves (see \aref{sec:ptc}) and measure linearity (see \aref{sec:lin}).
\end{itemize}

The cameras were tested using two different test setups. \aref{fig:dark_photo} shows the setup for taking dark frames: the cameras are face down and covered by a sheet. The long dark exposures required were taken overnight to minimise the background light reaching the detectors. For flat fields a computer monitor was used as a flat source, shown in \aref{fig:flat_photo}. Sheets of paper were placed between the camera and the monitor to reduce the illumination and diffuse the light. The LCD monitor will produce polarised light, however this should not affect the resulting images as long as the angle of the camera remains constant.

\begin{figure}[p]
    \begin{center}
        \includegraphics[width=0.75\linewidth]{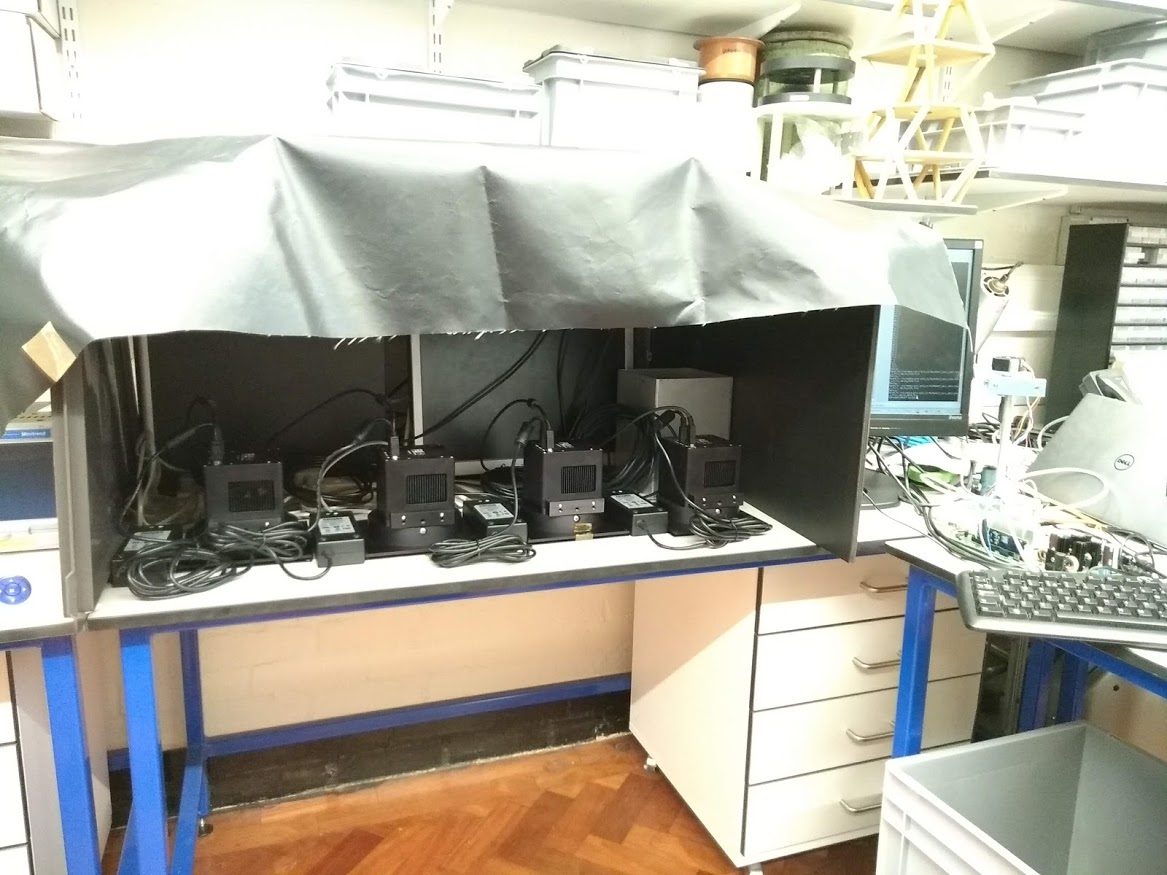}
    \end{center}
    \caption[The dark frame test setup]{
        A photo of the dark frame test setup in the lab in Sheffield. Dark frames were taken at night with the cover down to minimise the ambient light.
    }\label{fig:dark_photo}
\end{figure}

\begin{figure}[p]
    \begin{center}
        \includegraphics[width=0.75\linewidth]{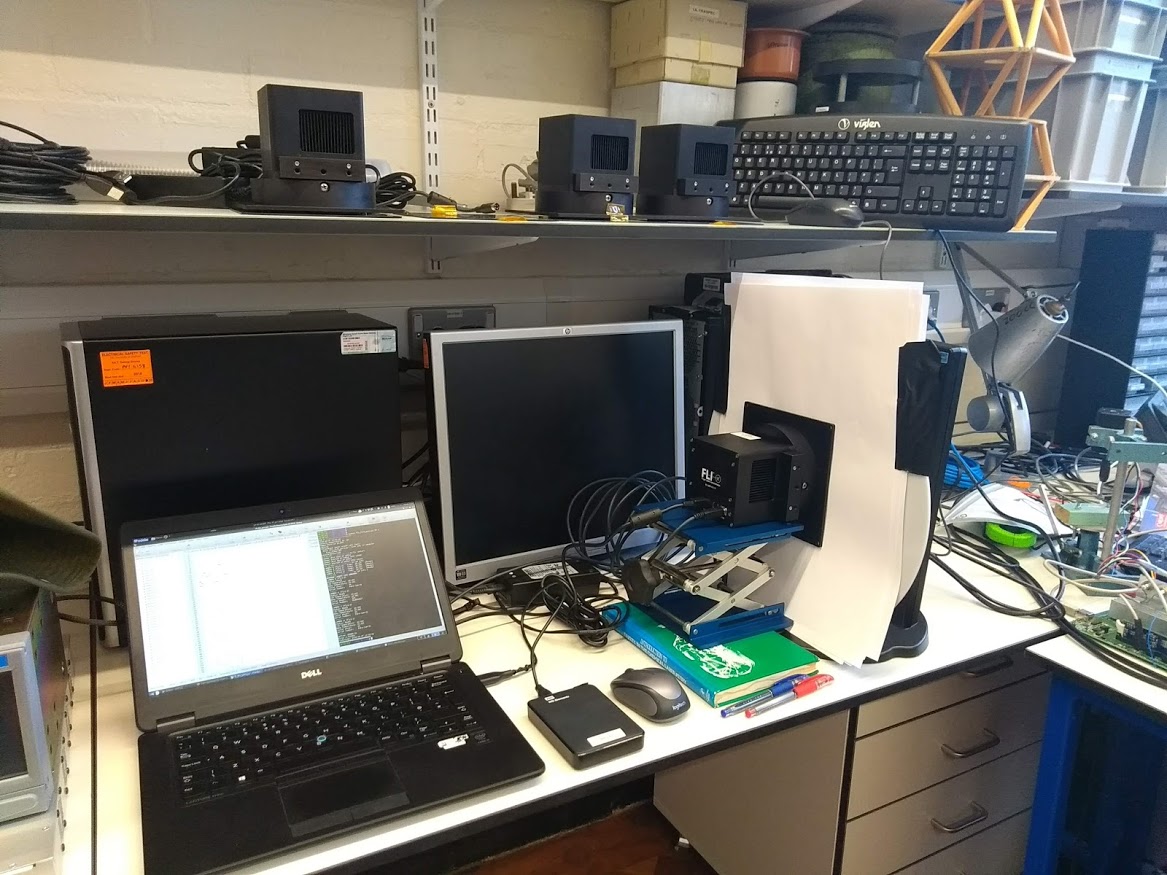}
    \end{center}
    \caption[The flat field test setup]{
        A photo of the flat field test setup. A spare computer monitor was used as a flat panel, with sheets of paper placed between it and the camera. The cover shown in \aref{fig:dark_photo} was also placed over the setup.
    }\label{fig:flat_photo}
\end{figure}

\clearpage
\newpage

\end{colsection}

\subsection{CCD sensors}
\label{sec:chip_layout}
\begin{colsection}

As mentioned previously, the MicroLine ML50100 cameras used by GOTO contain KAF-50100 \acro{ccd} sensors manufactured for FLI by ON Semiconductor\footnote{\url{http://www.onsemi.com}}. These are high-resolution, front-illuminated CCDs with two read-out channels. The \acro{qe} curve for the detector is shown in \aref{fig:qe} in \aref{sec:qe}. The detector is covered in a multilayer anti-reflective coating, and includes a microlens array to focus light onto each pixel and improve the quantum efficiency. The microlenses limit the acceptance angle of the detector to approximately $\SI{\pm20}{\degree}$, for a \SI{40}{\centi\metre} aperture this corresponds to a maximum focal ratio of 2.8 (the \SI{40}{\centi\metre} GOTO unit telescopes are f/2.5, see \aref{sec:optics}).

The KAF-50100 sensor consists of a 50-megapixel CCD with $\SI{6}{\micro\metre} \times \SI{6}{\micro\metre}$ square pixels. The layout of the sensor is shown in \aref{fig:chip}, adapted from the ON Semiconductor sensor specification sheet. The sensor has $8282 \times 6220$ pixels with an imaging area of $8176 \times 6132$ pixels; when taking data in full-frame mode the camera outputs an $8304 \times 6220$ array. Surrounding the image area on each edge are 16 \emph{active buffer pixels}, which are light-sensitive but not considered part of the primary active region (they are not tested for deformities by the manufacturer). Around the edge of the active area is a border of light-shielded \emph{dark reference pixels} which do not respond to light and therefore can be used as a dark current reference. At the beginning and end of each row there is also a test column with 4 blank columns either side, as well as a test row at the end of each frame; these are used to test charge transfer efficiency during the manufacturing process. Finally, at the start of each row the register reads out a test pixel, used in the readout process, followed by 10 \emph{dummy pixels} which do not correspond to physical pixels on the sensor. These form an overscan region which can be used to measure the bias level. A sample flat frame highlighting these areas is shown in \aref{fig:frame}.

\begin{figure}[p]
    \begin{center}
        \includegraphics[width=0.78\linewidth]{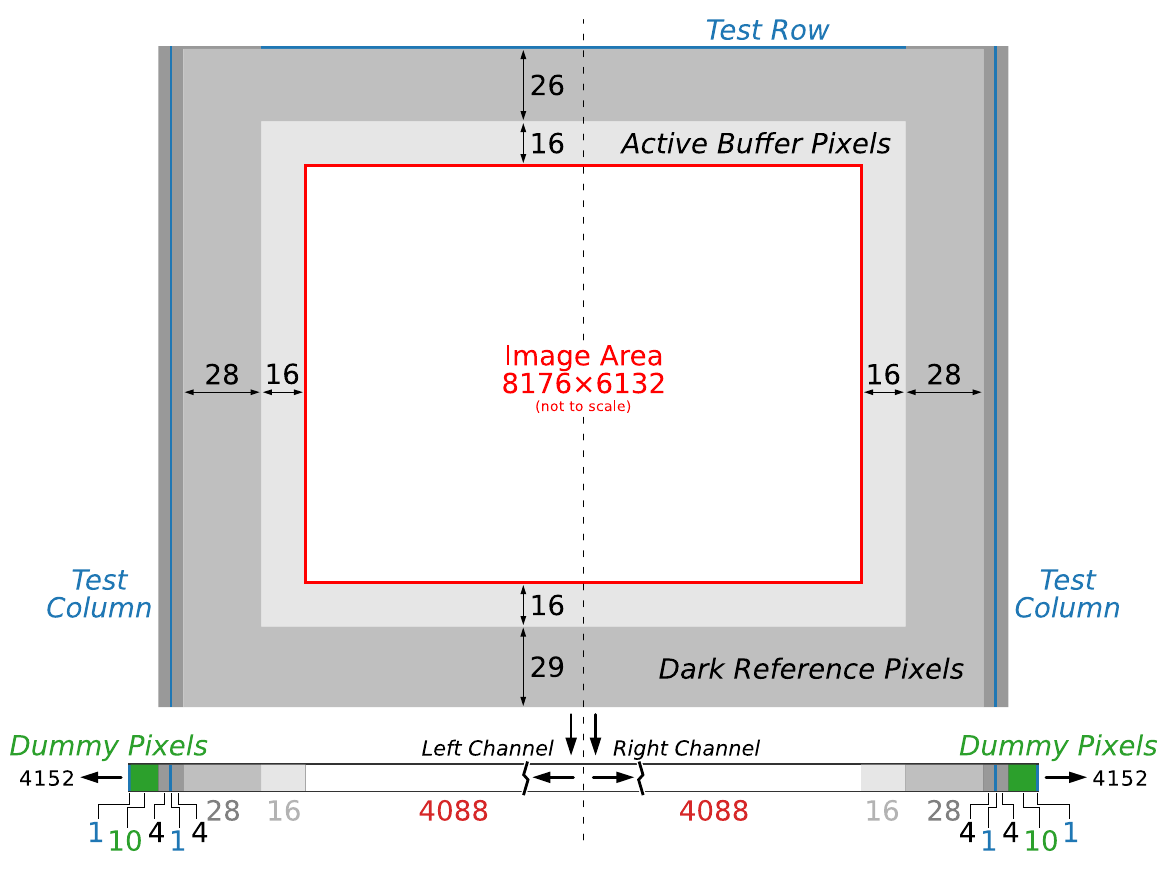}
    \end{center}
    \caption[The layout of the KAF-50100 CCD sensor]{
        The layout of the KAF-50100 CCD sensor. The central image area is not shown to scale, but the surrounding rows and columns are all in proportion.
    }\label{fig:chip}
\end{figure}

\begin{figure}[p]
    \begin{center}
        \includegraphics[width=0.78\linewidth]{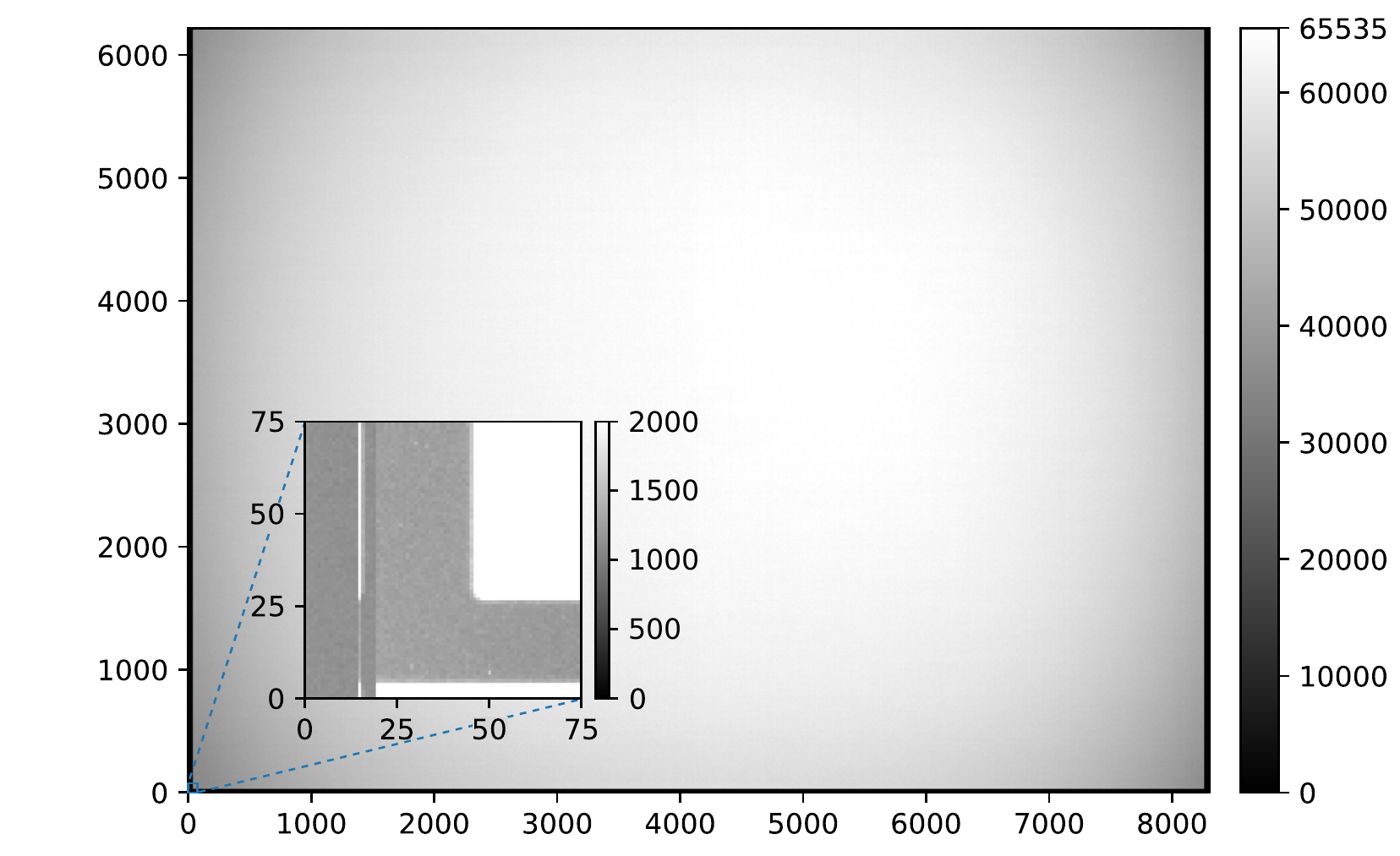}
    \end{center}
    \caption[A sample bright frame from one of the MicroLine cameras]{
        A sample bright frame from one of the MicroLine cameras. The highlighted corner shows some of the features described in \aref{fig:chip}. This image was taken as shown in \aref{fig:flat_photo} with no optical elements between the camera and the screen. As no optical elements were used the cause of the visible vignetting is unclear, but may come from the camera aperture.
    }\label{fig:frame}
\end{figure}

\clearpage
\newpage

\end{colsection}

\subsection{Bias}
\label{sec:bias}
\begin{colsection}

The bias level in each pixel can be measured by taking a dark, zero-second exposure image. This image will not include any electrons from a source or background, so the shot noise is zero, and as the dark current is proportional to the exposure time this will also be minimised (there will still be a small component due to the time taken to read out the sensor). To account for the read-out noise, multiple images are taken, 50 for each of the eight cameras, and combined to form a master bias frame by taking the median value in each pixel (to eliminate cosmic rays).

\aref{tab:bias} gives the median bias level from each master bias, measured within a 2000$\times$2000 pixel region in the centre of both channels, along with the standard deviation in the same region. The measured bias levels are all around 1000 counts, which would be typical for a bias level set by the manufacturer. Combining $n$ frames should reduce the noise by $\sqrt{n}$, and, as expected, the errors are equivalent to the read-out noise values given in \aref{tab:ptc} when converted into counts and reduced by a factor of $\sqrt{50}$ ($\approx 7$).

\begin{table}[t]
    \begin{center}
        \begin{tabular}{c|rr} %
             & \multicolumn{2}{c}{Bias} \\
             & \multicolumn{2}{c}{(ADU)} \\
             & \multicolumn{1}{c}{L} & \multicolumn{1}{c}{R} \\
            \midrule
            Camera 1 & $971\pm3.6$ & $969\pm3.7$ \\
            Camera 2 & $989\pm3.3$ & $983\pm3.3$ \\
            Camera 3 & $1004\pm3.2$ & $991\pm3.1$ \\
            Camera 4 & $974\pm3.4$ & $1008\pm3.8$ \\
        \end{tabular}
        \hspace{0.5cm}
        \begin{tabular}{c|rr} %
             & \multicolumn{2}{c}{Bias} \\
             & \multicolumn{2}{c}{(ADU)} \\
             & \multicolumn{1}{c}{L} & \multicolumn{1}{c}{R} \\
            \midrule
            Camera 5 & $994\pm2.9$ & $986\pm3.0$ \\
            Camera 6 & $984\pm2.7$ & $991\pm3.0$ \\
            Camera 7 & $992\pm3.1$ & $981\pm3.0$ \\
            Camera 8 & $1008\pm3.3$ & $1012\pm2.9$ \\
        \end{tabular}
    \end{center}
    \caption[Bias values]{
        Bias values for each camera.
    }\label{tab:bias}
\end{table}

\end{colsection}

\subsection{Gain, read-out noise and fixed-pattern noise}
\label{sec:ptc}
\begin{colsection}

The gain, read-out and fixed-pattern noise of a CCD camera can be measured using the \acro{ptc} method \citep{CCDs, PTC}. A photon transfer curve is a log-log plot of a signal value against the noise in the signal. To construct a photon transfer curve a series of bright exposures of a flat light source were taken with varying exposure times. For these images there is no background signal, the cameras were cooled meaning dark current noise is negligible (see \aref{sec:dc}), and the master bias frames described in \aref{sec:bias} were subtracted from each frame. The total noise per pixel in electrons is therefore given by \aref{eq:noise} as
\begin{equation}
    \sigma_\text{Total}^2 = N + R^2 + k_\text{FP}^2{(N)}^2.
    \label{eq:noise_2}
\end{equation}

The signal and total noise in \aref{eq:noise} are all in electrons (\elec), however the output of the camera's \acro{adc} is a digital signal, $S$, measured in counts or \acro{adu}. This signal is linearly related to the actual number of electrons detected, $N$, through the gain, $g$, in \elec/ADU, as
\begin{equation}
    N = g S.
    \label{eq:gain}
\end{equation}
The gain is an important parameter of a CCD, and is set by the manufacturer based on the properties of the detector. For example, if a CCD has the gain set to 3 \elec/ADU then pixels containing 0, 1 or 2 electrons would all have a measured value of 0 ADU;\@ this is a form of rounding error called quantisation error. If the same camera had a read-out noise of 1 \elec{} per pixel then the readout-noise would be under-sampled, and setting a lower gain would be required. However, setting the gain too low results in the full-well capacity of each pixel being under-utilised. The KAF-50100 detectors have a full well capacity of 40,300 electrons, and the cameras have a 16-bit ADC (meaning the signal from each pixel can vary from 0 to 65535 ($2^{16}-1$) ADU). If the gain is set to $0.5$ \elec/ADU then the ADC would saturate after reading 32,768 electrons, which is much less than the capacity of each pixel. Setting the gain therefore is a balance between these two effects.

As electrons and counts are proportional, the noise in both is also proportional (i.e.\ as $N=gS$ from \aref{eq:gain},  $\sigma_\text{Total} = g\sigma_S$). Using these relationships \aref{eq:noise_2} can be converted to give the noise in ADU,
\begin{equation}
    \sigma_S^2 = \frac{1}{g} S + \frac{R^2}{g^2} + k_\text{FP}^2 S^2.
    \label{eq:ptc}
\end{equation}
This is a quadratic equation which relates the measured total signal $S$ to the variance in the signal $\sigma_S^2$, and can be fitted to a photon transfer curve to determine values for the gain $g$ (in \elec/ADU), read-out noise $R$ (still in \elec{}) and fixed-pattern noise $k_\text{FP}$ (dimensionless).

\begin{figure}[t]
    \begin{center}
        \includegraphics[width=0.8\linewidth]{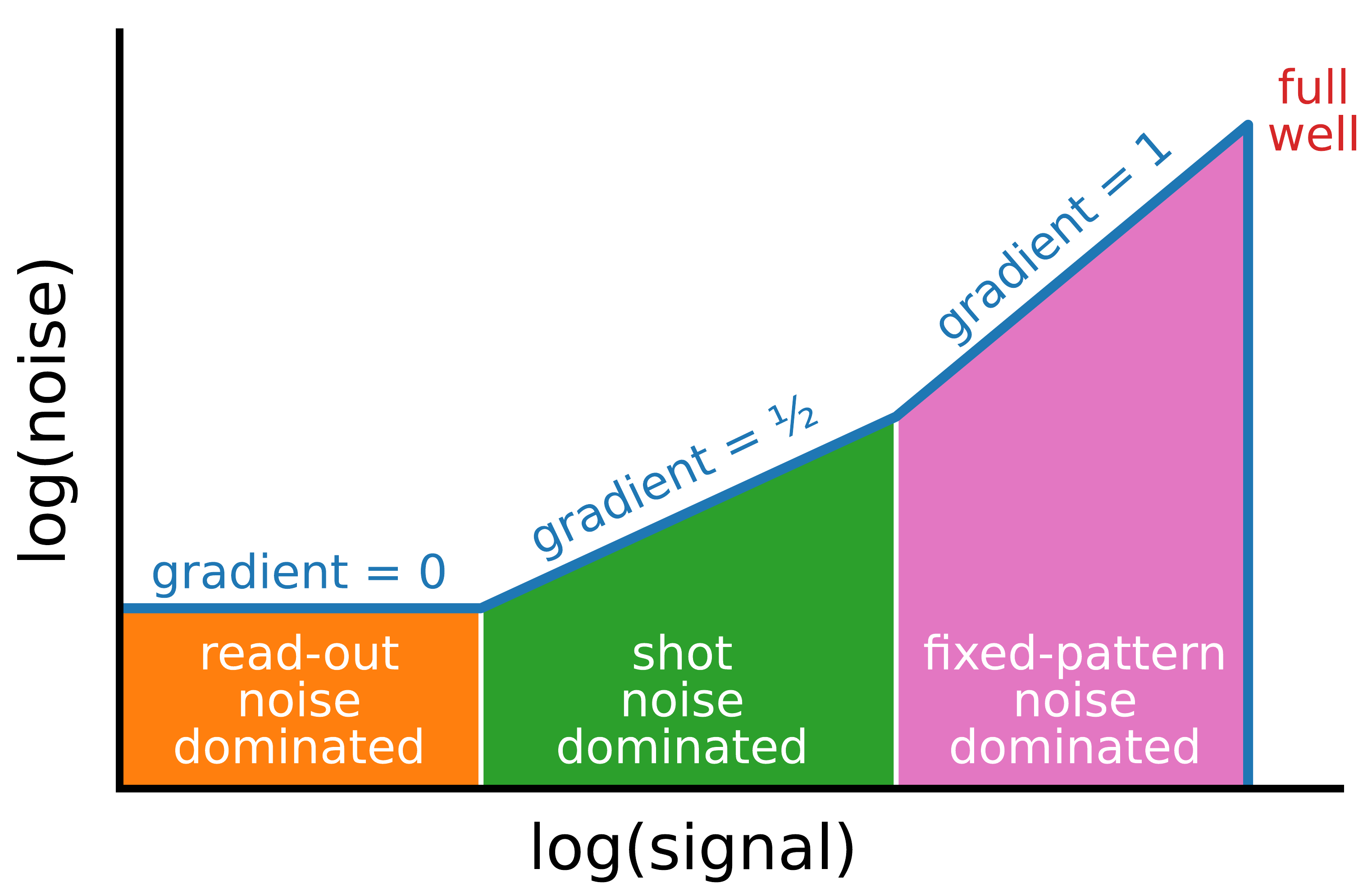}
    \end{center}
    \caption[Key features of the photon transfer curve]{
        The key features of a photon transfer curve, adapted from \citet{CCDs}.
    }\label{fig:ptc_cartoon}
\end{figure}

The key features of a photon transfer curve are common for all CCDs, and are shown in cartoon form in \aref{fig:ptc_cartoon}. The first noise regime is when the signal is small: from \aref{eq:ptc} for small $S$ the noise is constant and equal to $R^2/g^2$, or the read-out noise in ADU.\@ At higher signals the noise is dominated by the shot noise which is proportional to $\sqrt{S}$, so this region has a gradient of \sfrac{1}{2} when plotted on the log-log axis. As the signal increases further the fixed-pattern noise begins to dominate, and, as this noise is proportional to the signal, this produces a gradient of 1 in the PTC.\@ Finally, the pixel reaches its full well capacity (assuming the gain has been set so this occurs before the ADU saturates), so the noise drops to zero.

\begin{table}[t]
    \begin{center}
        \begin{tabular}{l|cc|cc|cc|cc} %
             &
            \multicolumn{2}{c|}{Gain} &
            \multicolumn{2}{c|}{RO noise} &
            \multicolumn{2}{c|}{FP noise} &
            \multicolumn{2}{c}{Saturation level} \\
            &
            \multicolumn{2}{c|}{(\elec/ADU)} &
            \multicolumn{2}{c|}{(\elec)} &
            \multicolumn{2}{c|}{(\%)} &
            \multicolumn{2}{c}{(ADU)} \\
             & L & R & L & R & L & R & L & R \\
            \midrule
            Camera 1 & 0.53 & 0.53 & 12.4 & 12.0 & 0.46 & 0.45 & 64568 & 64585 \\
            Camera 2 & 0.53 & 0.53 & 11.9 & 11.7 & 0.44 & 0.46 & 64552 & 64555 \\
            Camera 3 & 0.57 & 0.57 & 12.6 & 11.8 & 0.45 & 0.42 & 64540 & 64552 \\
            Camera 4 & 0.57 & 0.58 & 13.4 & 14.0 & 0.41 & 0.43 & 64577 & 64536 \\
            Camera 5 & 0.62 & 0.63 & 12.3 & 12.8 & 0.40 & 0.40 & 64544 & 64550 \\
            Camera 6 & 0.63 & 0.62 & 11.8 & 12.6 & 0.40 & 0.40 & 64554 & 64545 \\
            Camera 7 & 0.62 & 0.62 & 13.1 & 12.5 & 0.41 & 0.39 & 64544 & 64552 \\
            Camera 8 & 0.62 & 0.62 & 14.3 & 12.2 & 0.41 & 0.39 & 64529 & 64522 \\
        \end{tabular}
    \end{center}
    \caption[Gain, read-out noise, fixed-pattern noise and saturation values]{
        Gain, read-out noise, fixed-pattern noise and saturation values found by fitting photon transfer curves for each camera.
    }\label{tab:ptc}
\end{table}

Photon transfer curves were constructed for all eight cameras by taking flat fields of varying exposure times between \SI{0.01}{\second} and \SI{90}{\second}. Twelve 50$\times$50 pixel regions were selected across each image, and the mean and standard deviation of the pixel values within each region were plotted to form the PTC for each camera, shown in \aref{fig:ptcs}. \aref{eq:ptc} was then fitted to the data, and the resulting values for the gain ($g$), read-out noise ($R$) and fixed-pattern noise ($k_\text{FP}$) parameters are given in \aref{tab:ptc}. The saturation level for each channel was also measured as the maximum signal (the point where the PTC turns over and the noise drops), these values are also given in \aref{tab:ptc}.

The gain values are all around 0.6 \elec/ADU, and would have been set as such to maximise the dynamic range based on the full well capacity (65535\,$\times$\,0.6\,$\approx$\,40,000 \elec). The saturation levels are all around 64550 ADU, but note these images were bias-subtracted and \aref{sec:bias} found the bias levels were around 980--1000 ADU.\@ The read-out noise values match the FLI specification of 12 \elec{} for the MicroLine cameras, and also match the errors found in the master biases. Finally, the fixed-pattern noise is a very small fraction of the signal ($<0.5\%$) which suggests a low pixel non-uniformity.

\newpage

\begin{figure}[p]
    \begin{center}
        \begin{minipage}[t]{0.49\linewidth}\vspace{10pt}
            \includegraphics[width=\linewidth]{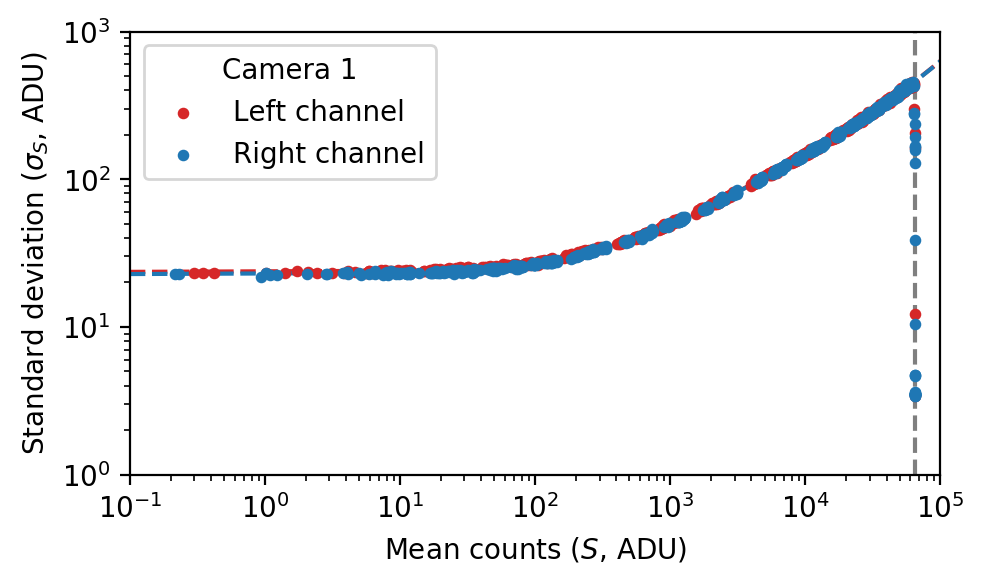}
        \end{minipage}
        \begin{minipage}[t]{0.49\linewidth}\vspace{10pt}
            \includegraphics[width=\linewidth]{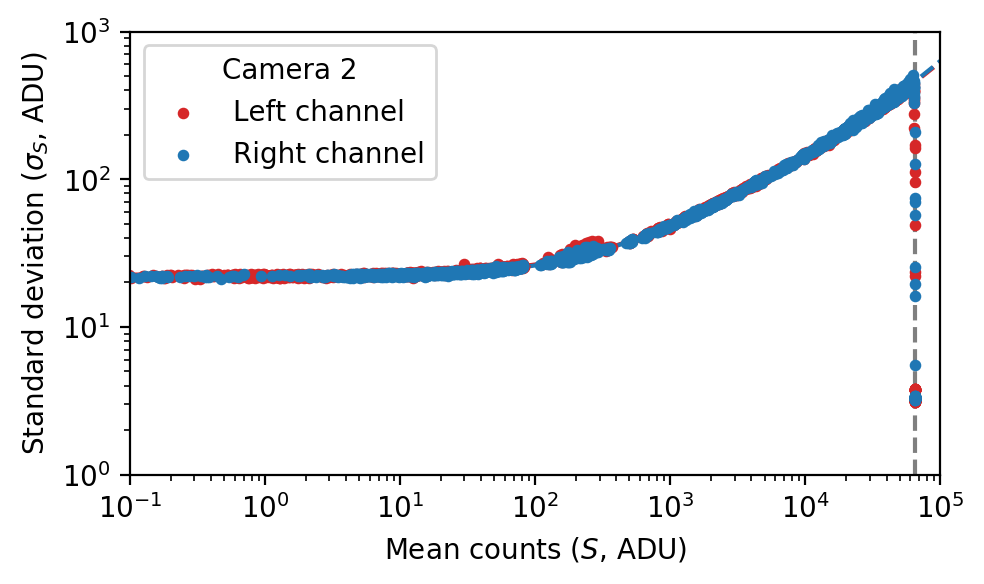}
        \end{minipage}

        \begin{minipage}[t]{0.49\linewidth}\vspace{10pt}
            \includegraphics[width=\linewidth]{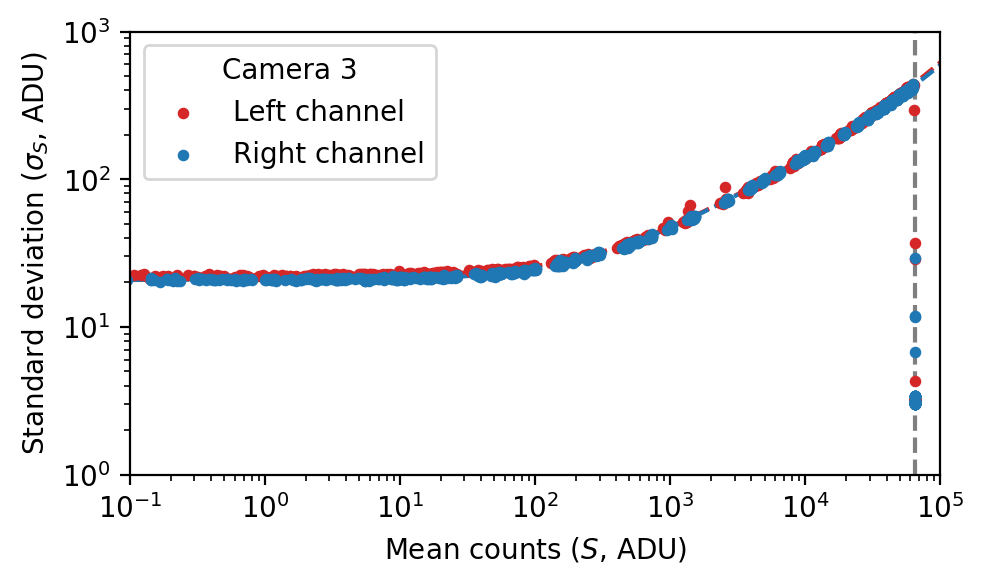}
        \end{minipage}
        \begin{minipage}[t]{0.49\linewidth}\vspace{10pt}
            \includegraphics[width=\linewidth]{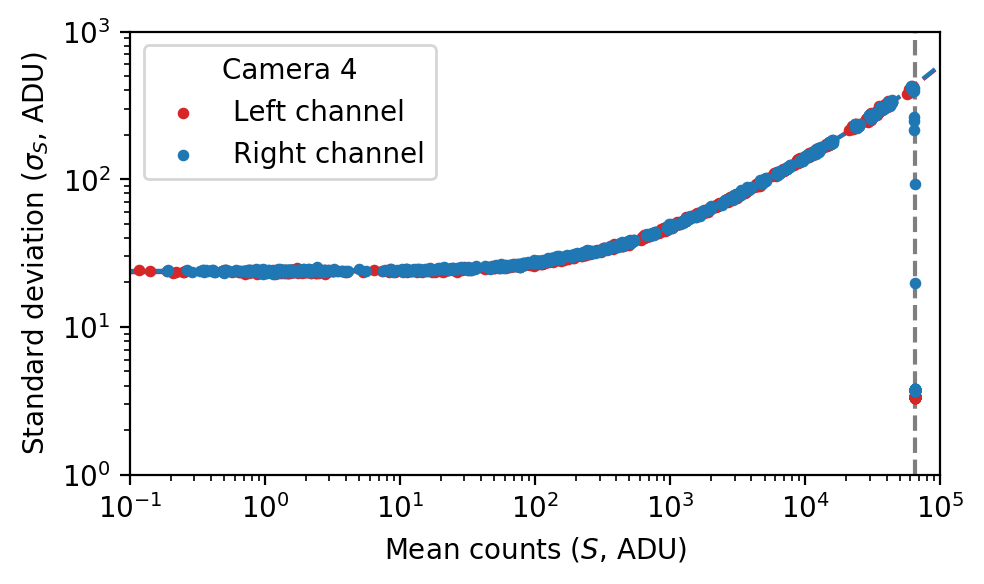}
        \end{minipage}

        \begin{minipage}[t]{0.49\linewidth}\vspace{10pt}
            \includegraphics[width=\linewidth]{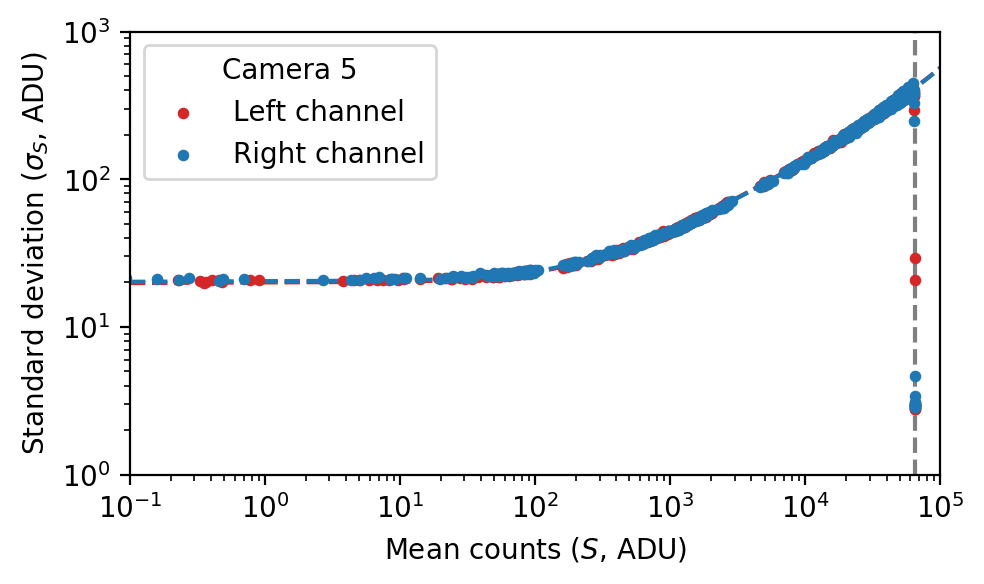}
        \end{minipage}
        \begin{minipage}[t]{0.49\linewidth}\vspace{10pt}
            \includegraphics[width=\linewidth]{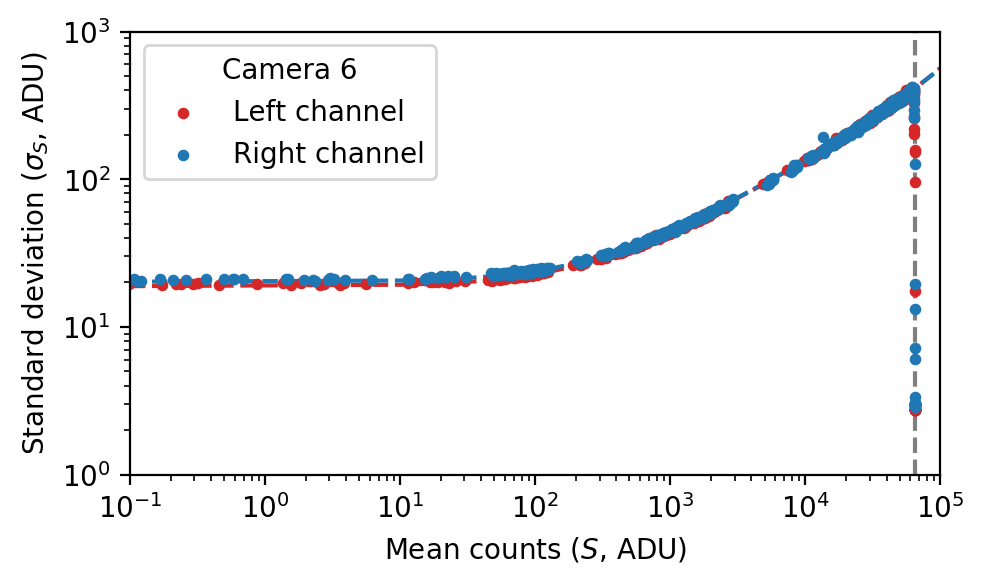}
        \end{minipage}

        \begin{minipage}[t]{0.49\linewidth}\vspace{10pt}
            \includegraphics[width=\linewidth]{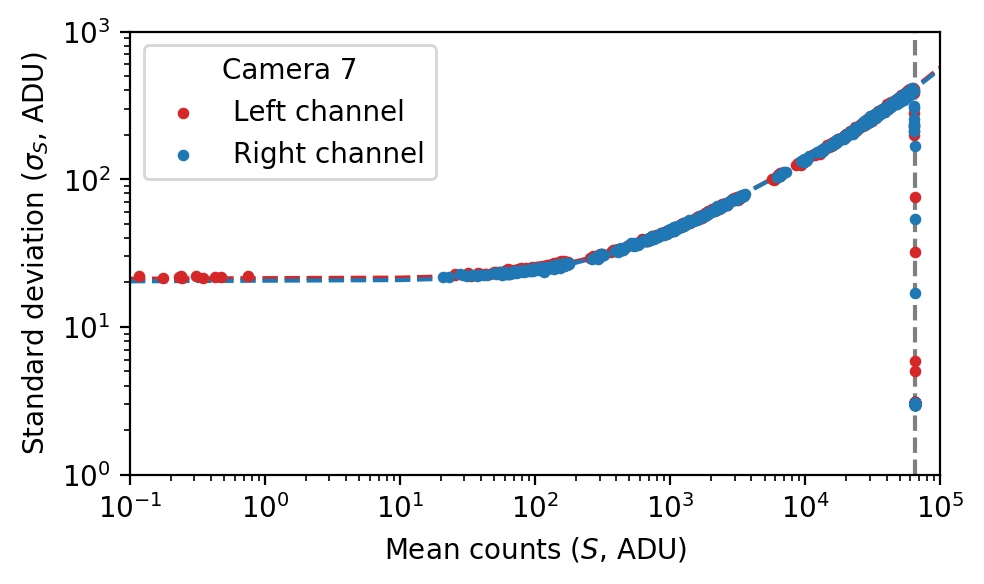}
        \end{minipage}
        \begin{minipage}[t]{0.49\linewidth}\vspace{10pt}
            \includegraphics[width=\linewidth]{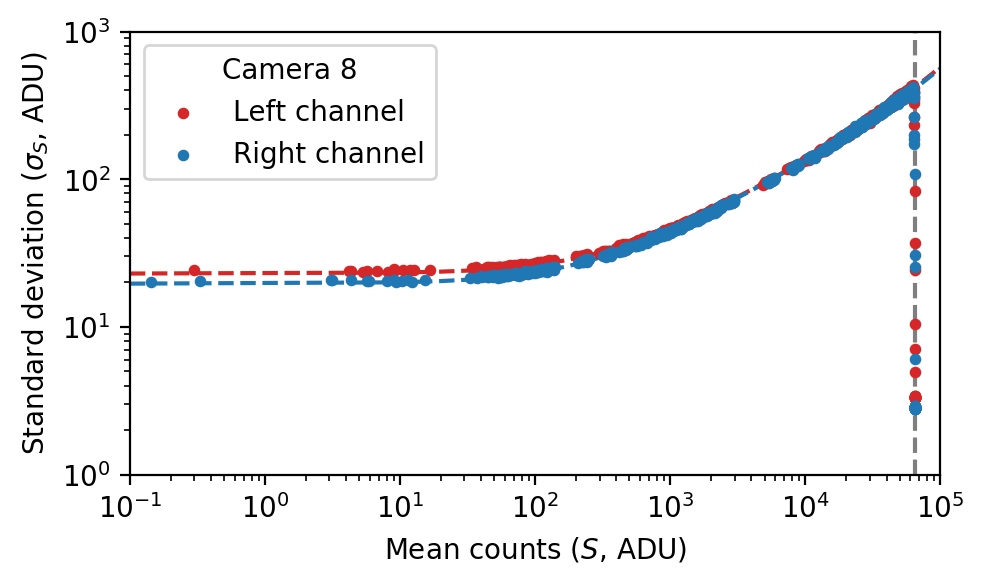}
        \end{minipage}
    \end{center}
    \caption[Photon transfer curve plots]{
        Photon transfer curve plots for each camera. The vertical dashed line shows the measured saturation level.
    }\label{fig:ptcs}
\end{figure}

\clearpage
\newpage

\end{colsection}

\subsection{Dark current}
\label{sec:dc}
\begin{colsection}

Dark current noise is independent of the incoming signal but depends on the exposure time of the image. It also increases as a function of temperature, $T$. The dark current per second, $D$, increases at an exponential rate and is usually parametrised as doubling after a fixed increase in temperature called the doubling temperature, $T_d$, so that if the dark current $D(T_0) = D_0$ then $D(T_0 + T_d) = 2D_0$. The dark current as a function of temperature is therefore defined as
\begin{equation}
    D(T) = D_0 e^{\frac{\ln2}{T_d}(T - T_0)}.
    \label{eq:dc}
\end{equation}
The choice of $T_0$ is arbitrary, and is usually decided as a reasonable operating temperature by the CCD manufacturer. The FLI specifications give a value for the typical dark current at \SI{-25}{\celsius}, so that is the value of $T_0$ used in this test.

In order to find values for the dark current $D_0$ and doubling temperature $T_d$, a series of long (30 minute) dark exposures were taken with each camera at varying temperatures. The MicroLine cameras have an in-built air-cooled Peltier cooler which can reach \SI{40}{\celsius} below the ambient temperature. The laboratory the tests were carried out in was air-conditioned, but only to a typical office level, and the cameras were unable to reach below \SI{-26}{\celsius} even when taking images in the middle of the night. The median dark signal was then measured in a 2000$\times$2000 pixel region in the centre of each channel, and divided by 1800 (as each exposure was 30 minutes) to get the dark current in ADU/second. This value was plotted against temperature, as shown in \aref{fig:dcs}. The points were fitted by \aref{eq:dc}, and the resulting values for the dark current and doubling temperature are given in \aref{tab:dc}.

\begin{table}[t]
    \begin{center}
        \begin{tabular}{c|cc|cc|rr} %
             &
            \multicolumn{4}{c|}{Dark current per pixel} &
            \multicolumn{2}{c}{Doubling} \\
             &
            \multicolumn{4}{c|}{at \SI{-25}{\celsius}} &
            \multicolumn{2}{c}{temperature} \\
             &
            \multicolumn{2}{c|}{(ADU/s)} &
            \multicolumn{2}{c|}{(e-/s)} &
            \multicolumn{2}{c}{(\SI{}{\celsius})} \\
             & L & R & L & R &
             \multicolumn{1}{c}{L} & \multicolumn{1}{c}{R} \\
            \midrule
            Camera 1 & 0.0022 & 0.0017 & 0.0012 & 0.0009 &  7.9 &  6.7 \\
            Camera 2 & 0.0030 & 0.0027 & 0.0016 & 0.0014 &  8.9 &  8.2 \\
            Camera 3 & 0.0034 & 0.0036 & 0.0019 & 0.0020 & 10.7 & 10.9 \\
            Camera 4 & 0.0026 & 0.0030 & 0.0015 & 0.0017 &  9.5 & 10.2 \\
            Camera 5 & 0.0015 & 0.0017 & 0.0009 & 0.0011 &  6.6 &  7.2 \\
            Camera 6 & 0.0020 & 0.0017 & 0.0013 & 0.0011 &  7.5 &  6.8 \\
            Camera 7 & 0.0017 & 0.0014 & 0.0011 & 0.0008 &  7.6 &  6.5 \\
            Camera 8 & 0.0019 & 0.0015 & 0.0012 & 0.0009 &  7.5 &  6.5 \\
        \end{tabular}
    \end{center}
    \caption[Dark current values]{
        Dark current values for each camera. The conversion from ADU/s to \elec/s used the gain values given in \aref{tab:ptc}.
    }\label{tab:dc}
\end{table}

The FLI specification for dark current changed between the two test periods; initially the company gave a typical per-pixel value of 0.002~\elec/s at \SI{-25}{\celsius}, for the second set of cameras this was increased to 0.008~\elec/s. All the cameras were found to have a dark current well within the revised specification value, and all except Camera 3 are comfortably below the original 0.002~\elec/s specification.

The KAF-50100 specification includes a value for the doubling temperature of \SI{5.7}{\celsius} but the measured values are all higher than this. In practice, the temperature dependence of the dark current is not important; the GOTO cameras are cooled to \SI{-20}{\celsius} in the evening and remain there through the night (\SI{-20}{\celsius} is used instead of \SI{-25}{\celsius} as during the summer on La Palma the ambient nightly temperature can reach higher than \SI{15}{\celsius}).

The dark current was also examined as a function of time since power on, as in some cameras there are a noticeable amount of free electrons left trapped in the lattice which take time to dissipate \citep{Liam}. No such trend was visible using the FLI cameras. Since the MicroLine cameras have the detector and cooler integrated into the same body there has to be some time spent waiting after power on for the camera to cool to the target temperature before any images can be taken, thus negating the effect.

\begin{figure}[p]
    \begin{center}
        \begin{minipage}[t]{0.49\linewidth}\vspace{10pt}
            \includegraphics[width=\linewidth]{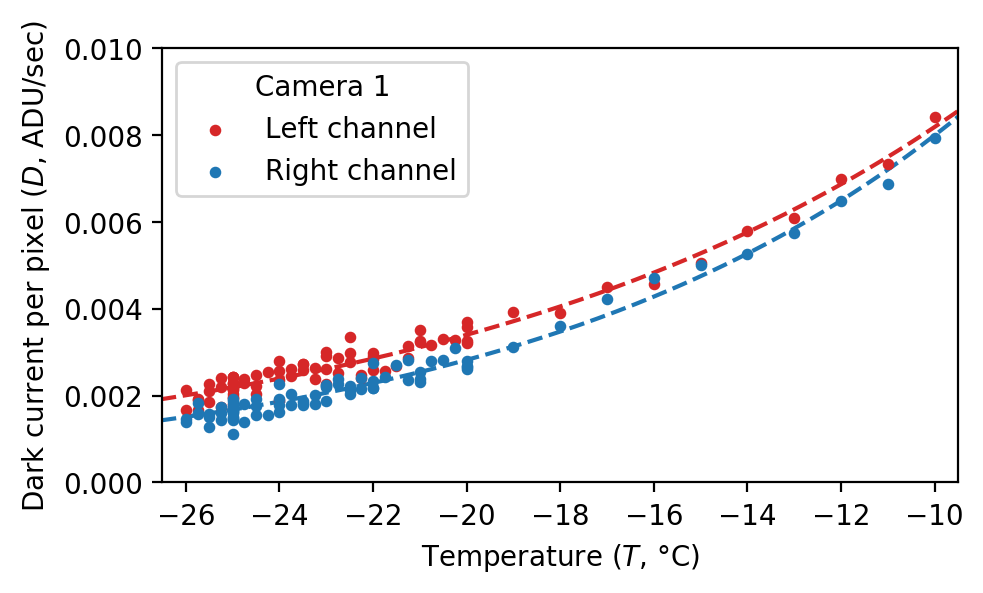}
        \end{minipage}
        \begin{minipage}[t]{0.49\linewidth}\vspace{10pt}
            \includegraphics[width=\linewidth]{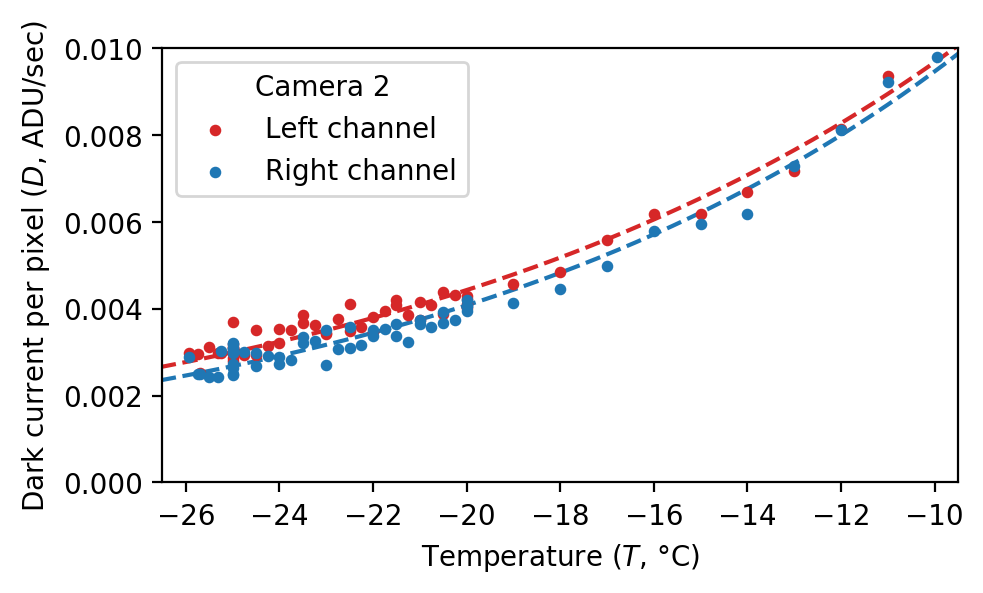}
        \end{minipage}

        \begin{minipage}[t]{0.49\linewidth}\vspace{10pt}
            \includegraphics[width=\linewidth]{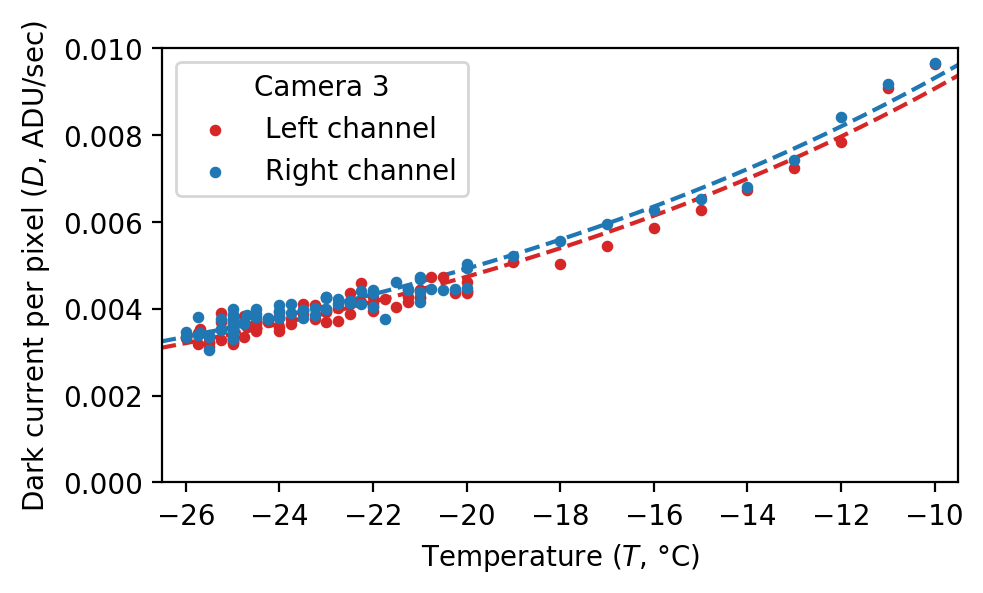}
        \end{minipage}
        \begin{minipage}[t]{0.49\linewidth}\vspace{10pt}
            \includegraphics[width=\linewidth]{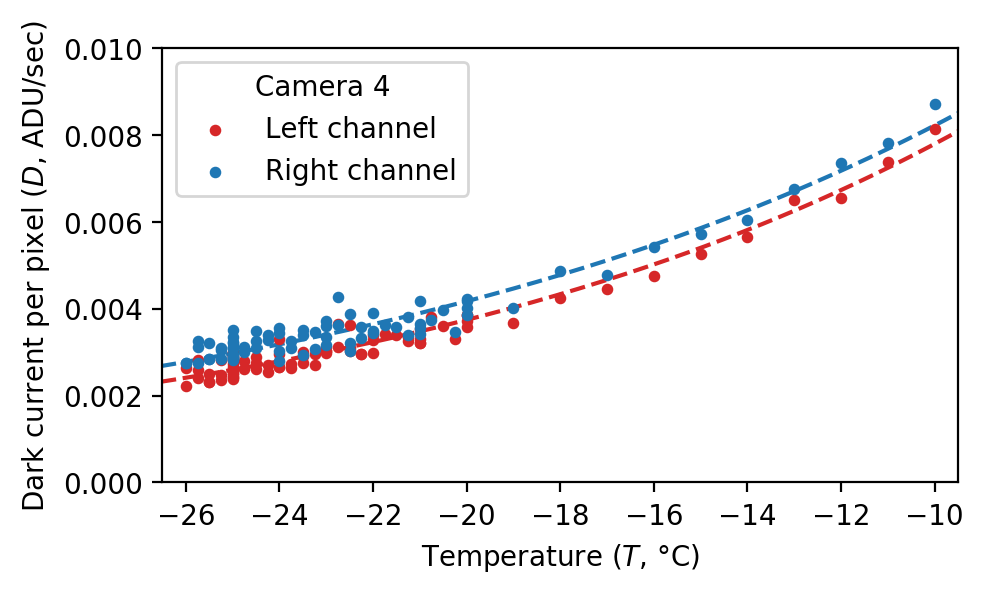}
        \end{minipage}

        \begin{minipage}[t]{0.49\linewidth}\vspace{10pt}
            \includegraphics[width=\linewidth]{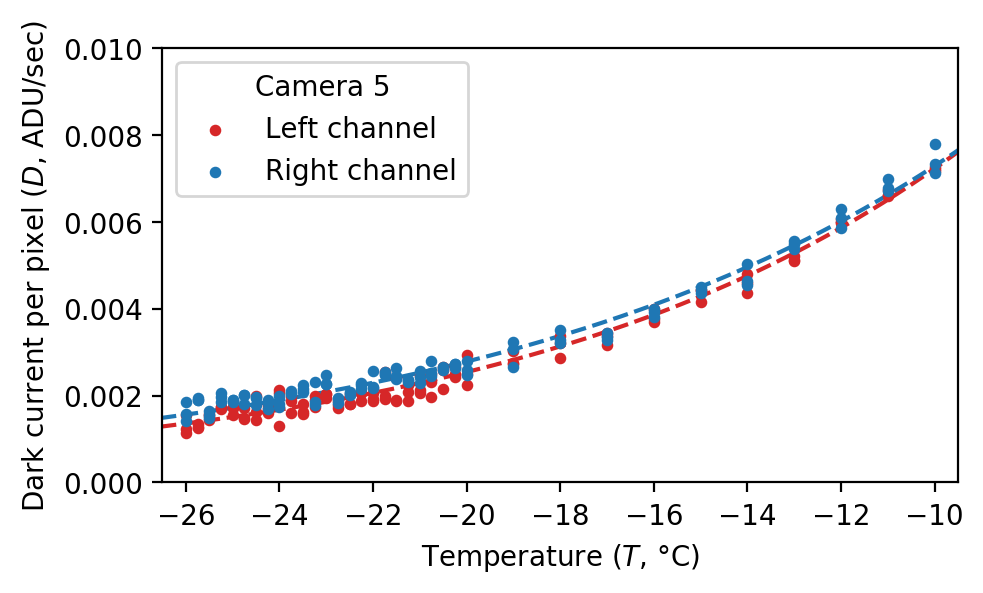}
        \end{minipage}
        \begin{minipage}[t]{0.49\linewidth}\vspace{10pt}
            \includegraphics[width=\linewidth]{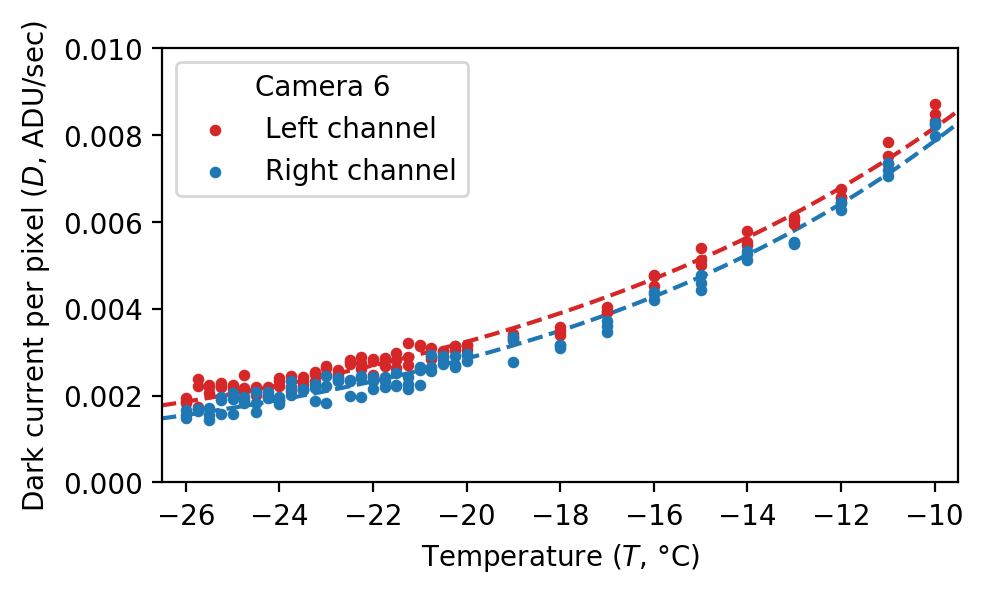}
        \end{minipage}

        \begin{minipage}[t]{0.49\linewidth}\vspace{10pt}
            \includegraphics[width=\linewidth]{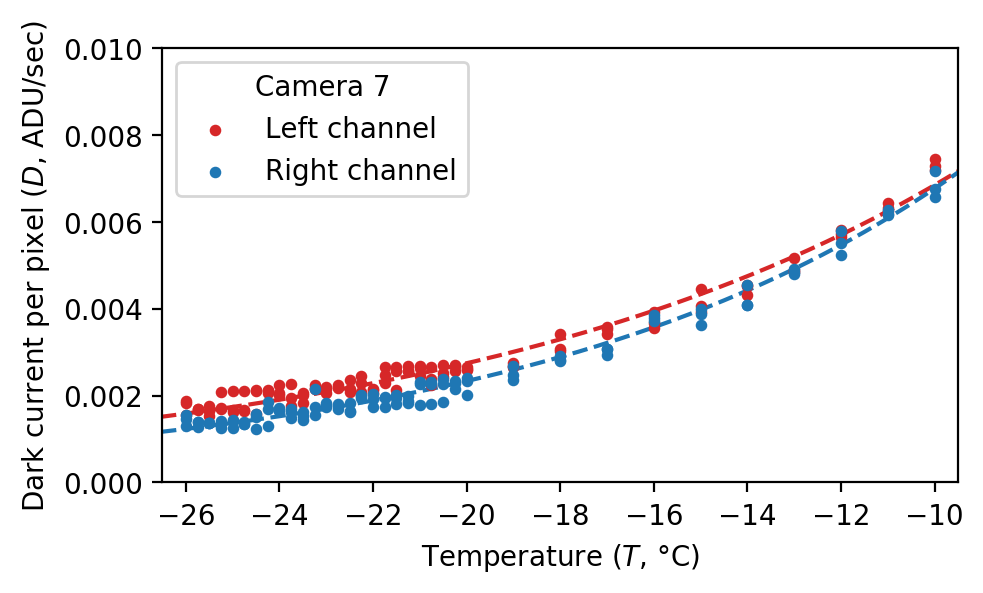}
        \end{minipage}
        \begin{minipage}[t]{0.49\linewidth}\vspace{10pt}
            \includegraphics[width=\linewidth]{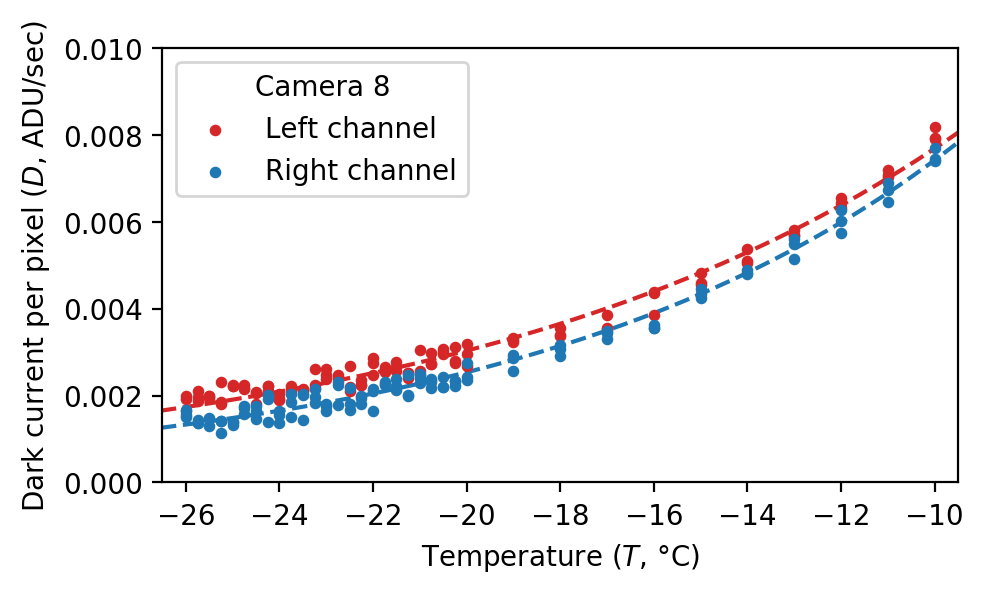}
        \end{minipage}
    \end{center}
    \caption[Dark current plots]{
        Dark current plots for each camera.
    }\label{fig:dcs}
\end{figure}

\clearpage
\newpage

\end{colsection}

\subsection{Linearity}
\label{sec:lin}
\begin{colsection}

Linearity is a measure of the response of the CCD over its dynamic range. The output counts should ideally be linearly related to the input photons, i.e.\@ if the target doubles in brightness then double the counts should be recorded.

The non-linearity of each camera was measured using the same images taken for the photon transfer curves in \aref{sec:ptc} --- bright images of a flat field with increasing exposure times. The images were bias-subtracted, and the median counts of a 2000$\times$2000 pixel region in the centre of each channel was plotted against the exposure time, shown in \aref{fig:lin}. A linear relation was fitted to the central potion of the data, excluding the upper and lower 10\% of the dynamic range. Residuals from this fit are also plotted in \aref{fig:lin}, and the mean absolute deviation from the linear fit is given in \aref{tab:lin}.

The values for non-linearity measured vary greatly between each camera, and several are over 1\%. If these values were true this would be a major problem when making accurate photometric measurements. However, the FLI specification advertises a non-linearity of <1\%, and FLI's own tests of the cameras consistently report non-linearity of 0.2\% or less. Accurately measuring the response of a CCD requires a stable, uniform light source, which I had to approximate with an LCD screen as described in a \aref{sec:camera_tests}. A better test would be to vary the screen brightness instead of the exposure time, which would prevent systematic effects such as the shutter closing time affecting the results.

\begin{table}[t]
    \begin{center}
        \begin{tabular}{c|cc} %
             & \multicolumn{2}{c}{Non-linearity} \\
             & \multicolumn{2}{c}{(\%)} \\
             & \multicolumn{1}{c}{L} & \multicolumn{1}{c}{R} \\
            \midrule
            Camera 1 & 2.29 & 2.00 \\
            Camera 2 & 0.76 & 0.65 \\
            Camera 3 & 0.34 & 0.39 \\
            Camera 4 & 0.18 & 0.22 \\
        \end{tabular}
        \hspace{0.5cm}
        \begin{tabular}{c|cc} %
             & \multicolumn{2}{c}{Non-linearity} \\
             & \multicolumn{2}{c}{(\%)} \\
             & \multicolumn{1}{c}{L} & \multicolumn{1}{c}{R} \\
            \midrule
            Camera 5 & 1.25 & 1.20 \\
            Camera 6 & 1.20 & 1.13 \\
            Camera 7 & 0.70 & 0.68 \\
            Camera 8 & 0.82 & 0.80 \\
        \end{tabular}
    \end{center}
    \caption[Non-linearity values]{
        Non-linearity values for each camera.
    }\label{tab:lin}
\end{table}

\begin{figure}[p]
    \begin{center}
        \begin{minipage}[t]{0.47\linewidth}\vspace{10pt}
            \includegraphics[width=\linewidth]{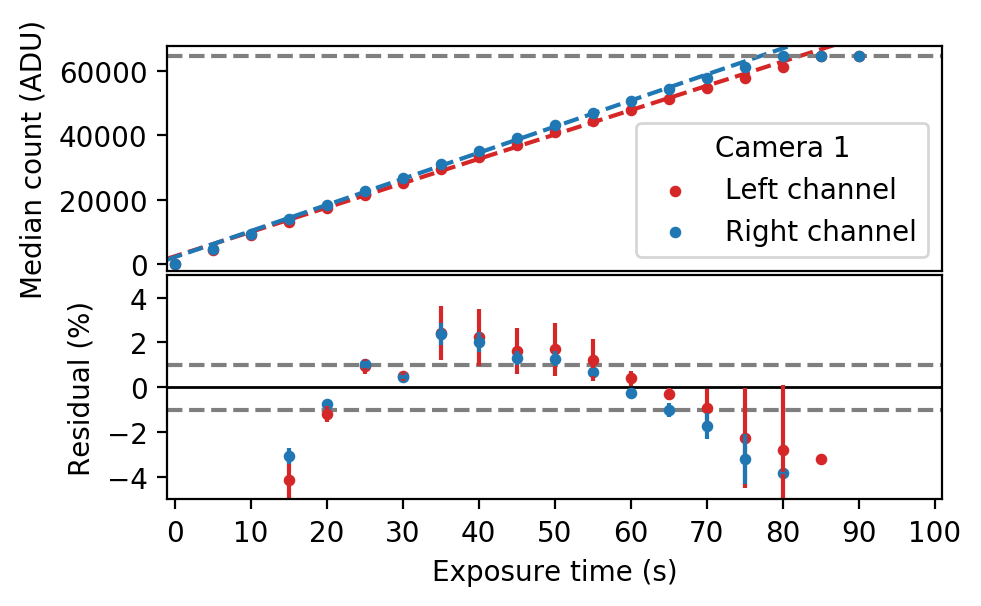}
        \end{minipage}
        \begin{minipage}[t]{0.47\linewidth}\vspace{10pt}
            \includegraphics[width=\linewidth]{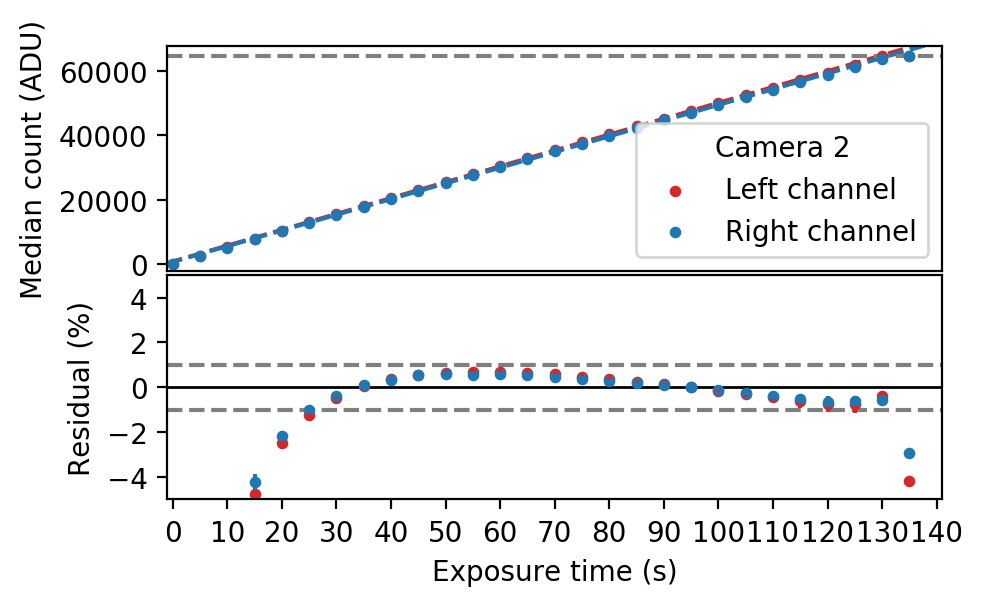}
        \end{minipage}

        \begin{minipage}[t]{0.47\linewidth}\vspace{10pt}
            \includegraphics[width=\linewidth]{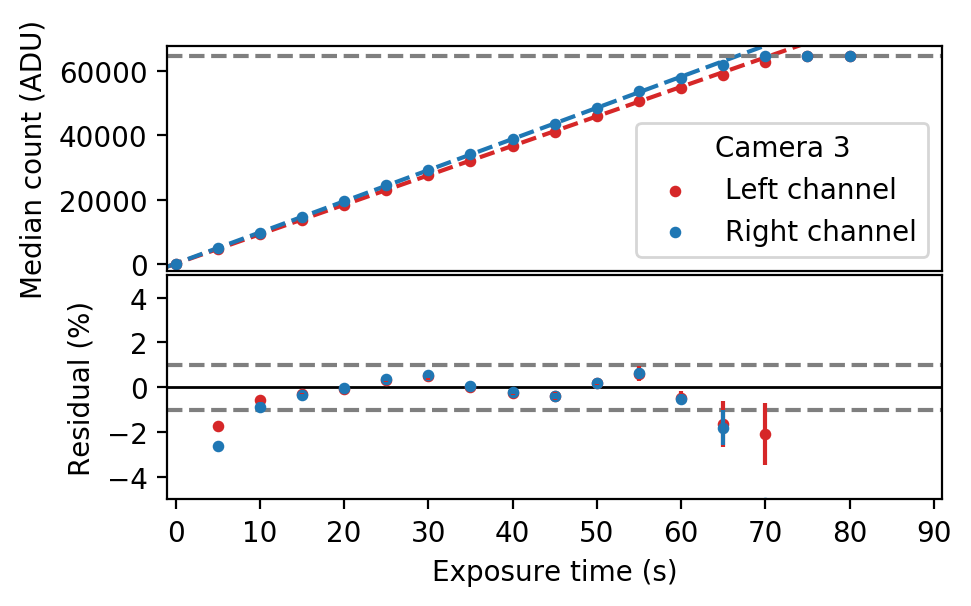}
        \end{minipage}
        \begin{minipage}[t]{0.47\linewidth}\vspace{10pt}
            \includegraphics[width=\linewidth]{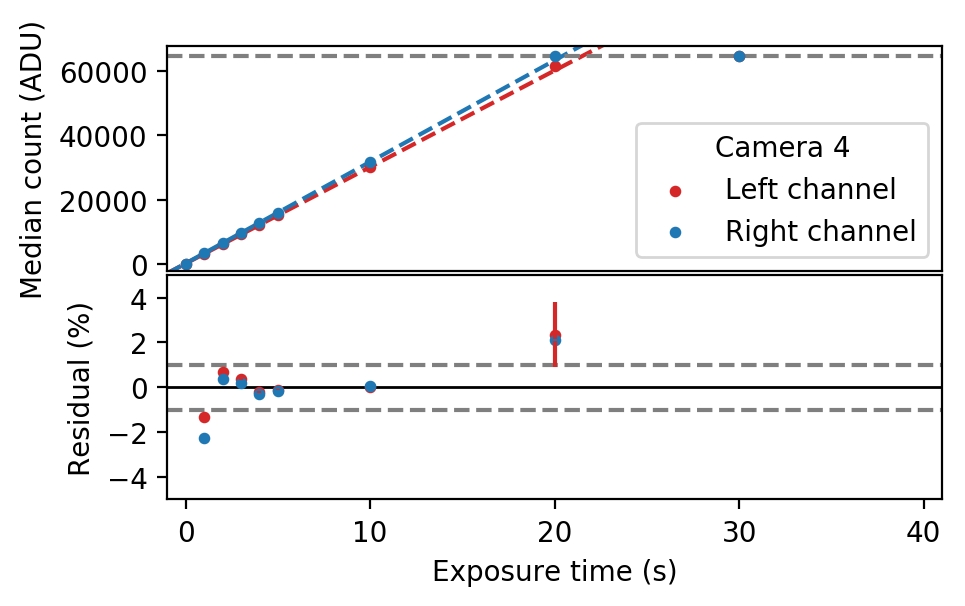}
        \end{minipage}

        \begin{minipage}[t]{0.47\linewidth}\vspace{10pt}
            \includegraphics[width=\linewidth]{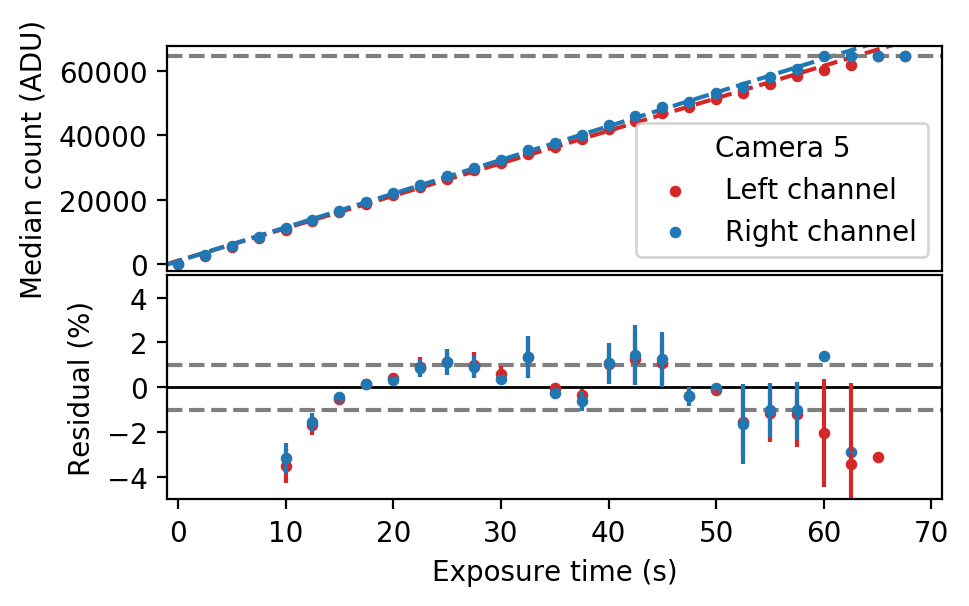}
        \end{minipage}
        \begin{minipage}[t]{0.47\linewidth}\vspace{10pt}
            \includegraphics[width=\linewidth]{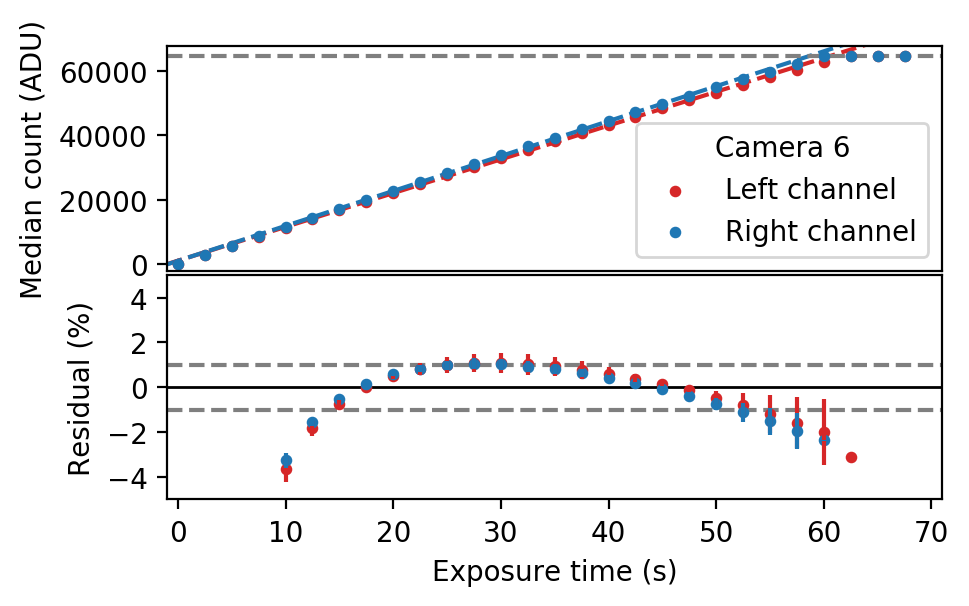}
        \end{minipage}

        \begin{minipage}[t]{0.47\linewidth}\vspace{10pt}
            \includegraphics[width=\linewidth]{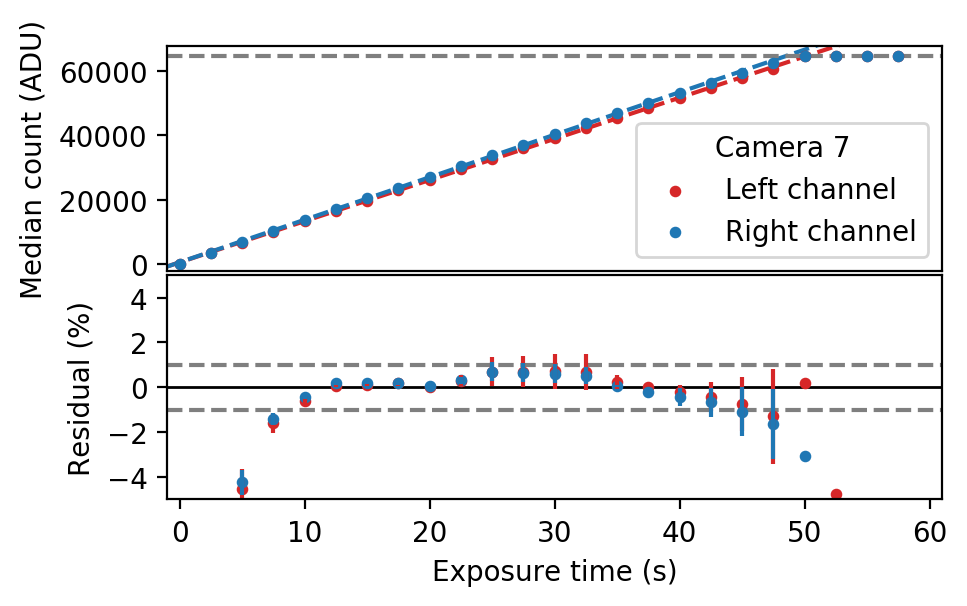}
        \end{minipage}
        \begin{minipage}[t]{0.47\linewidth}\vspace{10pt}
            \includegraphics[width=\linewidth]{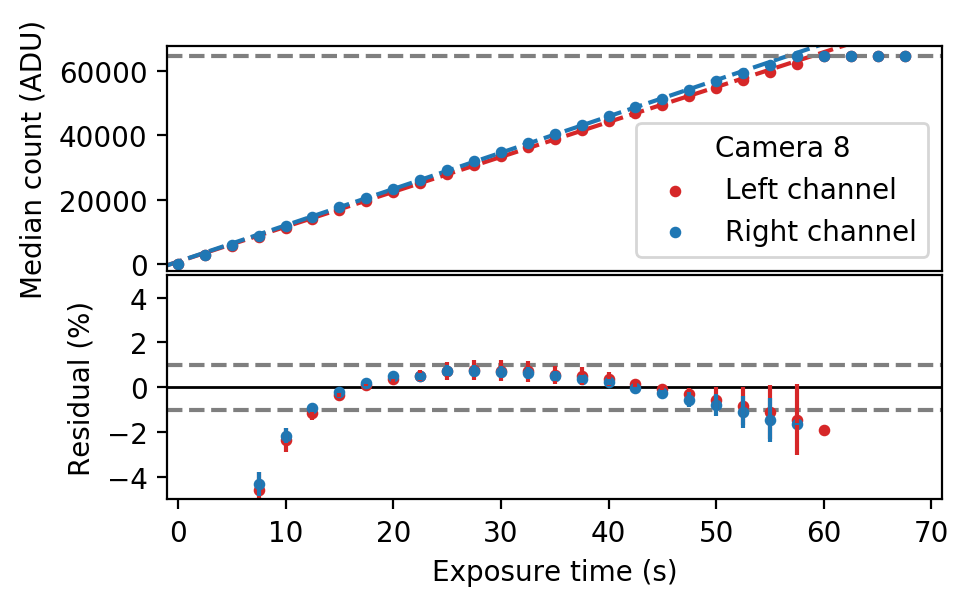}
        \end{minipage}
    \end{center}
    \caption[Linearity plots]{
        Linearity plots for each camera. The horizontal dashed line in the top panel shows the saturation level from \aref{tab:ptc}, and the dashed lines in the lower panels show the target $\pm$1\% non-linearity range.
    }\label{fig:lin}
\end{figure}

\clearpage
\newpage

\end{colsection}

\subsection{Defects}
\label{sec:defects}
\begin{colsection}

There are several possible defects in CCD sensors \citep{CCDs}: hot pixels, which have atypically high dark currents, dead pixels, which produce low or zero counts, and trap pixels, which ``trap'' electrons and prevent read out from it and any pixels above it in the column. It is important to identify any bad pixels so that the GOTOphoto pipeline (see \aref{sec:gotophoto}) can mask them when reducing the images from each camera.

Single hot or dead pixels can be removed to some extent by subtracting dark frames and flat fielding. Trap pixels are more of an issue, as they can potentially take out a large fraction of a column. For each camera, a defect mask was made by taking the ratio of two flat field images with different exposure times, making any bad columns easy to pick out by comparing to the surrounding pixels. An example of a bad column caused by a trap pixel is shown in \aref{fig:itsatrap}. The positions of bad columns for each camera are given in \aref{tab:traps}. The KAF-50100 chip specification gives an allowed limit of less than 20 column defects per device, which the GOTO cameras are well within.

\begin{table}[t]
    \begin{center}
        \begin{tabular}{c|ccc} %
             & \multicolumn{2}{c}{Trap location} & Height \\
             & x & y & \% \\
            \midrule
            Camera 1 & 7751 & 4361 & 30 \\
            Camera 2 & 1658 & ~172 & 97 \\ %
            Camera 3 & 1224 & 1844 & 70 \\
                     & 5058 & 5185 & 17 \\
            Camera 4 & 5406 & 2607 & 58 \\
            Camera 5 & 6293 & 1416 & 77 \\
            Camera 6 & 5455 & 5036 & 19 \\
        \end{tabular}
        \hspace{0.5cm}
        \begin{tabular}{c|ccc} %
            & \multicolumn{2}{c}{Trap location} & Height \\
            & x & y & \% \\
            \midrule
            Camera 7 & 1344 & 3037 & 51 \\
                     & 2326 & 2495 & 60 \\
                     & 2610 & 5688 & ~9 \\ %
                     & 7491 & 5120 & 18 \\
            Camera 8 & 1184 & 3043 & 51 \\
                     & 5659 & 2778 & 55 \\
            \multicolumn{4}{c}{} \\
        \end{tabular}
    \end{center}
    \caption[Locations of bad columns]{
        Locations and extent (as a percentage of the total column height) of bad columns for each camera.
    }\label{tab:traps}
\end{table}

\begin{figure}[p]
    \begin{center}
        \includegraphics[width=\linewidth]{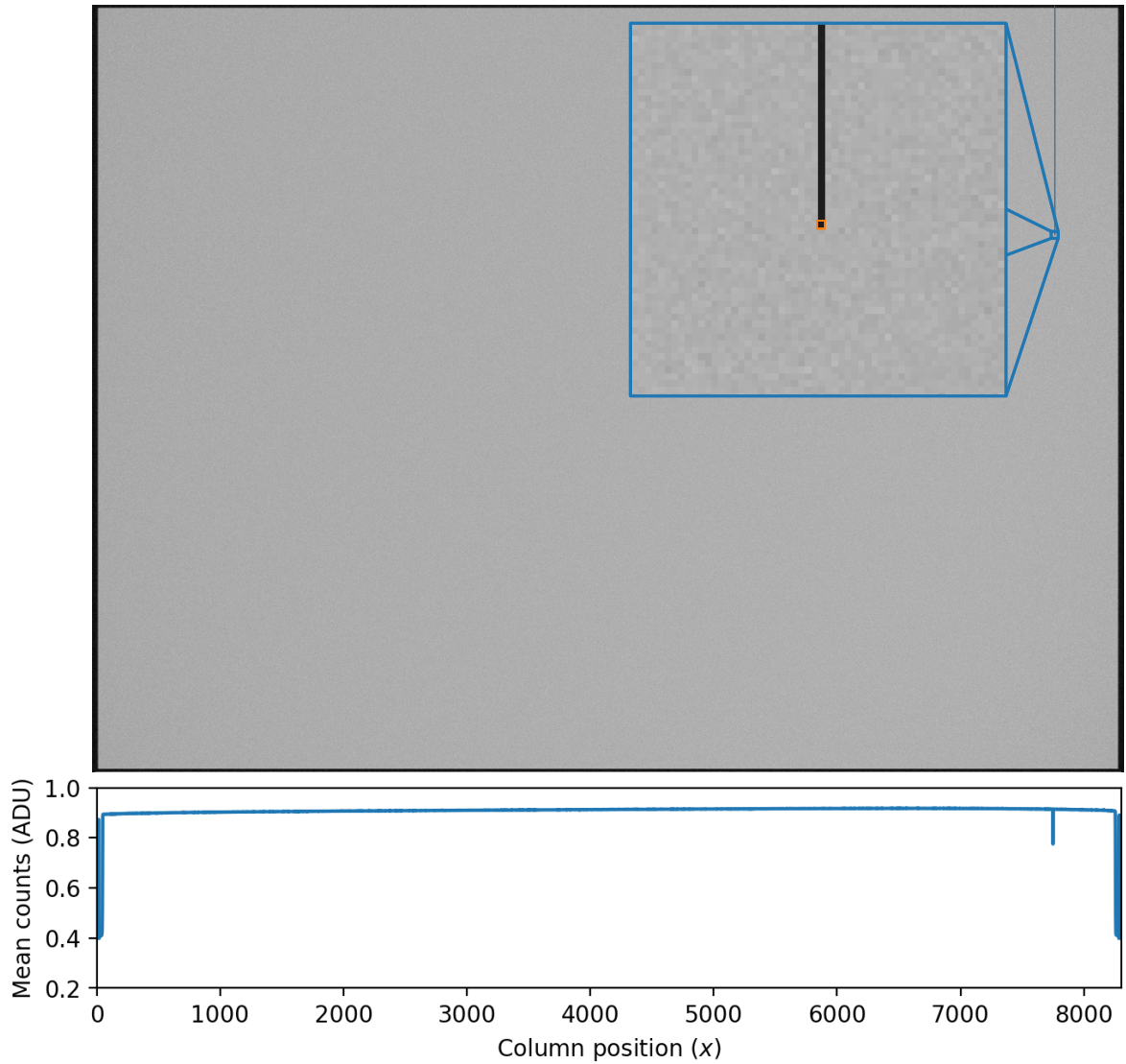}
    \end{center}
    \caption[An example of a column defect]{
        A flat field for Camera 1 showing a bad column. The location of the trap pixel is magnified, showing that the pixels in the column above the trap have been prevented from being read out. The bad column is also also clearly visible in the plot below, which shows the average counts in each column.
    }\label{fig:itsatrap}
\end{figure}

\clearpage

\end{colsection}

\section{System throughput}
\label{sec:throughput}

\begin{colsection}

Unfortunately, not every photon emitted by a target object will be recorded by a telescope: photons will be lost due to absorption and scattering within the telescope's optics and camera, as well as in the Earth's atmosphere (for ground-based telescopes). Understanding each of these factors is required in order to produce a complete throughput model, which can then be compared to the real system (in \aref{sec:onsky_comparison}) to see if the hardware is performing as expected.

\end{colsection}

\subsection{Optical elements}
\label{sec:optics}
\begin{colsection}

As described in \aref{sec:goto_design}, the GOTO unit telescopes are Wynne-Newtonian astrographs: fast (f/2.5) Newtonian telescopes with a \SI{40}{\centi\meter} primary mirror, a flat elliptical secondary (\SI{19}{\centi\metre} short axis) and a Wynne corrector between the secondary and the camera. A drawing of the \acro{ota} is shown in \aref{fig:ota}, and the five elements the light must pass through (the three corrector lenses, the filter in the filter wheel and the window in front of the detector) are shown in \aref{fig:wynne}. In order to model the throughput each element needed to be considered in turn.

\subsubsection{Mirrors}

The GOTO mirrors are were manufactured by Orion Optics\footnote{\url{https://www.orionoptics.co.uk}}. Orion used their own ``HiLux'' high reflectivity aluminium coating, and while individual reflectance curves were not available for the GOTO mirrors at the time this work was carried out, Orion does have a representative curve on their website (shown in \aref{fig:trans_ota}). As there are two mirrors this curve will be included twice in the final throughput model. The difference in the angle of incidence of light on the two mirrors is accounted for in the coating applied to each mirror, so the reflectivity curves of the two are assumed to be identical.

\newpage

\begin{figure}[p]
    \begin{center}
        \includegraphics[width=0.7\linewidth]{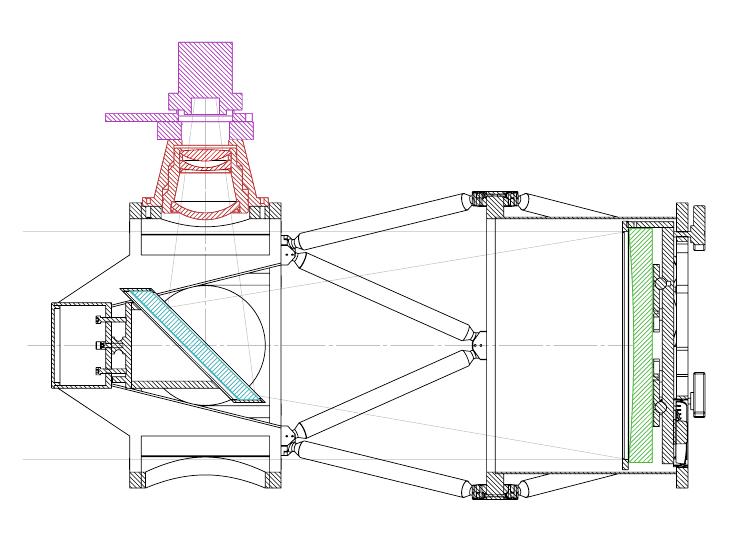}
    \end{center}
    \caption[GOTO optical telescope assembly]{
        The \acro{ota} design for one of the GOTO unit telescopes. Light enters from the left, and relevant elements have been highlighted: the primary mirror in \textcolorbf{Green}{green}, the secondary mirror in \textcolorbf{BlueGreen}{blue}, the Wynne corrector in \textcolorbf{Red}{red} and the FLI camera hardware in \textcolorbf{Purple}{purple}.
    }\label{fig:ota}
\end{figure}

\begin{figure}[p]
    \begin{center}
        \includegraphics[width=0.7\linewidth]{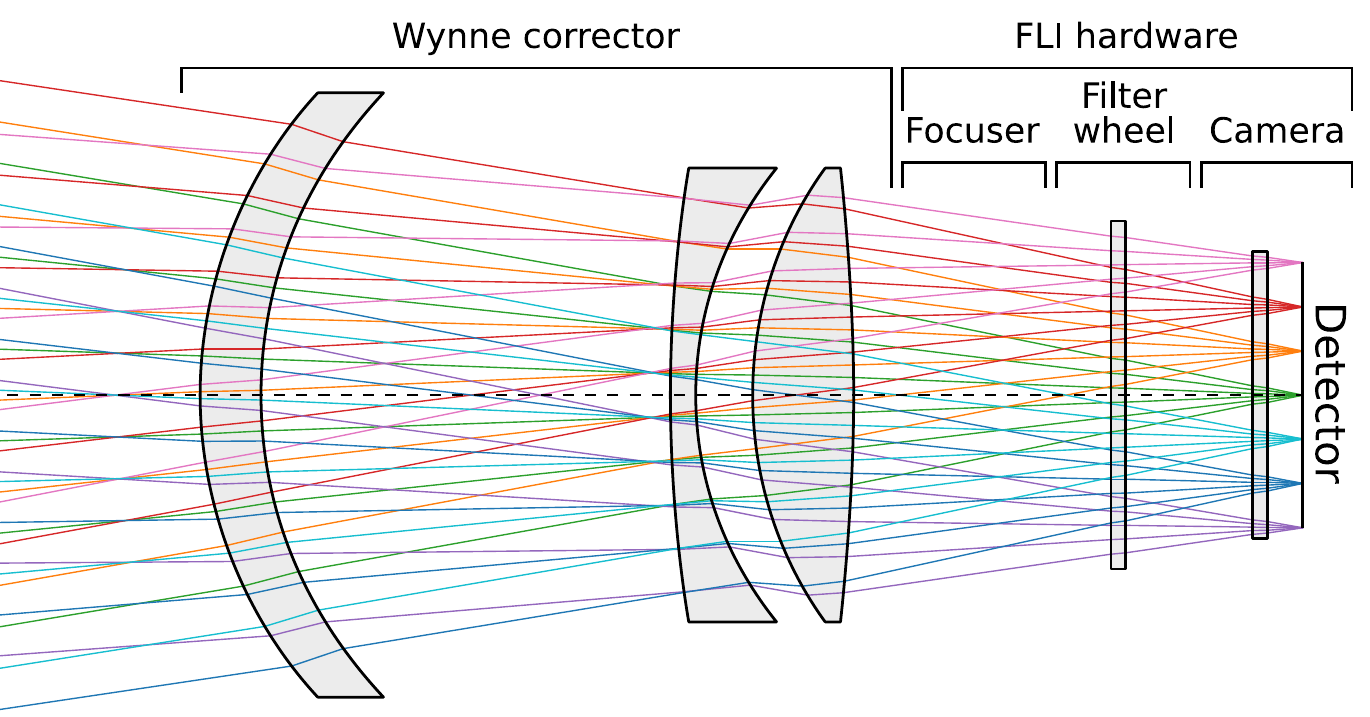}
    \end{center}
    \caption[Ray tracing the corrector elements]{
        A ray trace through the optical elements after the primary and secondary mirrors. From left-to-right light passes through the three Wynne corrector lenses, the filter, and the camera window before reaching the detector located in the focal plane.
    }\label{fig:wynne}
\end{figure}

\clearpage

\subsubsection{Lenses}

\begin{table}[t]
    \begin{center}
        \begin{tabular}{c|ccccc|c} %
                 &       &                        & \multicolumn{2}{c}{Radius of curvature}         & On-axis               & Glass \\
            Lens & Shape & Diameter               & Exterior               & Interior               & thickness             & type \\
            \midrule
            1 & Meniscus & \SI{120}{\milli\metre} &  \SI{89}{\milli\metre} &  \SI{86}{\milli\metre} & \SI{12}{\milli\metre} & H-K9L   \\
            2 & Meniscus &  \SI{90}{\milli\metre} & \SI{278}{\milli\metre} &  \SI{71}{\milli\metre} &  \SI{5}{\milli\metre} & H-K9L   \\
            3 & Biconvex &  \SI{90}{\milli\metre} &  \SI{77}{\milli\metre} & \SI{378}{\milli\metre} & \SI{20}{\milli\metre} & S-FPL53 \\
        \end{tabular}
    \end{center}
    \caption[Wynne corrector lens properties]{
        Properties of the three Wynne corrector lenses.
    }\label{tab:lenses}
\end{table}

Each Wynne corrector contains three lenses, as shown in \aref{fig:wynne} --- the details of each lens are given in \aref{tab:lenses}. No complete transmission data was available, so a model throughput curve had to be created.

For each lens the reflectivity of the front and rear surfaces and the internal transmittance of the glass needs to be considered. Each surface is coated with an anti-reflection coating, the profile of which was included in the GOTO optical report. Transmittance curves for each lens were not available, but the glass types were included in the report and are given in \aref{tab:lenses}. Transmittance data provided by the glass manufacturers were retrieved from the online Refractive Index Database\footnote{\url{https://refractiveindex.info}}. For simplicity, each lens was modelled as having a constant thickness, using their on-axis thickness. As shown in \aref{fig:wynne}, this is a good approximation for lens 1 but will underestimate the absorption within lens 2 and overestimate the absorption within lens 3.

Throughput curves for the anti-reflection coatings and the glass for the three lenses are shown in \aref{fig:trans_lenses} along with the total throughput of the corrector, found by multiplying the contributions from the glass transmission and the coating on both surfaces of each lens:
\begin{equation}
    T_\text{corrector} = T_\text{Lens1} \times
                         T_\text{Lens2} \times
                         T_\text{Lens3} \times
                         {(T_\text{coating})}^6.
    \label{eq:corrector}
\end{equation}

\newpage

\begin{figure}[t]
    \begin{center}
        \includegraphics[width=\linewidth]{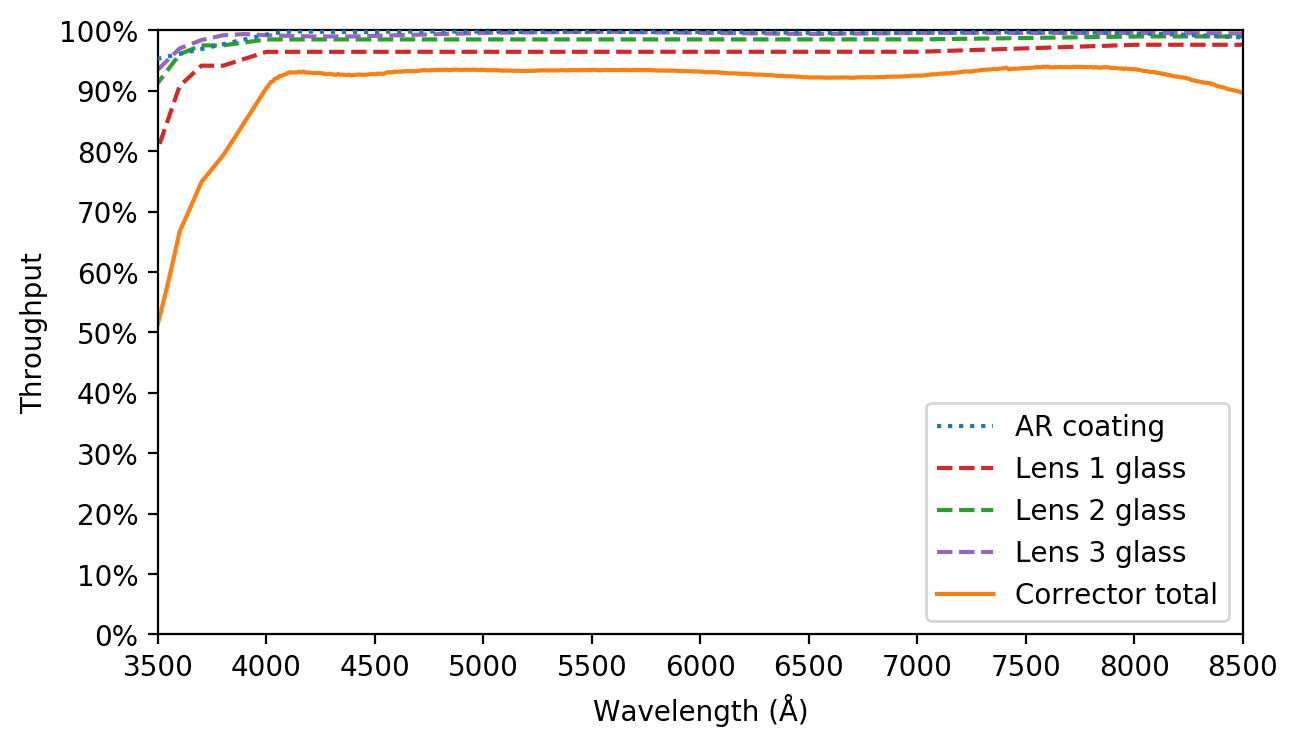}
    \end{center}
    \caption[Wynne corrector transmission curve]{
        Transmission curve for the Wynne corrector (the \textcolorbf{Orange}{orange} solid line), comprised of the glass throughput of each lens (\textcolorbf{Red}{red}, \textcolorbf{Green}{green} and \textcolorbf{Purple}{purple} dashed lines) and an anti-reflection (AR) coating on all six surfaces (\textcolorbf{NavyBlue}{blue} dotted line).
    }\label{fig:trans_lenses}
\end{figure}

\subsubsection{Filters}

The filter transmittance is included in their bandpass profiles, described below in \aref{sec:filters}. At this stage we will consider the OTA with no filter, so as to produce an unfiltered OTA transmission curve which can then by multiplied by the chosen filter bandpass in \aref{sec:total_throughput}.

\subsubsection{Camera window}

Finally, before reaching the detector, light must pass through a glass window in the camera which protects the CCD sensor. The window is made of F116 glass, and a transmission profile was provided by FLI.\@ This is shown in \aref{fig:trans_ota}.

\newpage

\subsubsection{Combined OTA throughput}

The combined throughput for the whole unfiltered OTA is shown in \aref{fig:trans_ota}. This was constructed by multiplying through the transmission curves for the two mirrors, the corrector (from \aref{eq:corrector}) and the camera window:
\begin{equation}
    T_\text{OTA} = {(T_\text{mirror})}^2 \times T_\text{corrector} \times T_\text{window}.
    \label{eq:ota}
\end{equation}
In the 4000--\SI{7000}{\angstrom} visible region used by GOTO the throughput is typically 60\% or above, although all the elements have a sharp cut-off towards the blue.

\begin{figure}[t]
    \begin{center}
        \includegraphics[width=\linewidth]{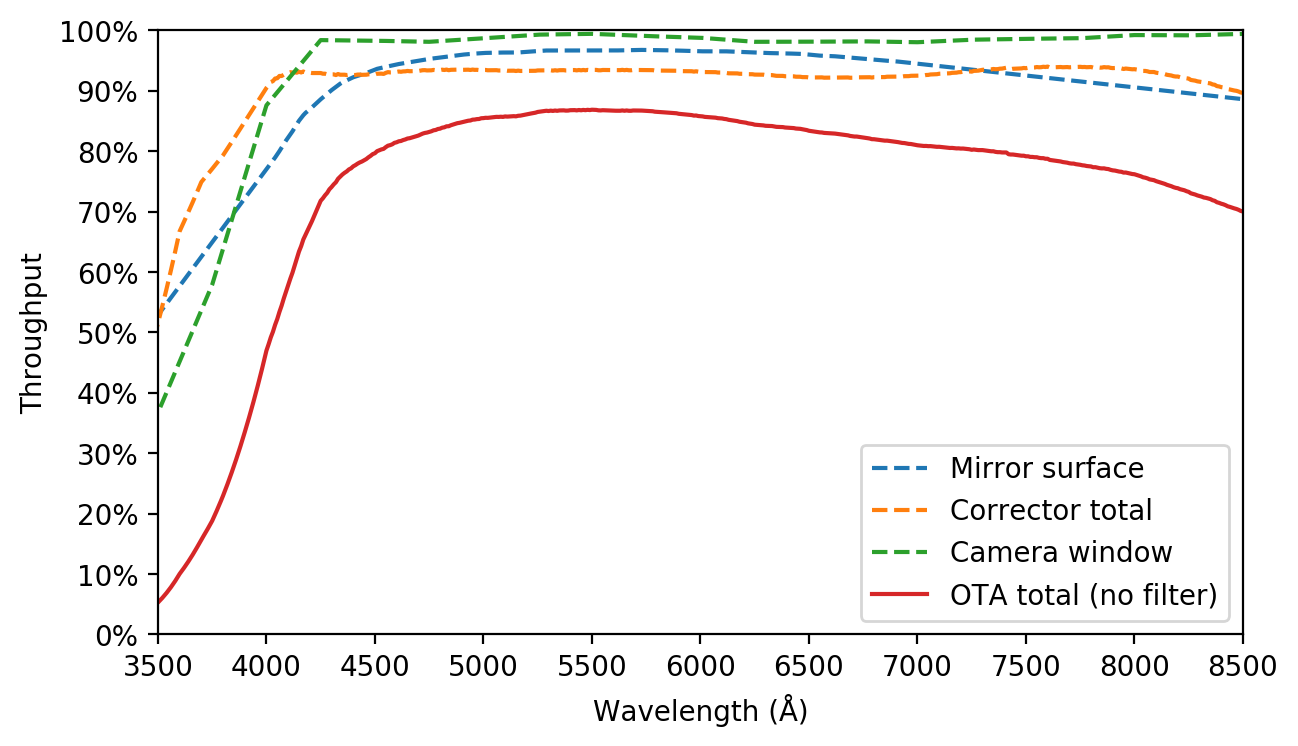}
    \end{center}
    \caption[Combined OTA transmission curve]{
        Transmission curve for the unfiltered OTA (the \textcolorbf{Red}{red} solid line), which includes the two mirrors (\textcolorbf{NavyBlue}{blue} dashed line), the combination of all three corrector lenses (\textcolorbf{Orange}{orange} dashed line, from \aref{fig:trans_lenses}) and the camera window (\textcolorbf{Green}{green} dashed line).
    }\label{fig:trans_ota}
\end{figure}

\newpage

\end{colsection}

\subsection{Filters}
\label{sec:filters}
\begin{colsection}

\begin{figure}[t]
    \begin{center}
        \includegraphics[width=\linewidth]{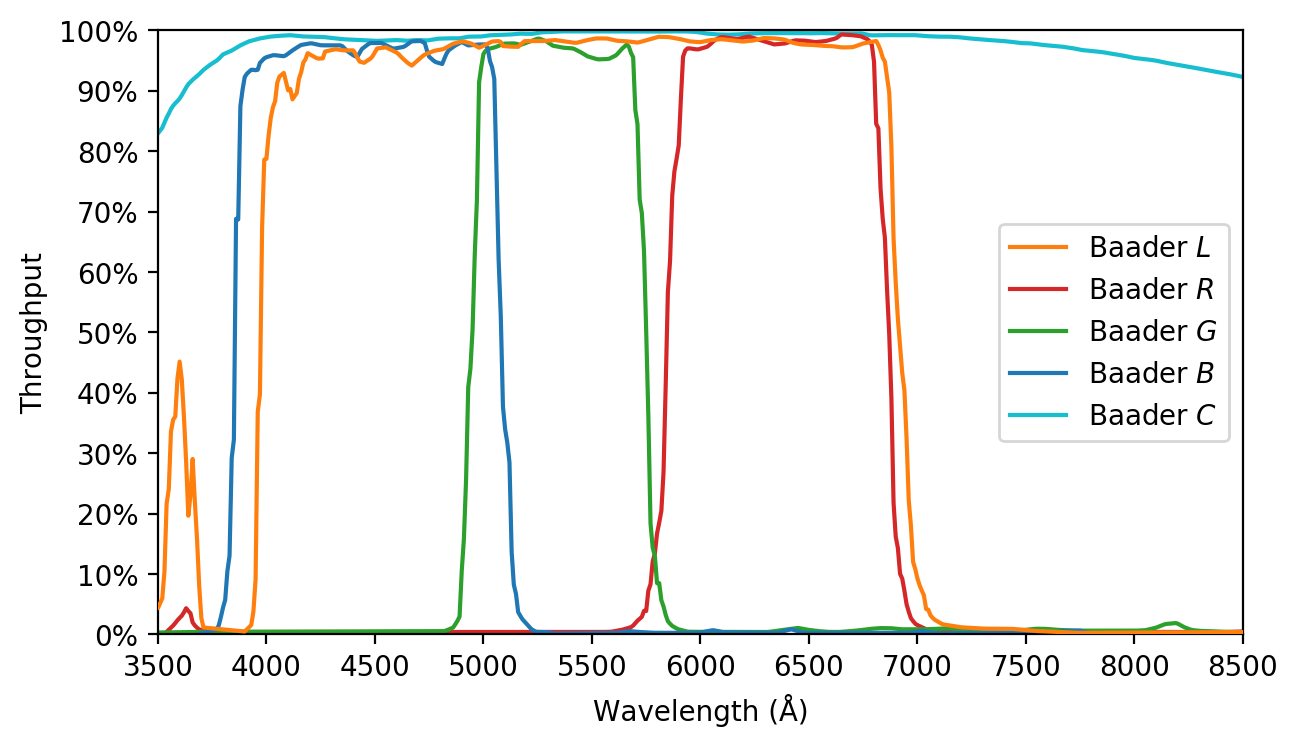}
    \end{center}
    \caption[Baader filter transmission curves]{
        Transmission curves for the Baader \textit{LRGBC} filter set used by GOTO.\
    }\label{fig:filters}
\end{figure}

Each GOTO unit telescope has a five-slot filter wheel containing a set of \SI{65}{\milli\metre} square filters from Baader Planetarium\footnote{\url{https://www.baader-planetarium.com}}: three coloured filters (\textit{R}, \textit{G}, \textit{B}), one wide ``luminance'' filter (\textit{L}) covering the whole visible range, and a clear glass filter ({\textit{C}}). Transmission curves for each filter are shown in \aref{fig:filters}. Each filter has a high throughput and steep cut-offs outside of the desired bandpasses. For the coloured filters the cut-offs were chosen so the [O\textsc{iii}] $\lambda 5007$ emission line falls within the overlap of the \textit{B} and \textit{G} filters and the region around \SI{5800}{\angstrom}, which contains emission lines from Mercury and Sodium vapour lamps, is excluded by the gap between the \textit{G} and \textit{R} filters.

Most GOTO observations are taken using the \textit{L} filter, however the \textit{RGB} filters have been used for manual follow-up observations. The clear filter is never used for scientific observations, so it is not considered as part of the throughput model going forward.

\begin{table}[t]
    \begin{center}
        \begin{tabular}{c|cccc} %
             & Effective wavelength & Effective bandwidth\\
            Filter & ($\lambda_\text{eff}$, \SI{}{\angstrom}) & ($\Delta\lambda$, \SI{}{\angstrom}) \\
            \midrule
            Baader \textit{L} & 5355 & 2942 \\
            Baader \textit{R} & 6573 &  979 \\
            Baader \textit{G} & 5373 &  813 \\
            Baader \textit{B} & 4509 & 1188 \\
        \end{tabular}
    \end{center}
    \caption[Baader filter properties]{
        Properties of the Baader \textit{LRGB} filters.
    }\label{tab:filters}
\end{table}

Properties of the \textit{LRGB} filters are given in \aref{tab:filters}. The effective wavelength ($\lambda_\text{eff}$) is the pivot wavelength as defined in \citet{HST_calibration} for HST filters:
\begin{equation}
    \lambda_\text{eff}^2 = \frac{\int T\lambda~d\lambda}{\int T/\lambda~d\lambda},
    \label{eq:pivot_wavelength}
\end{equation}
where $T$ is the transmission integrated over all wavelengths $\lambda$. The effective bandwidth ($\Delta\lambda$) is found by calculating the equivalent width, the width of a rectangle that has a height equal to the maximum transmission (unity) and the same area as the area under the filter transmission curve, i.e.
\begin{equation}
    \Delta\lambda = \int T~d\lambda.
    \label{eq:bandwidth}
\end{equation}

The Baader filters were designed for amateur astronomers and astro-photographers, and are less commonly used by professional instruments than other sets, such as the \textit{u'g'r'i'z'} set used by the Sloan Digital Sky Survey \acroadd{sdss} \citep{Sloan_filters} or the traditional Johnson-Cousins \textit{UBVRI} set redefined by \citet{Bessell_filters}. GOTO primarily uses Baader filters to reduce costs, as each unit telescope requires a full set. A comparison of the Baader \textit{LRGB} transmission curves with Sloan are shown in \aref{fig:filter_comparison1} and with Bessell \aref{fig:filter_comparison2}. The Baader \textit{L} filter approximately covers the Sloan \textit{g'} and \textit{r'} filters, the \textit{B} and \textit{G} filters cover \textit{g'} and \textit{R} roughly matches \textit{r'}. Colour terms to compare GOTO \textit{RGB} observations with Sloan \textit{g'} and \textit{r'} observations were calculated by \citet{Phaethon}.

\newpage

\makeatletter
\setlength{\@fptop}{0pt}
\makeatother

\begin{figure}[t]
    \begin{center}
        \includegraphics[width=\linewidth]{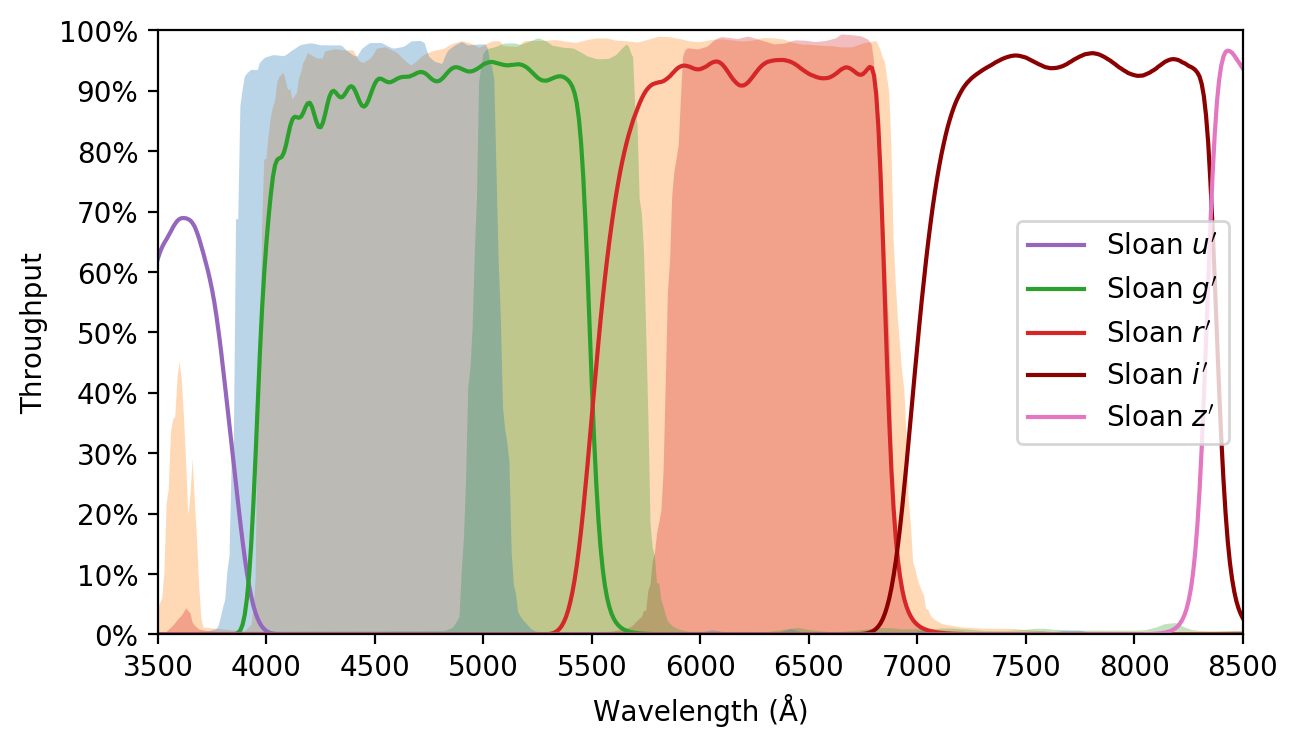}
    \end{center}
    \caption[Comparison of Baader and Sloan filters]{
        A comparison of the Baader \textit{LRGB} filters to the Sloan \textit{u'g'r'i'z'} set.
    }\label{fig:filter_comparison1}
\end{figure}

\begin{figure}[t]
    \begin{center}
        \includegraphics[width=\linewidth]{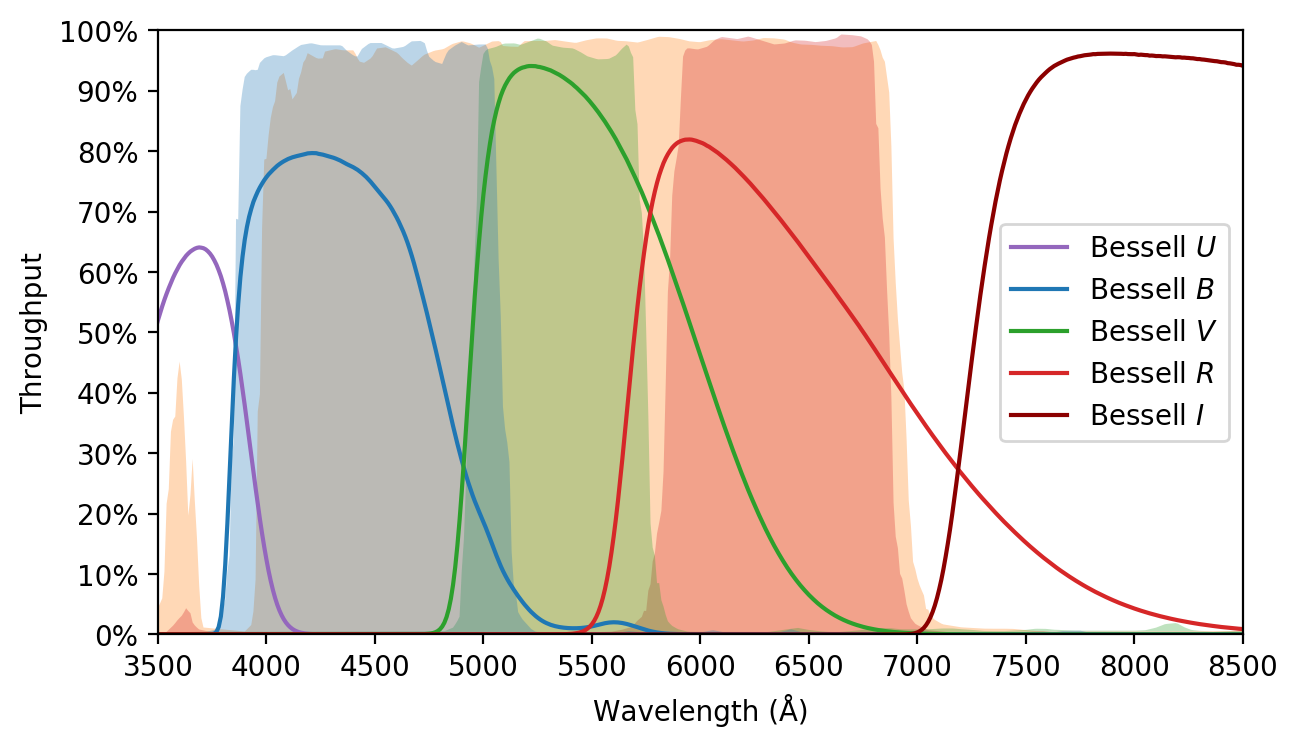}
    \end{center}
    \caption[Comparison of Baader and Bessell filters]{
        A comparison of the Baader \textit{LRGB} filters to the Bessell \textit{UBVRI} set.
    }\label{fig:filter_comparison2}
\end{figure}

\clearpage

\makeatletter
\setlength{\@fptop}{0\p@ \@plus 1fil} %
\makeatother

\newpage

\end{colsection}

\subsection{Quantum efficiency}
\label{sec:qe}
\begin{colsection}

\begin{figure}[t]
    \begin{center}
        \includegraphics[width=\linewidth]{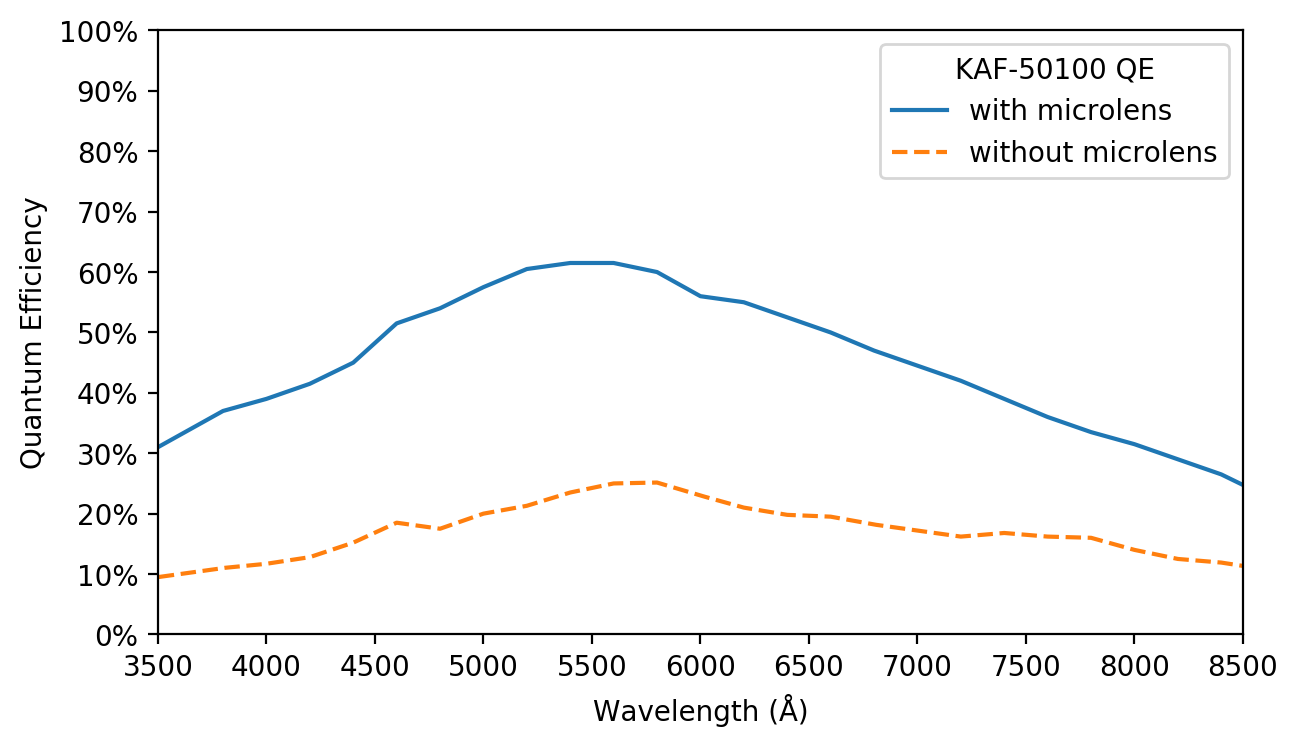}
    \end{center}
    \caption[CCD quantum efficiency curve]{
        QE curve for the KAF-50100 CCDs, both with and without microlensing.
    }\label{fig:qe}
\end{figure}

After passing through the telescope photons are focused onto the CCD, where they interact with the photosensitive layer and produce electrons which are recorded by the detector \citep{CCDs}. The conversion from photons to electrons is the \acro{qe} of the CCD, and is dependent on wavelength: short-wavelength photons will be absorbed before reaching the photosensitive layer, while long-wavelength photons will not have enough energy to create free electrons in the silicon. CCDs that are back-side illuminated have improved blue QE due to the photons not having to pass through the electrode layer and therefore having less chance of being absorbed, however these are more complicated and expensive to build. The QE can also change with temperature in the near IR, but this is negligible in the optical. The QE curve for the KAF-50100 CCDs is shown in \aref{fig:qe}. As described in \aref{sec:chip_layout}, these CCDs are front-illuminated, and include a microlens array in front of the sensor to improve the QE.\@

\end{colsection}

\subsection{Total throughput}
\label{sec:total_throughput}
\begin{colsection}

\begin{figure}[t]
    \begin{center}
        \includegraphics[width=\linewidth]{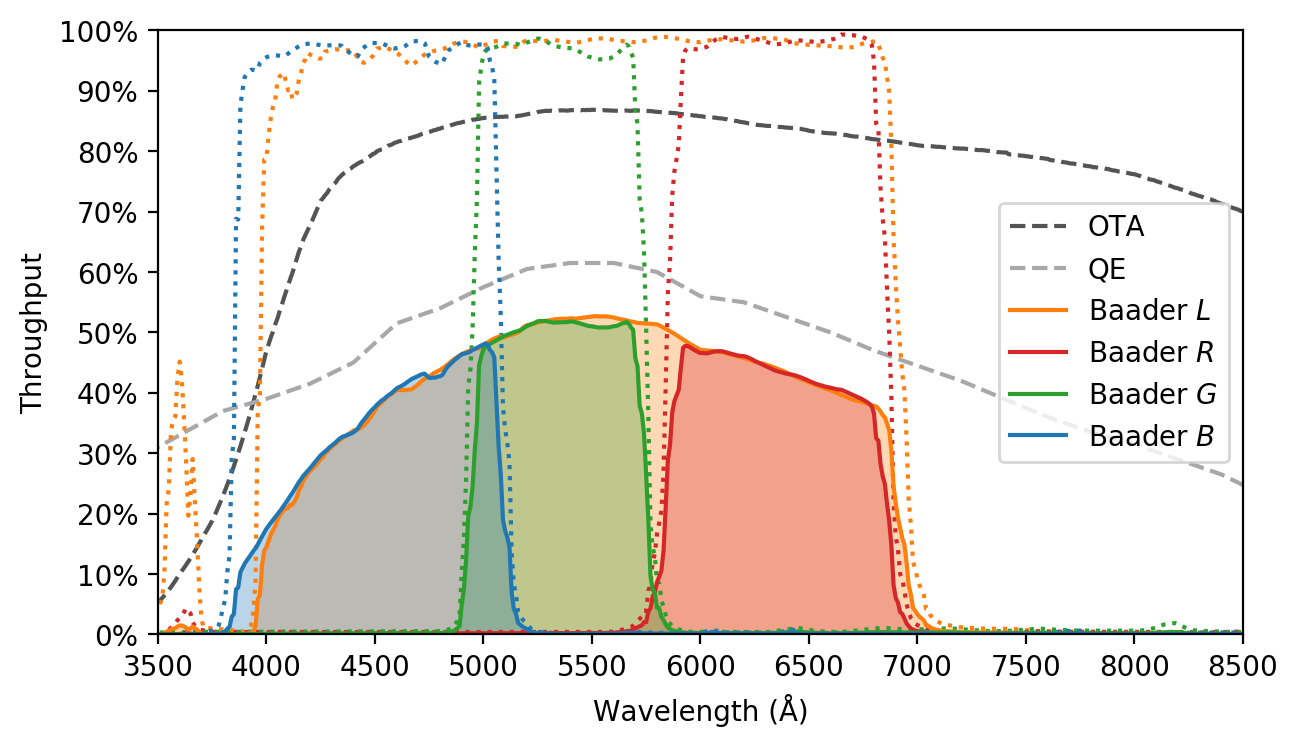}
    \end{center}
    \caption[Complete throughput model for the GOTO filters]{
        The complete GOTO throughput model. The model elements (dashed lines) are the combined throughput of the OTA elements (from \aref{fig:trans_ota}) and the quantum efficiency of the microlensed CCD (from \aref{fig:qe}); and the Baader \textit{LRGB} filter bandpasses (from \aref{fig:filters}) are shown by the coloured dotted lines. The filled lines shown the total throughput in each filter when the model is applied.
    }\label{fig:throughput}
\end{figure}

The complete GOTO throughput is a combination of all of the elements discussed in the previous sections. Each source profile was linearly interpolated to the same wavelength range (3500--\SI{8500}{\angstrom}) and multiplied together to produce the total GOTO throughput model, shown in \aref{fig:throughput}. Since the quantum efficiency has been included, the total throughput describes the conversion between photons to electrons detected in the CCD, and using the gain values given in \aref{tab:ptc} the full conversion between photons and output counts can be made (this does not include photons lost to extinction in the atmosphere, see \aref{sec:atmosphere}). The mean throughput in each filter can be found by dividing the filled areas in \aref{fig:throughput} by the area of the filter bandpass, and are given in electrons per photon in \aref{tab:throughput_extinction}.

\newpage

\end{colsection}

\subsection{Atmospheric extinction}
\label{sec:atmosphere}
\begin{colsection}

\begin{figure}[t]
    \begin{center}
        \includegraphics[width=\linewidth]{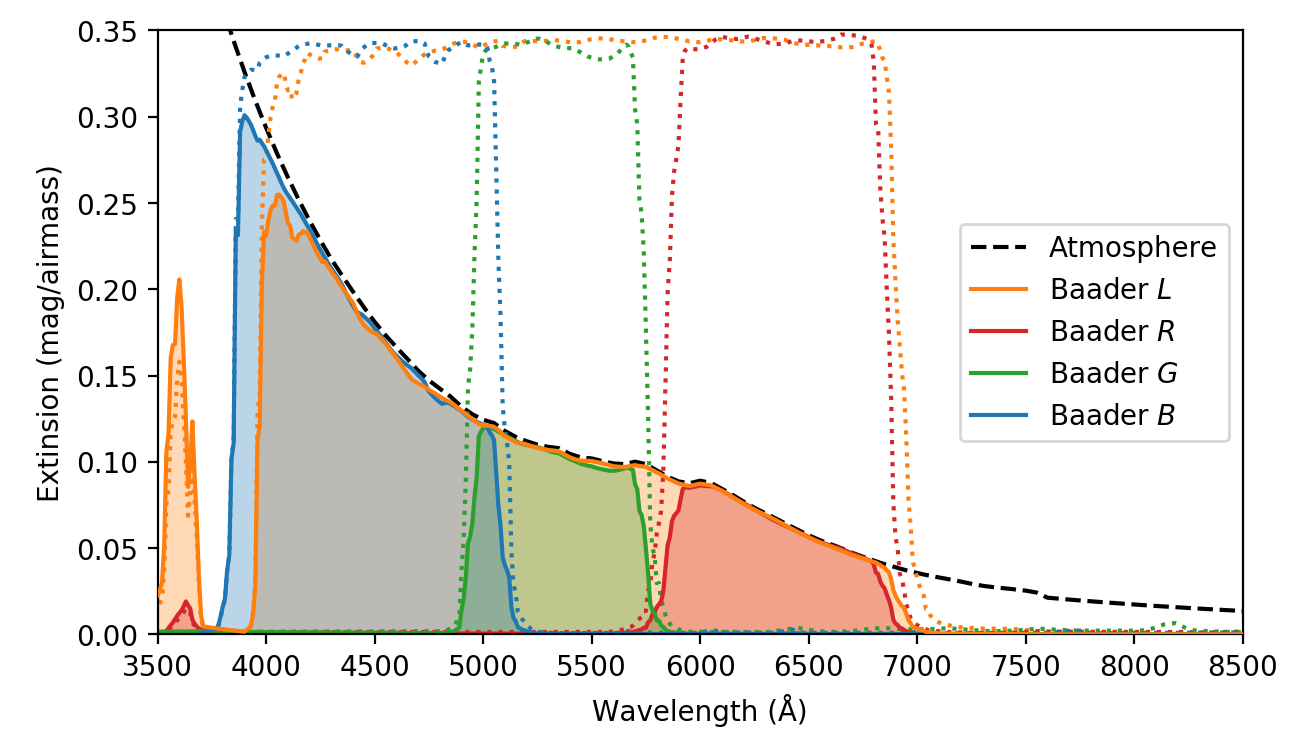}
    \end{center}
    \caption[Atmospheric extinction in the GOTO filters]{
        Atmospheric extinction in the GOTO filters, in magnitude per airmass. The measured extinction curve from \citet{tn31} is shown by the black dashed line, and the filled lines show the extinction curve multiplied by the filter bandpasses from \aref{fig:filters} (shown by the coloured dotted lines).
    }\label{fig:extinction}
\end{figure}

In order to model the entire light path from an astronomical source through to the CCD detector, the absorption of light by the Earth's atmosphere must also be considered. The atmosphere is close to transparent over most of the visible region, however losses due to Rayleigh scattering begin to dominate closer to the UV \citep{atmosphere}.

The amount of light lost due to absorption and scattering in the atmosphere will depend on the altitude of the source, as light from sources closer to the horizon will pass through a thicker layer of atmosphere. The extinction of the atmosphere above La Palma has been measured in terms of magnitude per source airmass by \citet{tn31}, and is shown in \aref{fig:extinction}.

\newpage

Atmospheric extinction can be treated like the throughput elements considered in \aref{sec:total_throughput}, except it is not included in the throughput model as it is not a function of the GOTO hardware. For the \textit{LRGB} filters, the mean extinction can be found by multiplying the extinction curve by the filter bandpasses, as shown in \aref{fig:extinction}, and then dividing the filled areas under each curve by the area of the unmodified filter bandpass. These extinction coefficients are given in magnitudes per airmass in \aref{tab:throughput_extinction}.

Another atmospheric factor unique to observing from La Palma is the \textit{calima}, large quantities of dust from the Sahara Desert which can be carried over the site by easterly winds. The calima occurs most often in the summer, and analysis of dust-affected images has shown that the additional extinction is not wavelength dependent \citep{ORM_dust}. The extinction curve measured by \citet{tn31} was based on observations taken on dust-free nights, and the values in \aref{tab:throughput_extinction} do not include any additional extinction to model the effects of the calima. Based on analysing archival images over 20 years, \citet{ORM_dust} suggests heavy calima could increase the extinction by up to 0.04 mag/airmass.

\begin{table}[t]
    \begin{center}
        \begin{tabular}{c|cc} %
                   & Throughput     & Extinction \\
            Filter & (\elec/photon) & (mag/airmass) \\
            \midrule
            Baader \textit{L} & 0.43 & 0.13 \\
            Baader \textit{R} & 0.46 & 0.07 \\
            Baader \textit{G} & 0.51 & 0.11 \\
            Baader \textit{B} & 0.36 & 0.20 \\
        \end{tabular}
    \end{center}
    \caption[Theoretical throughput and extinction coefficients for the GOTO filters]{
        Theoretical throughputs and extinction coefficients for the GOTO filters.
    }\label{tab:throughput_extinction}
\end{table}

\end{colsection}

\section{Photometric modelling}
\label{sec:photometry}

\begin{colsection}

Using the throughput model created in \aref{sec:throughput} it was possible to simulate photometric observations with GOTO before the telescope was commissioned. This section applies the theoretical throughput model to two important photometric properties: the magnitude zeropoint (the correction required to convert between instrumental and calibrated magnitude values) and the limiting magnitude (the faintest magnitude a source can be to still produce a detectable signal above a given noise threshold). These theoretical values are then compared to values calculated from real GOTO observations, in order to check that the hardware is performing to specification.

\end{colsection}

\subsection{Magnitude zeropoints}
\label{sec:zeropoints}
\begin{colsection}

The flux of a source, $F$, is related to its magnitude, $m$, by
\begin{equation}
    m = -2.5 \log_{10}(F).
    \label{eq:apparent_magnitude}
\end{equation}
In practice, magnitudes are usually measured relative to a reference star using
\begin{equation}
    m - m_\text{ref} = -2.5 \log_{10}\left(\frac{F}{F_\text{ref}}\right),
    \label{eq:magnitude_ref}
\end{equation}
which requires a reference star of known magnitude $m_\text{ref}$ and flux $F_\text{ref}$. Traditionally Vega is used as a reference star as it has a magnitude of very close to 0.

The instrumental magnitude measured from an image is related to the number of photo-electrons recorded, $N$, using the same magnitude definition
\begin{equation}
    \begin{split}
        m_\text{ins} & = -2.5 \log_{10}(N/t),
    \end{split}
    \label{eq:ins_mag}
\end{equation}
where $t$ is the exposure time. The number of photo-electrons recorded per second $N/t$ from a given source should be proportional to the source flux $F$ (assuming the camera has a low non-linearity, see \aref{sec:lin}). Relating the two through a constant $\kappa$ \aref{eq:ins_mag} becomes
\begin{equation}
    \begin{split}
        m_\text{ins} & = -2.5 \log_{10}\left(\kappa F\right) \\
                     & = -2.5 \log_{10}\left(F\right) - m_\text{ZP}    \\
                     & = m - m_\text{ZP},
    \end{split}
    \label{eq:ins_mag2}
\end{equation}
where the constant $m_\text{ZP}$ is defined as the instrumental \emph{zeropoint}.

The zeropoint is so called because observing an object with a true magnitude equal to the zeropoint ($m = m_\text{ZP}$) will produce an instrumental magnitude of 0, which corresponds to one electron per second on the detector. The zeropoint is usually defined based on the electron rate that would be measured above the atmosphere, which allows zeropoints to be compared between telescopes (i.e.\ not including an atmospheric profile as discussed in \aref{sec:atmosphere}). Each telescope and filter combination will have a unique zeropoint, and once determined it can be used to convert instrumental magnitudes measured using that telescope to a calibrated magnitude using
\begin{equation}
    m = m_\text{ins} + m_\text{ZP}.
    \label{eq:zp}
\end{equation}
Therefore, were it possible to observe a star with $m=0$ (without saturating the detector) the zeropoint can be calculated as
\begin{equation}
    \begin{split}
        m_\text{ZP} & = 0 - m_\text{ins} \\
                    & = 2.5 \log_{10}(N/t).
    \end{split}
    \label{eq:zp2}
\end{equation}

\newpage

\end{colsection}

\subsection{Calculating theoretical zeropoints}
\label{sec:model_zeropoints}
\begin{colsection}

Consider taking an observation of a zero magnitude star, such as Vega. From \aref{eq:zp}, the instrumental magnitude will be equal to the negative zeropoint. In the AB magnitude system a zero magnitude star has a fixed flux density $F_\nu = $ \SI{3631}{\jansky} \citep{Sloan_filters}. Therefore, passing this flux through the throughput model for each filter created in \aref{sec:throughput} will produce a predicted signal in photo-electrons, which can be used to calculate a theoretical zeropoint.

\subsubsection{Estimating predicted counts from a 0 mag star}

First, the zero-magnitude flux density needs to be converted into a flux in photons. \SI{3631}{\jansky} is equal to \SI{3.631e-20}{\erg\per\second\per\centi\metre\squared\per\hertz}. To convert from $F_\nu$ to $F_\lambda$ this needs to be multiplied by a factor of $c/\lambda_\text{eff}^2$, where $c$ is the speed of light and $\lambda_\text{eff}$ is the effective wavelength of the photon, in this case the effective wavelength of the filter in question\footnote{The $c/\lambda_\text{eff}^2$ conversion factor comes from differentiating the relationship $\nu = c/\lambda$.}. This will then give a flux in \si{\erg\per\second\per\centi\metre\squared\per\angstrom}, but to convert to a photon count it needs to be divided by the energy of each photon given by
\begin{equation}
    \begin{split}
        E_\lambda = \frac{hc}{\lambda_\text{eff}},
    \end{split}
    \label{eq:photon_energy}
\end{equation}
where $h$ is Planck's constant. Again, at this stage it is assumed that all of the photons have the effective wavelength of the filter. Therefore, the expected flux in photons from a 0 magnitude star is given by
\begin{equation}
    \begin{split}
        F_\lambda = 5.5 \times 10^{6}/\lambda_\text{eff}~\si{\photon\per\second\per\centi\metre\squared\per\angstrom}
    \end{split}
    \label{eq:zero-mag_photons}
\end{equation}
where $\lambda_\text{eff}$ is given in Angstroms.

\newpage

\begin{table}[t]
    \begin{center}
        \begin{tabular}{c|cc|c} %
                   & \multicolumn{2}{c|}{Zero-magnitude star} & \\
            Filter & flux       & predicted signal            & Zeropoint \\
                   & (photon/s) & (\elec/s)                   & (mag) \\
            \midrule
            Baader \textit{L} & \num{3.41e9} & \num{1.26e+09} & 22.75 \\
            Baader \textit{R} & \num{9.24e8} & \num{3.67e+08} & 21.41 \\
            Baader \textit{G} & \num{9.38e8} & \num{4.14e+08} & 21.54 \\
            Baader \textit{B} & \num{1.63e9} & \num{5.04e+08} & 21.76 \\
        \end{tabular}
    \end{center}
    \caption[Theoretical zeropoints for each of the GOTO filters]{
        The flux from a zero-magnitude star in each of the GOTO filters, along with the predicted signal and corresponding theoretical zeropoint found using the throughput model from \aref{sec:throughput}.
    }\label{tab:zeropoints}
\end{table}

Multiplying the value in \aref{eq:zero-mag_photons} by the effective filter bandwidth (in \si{\angstrom}) and the collecting area of the telescope (in \si{\centi\metre\squared}) will give the predicted photon flux in the detector. Each of GOTO's unit telescopes has a \SI{40}{\centi\metre} diameter primary mirror, with an area of \SI{1257}{\centi\metre\squared}. However not all of this is available to collect photons due to the shadow cast by the secondary mirror. The secondary mirror measures \SI{19}{\centi\metre} on its short axis (see \aref{sec:optics}), modelling this as circular gives the GOTO unit telescopes an effective collecting area of \SI{973}{\centi\metre\squared} --- meaning approximately 23\% of light is blocked. Using this area and the filter bandwidths given in \aref{tab:filters}, the theoretical flux in photons per second expected above the atmosphere from a zero-magnitude star can be calculated for each filter. These values are given in \aref{tab:zeropoints}.

Finally, multiplying these theoretical fluxes by the mean throughput values for each filter from \aref{tab:throughput_extinction} gives the predicted signal on the detector in photo-electrons per second (again still excluding atmospheric extinction), and using \aref{eq:zp2} gives the theoretical zeropoint. The predicted signals and zeropoints are also given in \aref{tab:zeropoints}.

\newpage

\subsubsection{Modeling observations with pysynphot}

The above method is the typical way to calculate a theoretical zeropoint, but it only approximates the bandpasses of each filter by using the effective wavelength and bandwidth, and only considers the mean throughput instead of over the whole bandwidth. To account for the full bandpass a more robust model was created using the pysynphot Python package (Python Synthetic Photometry, \pkg{pysynphot}\footnote{\url{https://pysynphot.readthedocs.io}}), which is based the IRAF \pkg{SYNPHOT} package\footnote{\url{http://www.stsci.edu/institute/software_hardware/stsdas/synphot}}. Each of the throughput elements described in \aref{sec:throughput} were imported to create throughput profiles for each filter, and observations were simulated of a flat spectrum of \SI{3631}{\jansky} (0 mag in the AB system) and the built-in Vega spectrum (0 mag in the Vega system) by multiplying the spectra with the bandpasses. The resulting spectra are shown in \aref{fig:pysynphot}; the area under each curve gives the predicted number of electrons produced by the zero-magnitude star in each system.

The predicted signals found using pysynphot, and the derived zeropoints, are given in \aref{tab:pysynphot_zeropoints}. The difference between the two photometric systems is visible, the AB spectrum gives more electrons in the red filter while the Vega spectrum is brighter in the blue (as shown in \aref{fig:pysynphot}).

\begin{table}[t]
    \begin{center}
        \begin{tabular}{c|cc|cc} %
                   & \multicolumn{2}{c|}{AB system} & \multicolumn{2}{c}{Vega system}\\
            Filter & Signal    & Zeropoint & Signal    & Zeropoint\\
                   & (\elec/s) & (mag)     & (\elec/s) & (mag) \\
            \midrule
            Baader \textit{L} & \num{1.25e+09} & 22.74 & \num{1.23e+09} & 22.72 \\
            Baader \textit{R} & \num{3.84e+08} & 21.46 & \num{3.25e+08} & 21.28 \\
            Baader \textit{G} & \num{4.19e+08} & 21.55 & \num{4.21e+08} & 21.56 \\
            Baader \textit{B} & \num{4.98e+08} & 21.74 & \num{5.47e+08} & 21.84 \\
        \end{tabular}
    \end{center}
    \caption[Zeropoints in the AB and Vega systems calculated using pysynphot]{
        Zeropoints in the AB and Vega systems calculated using pysynphot.
    }\label{tab:pysynphot_zeropoints}
\end{table}

\newpage

\makeatletter
\setlength{\@fptop}{0pt}
\makeatother

\begin{figure}[p]
    \begin{center}
        \includegraphics[width=\linewidth]{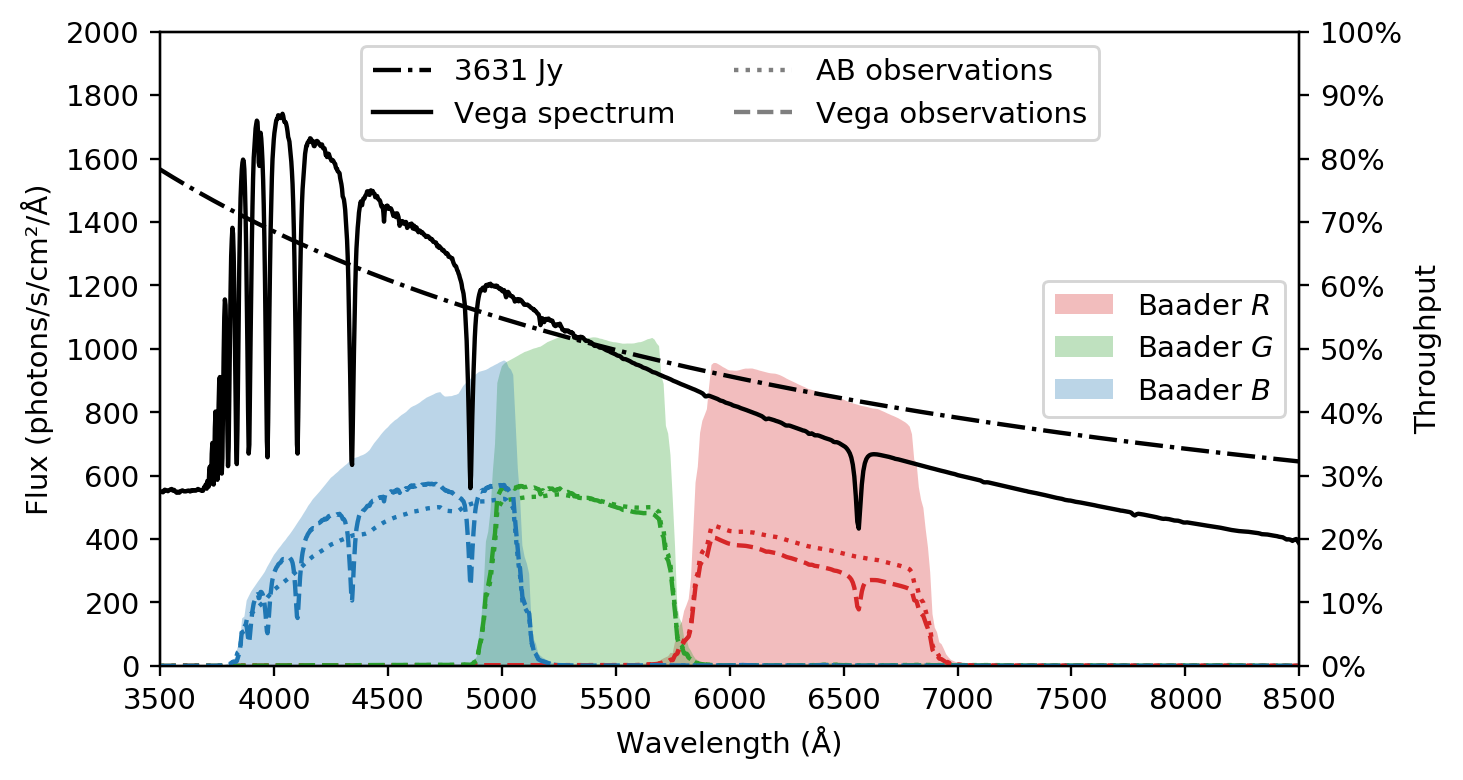}
        \includegraphics[width=\linewidth]{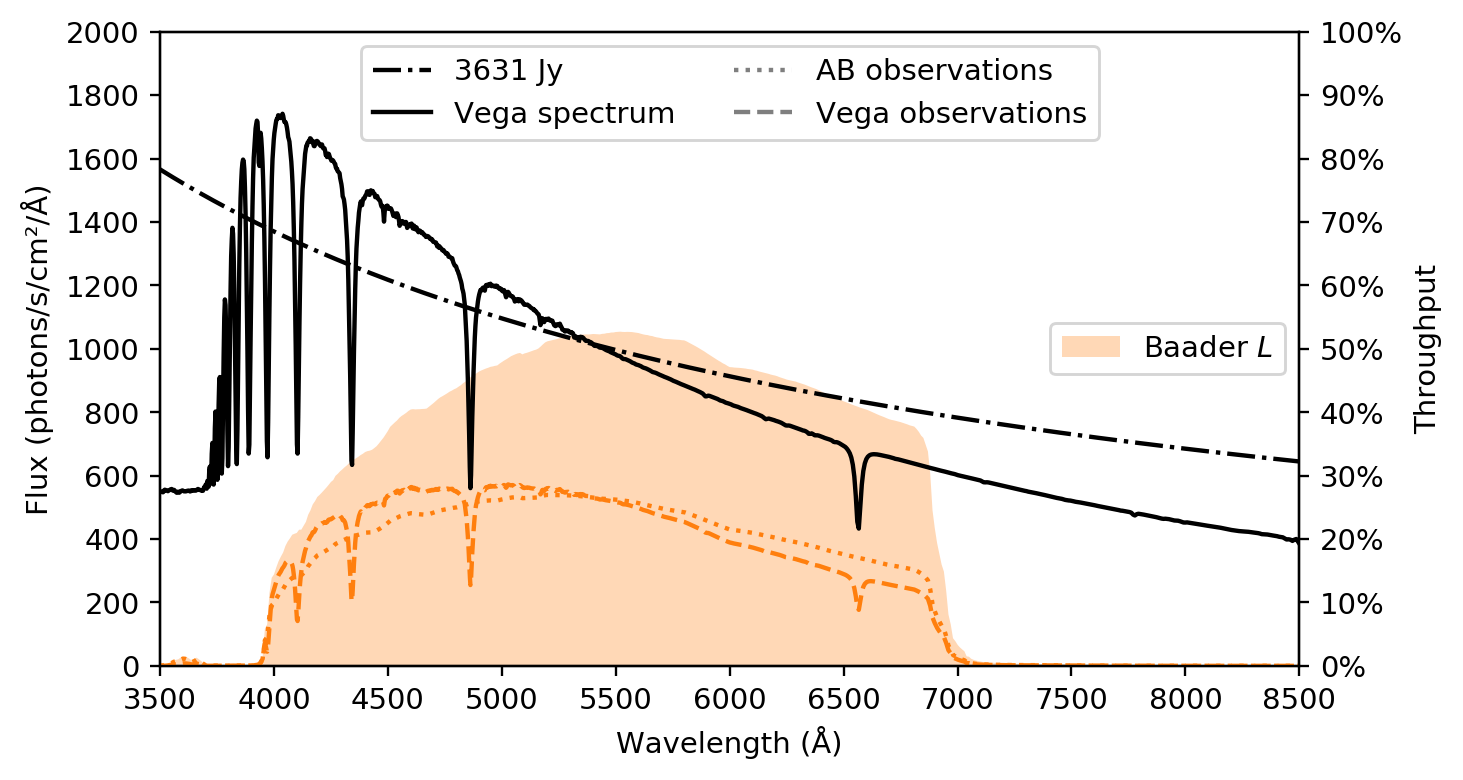}
    \end{center}
    \caption[Simulating photometric observations using pysynphot]{
        Simulating GOTO observations using pysynphot. The filled coloured areas show the theoretical throughputs from \aref{fig:throughput} for the \textit{RGB} filters in the upper plot and the \textit{L} filter in the lower plot. The coloured dotted lines show the throughputs multiplied by a flat \SI{3631}{\jansky} spectrum (dot-dashed black line), while the coloured dashed lines show the throughputs multiplied with the model Vega spectrum (solid black line).
    }\label{fig:pysynphot}
\end{figure}

\clearpage

\makeatletter
\setlength{\@fptop}{0\p@ \@plus 1fil} %
\makeatother

\newpage

\end{colsection}

\subsection{Limiting magnitude}
\label{sec:lim_mag}
\begin{colsection}

Using the CCD parameters determined in \aref{sec:detectors}, the throughput model created in \aref{sec:throughput} and the zeropoints calculated in \aref{sec:model_zeropoints}, a complete photometric model of the GOTO telescopes can be created. One use of this is to predict the system limiting magnitude for a target signal-to-noise ratio.

\subsubsection{Signal-to-noise}

The common sources of noise in CCDs are discussed in \aref{sec:noise}. Discounting the bias level and fixed-pattern noise, both properties of the detector that are easy to remove by subtracting a master bias and dividing by a flat field respectively, the major sources of noise in an astronomical image will be the dark current and read-out noise, as well as the shot noise from the target and the sky background. Accounting for these, the total noise in the image is given by
\begin{equation}
    \sigma_\text{Total} = \sqrt{N + N_\text{sky} + D + R^2},
    \label{eq:total_noise}
\end{equation}
where $N$ is the electron signal from the source object, $N_\text{sky}$ is the background signal from the sky, $D$ is the dark current and $R$ is the read-out noise. Noise is usually quantified as a fraction of the target signal $N$, known as the signal-to-noise ratio \acroadd{snr}:
\begin{equation}
    \text{SNR} = \frac{N}{\sigma_\text{Total}} = \frac{N}{\sqrt{N + N_\text{sky} + D + R^2}}.
    \label{eq:snr}
\end{equation}
To be confident of the detection of an astronomical source a signal-to-noise ratio of 5 or more is required, also known as a $5\sigma$ detection (as the signal will be more than 5 times the noise).

\newpage

\subsubsection{Sky background noise}

\begin{figure}[t]
    \begin{center}
        \includegraphics[width=\linewidth]{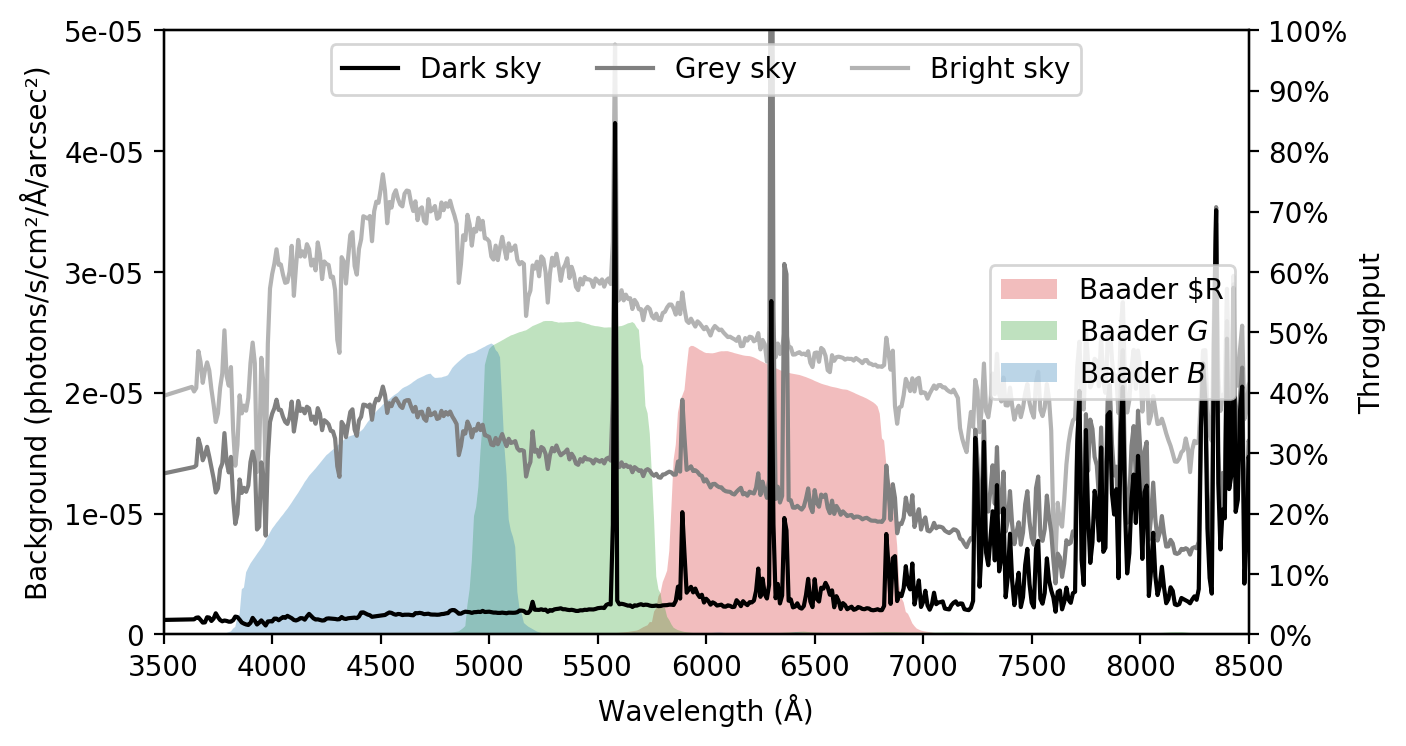}
    \end{center}
    \caption[Simulating sky background observations]{
        Simulating sky background observations using pysynphot.
    }\label{fig:background}
\end{figure}

The one value in \aref{eq:snr} that has not yet been considered is the sky background noise, $N_\text{sky}$. The brightness of the sky will change most noticeably as a function of the Moon phase, with a full Moon creating a background noise several magnitudes brighter than during a new Moon or when the Moon is below the horizon. In order to model the background, sky spectra were taken from \citet{sky_background}, which were obtained from 6 years of VLT observations on using the FORS1 instrument\footnote{Spectra available at \href{http://www.eso.org/~fpatat/science/skybright}{\texttt{http://www.eso.org/\raisebox{0.5ex}{\texttildelow}fpatat/science/skybright}}.} (no equivalent spectra taken from La Palma were available). Three sample spectra were selected to give a range of background signals: a ``Dark'' spectrum taken when the Moon was new and below the horizon, a ``Grey'' spectrum taken when the Moon was 60\% illuminated, and a ``Bright'' spectrum taken when the Moon was full. These spectra are shown in \aref{fig:background}.

In order to determine the sky background noise in the GOTO filters the same method using pysynphot can be used as in \aref{sec:model_zeropoints}. The spectra were again multiplied by the throughput curve for each filter from \aref{sec:total_throughput}, the area under the curve was measured and multiplied by the collecting area of the telescope in order to get an predicted signal in photo-electrons. These were converted into instrumental magnitudes using \aref{eq:ins_mag} and then into calibrated magnitudes using \aref{eq:zp} and the AB zeropoints given in \aref{tab:pysynphot_zeropoints}. The resulting signals are given in \aref{tab:pysynphot_background}. Note that the values are given per square arcsecond; when calculating the sky background flux the signal must be multiplied by the squared plate scale of the camera to get the signal per pixel (the plate scale of the GOTO CCDs is \SI[per-mode=symbol]{1.24}{\arcsec\per\pixel}).

\begin{table}[t]
    \begin{center}
        \begin{tabular}{c|ccc|ccc} %
                   & \multicolumn{6}{c}{Thoretical sky signal} \\
            Filter &
            \multicolumn{3}{c|}{(\elec/s/arcsec$^2$)} &
            \multicolumn{3}{c}{(mag/s/arcsec$^2$)} \\
                   & Dark & Grey & Bright & Dark & Grey & Bright \\
            \midrule
            Baader \textit{L} & 3.34 & 18.38 & 34.98 & 21.43 & 19.58 & 18.88 \\
            Baader \textit{R} & 1.52 &  5.64 & 10.58 & 21.00 & 19.58 & 18.90 \\
            Baader \textit{G} & 1.11 &  6.17 & 12.14 & 21.45 & 19.58 & 18.84 \\
            Baader \textit{B} & 0.68 &  7.24 & 13.50 & 22.16 & 19.59 & 18.92 \\
        \end{tabular}
    \end{center}
    \caption[Sky background signals calculated using pysynphot]{
        Sky background signals calculated using pysynphot for different Moon phases, in AB magnitudes.
    }\label{tab:pysynphot_background}
\end{table}

\subsubsection{Calculating limiting magnitudes}

The limiting magnitude of a telescope is defined as the signal which would be required to obtain a particular SNR, typically $5\sigma$. \aref{eq:snr} can be rearranged into a quadratic formula
\begin{equation}
    N_\text{lim}^2 - \text{SNR}^2 N_\text{lim} - \text{SNR}^2 (N_\text{sky} + D + R^2) = 0,
    \label{eq:snr2}
\end{equation}
and this can be solved to find $N_\text{lim}$ for a given SNR (e.g.\ setting $\text{SNR}=5$).

It is important to remember that $N_\text{lim}$, $N_\text{sky}$, $D$ and $R$ are usually given as a value per pixel. Each therefore needs to be multiplied by the number of pixels the source is spread across, which will be determined by the size of the seeing disk. A given seeing $s$ in arcseconds is defined as the \acro{fwhm} of the seeing disk in the image, which using a Gaussian profile is given by $ 2\sqrt{2 \ln 2}~\sigma$. Taking the $3\sigma$ radius, the number of pixels the source will be spread across is
\begin{equation}
    n = \pi {\left( \frac{3\sigma}{p} \right) }^2
      = \pi {\left( \frac{3s}{2\sqrt{2 \ln 2}~p} \right) }^2,
    \label{eq:seeing2}
\end{equation}
where $p$ is the plate scale in arcseconds/pixel.

Finally, the limiting magnitude in each filter can be calculated for a range of exposure times. These are plotted in \aref{fig:lim_mags} for dark and bright skies, for each GOTO filter and camera and using a seeing of \SI{1.5}{\arcsecond}. Note that it is almost impossible to distinguish between the curves for each camera, as the differences between their dark and read out noise values are very small. The limiting magnitudes for a \SI{60}{\second} image, the typical exposure time for GOTO observations, are given in \aref{tab:lim_mags}.

\begin{figure}[t]
    \begin{center}
        \includegraphics[width=\linewidth]{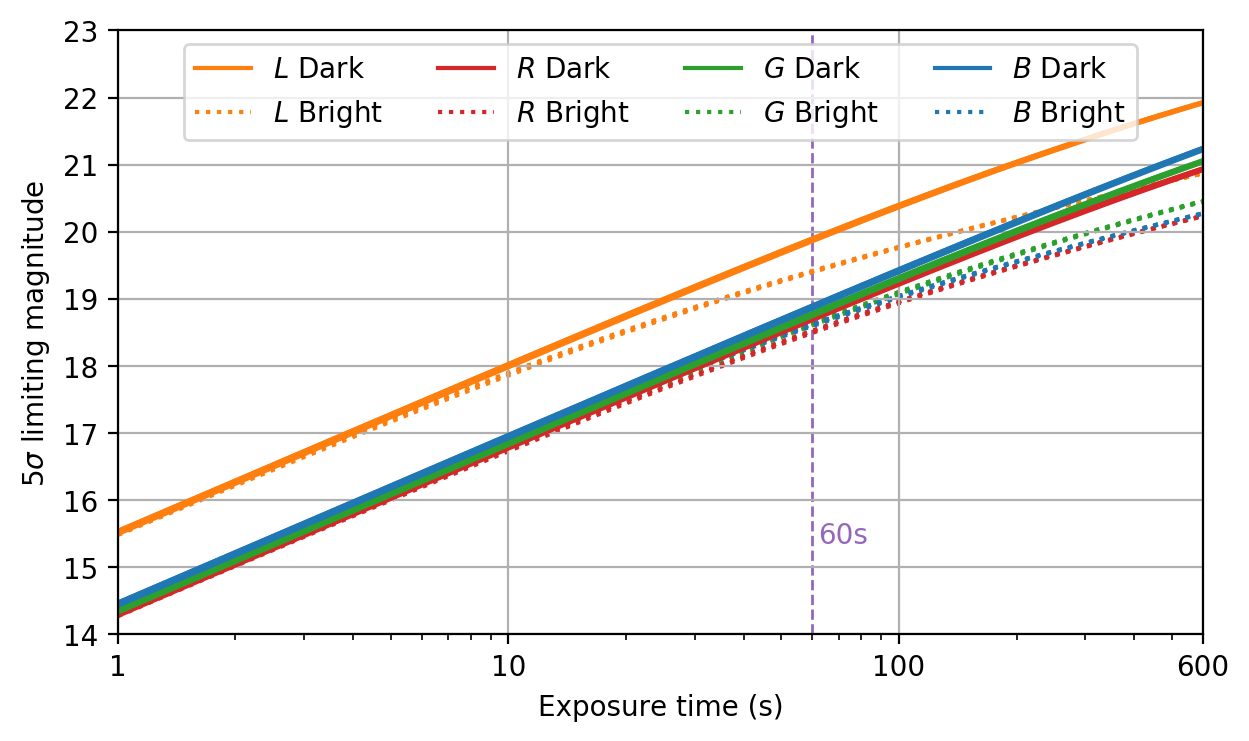}
    \end{center}
    \caption[$5\sigma$ limiting magnitudes for GOTO]{
        $5\sigma$ limiting magnitudes for GOTO plotted as a function of exposure time, assuming an airmass of 1 and seeing of \SI{1.5}{\arcsecond}. Solid lines give the limiting magnitude during dark time, dotted lines during bright time. The \textcolorbf{Purple}{purple} vertical line marks \SI{60}{\second}, the typical GOTO exposure time.
    }\label{fig:lim_mags}
\end{figure}

\begin{table}[t]
    \begin{center}
        \begin{tabular}{c|ccc} %
                   & \multicolumn{3}{c}{Limiting magnitude} \\
            Filter & \multicolumn{3}{c}{(mag)} \\
                   & Dark & Grey & Bright \\
            \midrule
            Baader \textit{L} & 19.88 & 19.61 & 19.41 \\
            Baader \textit{R} & 18.71 & 18.61 & 18.51 \\
            Baader \textit{G} & 18.77 & 18.65 & 18.63 \\
            Baader \textit{B} & 18.88 & 18.63 & 18.61 \\
        \end{tabular}
    \end{center}
    \caption[$5\sigma$ limiting magnitudes for a \SI{60}{\second} exposure]{
        $5\sigma$ limiting magnitudes for a \SI{60}{\second} exposure.
    }\label{tab:lim_mags}
\end{table}

\newpage

\end{colsection}

\subsection{Comparison to on-sky observations}
\label{sec:onsky_comparison}
\begin{colsection}

The GOTO prototype finally reached a stable 4-UT configuration in February 2019 (see \aref{sec:timeline}). In order to determine if it was performing to expectations, the theoretical zeropoints calculated in \aref{sec:model_zeropoints} and limiting magnitudes calculated in \aref{sec:lim_mag} can be compared to those found from on-sky observations. Since GOTO is a wide-field survey instrument there was no need to observe a particular standard star or field --- each frame contains thousands of sources that can be matched to a photometric catalogue. A set of sample observations were used: three \SI{60}{\second} exposures in each of the four filters (so 12 in total) of the Virgo Cluster, taken on the 16th of March 2019. These observations were taken during dark time when the field was at a high altitude (airmass 1.08).

Each image was processed using the standard GOTOphoto pipeline described in \aref{sec:gotophoto}, which corrected the frames for bias, dark and flat frames and extracted sky-subtracted source counts using Source Extractor \citep{SE}. These counts were converted into instrumental magnitudes using \aref{eq:ins_mag}, with $t=\SI{60}{\second}$, and using the gain values for each camera calculated in \aref{sec:ptc}. As GOTO uses the non-standard Baader filters (see \aref{sec:filters}) there are no catalogue magnitudes to compare to. The GOTO pipeline instead makes do with the best available catalogues: the Pan-STARRS PS1 catalogue \citep{Pan-STARRS} and APASS, the AAVSO Photometric All-Sky Survey \citep{APASS}. The \textit{L} and \textit{G} Baader filters are matched to Pan-STARRS \textit{g}, \textit{R} to Pan-STARRS \textit{r} and \textit{B} to APASS \textit{B}.

\newpage

\begin{figure}[t]
    \begin{center}
        \includegraphics[width=\linewidth]{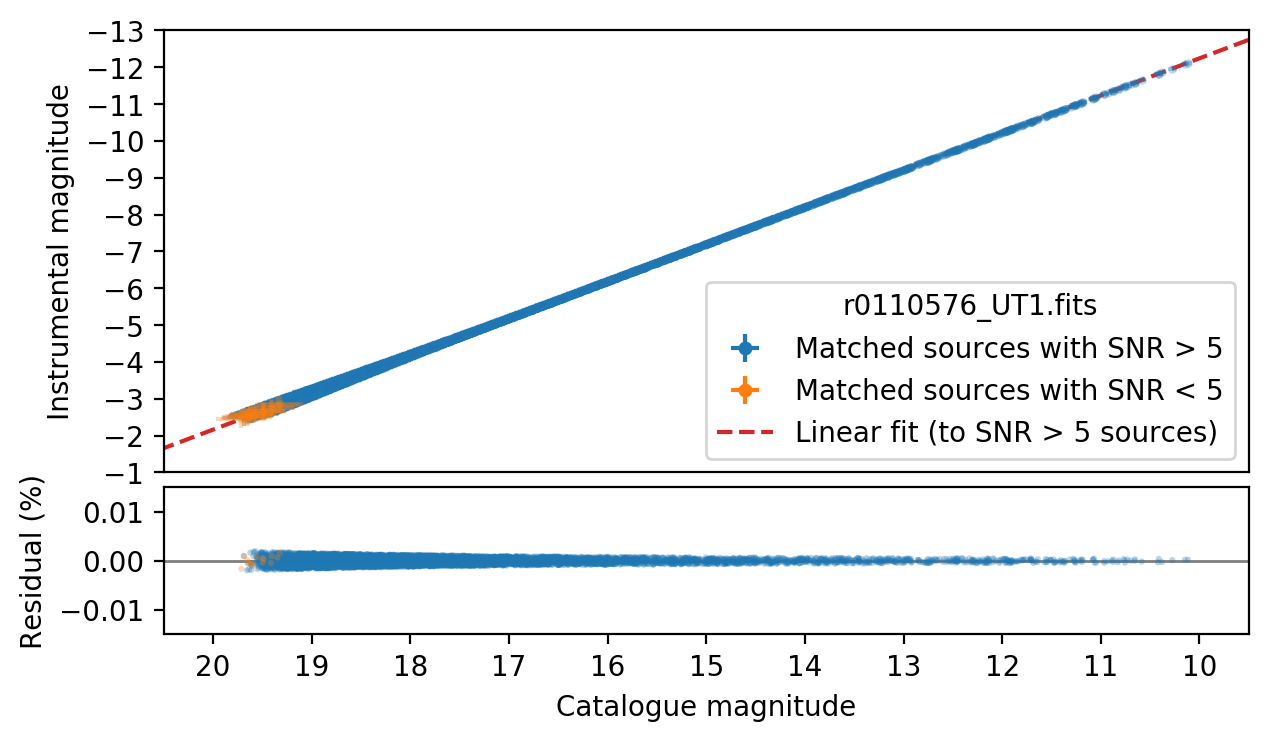}
    \end{center}
    \caption[Finding the observed zeropoint from a GOTO image]{
        Finding the observed zeropoint from a GOTO image. This particular frame was taken with UT1, the first of a set of three in the \textit{L} filter. It contains 6,117 matched sources, of which 6,126 have a signal-to-noise ratio (SNR) of 5 or better. Note the accuracy of the linear fit is much better than measured in \aref{sec:lin}.
    }\label{fig:zeropoint}
\end{figure}

In order to find the zeropoint for each image a linear function was fitted to the measured instrumental magnitudes of each source as a function of the catalogue magnitude of the star it was matched against, with the $y$-intercept being equal to the zeropoint for that image. This is shown in \aref{fig:zeropoint} for one of the \textit{L}-band images. To exclude faint sources with large errors, only sources with a signal-to-noise ratio of 5$\sigma$ or above were included in the fit. This was repeated for every image, and the zeropoints for each are given in \aref{tab:zps_lms}.

The theoretical zeropoints found in \aref{sec:model_zeropoints} were calculated for zero-magnitude stars above the atmosphere, i.e.\ not including the effects of atmospheric extinction described in \aref{sec:atmosphere}. Obviously the real zeropoints measured from GOTO images will include this effect, and so in order to compare to the observed zeropoints the extinction coefficients from \aref{tab:throughput_extinction} were subtracted (multiplied by 1.08, the airmass of the source at the time it was observed) from the theoretical values. This was done for each filter using the AB magnitude zeropoints from \aref{tab:pysynphot_zeropoints}, as both the PS1 and APASS catalogues use AB magnitudes. The new theoretical zeropoints are given in \aref{tab:zps_comparison}, along with the best observed zeropoint from each set of three images.

To measure the limiting magnitude from each image, the catalogue magnitude of each source was compared to the magnitude error measured by Source Extractor, plotted in \aref{fig:lim_mag}. A signal-to-noise ratio of 5 corresponds to a magnitude error of 0.198\footnote{An SNR of 5 means an error of $\pm20\%$, and $2.5\log(1.2)=0.198$. A common approximation is that the magnitude error $\approx 1/$SNR.}, and the limiting magnitude was taken as the lowest magnitude source with a magnitude error greater than 0.198. The best limiting magnitudes from each set of three are given in \aref{tab:lms_comparison}, along with the theoretical dark-time limiting magnitudes again accounting for the target airmass.

\begin{figure}[t]
    \begin{center}
        \includegraphics[width=\linewidth]{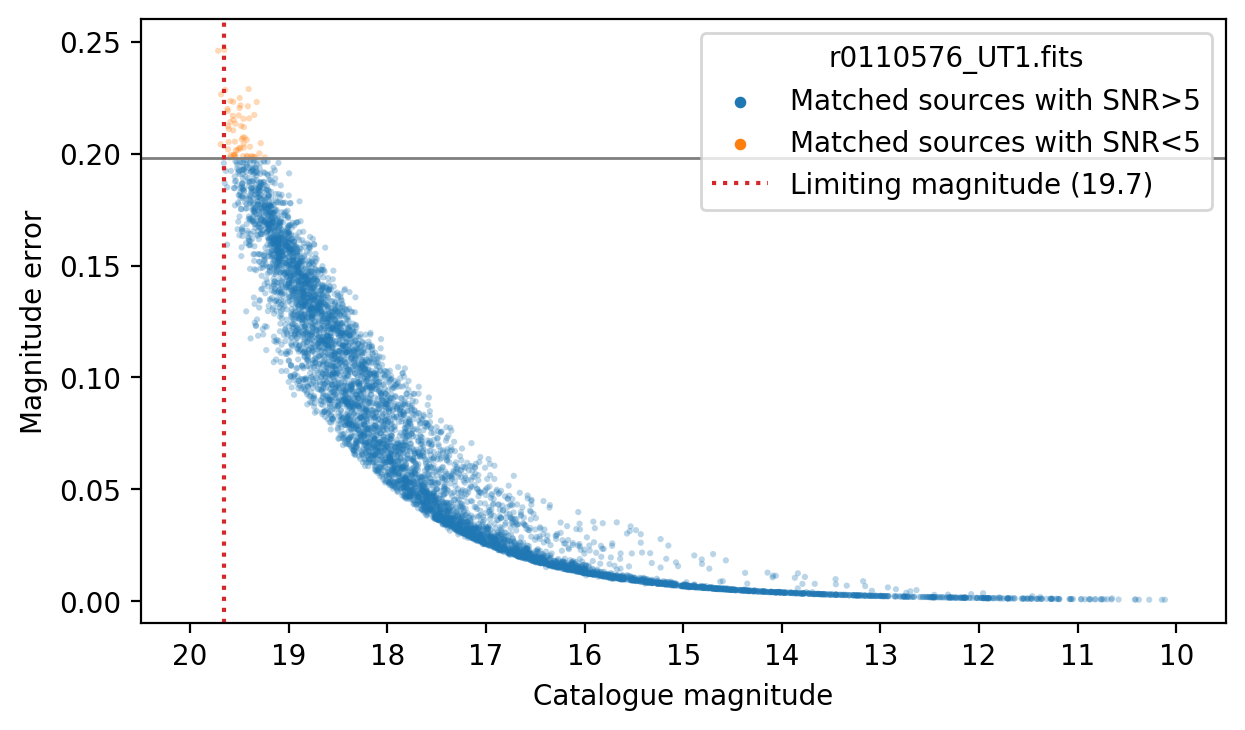}
    \end{center}
    \caption[Finding the limiting magnitude from a GOTO image]{
        Finding the limiting magnitude from a GOTO image.
    }\label{fig:lim_mag}
\end{figure}

\newpage

\begin{table}[p]
    \begin{center}
        \begin{tabular}{cc|cc|cc|cc|cc} %
             & &
            \multicolumn{2}{c|}{UT1} &
            \multicolumn{2}{c|}{UT2} &
            \multicolumn{2}{c|}{UT3} &
            \multicolumn{2}{c}{UT4}
            \\
             & &
            \multicolumn{2}{c|}{{\footnotesize(ML6094917)}} &
            \multicolumn{2}{c|}{{\footnotesize(ML0010316)}} &
            \multicolumn{2}{c|}{{\footnotesize(ML0420516)}} &
            \multicolumn{2}{c}{{\footnotesize(ML5644917)}}
            \\
            \multicolumn{2}{c|}{Filter} &
            ZP & LM & ZP & LM & ZP & LM & ZP & LM \\
            \midrule
            \textit{L} & 1 &
            22.32 & 19.7 &
            22.31 & 19.7 &
            22.40 & 19.8 &
            22.32 & 19.6
            \\
            & 2 &
            22.26 & 19.6 &
            22.27 & 19.7 &
            22.44 & 19.7 &
            22.37 & 19.7
            \\
            & 3 &
            22.39 & 19.6 &
            22.42 & 19.6 &
            22.45 & 19.7 &
            22.40 & 19.7
            \\
            \midrule
            \textit{R} & 1 &
            20.83 & 18.3 &
            21.05 & 18.3 &
            21.10 & 18.4 &
            21.05 & 18.2
            \\
            & 2 &
            20.84 & 18.4 &
            21.11 & 18.4 &
            21.13 & 18.5 &
            21.06 & 18.2
            \\
            & 3 &
            20.91 & 18.4 &
            21.01 & 18.4 &
            21.04 & 18.5 &
            20.94 & 18.2
            \\
            \midrule
            \textit{G} & 1 &
            21.20 & 18.7 &
            21.39 & 18.8 &
            21.40 & 18.8 &
            21.27 & 18.7
            \\
            & 2 &
            21.16 & 18.7 &
            21.43 & 18.8 &
            21.46 & 18.7 &
            21.36 & 18.6
            \\
            & 3 &
            21.26 & 18.6 &
            21.44 & 18.8 &
            21.45 & 18.8 &
            21.32 & 18.6
            \\
            \midrule
            \textit{B} & 1 &
            21.22 & 19.0 &
            21.35 & 18.9 &
            21.43 & 19.1 &
            21.27 & 19.1
            \\
            & 2 &
            21.22 & 19.0 &
            21.32 & 19.2 &
            21.44 & 19.1 &
            21.22 & 19.0
            \\
            & 3 &
            21.20 & 18.9 &
            21.35 & 19.0 &
            21.44 & 19.1 &
            21.26 & 19.0
            \\
        \end{tabular}
    \end{center}
    \caption[Observed zeropoints and limiting magnitudes]{
        Observed zeropoints (ZP) \acroadd{zp} and limiting magnitudes (LM) \acroadd{lm} from three \SI{60}{\second} exposures taken in each filter. The camera serial numbers can be matched to \aref{tab:cameras}.
    }\label{tab:zps_lms}
\end{table}

\begin{table}[p]
    \begin{center}
        \begin{tabular}{c|c|cccc|cccc} %
             &
            Theoretical &
            \multicolumn{4}{c|}{Best observed zeropoint} &
            \multicolumn{4}{c}{Difference (obs-theory)}
            \\
            Filter & zeropoint & UT1 & UT2 & UT3 & UT4 & UT1 & UT2 & UT3 & UT4 \\
            \midrule
            \textit{L} & 22.60 & 22.39 & 22.42 & 22.45 & 22.40 & -0.21 & -0.18 & -0.15 & -0.20 \\
            \textit{R} & 21.38 & 20.91 & 21.11 & 21.13 & 21.06 & -0.47 & -0.27 & -0.25 & -0.32 \\
            \textit{G} & 21.43 & 21.20 & 21.44 & 21.46 & 21.36 & -0.23 & +0.01 & +0.03 & -0.07 \\
            \textit{B} & 21.52 & 21.22 & 21.35 & 21.44 & 21.27 & -0.30 & -0.12 & -0.08 & -0.25 \\
        \end{tabular}
    \end{center}
    \caption[Comparison between theoretical and observed zeropoints]{
        Comparison between the theoretical zeropoints (accounting for extinction) and the best observed zeropoints.
    }\label{tab:zps_comparison}
\end{table}

\begin{table}[p]
    \begin{center}
        \begin{tabular}{c|c|>{\centering\arraybackslash}p{1.2cm}>{\centering\arraybackslash}p{1.2cm}>{\centering\arraybackslash}p{1.2cm}>{\centering\arraybackslash}p{1.2cm}} %
             &
            Theoretical &
            \multicolumn{4}{c}{Best observed limiting magnitude}
            \\
            Filter & limiting magnitude & UT1 & UT2 & UT3 & UT4 \\
            \midrule
            \textit{L} & 19.87 & 19.7 & 19.7 & 19.8 & 19.7 \\
            \textit{R} & 18.70 & 18.4 & 18.4 & 18.5 & 18.2 \\
            \textit{G} & 18.76 & 18.7 & 18.8 & 18.8 & 18.7 \\
            \textit{B} & 18.87 & 19.0 & 19.2 & 19.1 & 19.1 \\
        \end{tabular}
    \end{center}
    \caption[Comparison between theoretical and observed limiting magnitudes]{
        Comparison between the theoretical limiting magnitudes (for dark time) and the best observed limiting magnitudes.
    }\label{tab:lms_comparison}
\end{table}

\clearpage

From \aref{tab:zps_comparison} in most cases the theoretical zeropoints are 0.2--0.3 magnitudes higher than those measured from the sample images, which might suggest that the theoretical model is overestimating the throughput of the system. There is a clear difference between the four unit telescopes, with UT3 consistently performing better than the others. This might be because its mirrors were re-aluminised and returned to La Palma only a month before the images were taken (see \aref{sec:timeline}), and therefore the difference in the observed and theoretical zeropoints may be due to a lower mirror reflectivity than assumed in the throughput model in \aref{sec:optics} (e.g.\ due to dust build up on the mirrors).

There is also a noticeable difference between filters, with all the unit telescopes performing worse in the \textit{R} filter but surpassing the predicted limiting magnitudes in \textit{B}. One limitation in the method used was having to match sources to existing catalogues taken in other filters, without correcting for colour terms i.e.\ the differences in the filter bandpasses (see \aref{sec:filters}). This is something that should be integrated into the GOTOphoto pipeline. Further images taken over more nights would be needed to make any firm conclusions on the performance of the hardware.

\newpage

\end{colsection}

\section{Summary and Conclusions}
\label{sec:hardware_conclusion}

\begin{colsection}

In this chapter I have presented a description and analysis of the GOTO optical hardware.

I first detailed a series of in-lab tests I carried out on the GOTO cameras to determine their key characteristics. I confirmed that the camera properties met the manufacturer's specifications, and was able to independently calculate many of the key properties including the gain and noise levels. I then created a full throughput model of the GOTO unit telescopes, including the contribution of the optics, filters and atmospheric extinction.

I then used the throughput model and camera characteristics to predict values showing the performance of the GOTO telescopes, and then compared them to real on-sky observations. I confirmed that the model does a reasonable job at predicting the zeropoints and limiting magnitudes of real images, and it appears that the GOTO hardware is performing as expected. Future enhancements to the model would include a more detailed comparison of the Baader filters to other sets and calculation of colour terms, to better compare GOTO observations to existing catalogues.

\end{colsection}

\chapter[The GOTO Telescope Control System]{%
    \protect\scalebox{0.96}[1.0]{\mbox{The GOTO Telescope Control System}}
}
\label{chap:gtecs}

\chaptoc{}

\section{Introduction}
\label{sec:gtecs_intro}

\begin{colsection}

Over the next three chapters I detail my work creating a software control system for GOTO.\@ This chapter includes the initial requirements and outline of the control system, and then focuses on the core programs to control the telescope hardware.
\begin{itemize}
    \item In \nref{sec:control_systems} I go through the requirements for the GOTO control system and describe the different options considered.
    \item In \nref{sec:gtecs} I give an overview of the software that makes up the GOTO Telescope Control System and how it was implemented.
    \item In \nref{sec:hardware_control} I go through each category of hardware and describe how the G-TeCS programs were written to control them.
\end{itemize}
All work described in this chapter is my own unless otherwise indicated. This and the following two chapters have been expanded from my SPIE conference paper on G-TeCS, \citet{Dyer}. G-TeCS is based on the pt5m control system \citep{pt5m}, written primarily by Tim Butterly at Durham, and Stu Littlefair and Vik Dhillon at Sheffield.

\newpage

\end{colsection}

\section{The telescope control system}
\label{sec:control_systems}

\begin{colsection}

The term \acro{tcs} describes the various software packages and scripts required to operate a telescope. As described in \aref{sec:goto_design}, GOTO was designed to use standard, off-the-shelf hardware, the type used by high-end amateur astronomers. The control software for this hardware has increasingly been standardised, and there are many TCS software packages available on the market that can be used to operate all aspects of an observatory. However, GOTO has an unusual multi-telescope design and strict requirements for target scheduling, meaning a more customised control system was required. In this section I detail the requirements of the GOTO project and how the choice of TCS was made.

\end{colsection}

\subsection{Requirements}
\label{sec:control_requirements}
\begin{colsection}

My first task as part of the GOTO collaboration, in the summer of 2015 before I started my PhD in Sheffield, was to decide on what control system software to use. There were several requirements to consider.

First, the chosen system had to allow for remote and, most importantly, robotic operation of GOTO.\@ There are many telescope control software packages available, but the majority are designed for a human observer to operate. GOTO, however, was to be a fully autonomous telescope, which meant operating nightly with no human intervention. This meant the control system had to contain routines for observing targets and standard tasks like taking calibration frames. On top of that it was desirable for the system to be able to monitor itself to detect and fix any errors as much as possible without the need for human intervention. Finally it had to be able to monitor and react to external conditions, for example closing the dome if rain was detected.

Second, the system had to include an observation scheduler, which could decide what the telescope should observe during the night. A basic scheduler might be run in the evening to create a night plan, as observers typically do when operating a telescope. However that function alone would not meet the expected operations required from GOTO:\@ normally carrying out an all-sky survey but with a robust interrupt protocol for gravitational-wave follow-up. The system therefore had to be able to recalculate what to observe on-the-fly, and be able to react immediately to transient \acro{too} events.

Furthermore, although the project was still at an early stage the idea of linking together multiple telescopes into a global network was also considered, and the chosen control system would ideally be expandable to facilitate this in the future.

There were also several physical considerations when it came to choosing between software systems. The telescope hardware described in \aref{sec:goto_design} had already been decided on: a clamshell dome from AstroHaven Enterprises, a custom mount with a \acro{sitech} servo controller and multiple unit telescopes all equipped with \acro{fli} cameras, focusers and filter wheels. Any control system would need to communicate with all of this hardware, so any software package with existing drivers would be desirable.

Two particular hardware-related challenges faced the control system project. The first was that the SiTech controller software, SiTechEXE, only ran on Microsoft Windows. The software did have an accessible \acro{api} through the ASCOM standard\footnote{\url{www.ascom-standards.org}}, but that still required some form of the mount control system to be running on Windows. As most professional scientific software in astronomy runs on Linux systems, this led to two options: either have just a small interface running on the Windows machine and the rest of the system on Linux, or have the entire system run on Windows.

The second hardware-related challenge was to deal with the multiple-unit telescope design of GOTO.\@ A full array of eight unit telescopes (UTs) would require eight cameras, focusers and filter wheels. These would all need to be run in parallel, most importantly there needed to be no delay between the exposures starting and finishing on each camera. The physical construction of the telescope also came into play. The FLI units all require a USB connection to the control computer. A single computer situated in the dome would therefore require 24 extra-long USB cables to run up the mount. The suggested solution was to have small computers attached to the mount boom arms next to the unit telescopes, to act as intermediate interfaces to the hardware. The control system therefore needed to be able to run in a distributed manner across multiple computers, potentially even running different operating systems.

There were also practical details to consider when choosing the control software. GOTO was designed as a relatively inexpensive project that could be built quickly and copied across multiple sites, therefore any costly software licenses should ideally be avoided. Experience and support requirements should also be considered, and reusing a software system that members of the collaboration had experience with would provide benefits compared to a completely new system.

\end{colsection}

\subsection{Existing software options}
\label{sec:control_options}
\begin{colsection}

Four possible options for the GOTO control system were considered: the existing software packages ACP Expert, Talon and RTS2, or a custom system based on the code written at Durham and Sheffield for their \acro{pt5m}. At the July 2015 GOTO meeting at Warwick University I gave a talk outlining the control system requirements and presenting the four options, and the decision taken was to adapt the pt5m system for use by GOTO.\@ The three rejected systems are described below, while the pt5m system is described in more detail in \aref{sec:pt5m}.

\subsubsection{ACP Expert}

ACP Expert\footnote{\url{http://acp.dc3.com}} is a commercial observatory control software system by DC3-Dreams. It is used by some advanced amateur astronomers and a few scientific and university telescopes, such as the Open University's PIRATE telescope \citep{PIRATE}. As a complete Windows software package with a web interface it is marketed as being straightforward to use, in either remote or fully robotic modes. It uses the ASCOM standard library and DC3-Dreams also provide professional support and updates. This however came at a cost: \$2495 for the base software, plus an additional \$599 for Maxim DL camera control and \$650 per year for continued support. At the time, GOTO was anticipated to be deployed in a matter of months, so the quick and simple pre-existing commercial solution was tempting. However it was unclear if the ACP software would be able to cope with GOTO's unusual design, and its closed-source model would restrict our ability to make modifications.

\subsubsection{Talon}

The Talon observatory control system\footnote{\url{https://sourceforge.net/projects/observatory}} is a Linux-based, open-source system created by \acro{omi}. It was included as an option primarily as at the time it was the control system of choice for the other observatories operated by Warwick University, such as SuperWASP \citep{SuperWASP}. OMI had built the SuperWASP mount and developed Talon alongside it, before later making it open source. However development of Talon has been almost non-existent over the past decade, and when building the \acro{ngts} a large amount of custom software was needed to allow Talon to work with its multiple telescopes \citep{ngts}. Warwick were already looking at replacing Talon for their \acro{w1m} and when upgrading SuperWASP.\@ Therefore adopting it for GOTO would be unlikely and even counter-productive, as whatever was chosen for GOTO was expected to (and ultimately did) influence and benefit the concurrent development of a new control system for W1m.

\subsubsection{RTS2}

The \acro{rts2}\footnote{\url{https://rts2.org}} \citep{RTS2, RTS2b} is another free and open-source Linux software package. Unlike Talon, RTS2 is under active development and is used by telescopes and observatories around the world \citep{BORAT, BOOTES-3, antarctic, ARTN}. There is a small but active user community and drivers for the hardware GOTO would use had already been developed. The first version of RTS was written in Python, while the second version was rewritten in C++ but with a Python interface available. RTS2 was an attractive choice, however like the others it was unclear if it could be easily modified to meet the requirements for GOTO's multiple telescopes and Windows-controlled mount and no one in the collaboration had prior experience of using or implementing it.

\end{colsection}

\subsection{The pt5m control system}
\label{sec:pt5m}
\begin{colsection}

Built and operated by Sheffield and Durham Universities, \emph{pt5m} is a \SI{0.5}{\metre} telescope located on the roof of the \SI{4.2}{\metre} \acro{wht} on La Palma \citep{pt5m}. The telescope was originally developed as a \acro{slodar} system for atmospheric turbulence profiling in support of the CANARY laser guide star project on the WHT \citep{SLODAR_LaPalma, CANARY}. There are several SLODAR telescopes around the world operated by Durham, including one in Korea that had just been commissioned at the time I joined the GOTO project \citep{SLODAR_Korea}. In order to make the most of the telescope when not being used for SLODAR observations, a science camera was added to pt5m by Sheffield, and in-house control software was written to enable robotic observations. It has successfully been used for automatic observations of transient events since 2012, as well as being used for undergraduate teaching at Sheffield and Durham. All of the SLODAR telescopes used a custom telescope control system developed at Durham, and a similar system was used by the teaching telescopes of the Durham Department of Physics that I worked with during my undergraduate degree. The pt5m control system had been modified by the team at Durham and Sheffield for robotic operation, which matched well with what we needed for GOTO.\@ For this reason, on top of the Sheffield group's existing experience, the pt5m software was chosen to be the base for the GOTO control system.

An overview of the pt5m control system architecture from \citet{pt5m} is shown in \aref{fig:pt5m_software}. The software is written in Python and was built around multiple independent background programs called \emph{daemons}. Each daemon controls one category of hardware, for example the dome, mount or CCD controller. A script called the \emph{pilot} sends commands to the daemons when the system is operating in robotic mode, and the decision of what to observe is taken by the \emph{scheduler} which picks targets out of a database and returns the highest priority to the pilot. Finally a separate script called the \emph{conditions monitor} checks the local weather conditions and tells the pilot to close the dome in bad weather.

The basic framework of the pt5m control system was adopted for GOTO, but with several changes. The major difference between pt5m and GOTO are the multiple unit telescopes, but the same software control system could be adapted through the creation of interface daemons which allow communication to the unit telescopes over the internal dome network. In fact the independent, distributed nature of the daemon system made it very easy to expand to have daemons running on physically separate machines but still communicating over the same local network, including on both Linux and Windows computers.

\begin{figure}[p]
    \begin{center}
        \includegraphics[width=\linewidth]{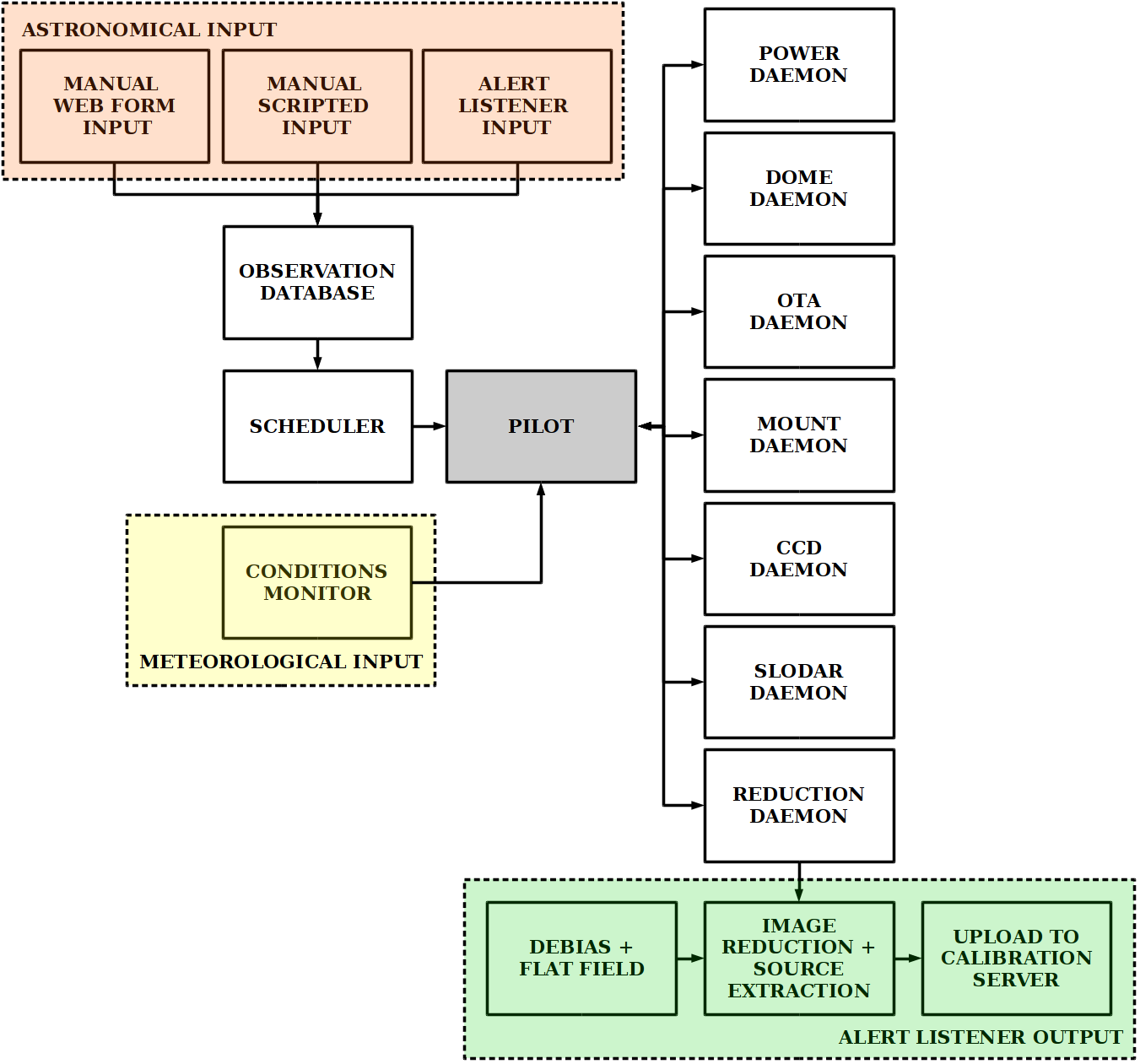}
    \end{center}
    \caption[The pt5m control system architecture]{
        The pt5m control system architecture, taken from \citet{pt5m}. The hardware daemons are shown on the right; they communicate with the pilot which receives information from the observation scheduler and the conditions monitor. This basic framework was adapted for the GOTO control system, c.f. \aref{fig:flow}.
    }\label{fig:pt5m_software}
\end{figure}

\end{colsection}

\section{Overview of G-TeCS}
\label{sec:gtecs}

\begin{colsection}

The \acro{gtecs} is the name given to the collection of programs that have been developed to fulfil the requirements of the GOTO project given in \aref{sec:control_requirements}. The pt5m control system as described in the previous section formed the basis for G-TeCS.\@ Its structure of multiple independent daemons was developed into the core system architecture of G-TeCS, shown in \aref{fig:flow}. This section gives an overview of the system and its implementation. There are two core branches of G-TeCS:\@ the base hardware control programs, described in \aref{sec:hardware_control}, and the autonomous systems built on top of them. The latter software is described in \aref{chap:autonomous} and \aref{chap:scheduling}.

\begin{figure}[p]
    \begin{center}
        \includegraphics[width=\linewidth]{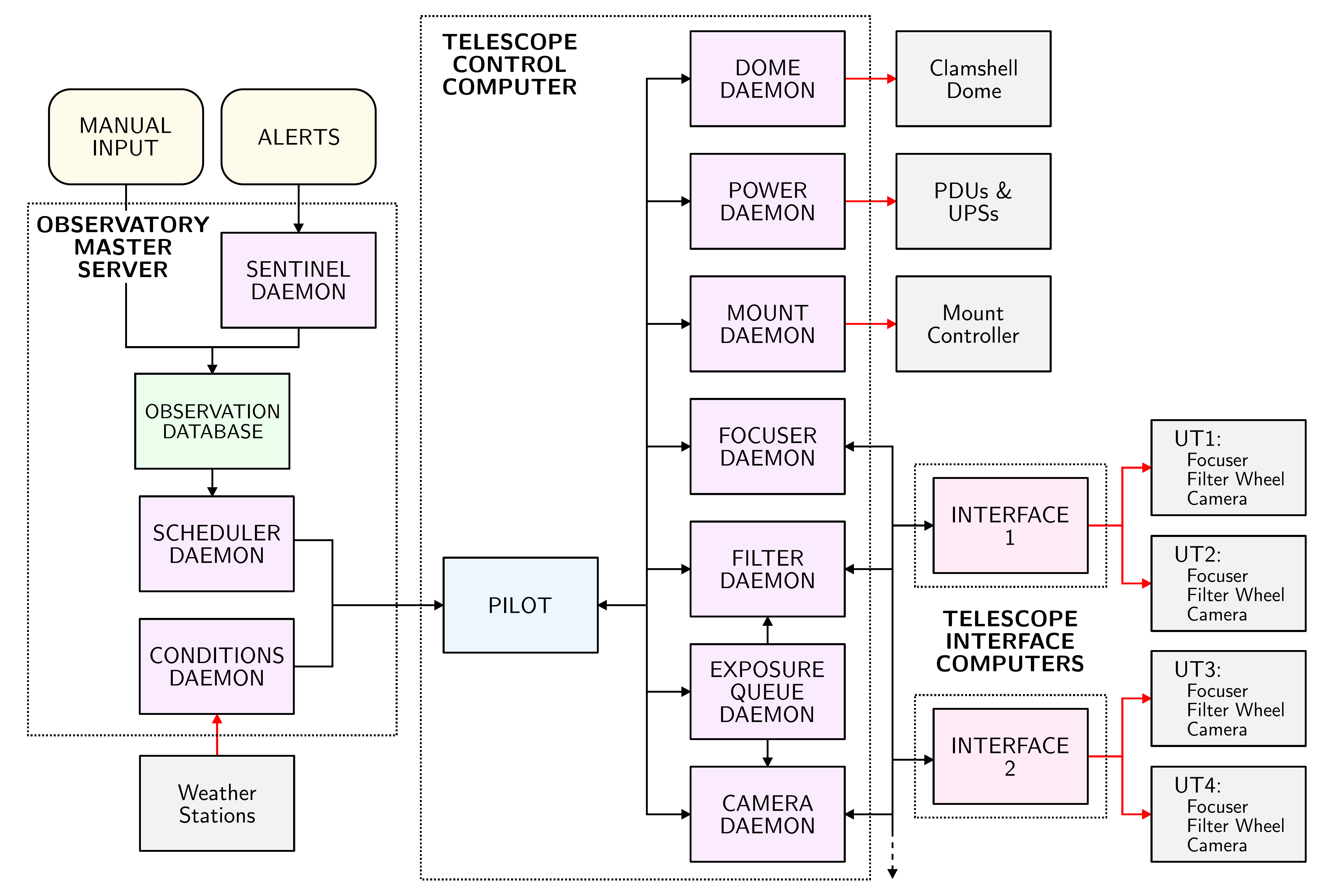}
    \end{center}
    \caption[The G-TeCS system architecture]{
        The G-TeCS system architecture as deployed on La Palma, taken from \citet{Dyer}. The observation database as well as the sentinel, scheduler and conditions daemons shown to the left run on a central observatory-wide server located in the SuperWASP building next to the GOTO domes, while the pilot and hardware daemons are located on the telescope control computer within the dome. Control for the unit telescope hardware (focuser, filter wheel and camera) is sent via an interface daemon for each pair of UTs, running on computers attached to the mount. Only the system for the prototype instrument (one mount with four unit telescopes) is shown.
    }\label{fig:flow}
\end{figure}

\end{colsection}

\subsection{Implementation}
\label{sec:implementation}
\begin{colsection}

The core G-TeCS code is contained in a Python package (\pkg{gtecs}\footnote{\url{https://github.com/GOTO-OBS/g-tecs}}). This includes all of the core daemons, scripts, associated modules and functions. One important module, containing code and functions to interact with the observation database (see \aref{sec:obsdb}), was split off into a separate Python package ObsDB;\@ this was done to allow other users to interact with the database without the need to install the entire G-TeCS package. In addition, the code for alert processing within the sentinel (see \aref{sec:sentinel}) is in a separate package, GOTO-alert. This is because it originated as a separate coding project written by Alex Obradovic at Monash, that I then took over and integrated with G-TeCS.\@ GOTO-alert is described in detail in \aref{chap:alerts}.

G-TeCS and the associated packages are written almost entirely in Python \citep{Python}. Python is a versatile programming language that is increasingly common in astronomy, helped by the popular open-source Astropy Project \citep{astropy}. Python version 3.0 was released in 2008 and was infamously not backwards-compatible with Python2. The code for pt5m was written in Python2, and therefore initially so was G-TeCS.\@ Over the subsequent years the G-TeCS code was re-written to be compatible with both Python2 and Python3, which was possible due to the standard \pkg{\_\_future\_\_} library in Python2 and the Six package (\pkg{six}\footnote{\url{https://six.readthedocs.io}}). Eventually, the addition of new features added to Python3, such as the AsyncIO library used heavily by the pilot (see \aref{sec:async}) in version 3.5, and the imminent end-of-life of Python2 in 2020, led to the dropping of Python2 support. This is in line with most other scientific Python packages including Astropy, which is no longer developed for Python2. %

The core G-TeCS packages have multiple dependencies. Some of the most critical external packages (not included in the Python standard library) are NumPy for mathematical and scientific structures \citep{NumPy}, Astropy for astronomical functions \citep{astropy}, Pyro for communicating between daemons (see \aref{sec:daemons} below), SQLAlchemy for database management (see \aref{sec:obsdb}), Astroplan for scheduling \citep[][see \aref{sec:scheduler}]{astroplan}, VOEvent-parse for handling VOEvents \citep[][see \aref{sec:voevents}]{voevent-parse} and GOTO-tile, a custom package written for GOTO (described in \aref{chap:tiling}).

\end{colsection}

\subsection{Daemons}
\label{sec:daemons}
\begin{colsection}

The core elements of the control system are the daemons. A \emph{daemon} is a type of computer program that runs as a background process, continually cycling and awaiting any input from the user. This is in contrast to a \emph{script} which is run once (either by the system or a user), carries out a series of tasks in the foreground and then exits once it is completed. Common examples of daemons on a Unix-based system are sshd, which listens for and creates \acro{ssh} connections, and cron, which runs commands at predefined times. Incidentally both are used by G-TeCS:\@ SSH is used to execute commands on remote machines, and cron is used to run scripts like the pilot at a set time of day.

Daemons are an ideal model for hardware control software. Once started, each daemon runs continually as a background process, with a main control loop that repeats on a regular timescale (for the G-TeCS hardware daemons this is usually every \SI{0.1}{\second}). There are two primary tasks that are carried out within the loop by every daemon: monitoring the status of the hardware, and listening for and carrying out commands. The former is typically not carried out every time the loop runs, because attempting to request and process the hardware status every \SI{0.1}{\second} would overwhelm the daemon and delay the loop. Instead the status checks are typically carried out every \SI{2}{\second}, or sooner if requested. By continually requesting the status of the hardware the daemon will detect very quickly if there are any problems, and should it be unable to reach its hardware it will enter an error state. However the daemons themselves will not attempt to self-diagnose and fix any problems that are detected, with the notable exception of the dome daemon (see \aref{sec:dome}). Instead, that is the job of the hardware monitors (see \aref{sec:monitors}); the daemons themselves will just report any problems to the pilot or user. The second reason for a control loop within the daemons is to listen for and carry out any commands issued to them. As these commands are dealt with within the loop it ensures only one command is carried out at a time; the alternative of user input going directly to the hardware could cause problems with overlapping commands. These commands can be as simple as querying the cameras for how long left until an exposure finishes, or the mount for the current position, to opening the dome, taking and saving an image, or calculating the current highest priority pointing to observe.

Within G-TeCS each category of hardware has a dedicated control daemon that acts as an interface to the hardware. For example, the mount daemon communicates with the SiTech mount controller, sending commands and reading the current status, while the camera daemon does the same for every camera attached to the telescope. Therefore there is not necessarily a one-to-one correspondence between daemons and pieces of hardware. Having separate daemons for each hardware type allows them to operate independently and allows the pilot, or a human operator, to send commands to each in turn without needing the other one to complete. It also means that a failure in one daemon or its hardware is isolated from the others, should the mount develop a fault, for example, the dome daemon will still be able to communicate with, and close, the dome. Not every daemon within G-TeCS interacts with external hardware: there is the sentinel daemon which monitors alert channels and adds pointings to the observation database, and the scheduler daemon which selects which pointing should be observed at a given time.

Functionally, each daemon is built around a Python class which contains hardware control functions and a main loop. When the daemon starts, the loop is set running in its own thread, and when a control function is called it sets a flag within this loop to carry out the requested commands. The daemons are created using the Pyro Python package (Python Remote Objects, \pkg{Pyro4}\footnote{\url{https://pythonhosted.org/Pyro4}}). Each daemon is run as a Pyro server, so any client script can then access its functions and methods across the network using the associated server ID.\@ This system allows complicated interactions across the network between daemons and scripts with very simple code, and was one of the major benefits of adopting the pt5m system.

\end{colsection}

\subsection{Scripts}
\label{sec:scripts}
\begin{colsection}

As well as the daemons, the G-TeCS package includes multiple Python scripts. These scripts can be run on the control computer from the command line by a human user, called from within other scripts like the pilot, or started through utilities like cron.

In order to send commands to the daemons, each has an associated control script that can be called by a user from a terminal, or by the pilot in robotic mode (see \aref{sec:pilot}). The commands follow a simple format which was inherited from pt5m, first the short name of the daemon, then the command, and finally any arguments. There are several commands that are common to all daemons: \code{start}, \code{shutdown} and \code{restart} to control if the daemon is running; \code{ping} to see the current status of the daemon; \code{info} to see the current status of the hardware; \code{log} to print the daemon output log. Examples of daemon-specific commands include ``\code{dome~close}'' to close the dome, ``\code{mnt~slew~30.54~+62}'' to slew the mount to the given coordinates, ``\code{cam~image~60}'' to take a \SI{60}{\second} exposure with all connected cameras and ``\code{cam~image~2~60}'' to take a \SI{60}{\second} exposure with camera 2 only.

Every daemon can also be controlled in ``interactive mode'', which is a user-friendly way to save time sending multiple commands to the same daemon. Interactive mode is entered with \code{i} and exited with \code{q}.

There is also a utility script, \code{lilith.py}, which can send the same command to all the daemons. For example, to shutdown every daemon it is possible to call each directly (\code{cam~shutdown}, \code{foc~shutdown}, \code{mnt~shutdown} etc\ldots) but it is instead much easier to run \code{lilith~shutdown}. The name ``Lilith'' comes from the biblical ``mother of demons''.

The most important script to the robotic operation of the telescope is the pilot, detailed in \aref{sec:pilot}. The pilot is started every night using cron at 5pm, but can also be started manually with the command ``\code{pilot~start}'' (note although this uses the same syntax as a daemon it simply runs the pilot script in the current terminal instead of starting a background process). There is also a daytime counterpart to the pilot, called the day marshal, which is run in the same way (see \aref{sec:day_marshal}). Finally, several of the more common observing tasks are separated off into ``observation scripts''. These contain lists of commands to send to the daemons to carry out tasks such as focusing the telescope, taking flat fields or starting/shutting down the hardware in the evening/morning respectively. These are run at specific times each night by the pilot night marshal routine (see \aref{sec:night_marshal}), but they can also be run by human observers through the command line (for example ``\code{obs\_script~startup}'' to run the startup script, or ``\code{obs\_script~autofocus}'' to start the autofocus routine).

\end{colsection}

\section{Hardware Control}
\label{sec:hardware_control}

\begin{colsection}

The core programs of G-TeCS are the hardware daemons. There are seven primary daemons, as shown in the centre of \aref{fig:flow}. This section provides a summary of each of the hardware categories, describing how the daemons interact with them and the particular challenges and features unique to each.

\end{colsection}

\subsection{FLI interfaces}
\label{sec:fli}
\begin{colsection}

As described previously, GOTO uses off-the-shelf camera, focuser and filter-wheel hardware from \acro{fli}. Each GOTO unit telescope has a MicroLine ML50100 camera, an Atlas focuser and a CFW9--5 filter wheel, these are connected to a small Intel \acro{nuc} attached to the boom arm (one per pair of UTs, shown in \aref{fig:boomarm}). These NUCs run very basic daemons called the FLI interfaces, shown in \aref{fig:flow}. Barely daemons by the definition given in \aref{sec:daemons}, these interfaces have no control loop and exist only as a way to expose the serial connection of the hardware to the wider Pyro network. By using these interface daemons, the primary control daemons for the FLI hardware can run on the main control computer without being physically connected to the hardware (aside from via ethernet).

Communicating with the hardware has to be done using the \acro{sdk} provided by FLI, which is written in C. In order to use this SDK with the control system written in Python, a separate wrapper package FLI-API (\pkg{fliapi}\footnote{\url{https://github.com/GOTO-OBS/fli-api}}) was written by Stu Littlefair in Cython, a programming language that provides a way for C code to be imported and run in Python.

\begin{figure}[t]
    \begin{center}
        \includegraphics[width=\linewidth]{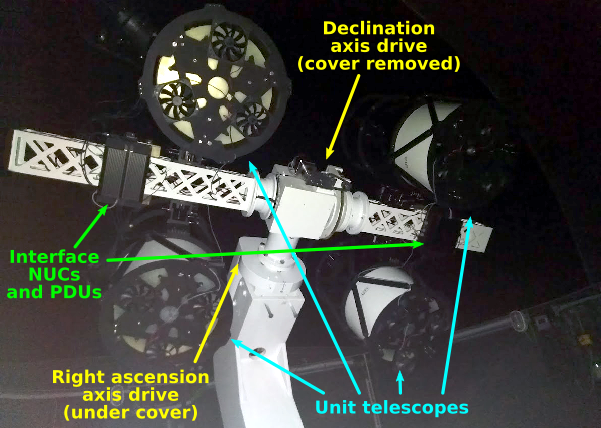}
    \end{center}
    \caption[Photo of GOTO with hardware elements labelled]{
        A photo of GOTO observing at night, with key hardware elements labelled. The back of the four unit telescopes is visible, attached to the central boom arm. Each pair of UTs is connected to a NUC interface computer and a \acro{pdu}. The mount declination drive is also visible with the cover removed, the right ascension drive is on the other side of the mount.
    }\label{fig:boomarm}
\end{figure}

The interfaces and the Pyro network allow a single daemon to interact with multiple pieces of hardware across multiple computers. This means that the single camera daemon running on the primary control computer can interface with all of the cameras attached to the mount, and instead of sending commands to each camera individually the user can speak to all of them together through the daemon. As an example, the command ``\code{cam~image~60}'' will take a 60 second exposure on every attached camera simultaneously. Including a specific number (``\code{cam~image~2~60}'') will only start the exposure on the camera attached to UT2. Multiple selections can also be made using a simple comma-separated syntax, such as ``\code{cam~image~1,2,4~60}''. This notation and functionality is one of the major differences between G-TeCS and the pt5m control system, and in fact all of the other control systems considered in \aref{sec:control_options}, which typically can only communicate with a single telescope at a time.

\newpage

There are three control daemons that interact with the FLI interfaces: the camera, filter wheel and focuser daemons. There is also a fourth, the exposure queue daemon, which coordinates sets of exposures and communicates with both the cameras and filter wheels through their daemons, not the interfaces directly. Each of these four daemons are described in the following sections.

\end{colsection}

\subsection{Camera control}
\label{sec:cam}
\begin{colsection}

The camera daemon interacts with all of the FLI cameras on the GOTO mount, making it the most complicated daemon to design. The commands to the camera daemon, however, are fairly straightforward. There are four types of exposures that can be taken:

\begin{itemize}
    \item Normal images, with the shutter opening and closing for the given exposure time.
    \item ``Glance'' images, which are the same as normal images but are saved to a separate file that is overwritten each time a glance is taken.
    \item Dark images, where the shutter remains closed during the exposure time.
    \item Bias images, where the shutter remains closed and a zero-second exposure is taken.
\end{itemize}

The FLI-API interface also gives other options for exposures aside from just the exposure time, including different binning factors and windowing the active area of the chip to read out. Although the camera daemon does offer commands to set these they are never used during normal operations, and the GOTOphoto image processing pipeline is set up to only expect full-frame, unbinned images.

Once an exposure is completed, the image data needs to be downloaded from the cameras and sent through the interfaces to the camera daemon, before the frames can be saved as \acro{fits} files. This is a disadvantage of the interface system, and consideration was given to instead having the interfaces write out the files to their local NUCs. Although this would have been faster to save the raw images, they would still need to be copied down from the NUCs to the primary archive on the control computer. Having the interfaces send the raw count arrays to the camera daemon for processing proved to save more time in the long run. The camera daemon also queries all the other hardware daemons at the start of the exposure, to get their current statuses to add to the FITS headers (for example, getting the current pointing position from the mount daemon).

The time taken by each exposure, from the command being received to the FITS images being written to disk, has been optimised to minimise the amount of ``dead time'' between exposures. One of the primary ways to save time was to have the two most time-dependent processes, downloading the images from the interfaces and writing them to disk, run as separate threads for each camera independently of the main daemon control loop. Other time-saving improvements included only fetching the status information from the other daemons once, just after starting the exposures (so it does not take any extra time in addition to the exposure time).

Images are written to FITS files by the camera daemon and are archived in different directories by date (e.g. \code{2019--09--30}). Each camera output is saved as a separate file, named by the current run number and the name of the unit telescope it originated from (e.g. \code{r000033\_UT2.fits} is the image from camera 2 for run 33). The run number is increased whenever a non-glance exposure is taken, even if the exposure is subsequently aborted. After being saved the images are copied at regular intervals from La Palma to Warwick University via a dedicated fibre link, where the GOTOphoto photometry pipeline is run (as described in \aref{sec:gotophoto}). GOTOphoto has been developed at Warwick and Monash separately from the control system, which means image calibration, astrometry and photometry are all out of the scope of this thesis.

\newpage

\end{colsection}

\subsection{Filter wheel control}
\label{sec:filt}
\begin{colsection}

The filter wheel daemon (sometimes shortened to just the filter daemon) controls the filter wheels on the GOTO unit telescopes. The FLI CFW9--5 filter wheels are fairly standard pieces of hardware, with 5 slots that contain the \SI{65}{\milli\metre} square Baader \textit{R}, \textit{G}, \textit{B}, \textit{L} and \textit{C} filters (see \aref{sec:filters}). Moving the filter wheel is usually done via the exposure queue daemon (see \aref{sec:exq}) but can be done individually. When powered-on the filter wheels must be homed to position 0 before moving. The \textit{L} filter was placed in the home position as the vast majority of GOTO observations are taken in this filter.

\end{colsection}

\subsection{Focuser control}
\label{sec:foc}
\begin{colsection}

The focuser daemon is the third of the three FLI hardware daemons. Each connected focuser can be set to a specific position or moved by a given offset by the daemon. The focuser daemon is usually only used when the pilot runs the autofocus routine at the start of the night (see \aref{sec:night_marshal} and \aref{sec:autofocus}).

\end{colsection}

\subsection{Exposure queue control}
\label{sec:exq}
\begin{colsection}

The exposure queue daemon (often abbreviated to `ExQ' or `exq') does not directly talk to hardware; instead it is the only daemon with the primary purpose of communicating with other daemons, specifically the camera and filter wheel daemons. The exposure queue daemon coordinates taking frames in sequence and setting the correct filters before each exposure starts. For example, consider needing a series of three \SI{30}{\second} exposures, one each in the \textit{R}, \textit{G} and \textit{B} filters. Through the camera and filter wheel daemons this would require six commands: \code{filt~set~R}, \code{cam~image~30}, \code{filt~set~G}, \code{cam~image~30}, \code{filt~set~B}, \code{cam~image~30}. The exposure queue daemon gives shorter method to carry out the same commands, and these same exposures can be requested with a single command: ``\code{exq~mimage~30~R,G,B}'' (\code{mimage} is short for multiple-image).

\begin{figure}[t]
    \begin{center}
        \vspace{1cm}
        \code{1111;30;R;1;normal;M101;SCIENCE;0;1;3;545}\\
        \code{1111;30;G;1;normal;M101;SCIENCE;0;2;3;545}\\
        \code{1111;30;B;1;normal;M101;SCIENCE;0;3;3;545}\\
        \vspace{0cm}
    \end{center}
    \caption[A sample exposure queue file]{
        A sample of an exposure queue file. Each line is a new exposure, and details of the exposure are separated by semicolons. In order, these are: the binary UT mask, exposure time in seconds, filter, binning factor, frame type, object name, image type, glance flag, set position, set total and database set ID number.
    }\label{fig:exq_file}
\end{figure}

When a set of exposures is defined and passed to the exposure queue daemon they are added to the queue, which is stored in a text file written to and read by the daemon. An example of the contents of the file is given in \aref{fig:exq_file}. The details of each exposure are saved in this file, and adding more using the \code{exq} command adds more exposures to the end of the queue. When the queue is running (it can be paused and resumed, for example to allow slews between exposures) the daemon will select the first exposure in the queue, tell the filter wheel daemon to change filter if necessary and then tell the camera daemon to start the exposure.

As shown in \aref{fig:exq_file}, extra meta-data can be written for each exposure. The UT mask is simply a binary representation of the unit telescopes to use for this exposure, so \code{0101} would be exposing on UTs 1 and 3 only (counting from the right starting with UT1), while \code{1111} will be on all four. The frame type is a variable used within the FLI API, it is either \code{normal} or \code{dark} depending on if the shutter will open or not. Exposures taken through the exposure queue can also have a target name (e.g.\ the galaxy M101 in \aref{fig:exq_file}) and an image type (used to define the type of image, either SCIENCE, FOCUS, FLAT, DARK or BIAS). The glance flag is a boolean value, set to \code{1} (True) if the exposure is a glance or \code{0} (False) otherwise.

When multiple exposures are defined using the \code{exq} commands, as in the previous \code{mimage} example, they are grouped into a ``set''. The set position and set total values shown in \aref{fig:exq_file} denote those exposures as 1 of (a set of) 3, 2 of 3, and 3 of 3. Including this information in the exposure metadata is necessary so the photometry pipeline knows if an exposure is part of a set and, if they are all in the same filter, whether they should be co-added to produce reference frames. Exposure sets are defined in the observation database (see \aref{sec:obsdb}), with each pointing having at least one or more sets to be added to the exposure queue by the pilot when that pointing is observed.

Similar to the camera daemon, the timing of code and functions within the exposure queue daemon has been optimised to minimise the ``dead time'' between exposures. However, the commands also need to be timed correctly to ensure that, for example, the exposure does not start while the filter wheel is still moving. This was one of the major reasons for having a separate exposure queue daemon to handle these timing concerns, while the camera and filter wheel daemons dealt only with individual commands. Incidentally, pt5m uses a QSI camera with an integrated filter wheel \citep{pt5m}, so what in G-TeCS are separate camera, filter wheel and exposure queue daemons are all combined into a single ``CCD'' daemon in \aref{fig:pt5m_software}.

\end{colsection}

\subsection{Dome control}
\label{sec:dome}
\begin{colsection}

The dome daemon is the primary interface to the dome. It is in effect the most critical of all of the hardware control systems, because a failure in the software resulting in the dome opening in bad weather could be catastrophic to the hardware inside. As such, the dome daemon includes multiple levels of internal checks and backup systems. It is also the only daemon with a small amount of autonomy built in, and therefore blurs the line between a pure hardware control daemon and the more complicated autonomous systems described in \aref{chap:autonomous}.

\begin{figure}[t]
    \begin{center}
        \includegraphics[width=\linewidth]{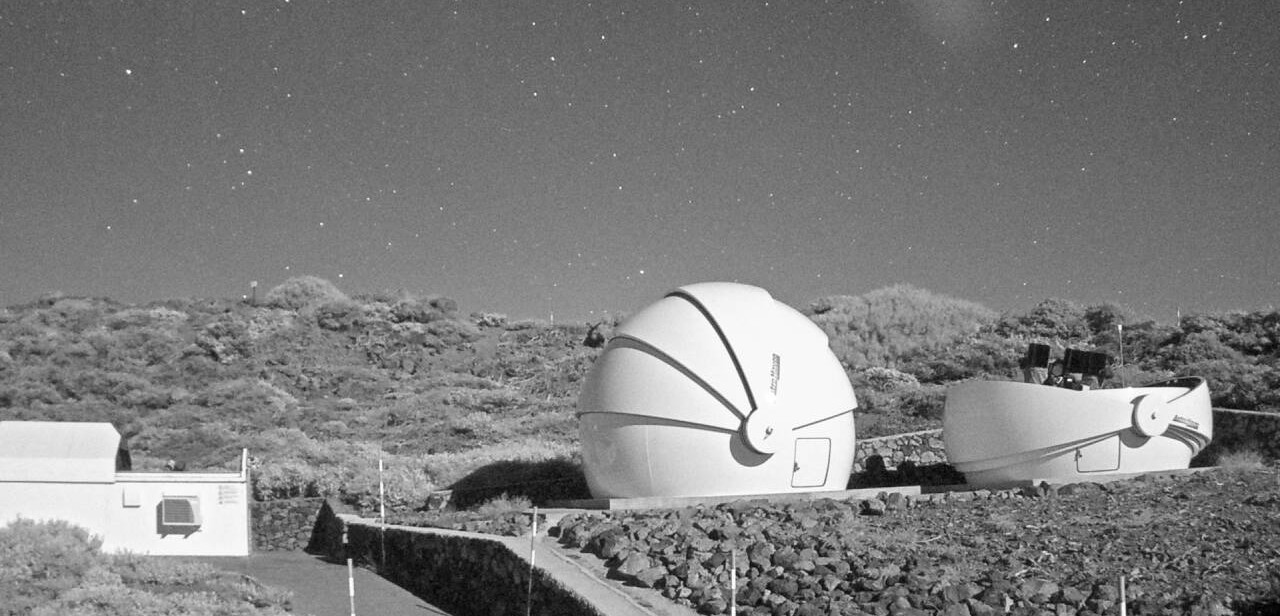}
    \end{center}
    \caption[Webcam image of the GOTO site at night]{
        A webcam image of the GOTO site taken in June 2018. The two GOTO clamshell domes are on the right; the northern, empty dome is closed while the other is open for observing. Note the SuperWASP shed roof on the left is also open.
    }\label{fig:webcam}
\end{figure}

GOTO uses an Astrohaven clamshell dome, shown in \aref{fig:webcam}. The dome daemon communicates with the \acro{plc} that comes with the dome through a simple serial (RS-232) connection. Moving the dome is achieved by sending a single character to the PLC:\@ \code{a} to open the south side, \code{A} to close it; \code{b}/\code{B} for the north side. The PLC will respond with another character: either returning the input while the dome is moving, \code{x}/\code{X} when the south side is fully open/closed and \code{y}/\code{Y} for the north side. This is a simplistic and quite limited interface. For example, while one side is moving there is no way to know the status of the other. Therefore, when commissioning it was decided to add additional independent limit switches, described in \aref{sec:arduino}. The Arduino system detailed in that section adds four additional inputs: one at the intersection of the two inner-most shutters to confirm the dome is fully closed, two on either side to confirm if either side is fully open, and one on the dome entrance hatch. Using all of these sensors, and the feedback from the dome PLC, it is possible to build up a complete picture of the current dome status. Inside the dome daemon each side has five possible statuses: \code{closed}, \code{part\_open}, \code{full\_open}, \code{opening} and \code{closing}. The dome as a whole is only considered confirmed closed if both sides report \code{closed}.

As the interface functions of the Astrohaven PLC are very limited, any more advanced functionality had to be coded from scratch. The commands to the dome are contained within a custom Python class \code{Astrohaven}, which also returns the status of the dome and the additional sensors. The class has functions to open and close the dome, which include being able to move a specific side or both. Due to the five-shutter design of the GOTO dome (shown in \aref{fig:webcam}) the overlapping side (south) is always opened before the north and closed after it, as it is easier for the shutter casters to roll over the lower shutter than for the lower shutter to force itself under the casters. When opening the south side the motion is deliberately stepped (i.e.\ moving in short bursts) rather than in one smooth motion. This was added due to the design of the top overlapping shutter: if the move command is sent too quickly slack will appear in the drive belts and the upper shutter will end up ``jerking'' the lower one, putting more stress on the belts. This sort of functionality is not included in the default Astrohaven software but is easy to do within the Python code by increasing the time between sending command characters.

As described in \aref{sec:arduino}, along with the extra dome sensors a small siren was attached to the Arduino to give an audible warning before the dome starts moving. This siren can be activated for a given number of seconds through a HTML request to the Arduino, and this is called by the dome daemon whenever the dome is moved automatically. The siren can be disabled in manual mode and is automatically off in engineering mode (see \aref{sec:mode}). One slight complexity is if the system is in manual mode with the alarm disabled but the autoclose feature still enabled. In this case the dome alarm will not sound when manually sending move commands to the daemon, however if the dome is due to close automatically in bad conditions it will re-enable the alarm and make sure it sounds before moving. Forcing the siren to sound whenever the dome moves autonomously is an important safety feature when operating a robotic telescope such as GOTO, and pt5m also has a similar alarm.

As mentioned above, the dome daemon has an ``autoclose'' feature that is unlike any feature of the other daemons. The normal design philosophy of the daemons is that they should not take any action without explicit instructions, which could come from a user or a script like the pilot. The dome, however, is an exception since in the case of bad weather the survival of the hardware is considered to be of higher importance. Therefore, in addition to checking for input commands, the dome daemon control loop also monitors the output of the conditions daemon (see \aref{sec:conditions}). If any conditions flag is set to bad, and the dome autoclose option is enabled, then the dome daemon will automatically enter a ``lockdown'' state. In this state if the dome is currently open it will immediately send itself a close command. Once it is closed the lockdown will prevent any open commands, until either the lockdown is cleared or autoclose is disabled. Another of the hardware additions during commissioning (see \aref{sec:arduino}) was a ``quick-close'' button directly attached to the serial port of the control computer in the dome. The dome daemon automatically sends a signal through the serial connection every time the control loop runs, and if the signal is broken (i.e.\ the button has been pressed, breaking the circuit) then it will immediately trigger a lockdown.

The other custom hardware device added to the GOTO dome was a small backup ``heartbeat'' system, developed by Paul Chote at Warwick. A recognised flaw of the G-TeCS dome control architecture was that it was entirely reliant on the dome daemon, and by extension the master control computer, to close the dome in an emergency. Should the dome daemon or the control computer crash for any reason the dome would be completely disabled. This therefore presented a single point of failure, and a system was designed at Warwick to mitigate against this. An extra circuit, also powered by an Arduino, is connected over the serial port to the dome PLC, and the dome daemon has a separate thread which continuously sends a ping byte to this port. Should the Arduino not receive a signal from the dome daemon after a given timeout period (the default is \SI{5}{\second}) it will automatically start sending the close characters (\code{A}/\code{B}) to the dome PLC.\@ This system therefore provides a secure secondary backup to the other dome software, and although it has so far not been needed it is an important insurance policy.

The dome daemon is also the hardware interface to the dehumidifier located within the dome. Like the dome, the dehumidifier requires automated control, as the unit uses a lot of power and can get clogged with dust if used excessively. The dome daemon will turn the dehumidifier on if the internal humidity gets too high or temperature gets too low, and will turn it off when they reach normal levels or if the dome is opened. This behaviour can also be overridden and, like all the automated G-TeCS systems, is disabled in engineering mode.

\end{colsection}

\subsection{Mount control}
\label{sec:mount}
\begin{colsection}

The mount daemon sends commands to the GOTO mount through the \acro{sitech} servo controller. As discussed in \aref{sec:control_requirements}, the software for the servo controller is a Windows program called SiTechEXE.\@ Therefore, enabling communication between SiTechEXE and the rest of the control system was a key requirement of G-TeCS.\@

Initially the only way to communicate with SiTechEXE was via the ASCOM software interface. It was possible to communicate directly with the servo controller through a serial interface, however this was a very low-level interface and would have required a lot of work to re-implement the array of commands and functions within SiTechEXE.\@ In particular, the PointXP pointing model software was essential to make a pointing model for the mount (see \aref{sec:pointxp}), and it would have been very difficult to implement using serial commands. ASCOM is so called because it uses the Microsoft \acro{com} interface standard to provide a unified \acro{api} for astronomical hardware. SiTEch provides their own ASCOM driver for their servo controller, and through the Python for Windows package (\pkg{pywin32}\footnote{\url{https://github.com/mhammond/pywin32}}) Python code could interact with ASCOM and therefore SiTechEXE.\@ The ASCOM API gave access to a wide variety of commands and status functions, including being able to slew the telescope, start and stop tracking, parking and setting and clearing targets.

The ASCOM method did however require the Python daemon to be running on the Windows computer. The solution to this was to write a \code{sitech} interface in the same manner that the FLI hardware connected to the boom arm computers use an interface daemon running on the NUCs. The \code{sitech} interface acted purely as a way of routing commands sent through the Pyro network to the ASCOM equivalent. However, as it had to run on the Windows machine, it differed slightly in implementation to the FLI interfaces and other daemons, as Windows and Linux have different ways of defining ``daemon'' processes (Windows generally does not call them daemons, instead using terms like ``background processes''). Furthermore, the interface had to be able to be started, stopped and killed from the remote control computer using a \code{sitech} control script, which meant G-TeCS needed to include functions specifically to interact with Windows processes. This meant the G-TeCS package needed to be installable on Windows and deal with configuration file paths and parameters (compare Windows \code{C:\textbackslash{}Users\textbackslash{}goto\textbackslash{}} to Linux \code{/home/goto/}). This was simplified by the use of the Cygwin package\footnote{\url{https://www.cygwin.com}}, which provides Unix-like commands and behaviour on Windows including mapping directories into Unix format. Once this was developed the system was reliable enough to correctly control the mount during commissioning.

In July 2017 the author of SiTechEXE, Dan Gray of Sidereal Technology, released an update to the software that enabled communication over a network using \acro{tcpip} commands. This meant the mount daemon running on the control computer could communicate directly with SiTechEXE without the need for the \code{sitech} interface, ASCOM, Cygwin or maintaining any Windows-compatible code. Although the existing code was functioning reliably, removing the need for compatibility with ASCOM enabled the addition of several new features, such as more error feedback, whether the limit switches have been triggered, and turning on and off ``blinky mode'' (the error state the mount automatically enters when drawing too much current or one of the inbuilt limit switches is triggered). As such it was seen as a worthwhile update, and therefore the \code{sitech} interface and any Windows code were removed from G-TeCS when the La Palma system was updated in August 2017. The TCP/IP interface provides a much simpler way to communicate with the mount than the previous ASCOM commands. Commands are sent as binary strings of characters; for example to get the current mount status information you send `\code{ReadScopeStatus}', and to slew to given coordinates the command is `\code{GoTo~<ra>~<dec>}'.

One catch is that the SiTech software expects coordinates in the JNow epoch, where the right ascension and declination coordinate system is defined for the current time rather than a fixed date in the past such as used for the J2000 equinox. Conversion from J2000 coordinates, which most professional astronomers use and is used everywhere else in G-TeCS, to the JNow epoch required by SiTechEXE is done using Astropy's coordinates module.

\end{colsection}

\subsection{Power control}
\label{sec:power}
\begin{colsection}

Similar to the camera, focuser and filter wheel daemons, the power daemon acts as an interface to multiple pieces of hardware. In this case, the daemon is connected to three types of power unit in two locations within the GOTO dome:

\begin{itemize}
    \item Two Power Distribution Units (PDUs)\acroadd{pdu} are located in the main computer rack within the dome. These are used to control and distribute power to a variety of sources, including the primary control computer and ethernet switches in the rack, the mount controller and Windows control NUC on the mount, the rack monitor, Wi-Fi router and LED lights within the dome.
    \item Two additional power relay boxes are attached to the mount boom arms. In the same way that the boom-arm NUCs are used to provide control interfaces instead of running multiple USB cables down the mount, these relays are used to provide and control power to the NUC and hardware (cameras, focusers and filter wheels).
    \item Two Uninterruptible Power Supplies (UPSs)\acroadd{ups} are also located in the rack. These are battery devices that provide backup power in the event of mains supply failure. The first of these is connected directly to the dome, so in case of a power failure the dome has its own supply to enable it to close. The second is connected to the other power units described above.
\end{itemize}

Each power outlet in any of the above units can be turned on, off or rebooted (switched off and then back on again after a short delay). Each outlet has a unique name assigned, and multiple outlets can be grouped together to be controlled using a single command similar to the commands for the exposure queue daemon (for example \code{power~off~cam1,cam2,cam3}). The FLI hardware (cameras, focusers and filter wheels) are usually powered down during the day, all other hardware including the dome and mount is left on. Power to the dehumidifier unit is controlled by the dome daemon as described in \aref{sec:dome}.

The rack PDUs and UPSs used by GOTO are manufactured by Schneider Electric (previously APC)\footnote{\url{https://www.apc.com}}, and are communicated with using \acro{snmp} commands over the network using the Linux snmpget and snmpset utilities. The relay boxes were manufactured for GOTO using Devantech ETH8020 ethernet boards\footnote{\url{https://www.robot-electronics.co.uk}}, controlled through simple TCP/IP commands. All of these are surrounded by Python wrappers within the power daemon.

\end{colsection}

\section{Summary and Conclusions}
\label{sec:gtecs_conclusion}

\begin{colsection}

In this chapter I have described the key elements of the GOTO Telescope Control System (G-TeCS).

Several options were considered for the GOTO control system, and based on the unique requirements of GOTO a custom system based on the pt5m software was ultimately decided on. I described the fundamental features of this new system, G-TeCS, in particular how the control system is built around standalone daemon programs. Adopting this system proved to be a successful decision, as it provided the flexibility required for GOTO's multi-telescope design. In the future the daemon-based system will provide the basis to expand the control system to multiple independent mounts (see \aref{sec:gtecs_future}).

I went on to describe the hardware control functionality of G-TeCS, and how each type of hardware (cameras, mount, dome etc) is controlled through the associated software daemons. This provides the foundations of the control system, but on its own it still requires an observer to operate the telescope. In the following chapter (\aref{chap:autonomous}) I describe the higher-level software within G-TeCS that replaces the human operator, and allows GOTO to function as a fully-robotic telescope.

\end{colsection}

\chapter{Autonomous Observing}
\label{chap:autonomous}

\chaptoc{}

\section{Introduction}
\label{sec:autonomous_intro}

\begin{colsection}

Continuing the description of the GOTO Telescope Control System from \aref{chap:gtecs}, in this chapter I describe the higher level programs written to enable GOTO to operate as a robotic observatory.
\begin{itemize}
    \item In \nref{sec:auto} I outline the additional functionality added to G-TeCS in order to allow the telescope to operate autonomously.
    \item In \nref{sec:pilot} I describe the master control program that operates the telescope when in robotic mode.
    \item In \nref{sec:conditions} I detail how G-TeCS monitors the local conditions, and list the different flags used to judge if it is safe to observe.
    \item In \nref{sec:observing} I give an outline of how targets are observed by the robotic system, and introduce the scheduling system that is expanded further in \aref{chap:scheduling}.
\end{itemize}
All work described in this chapter is my own unless otherwise indicated. As noted before, a description of the G-TeCS control system has previously been published as \citet{Dyer}.

\end{colsection}

\section{Automating telescope operations}
\label{sec:auto}

\begin{colsection}

The hardware control systems described in \aref{sec:hardware_control} provide the basic functions to control and operate GOTO.\@ A human observer could run through a series of simple commands to open the dome, slew the mount to a given target, take exposures once there, and then repeat with other targets for the rest of the night. There is a limited level of autonomy provided by the dome daemon, so the dome will close in bad weather without the delay from a human sending the command, but even that can be disabled if desired. Fundamentally, the software described in \aref{chap:gtecs} provides a perfectly usable human-operated telescope control system.

GOTO, however, was always designed as a fully robotic installation, as described in \aref{sec:goto_motivation}. Therefore an additional level of software was required, to take the place of the observer as the source of commands to the daemons.

\end{colsection}

\subsection{Robotic telescopes}
\label{sec:robotic_telescopes}
\begin{colsection}

One of the first robotic telescopes was the Wisconsin Automatic Photoelectric Telescope \citep{Wisconsin_APT}. Built in 1965, it could take routine observations unattended for several days. Today a huge number and variety of automated telescopes now regularly take observations of the night sky with limited or no human involvement, ranging from wide-field survey projects like the All-Sky Automated Survey for Supernovae \acroadd{asassn} \citep[ASAS-SN,][]{ASAS-SN}, large robotic telescopes like the \SI{2}{\metre} Liverpool Telescope \citep{Liverpool}, to countless small automated observatories around the world\footnote{A list of over 130 active robotic telescopes is available at \href{http://www.astro.physik.uni-goettingen.de/~hessman/MONET/links.html}{\texttt{http://www.astro.physik.uni-}} \\ \href{http://www.astro.physik.uni-goettingen.de/~hessman/MONET/links.html}{\texttt{goettingen.de/\raisebox{0.5ex}{\texttildelow}hessman/MONET/links.html}}.}. While larger facilities still tend to be manually operated, they often have multiple instances of automation in their hardware control or scheduling system; the planned conversion of the 50-year-old Isaac Newton Telescope for the automated HARPS3 survey \citep{INT_robotic} is a recent example of large, established telescopes exploiting the benefits of automation. Larger purely robotic telescopes are also being developed, such as the proposed \SI{4}{\metre} successor to the Liverpool telescope \citep{Liverpool2}. The opportunity for multiple robotic telescopes to be networked together into global observatories has also been exploited by projects like the Las Cumbres Observatory Global Telescope Network \citep{LCO} and the MASTER network \citep{MASTER}.

In G-TeCS, as in the pt5m system before it, the role of the observer is filled by a master control program called the pilot. The pilot sends commands to the daemons, monitors the hardware and attempts to fix any problems that arise. The intention is that the pilot will fully replicate anything a trained on-site observer would be required to do. In order to manage this there are several auxiliary systems and additional support daemons that the pilot confers with: the conditions daemon monitors weather and other system conditions, the sentinel daemon listens for alerts and enters new targets into the observation database and the scheduler daemon reads the database and calculates which target the pilot should observe. Each of these systems are described in this chapter.

\end{colsection}

\subsection{System modes}
\label{sec:mode}
\begin{colsection}

Although GOTO is a robotic telescope, sometimes it is necessary for a human operator to take control if one of the automated scripts fails or a situation arises that is easier to deal with manually. One example was taking observations of the asteroid Phaethon \citep{Phaethon}: G-TeCS was not designed to observe solar system objects, and although the mount allows non-sidereal tracking there was no way to add a pointing into the database without fixed coordinates. Therefore, it was necessary for a human observer to determine and slew to the coordinates of the asteroid as it moved past the Earth. There are also cases when it is important that the automated systems are disabled: if work is being done to the hardware on-site it could be dangerous if the system still tried to move the mount or dome autonomously.

\begin{table}[t]
    \begin{center}
        \begin{tabular}{c|ccccc} %
            mode &
            pilot &
            day marshal &
            dome autoclose &
            dome alarm &
            \code{hatch} flag
            \\
            \midrule
            \code{robotic} &
            \textcolor{Green}{active} &
            \textcolor{Green}{active} &
            \textcolor{Green}{enabled} &
            \textcolor{Green}{enabled} &
            \textcolor{Green}{active}
            \\[5pt]
            \code{manual} &
            \textcolor{Orange}{paused} &
            \textcolor{Green}{active} &
            \textcolor{Orange}{adjustable} &
            \textcolor{Orange}{adjustable} &
            \textcolor{Red}{ignored}
            \\[5pt]
            \code{engineering} &
            \textcolor{Red}{disabled} &
            \textcolor{Red}{disabled} &
            \textcolor{Red}{disabled} &
            \textcolor{Red}{disabled} &
            \textcolor{Red}{ignored}
            \\
        \end{tabular}
    \end{center}
    \caption[System mode comparison]{
        A comparison of the three G-TeCS system modes. In \code{robotic} mode all automated systems are enabled, in \code{engineering} mode they are all disabled, and in \code{manual} mode the pilot is paused and the observer can disable other systems if desired.
    }\label{tab:modes}
\end{table}

G-TeCS deals with manual operation by having an overall system mode flag stored in a datafile, which is checked by the automated systems before activating. There are three possible modes, outlined below and summarised in \aref{tab:modes}.

\begin{itemize}
    \item \code{robotic} mode is the default. In this case it is assumed that the system is completely automated and therefore could move at any time. In this mode during the night the pilot will be in complete control of the telescope, and the dome will automatically close in bad weather. The dome entry hatch being open is also treated as a critical conditions flag (see \aref{sec:conditions_flags}).

    \item \code{manual} mode is designed for manual observing, either on-site or remotely. In this mode the pilot will be paused and so will not interrupt commands sent by the observer. The dome will still sound the alarm when moving and autoclose in bad weather by default, but both can be disabled. It is intended that they should only be disabled if there is an observer physically present in the dome, otherwise the dome should still be able to close automatically when observing remotely.

    \item \code{engineering} mode is designed to be used if there are workers on site, when the hardware moving automatically could be dangerous. All of the dome systems are automatically disabled, and the pilot and day marshal will refuse to start. Leaving the system in this state for long periods of time is undesirable, and so it should only be used while work is ongoing or the telescope is completely deactivated.
\end{itemize}

\newpage

\end{colsection}

\subsection{Slack alerts}
\label{sec:slack}
\begin{colsection}

Although when in robotic mode GOTO is a completely autonomous system, it is still important that it does not operate completely unsupervised. As the GOTO collaboration has adopted the Slack messaging client\footnote{\url{https://slack.com}} for instant messaging and collaboration it was decided that the telescope control system should send reports automatically to a dedicated Slack channel. This was implemented through the Python Slack API package (\pkg{slackclient}\footnote{\url{https://python-slackclient.readthedocs.io}}), and has been widely adopted throughout G-TeCS.\@

The two most detailed Slack messages are the startup report, sent by the night marshal within the pilot (see \aref{sec:night_marshal}), and the morning report sent by the day marshal (see \aref{sec:day_marshal}). Examples of both are shown in \aref{fig:pilot_slack}. The startup report includes a summary of the current condition flags (see \aref{sec:conditions_flags}), links to the site weather pages, the external webcam view and the latest IR satellite image over La Palma. The morning report includes the internal webcam view and automatically-generated plots showing what the pilot observed last night and the current status of the all-sky survey.

Several other functions within the pilot send short messages to Slack when called. For example, as shown in \aref{fig:pilot_slack}, the pilot sends a message when the script starts and completes and when the dome opens or closes. A message will also be sent if the conditions turn bad, if the system mode changes, or if a hardware error is being fixed (see \aref{sec:monitors}). The pilot sends the majority of messages to Slack, but other daemons can also send their own alerts if nessesary. For example, the dome daemon sends a message when it enters lockdown (see \aref{sec:dome}), and the sentinel sends a series of messages whenever it processes an interesting alert through GOTO-alert (see \aref{sec:sentinel} and \aref{sec:event_slack}).

\begin{figure}[p]
    \begin{center}
        \includegraphics[width=\linewidth]{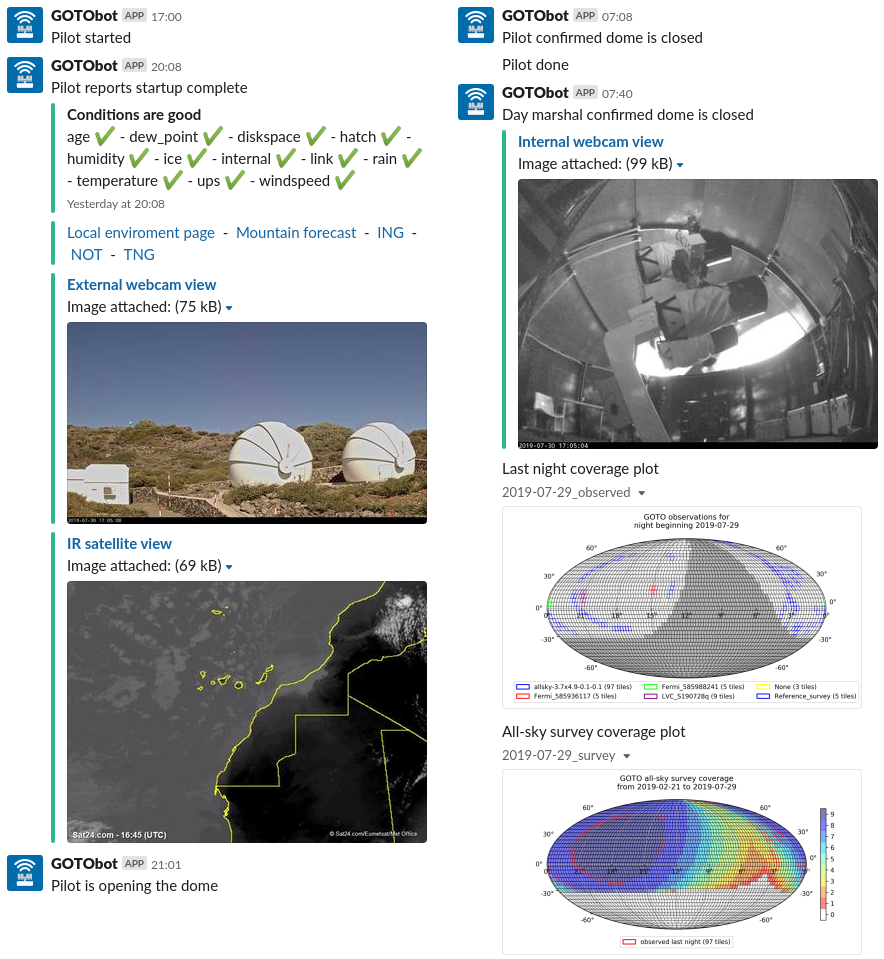}
    \end{center}
    \caption[Slack messages sent by the pilot and day marshal]{
        Slack messages sent by the pilot and day marshal on a typical night. The pilot reports when it starts automatically at 5pm, then the night marshal sends out the startup report when the STARTUP task has completed. The pilot also sends out messages when it is opening and closing the dome, and when it finishes in the morning. The day marshal later independently confirms the dome is closed and sends out its own morning report.
    }\label{fig:pilot_slack}
\end{figure}

\end{colsection}

\section{The pilot}
\label{sec:pilot}

\begin{colsection}

The pilot is a Python script, \code{pilot.py}, not a daemon. It is run once each night; started automatically at 5pm by the Linux cron utility, it runs through to the morning, quits, and then is started again in the afternoon. This happens every day, unless the system is in engineering mode.

\end{colsection}

\subsection{Asynchronous programming}
\label{sec:async}
\begin{colsection}

The pilot is written as an \textit{asynchronous} program, using the AsyncIO package from the Python standard library (\pkg{asyncio}). An asynchronous program is one where its code runs in separate parallel routines, which are switched between as required, and should not be confused with programs that are written utilising multiple processes or threads that run in parallel. See \aref{fig:async} for a graphical comparison between the two methods.

\begin{figure}[p]
    \begin{center}
        \includegraphics[width=0.89\linewidth]{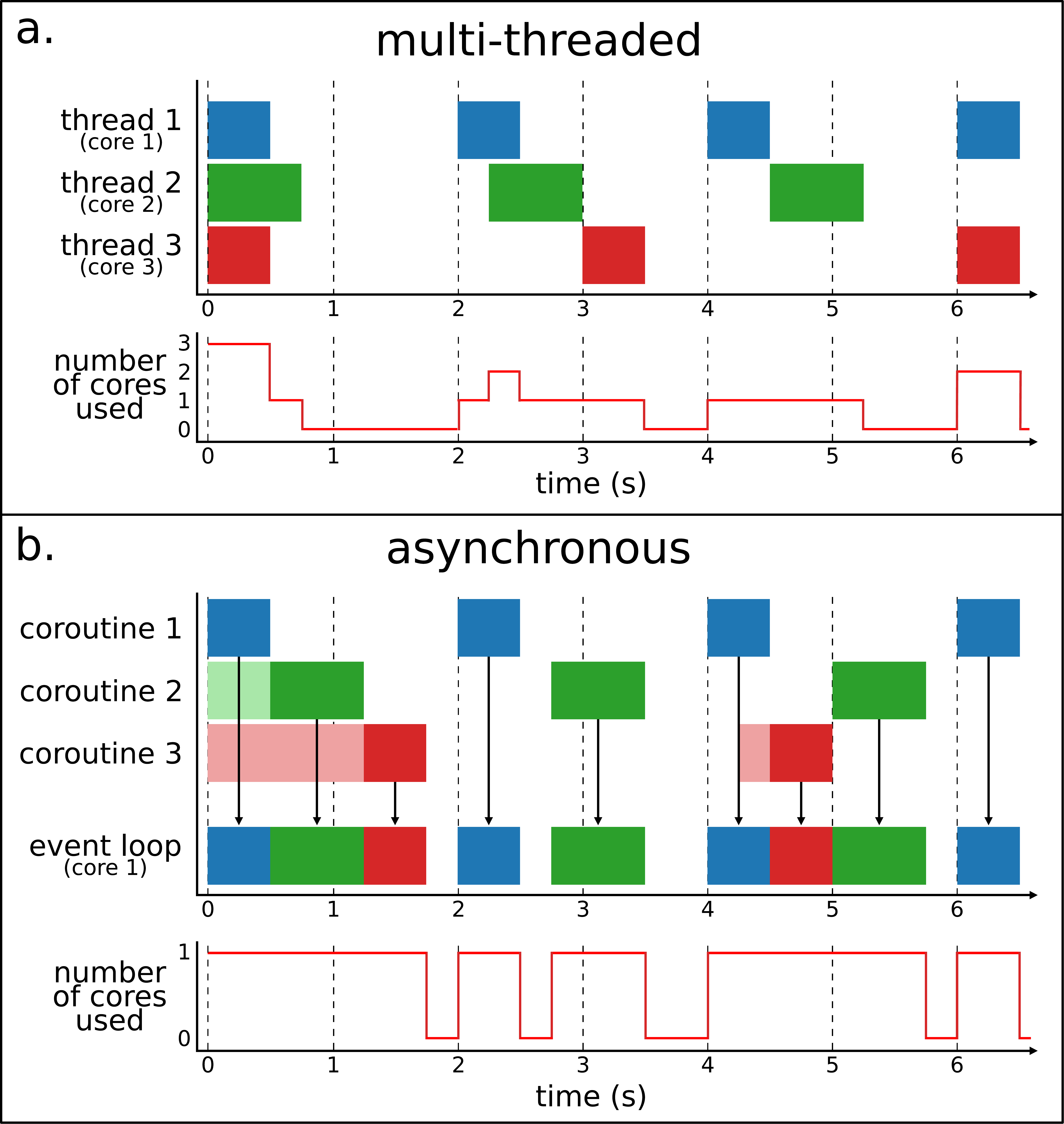}
    \end{center}
    \caption[Multi-threaded verses asynchronous programming]{
        A comparison of multi-threaded verses asynchronous programming. This example uses a \textcolorbf{NavyBlue}{blue} task that takes \SI{0.5}{\second} to execute and then waits for \SI{1.5}{\second}, a \textcolorbf{Green}{green} task that takes \SI{0.75}{\second} and waits for \SI{1.5}{\second} and a \textcolorbf{Red}{red} task that takes \SI{0.5}{\second} and waits for \SI{2.5}{\second}. These times are exaggerated, typically pilot tasks wait for between 10--\SI{60}{\second}.
        The upper plot shows three tasks with different execution periods (solid blocks) and wait times (blank) running in a multi-threaded program. Each task is being run in an independent parallel thread on its own core, even though they rarely overlap and it is uncommon for multiple cores to be in use at the same time.
        The lower plot shows the same three tasks running as coroutines in an asynchronous program. The event loop decides which coroutine to run on the single core, represented by the black arrows. This does lead to some coroutines being left waiting (lighter blocks) until the current one finishes, and, as routines can be delayed, it is not suitable for checks that need to happen at exact frequencies. However, the overall core usage is much more efficient.
    }\label{fig:async}
\end{figure}

An example of a simple task might be monitoring a particular source of data, like a weather station. It would contain a function to download the current weather information from the external mast, and then a \code{sleep} command to wait for 10 seconds, which when put inside a loop will ensure that the weather information is queried and updated every 10 seconds. If this loop was called in a multi-threaded program then the thread will be held up for a majority of the time not doing anything between checks. If there were multiple threads, for example checking different masts, then there could be no coordination between them and the whole program would end up being very inefficient. There are also other issues with multi-threaded programs, including input/output and sharing data between threads.

Asynchronous code contains multiple parallel \textit{coroutines}. The program itself runs an \textit{event loop}, which is a function with the job of choosing between the different coroutines to execute in the main thread. In an asynchronous version of the weather-monitoring program instead of a \code{sleep} function each coroutine would include an \code{await} function. When a routine reaches an \code{await} command it is suspended for the given time period and control is passed back to the event loop, which then chooses which of the other suspended routines should be run. Importantly, when the coroutine resumes it remembers where it stopped and continues from that point. The asynchronous style of writing code is ideally used with multiple coroutines that contain short functions with wait periods between when they need to be called again, and the pilot is a good example of this. The pilot runs a single-threaded event loop with multiple coroutines, which execute commands and then pause using the \code{await} command to allow other routines to be run.

One complication with the asynchronous model is handling errors within a coroutine. Should one coroutine block the thread control will never be returned to the event loop, meaning tasks in other coroutines will never be carried out. Within the G-TeCS pilot each task has a set timeout which will trigger an error should it take too long to complete, and a failure in any of the coroutines will raise an error within the pilot.

\end{colsection}

\subsection{Check routines}
\label{sec:checks}
\begin{colsection}

The coroutines within the pilot can be separated into two types: the check routines and the night marshal. Most of the coroutines are designed as monitors to regularly check different parts of the system, which fits well into the asynchronous model. These check routines are as follows:

\begin{itemize}

\item \code{check\_flags} is a routine that monitors the system flags, most notably those created by the conditions daemon (see \aref{sec:conditions}). If any of the conditions flags are bad then the dome daemon will enter lockdown and close the dome on its own (see \aref{sec:dome}), but the \code{check\_flags} routine will abort exposures, pause the pilot and ensure it is not resumed until the flag is cleared. When the pilot is paused the dome will close, the mount will park, and the night marshal (see below) will not trigger any more tasks. When conditions are clear again the pilot will reopen the dome and allow normal operations to be restored. The \code{check\_flags} routine also monitors the system mode and will pause the pilot if it is set to manual mode or exit if set to engineering mode (see \aref{sec:mode}).

\item \code{check\_scheduler} is a routine that queries the scheduler daemon (see \aref{sec:scheduler}) every 10 seconds to find the best job to observe. If the pilot is currently observing the scheduler will either return the database ID of the current pointing, in which case the pilot will continue with the current job, or a new ID which will lead to the pilot interrupting the current job and moving to observe the new one. If the pilot is not currently observing (either it is the start of the night, resuming from being paused or the previous pointing has just completed) then it will begin observing whatever the scheduler returns. The details of how the scheduler decides which target to observe are given in \aref{chap:scheduling}. The ID returned is then passed to the observe (OBS) task run by the night marshal.

\item \code{check\_hardware} monitors the hardware daemons (see \aref{sec:hardware_control}), checking every 60 seconds that they are all reporting their expected statuses. It does this using the hardware monitor functions described in \aref{sec:monitors}. If an abnormal status is returned then the pilot will pause, and a series of pre-set recovery commands generated by the monitor are executed in turn. While in recovery mode the pilot will check the monitors more frequently. If the commands work and the status returns to normal the pilot is resumed, but if the commands are exhausted without the problem being fixed then a Slack alert is issued reporting that the system requires human intervention and the pilot triggers an emergency shutdown.

\item \code{check\_dome} is a backup to the primary hardware check routine. \code{check\_hardware} does monitor the dome along with the other hardware daemons, but \code{check\_dome} provides a simple, dedicated backup to ensure the dome is closed when it should be and to raise the alarm if it is not.

\end{itemize}

\newpage

\end{colsection}

\subsection{Monitoring the hardware}
\label{sec:monitors}
\begin{colsection}

One of the important tasks that the pilot is required to do is monitoring the status of the various system daemons, and therefore the hardware units they are connected to. If any problems are detected (e.g.\ hardware not responding) the easiest automated response would be to shut down everything and send a message for a human to intervene. However this would be unnecessary in the case of small problems that could be easily fixed with one command, and it would be much better if the pilot could identify the problem and issue the command itself. The other benefit of this is a much faster reaction time than potentially needing to wake a human operator in the middle of the night, this is important both to minimise observing time lost and also potentially save the hardware by, for example, making sure the dome is closed in bad weather.

Therefore, a system was created to enable the pilot to attempt to respond and fix any errors that occur itself. This is done within the \code{check\_hardware} coroutine though a series of hardware monitor Python classes, one for each of the daemons (i.e. \code{DomeMonitor}, \code{CamMonitor} etc.). Each daemon has a set of recognised statuses, representing the current hardware state, and a set of valid modes which represent the expected state. The current status is fetched from the hardware daemon, the mode is set by the pilot, and the hardware checks consist of comparing the two to discover if there are any inconsistencies. For example, the dome daemon can have current statuses of \code{OPEN}, \code{CLOSED} or \code{MOVING} (or \code{UNKNOWN}), and its valid modes are just \code{OPEN} and \code{CLOSED}. At the start of the night when the pilot starts the dome should be in \code{CLOSED} mode, and the pilot only switches it to \code{OPEN} mode when it is ready to open the dome. If when a check is carried out the dome is in \code{CLOSED} mode but the current status is reported as not \code{CLOSED} then that is a problem, and the hardware check function returns that it has detected an error with the dome. These checks can have timeouts associated with each status. For example, if the dome is in \code{CLOSED} mode and is reported as \code{MOVING} that is not necessarily an error, as it might be currently closing. The hardware monitor stores the time since the hardware status last changed, so if the dome reports that it has been in the \code{MOVING} state for longer than it should normally take to close ($\sim$\SI{90}{\second}) then that raises an error. This example used states specific to that hardware, but every daemon also has various other possible states and errors --- for example if the daemon is not running, or is running and not responding.

When one of the monitor checks returns an error then the pilot will take action as described within the \code{check\_hardware} routine: pause night marshal (see below), stop any current tasks and send a Slack alert to record the error. But instead of stopping there, the monitor goes on to attempt to recover from the error and fix the problem. In the same way that a human observer would run though a series of commands in order to solve the problem, each monitor has a defined set of recovery steps to be run through depending on the error reported. Continuing with the previous example, if the dome reports \code{OPEN} when in \code{CLOSED} mode then the first recovery step is simple: execute the command \code{dome~close}. Each step then has a timeout value and an expected state if the recovery command worked. If after 10 seconds the status of the dome has not changed from \code{OPEN} to \code{MOVING}, then the error is persisting and more actions need to be taken. If however the dome daemon reports that the dome is moving then the error is not cleared immediately, only when the status finally reaches \code{CLOSED}. As mentioned previously, should a monitor run out of recovery steps then the pilot will send out an alert that there is nothing more that it can do and will attempt an emergency shutdown.

Using the above method, the vast majority of minor hardware issues can be solved by the pilot without the need for human intervention. Every time the recovery steps are triggered a message is sent to Slack (see \aref{sec:slack}) containing the error code and the steps required to fix it, so it is easy to then go back and examine why the error occurred and how to prevent it in the future.

\newpage

\end{colsection}

\subsection{The night marshal}
\label{sec:night_marshal}
\begin{colsection}

The check routines described in \aref{sec:checks} are support tasks for the primary routine, which is called the \code{night\_marshal}. Unlike the check routines, the night marshal does not contain a loop, instead it runs through a list of tasks as the night progresses, based on the altitude of the Sun. Each task is contained in a separate Python observation script, which contains the commands to send to the hardware daemons (see \aref{sec:scripts}). Each is run by spawning a new coroutine, meaning that while they are running the other routines --- such as the check tasks --- can continue. In the order they are performed during the night, the night marshal tasks are:

\begin{enumerate}

\item STARTUP, run immediately when the pilot starts. The \code{startup.py} script powers on the camera hardware, unparks the mount, homes the filter wheels and cools the CCDs down to their operating temperature of \SI{-20}{\celsius}. Once startup has finished the pilot will send a report of the current conditions to Slack (see \aref{fig:pilot_slack}).

\item DARKS, run after the system start up is complete before opening the dome. This executes the \code{take\_biases\_and\_darks.py} script to take bias and dark frames at the start of the night.

\item OPEN, run once the Sun reaches \SI{0}{\degree} altitude. It simply executes the \code{dome~open} command. If the pilot is paused due to bad weather or a hardware fault then the night marshal will wait and not open until the weather improves or the fault is fixed. If it is never resolved then the night marshal will remain at this point until the end of the night and the shutdown timer runs out (see below).

\item FLATS, run once the dome is open and the Sun reaches \SI{-1}{\degree}. This executes the \code{take\_flats.py} script, which moves the telescope into a position pointing away from the Sun and then takes flat fields in each filter, stepping in position between each exposure and automatically increasing the exposure time as the sky darkens. See \aref{sec:flats} for details of the flat field routine.

\item FOCUS, run once the Sun reaches \SI{-11}{\degree}. This executes the \code{autofocus.py} script, which finds the best focus position for each of the unit telescopes. See \aref{sec:autofocus} for details of the autofocus routine. If the routine fails for any reason the previous nights' focus positions are restored.

\item OBS (short for ``observing''), begun once autofocusing is finished and continuing for the majority of the night until the Sun reaches \SI{-12}{\degree} in the morning. When a database ID is received from the scheduler via the \code{check\_schedule} routine the \code{observe.py} script is executed. The script queries the observation database (see \aref{sec:obsdb}) to get the coordinates and exposure settings for that pointing and then sends the commands to the mount and exposure queue daemons. Once a job is finished, either through completing all of its exposures or being interrupted, the entry in the database is updated and the routine starts observing the next job from the scheduler, starting the \code{observe.py} script again with the new pointing ID.\

\item FLATS is repeated once the Sun reaches \SI{-10}{\degree} in the morning, using the same script but this time decreasing the exposure times as the sky brightens.

\end{enumerate}

Once the night marshal has completed all of its tasks it exits and triggers the \code{shutdown.py} script, which powers off the cameras, parks the mount and ensures the dome is closed. Once this is finished the pilot quits. In addition there is a separate night countdown timer within the pilot, which will trigger the shutdown once the Sun reaches \SI{0}{\degree} in the morning. Normally the night marshal will have finished and triggered the shutdown long before that point, but the countdown acts as a backup ensuring that if there is a problem with the night marshal the pilot will still trigger a shutdown.

\newpage

It is also possible for the pilot to trigger an emergency shutdown during the night. This triggers the same \code{shutdown.py} script observing script, with the only difference being that it ensures the dome is closed first. An emergency shutdown will be triggered by the pilot only in situations that it could not recover from without human intervention. Notably, this occurs if the hardware monitors called by the \code{check\_hardware} routine reach the end of a daemon's recovery steps without fixing the problem.

\end{colsection}

\subsection{The day marshal}
\label{sec:day_marshal}
\begin{colsection}

The day marshal is a completely separate script (\code{day\_marshal.py}) which provides a counterpart and backup to the pilot (the name is the mirror of the night marshal). The script is run as a cron job like the pilot, but starts in the early morning rather than the late afternoon. The day marshal is a much simpler script, with only one key task --- to wait until dawn and then check that the dome is closed. In this sense it is specifically a backup for the pilot's inbuilt night countdown timer, and as it is completely independent of the pilot it will run even if the pilot script has frozen or crashed during the night.

If the day marshal finds that the dome is still open when it runs it will send out Slack alerts that the system has failed, and then try closing the dome itself by sending commands to the dome daemon. So far this has not occurred, aside from deliberately during on-site tests. If all is well the day marshal will send out a Slack report as shown in \aref{fig:pilot_slack}, again as a mirror of the report that the night marshal sends after startup. This report will confirm that the dome is closed, and also contains some simple plots showing the targets the pilot observed last night.

\end{colsection}

\section{Conditions monitoring}
\label{sec:conditions}

\begin{colsection}

Perhaps the most important role of the autonomous systems is monitoring the on-site conditions. The weather at the site on La Palma is typically very good, however storms can affect the mountain-top observatory, especially in the winter months (see \aref{sec:challenges}). It is vital that the dome is closed whenever the weather turns bad, or in any other abnormal circumstances. For example, if the site loses power or internet connection it is better to stop observing and close the dome in case they are not restored quickly. The system had to be trusted to close in an emergency before it was allowed to run completely without human supervision.

\end{colsection}

\subsection{The conditions daemon}
\label{sec:conditions_daemon}
\begin{colsection}

The conditions daemon is a support daemon that runs on the central observatory server in the SuperWASP building on La Palma alongside the observation database (see \aref{fig:flow}). The daemon is run on the central server because it deals with site-wide values, so when the second GOTO telescope on La Palma is built it is envisioned that they will both share the same conditions daemon (see \aref{sec:gtecs_future}).

The daemon takes in readings from the three local weather stations next to the GOTO dome on La Palma (shown in \aref{fig:conditions}) every 10 seconds, as well as other sources such as internal sensors. The daemon processes these inputs into a series of output flags, which have a value of \code{0} (good), \code{1} (bad) or \code{2} (error). If any of the flags are marked as not good (i.e.\ the sum of all flags is $>0$) then the overall conditions are bad. The output flags are monitored by the dome daemon and the pilot, if the conditions are bad the dome will enter lockdown and automatically close if it is open (see \aref{sec:dome}) and the pilot \code{check\_flags} routine will trigger the pilot to pause observations (see \aref{sec:pilot}).

\begin{figure}[t]
    \begin{center}
        \includegraphics[height=0.5\linewidth]{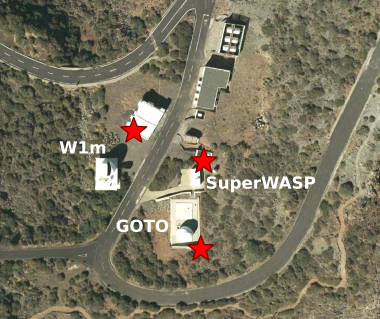}
        \includegraphics[height=0.5\linewidth]{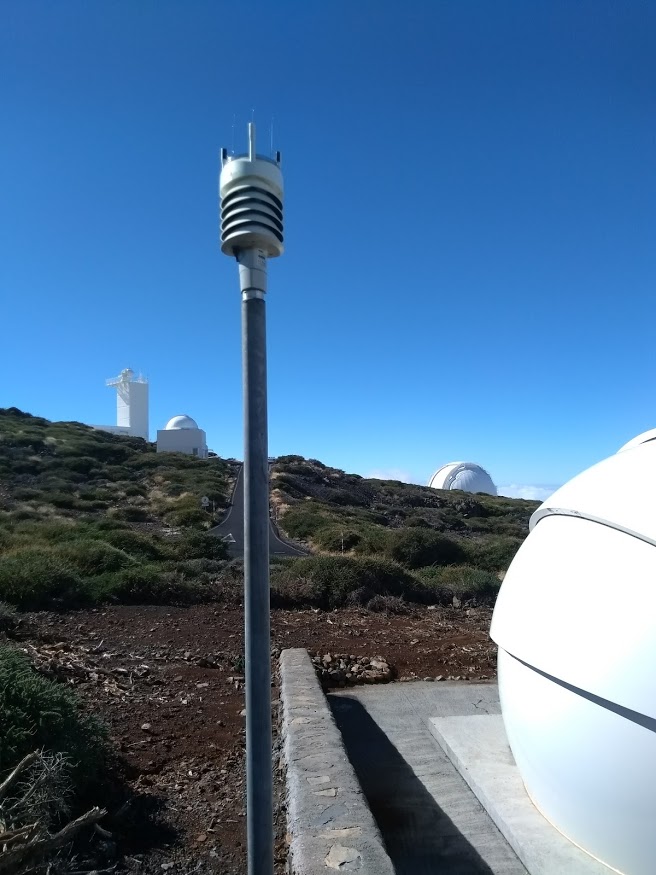}
    \end{center}
    \caption[Locations of the three weather masts on La Palma]{
        On the left the locations of the three local weather masts on La Palma are marked by the \textcolorbf{Red}{red} stars (see \aref{fig:orm} for the context of the site). There are three masts around GOTO:\@ one next to the GOTO platform (shown on the right), one on the SuperWASP shed and a third on the liquid nitrogen plant next to W1m.
    }\label{fig:conditions}
\end{figure}

\newpage

\end{colsection}

\subsection{Conditions flags}
\label{sec:conditions_flags}
\begin{colsection}

Each conditions flag has a limit below or above which the flag will turn from good to bad. For categories with multiple sources (for example the three local weather stations each give an independent external temperature reading) then the limit will be applied to each, and if \textit{any} are found to be bad then the flag is set. It follows therefore that \textit{all} the conditions sources must be good for the flag to be set to good. Each category also has two parameters: the bad delay and the good delay. These are the time the conditions daemon waits between an input going bad/good and setting the flag accordingly, which has the effect of smoothing out any sudden spikes in a value and ensures the dome will not be opening and closing too often.

The conditions flags can be grouped into three categories, divided according to severity. The latest version of G-TeCS contains 13 flags, listed in \aref{tab:conditions_flags}. An explanation of the different categories, and the flags within each, is given below.

\begin{table}[p]
    \begin{center}
        \begin{tabular}{c|cccc} %
            Flag name           & Criteria measured & Bad criteria      & Good criteria     & Category    \\
            \midrule
            \code{dark}         & Sun altitude
                                & > \SI{0}{\degree}
                                & < \SI{0}{\degree}
                                & info
            \\[20pt]
            \code{clouds}       & IR opacity
                                & > 40\%
                                & < 40\%
                                & info
            \\[20pt]
            \code{rain}         & Rain detectors
                                & \makecell{\code{True} \\ for \SI{30}{\second}}
                                & \makecell{\code{False} \\ for \SI{60}{\minute}}
                                & normal
            \\[20pt]
            \code{windspeed}    & Wind speed
                                & \makecell{> \SI[per-mode=symbol]{35}{\kilo\metre\per\hour} \\ for \SI{2}{\minute}}
                                & \makecell{< \SI[per-mode=symbol]{35}{\kilo\metre\per\hour} \\ for \SI{10}{\minute}}
                                & normal
            \\[20pt]
            \code{humidity}     & Humidity
                                & \makecell{> 75\% \\ for \SI{2}{\minute}}
                                & \makecell{< 75\% \\ for \SI{10}{\minute}}
                                & normal
            \\[20pt]
            \code{dew\_point}   & \makecell{Dew point \\ above ambient \\ temperature}
                                & \makecell{< +\SI{4}{\degree} \\ for \SI{2}{\minute}}
                                & \makecell{> +\SI{4}{\degree} \\ for \SI{10}{\minute}}
                                & normal
            \\[20pt]
            \code{temperature}  & Temperature
                                & \makecell{< \SI{-2}{\degree} \\ for \SI{2}{\minute}}
                                & \makecell{> \SI{-2}{\degree} \\ for \SI{10}{\minute}}
                                & normal
            \\[20pt]
            \code{ice}          & Temperature
                                & \makecell{< \SI{0}{\degree} \\ for \SI{12}{\hour}}
                                & \makecell{> \SI{0}{\degree} \\ for \SI{12}{\hour}}
                                & critical
            \\[20pt]
            \code{internal}     & \makecell{Internal \\ temperature \\ \& humidity}
                                & \makecell{< \SI{-2}{\degree} or > 75\% \\ for \SI{1}{\minute}}
                                & \makecell{> \SI{-2}{\degree} and < 75\% \\ for \SI{10}{\minute}}
                                & critical
            \\[30pt]
            \code{link}         & \makecell{Network \\ connection}
                                & \makecell{ping fail \\ for \SI{10}{\minute}}
                                & \makecell{ping okay \\ for \SI{1}{\minute}}
                                & critical
            \\[20pt]
            \code{diskspace}    & \makecell{Free space \\ remaining}
                                & < 5\%
                                & > 5\%
                                & critical
            \\[20pt]
            \code{ups}          & \makecell{Battery power \\ remaining}
                                & < 99\%
                                & > 99\%
                                & critical
            \\[20pt]
            \code{hatch}        & Hatch sensor
                                & \makecell{\code{open} \\ for \SI{30}{\minute}}
                                & \makecell{\code{closed} \\ for \SI{30}{\minute}}
                                & critical
            \\
        \end{tabular}
    \end{center}
    \caption[List of conditions flags and change criteria]{
        A list of all the conditions flags, and the criteria for them to switch from good to bad and bad to good.
    }\label{tab:conditions_flags}
\end{table}

\clearpage

\subsubsection{Information flags}

The first category are the `information' flags. These are assigned values like the other flags, however they are purely for information purposes and do not contribute to the overall decision of whether the conditions are bad or not. In other words, an information flag can be bad, but the overall system conditions still considered good because the flag is not included in the final calculation. The information flag being being bad is not a reason to send the dome into lockdown, however it is still useful information to record. The two current information flags are described below:

\begin{itemize}
    \item \code{dark}: A simple information flag that is bad when the Sun is above the \SI{0}{\degree} horizon and good when it is below. This has no effect on the robotic system, but is useful for human observers.

    \item \code{clouds}: This information flag uses free IR satellite images downloaded from the sat24.com website\footnote{\url{https://en.sat24.com}} to measure a rough cloud coverage value, based on the methods of \citet{clouds}. Although initially trialled as a normal flag, meaning the dome would close when high cloud was detected, the results were not consistent enough and the presence of clouds was more reliably calculated by the zero point measured by the data processing pipeline. The flag remains a useful information source however, and the satellite cloud opacity is added to the image headers to assist in later data quality control checks.
\end{itemize}

Other information flags that have been proposed include seeing (from the ING or TNG seeing monitors on La Palma) and dust (from the TNG aerosols monitor). Both would be useful information to gather and store in image headers, but as they pose little threat to the hardware they are not valid reasons to close the dome unlike the other conditions flags described below.

\newpage

\subsubsection{Normal flags}

The second category contains the `normal' flags, and makes up the conditions flags relating to the external weather conditions. These flags going bad are valid grounds to close the dome, however as they relate to natural events they are not in any way unusual and the pilot can happily remain paused and wait for the flags to clear. The normal flags are described below:

\begin{itemize}
    \item \code{rain}: This flag is set to bad if any of the weather stations report rain, and will only be cleared after 60 minutes of no more rain being reported. In practice rain usually coincides with high humidity, meaning the \code{rain} and \code{humidity} flags often overlap.

    \item \code{windspeed}: This flag gets set if the windspeed is above \SI[per-mode=symbol]{35}{\kilo\meter\per\hour}, with a bad delay of two minutes and a good delay of ten minutes.

    \item \code{humidity}: The humidity limit is 75\%, with a bad delay of two minutes and a good delay of ten minutes.

    \item \code{dew\_point}: The dew point is related to the humidity, and has a limit of \SI{4}{\celsius} above the ambient external temperature (so if the external temperature is \SI{2}{\celsius} then the flag is set to bad if the dew point is \SI{6}{\celsius} or below).

    \item \code{temperature}: The \code{temperature} flag is set if the temperature drops below \SI{-2}{\celsius} for two minutes, and also has a good delay of ten minutes. The telescope can operate in below-freezing temperatures for short amounts of time, but for longer cold periods when ice build-up is a concern see the critical \code{ice} flag below.
\end{itemize}

The limits for some of these flags and how they were determined is discussed further in \aref{sec:conditions_limits} below.

\subsubsection{Critical flags}

The final category are the `critical' flags, for more serious situations that might arise. In early versions of G-TeCS any of these flags turning bad was enough to trigger an emergency shutdown and stop the pilot for the night. However this proved to be an over-reaction, and there were no issues with having the pilot continue, although remaining paused while the flag was bad. The only difference now between `normal' and `critical' flags is that when a critical flag changes a Slack alert is sent out to ensure it is brought to the attention of the human monitors. The critical flags are described below:

\begin{itemize}
    \item \code{ice}: A critical flag which uses the same input as the \code{temperature} flag, but is set to bad if the temperature is below \SI{0}{\celsius} for 12 hours and will only clear if it is constantly above freezing for another 12 hours. These longer timers mean this flag prevents the dome opening after a serious cold period until the temperature is regularly back above freezing, and also gives time for a manual inspection to be carried out to ensure the dome is free of ice.

    \item \code{internal}: A combination flag for the two internal temperature and humidity sensors within the dome. These have very extreme limits, a humidity above 75\% or a temperature below \SI{-2}{\celsius}, which should never be reached inside under normal circumstances due to the internal dehumidifier. This flag therefore is a backup for an emergency case, when either the dehumidifier is not working or the dome has somehow opened in bad conditions (see \aref{sec:challenges}).

    \item \code{link}: The conditions daemon also monitors the external internet link to the site, by pinging the Warwick server and other public internet sites. After 10 minutes of unsuccessful pings the flag is set to bad. It is technically possible for the system to observe without an internet link, and there is a backdoor into the system through the separate SuperWASP network, but it is an unnecessary risk: in an emergency alerts could not be sent out and external users would not be able to log in.

    \item \code{diskspace}: The amount of free disk space on the image data drive is also monitored, with the flag being set to bad if there is less than 5\% of free space available. As images are immediately sent to Warwick and then regularly cleared from the local disk this should never be an issue, but this is a critical conditions flag as if the local disk was full it would prevent any more data being taken.

    \item \code{ups}: The conditions daemon will set the \code{ups} flag if the observatory has lost power and the system UPSs are discharging (see \aref{sec:power}). Brief power cuts do occur on La Palma, but rarely for more than a few minutes as there are on-site backup generators that take over.

    \item \code{hatch}: A critical flag to detect if the access hatch into the dome has been left open. This flag is unique in that it is only valid in robotic mode (see \aref{sec:mode}); when in manual or engineering mode it is assumed that the hatch being opened is a result of someone operating the telescope. But when the system is observing robotically the hatch being open is a problem, as there is no way to close it remotely and in bad weather damage could be caused to the telescope.
\end{itemize}

\subsubsection{Age flag}

There is a 13th pseudo-flag that is not set by the conditions monitor: the \code{age} flag. The output of the conditions daemon is saved in a datafile with a timestamp, and the dome daemon and the pilot monitors this file for changes, instead of querying the conditions daemon directly, to prevent any errors if the conditions daemon freezes or crashes. If the timestamp is out of date compared to the current time (2 minutes by default) then something must have happened to the conditions daemon and the flags are not reliable. The \code{age} flag is then created and set to bad, and it is then treated identically to the other 11 non-info flags when checking the conditions. Note that the \code{age} flag is included in the startup report sent to Slack in \aref{fig:pilot_slack} alongside the others given in \aref{tab:conditions_flags}.

\end{colsection}

\subsection{Determining conditions limits}
\label{sec:conditions_limits}
\begin{colsection}

Setting the conditions flags based on the external weather is a balance between the loss of sky time and the potential risk to the hardware.

\subsubsection{Temperature}

\begin{figure}[t]
    \begin{center}
        \includegraphics[width=\linewidth]{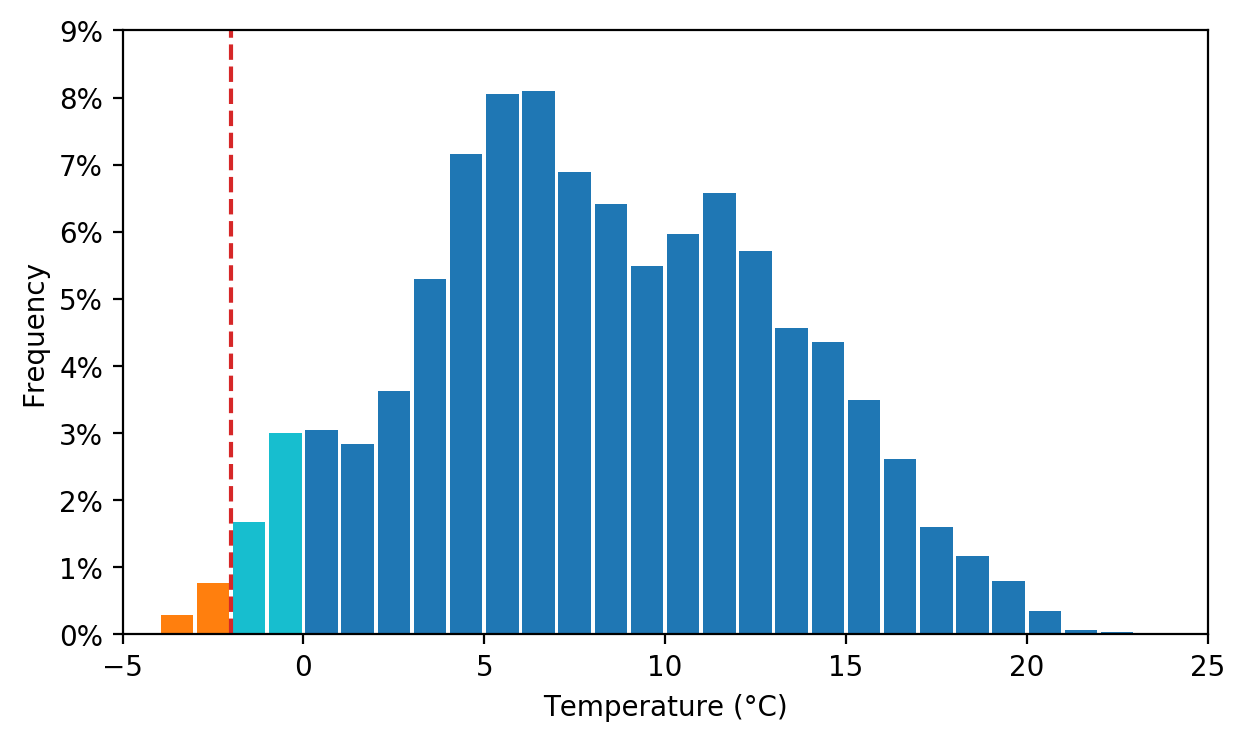}
    \end{center}
    \caption[Histogram of temperature readings]{
        Histogram of temperature readings recorded for all nights in 2018. The temperature was recorded as below the \SI{-2}{\celsius} limit 1.1\% of this period (the \textcolorbf{Orange}{orange} bars) and below \SI{0}{\celsius} (the \textcolorbf{Orange}{orange} and \textcolorbf{Cyan}{cyan} bars) 5.7\% of the time.
    }\label{fig:temperature}
\end{figure}

Originally the temperature limit was set to \SI{0}{\celsius}, meaning the dome would close if any of the conditions masts recorded the temperature as below freezing for more than two minutes (the standard `bad delay' for the normal flags). However the temperature on its own is not a risk to the hardware unless coupled with high humidity, meaning that the limit was later lowered to \SI{-2}{\celsius}. As shown in \aref{fig:temperature} this gained up to an extra 4.6\% of observing time that was previously lost, although note that is an upper limit as other flags might be bad during that period.

\newpage

\subsubsection{Humidity}

\begin{figure}[t]
    \begin{center}
        \includegraphics[width=\linewidth]{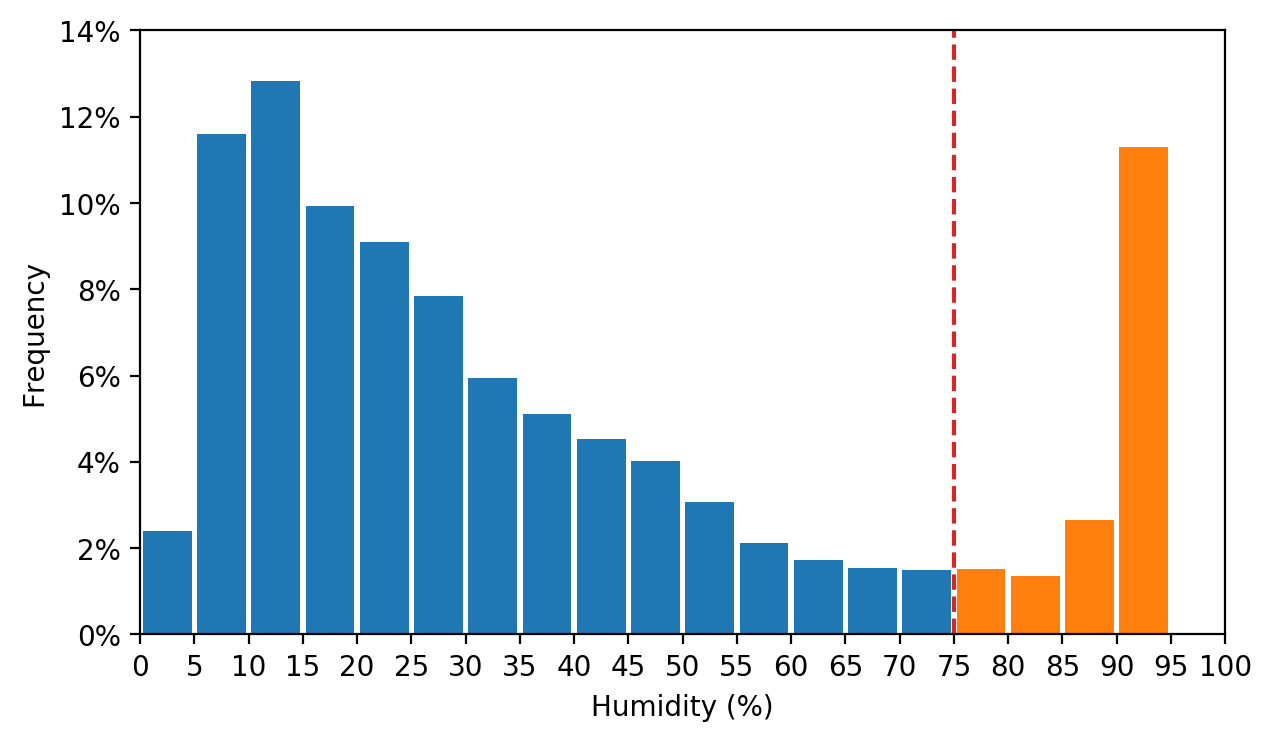}
    \end{center}
    \caption[Histogram of humidity readings]{
        Histogram of humidity readings recorded for all nights in 2018. The humidity was recorded as above the 75\% limit 16.8\% of this period (\textcolorbf{Orange}{orange} bars).
    }\label{fig:humidity}
\end{figure}

\end{colsection}

High humidity at the GOTO site on La Palma is associated with clouds forming in the caldera and spilling over the edge to cover the telescopes. There is a clear bimodal distribution in the site humidity readings shown in \aref{fig:humidity}, a result of fairly distinct high-humidity periods (mainly in the winter) and a range of lower humidities that pose no threat to the hardware. The humidity limit of 75\% is semi-arbitrary, in that if it is surpassed that acts as a forewarning of a high humidity period. Changing the limit to 80\% would not in practice gain much on-sky time as the humidity tends to rise rapidly when clouds are forming, and it would come at a risk of condensation on the hardware (the related dew point measurement is also an important measurement of this).

\newpage

\subsubsection{Windspeed}

\begin{figure}[t]
    \begin{center}
        \includegraphics[width=\linewidth]{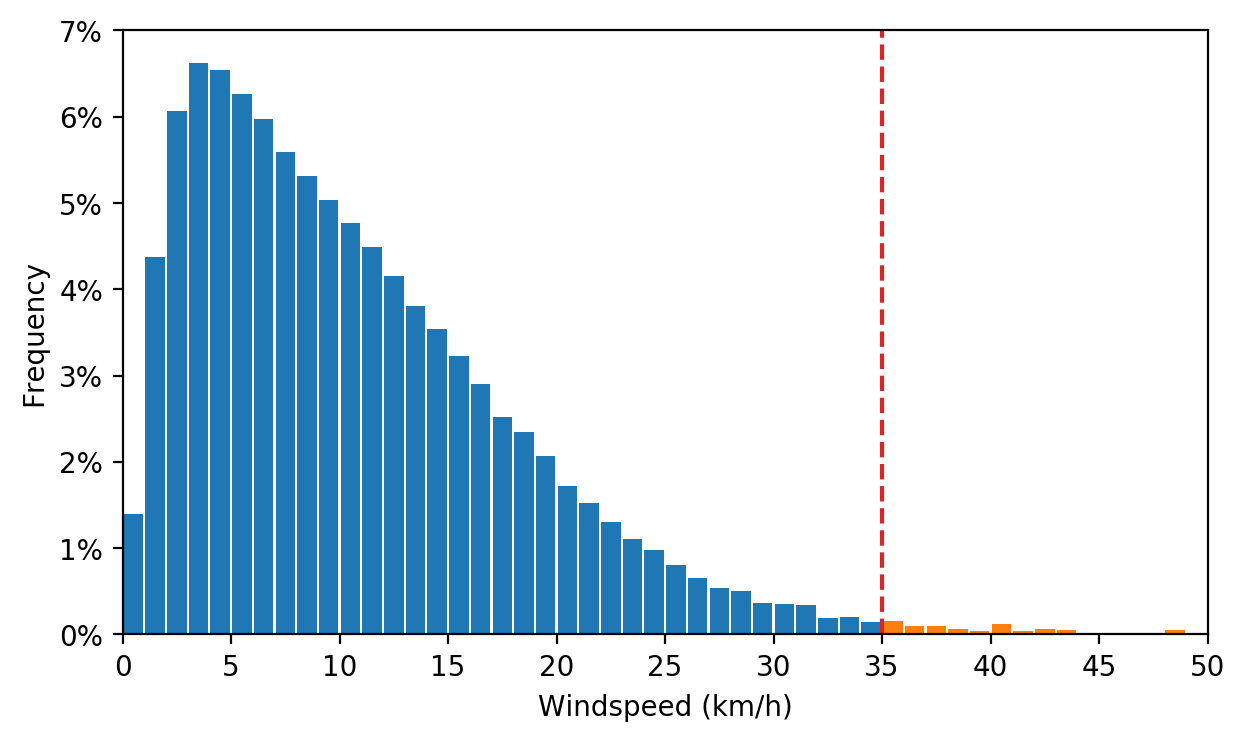}
    \end{center}
    \caption[Histogram of windspeed readings]{
        Histogram of windspeed readings recorded for all nights in 2018. The windspeed was recorded as above the \SI[per-mode=symbol]{35}{\kilo\metre\per\hour} limit 2.3\% of this period (\textcolorbf{Orange}{orange} bars).
    }\label{fig:windspeed}
\end{figure}

The \code{windspeed} flag is set to bad if the wind speed from any of the local masts is recorded as above \SI[per-mode=symbol]{35}{\kilo\meter\per\hour} for more than two minutes. This limit is well below the winds needed to damage the GOTO hardware, but is more a factor of the effect of wind shake on image quality. This is rarely the case, as shown in \aref{fig:windspeed}, and only tends to occur when storms are passing over the island; otherwise the conditions are fairly stable.

The wind limit was previously \SI[per-mode=symbol]{40}{\kilo\metre\per\hour}, but when the full four unit telescope array was installed, with the addition of the light shields, the wind sensitivity of the mount was increased and the high wind limit had to be lowered. It is possible the wind limit will also need to be revisited when the next four unit telescopes are added to the mount.

\section{Observing targets}
\label{sec:observing}

\begin{colsection}

In order for the pilot to function during the night it needs to know what targets to observe. This section describes the architecture within G-TeCS to allow targets to be defined, selected and passed to the pilot, while the details of the scheduling algorithms are discussed in more detail in \aref{chap:scheduling}.

\end{colsection}

\subsection{The observation database}
\label{sec:obsdb}
\begin{colsection}

\begin{sidewaysfigure}[p]
    \begin{center}
        \includegraphics[width=\linewidth]{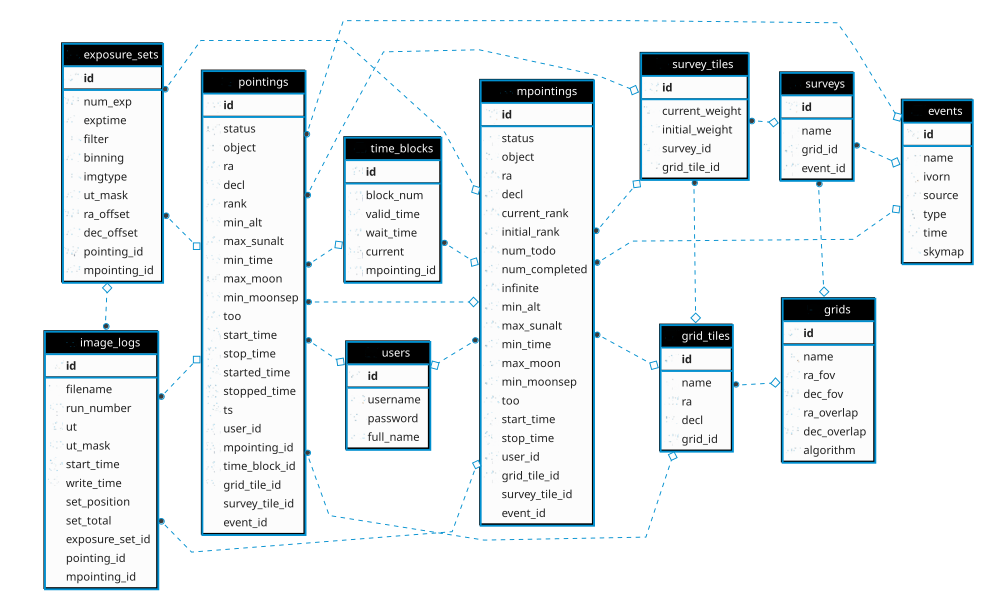}
    \end{center}
    \caption[Relationship diagram for the G-TeCS observation database]{
        Relationship diagram for the G-TeCS observation database.
    }\label{fig:schema}
\end{sidewaysfigure}

The scheduling system for G-TeCS is based around a database known as the observation database or ``ObsDB''. This database is located on the central observatory server hosted by SuperWASP, which not only is a faster machine than the control computer in the dome but in the future will allow a single database to be shared between mounts (see \aref{sec:multi_tel} and \aref{sec:gtecs_future}). The database is implemented using the MariaDB database management system\footnote{\url{https://mariadb.com}}, and is queried and modified using \acro{sql} commands. In order to interact easily with the database within G-TeCS code a separate Python package, ObsDB (\pkg{obsdb}\footnote{\url{https://github.com/GOTO-OBS/goto-obsdb}}), was written as an \acro{orm} package utilising the SQLAlchemy package (\pkg{sqlalchemy}\footnote{\url{https://sqlalchemy.org}}). An entity relationship diagram for the database schema is shown in \aref{fig:schema}.

The primary table in the database is for individual \code{pointings}. These each represent a single visit of the telescope, with defined RA and Dec coordinates and a valid time range for it to be observed within, as well as other observing constraints. Each pointing has a status value which is either \code{pending}, \code{running}, \code{completed} or some other terminal status (\code{aborted}, \code{interrupted}, \code{expired} or \code{deleted}). Ideally a pointing passes through three stages: it is created as \code{pending}, the scheduler selects it and the pilot marks it as \code{running}, then if all is well when it is finished it is marked as \code{completed}. If it stays in the database and never gets observed it will eventually pass its defined stop time (if it has one) and will be marked as \code{expired}. If the pointing is in the middle of being observed but is then cancelled before being completed it will be marked either \code{interrupted} (if the scheduler decided to observe another pointing of a higher priority) or \code{aborted} (in the case of a problem such as having to close for bad weather). The \code{deleted} status is reserved for pointings being removed from the queue before being observed, such as updated pointings being inserted by the sentinel and overwriting the previous ones (see \aref{sec:event_insert}). A representation of the relationship between the pointing statuses and how they progress is shown in \aref{fig:pointings}.

\begin{figure}[p]
    \begin{center}
        \includegraphics[width=\linewidth]{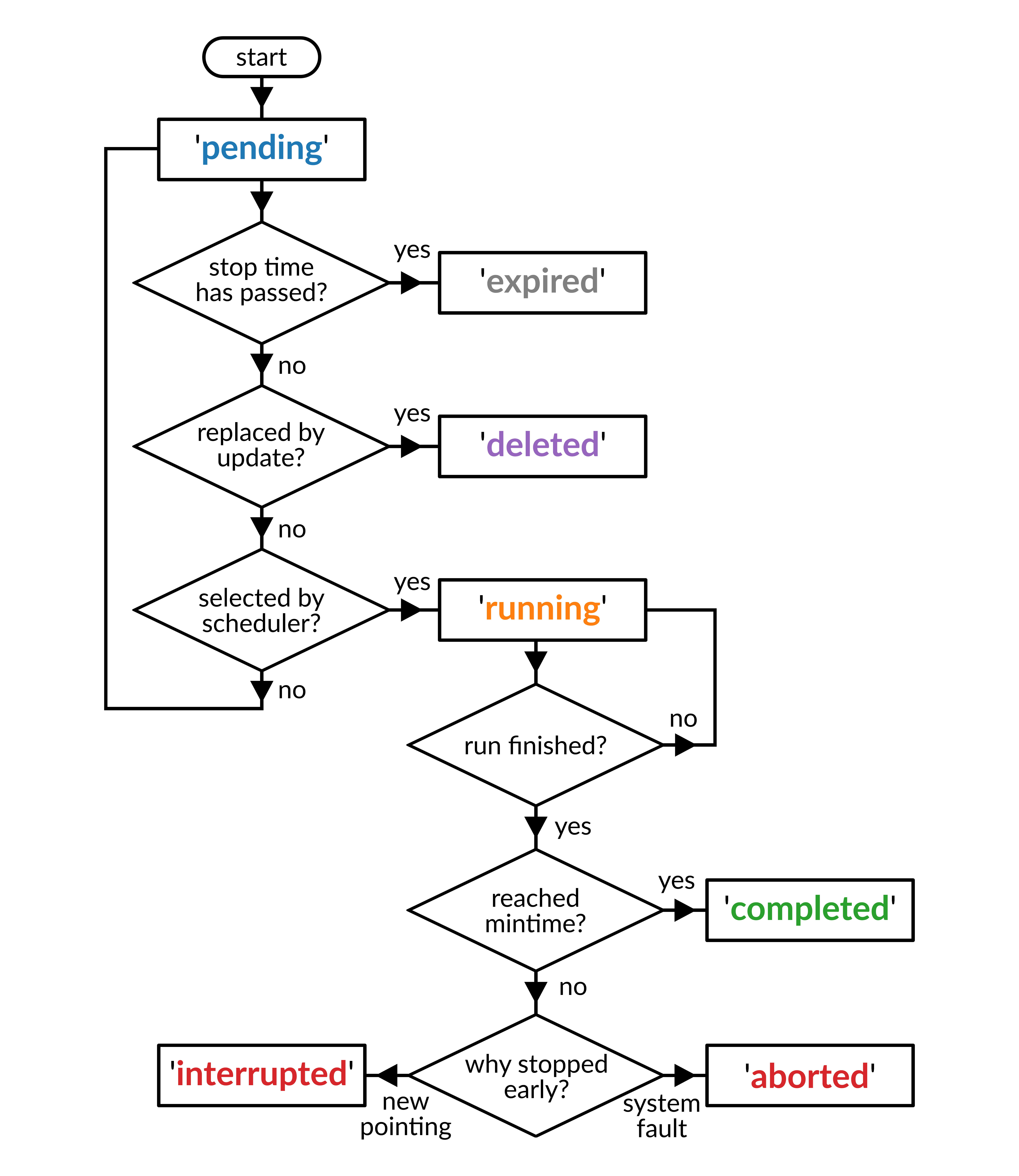}
    \end{center}
    \caption[Pointing status progression flowchart]{
        A flowchart showing how the status of an entry in the pointings table can change.
    }\label{fig:pointings}
\end{figure}

As well as the target information (RA, Dec, name) a pointing entry contains constraints on when they can be observed. Each pointing can have set start and stop times; the scheduler will only select pointings where the current time is within their valid range (and once the stop time has passed they will be marked as \code{expired}). Limits can also be set on minimum target altitude, minimum distance from the Moon, maximum Moon brightness (in terms of Bright/Grey/Dark time) and maximum Sun altitude. These constraints are applied by the scheduler to each pointing when deciding which to observe (see \aref{sec:constraints}), and unless they all pass the pointing is deemed invalid. When created, a pointing is also assigned a rank, usually from 0--9, as well as a True/False flag marking it as a time-critical \acro{too}. These are used when calculating the priority of the pointing, to compare with others in order to determine which is the highest priority to observe (see \aref{sec:ranking}).

The commands to be executed once the telescope is in position are stored in a separate \code{exposure\_sets} table. The table contains the number of exposures to take, the exposure time and the filter to use; so an observation requiring three \SI{60}{\second} exposures in the \textit{L} filter only requires one entry in the table. When the pointing is observed the pilot will read the set information and issue the appropriate commands to the exposure queue daemon (see \aref{sec:exq}), where each exposure is added to the queue and observed in turn.

\newpage

Each entry in the \code{pointings} table can only be observed once, after which it is marked as \code{completed} and is therefore excluded from future scheduler checks (which only consider \code{pending} pointings). For observing a target more than once there also exists the \code{mpointings} table, which contains information to dynamically re-generate pointings for a given target. An mpointing entry is defined with three key values: the requested number of observations, the time each should be valid in the queue and the minimum time to wait between each observation. Each time the database caretaker script is run it looks for any entries in the \code{mpointings} table that still have observations to do and it creates another entry in the \code{pointings} table for that target (this is tracked in a separate \code{time\_blocks} table). Setting the time values allows a lot of control over when pointings can be valid; for example, scheduling follow-up observations a set number of hours or days after an initial pointing is observed (see \aref{sec:event_strategy}).

The three tables described above (\code{pointings}, \code{exposure\_sets} and \code{mpointings}) are the core tables required for observation scheduling. However, there are several other tables defined in the database which are used to group pointings together and relate to GOTO's purpose as a survey instrument. As described in more detail in \aref{sec:gototile}, GOTO observes the sky divided into a fixed grid of individual tiles. The database therefore also contains a \code{grids} table and a \code{grid\_tiles} table, which define the current grid based on the field of view of the telescope. Mapping pointings to the grid is achieved through two more tables, \code{surveys} and \code{survey\_tiles}. A \textit{survey} in this context is a group of tiles that are being observed for a specific reason, one example are the pointings comprising the all-sky survey that GOTO carries out every night. Events that are processed by the alert sentinel might have a skymap that covers multiple tiles, and therefore the set of pointings required to cover it forms a survey within the database (the details of adding event pointings to the database are described in \aref{sec:event_insert}). Each pointing within the survey is linked to a survey tile, and each survey tile is linked to a grid tile of the current grid. The additional field added by the survey tile is a `weighting' column, which allows tiles within a survey to be weighted relative to each other. In the all-sky survey each tile is weighted equally, but in a survey coming from an event skymap the tiles will be weighted by the contained probability within that tile. The scheduler takes this weighting into account when deciding which pointing to observe (see \aref{sec:scheduler_tiebreaker}).

There are two additional tables in the database that are used to contain supporting information: the \code{events} and \code{users} tables. The \code{events} table contains fields such as the event type and source, and is filled by the sentinel when events are processed (see \aref{sec:event_insert}). The \code{users} table connects each pointing to the user who added it to the database. At the moment this is unused, and every pointing is linked to the single generic ``GOTO'' user, but in the future individuals might wish to insert and keep track of their own targets. Finally there is an \code{image\_logs} table that is populated by the camera daemon whenever an image is taken, this builds a record and allows individual images to be traced back to other database entries if required (all connected database IDs are also stored in the image FITS header).

\end{colsection}

\subsection{The sentinel}
\label{sec:sentinel}
\begin{colsection}

In order for targets to be observed by the pilot they must have entries defined in the \code{pointings} table in the observation database. These can be added manually, but for automated follow-up observations they have to be inserted whenever an alert is received. As shown in \aref{fig:flow} this is the job of the sentinel daemon.

In addition the normal control loop, the sentinel daemon includes an independent alert listener loop that is continuously monitoring the transient alert stream output by the 4 Pi Sky event broker \citep{4pisky}, using functions from the PyGCN Python package (\pkg{pygcn}\footnote{\url{https://pypi.org/project/pygcn}}). Should the link to the server fail the daemon will automatically attempt to re-establish the connection every few seconds until it is restored. Alerts come in to the listener and are are appended to an internal queue, and the sentinel also has an additional \code{ingest} command which can be used to manually insert test events or bypass the alert listener. Alerts are then removed from the queue and processed using the handler from the GOTO-alert Python package. The details of how alerts are processed are described in \aref{chap:alerts}, which includes how events are defined, processed, mapped onto the all-sky grid and ultimately added to the database.

The alert listener is a key part of the automated system but was not initially planned to be assigned to its own independent daemon. The pt5m system uses the Comet software \citep{comet} in a separate script independent of any daemons. The advantages to including a dedicated alert listener daemon in G-TeCS, which became known as the sentinel, come from it being integrated into the pilot monitoring systems like the other daemons (described in \aref{sec:monitors}). Should the sentinel daemon crash or not respond to checks the pilot will notice and restart it like any other daemon.

\end{colsection}

\subsection{The scheduler}
\label{sec:scheduler}
\begin{colsection}

All entries in the observation database \code{pointings} table with status ``\code{pending}'' form the current queue, and the task of selecting which of these pointings the system should observe is the role of the scheduler. Within G-TeCS the \emph{scheduler} can refer to two linked concepts: the scheduling functions or the scheduler daemon itself. This section describes how the scheduler daemon operates; for how the scheduling functions chose which pointing to observe see \aref{chap:scheduling}.

The pt5m control system has no independent scheduler daemon; when the pilot needs to know what to observe it simply calls the scheduling functions to read the current queue, rank the pointings and find the one with the highest priority. When expanding the system for GOTO it was decided to farm these calculations off to a separate daemon, which the pilot queries just like the other hardware daemons. There are several advantages to this method. Firstly, just like with the sentinel alert monitor, having a dedicated daemon means it can be monitored by the pilot using the functions described in \aref{sec:monitors}. Furthermore, the scheduling commands can take a significant amount of time to run (several seconds), so splitting them out to a separate program saves time and frees up the pilot thread for other routines (recall the pilot is asynchronous but not multi-threaded). Also, having an independent scheduler daemon allows it to be run on the faster central server in SuperWASP that hosts the observation database, as shown in \aref{fig:flow}. Having the database queries run on the same, machine as the database, instead of over the network, improves the speed of the scheduling functions. Finally, when GOTO moves to a multi-telescope system it is anticipated that the scheduler will be one of the common systems shared between telescopes (see \aref{sec:multi_tel_scheduling}), so it makes sense to have the daemon on the central server alongside the other shared systems.

The scheduler daemon contains the usual control loop, which runs the scheduling functions described in \aref{chap:scheduling} and internally stores the returned highest-priority pointing. The daemon exposes a single command, \code{check\_queue}, which returns the ID of that pointing. The pilot \code{check\_schedule} coroutine queries the daemon every 10 seconds using this command, and the scheduler returns one of three results: carry on with the current observation, switch to a new observation, or park the telescope (in the case that there are no valid targets). Most of the time the pilot will be observing a pointing previously given by the scheduler, and on the next check the scheduler will return the same pointing as it is still the highest priority --- in which case pilot will continue observing it. Even if the scheduler finds that a different pointing now has a higher priority it will not tell the pilot to change targets whilst observing the current target, unless the new pointing has the \acro{too} flag set. Otherwise the pilot will wait until it has finished the current job, mark it as complete in the database and ask the scheduler for the next target to observe. The different possible cases are summarised in \aref{tab:sched}.

\begin{sidewaystable}[p]
    \begin{center}
        \begin{tabular}{cc|cccc} %
            &
            & \multicolumn{4}{c}{Highest priority pointing is\ldots}
            \\[0.5cm]
            &
            & \makecell{\ldots same as \\ current pointing}
            & \makecell{\ldots a new, \\ valid pointing}
            & \makecell{\ldots a new, \\ invalid pointing}
            & \ldots None
            \\[0.5cm]
            \midrule
            & & & & &
            \\
            \multirow{8}{*}{\rotatebox[origin=c]{90}{Current pointing is\ldots}}
            & \ldots valid
            & \makecell{\textcolor{Green}{Continue} \\ \textcolor{Green}{current pointing}}
            & \makecell{\textcolor{BlueGreen}{Interrupt and start new pointing} \\ \textcolor{BlueGreen}{if it is a ToO and the current pointing is not,} \\ \textcolor{BlueGreen}{otherwise continue current pointing}}
            & \textcolor{Red}{Park}
            & \textcolor{Red}{Park}
            \\[1.5cm]
            & \ldots invalid
            & \textcolor{Red}{Park}
            & \textcolor{NavyBlue}{Interrupt and start new pointing}
            & \textcolor{Red}{Park}
            & \textcolor{Red}{Park}
            \\[1.5cm]
            & \makecell{\ldots N/A}
            & ---
            & \textcolor{NavyBlue}{Start new pointing}
            & \textcolor{Red}{Park}
            & \textcolor{Red}{Park}
            \\[0.5cm]
        \end{tabular}
    \end{center}
    \caption[Actions to take based on scheduler results]{
        Actions the pilot will take based on the scheduler results. The scheduler can return one of three options as the highest priority pointing: either the current pointing, a different pointing or None (meaning the current queue is empty). Each pointing can either be valid or invalid. The pilot will either continue with the current pointing (\textcolorbf{Green}{green}), switch to the new pointing depending on the ToO flag (\textcolorbf{NavyBlue}{blue}, \textcolorbf{BlueGreen}{blue-green}) or park the telescope (\textcolorbf{Red}{red}).
    }\label{tab:sched}
\end{sidewaystable}

\clearpage

\end{colsection}

\section{Summary and Conclusions}
\label{sec:autonomous_conclusion}

\begin{colsection}

In this chapter I have described the autonomous systems that allow GOTO to function as a robotic telescope.

The purpose of the programs described in this chapter is to add onto the core G-TeCS software as described in \aref{chap:gtecs}, and to replicate and replace every role of a human telescope operator. The core of the robotic control system is the pilot master control program, and I described how the pilot operates as an asynchronous program with multiple coroutines dedicated to monitoring or carrying out specific tasks throughout the observatory.

There are also several additional daemons added to support the pilot. The conditions daemon performs a vital role in the control system, and I described the different conditions flags and weather limits used for the telescope on La Palma. I then gave an outline of how targets are observed by the robotic system, with the sentinel daemon adding pointings to the observation database and the scheduler daemon selecting which ones to observe. How the scheduling functions determine which target is the highest priority is examined in the following chapter (\aref{chap:scheduling}), and how the sentinel daemon processes alerts is described in \aref{chap:alerts}.

\end{colsection}

\chapter{Scheduling Observations}
\label{chap:scheduling}

\chaptoc{}

\section{Introduction}
\label{sec:scheduling_intro}

\begin{colsection}

Completing the chapters describing the core functions of the GOTO Telescope Control System, in this chapter I detail how the robotic system decides which targets to observe.
\begin{itemize}
    \item In \nref{sec:ranking} I describe the functions used by the G\nobreakdash-TeCS scheduler to chose between targets and decide which is the highest priority.
    \item In \nref{sec:scheduler_tiebreaker} I examine how the ``tiebreak'' value is calculated to sort between equally-ranked targets.
    \item In \nref{sec:scheduler_sims} I describe how optimal tiebreak weighting parameters were determined, by running simulations of the G-TeCS system observing gravitational-wave events.
\end{itemize}
All work described in this chapter is my own unless otherwise indicated. The first two sections are based on the description of the scheduling functions in \citet{Dyer}.

\end{colsection}

\section{Determining target priorities}
\label{sec:ranking}

\begin{colsection}

GOTO operates under a ``just-in-time'' scheduling model \citep[see, for example,][]{LCO_scheduling}, rather than creating a plan at the beginning of the night of what to observe \citep[see, for example,][]{ZTF_scheduler}. Each time the pilot queries the scheduler the current queue of pointings is imported and the priority of each is calculated, with no explicit consideration for the past or future (aside from the ``mintime'' constraints, as described below). The highest priority pointing is then returned, as described in \aref{sec:scheduler}.

This system is very reactive to any incoming alerts, as the new pointings will immediately be included in the queue at the next check. This method also naturally works around any delay in observations due to poor conditions, unlike a fixed night plan. The just-in-time method can be less efficient than a night plan when observing predefined targets which can be deliberately optimised before the night starts. However the just-in-time system is perfectly reasonable for the all-sky survey GOTO is normally observing, and any other observations will be alerts entered by the sentinel daemon which could not be planned for, so it was determined to be the best option for GOTO.\@

Each time the scheduler functions are called several steps need to be carried out. The first of these is to fetch the current queue from the observation database (see \aref{sec:obsdb}). This is done by querying the database \code{pointings} table for any entries that have the \code{pending} status. Additional filters are also applied in order to reduce the number of invalid pointings imported: restricting the query to pointings within the visible region of the sky (based on the time and observatory location) and within the pointing's valid period (the start time has passed and stop time has not yet been reached). Any entries in the table that pass these filters make up the \emph{pointings queue}.

In order to find which pointing is the highest priority, the queue is sorted using a variety of parameters, with the pointing sorted at the top being returned by the scheduler. The sorting criteria are outlined in the following sections.

\end{colsection}

\subsection{Applying target constraints}
\label{sec:constraints}
\begin{colsection}

The first consideration is determining which pointings are currently valid. As described in \aref{sec:obsdb}, pointings have limits defined for physical constraints (minimum target altitude, minimum Moon separation, maximum Moon illumination, maximum Sun altitude). These constraints are calculated and applied to the pointings using the Astroplan Python package \citep[\pkg{astroplan}\footnote{\url{https://astroplan.readthedocs.io}},][]{astroplan}. The target altitude and Moon separation constraints depend on the position of the target, both the altitude constraints depend on the site the observations are being taken from, and all four constraints depend on the current time. Each constant is applied both at the current time and after the minimum observing time defined for each pointing. This ensures that, for example, targets that are setting are visible throughout their observing period by checking the altitude is above the minimum both at the beginning and end of the observation. The minimum time constraints are not applied to the pointing currently being observed (if any), as the pointing will already be part way through and will have already been passed as valid. The validity of the pointings is a simple boolean flag (True or False), and invalid pointings are naturally sorted below valid ones.

\end{colsection}

\subsection{Effective rank}
\label{sec:rank}
\begin{colsection}

The next order pointings are sorted by is the effective rank of the pointing, which is a combination of the integer starting rank the pointing was inserted with and the number of times it has since been observed.

The starting rank is fixed when the pointing is created in the observation database: every pointing is given an integer rank between 0 and 999. The highest and lowest ranks are reserved for particular classes of targets. Rank 0 is not intended to be used under normal circumstances: it is reserved for exceptional events, such as a local galactic supernova, as a pointing with rank 0 would outrank all other pointings including even gravitational-wave events. At the other end of the scale, rank 999 is reserved for the all-sky survey tiles, so that they are sorted below all other pointings. These pointings act as ``queue fillers'' in the system, ensuring there is always something for the telescope to observe. All other ranks are otherwise available, although by convention ranks ending in 1--5 are used for gravitational-wave events, 6--8 for other transient events (e.g. GRBs) and 9 for other fixed targets. See \aref{sec:event_strategy} for the details of determining the rank for different transient events.

Added to the starting rank is a count of the number of times that a target has been observed, based on the number of pointings previously associated with a given mpointing (see \aref{sec:obsdb}). This count only includes successful observations, so pointings that were interrupted or aborted are not included. The starting rank ($R_s$) and observation count ($n_\text{obs}$) are added to create the effective rank $R$ given by
\begin{equation}
    R = R_s + 10\times n_\text{obs}.
    \label{eq:effective_rank}
\end{equation}
This formula means a pointing with a starting rank of 2 that has been observed five times will have an effective rank of 52. Effective ranks are sorted in reverse order, so a rank-5 pointing that has been observed once (an effective rank of 15) will be a higher priority target than a rank-4 pointing that has been observed twice (an effective rank of 24). This system allows for a natural filtering of targets, as targets will move down the queue as they are observed. For example, pointings from a gravitational-wave event might be inserted into the database at rank 2, so will first appear in the queue with effective rank of $R=2$ ($n_\text{obs}=0$). The first pointing that is observed will reappear with $R=12$, and therefore be sorted below those tiles that have not yet been observed. Once all the pointings have been observed once they will all have effective rank 12 and the process repeats, with each pointing falling to effective rank 22, 32 etc. As the increase is by 10 each time pointings from other events, or which were manually inserted, might also be in the queue and interweave between the event follow-up pointings. For example, a manual observation might be inserted at rank 9, meaning it will fall below the first observation of the gravitational-wave tiles at $R=2$ but will take priority over subsequent observations. To prevent this, the manual observation could be inserted at rank 19 to come after two gravitational-wave observations, or even 509 to completely ensure it does not interfere with the gravitational-wave follow-up targets.

\end{colsection}

\subsection{Targets of Opportunity}
\label{sec:toos}
\begin{colsection}

For pointings with the same effective rank the next sorting parameter is the \acro{too} flag assigned to the pointing when it was inserted into the database. The flag is simply a boolean value that is true if the target is a ToO and false if it is not, and pointings that have the flag as true are sorted higher than those of the same rank that are not ToOs. This ensures that time-sensitive targets are prioritised ahead of other targets at the same rank, although it is important to remember that the effective rank does still take priority (this means a ToO at rank 4 will be sorted above any other rank 4 pointings, but will still be a lower priority than a non-ToO at rank 3).

\end{colsection}

\subsection{Breaking ties}
\label{sec:breaking_ties}
\begin{colsection}

Finally, if there are multiple pointings with the same values for the above parameters then a single \emph{tiebreak} value is calculated for each. This value is based on the current airmass of the pointing and the weighting of the survey tile the pointing is linked to, if any; pointings at lower airmass (closer to the zenith) and higher tile weightings are the higher priority. For example, if two new gravitational-wave pointings with equal ranks both contain the same skymap probability (see \aref{sec:skymaps}), then the one at the lower airmass at the time of the check will be prioritised.

\newpage

To calculate the tiebreak value, both the tile weighting ($W$) and airmass ($X$) values need to be scaled between 0 and 1. This is true by definition for the tile weights, while the airmass is scaled so airmasses 1 and 2 are set to 1 and 0 respectively (airmasses greater than 2 are set to zero). The parameters are then combined to form the tiebreak value $V$ in a ratio 10:1 using
\begin{equation}
    V = \frac{10}{11}~W + \frac{1}{11}~(2 - X).
    \label{eq:tiebreak}
\end{equation}
This ensures the tiebreak value is also between 0 and 1, with higher values being preferred. The best possible scenario is a tile which contains 100\% of the skymap localisation probability ($W=1$) and is exactly at zenith ($X=1$) which gives a tiebreak value $V=1$. How this tiebreak formula was determined is described in \aref{sec:scheduler_tiebreaker}. Note that \aref{eq:tiebreak} is just \aref{eq:wa_ratio} using a ratio of 10:1, which was determined based on the scheduling simulations described in \aref{sec:scheduler_sims}.

In the unlikely event that two pointings are still tied, all other parameters (rank, ToO flag) being otherwise equal, and they have exactly the same tiebreak value, then whichever was inserted into the database first (and therefore has a lower database ID) by default comes first in the queue.

\end{colsection}

\subsection{Queue sorting example}
\label{sec:sorting_example}
\begin{colsection}

\begin{table}[t]
    \begin{center}
        \begin{tabular}{c|l|c|ccc|c|ccc} %
            & Name & Valid & $R_s$ & $n_\text{obs}$ & Eff.\ rank & ToO & $W$ & $X$ & Tiebreaker \\
            \midrule
            1 & GW191202 P3   & \textcolor{Green}{Y} &  2  & 0 &   2 & \textcolor{Green}{Y} & 0.10 & 1.1 & 0.173 \\
            2 & GW191202 P4   & \textcolor{Green}{Y} &  2  & 0 &   2 & \textcolor{Green}{Y} & 0.05 & 1.1 & 0.127 \\
            3 &         M101  & \textcolor{Green}{Y} &  9  & 0 &   9 &   \textcolor{Red}{N} &    1 & 1.5 & 0.955 \\
            4 & GW191202 P2   & \textcolor{Green}{Y} &  2  & 1 &  12 & \textcolor{Green}{Y} & 0.30 & 1.1 & 0.355 \\
            5 &   AT 2019xyz  & \textcolor{Green}{Y} &  6  & 2 &  26 & \textcolor{Green}{Y} &    1 & 1.4 & 0.964 \\
            6 &          M31  & \textcolor{Green}{Y} &  16 & 1 &  26 &   \textcolor{Red}{N} &    1 & 1.2 & 0.982 \\
            7 & All-sky T0042 & \textcolor{Green}{Y} & 999 & 0 & 999 &   \textcolor{Red}{N} &    1 & 1.0 & 1.000 \\
            \vdots & & & & & & \\
              &  GW191202 P1  &   \textcolor{Red}{N} &   2 & 0 &   2 & \textcolor{Green}{Y} & 0.55 & 1.1 & 0.582 \\
              & All-sky T0123 &   \textcolor{Red}{N} & 999 & 0 & 999 &   \textcolor{Red}{N} &    1 & 2.0 & 0.909 \\
        \end{tabular}
    \end{center}
    \caption[Examples of sorting pointings by priority]{
        Some examples of a queue of pointings sorted by priority. Pointings are first sorted by validity, with invalid pointings shown at the bottom of the queue. Then pointings are sorted by effective rank, which is comprised of the starting rank ($R_s$) and the observation count ($n_\text{obs}$). Pointings with the same effective rank are sorted based on if they are targets of opportunity or not, with ToOs being ranked higher. Finally pointings with all other factors being equal are ranked by the tiebreaker value, combining tile weighting ($W$) and the current airmass ($X$) using \aref{eq:tiebreak}.
    }\label{tab:priority}
\end{table}

In order to show how the above sorting methods are applied in practice, an example queue of pointings is shown in \aref{tab:priority}. The current highest-priority pointing is one of four pointings from a fictional gravitational-wave event, GW191202. At the top of the queue are two of these pointings, marked as P3 and P4. Both are valid, both have the same starting rank (2) and neither have been observed yet ($n_\text{obs}=0$). They are also both at the same airmass (1.1), but as P3 has a higher tile weighting (containing 10\% of the skymap probability compared to 5\% for P4) it has a higher tiebreak value and is therefore sorted higher. Therefore P3 would be returned by the scheduler.

As a demonstration, the rest of the queue is also shown. The gravitational-wave pointing containing the highest probability, P1, is unfortunately not valid and is therefore at the bottom of the queue (but still above other invalid pointings). The second highest, P2, has already been observed once and therefore has an effective rank of 12. This puts it below a non-ToO pointing of M101 which has a lower starting rank, 9 compared to 2, but has not yet been observed and is therefore sorted higher. The other non-survey pointings in the queue are a pointing of a transient event, AT 2019xyz, and one of M31. Both are valid and have the same effective rank of 26, but the transient is a target of opportunity and therefore is sorted higher. This is true even though it is at a worse airmass, as the ToO sorting takes priority over the tiebreak; had they both (or neither) been ToOs then the M31 pointing would have been higher. Finally below those pointings is the first of the all-sky survey pointings. These will only be the highest priority if there are no other valid pointings above them, which for GOTO is actually most of the time.

\end{colsection}

\section{Calculating the tiebreaker}
\label{sec:scheduler_tiebreaker}

\begin{colsection}

The scheduler weights pointings in the current queue by several parameters, as described in the previous section: the assigned rank, the number of times it was previously observed, if it is a target of opportunity or not. But in practice most of the time the queue will contain a large number of pointings where these values are all the same. For example, when a new gravitational-wave event is processed by the sentinel the GOTO-alert event handler adds in a large number of pointings based on tiles from the skymap (see \aref{sec:event_insert}). On the next scheduler check the queue will be populated by a large number of pointings each with the same rank and ToO flag which have never been observed. Likewise, when observing the all-sky survey the queue will be filled with tiles that have all been observed the same number of times. This is why the scheduler then needs a further way to distinguish between pointings, which is known as the tiebreaker.

The older pt5m scheduling code that G-TeCS is based on (see \aref{sec:pt5m}) used only a single tiebreak parameter to decide between equally-ranked pointings: the airmass each target would be at at the midpoint of the observation. This works well to prioritise getting the best data quality, assuming the two targets are otherwise identical. However when adapting the pt5m system for GOTO it was clear there was the need for an additional parameter in the scheduling functions, to encode the relative weights of a set of pointings.

\end{colsection}

\subsection{Tile weighting}
\label{sec:weights}
\begin{colsection}

All the pointings added from a gravitational-wave skymap (or similar event such as a gamma-ray burst) will have the same rank, but they will have different weights from the amount of skymap probability they each contain (how event skymaps are mapped onto the tile grid is described in \aref{chap:tiling}). It makes sense that, of all the tiles from a given event, the ones with higher probability should be the ones to prioritise and observe first.

However, unlike an integer parameter such as the rank or the True/False ToO flag, the tile weights cover a wide range and often there will only be a small amount of difference between the values for neighbouring tiles. Prioritising a tile that contains 2.71\% of the skymap over another that contains 2.70\% in all cases is not the best strategy, especially if the latter is close to zenith while the former is low down close to the horizon. Observing a high-airmass tile over a low-airmass one for a gain of only 0.01\% probability is a poor choice, especially if the former tile is currently rising and will be at a better altitude in a few hours. For these reasons it was decided that the skymap probability weighting should be considered at the same level as the airmass tiebreaker, meaning a lower-airmass tile with only a slightly lower probability will be prioritised over one further from zenith. This should only be true up to a reasonable limit, however, as in the case of two tiles where one contains a probability of 95\% and the other 3\% it should always be true that observing the former is the better choice, even if it has a slightly worse airmass.

It should be noted that skymap probability is not necessarily the only way for a group of tiles to be weighted. In the past GOTO has carried out more focused surveys: when carrying out a galaxy-focused survey, for example, tiles were weighted by the sum of the magnitudes of all galaxies within them. The only requirement is that each tile has a weighting of between 0 and 1, relative to the other tiles added in that survey. For non-survey pointings, such as a single observation of a particular target, the weight is set to 1 (this can be considered as that tile having a 100\% chance of containing the target). This is also true for survey pointings where all tiles are weighted equally, such as the all-sky survey. This can be seen in the example \aref{tab:priority} in the previous section, as all the non-gravitational-wave pointings have $W=1$.

\newpage

\end{colsection}

\subsection{Combining tile weight and airmass}
\label{sec:wa}
\begin{colsection}

As mentioned previously, the pt5m system uses airmass as the sole tiebreak parameter. For the G-TeCS scheduler it was instead decided to create a new tiebreak value, which would combine both the tile weight, as described above, and the airmass of the target at the time the scheduler check was carried out. This allows the scheduler to take into account both parameters when deciding between otherwise-equal pointings.

Airmass is usually modelled using a plane-parallel atmosphere, which gives
\begin{equation}
    X = \sec{z},
    \label{eq:airmass}
\end{equation}
where $X$ is airmass and $z$ is the zenith distance ($z=90-h$ where $h$ is the altitude of the target). Targets are best to observe at low airmasses to get the best data quality.

In order to combine both weight and airmass into a single tiebreak value it was decided to scale both between 0 and 1, and then combine them such that the final tiebreak value $V$ was also between 0 and 1. This would then be sorted so that higher values are prioritised, as described previously in \aref{sec:ranking}. Helpfully, the tile weights are already defined as being between 0 and 1. In order to scale the airmasses it was decided that the airmass of a target that is at or below the horizon limit should be set to 0, and any target that was at the zenith ($h=90$, so $X=1$) should be set to 1. For GOTO the horizon limit is \SI{30}{\degree}, which corresponds to an airmass limit of 2. The final tiebreak value is defined using
\begin{equation}
    V = \frac{w}{w+a}~W + \frac{a}{w+a}~(2-X),
    \label{eq:wa_ratio}
\end{equation}
where the balance of tile weight $W$ to airmass $X$ is described by the ratio $w$:$a$. Using a ratio of 10:1, i.e.\ prioritising the tile weight 10 times more than the airmass, produces \aref{eq:tiebreak}.

\newpage

\end{colsection}

\subsection{Time-to-set}
\label{sec:tts}
\begin{colsection}

The definition of airmass given in \aref{eq:airmass} is, as would be expected, symmetric around the zenith. Scheduling using this parameter is therefore a problem for the following reason: consider two targets with equal or similar contained skymap probabilities, but one is \SI{5}{\degree} above the horizon in the west and the other is \SI{5}{\degree} above the horizon in the east. They have the same airmass value, but due to the rotation of the Earth the one in the west will be setting while the one in the east is rising. A good scheduling system could prioritise observing the target in the west, as unless it is observed quickly it will pass below the horizon and no longer be visible for the remainder of the night (assuming it is not circumpolar, see below).

In order to address this problem, a new parameter was required that prioritises targets that are about to set. This new parameter is called `time-to-set', and is simply the time until the target sets below the defined horizon. The units are arbitrary, but as it will repeat with a period of 24 hours the time-to-set value is normalised between 0 and 1 so that it is 0 when the target is at the horizon, 0.5 when it is 12 hours from setting and 1 when it is 24 hours from setting (there is therefore a degeneracy at 0 and 1). \aref{fig:airmass_tts} shows how the airmass and time-to-set values change between 0 and 1 over the course of a day for any non-circumpolar target. Circumpolar targets are ones that never set below the horizon, and therefore for these targets the time-to-set is an illogical value. However, the North Celestial Pole is just below the \SI{30}{\degree} horizon limit of GOTO from La Palma, so this is not a concern at present.\@ This may need to be reconsidered depending on which site is picked for GOTO's future southern node, see \aref{sec:multi_site_scheduling}.

\begin{figure}[t]
    \begin{center}
        \includegraphics[width=\linewidth]{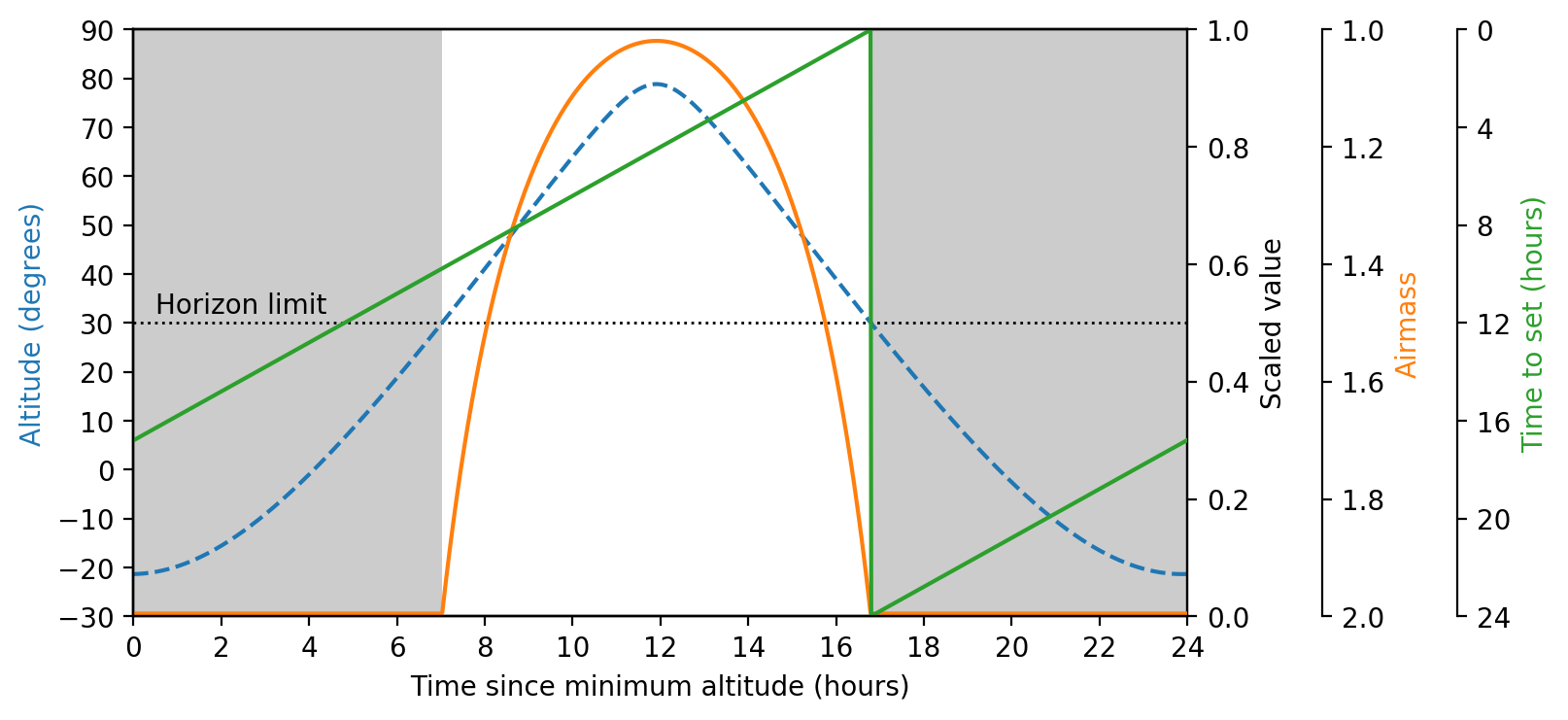}
    \end{center}
    \caption[Plotting scaled airmass and time-to-set values for a target]{
        Plotting scaled airmass and time-to-set values for a target over 24 hours.
        The altitude of the target is shown by the \textcolorbf{NavyBlue}{blue} dashed line, and the \textcolorbf{Gray}{grey} regions show the times when the target is below the \SI{30}{\degree} horizon altitude limit (the black dotted line).
        The two tiebreak values are also plotted, scaled between 0 and 1: airmass (in \textcolorbf{Orange}{orange}) is set to 0 when the target is at airmass 2 or below, and the time-to-set (in \textcolorbf{Green}{green}) linearly increases to 1 until the target passes below the horizon and it is reset to 0.
    }\label{fig:airmass_tts}
\end{figure}

Just replacing airmass in the tiebreaker calculation with the time-to-set would not produce good results, as the telescope would be prioritised to observe the western horizon continuously (as by design targets that are just about to set have the highest scaled time-to-set values). As when including the skymap probability, a weighted combination of both airmass and time-to-set would be best in order to take both parameters into account. This new parameter, $Z$, can be considered using the equation
\begin{equation}
    Z = \frac{a}{a+t}~(2-X) + \frac{t}{a+t}~T,
    \label{eq:at_ratio}
\end{equation}
where again the weighting factors $a$ and $t$ describe the relative ratio between the airmass ($X$) and time-to-set ($T$) in the ratio $a$:$t$. \aref{fig:at_ratio} shows how different ratios produces different distributions: a ratio of 1:0 only considers the airmass, a ratio of 1:1 has to two equally weighted and a ratio of 0:1 only considers the time-to-set. As the ratio is increased in favour of time-to-set (i.e.\ $t$ is larger than $a$) the peak of the distribution shifts to the right, which will favour targets that are setting over those at the zenith. By using both airmass and time-to-set values in the tiebreaker formula the scheduler should prioritise observing targets that are about to set but are still at a reasonable airmass.

\begin{figure}[t]
    \begin{center}
        \includegraphics[width=\linewidth]{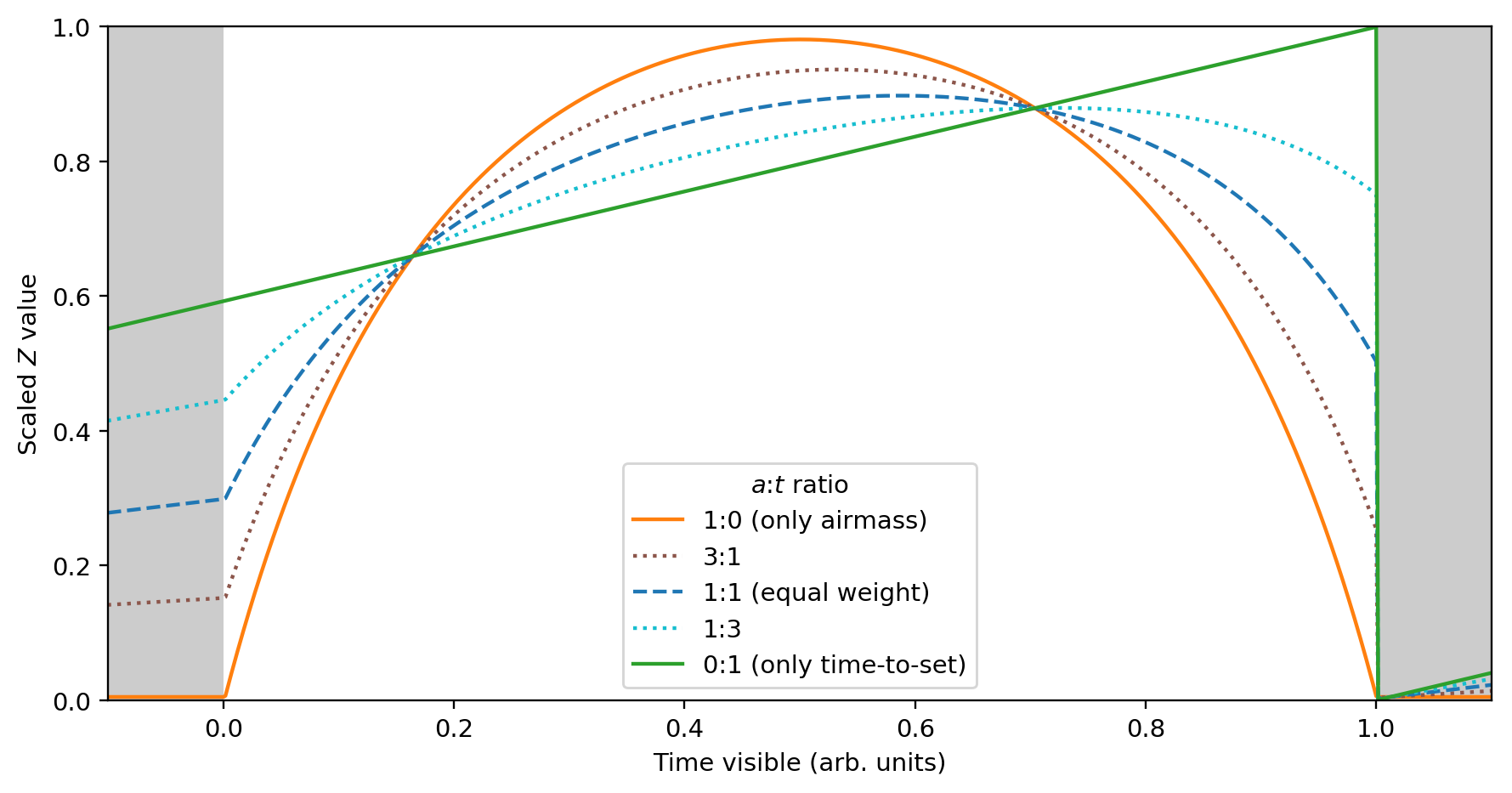}
    \end{center}
    \caption[Combining airmass and time-to-set values]{
        Combining airmass and time-to-set values in different ratios using \aref{eq:at_ratio} to create a new distribution. This plot uses the same target as \aref{fig:airmass_tts}, but focusing the $x$-axis on the time when the target is above the horizon.
    }\label{fig:at_ratio}
\end{figure}

\newpage

Previously, the scheduler tiebreak value $V$ was calculated using equation \aref{eq:wa_ratio} with just the tile weight ($W$) and the airmass ($X$). In order to determine how including the time-to-set ($T$) would affect the scheduler this equation can be rewritten as
\begin{equation}
    V = \frac{w}{w+a+t}~W + \frac{a}{w+a+t}~(2-X) + \frac{t}{w+a+t}~T,
    \label{eq:wat}
\end{equation}
where $w$, $a$ and $t$ are the relative weightings factors for the tile weight, airmass and time-to-set respectively (usually written in the form $w$:$a$:$t$). Using this equation a series of simulations were carried out using the GOTO scheduling code to determine optimal values for $w$, $a$ and $t$, as described in the \aref{sec:scheduler_sims}. Ultimately a ratio of 10:1:0 was selected; by setting $t=0$ the time-to-set value is ignored, as it was found to only hinder the scheduler performance. This is why \aref{eq:tiebreak} in \aref{sec:breaking_ties} only includes the tile weight and airmass values.

\end{colsection}

\section{Scheduler simulations}
\label{sec:scheduler_sims}

\begin{colsection}

When the scheduling functions described in the previous sections were written, it was not clear how best to combine the selected parameters (tile weight, airmass and time-to-set) to create a single tiebreak value. In order to examine how different weightings of the three parameters affected the scheduler performance, a series of simulations were carried out. These simulations, and their results, are described in this section.

\end{colsection}

\subsection{Simulating GOTO}
\label{sec:goto_sims}
\begin{colsection}

The G-TeCS control system described in \aref{chap:gtecs} and \aref{chap:autonomous} contains all of the code used to operate the GOTO telescope, as well as test code down to the level of the individual daemons and hardware units. This makes it possible to simulate, for example, the camera daemon taking exposures using fake hardware code that waits in real time until the exposure time is completed (plus some readout time) and then creates a blank FITS image file with all of the expected header information. On top of this the real pilot can run without knowing these daemons are fake, and the real sentinel daemon can add real or simulated events into a copy of the observation database for the real scheduler to choose between. In this way the entire control system can be run without connecting it to any real hardware.

However, the fully-featured test suite described above is not necessary for the simulations described in this section: ideally they would run faster than real time, and it is not necessary to simulate the full hardware system down to fake images being created, but the intention is still to model the response of the real control system as much as possible. For these simulations anything below the pilot (i.e.\ the hardware daemons, as shown in \aref{fig:flow}) is abstracted away, and the pilot itself is replaced by a new specialised script simply called the fake pilot. This mirrors the real code described in \aref{sec:pilot} in most ways, however there are several important simplifications:

\begin{itemize}
    \item The fake pilot does not call the scheduler daemon in order to find what pointing to observe, but instead imports and runs the scheduling functions itself. This was the original way the pilot ran before the scheduler was split into a separate daemon, as described in \aref{sec:scheduler}.
    \item The fake pilot does not include the full conditions monitoring system described in \aref{sec:conditions}. The \code{check\_conditions} routine still exists in order to stop observations when the Sun has risen, but is just a single function check. Code was written to simulate weather closing the dome using random Gaussian processes, however for the scheduling simulations described here only a single night of observations is considered, and including random weather effects in the simulations only distracted from their purpose to model the scheduler response.
    \item The night marshal and any observing tasks other than actually observing the scheduler target (e.g.\ autofocusing) have been removed from the fake pilot. Observations start immediately after sunset and continue to sunrise, and then when the dome is closed the pilot loop continues until the simulation has completed.
    \item While the real pilot works using loops that sleep until a given time has passed, the fake pilot contains an internal time which is increased for each `step' in the simulation. At each step the script checks the scheduler using the normal commands. One important factor in speeding the simulations up is to increase this step size, up to the point that each observation takes a single step. This means that at each step the pilot will observe a new target, and increase the internal simulation time by the appropriate amount. If, for example, the target pointing asks for three \SI{60}{\second} exposures the simulations would increase by 3 minutes, plus extra time for readout and slewing to the target before the exposures start.
\end{itemize}

\newpage

\end{colsection}

\subsection{Simulation results}
\label{sec:scheduler_sim_results}
\begin{colsection}

A series of simulations using the fake pilot code described above were carried out in order to find optimal values for the weights ($w$, $a$ and $t$) given in \aref{eq:wat}, and to see how the telescope response changes depending on the values used.

It should be noted that these simulations were carried out in 2016, early in the development of G-TeCS and before the first GOTO telescope had been commissioned on La Palma. They therefore included assumptions for values such as the field of view of the telescope (affecting the tile size), mount slew speed and readout time. In addition, most of the code to handle gravitational-wave skymaps detailed in \aref{chap:alerts} had not yet been written and the event follow-up strategy had not been fully defined.

In order to run the simulations, a selection of model skymaps from the LIGO First Two Years project \citep{First2Years} were manually processed to generate a series of tiled pointings, which were then added to the observation database. The fake pilot script was then run to simulate one night of observations using set values for the $w$:$a$:$t$ ratio. Once completed the tiles observed were recorded, and then the database was reset, the ratio changed and the simulations repeated.

Two metrics were used to judge the effectiveness of the scheduler response: the mean airmass of each tile when observed, and the fraction of the skymap probability covered (i.e.\ the total contained probability within all observed tiles). As simulated skymaps were being used the location of the source of the gravitational-wave signal was known, so it was possible to record if the tile containing the source was observed or not. However, for these simulations the overall response, in terms of skymap coverage, was deemed a better indicator of the scheduler performance than just if the source was observed or not (for example, the source might not have even been visible from La Palma). The later, more advanced simulations described in \aref{chap:multiscope} go into more detail about the source location and the probability that the source position is observed.

\newpage

\end{colsection}

\subsection{Analysis of simulation results}
\label{sec:scheduler_sim_analysis}
\begin{colsection}

\begin{figure}[t]
    \begin{center}
        \includegraphics[width=\linewidth]{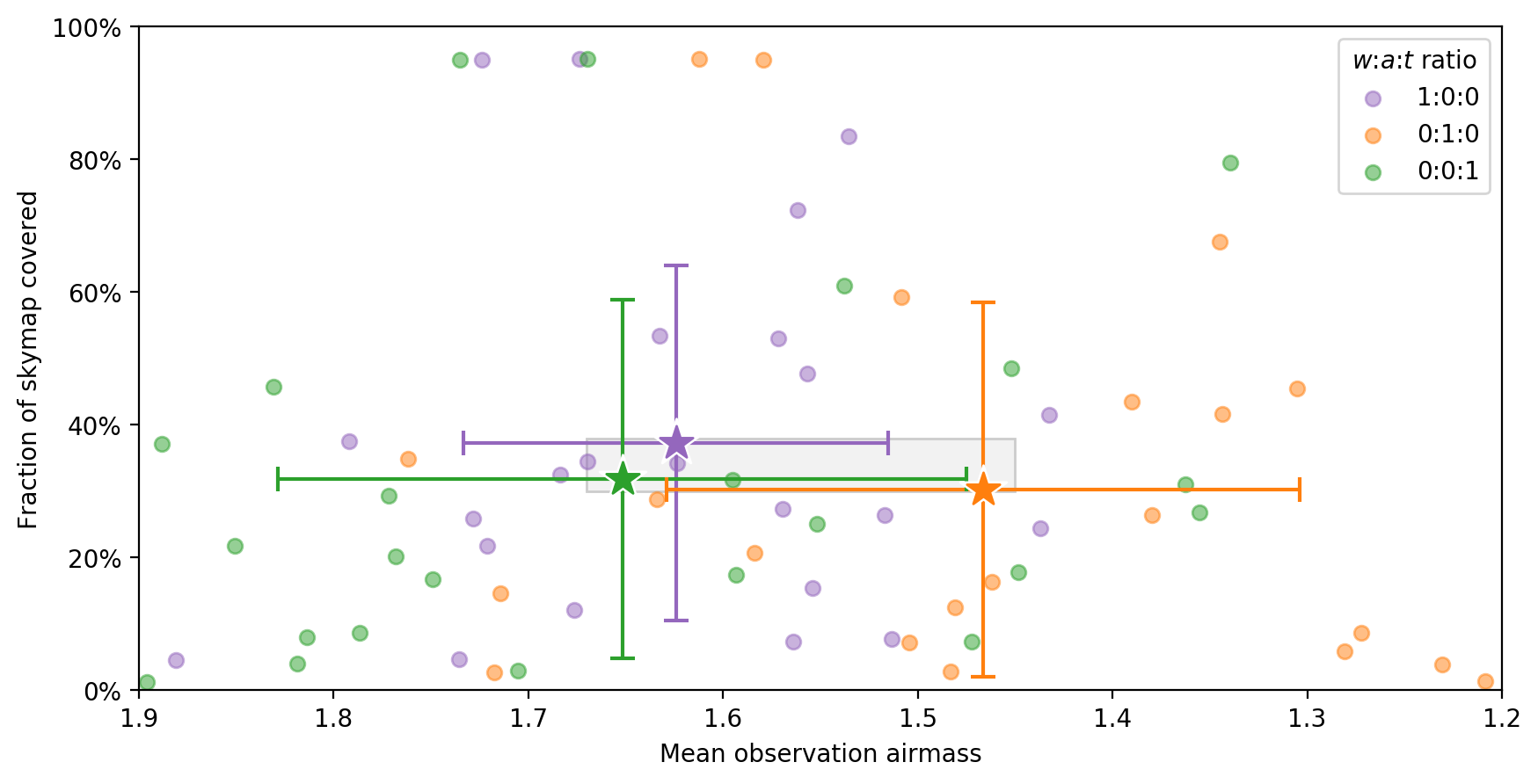}
    \end{center}
    \caption[Skymap coverage verses mean airmass for different $w$:$a$:$t$ ratios]{
        The fraction of the event skymap covered verses the mean observation airmass for three different $w$:$a$:$t$ ratios: 1:0:0 (\textcolorbf{Purple}{purple}), 0:1:0 (\textcolorbf{Orange}{orange}) and 0:0:1 (\textcolorbf{Green}{green}). Each point represents a simulation using one of the First 2 Year skymaps. The stars show the average position for each ratio, with the error bars showing the standard deviation. The shaded region shows the area of \aref{fig:scheduler_sim_results2}.
    }\label{fig:scheduler_sim_results1}
\end{figure}

The simulation results for three different ratios are plotted in \aref{fig:scheduler_sim_results1}. Each coloured point represents a single simulation of a night observing a single skymap. There is a large range of results: some simulations covered close to 100\% of the skymap while others covered almost none. Only simulations that included observing at least one skymap tile are included, otherwise the mean observed airmass would be undefined. The stars show the average position for each $w$:$a$:$t$ ratio. Although the errors are clearly large, some distributors are clear. For example, the 0:1:0 ratio (only including the airmass weighting) on average produces a better mean airmass than the others, which would be expected. Likewise the 1:0:0 case (only including the tile weight) on average results in a slightly higher fraction of the skymap being observed.

\begin{figure}[t]
    \begin{center}
        \includegraphics[width=\linewidth]{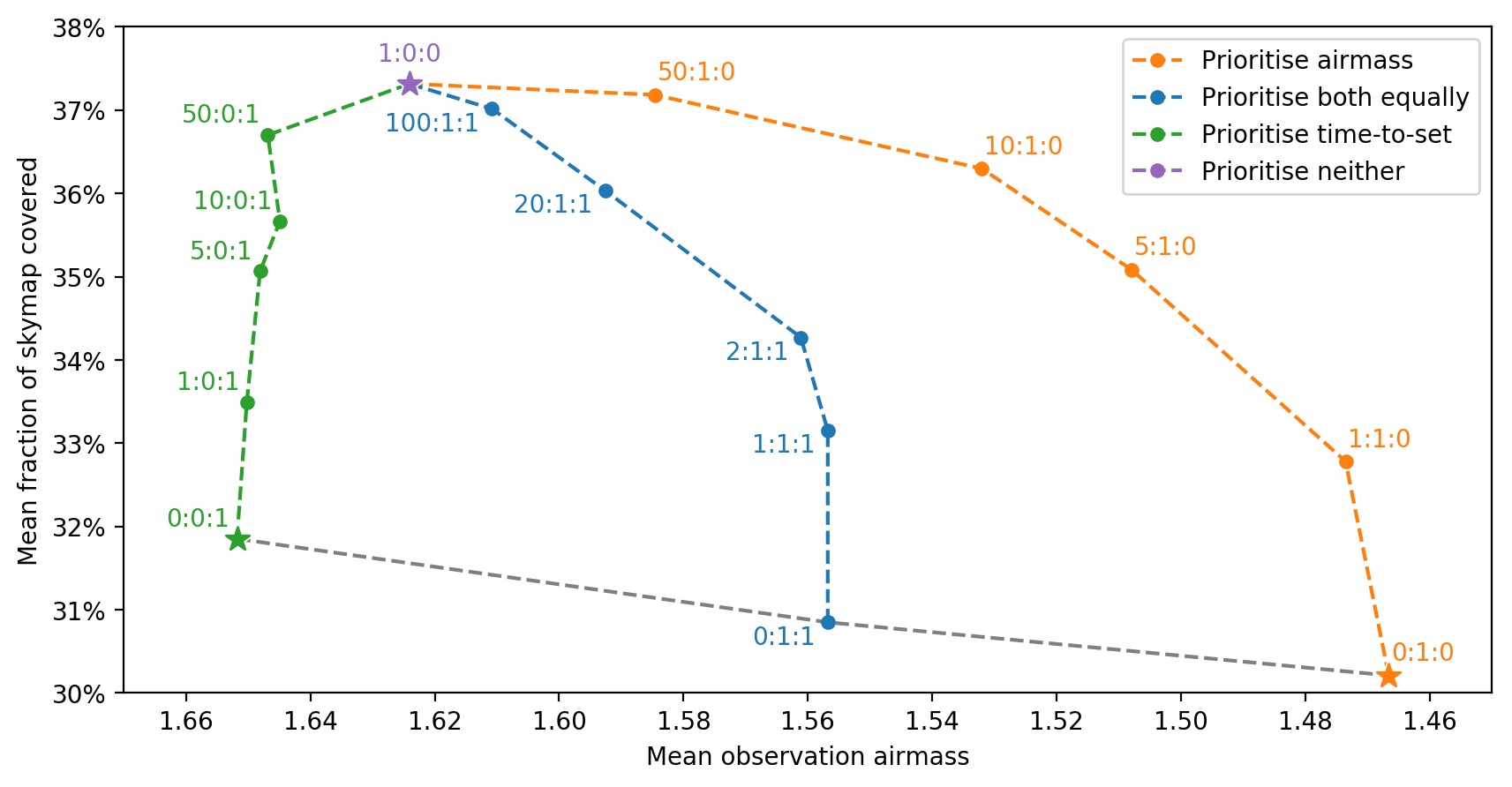}
    \end{center}
    \caption[Scheduler simulation results for different $w$:$a$:$t$ ratios]{
        The fraction of the event skymap covered verses airmass for multiple different $w$:$a$:$t$ ratios. Each point here shows the average over multiple simulations of different skymaps, and the ratios shown by stars are the same as plotted in \aref{fig:scheduler_sim_results1}. The dashed coloured lines join results with the same $a$:$t$ ratio; 1:0 in \textcolorbf{Orange}{orange}, 1:1 in \textcolorbf{NavyBlue}{blue}, and 0:1 in \textcolorbf{Green}{green}; and the \textcolorbf{Gray}{grey} dashed line joins ratios with $w=0$.
    }\label{fig:scheduler_sim_results2}
\end{figure}

Simulations were repeated for multiple different $w$:$a$:$t$ ratios, and the averages for each are plotted in in \aref{fig:scheduler_sim_results2}. This plot shows that, while there is a lot of underlying scatter between different skymaps as shown in \aref{fig:scheduler_sim_results1}, the mean positions for each ratio follow remarkably smooth trends (error bars are omitted from \aref{fig:scheduler_sim_results2} as they would spread off the page in both axes). For example, the higher the tile weight value $w$ relative to the other two parameters consistently increases the mean fraction of the skymap covered, however this is at a cost to the mean airmass of the observations.

Changing the relative ratios of airmass to time-to-set (from 1:0 through 1:1 to 0:1) shows an unexpected result. It is true the observed airmasses would be lower as less weight is put on the airmass parameter, however it was intended that introducing the time-to-set parameter would compensate by catching more setting tiles that would be missed by purely looking around the zenith, and therefore the skymap fraction covered would be higher. This is true if the tile weight is not included as a factor, as seen by the grey 0:0:1--1:0:0 line at the bottom of \aref{fig:scheduler_sim_results2}: when only airmass is considered (in the 0:1:0 case) the results produce the best average airmass per observation but the worse skymap coverage, and on the other hand only considering time-to-set (0:0:1) results in worse airmasses but better coverage. However, when the tile weight value $w$ is included this trend is counteracted, to the extent that the 50:0:1 point has both a higher mean airmass and a lower fraction of the skymap covered than the 50:1:0 point. In fact, based on the results which do not include the airmass parameter (the green line on the left of \aref{fig:scheduler_sim_results2}) the time-to-set value just suppresses the fraction of the skymap observed, while making almost no difference to the mean airmass.

The optimal scheduler result would, on average, produce the highest possible skymap coverage for the lowest average airmass. This would fall in the top-right region of \aref{fig:scheduler_sim_results2}. Based on these simulation results the best solution is to ignore the time-to-set value, and as described previously in \aref{sec:breaking_ties} the G-TeCS scheduler has been operating with a ratio of 10:1:0.

\end{colsection}

\subsection{Further simulations}
\label{sec:scheduler_sim_future}
\begin{colsection}

The conclusions from these scheduler simulations are not unreasonable if all that is needed is a one-size-fits-all set of weights that are hard-coded into the scheduler, as used by the current G-TeCS system. However, future work on these simulations should look in detail at other trends that might be hidden in the averages. For example, different $w$:$a$:$t$ ratios might be better suited for large skymaps versus smaller ones, or in cases where the whole skymap is visible at once compared to it slowly rising above the horizon during the night.

Although the time-to-set value seemed to only hinder the scheduler, other possible parameters could be considered. One idea is to convert time-to-set to time-visible, by including not only the time when the target sets below the horizon but also the time that the Sun rises. The existing time-to-set ratio prioritises observing targets later in the night when they are about to set, whereas airmass prioritises tiles near the zenith. Neither however considers the time remaining in the night, and while this is the same for every target including it as a factor in the scheduler could be a relatively-straightforward way to attempt to prioritise observations in the limited time available.

As mentioned previously, the simulations presented in this section were carried out before a lot of the G-TeCS code was finalised. Therefore it would also make sense to revisit the scheduler simulations with the newer simulation code, as used by the simulations in \aref{chap:multiscope}, in order to confirm if the 10:1:0 ratio is still found to be the best case. This would also allow the real parameters from the commissioned telescope to be included. When the next four unit telescopes are added to GOTO the effective field of view will be doubled, and the results in this section based on the 4-UT telescope might not necessarily be the same in the 8-UT case.

Finally, GOTO is ultimately planned to expand to multiple telescopes at several sites, as described in \aref{sec:goto_expansion}. Adapting the scheduling system described into this chapter to deal with multiple telescopes will be a major future project, which is considered in more detail in \aref{chap:multiscope}.

\end{colsection}

\section{Summary and Conclusions}
\label{sec:scheduling_conclusion}

\begin{colsection}

In this chapter I described how the G-TeCS automated scheduling system decides which pointings to observe.

The scheduler daemon is one of the autonomous support daemons as defined in \aref{chap:autonomous}. It has the task of taking the queue of possible targets from the observation database and sorting them based on several parameters to determine which is the highest priority. First each pointing is tested using a series of physical constraints to remove any which are invalid. The queue sorting then depends on the properties of each pointing: its rank, the number of times it has already been observed, whether it is defined as a target of opportunity or not. If there are still multiple pointings in the same position in the queue then a final tiebreaker value is calculated to chose between them, based on the skymap tile weighting (defined in \aref{chap:tiling}) and target airmass.

Determining how to calculate the tiebreaker and which parameters to base it on required a series of simulations of GOTO observations, to see how the choice of parameter weightings affected which pointings were observed. These simulations showed that a 10:1 ratio of tile weight to scaled airmass was closest to the preferred outcome, and that the introduction of a third parameter, time-to-set, could not improve on this. GOTO has been observing successfully using this ratio ever since. However, since those simulations were carried out further development of the control system and simulation code means that it would be worth revisiting them to see if the conclusions still hold.

\end{colsection}

\chapter{Tiling the Sky}
\label{chap:tiling}

\chaptoc{}

\section{Introduction}
\label{sec:tiling_intro}

\begin{colsection}

In this chapter I describe the software used by GOTO to create an all-sky survey grid, which gravitational-wave events are then mapped onto.
\begin{itemize}
    \item In \nref{sec:gototile} I describe the GOTO-tile Python package, and the algorithms it uses to define the GOTO all-sky survey grid.
    \item In \nref{sec:skymaps} I describe how transient alert localisations are defined using skymaps and how they are mapped onto the GOTO-tile grid.
    \item In \nref{sec:custom_skymaps} I give some examples of how other skymaps can be used to direct GOTO observations.
\end{itemize}
All work described in this chapter is my own unless otherwise indicated, and has not been published elsewhere. The original GOTO-tile package was written by Darren White at Sheffield and Evert Rol at Monash, before I took over development and made substantial changes as described in this chapter.

\end{colsection}

\section{Defining the sky grid}
\label{sec:gototile}

\begin{colsection}

GOTO-tile is a Python package (\pkg{gototile}\footnote{\url{https://github.com/GOTO-OBS/goto-tile}}) created for the GOTO project to contain all of the functions related to tiling the sky and processing skymaps. It was originally developed by Darren White as a way to process gravitational-wave skymaps for GOTO, and then maintained by Evert Rol who rearranged it into a package usable for some other telescopes, including SuperWASP on La Palma and a proposed southern GOTO node. My contributions to the package have been extensive: reworking the foundations to improve how the sky grid is defined and how skymaps are applied, as well as adding additional code to create new skymaps (described in \aref{sec:custom_skymaps}).

\end{colsection}

\subsection{Creating sky grids}
\label{sec:grids}
\begin{colsection}

The core of GOTO-tile as it now exists is the \code{SkyGrid} Python class, which is used to define a sky grid: a collection of regularly-spaced points on the celestial sphere. These points are used as the centre of rectangular `tiles' aligned to the equatorial right ascension/declination coordinate system, which create a framework for survey observations to be mapped on to.

The most important parameter required when defining a sky grid is the field of view of the telescope, which is taken as the size of the tiles that make up the grid. The field of view is defined within GOTO-tile by giving a width and height value in degrees, meaning the tiles can only be square or rectangular. This is typically fine for the GOTO array, which has a total field of view comprising of overlapping rectangles from each unit telescope (see \aref{fig:fov}). There was a period when having three unit telescopes in an `L'-shape was considered, but this was abandoned due mainly to the complexity of tiling the grid based on abstract shapes. For the prototype 4-UT GOTO system currently on La Palma a rectangular 18 square degree tile (\SI{3.7}{\degree} $\times$ \SI{4.9}{\degree}) was defined during the commissioning period (see \aref{sec:timeline}).

The second parameter required to define a sky grid is the desired overlap between the tiles. This is given as a value between zero and one in both the right ascension and declination axes, with zero meaning no overlap and one meaning all the tiles are completely overlapping (in practice the overlap is restricted to no more than $0.9$). The overlap is used to define the spacing between the tile centres, depending on the algorithm used. The current 4-UT grid uses an overlap of 0.1 (10\%) in both axes.

As GOTO-tile has developed, the algorithm used to define the grid has changed (see \aref{sec:algorithms}), but the basic method remained the same:

\begin{enumerate}
    \item On the celestial sphere (\aref{fig:sphere}) equally spaced lines of constant declination are defined, separated by the value $\Delta\delta$ (\aref{fig:deltadelta}). These ``declination strips'' are the basis for the grid points, which the tiles are centred on.
    \item Each declination strip is then filled with equally spaced points, separated by the value $\Delta\alpha$ (\aref{fig:deltaalpha}). This value is constant within each strip but is (in most algorithms) a function of declination, $\Delta\alpha(\delta)$, meaning that as one moves away from the equator towards the poles each strip will contain a fewer number of points.
    \item These points are then defined as the centres of the tiles, the size of which is given by the field of view (\aref{fig:tiledsphere}).
\end{enumerate}

Once the grid has been created it is encapsulated within the GOTO-tile \code{SkyGrid} class. Each tile is defined by a coordinate at its centre, and each is also given a unique name of the form \code{T0001}. The grid itself is also given a name formed using the input field of view and overlap parameters, so the current grid (with a field of view of \SI{3.7}{\degree}$\times$\SI{4.9}{\degree} and overlap factor of 0.1 in both axes) is given the name \code{allsky-3.7x4.9--0.1--0.1}. In this way a given tile in a given grid can be recreated just from the grid and tile name, which is used when storing the details in the \code{grids} and \code{grid\_tiles} tables in the observation database (see \aref{sec:obsdb}). %

\newpage

\makeatletter
\setlength{\@fptop}{0pt}
\makeatother

\begin{figure}[t]
    \begin{center}
        \includegraphics[width=\linewidth]{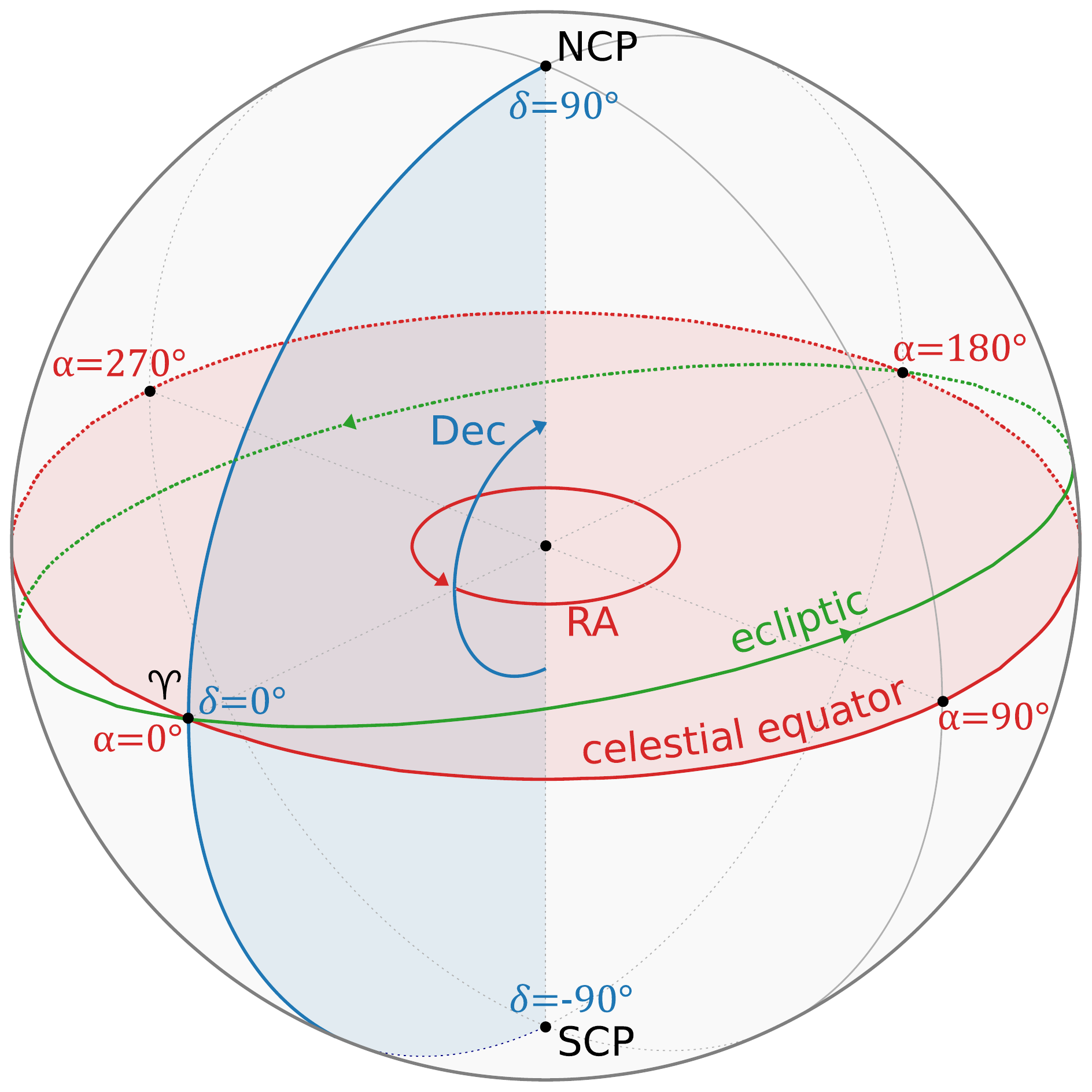}
    \end{center}
    \caption[The celestial sphere]{
        The celestial sphere. The northern and southern celestial poles are marked as \textbf{NCP} \acroadd{ncp} and \textbf{SCP} \acroadd{scp} respectively, and the celestial equator is marked in \textcolorbf{Red}{red}. The ecliptic (the path of the Sun) is marked in \textcolorbf{Green}{green}, the point where the ecliptic rises above the celestial equator (the vernal equinox) is marked with the symbol \Aries{}, and the meridian that intercepts the poles and the vernal equinox is marked in \textcolorbf{NavyBlue}{blue}. Traditionally the equatorial coordinate system is defined as shown: declination (\textcolorbf{NavyBlue}{Dec}, $\delta$) is the angle from the equator (between \SI{-90}{\degree} at the SCP to \SI{90}{\degree} at the NCP) and right ascension (\textcolorbf{Red}{RA}, $\alpha$) \acroadd{ra} is the angle east of the vernal equinox (between \SI{0}{\degree} and \SI{360}{\degree}). The modern \acro{icrs} defines coordinates based on radio sources which approximately match the ones described here \citep{ICRF}.
    }\label{fig:sphere}
\end{figure}

\clearpage

\begin{figure}[t]
    \begin{center}
        \includegraphics[width=\linewidth]{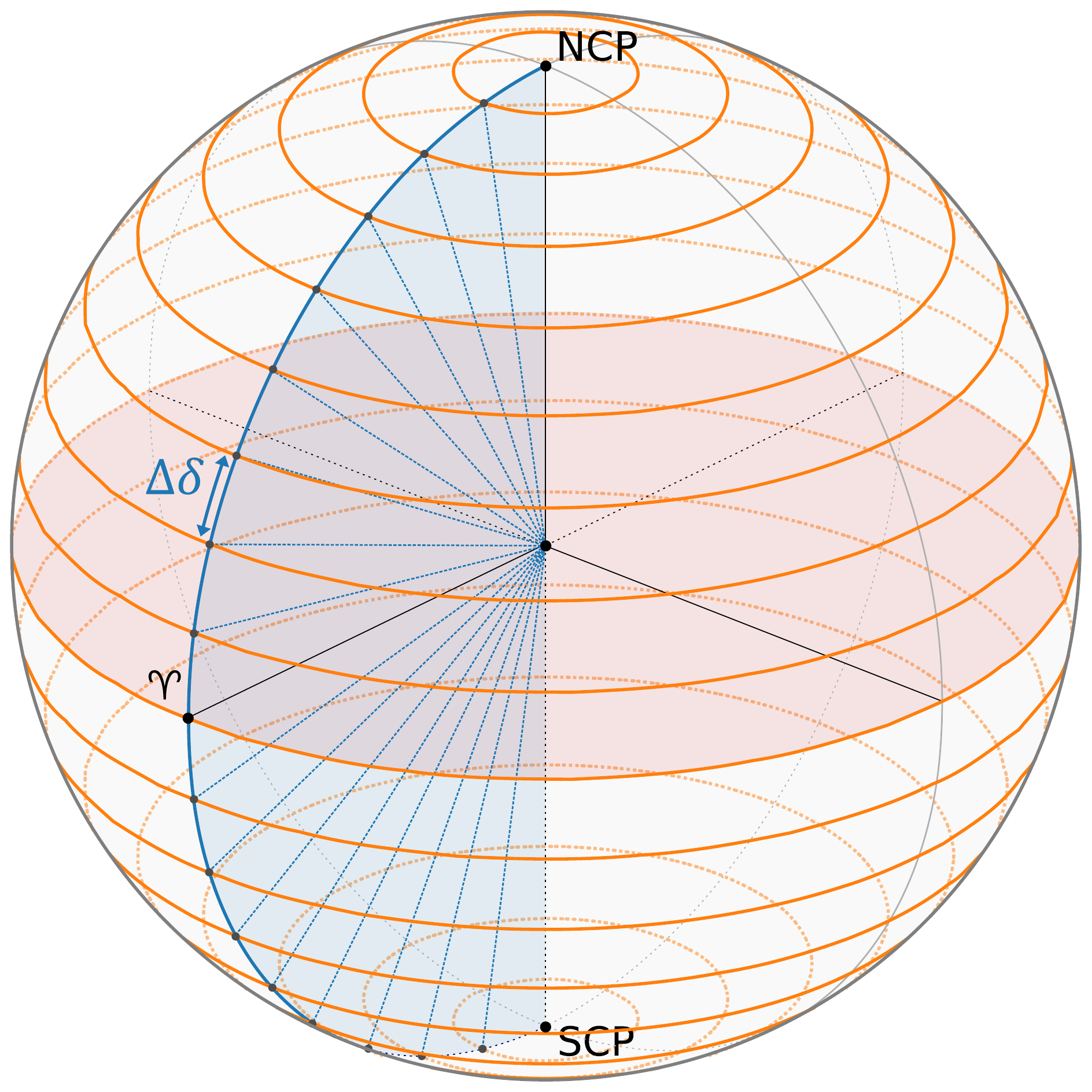}
    \end{center}
    \caption[Defining declination strips]{
        Defining the declination strips (shown in \textcolorbf{Orange}{orange}). The full declination range (\SI{-90}{\degree} to \SI{90}{\degree}) is divided equally by a constant spacing value $\Delta\delta$ (in \textcolorbf{NavyBlue}{blue}). In this example $\Delta\delta =$ \SI{10}{\degree}, and so the centre of each strip is set at $\delta=$ \SI{0}{\degree}, $\pm$\SI{10}{\degree}, $\pm$\SI{20}{\degree} etc. This gives 19 strips, 9 in each hemisphere and one on the equator. There is always a strip centred on $\delta=0$, and using the ``minverlap'' algorithm (see \aref{sec:algorithms}) there is always a `strip' of tiles at the poles which will include a single point.
    }\label{fig:deltadelta}
\end{figure}

\clearpage

\begin{figure}[t]
    \begin{center}
        \includegraphics[width=\linewidth]{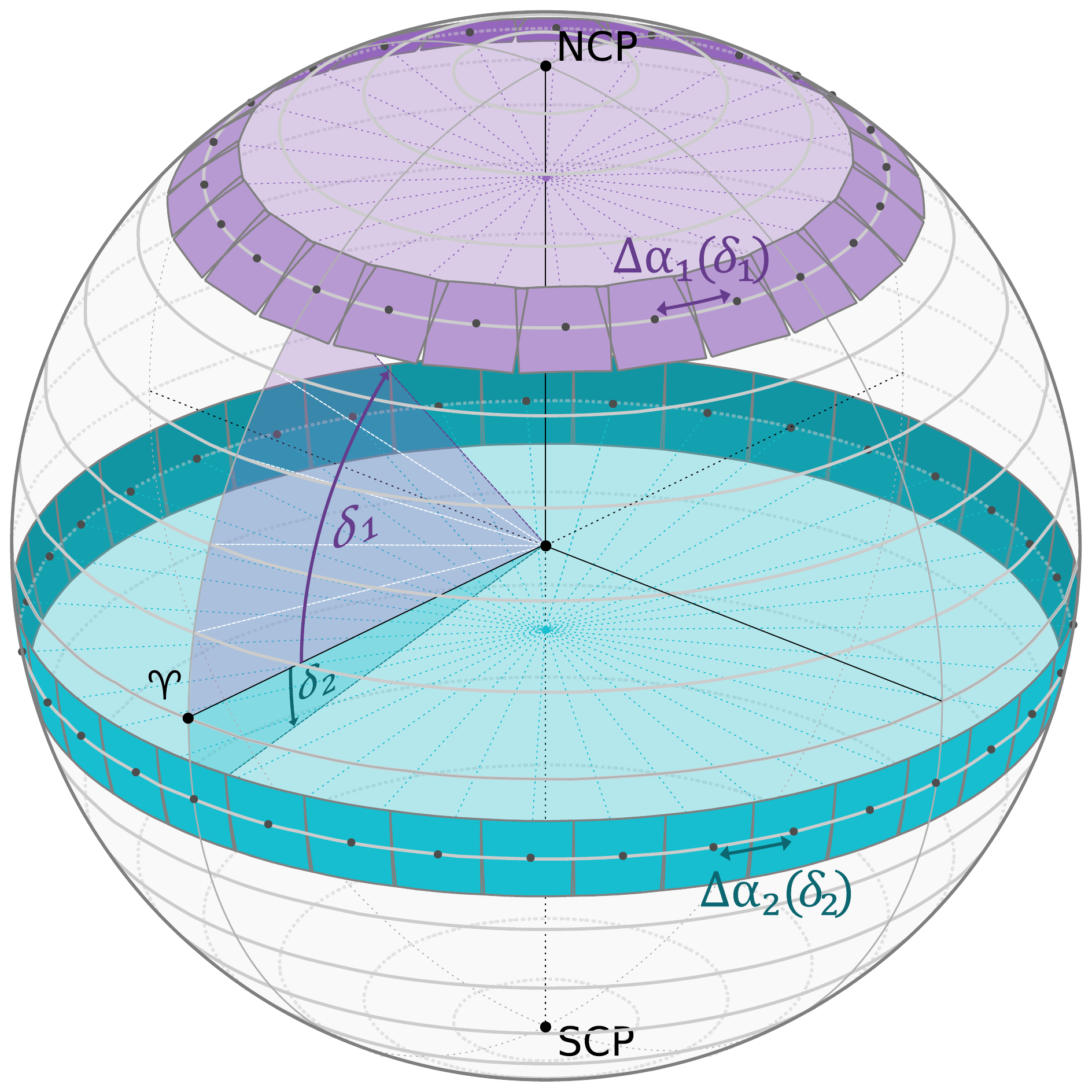}
    \end{center}
    \caption[Defining grid points]{
        Defining the grid points, based on the declination strips from \aref{fig:deltadelta}. Using the ``minverlap'' algorithm (see \aref{sec:algorithms}) points are uniformly distributed on each declination strip with a spacing $\Delta\alpha(\delta)$. Unlike $\Delta\delta$, which is fixed across the sphere, $\Delta\alpha$ varies as a function of declination, meaning strips closer to the poles will contain fewer points (and therefore fewer tiles). Two examples of defining grid points are shown, one at declination $\delta_1=+$\SI{50}{\degree} (in \textcolorbf{Purple}{purple}) in the northern hemisphere and another at $\delta_2=-$\SI{10}{\degree} (in \textcolorbf{BlueGreen}{cyan}) in the southern hemisphere. The survey tiles are then centred on the grid points, as shown for these two strips.
    }\label{fig:deltaalpha}
\end{figure}

\clearpage

\begin{figure}[t]
    \begin{center}
        \includegraphics[width=\linewidth]{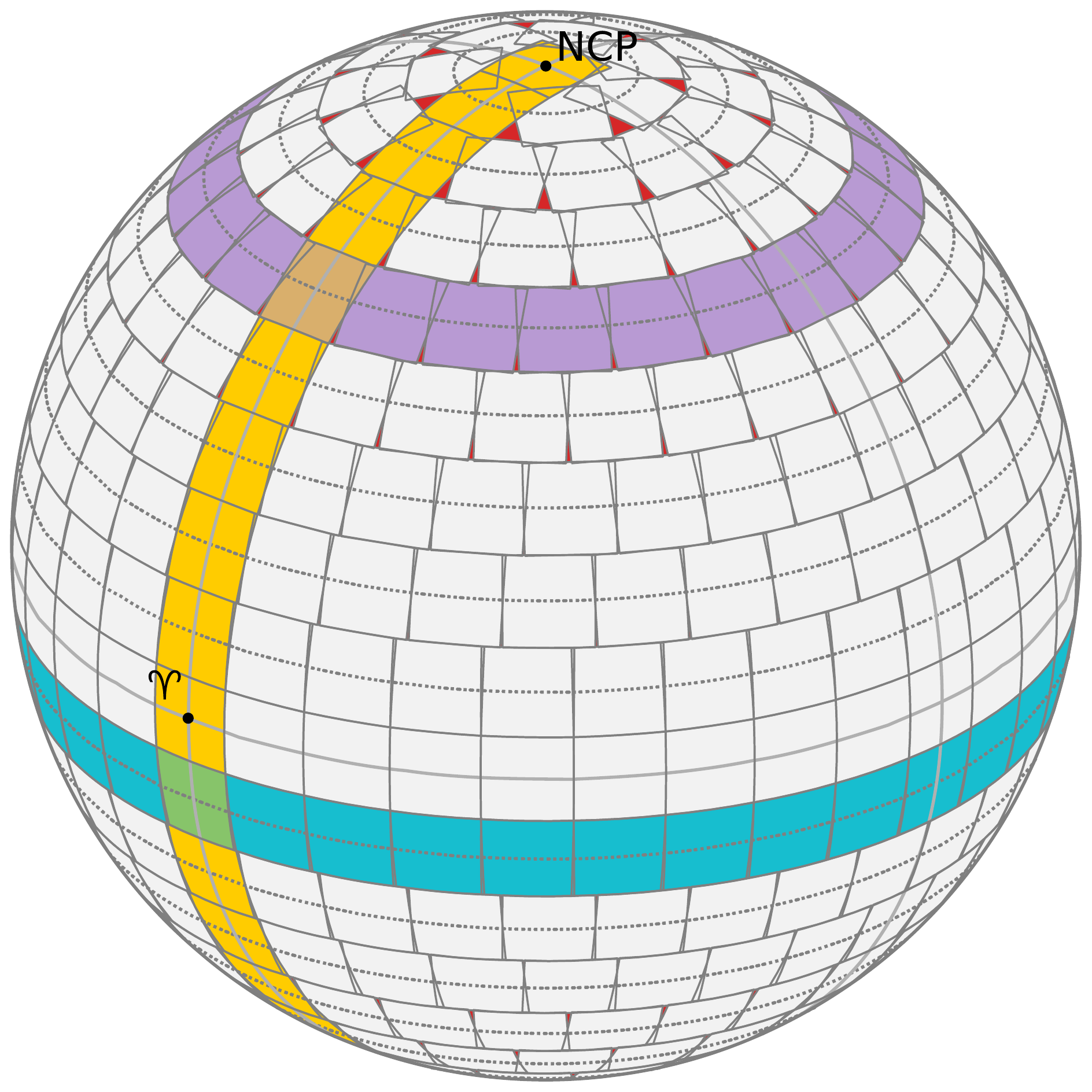}
    \end{center}
    \caption[A fully tiled celestial sphere]{
        A fully tiled celestial sphere. The same two strips of tiles are coloured in \textcolorbf{Purple}{purple} and \textcolorbf{BlueGreen}{cyan} as in \aref{fig:deltaalpha}. As each strip starts with a tile at RA$=0$ there is a fully aligned column of tiles along the meridian through the vernal equinox, coloured in \textcolorbf{YellowOrange}{yellow}, and there is always a tile centred on the vernal equinox and each pole. This grid was defined using the ``minverlap'' algorithm (see \aref{sec:algorithms}), with each tile having a field of view of \SI{10}{\degree} $\times$ \SI{10}{\degree} and a overlap of zero for clarity. Note zero overlap can lead to gaps between tiles towards the poles, shown by the \textcolorbf{Red}{red} patches. In this case the complete grid contains 424 tiles.
    }\label{fig:tiledsphere}
\end{figure}

\clearpage

\makeatletter
\setlength{\@fptop}{0\p@ \@plus 1fil} %
\makeatother

\end{colsection}

\subsection{Different gridding algorithms}
\label{sec:algorithms}
\begin{colsection}

There have been three different algorithms used by GOTO-tile to define the grid.

\subsubsection{The product algorithm}

The first has since retroactively been called the ``\textbf{product}'' algorithm, and was used when Darren White first wrote GOTO-tile. It first defines the declination step size as
\begin{equation}
    \Delta\delta = f_\text{dec}(1-v_\text{dec}),
    \label{eq:product_deltadelta}
\end{equation}
where $f_\text{dec}$ and $v_\text{dec}$ are respectively the field of view in degrees and the fractional overlap parameters in the declination direction. The declination strips are then defined by taking steps of this size from the equator towards the poles, stopping when $|\delta| > 90$. An equivalent formula is used to calculate the steps in right ascension

\begin{equation}
    \Delta\alpha = f_\text{RA}(1-v_\text{RA}).
    \label{eq:product_deltaalpha}
\end{equation}

The clear downside of this method is that $\Delta\alpha$ does not vary with declination. In effect this algorithm attempts to define the grid as if it was on a flat plane, where the tiles could be arranged in orthogonal rows and columns. In practice when applied to a sphere this leads to a vast number of redundant tiles at the poles, as shown in \aref{fig:product}.

\subsubsection{The cosine algorithm}

Due to the obvious problems with the product algorithm, a replacement was written by Evert Rol, which I have since called the ``\textbf{cosine}'' algorithm. It is a more refined version of the product algorithm, and the declination strips are calculated in the same manner using \aref{eq:product_deltadelta}. However \aref{eq:product_deltaalpha} is modified to depend on declination:
\begin{equation}
    \Delta\alpha(\delta) = \frac{f_\text{RA}(1-v_\text{RA})}{\cos \delta}.
    \label{eq:cosine_deltaalpha}
\end{equation}

This produces a more sensible grid with fewer redundant tiles at the poles, as shown in \aref{fig:cosine}. However, there remained an issue of asymmetry: the strips are arranged increasing and decreasing from $\delta=0$ and the tiles are then arranged within the strips starting from $\alpha=0$. This leads to varying levels of overlap when the tiles within the strips meet as $\alpha$ approaches \SI{360}{\degree}, as visible in \aref{fig:cosine}. Although more subtle, there are similar issues at the north and south celestial poles. It is also common for there to be small gaps between the tiles at high and low declinations.

\subsubsection{The minverlap algorithm}

Due to these problems I created a new method to create the grid, called the ``\textbf{minverlap}'' (\emph{min}imum o\emph{verlap}) algorithm. A grid created with this algorithm is shown in \aref{fig:minverlap}. The intention of the new algorithm was to solve the issues with the product and cosine algorithms by dynamically adjusting the spacing between tiles. The previous two algorithms both treated the user-specified overlap parameter as fixed, and if the resulting spacings did not give an integer number of tiles within the ranges available then they produced uneven gaps at the edges. This is shown more clearly in \aref{fig:cosine_spacing}, where a particular spacing results in gaps at the celestial poles and variable overlaps across the RA$=0$ meridian.

The minverlap algorithm solves these problems by treating the overlap parameter not as fixed but as the \textit{minimum} required overlap between tiles. For example, if a grid is requested with an overlap of $0.2$ (20\%) but the field of view of the tiles does not neatly divide by \SI{90}{\degree} then the overlap can be increased until an integer number of tiles fit, as shown in \aref{fig:minverlap_spacing}.

\newpage

\begin{figure}[p]
    \begin{minipage}[c]{0.46\linewidth}
        \includegraphics[width=\linewidth]{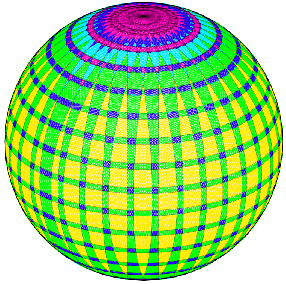}
    \end{minipage}
    \hfill
    \begin{minipage}[c]{0.50\linewidth}
        \caption[The product gridding algorithm]{
            A sky grid of tiles defined using the ``product'' gridding algorithm. The inputs were a field of view of \SI{13}{\degree} $\times$ \SI{13}{\degree} and an overlap factor of $0.2$ in both axes. The colours show overlapping coverage: \textcolorbf{YellowOrange}{yellow} areas are within only one tile, \textcolorbf{ForestGreen}{green} two, \textcolorbf{cyan}{cyan} three, \textcolorbf{blue}{blue} four and \textcolorbf{RubineRed}{pink} five or more. This grid contains 595 tiles. Note the constant spacing of tiles in RA and the huge number of redundant tiles at the pole.
        }\label{fig:product}
    \end{minipage}
\end{figure}

\begin{figure}[p]
    \begin{minipage}[c]{0.50\linewidth}
        \caption[The cosine gridding algorithm]{
            A sky grid of tiles defined using the ``product'' gridding algorithm. The input parameters and colours are the same as in \aref{fig:product}. This grid contains 393 tiles. Note the asymmetric ``seam'' along the $\alpha=0$ meridian, and the \textcolorbf{red}{red} areas near the pole that are not within the area of any tiles.
        }\label{fig:cosine}
    \end{minipage}
    \hfill
    \begin{minipage}[c]{0.46\linewidth}
        \includegraphics[width=\linewidth]{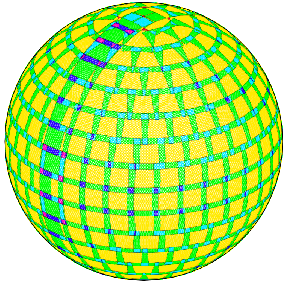}
    \end{minipage}
\end{figure}

\begin{figure}[p]
    \begin{minipage}[c]{0.46\linewidth}
        \includegraphics[width=\linewidth]{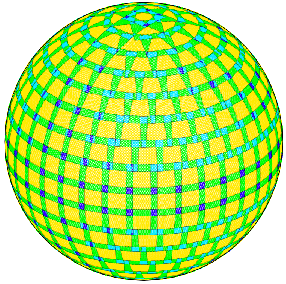}
    \end{minipage}
    \hfill
    \begin{minipage}[c]{0.50\linewidth}
        \caption[The minverlap gridding algorithm]{
            A sky grid of tiles defined using the ``minverlap'' gridding algorithm. The input parameters and colours are the same as in \aref{fig:product}. This grid contains 407 tiles. Note the even spacing of tiles even over the $\alpha=0$ meridian, and the better coverage at the pole.
        }\label{fig:minverlap}
    \end{minipage}
\end{figure}

\newpage

\begin{figure}[p]
    \begin{center}
        \includegraphics[width=\linewidth]{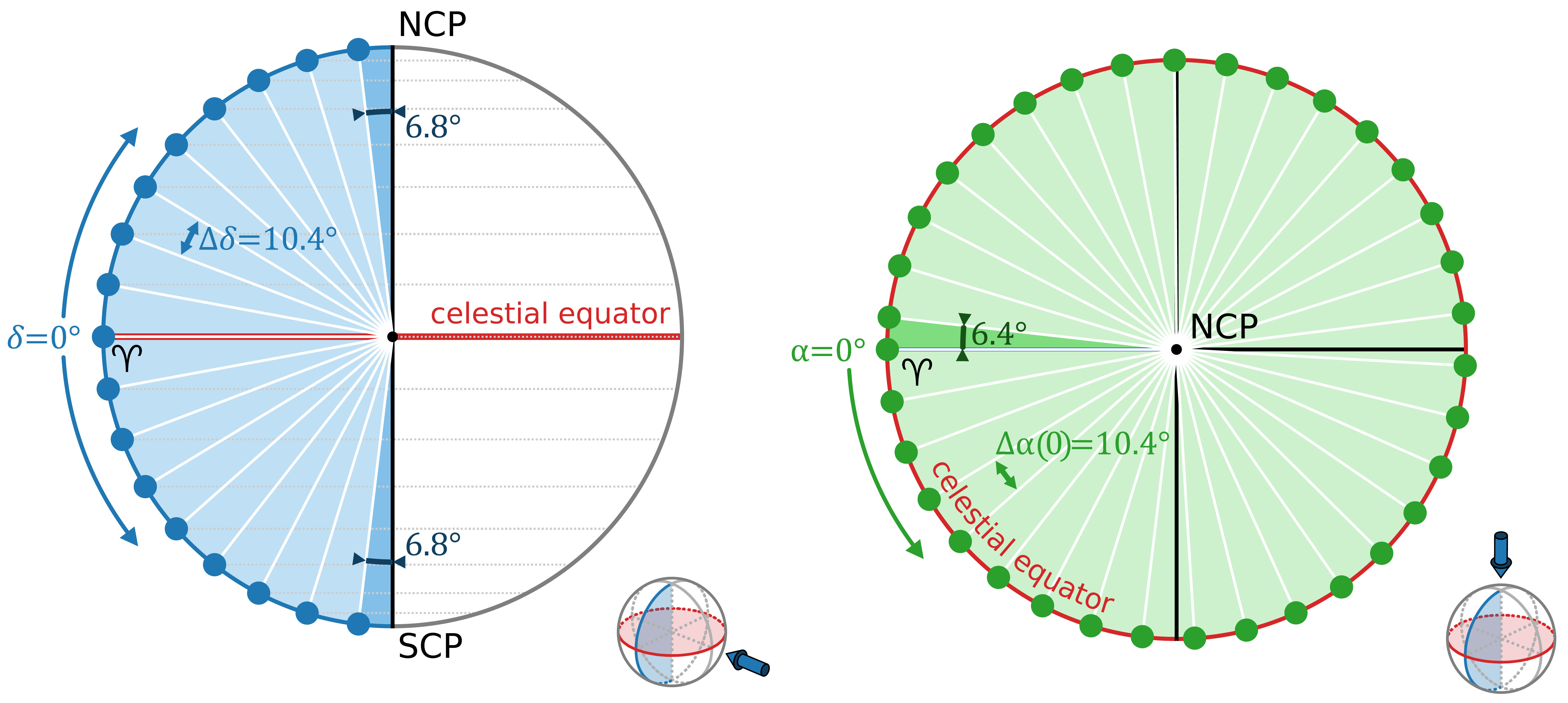}
    \end{center}
    \caption[Grid spacing with the cosine algorithm]{
        Grid spacing with the cosine algorithm. Using a \SI{13}{\degree}$\times$\SI{13}{\degree} field of view and an overlap of $0.2$ \aref{eq:product_deltadelta} gives $\Delta\delta = $ \SI{10.4}{\degree}. 17 declination strips are defined moving away from $\delta=0$, as shown in the equatorial view on the left. The final strips are \SI{6.8}{\degree} from the poles, as this is more than half of the field of view (\SI{6.5}{\degree}) the poles themselves will not by within the area of any tile. \aref{eq:cosine_deltaalpha} gives $\Delta\alpha = $ \SI{10.4}{\degree} on the equator ($\delta=0$). This results in 35 points arranged as shown in the polar view on the right, and a reduced spacing of \SI{6.4}{\degree} to the west of the $\alpha=0$ meridian. This remainder will be different for each strip, as shown in \aref{fig:cosine}.
    }\label{fig:cosine_spacing}
\end{figure}

\begin{figure}[p]
    \begin{center}
        \includegraphics[width=\linewidth]{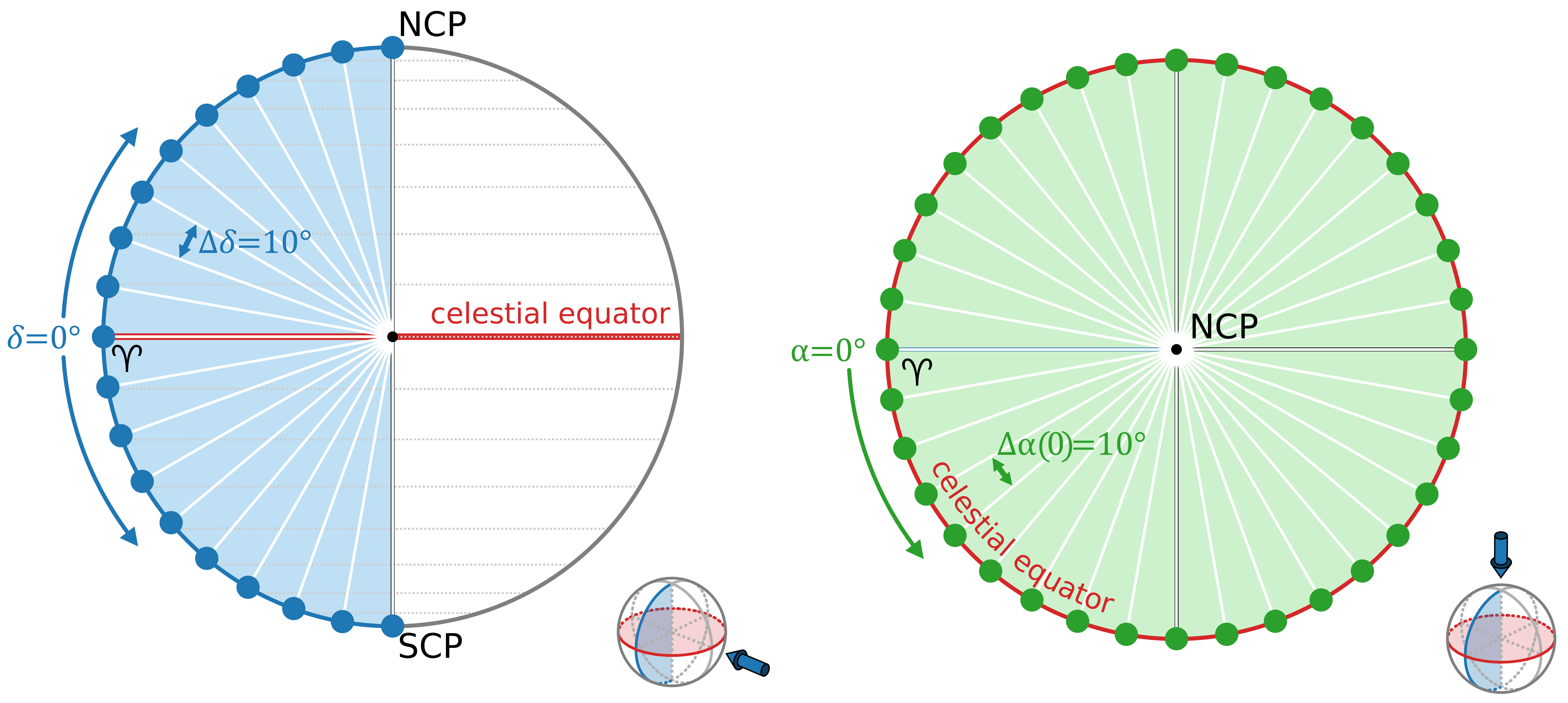}
    \end{center}
    \caption[Grid spacing with the minverlap algorithm]{
        Grid spacing with the minverlap algorithm. Using the same parameters as \aref{fig:cosine_spacing}, \aref{eq:minverlap_deltadelta} gives $\Delta\delta = $ \SI{10}{\degree}, therefore neatly arranging 19 declination strips between \SI{-90}{\degree} and \SI{90}{\degree}. \aref{eq:minverlap_deltaalpha} also gives $\Delta\alpha = $ \SI{10}{\degree} on the equator ($\delta=0$), so 36 points are uniformly arranged around the circumference.
    }\label{fig:minverlap_spacing}
\end{figure}

\clearpage

To define the grid points using the minverlap algorithm, it is first necessary to find the number of tiles $n$ that would fit into the available range using the cosine algorithm spacing; if this is not an integer number of tiles then round it up to the next whole number. In declination this is calculated as
\begin{equation}
    n_\text{dec} = \left \lceil \frac{90}{f_\text{dec}(1-v_\text{dec})} \right \rceil,
    \label{eq:minverlap_ndec}
\end{equation}
where $\lceil x \rceil$ is the mathematical ceiling function. This is a modification of \aref{eq:product_deltadelta}, but one that will always find an integer number of tiles. For example, for a tile with a declination field of view $f_\text{dec} = $ \SI{13}{\degree} and overlap $v_\text{dec} = 0.2$ \aref{eq:product_deltadelta} gives $\Delta\delta = 13 \times (1-0.2) = $ \SI{10.4}{\degree}. This clearly does not divide into \SI{90}{\degree} without a remainder, \SI{6.8}{\degree}, as shown in \aref{fig:cosine_spacing}. The problem is that \SI{90}{\degree} $/$ \SI{10.4}{\degree} $= 8.65$, so the product and cosine algorithms will fit in 8 declination strips and then have over half the height of a tile remaining at the poles. Instead, the minverlap algorithm rounds this up to $n_\text{dec} = 9$ and then calculates the spacing using
\begin{equation}
    \Delta\delta = \frac{90}{n_\text{dec}}.
    \label{eq:minverlap_deltadelta}
\end{equation}

In this case the new $\Delta\delta = $ \SI{10}{\degree}, which gives an even arrangement of tiles from the equator to the poles, as shown in \aref{fig:minverlap_spacing}. The other benefit of this method is that, in addition to there always being a declination strip at $\delta=0$, there will always be ``strips'' at \SI{+90}{\degree} and \SI{-90}{\degree}, which results in a single tile being located over the celestial poles and ensuring there are no major gaps in coverage.

In the minverlap algorithm the spacing in right ascension is treated in a similar way. The integer number of tiles that can fit into a given declination strip is given by
\begin{equation}
    n_\text{RA}(\delta) = \left \lceil \frac{360}{f_\text{RA}(1-v_\text{RA})/\cos \delta} \right \rceil + 1,
    \label{eq:minverlap_nra}
\end{equation}
where the $+1$ is to account for tiles being located both at $\alpha=$\SI{0}{\degree} and $\alpha=$\SI{360}{\degree}. The logic is exactly the same as with declination, and the revised spacing is given by
\begin{equation}
    \Delta\alpha(\delta) = \frac{360}{n_\text{RA}(\delta)}.
    \label{eq:minverlap_deltaalpha}
\end{equation}

This spacing is also shown in \aref{fig:minverlap_spacing}, with the grid points uniformly spaced around the celestial equator. Note that the ceiling function means that, in some cases, strips at different declinations can have the same number of tiles. For example, using the same parameters as previously $\Delta\delta=$\SI{10}{\degree}, so declination strips start at \SI{0}{\degree} and continue to \SI{\pm10}{\degree}, \SI{\pm20}{\degree} \ldots (mirrored in both hemispheres). From \aref{eq:minverlap_nra} the number of tiles on the equator is $n_\text{RA}(\delta=\SI{0}{\degree}) = \lceil 360/(10.4/\cos \SI{0}{\degree}) \rceil + 1 = \lceil 34.6 \rceil + 1 = 36$. But on the next strip up (or down) $n_\text{RA}(\delta=\SI{\pm10}{\degree}) = \lceil 360/(10.4/\cos(\pm\SI{10}{\degree})) \rceil + 1 = \lceil 34.1 \rceil + 1 = 36$ as well. This occurs because there are only a limited number of ways to fit an integer number of fixed tiles into a given declination strip, and so, as shown in \aref{fig:minverlap}, the three strips around the equator align perfectly with the same number of tiles.

\subsubsection{Limitations of the minverlap algorithm}

The new minverlap algorithm is a significant improvement on the previous gridding algorithms. In particular it reduces the occurrences of gaps in coverage close to the poles which occur when using the cosine algorithm. However, gaps can still occur when using the minverlap with a particularly low overlap parameter. For example, \aref{fig:tiledsphere} shows a sphere tiled using the minverlap algorithm with an overlap parameter of 0 and in this case gaps are visible just below the northern celestial pole.

A proposed solution to this problem would be to force tiles to meet at their lower corners (in the northern hemisphere; upper corners in the south), therefore overlapping further and removing the possibility of gaps forming due to the angle between the tiles. An attempt to make this change and create an ``enhanced minverlap'' algorithm was tested, however ultimately it proved unnecessary. Although the current minverlap algorithm is deficient at low overlap values, this is only an issue when used with large tiles. The \SI{10}{\degree} $\times$ \SI{10}{\degree} tiles and 0 overlap used for \aref{fig:tiledsphere} are extreme values, and even for the roughly \SI{8}{\degree} $\times$ \SI{5}{\degree} full field of view of GOTO with 8 unit telescopes the overlap has to be less than 0.1 before noticeable gaps start appearing.

A further possible improvement to the minverlap algorithm has also been identified since it was implemented. Instead of locating two grid points precisely at the celestial poles (i.e.\ using declination strips at $\pm$\SI{90}{\degree}) as shown in \aref{fig:minverlap_spacing}, it would instead be enough to have the highest/lowest declination strips exactly half of the field of view away from the poles (e.g.\ at $\pm$\SI{85}{\degree} if the tile was \SI{10}{\degree} tall). This would ensure the poles were still included within the tiled area, at the top/bottom of the highest/lowest strip, but could reduce the number of strips needed to cover the entire sphere.

As it happens, when viewed from La Palma, the northern celestial pole is below the GOTO altitude limit of \SI{30}{\degree}. This means that tiles closest to the pole are not visible, and so the issues described above are irrelevant. Should GOTO-tile be applied in the future to other telescopes at other sites then this issue would need to be revisited, but it was not a priority to fix within the context of this work.

\end{colsection}

\section{Probability skymaps}
\label{sec:skymaps}

\begin{colsection}

When identifying a particular target in the sky its coordinates can be given in right ascension and declination, and if there is some uncertainty in the position, then errors can be given on the coordinate values. For example, a \acro{grb} event detected by the \textit{Fermi} \acro{gbm} might have a central position and an error radius ranging from arcseconds to tens of degrees. However, multiple gravitational-wave detectors produce large and distinctly asymmetric localisation areas (see \aref{sec:gw_localisation}). For these cases, the \acro{lvc} produce probability skymaps which map the localisation area onto the celestial sphere. This section describes how these skymaps are defined and how they are mapped onto the GOTO all-sky grid as described in \aref{sec:gototile}.

\end{colsection}

\subsection{Defining skymaps with HEALPix}
\label{sec:healpix}
\begin{colsection}

\acro{healpix} is a system used to define pixelised data on the surface of a sphere~\citep{HEALPix}. Developed at NASA JPL for microwave background data, it is now widely used for other applications including for gravitational-wave skymaps produced by the LVC.\@ HEALPix divides the sphere into a series of nested (hierarchical), equal-area (although not equal-shape) pixels arranged in declination strips (``isoLatitude''). The first four orders of spheres are shown in \aref{fig:healpix}, starting from a base resolution with 12 pixels and increasing as each pixel is split into four.

\begin{figure}[t]
    \begin{center}
        \includegraphics[width=0.7\linewidth]{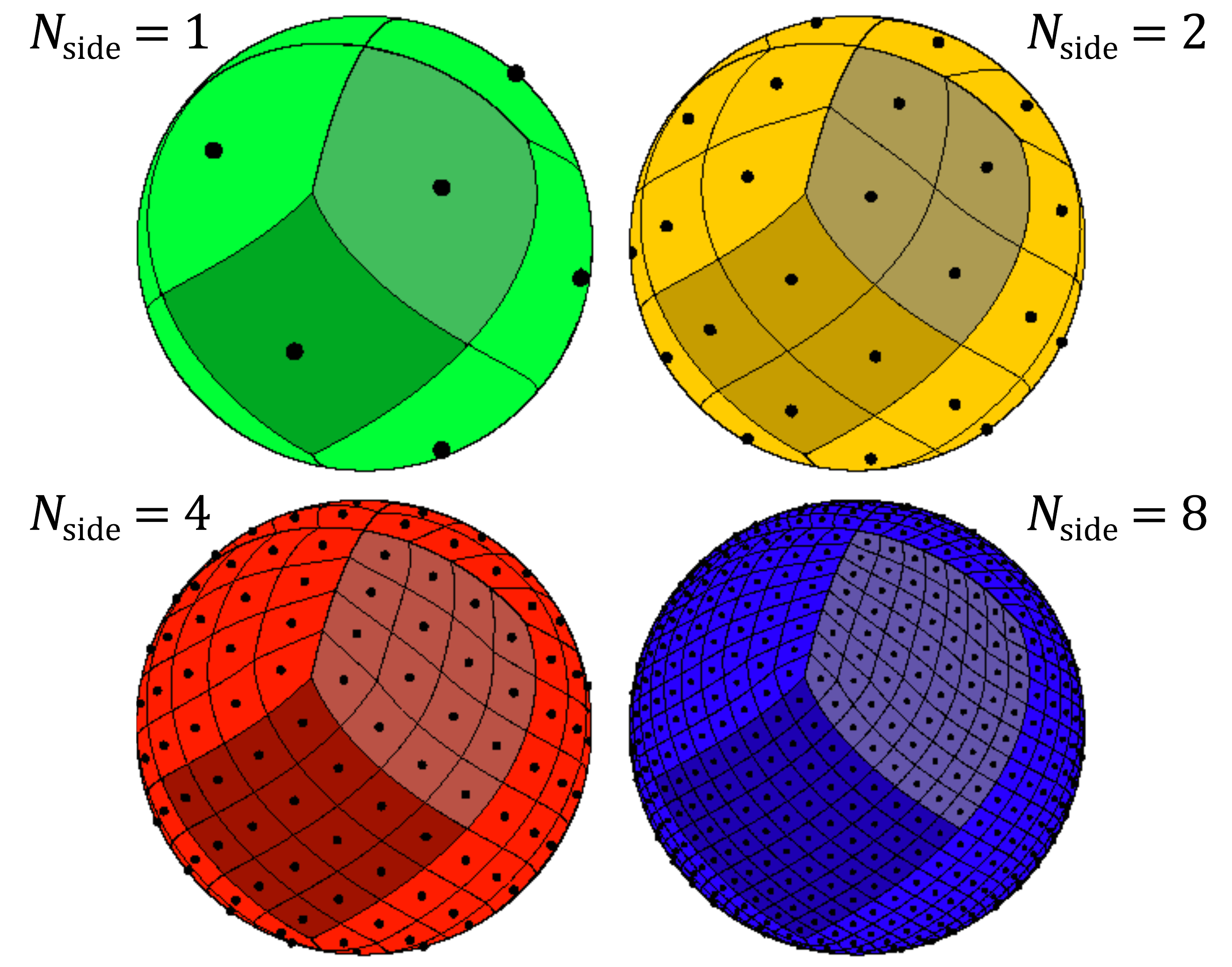}
    \end{center}
    \caption[HEALPix partitions of a sphere]{
        The first four orders of the HEALPix partition of a sphere, with increasing $N_\text{side}$ resolution parameter. Note that as the resolution doubles each pixel on the previous sphere is split into four, and $N_\text{side}$ is the number of pixels along the side of a first-order pixel (two of which are highlighted). Adapted from \citet{HEALPix}.
    }\label{fig:healpix}
\end{figure}

The resolution of the HEALPix grid is defined using the $N_\text{side}$ parameter, which for the given resolution is the number of pixels along each side of one of the 12 base pixels. At every resolution each first-order pixel contains $N_\text{side}^2$ pixels, so the total number of pixels in a sphere is given by
\begin{equation}
    N_\text{pix} = 12 N_\text{side}^2.
    \label{eq:healpix_npix}
\end{equation}
Each pixel therefore has an equal area of
\begin{equation}
    \Omega_\text{pix} = \frac{4\pi}{12 N_\text{side}^2} = \frac{\pi}{3 N_\text{side}^2},
    \label{eq:healpix_area}
\end{equation}
on a unit sphere where the radius $r=1$. Taking the celestial sphere, the circumference in degrees is $\SI{360}{\degree} = 2 \pi r$ meaning the area of the whole sky is given by
\begin{equation}
    A_\text{sky} = 4 \pi r^2 = 4 \pi \left ( \frac{\SI{360}{\degree}}{2 \pi} \right )^2 = \frac{129600}{\pi}~\text{sq deg} \approx 41252~\text{sq deg} , %
    \label{eq:sky_area}
\end{equation}
and therefore the area of each HEALPix pixel is
\begin{equation}
    A_\text{pix} = \frac{129600}{12 \pi N_\text{side}^2}~\text{sq~deg} \approx \frac{3438}{N_\text{side}^2}~\text{sq~deg}.
    \label{eq:healpix_area_degrees}
\end{equation}

\aref{fig:healpix} shows only the first four orders of HEALPix pixelisation, up to $N_\text{side} = 8$ where the sphere is split into 768 pixels each with an area of 53.7~sq deg. An initial, low-resolution LVC skymap might use a grid with $N_\text{side} = 64$ (approximately 49 thousand pixels, each with an area of 0.84~sq~deg), whereas a final output skymap will have $N_\text{side} = 1024$ (12.5 million pixels, and a pixel size resolution of $3.27 \times 10^{-3}$~sq~deg or 11.7~square~arcminutes).

In addition to being a way to divide the sphere, each HEALPix pixel has a unique index from one of two different numbering schemes: either the ring (counting around each ring from the north to the south) or nested (based on the sub-pixel tree) system.

HEALPix is used to provide localisation of sky probabilities for transient astronomical events, in the form of ``skymaps''. Each point on the HEALPix grid is assigned a probability between 0 and 1 that the counterpart object is located within that pixel, and the whole sphere should sum to unity. \aref{fig:skymap_regrade} shows a typical LVC skymap, for the gravitational-wave event S190521r \citep{S190521r}, at various HEALPix $N_\text{side}$ parameters.

\begin{figure}[p]
    \begin{center}
        \begin{tabular}{cc}
            $N_\text{side} = 1$ &
            $N_\text{side} = 2$ \\
            \includegraphics[width=0.45\linewidth]{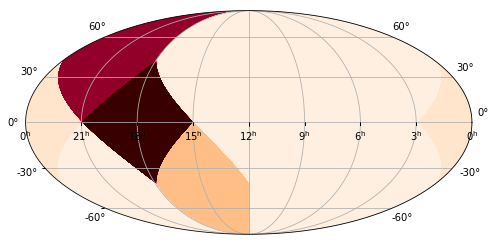} &
            \includegraphics[width=0.45\linewidth]{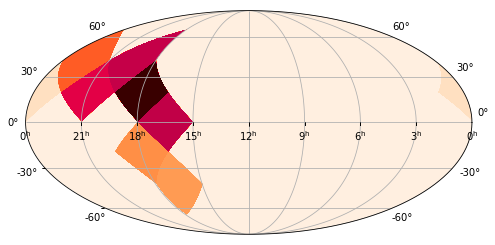} \\
            \\
            $N_\text{side} = 4$ &
            $N_\text{side} = 8$ \\
            \includegraphics[width=0.45\linewidth]{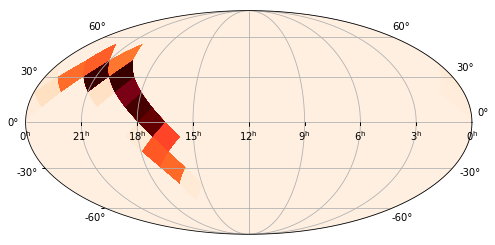} &
            \includegraphics[width=0.45\linewidth]{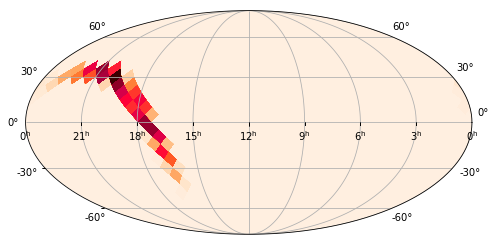} \\
            \\
            $N_\text{side} = 16$ &
            $N_\text{side} = 32$ \\
            \includegraphics[width=0.45\linewidth]{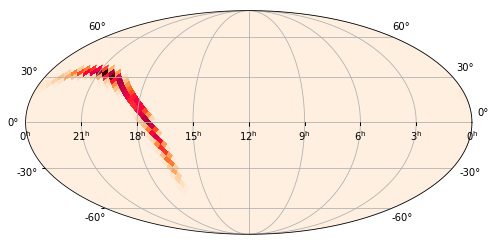} &
            \includegraphics[width=0.45\linewidth]{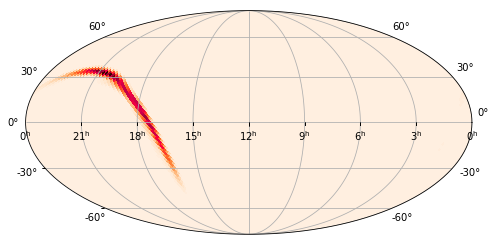} \\
            \\
            $N_\text{side} = 64$ &
            $N_\text{side} = 128$ \\
            \includegraphics[width=0.45\linewidth]{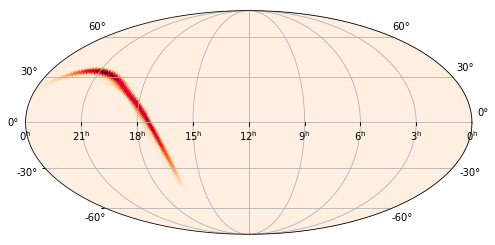} &
            \includegraphics[width=0.45\linewidth]{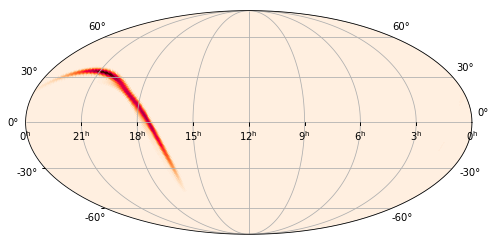} \\
        \end{tabular}
    \end{center}
    \caption[Regrading a gravitational-wave skymap]{
        Changing the HEALPix resolution of a gravitational-wave skymap (also known as regrading). At every stage each pixel is assigned a probability value which indicates the probability the source is located within that pixel; here, darker colours denote higher probabilities. At lower $N_\text{side}$ values individual pixels are visible, but as the resolution increases the HEALPix structure is less visible.
    }\label{fig:skymap_regrade}
\end{figure}

As well as the individual probabilities assigned to each pixel, it is also useful to consider the overall spread of the probability. This is done by considering the probability contour areas, typically at the 50\% and 90\% levels. The 50\% contour area of a skymap is defined by encircling the smallest number of pixels so that the total probability within the area is 50\% of the overall skymap probability. When a skymap is processed using GOTO-tile each pixel is assigned a contour value as well as its individual probability value. This is calculated by sorting all of the pixels by probability from highest to lowest, and the contour value for each pixel is then the cumulative sum of the probability within the pixels above it. This contour value can be considered as the lowest contour area that each pixel is within, meaning the pixels that are contained within the 50\% contour area are those with contour values of less than 50\%. \aref{fig:sim_skymap_probs} shows a cartoon 2-dimensional skymap, and \aref{fig:sim_skymap_conts} illustrates how the 50\% and 90\% contours are calculated.

\makeatletter
\setlength{\@fptop}{1cm}
\makeatother

\begin{figure}[t]
    \begin{center}
        \includegraphics[width=0.95\linewidth]{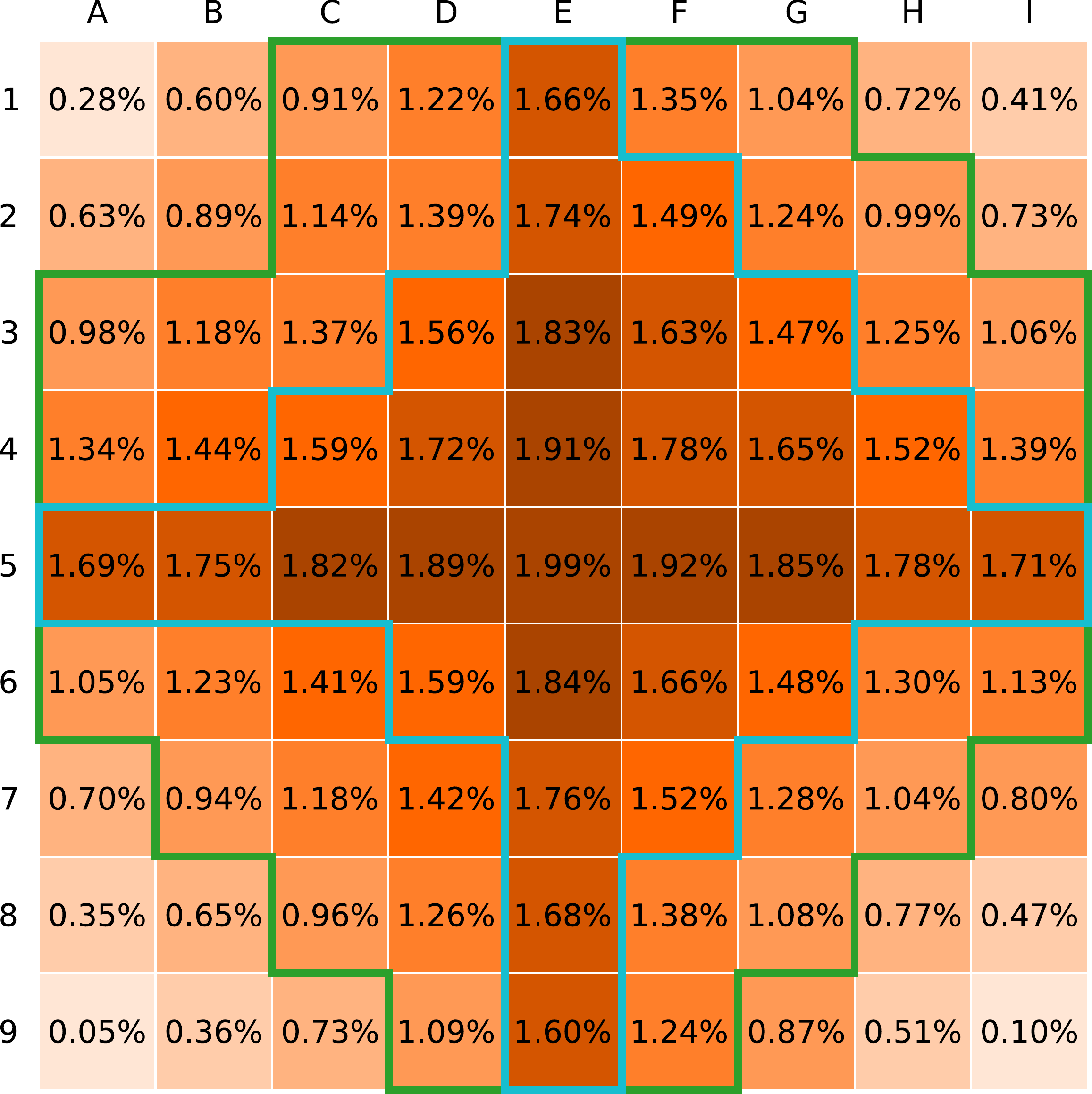}
    \end{center}
    \caption[An example 2D probability skymap]{
        A cartoon 2-dimensional skymap. Each pixel (represented by one of the 81 squares) has an assigned probability, and together they all sum to 100\%. The \textcolorbf{BlueGreen}{blue} inner contour contains 50\% of the probability, while the \textcolorbf{Green}{green} outer contour contains 90\% of the probability. These contours are created based on the values shown in \aref{fig:sim_skymap_conts}.
    }\label{fig:sim_skymap_probs}
\end{figure}

\clearpage

\begin{figure}[t]
    \begin{center}
        \includegraphics[width=0.95\linewidth]{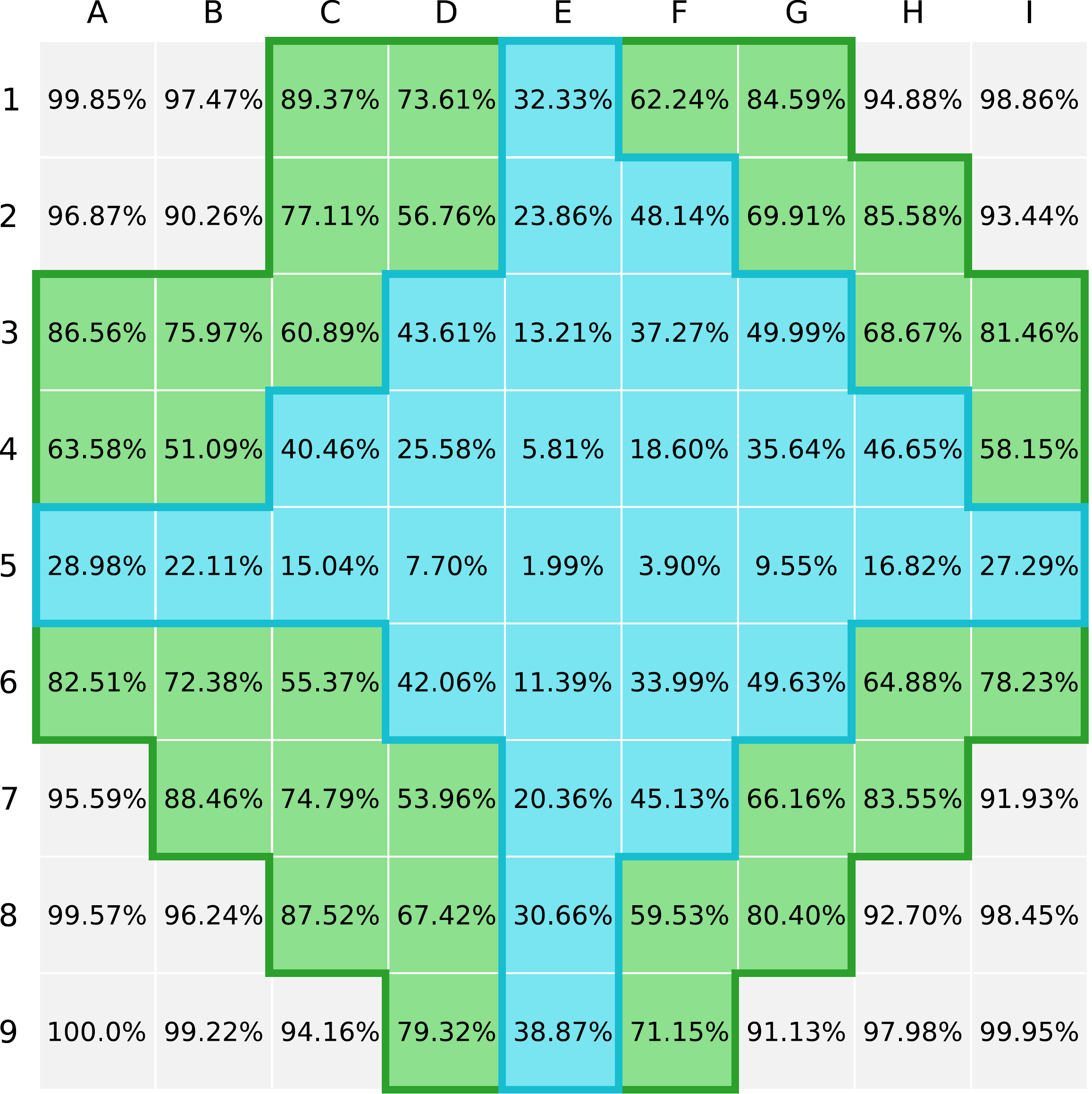}
    \end{center}
    \caption[An example 2D skymap with pixel contour values]{
        The same cartoon skymap as in \aref{fig:sim_skymap_probs}, but now each pixel contains its contour value calculated by sorting the pixels by probability and assigning each pixel the value of the cumulative sum. The pixel with the highest probability (E5) has the same contour value as its probability value in \aref{fig:sim_skymap_probs}, the second highest (F5) has the sum of the probability values of both E5 and F5, and this continues to the pixel with the lowest probability value (A9) which has a contour value of 100\%. The \textcolorbf{BlueGreen}{blue} area is the 50\% probability contour, which encloses all pixels with a contour value of less than 50\%. The \textcolorbf{Green}{green} area likewise encloses all pixels with a contour value of less than 90\%. Note in this example the contours are continuous, but it is possible to have multiple `islands' of probability within a single skymap.
    }\label{fig:sim_skymap_conts}
\end{figure}

\clearpage

\makeatletter
\setlength{\@fptop}{0\p@ \@plus 1fil} %
\makeatother

\newpage

\end{colsection}

\subsection{Mapping skymaps onto the grid}
\label{sec:mapping_skymaps}
\begin{colsection}

When a gravitational-wave signal is detected the LVC analysis pipelines create HEALPix skymaps to describe the sky localisation, and these are then distributed with the public alert (see \aref{sec:voevents}). GOTO-tile is used to map the skymaps onto the grid used for the all-sky survey (defined in \aref{sec:gototile}). This requires finding which HEALPix pixels fall within each tile, which is done by defining polygons that match the projected tile areas and using the \code{query\_polygon} function from the healpy Python package (\pkg{healpy}\footnote{\url{https://healpy.readthedocs.io}}). For each tile it is then simple to sum the probability of all the HEALPix pixels within it, which gives the total contained probability. This is shown for a cartoon skymap in \aref{fig:sim_skymap_tiles}. In cases where grid tiles overlap a given HEALPix pixel could fall within the area of multiple tiles, and therefore that pixel would contribute to the total probability of more than one tile. This means the total contained probability within all tiles can add to more than 100\%.

\begin{figure}[t]
    \begin{center}
        \includegraphics[width=0.46\linewidth]{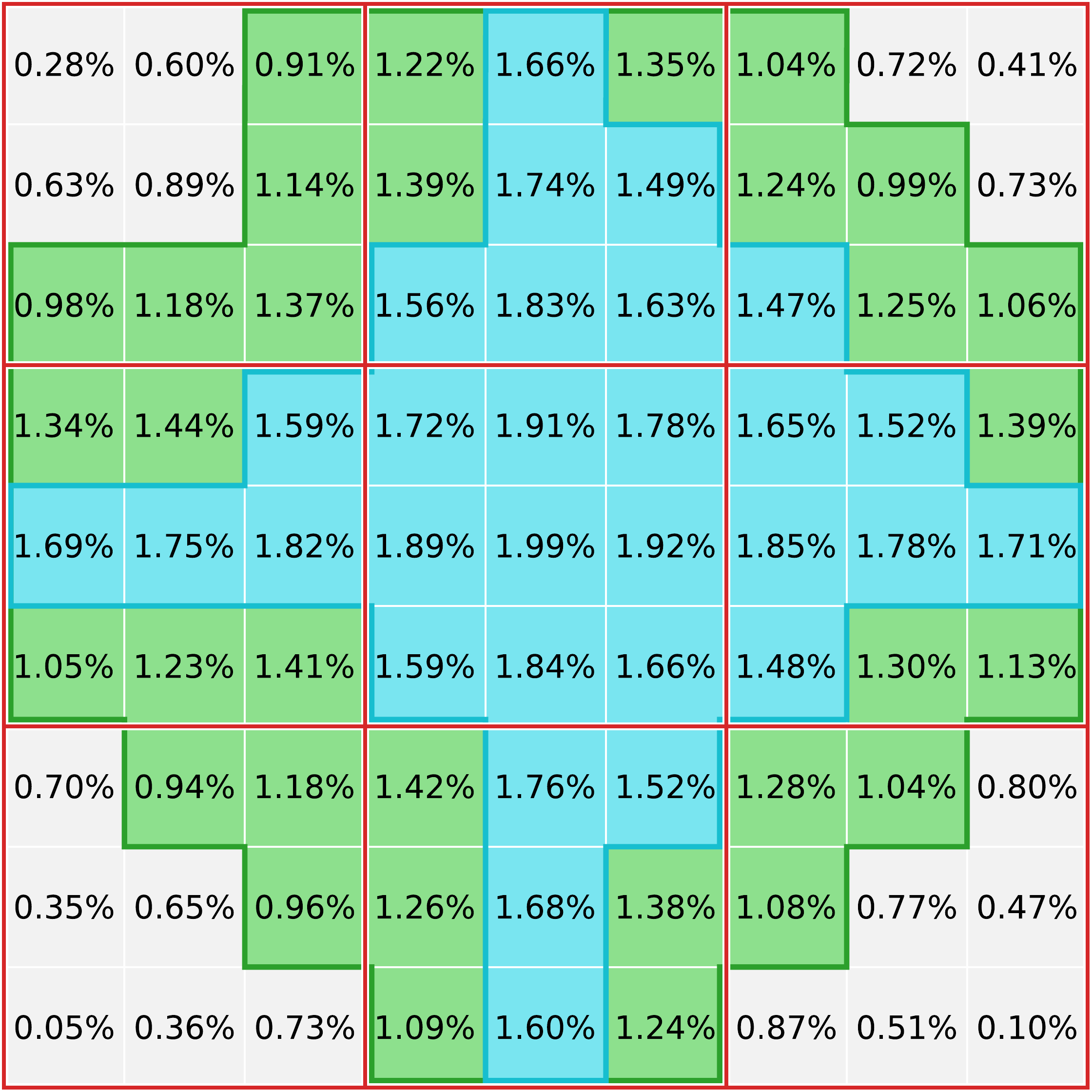}
        \includegraphics[width=0.46\linewidth]{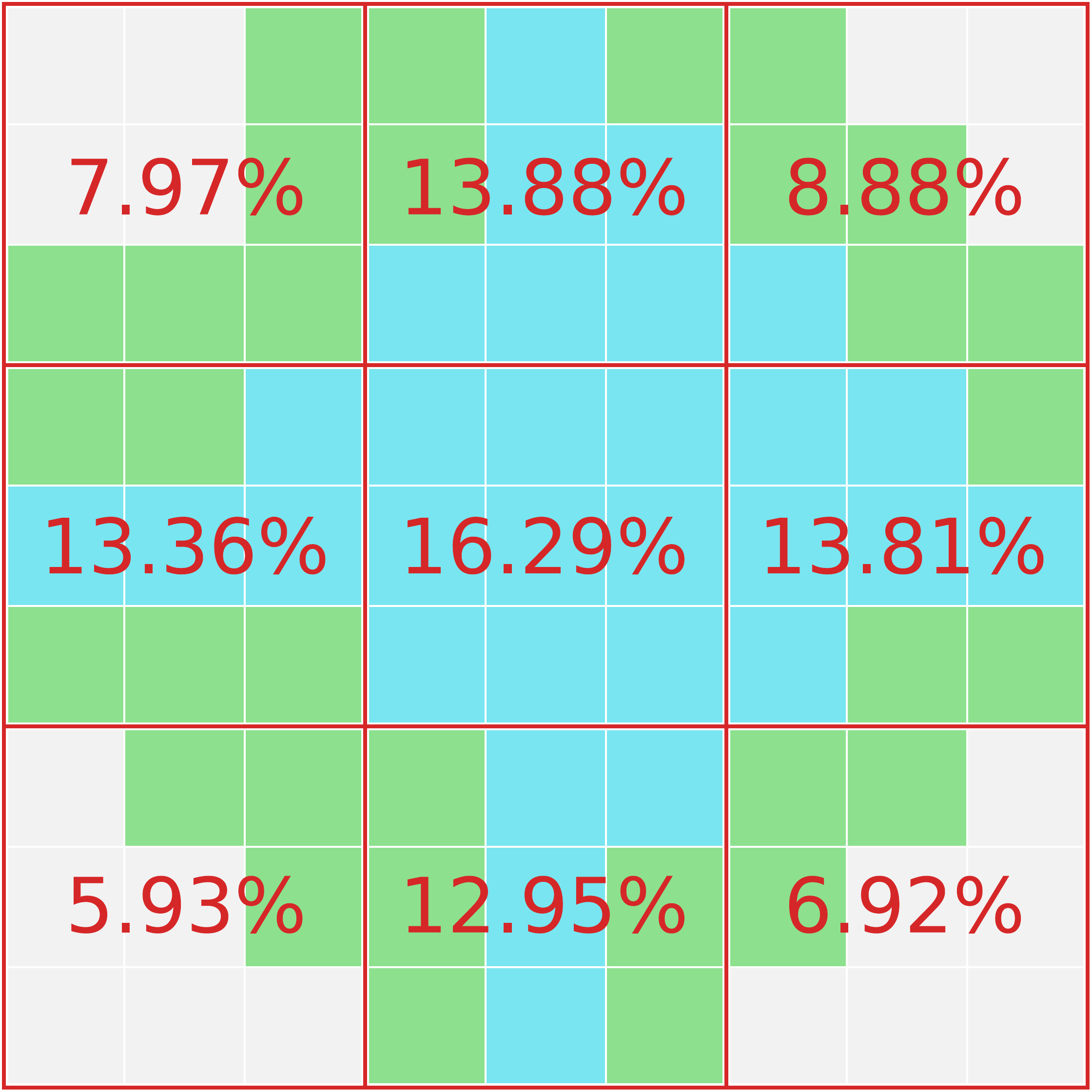}
    \end{center}
    \caption[Mapping a 2D probability skymap onto grid tiles]{
        On the left, the cartoon skymap from \aref{fig:sim_skymap_probs} has been divided into nine survey grid tiles (outlined in \textcolorbf{Red}{red}). On the right, the total contained probability for each tile is found by summing the probability of nine pixels within them.
    }\label{fig:sim_skymap_tiles}
\end{figure}

\newpage

\end{colsection}

\subsection{Selecting tiles}
\label{sec:selecting_tiles}
\begin{colsection}

When a gravitational-wave event is processed, event pointings are added into the observation database as described in \aref{sec:event_insert}. However, only a certain number of pointings should be added to prevent GOTO wasting too much time observing low-probability areas. Each pointing is mapped to a grid tile, and only tiles with a reasonably high contained probability are worth observing. The GOTO-alert event handling code described in \aref{chap:alerts} selects tiles based on their contour level, meaning GOTO could, for example, chose to observe the 90\% contour of each skymap. However, determining which contour level each tile is within is not as simple as calculating the contained probability, as there are multiple ways to define the contour value for each tile.

For example, a tile could be defined as being within a given contour area if \textit{every} pixel contained within that tile is within that contour. However this is unreasonable for large tiles, such as GOTO's, as the tile areas are often wider than the long, stretched out probability areas seen in typical gravitational-wave typical skymaps. An alternative then would be say that a tile is within a contour if \textit{any} of its contained pixels are within the contour. However, this will find every tile covering the contour area even if only the smallest fraction of the tile's area is within that region, which leads to over-selecting tiles. Several alternative methods were considered, including taking the median or mean of the contained pixel contour levels within each tile. Some different selection methods applied to the S190521r skymap are shown in \aref{fig:selecting_tiles}.

The method used within the GOTO-alert event handler is to select all tiles which have a mean contour value within 90\%. However, more quantitative simulations of different skymaps could be used to determine if this is the optimal choice for all cases. For example, the selection level could be modified depending on the size of the skymap, and the event strategy might need to be modified as more GOTO telescopes are built. These possibilities are discussed in \aref{sec:event_insert}.

\begin{figure}[p]
    \begin{center}
        \includegraphics[width=0.49\linewidth]{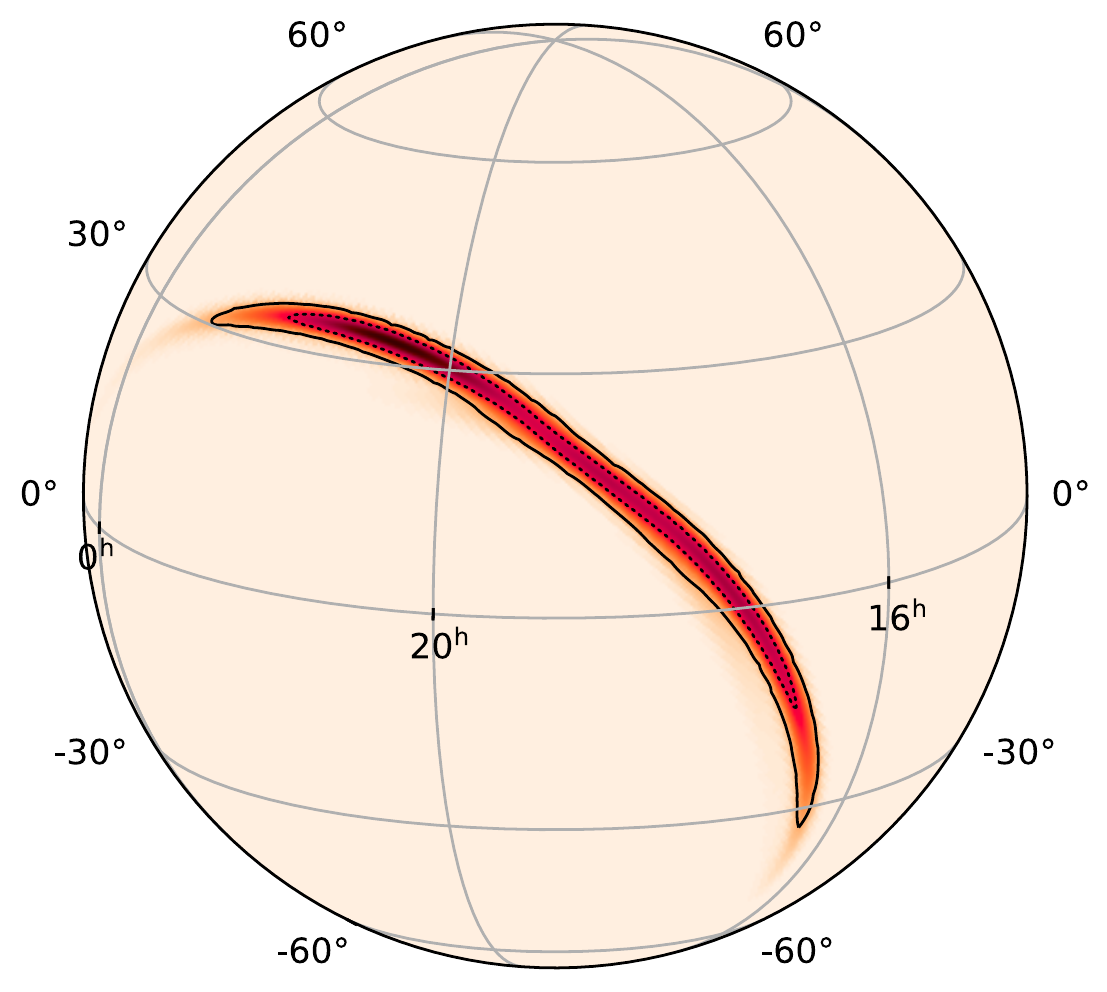}
        \includegraphics[width=0.49\linewidth]{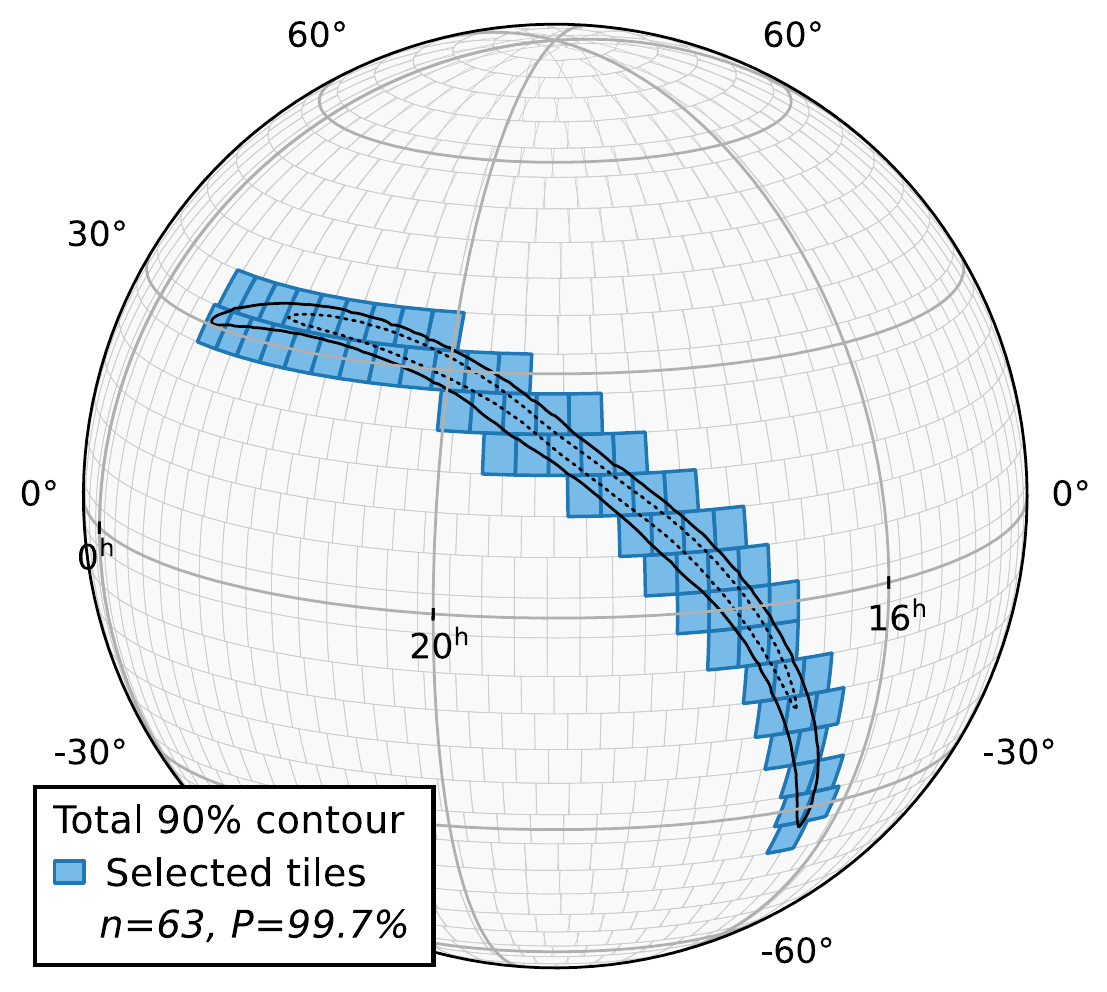}

        \vspace{0.5cm}

        \includegraphics[width=0.49\linewidth]{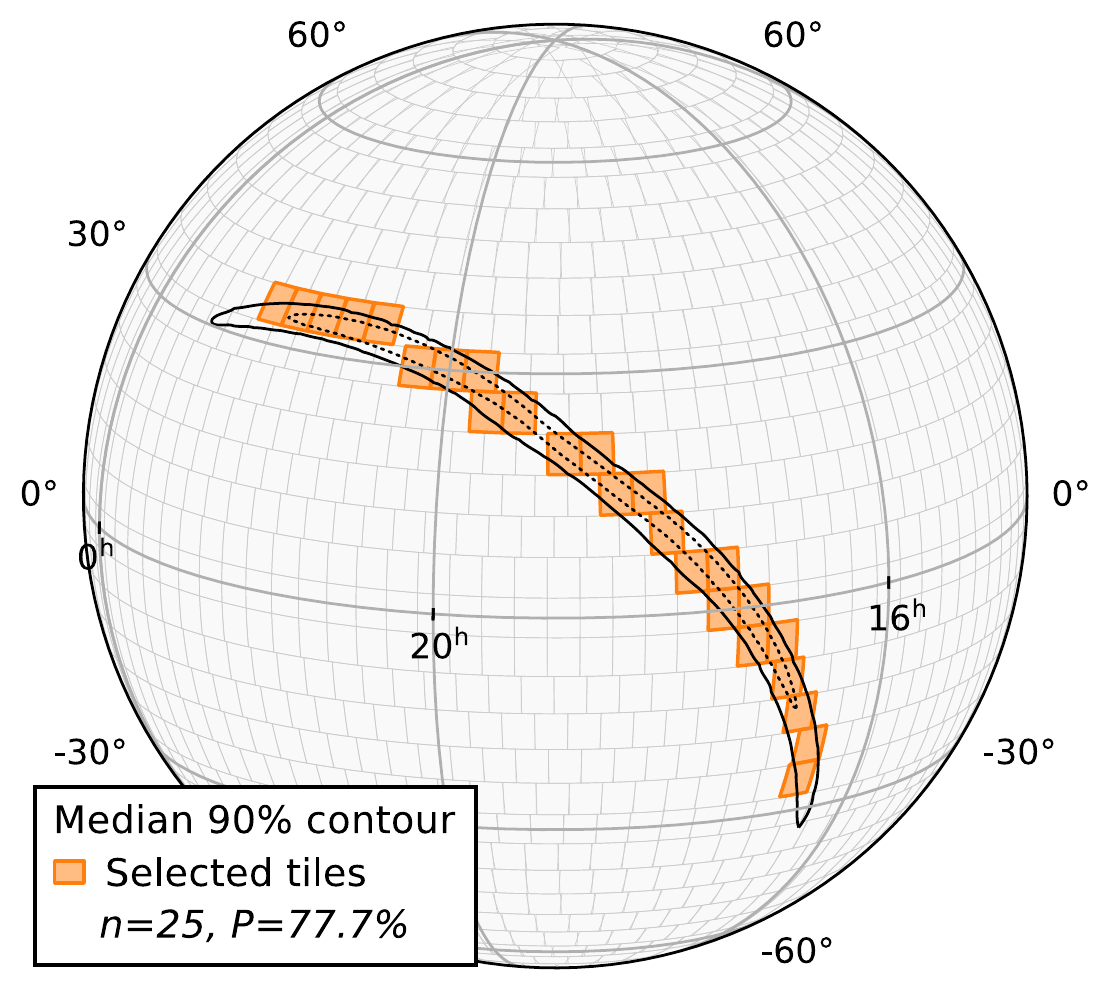}
        \includegraphics[width=0.49\linewidth]{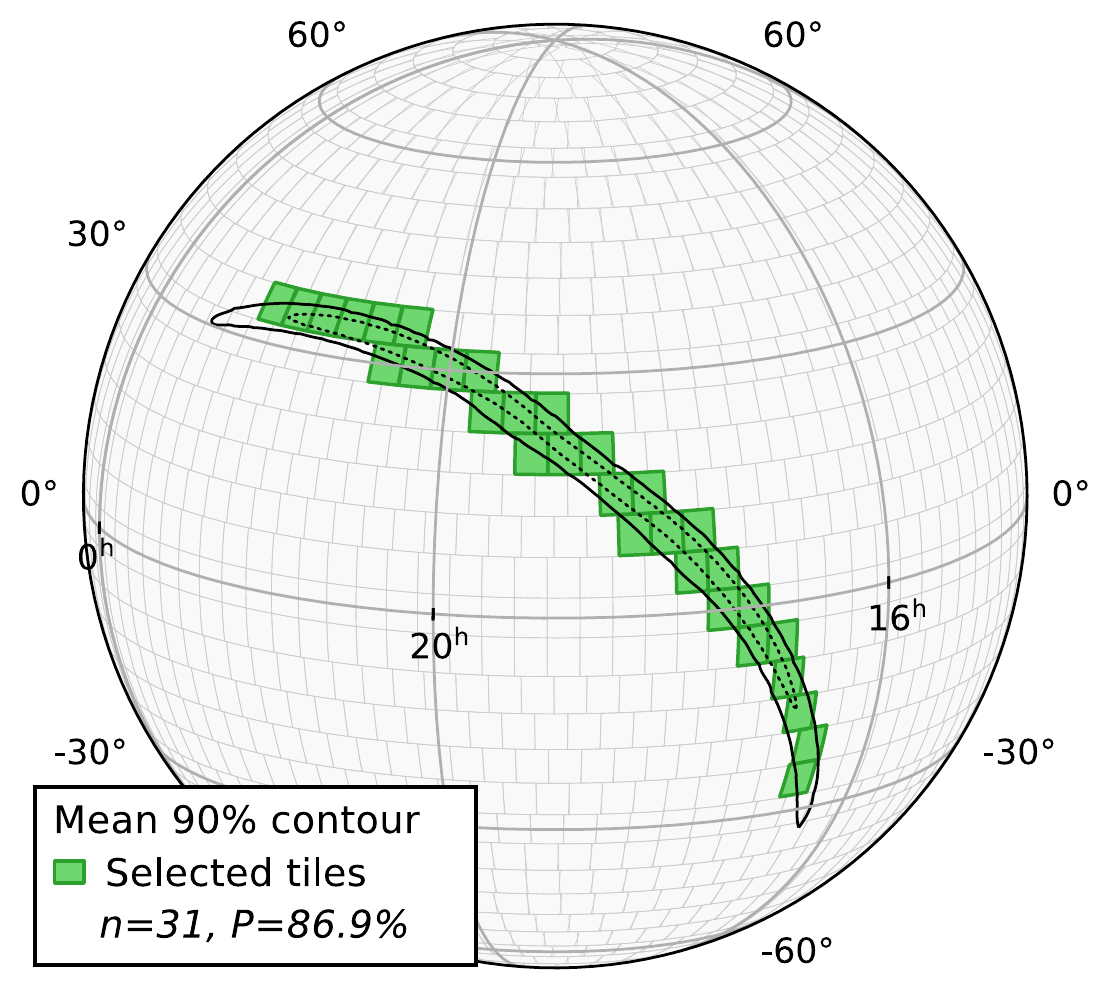}
    \end{center}
    \caption[Selecting tiles for a gravitational-wave skymap]{
        Selecting tiles for the S190521r skymap \citep[also shown in \aref{fig:skymap_regrade}]{S190521r}.
        In the upper left the skymap is plotted on the celestial sphere, forming the ``banana'' shape typical of gravitational-wave localisations, and the 50\% and 90\% probability contours are shown.
        The other three plots show grid tiles selected using one of three methods: selecting tiles to cover the entire 90\% contour (\textcolorbf{NavyBlue}{blue}), selecting tiles with a median contour value of 90\% (\textcolorbf{Orange}{orange}) and selecting tiles with a mean contour value of 90\% (\textcolorbf{Green}{green}). The number of tiles selected ($n$) and total probability within them ($P$) is given.
        The mean contour method provides a good compromise, selecting fewer than half of the tiles needed to cover the whole 90\% contour (31 compared to 63), but together they still contain nearly 87\% of the total probability.
    }\label{fig:selecting_tiles}
\end{figure}

\newpage

\begin{figure}[t]
    \begin{center}
        \includegraphics[width=\linewidth]{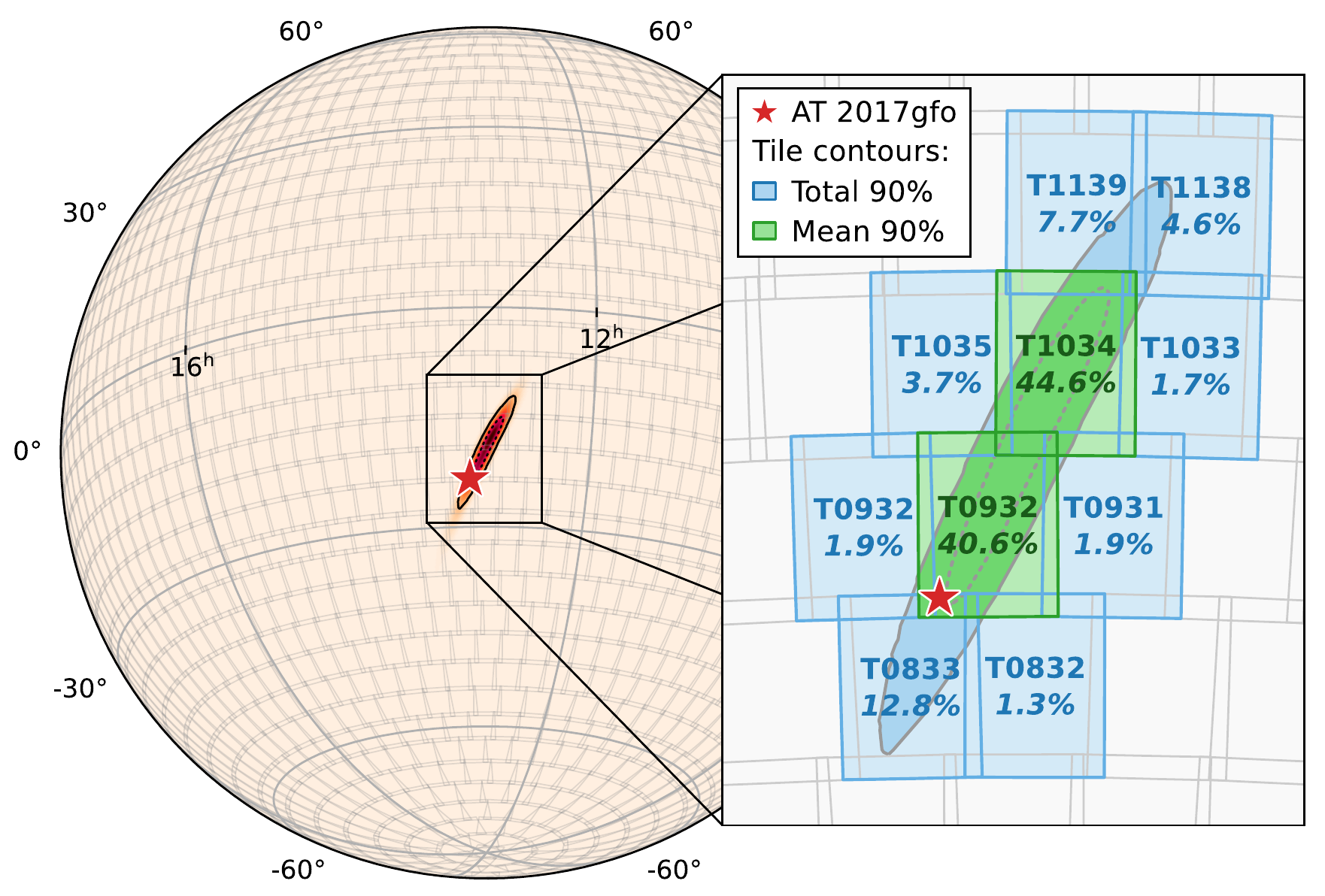}
    \end{center}
    \caption[GOTO tile probabilities for GW170817]{
        GOTO tiling applied to the final GW170817 skymap \citep{GW170817}.
        The grid shown is for the 4-UT GOTO field of view, with tiles of \SI{3.7}{\degree}$\times$\SI{4.9}{\degree} and an overlap of $0.1$. The inset shows the 10 tiles within the total 90\% contour (containing 97.4\% of the total probability) in \textcolorbf{NavyBlue}{blue} and the 2 selected by the 90\% mean contour (containing 79.5\%) in \textcolorbf{Green}{green}. Compare to \aref{fig:swope_decam}, which shows follow-up observations of GW170817 by the Swope and DECam projects.
    }\label{fig:170817_gw}
\end{figure}

\aref{fig:170817_gw} shows the GOTO-tile selection code applied to the skymap for GW170817 \citep{GW170817}. As described in \aref{sec:followup} this is the only gravitational-wave detection so far with an identified counterpart, AT~2017gfo \citep{GW170817_followup}. In this case the mean contour selection method is perhaps too restrictive, only adding two tiles to the database, and for similarly well-localised events adding more tiles would be better. Still, based on the performance during O3 (see \aref{sec:gw_results}), had GOTO been able to observe this event then the counterpart could have been observed in tile T0932 within minutes of the alert notice being received.

\end{colsection}

\section{Creating and modifying skymaps}
\label{sec:custom_skymaps}

\begin{colsection}

The GOTO-tile skymap processing system described in \aref{sec:skymaps} provides a framework which allows any gravitational-wave skymap to be mapped onto the GOTO survey grid, from which pointings can be generated for the pilot to observe. However, there is no particular reason that the system has to be restricted to just the gravitational-wave skymaps produced by the LVC.\@ This section describes three further projects based on the GOTO-tile skymap code, from when I was working with Yik Lun Mong at Monash.

\end{colsection}

\subsection{Creating Gaussian skymaps for GRB events}
\label{sec:grb_skymaps}
\begin{colsection}

As part of the GOTO commissioning observations when the LIGO-Virgo detectors were not operating (see \aref{sec:timeline}), GOTO followed up \acro{grb} events from the \textit{Fermi} satellite Gamma-ray Burst Monitor \citep[GBM;][]{Fermi_GBM}. At the time, the alert notices for GBM events did not include probability skymaps, only right ascension, declination and an error radius, and so code was developed in order to create a skymap from these details based on a 2D Gaussian profile; therefore allowing them to be processed by GOTO-tile using the same methods already created for gravitational-wave events.

Taking the radius $r$ as half the full-width at half-maximum, the standard deviation of a 2D Gaussian distribution $\sigma$ is given by
\begin{equation}
    \sigma = \frac{r}{\sqrt{2 \ln 2}}.
    \label{eq:gaussian_sigma}
\end{equation}
The distance $d$ between a given point on the sphere ($\alpha, \delta$) and the central coordinates of the distribution ($\alpha_c, \delta_c$) is given by
\begin{equation}
    \sin^2 \left ( \frac{1}{2} d \right )
    = \sin^2 \left ( \frac{\delta-\delta_c}{2} \right)
      + \cos \delta \cos \delta_c \sin^2 \left ( \frac{\alpha-\alpha_c}{2} \right),
    \label{eq:gaussian_distance}
\end{equation}

\noindent and the probability at each point for a 2D Gaussian is given by
\begin{equation}
    P(\alpha, \delta) = \frac{1}{2\pi\sigma} \exp \left ( \frac{d^2}{2\sigma^2} \right ).
    \label{eq:gaussian_prob}
\end{equation}
This probability is calculated for the location of every HEALPix pixel on a sphere, which produces a skymap array that can then be processed using GOTO-tile.

Using the above method, skymaps can be created for any single-target alert that has a given error radius. Several sources of transient events, such as \textit{Gaia} and \textit{Swift}, produce well-localised events with error circles much smaller than the GOTO tiles, so creating skymaps is less important. \textit{Fermi} GRB skymaps however cover much larger areas. For example, the GBM detection of GRB~170817A that helped localise the GW170817 gravitational-wave detection produced an initial alert with an error radius of \SI{17.45}{\degree}, later reduced to \SI{11.58}{\degree} in the final alert\footnote{GCN Notices available at \url{https://gcn.gsfc.nasa.gov/other/524666471.fermi}.}, which corresponded to a 50\% confidence region of $\sim$500~square~degrees \citep{GW170817_Fermi}.

The error values given in GBM notices only account for statistical errors for that event, not systematic errors. The GBM systematic errors are described in \citet{Fermi_localisation} to be well modelled by a core Gaussian with a radius (FWHM) of \SI{3.71}{\degree} and a non-Gaussian tail extending to \SI{14}{\degree}. For the purposes of GOTO tiling only the Gaussian portion is considered, with a radius obtained by combining the statistical radius ($r_\text{notice}$) and the systematic error in quadrature as
\begin{equation}
    r = \sqrt{r_\text{notice}^2 + {(\SI{3.71}{\degree})}^2}.
    \label{eq:fermi_radius}
\end{equation}
This is then used with the previous method to create a Gaussian skymap, which can be processed by GOTO-tile as described in \aref{sec:mapping_skymaps}. The skymap generated using this method for GRB~170817A is shown in \aref{fig:170817_grb}. Note that the location of AT~2017gfo falls quite far from the reported peak of the GRB skymap, and had there not been the coincident gravitational-wave detection it would have been unlikely that the source of the gamma-ray burst would have been observed.

\begin{figure}[t]
    \begin{center}
        \includegraphics[width=\linewidth]{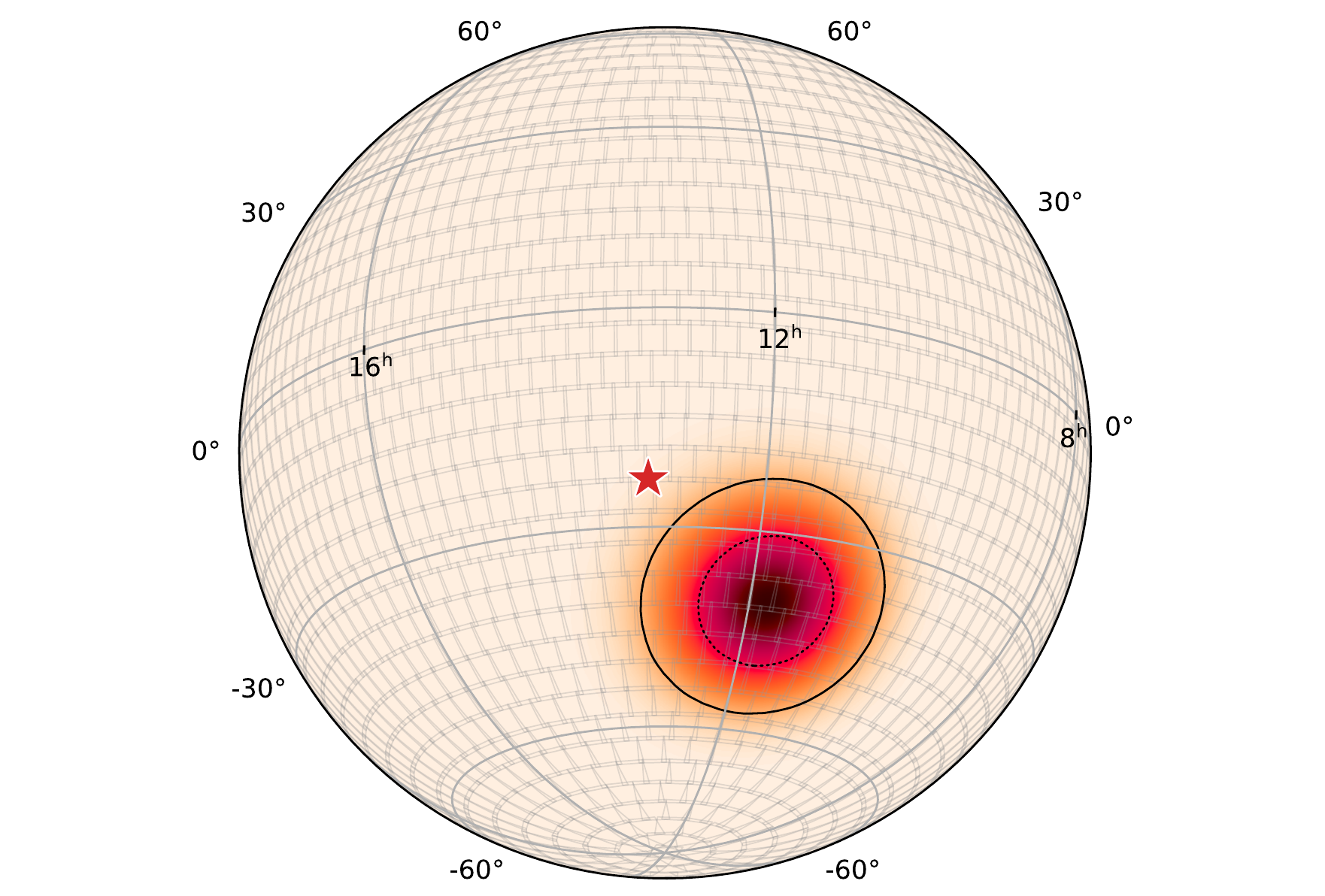}
    \end{center}
    \caption[Gaussian skymap for GRB 170817A]{
        A Gaussian skymap generated from the initial GBM alert for GRB 170817A, shown on the GOTO tile grid. The \textcolorbf{Red}{red} star shows the location of the counterpart AT~2017gfo. The final GRB skymap for this event is shown in \aref{fig:170817_skymaps}.
    }\label{fig:170817_grb}
\end{figure}

It should be noted that the GBM localisation areas are actually not perfectly symmetric, but the above procedure works as a reasonable approximation. Since May 2019 \textit{Fermi} has started including HEALPix skymaps with alert notices \citep{Fermi_skymaps}, however, unlike the LVC, the GBM team do not guarantee that the skymap will have been generated by the time the alert notice is issued. Therefore the above procedure is still used when processing alerts if the official skymap is not yet available.

\newpage

\end{colsection}

\subsection{Weighting GW skymaps using galaxy positions}
\label{sec:galaxy_skymaps}
\begin{colsection}

As described in \aref{sec:followup}, telescopes with small fields of view can focus on observing possible host galaxies instead of covering an entire GW probability region \citep{GW_weighting}. The most recent catalogue of potential host galaxies is the \acro{glade} catalogue \citep{GLADE}, which combines multiple prior catalogues including the Gravitational Wave Galaxy Catalogue \citep[GWGC,][]{GWGC} used by Swope to successfully find the GW170817 counterpart.

GOTO does not use a galaxy-focused strategy; due to its large field of view each GOTO pointing will contain tens of possible host galaxies. However, for skymaps that cover large numbers of tiles, the order in which GOTO observes could potentially be optimised by focusing on the tiles that contain the most potential host galaxies. One way of doing this using the existing G-TeCS scheduling framework (described in \aref{chap:scheduling}) is to adjust the weighting factor assigned to each tile, which can be done by multiplying the LVC localisation skymap with another skymap, containing the position of possible host galaxies, before applying the result to the tile grid as described in \aref{sec:mapping_skymaps}. Constructing weighted skymaps like this is a strategy used by several smaller field-of-view instruments, such as \textit{Swift} \citep{GW_Swift}.

In order to create such a weighted skymap the GLADE catalogue can be queried for the position of each galaxy within the event distance limits (each LVC event notice contains an estimate for the distance to the source, see \aref{sec:event_strategy} for how this is used to determine the follow-up strategy for each event). This is only possible for events within a few \SI{100}{\mega\parsec}, beyond which the GLADE catalogue is increasingly incomplete \citep{GLADE}. Once found, a HEALPix skymap can be constructed by weighting each pixel by the number of possible host galaxies located within it. As galaxies are not being point sources, the resulting skymap is then passed through a Gaussian smoothing function with a default standard deviation of 15~arcseconds. This new skymap can then be normalised and multiplied with the gravitational-wave position skymap, to produce a new skymap which contains the information from both.

\aref{fig:galaxy_skymap} shows this method applied to the large skymap for event S190425z \citep{S190425z}, which included a reported luminosity distance of $155\pm\SI{45}{\mega\parsec}$. The underlying pattern of the gravitational-wave localisation regions is still clearly visible in the final skymap, but by including the galaxy information the resulting tile pointings will be weighted towards regions with larger numbers of possible host galaxies. This method still needs more work before being implemented, in particular the relative weighting to apply to each skymap needs to be considered. However, it could prove beneficial by further prioritising GOTO observations towards regions more likely to include a counterpart source.

\begin{figure}[p]
    \begin{center}
        \includegraphics[width=0.8\linewidth]{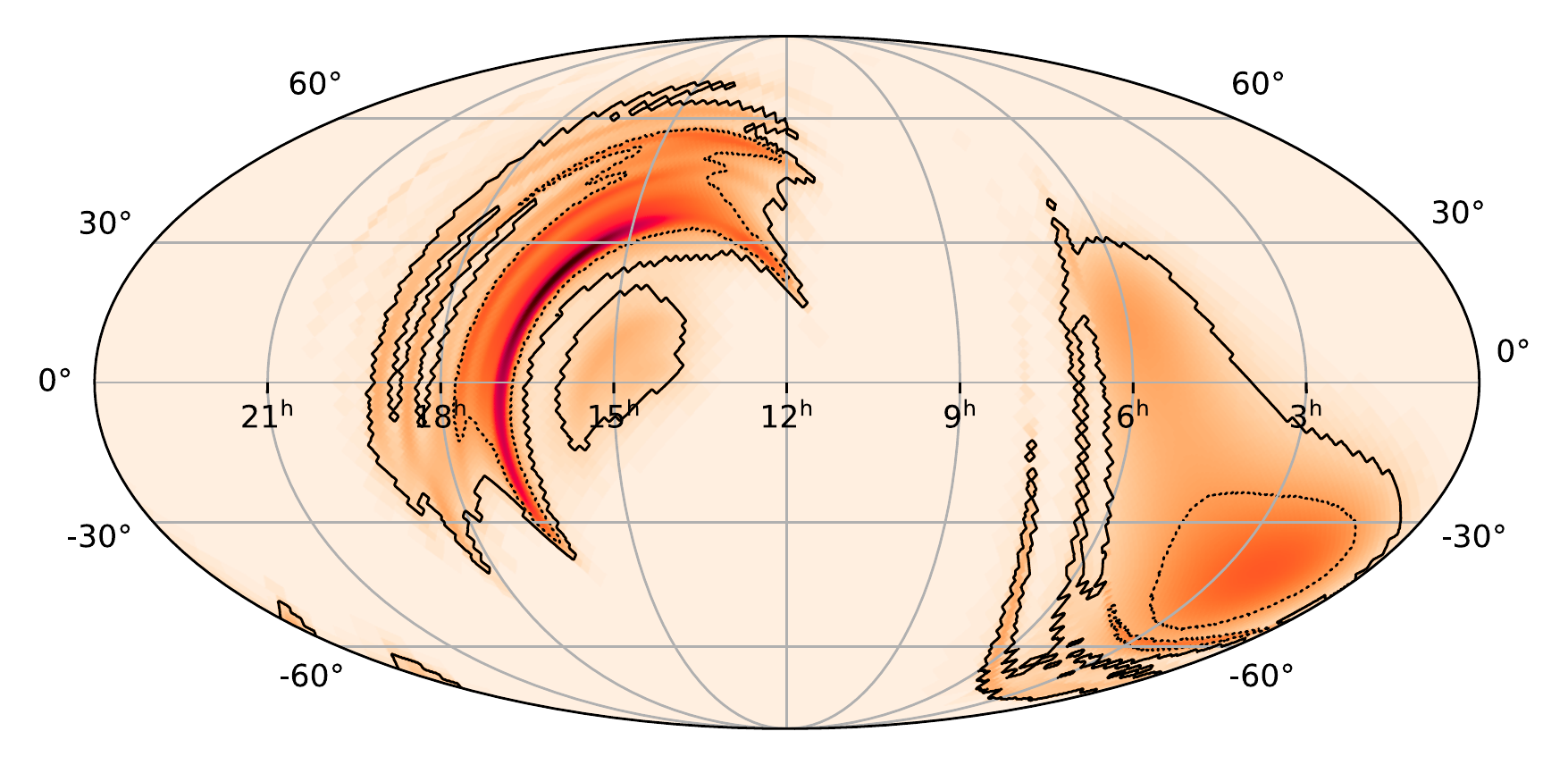}
        \includegraphics[width=0.8\linewidth]{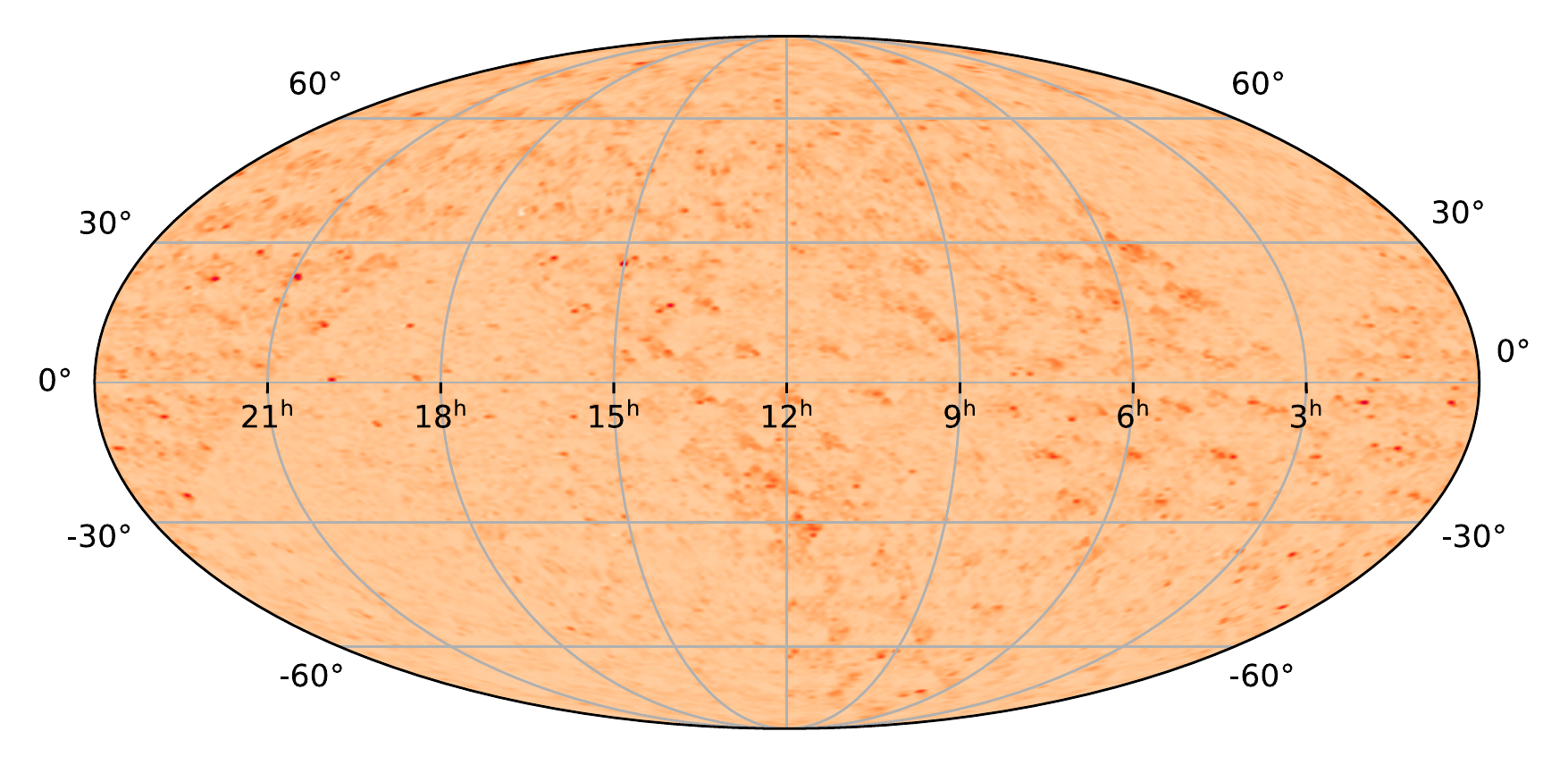}
        \includegraphics[width=0.8\linewidth]{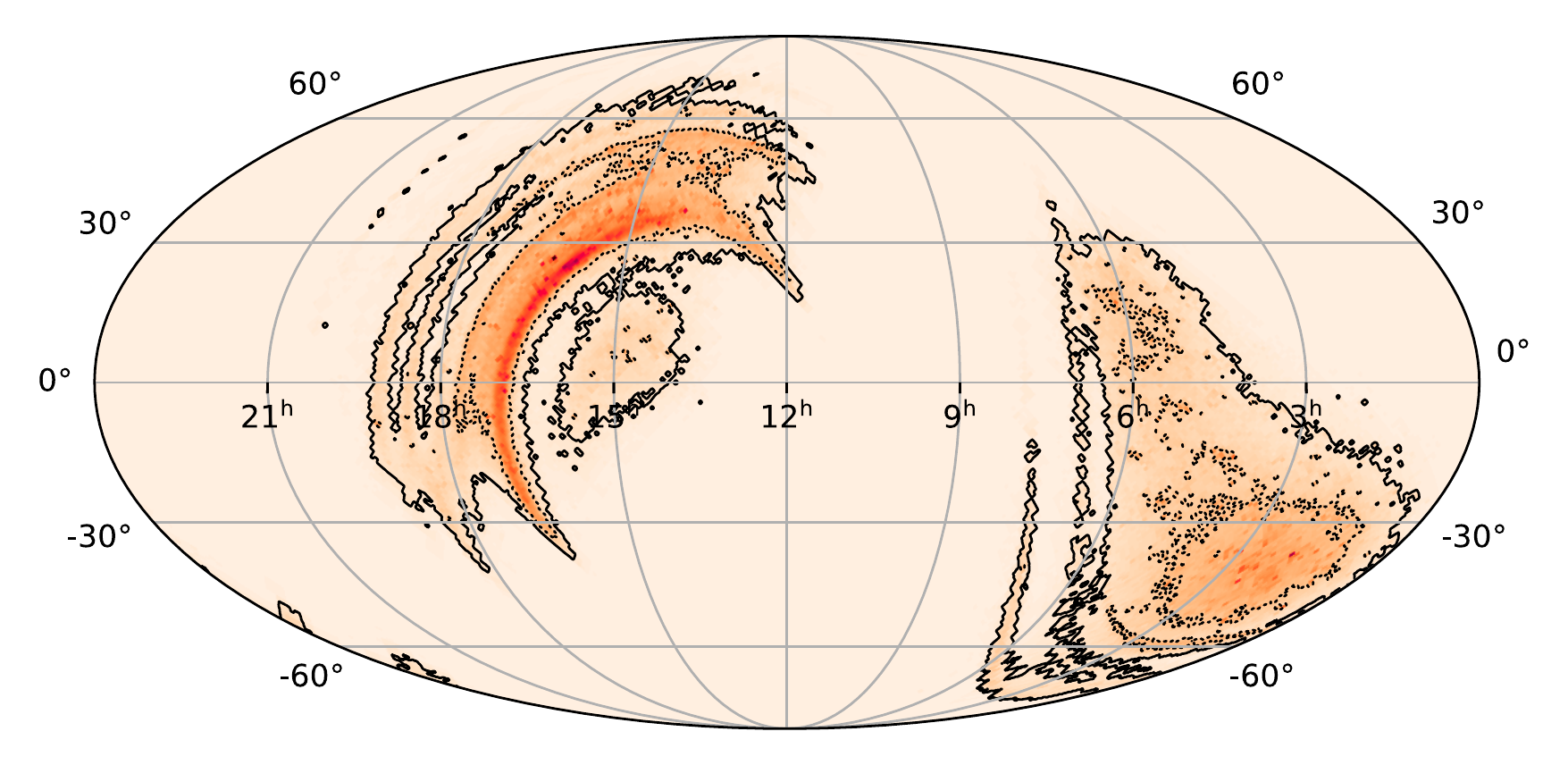}
    \end{center}
    \caption[Weighting a GW skymap using galaxy positions]{
        Weighting a GW skymap (for event S190425z) using galaxy positions. The top plot shows the skymap distributed by the LVC, the central plot shows a skymap using GLADE galaxies within the estimated distance to the source ($155\pm\SI{45}{\mega\parsec}$), and the lower plot shows the result of multiplying the two together.
    }\label{fig:galaxy_skymap}
\end{figure}

\end{colsection}

\subsection{Prioritising observations using dust extinction skymaps}
\label{sec:extinction_skymaps}
\begin{colsection}

In very crowded fields a single GOTO image can contain tens of thousands of sources, which makes it difficult for the GOTOphoto photometry pipeline (see \aref{sec:gotophoto}) to identify potential counterpart candidates. Observing high-interstellar-extinction areas, i.e.\ through the galactic plane, also makes it harder to observe extra-galactic sources such as counterparts to gravitational-wave events. Therefore another possible reason to modify localisation skymaps is to de-prioritise observations of the galactic plane.

One method to do this is shown in \aref{fig:extinction_skymap} --- multiplying the gravitational-wave skymap by an inverted thermal dust emission skymap from the \textit{Planck} observatory \citep{Planck_dust}. The effect is intended to be subtle, not enough to completely wipe out the skymap probability in the high-extinction regions, but instead to just reduce the weighting of those tiles so that GOTO first prioritises other areas. Again, this is not currently implemented in the scheduling system, but it is presented as another example of using skymaps to optimise observation priorities.

\begin{figure}[p]
    \begin{center}
        \includegraphics[width=0.8\linewidth]{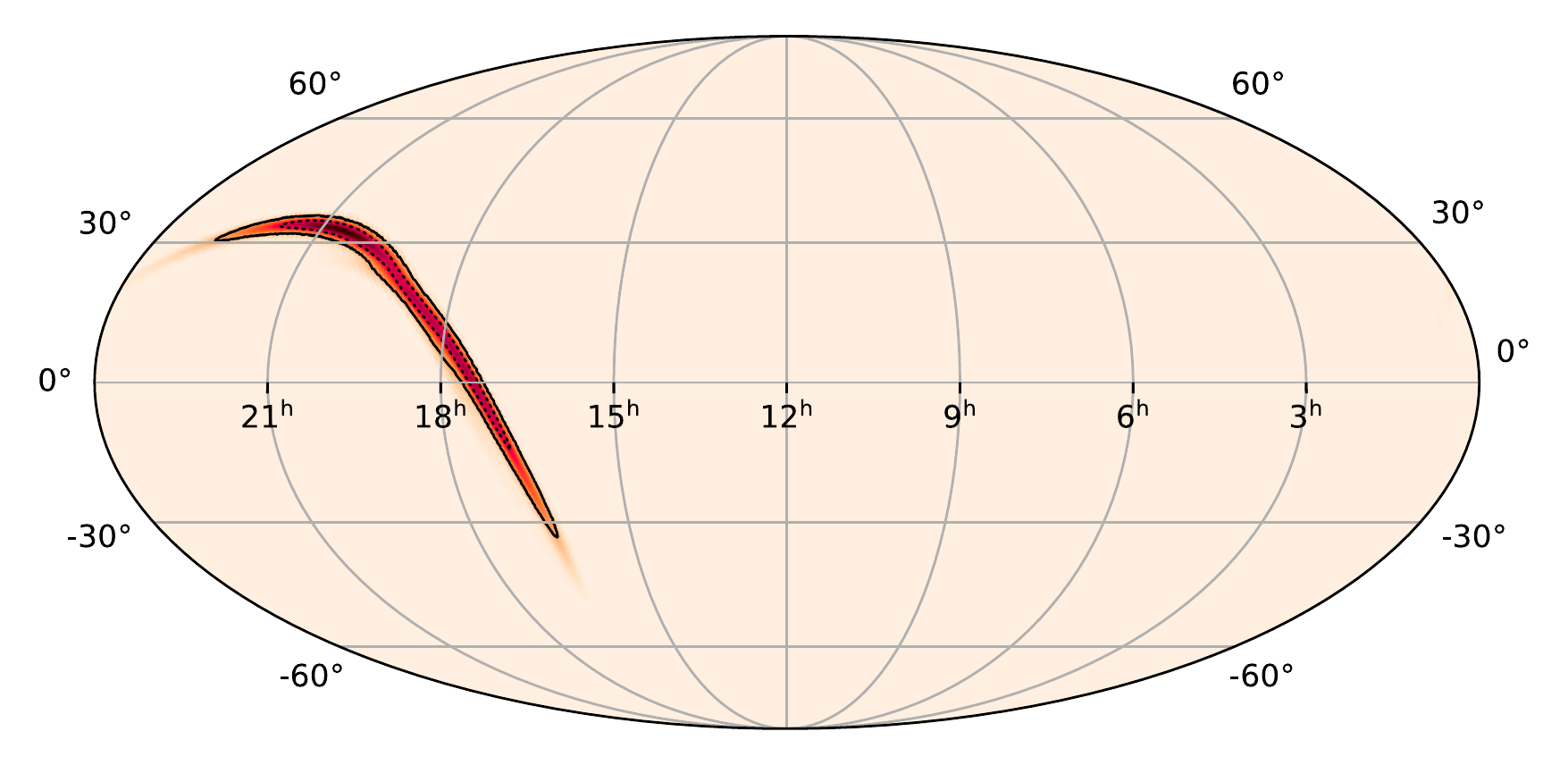}
        \includegraphics[width=0.8\linewidth]{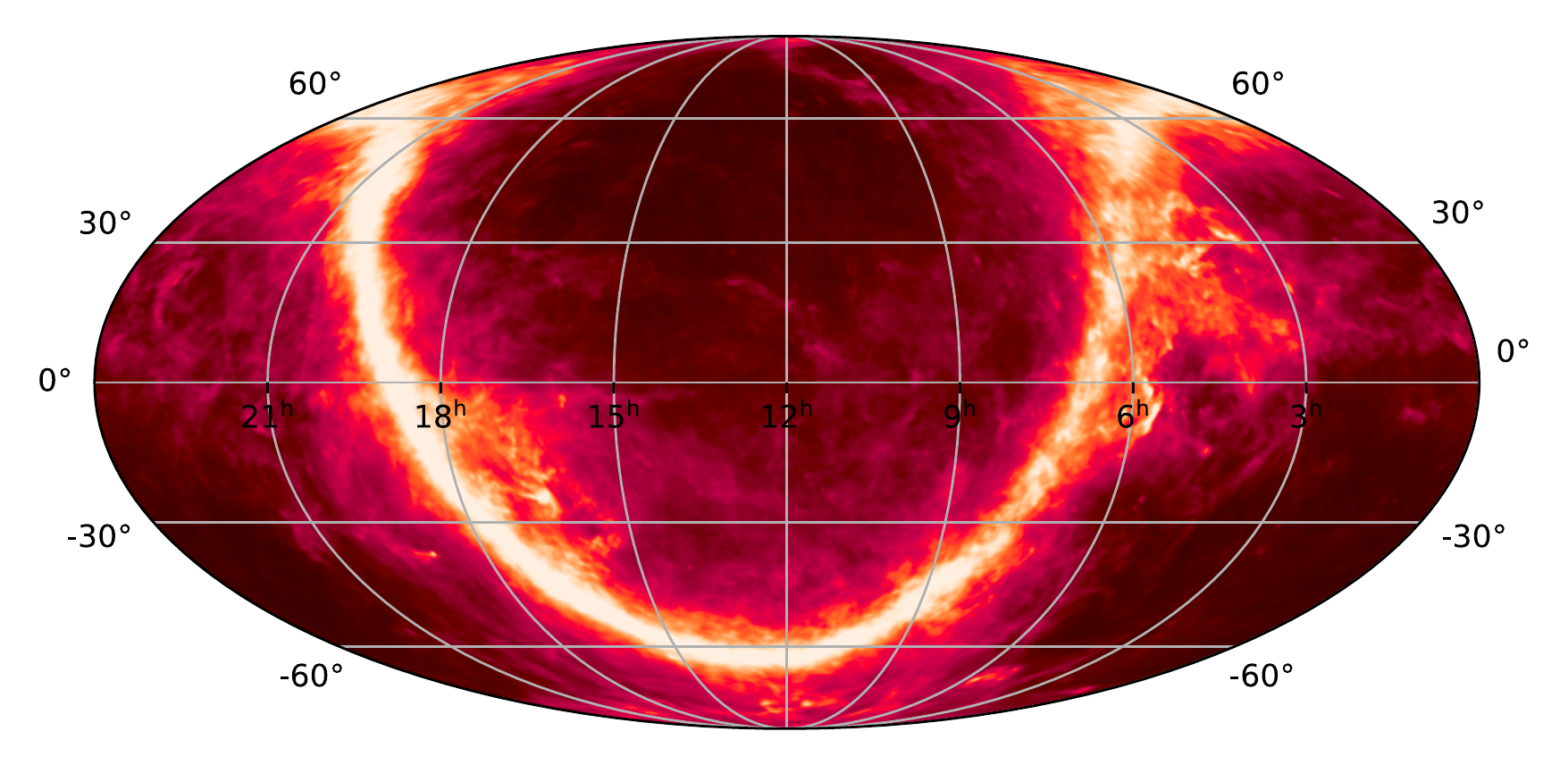}
        \includegraphics[width=0.8\linewidth]{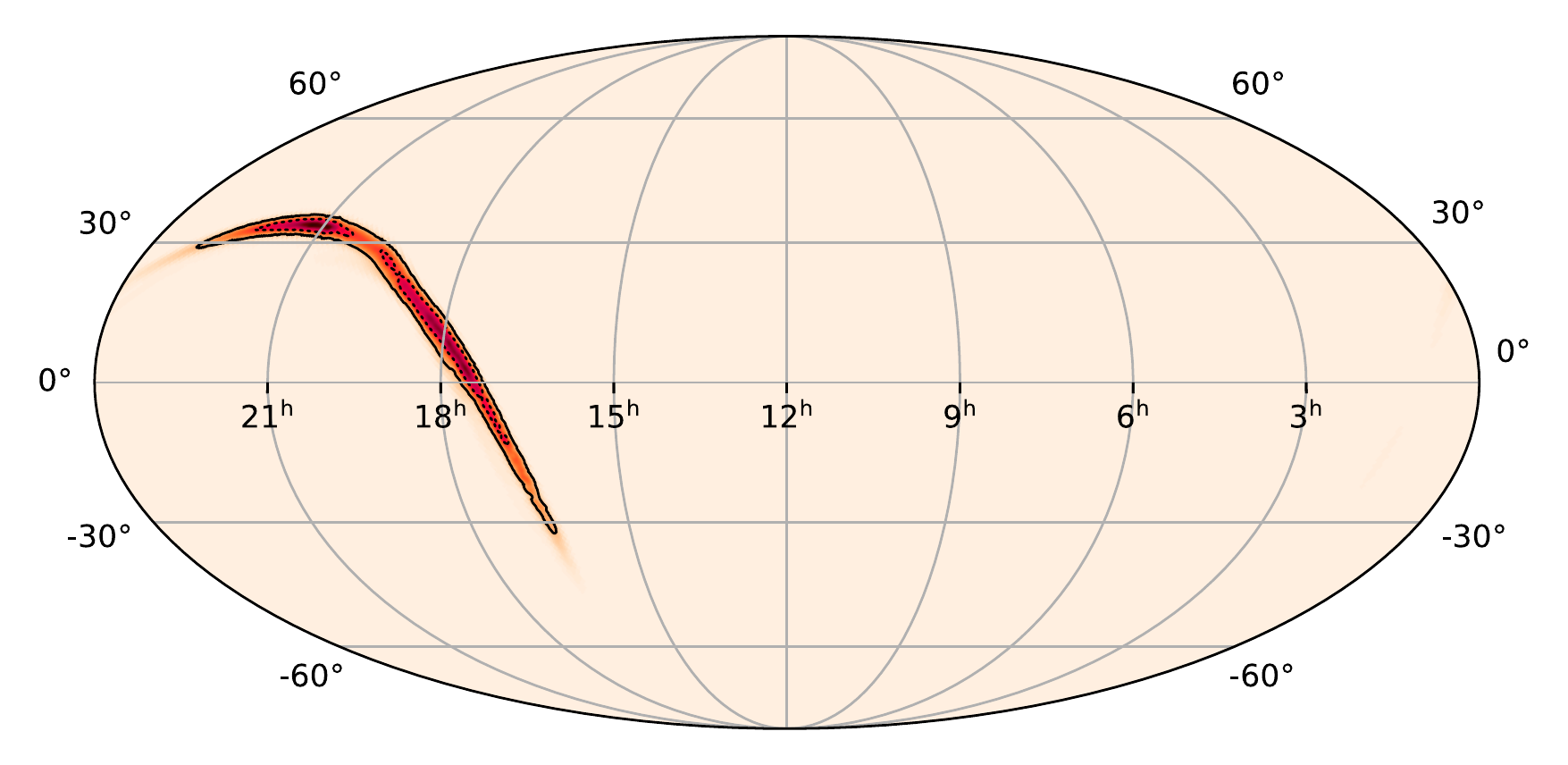}
    \end{center}
    \caption[Weighting a GW skymap using galactic extinction]{
        Weighting a GW skymap (for event S190521r) using galactic extinction. The top plot shows the skymap distributed by the LVC, the central plot shows an inverted thermal dust emission skymap from \textit{Planck}, and the lower plot shows the result of multiplying the two together.
    }\label{fig:extinction_skymap}
\end{figure}

\end{colsection}

\section{Summary and Conclusions}
\label{sec:tiling_conclusion}

\begin{colsection}

In this chapter I described how the GOTO all-sky survey grid is defined, and how it is used for targeting gravitational-wave follow-up observations.

As a survey-based project all GOTO observations are taken aligned to a fixed grid. This all-sky grid is defined using the ``minverlap'' algorithm I developed for the GOTO-tile Python module, which produces a much more even grid compared to the previous algorithms used. Gravitational-wave localisation areas are produced in the form of HEALPix skymaps, which need to be mapped onto the grid used by GOTO before observations can be carried out. This is carried out by the sentinel alert-listening daemon described in \aref{chap:autonomous}, and the resulting pointings are sorted and prioritised by the scheduler as described previously in \aref{chap:scheduling}.

The functions used by the sentinel to process transient alerts are described in the following chapter (\aref{chap:alerts}). It is at this stage that the relative weighting of each tile is determined based on the transient skymap, for both LVC gravitational-wave detections and other alerts such as gamma-ray bursts. In this chapter I also outlined some further possible additions to this weighting by including skymaps based on host galaxy catalogues or galactic extinction. By including these extra weightings it may be possible to further optimise the GOTO follow-up observations and increase the chance of taking a quick observation of a gravitational-wave counterpart.

\end{colsection}

\chapter{Processing Transient Alerts}
\label{chap:alerts}

\chaptoc{}

\section{Introduction}
\label{sec:alerts_intro}

\begin{colsection}

In this chapter I describe the software used by GOTO to process alerts generated from transient astronomical events, including gravitational-wave detections.
\begin{itemize}
    \item In \nref{sec:transient_alerts} I give an overview of the established systems through which the LVC, NASA and other organisations publish and distribute astronomical alerts.
    \item In \nref{sec:gotoalert} I describe the GOTO-alert Python package, and how transient alerts are received and processed.
    \item In \nref{sec:strategy} I describe how the optimal follow-up strategy is determined for different types of alert, and how the targets are defined for GOTO to observe.
\end{itemize}
All work described in this chapter is my own unless otherwise indicated, and has not been published elsewhere.

\end{colsection}

\section{Transient event alerts}
\label{sec:transient_alerts}

\begin{colsection}

For the last few decades the number of detections of transient astronomical sources has been rapidly increasing. From space-based gamma-ray burst monitors such as \textit{Fermi} and \textit{Swift} to wide-field survey telescopes such as the \acro{ptf} and \acro{asassn}, increasing numbers of time-critical events have been detected and rapidly sent to other observing partners for follow-up campaigns. Historically this was done, at a much slower pace, through physical post and telegrams --- hence the names of some of the current services offering modern, email-based alternatives: the Astronomer's Telegram \citep{ATel} and the \acro{gcn} Circulars and Notices \citep{GCN}.

Today the global system has evolved to remove the human factor entirely, in order to reduce the delay between events being detected and follow-up observations being taken. Robotic telescopes are now common and can be triggered automatically by machine processing of alerts, which are themselves generated automatically by the detection instruments. The \acro{ivoa} VOEvent protocol \citep{voevent} has become the standard language for such robotic communications, allowing telescopes around the world to respond within seconds to transient detections. New projects such as the \acro{ztf}, itself paving the way for the forthcoming \acro{lsst}, will produce millions of events per night requiring even faster and more efficient alert systems \citep{ZTF_alerts}.

GOTO's priority is, of course, detecting optical counterparts to gravitational-wave events. Such events are published by the LIGO-Virgo Collaboration as VOEvents through the GCN Notice system.

\newpage

\end{colsection}

\subsection{GCN alerts}
\label{sec:gcn}
\begin{colsection}

The Gamma-ray burst Coordinates Network, also known as the Transient Astronomy Network or together GCN/TAN, is a system hosted by NASA originally to publish alerts relating to gamma-ray burst detections \citep{GCN}. It publishes events from a variety of telescopes including \textit{INTEGRAL}, \textit{Fermi} and \textit{Swift}, and more recently has expanded to neutrino and gravitational-wave alerts --- including publishing alerts from the LIGO-Virgo Collaboration.

Alerts are produced by the various facilities in the form of GCN Notices, standard machine-readable text messages that are distributed by the network. Notices are designed to be written and sent out automatically by the facility without the need for human intervention, and likewise can be received and acted on by automated systems run by follow-up projects such as GOTO's sentinel (see \aref{sec:sentinel}). Transmitting notices is only done by the projects that are part of the network.

There is a second form of alerts distributed by the network called GCN Circulars. Unlike notices, circulars are intended to be written by humans and sent out by email to anyone subscribed to the distribution list. They act as a formal, citable way to share information about events, both the initial detection by the facility any any follow-up activity by other groups.

\end{colsection}

\subsection{VOEvents}
\label{sec:voevents}
\begin{colsection}

The GCN/TAN system broadcasts notices in multiple ways over many different channels, but the most useful for automated telescopes such as GOTO uses the IVOA VOEvent standard \citep{voevent}. VOEvents are a standard way to transmit information about transient astronomical events, in a structured format to make the reports easily machine-readable. Each event is assigned an \acro{ivorn} and follows a defined schema. By defining a standard template to follow, diverse events can be automatically processed and robotic telescopes triggered without the need for human interpretation or vetting.

The structure to transmit these events is fairly flexible, but there are certain common roles. The names below are taken from \citet{voevent}:

\begin{itemize}
    \item \textbf{Authors} are the projects, facilities or institutions that create the original data worthy of reporting in the VOEvent.
    \item \textbf{Publishers} take the information about the astronomical event, put it into the VOEvent format and broadcast it from their servers.
    \item \textbf{Brokers} act as nodes in the communication web that can take in events from multiple publishers and rebroadcast them in a single stream.
    \item \textbf{Subscribers} are the end users that listen to VOEvent servers, either directly to the publishers or to a broker.
\end{itemize}

In some cases the above roles can be combined, but for the case of GOTO receiving gravitational-wave events there are distinct actors: the LIGO-Virgo Collaboration is the event's author, NASA and the GCN/TAN system are the publishers and the GOTO sentinel is the subscriber.

It is possible to listen directly to the GCN/TAN servers, in which case the sentinel would receive VOEvents from the NASA missions and other projects like LIGO.\@ However, there are other groups publishing their own VOEvents, separate to the GCN system, which we might want to receive. It would be possible to run multiple event listeners within the sentinel, each listening to a different server, but it is much easier to listen to a broker that already does that and provides a single point of access to these pipelines.

The broker listened to by the G-TeCS sentinel is the 4 Pi Sky VOEvents Hub \citep{4pisky}. 4 Pi Sky combines alerts from the GCN system\footnote{\url{https://gcn.gsfc.nasa.gov/burst_info.html}} as well as from the \textit{Gaia}\footnote{\url{http://gsaweb.ast.cam.ac.uk/alerts/alertsindex}} and ASAS-SN\footnote{\href{http://www.astronomy.ohio-state.edu/~assassin/transients}{\texttt{http://www.astronomy.ohio-state.edu/\raisebox{0.5ex}{\texttildelow}assassin/transients}}} projects. At the time of writing GOTO only follows up LVC gravitational-wave events and \textit{Fermi} and \textit{Swift} gamma-ray burst detections, all of which are published through GCNs, meaning there is technically no benefit of listening to 4 Pi Sky over listening directly to NASA.\@ However pt5m uses the 4 Pi Sky broker to receive and automatically follow up \textit{Gaia} transient detections, and it has been suggested GOTO could do the same in the future.

In order to receive VOEvents from any source it is necessary to set up a VOEvent client. The most common way to do this is using the Comet software \citep{comet}, which allows both sending and receiving of events. For the G-TeCS sentinel (see \aref{sec:sentinel}) all that was required was a simple way to listen to and download alerts, which is why it instead uses code based on the PyGCN Python package (\pkg{pygcn}\footnote{\url{https://github.com/lpsinger/pygcn}}). Despite the name, PyGCN can receive any VOEvents, not just those from the GCN servers. The sentinel uses PyGCN to open a socket to the 4 Pi Sky server and ingest binary packets, as well as sending the required receipt and ``\code{iamalive}'' responses to the server to ensure it keeps receiving events.

VOEvents take the form of a structured \acro{xml} document. XML is a ``markup'' language similar to HTML, JSON or \LaTeX, meaning it is understandable by humans but follows a set schema and so can be easily read and processed by computers. A sample of a VOEvent is given in \aref{fig:voevent_xml}.

\begin{figure}[p]
\begin{lstlisting}[language=xml,
       tabsize=2,
       breaklines=true,
       keywordstyle={},
       stringstyle=\color{red},
       showstringspaces=false,
       basicstyle=\ttfamily\scriptsize,
       emph={voe,Who,What,WhereWhen,How,Citations},
       emphstyle={\color{magenta}},
       columns=fullflexible
       ]
<?xml version="1.0" encoding="UTF-8"?>
<voe:VOEvent
 xmlns:xsi="http://www.w3.org/2001/XMLSchema-instance"
 xmlns:voe="http://www.ivoa.net/xml/VOEvent/v2.0"
 xsi:schemaLocation="http://www.ivoa.net/xml/VOEvent/VOEvent-v2.0.xsd"
     role="observation" 
     ivorn="ivo://gwnet/LVC#S190408an-2-Initial">
<Who>
 <Date>2019-04-08T20:21:42</Date>
 <Author>
  <contactName>LIGO Scientific Collaboration and Virgo Collaboration</contactName>
 </Author>
</Who>
<What>
 <Param name="Packet_Type" dataType="int" value="151"></Param>
 <Param name="internal" dataType="int" value="0"></Param>
 <Param name="Pkt_Ser_Num" dataType="string" value="2"></Param>
 <Param name="GraceID" dataType="string" value="S190408an" ucd="meta.id"></Param>
 <Param name="AlertType" dataType="string" value="Initial" ucd="meta.version"></Param>
 <Param name="HardwareInj" dataType="int" value="0" ucd="meta.number"></Param>
 <Param name="OpenAlert" dataType="int" value="1" ucd="meta.number"></Param>
 <Param name="EventPage" dataType="string" value="https://gracedb.ligo.org/superevents/S190408an/view/" ucd="meta.ref.url"></Param>
 <Param name="Instruments" dataType="string" value="H1,L1,V1" ucd="meta.code"> </Param>
 <Param name="FAR" dataType="float" value="2.81096164616e-18" ucd="arith.rate;stat.falsealarm" unit="Hz"></Param>
 <Param name="Group" dataType="string" value="CBC" ucd="meta.code"></Param>
 <Param name="Pipeline" dataType="string" value="gstlal" ucd="meta.code"></Param>
 <Param name="Search" dataType="string" value="AllSky" ucd="meta.code"></Param>
 <Group type="GW_SKYMAP" name="bayestar">
  <Param name="skymap_fits" dataType="string" value="https://gracedb.ligo.org/api/superevents/S190408an/files/bayestar.fits.gz" ucd="meta.ref.url"></Param>
 </Group>
 <Group type="Classification">
  <Param name="BNS" dataType="float" value="0.0" ucd="stat.probability"></Param>
  <Param name="NSBH" dataType="float" value="0.0" ucd="stat.probability"></Param>
  <Param name="BBH" dataType="float" value="0.99999999999" ucd="stat.probability"></Param>
  <Param name="MassGap" dataType="float" value="0.0" ucd="stat.probability"></Param>
  <Param name="Terrestrial" dataType="float" value="9.82357724531e-12" ucd="stat.probability"></Param>
 </Group>
 <Group type="Properties">
  <Param name="HasNS" dataType="float" value="0.0" ucd="stat.probability"></Param>
  <Param name="HasRemnant" dataType="float" value="0.12" ucd="stat.probability"></Param>
 </Group>
</What>
<WhereWhen>
 <ObsDataLocation>
  <ObservatoryLocation id="LIGO Virgo"/>
  <ObservationLocation>
   <AstroCoordSystem id="UTC-FK5-GEO"/>
   <AstroCoords coord_system_id="UTC-FK5-GEO">
   <Time><TimeInstant><ISOTime>2019-04-08T18:18:02.288180</ISOTime></TimeInstant></Time>
   </AstroCoords>
  </ObservationLocation>
 </ObsDataLocation>
</WhereWhen>
...
\end{lstlisting}
    \caption[VOEvent XML sample]{
        A sample of VOEvent text from an LVC event, formatted so the core XML structure is visible. Some of the key pieces of information are the role and IVORN defined in the header, the skymap URL, and the event classification probabilities.
    }\label{fig:voevent_xml}
\end{figure}

\newpage

\end{colsection}

\section{Processing alerts}
\label{sec:gotoalert}

\begin{colsection}

Once a VOEvent is received by the G-TeCS sentinel, the task of parsing and processing the event uses another Python package, GOTO-alert (\pkg{gotoalert}\footnote{\url{https://github.com/GOTO-OBS/goto-alert}}), which contains functions related to processing transient alerts. GOTO-alert was originally written by Alex Obradovic at Monash to listen for GRB alerts, when I took the code over I rewrote it to integrate it into G-TeCS, as well as adding the capability to process gravitational-wave alerts.

\end{colsection}

\subsection{Event classes}
\label{sec:event_classes}
\begin{colsection}

\begin{table}[t]
    \begin{center}
        \begin{tabular}{clll}
            Packet type & Source              & Notice type                  & Event subclass           \\
            \midrule
            \code{61}   & NASA/\textit{Swift} & \code{SWIFT\_BAT\_GRB\_POS}  & \code{GRBEvent}          \\
            \code{115}  & NASA/\textit{Fermi} & \code{FERMI\_GBM\_FIN\_POS}  & \code{GRBEvent}          \\
            \code{150}  & LVC                 & \code{LVC\_PRELIMINARY}      & \code{GWEvent}           \\
            \code{151}  & LVC                 & \code{LVC\_INITIAL}          & \code{GWEvent}           \\
            \code{152}  & LVC                 & \code{LVC\_UPDATE}           & \code{GWEvent}           \\
            \code{164}  & LVC                 & \code{LVC\_RETRACTION}       & \code{GWRetractionEvent} \\
        \end{tabular}
    \end{center}
    \caption[GCN notices recognised by the GOTO-alert event handler]{
        GCN notices and corresponding classes recognised by the GOTO-alert event handler. The packet type is used by the GCN system to identify the class of event.
    }\label{tab:events}
\end{table}

At the core of the GOTO-alert code is the \code{Event} object class. Events are Python classes created from the raw VOEvent XML payload received by the PyGCN listener, containing the basic event information (IVORN, type, source etc). Once the basic Event is created it is checked against an internal list of so-called ``interesting'' event packet types --- the ones we care about processing for GOTO.\@ At the time of writing these are \code{SWIFT\_BAT}, \code{FERMI\_GBM} and \code{LVC} events, as listed in \aref{tab:events}. If the event matches any of the recognised packet types then the Event is subclassed into a new object, which allows more specific properties and methods. The current subclasses are as follows:

\subsubsection{GRB Events}

The \code{GRBEvent} class is used for events relating to gamma-ray burst detections, specifically from \textit{Fermi} and \textit{Swift}. The VOEvents for these events contain a sky position in right ascension and declination as well as an error radius, so a HEALPix skymap is produced using the Gaussian method described in \aref{sec:grb_skymaps}. For \textit{Fermi} events the class also has an additional attribute extracted from the VOEvent: the duration of the burst (Long or Short).

\subsubsection{GW Events}

The \code{GWEvent} class is used for LVC gravitational-wave events. LVC events have several stages: a ``Preliminary'' alert is released as soon as the signal is detected, then an ``Initial'' alert is issued once it has been human-vetted. From then on future versions are marked as ``Update'' alerts, unless the event itself is found to be non-physical or below certain thresholds in which case a ``Retraction'' alert is issued. As these are events produced by LIGO-Virgo they should contain a \code{skymap\_fits} parameter (as shown in \aref{fig:voevent_xml}), which gives a URL pointing to where the skymap can be downloaded from GraceDB, the LIGO event database\footnote{\url{https://gracedb.ligo.org}}. The gravitational-wave VOEvents also contain a variety of properties that are stored in the event class and which can be used to determine the observing strategy (see \aref{sec:event_strategy}) to use. These include:
\begin{itemize}
    \item \textbf{\acro{far}}: an estimate of the probability that this event is a false alarm, i.e.\ not from a real astronomical source. Given in the form of an expected frequency or rate, so an event with a false alarm rate of 1~per~year is much less significant than one with a FAR of 1~per~10,000 years.
    \item \textbf{Instruments}: which of the active gravitational-wave detectors (currently LIGO-Livingston, LIGO-Hanford and Virgo) detected the signal. A non-detection in one or more instruments is also accounted for in the false alarm rate.
    \item \textbf{Group}: which type of GW pipeline detected the event signal, either ``CBC'' (Compact Binary Coalescence)\acroadd{cbc} or ``Burst'' \citep[other, unmodelled detections, see][]{GW_burst}. The following parameters only apply to CBC events.
    \item \textbf{Classification}: the VOEvents for CBC events include probabilities that the source falls into one of five categories: \acro{bns} mergers, \acro{nsbh} mergers, \acro{bbh} mergers, ``MassGap'' mergers \citep[one or other of the components is in the hypothetical ``mass gap'' between neutron stars and black holes, defined as 3--\SI{5}{\solarmass};][]{GW_MassGap}, or ``Terrestrial'' (a non-astronomical source).
    \item \textbf{Properties}: CBC events also contain two important properties: ``HasNS'', the probability that the mass of one or both of the components is consistent with a neutron star ($<$\SI{3}{\solarmass}); and ``HasRemnant'', the probability that a non-zero amount of material was ejected during coalescence and therefore an electromagnetic signal might be expected \citep{LVC_userguide}.
    \item \textbf{Distance}: The skymaps produced by the LVC contain three-dimensional localisation information \citep{GW_distance}. For deciding on event strategy the mean distance, in megaparsec, is read from the skymap FITS header, along with the standard deviation.
\end{itemize}

\subsubsection{GW Retraction Events}

\code{GWRetractionEvent} is a special event subclass used to handle LVC notices that are retractions of earlier events. They are effectively just a more limited version of the \code{GWEvent} class, as the retraction VOEvents do not contain a skymap or any of the additional parameters listed above. Having retraction events occupy their own subclass makes it easier to identify and process them when sent through the event handler.

\newpage

\end{colsection}

\subsection{The event handler}
\label{sec:event_handler}
\begin{colsection}

Once the correct \code{Event} class has been created the sentinel passes it to the GOTO-alert \code{event\_handler} function. Before processing the event, the handler first filters out any unwanted events which do not come under any of the above subclasses. These are primarily alerts from other facilities (\textit{INTEGRAL}, \textit{Gaia} etc) or other event classes that are not ``interesting'' to GOTO (\textit{Fermi} releases several types of alerts, but only the final GBM positions are processed). If the event passed to the event handler is not marked as ``interesting'' then it is rejected at this stage and the handler exits. The handler will also intentionally reject an event that is ``interesting'' if it has an incorrect role. LVC sends out test VOEvents to allow full testing of any follow-up systems; these are identical to real events (even including simulated skymaps) but are explicitly marked with the role of \code{test} rather than \code{observation}. These can be optionally processed by the event handler, but in the live sentinel system they are rejected at this point.

If the event passes the above filter the next step is to download the event's skymap (for GW events, see \aref{sec:skymaps}) or create a corresponding Gaussian skymap (for GRB events, see \aref{sec:grb_skymaps}). Doing this after filtering-out uninteresting events saves time and space when downloading or creating skymaps that are not used, for example for LVC test events.

Once the event has a skymap the observing strategy for the event can be generated. This can only happen after the skymap is downloaded as some parameters, notably the distance for GW events, are only stored within the skymap headers instead of in the VOEvent XML.\@ The details of the different event strategies and how they are defined are given in \aref{sec:event_strategy}.

Finally, the event handler inserts pointings and other information into the observation database, with parameters depending on the strategy determined in the previous stage. This is detailed in \aref{sec:event_insert}.

\newpage

\end{colsection}

\subsection{Event reports}
\label{sec:event_slack}
\begin{colsection}

Throughout the event handling process, GOTO-alert has the option of sending confirmation messages to Slack in a similar way to the G-TeCS pilot (see \aref{sec:slack}). This allows human observers to be informed of any new alerts processed by the sentinel, as well as the expected outcome of upcoming observations. Alerts are deliberately spaced out at different stages within the event handler, rather than all sent at the end, to make it obvious if a problem occurs and one or more alerts are missing.

Four alerts are sent out in total:
\begin{enumerate}
    \item An \textbf{initial} alert is sent out by the sentinel as soon as an interesting event is received, and contains only one line reporting the IVORN of the event to be processed. Sending this first means there is a record should an error subsequently occur.
    \item Next, the \textbf{event} alert is sent out by GOTO-alert once the event has been created and the skymap downloaded. It contains the key information and properties of the event, as well as a plot of the skymap produced by GOTO-tile.
    \item The \textbf{strategy} alert is sent out after the event observing strategy has been retrieved, and reports the contents of the strategy dictionaries (see \aref{sec:event_strategy}).
    \item Finally, the \textbf{visibility} alert is sent out after the event tiles have been added to the observation database (see \aref{sec:event_insert}). As well as showing the number of tiles selected and their combined skymap coverage, this alert also predicts which tiles will be visible from the GOTO site on La Palma over the valid observing period. This is only based on tile altitudes during the night, and excludes other factors (weather conditions, the position of the Moon, any higher-priority pointings that would take precedence) considered by the G-TeCS just-in-time scheduler (see \aref{sec:ranking}).
\end{enumerate}
An example of alerts generated for a gravitational-wave event are shown in \aref{fig:gotoalert_slack}.

\begin{sidewaysfigure}[p]
    \begin{center}
        \includegraphics[width=\linewidth]{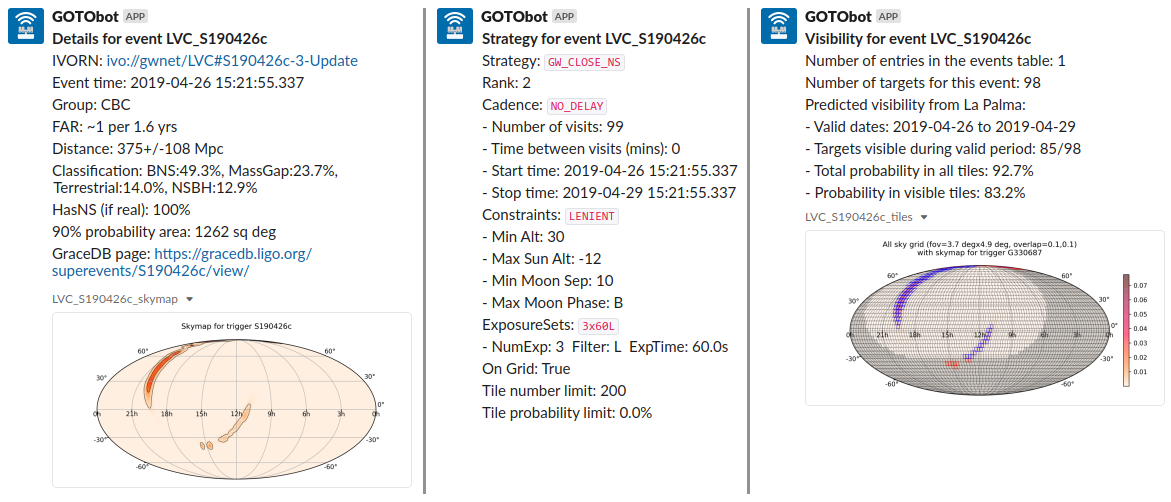}
    \end{center}
    \caption[Slack alerts created by GOTO-alert for a GW event]{
        Slack alerts generated by GOTO-alert for GW event S190426c \citep{S190426c}.
        The event alert (left) includes key information about the event gathered from the GCN notice. For GW events that includes FAR, classification and distance information, a link to the GraceDB page and a skymap generated by GOTO-tile.
        The strategy alert (centre) details the event observing strategy (see \aref{sec:event_strategy}).
        The visibility alert (right) reports how many tiles were inserted into the observation database and what skymap probability they cover. The attached plot shows a prediction of which tiles will be visible during the valid observing period.
    }\label{fig:gotoalert_slack}
\end{sidewaysfigure}

\end{colsection}

\section{Strategies for follow-up observations}
\label{sec:strategy}

\begin{colsection}

An important function of the GOTO-alert event handler is to determine the specific strategy to be used for follow-up observations of each interesting event. Through the G-TeCS observation database (see \aref{sec:obsdb}), observations can be tailored to the properties of the triggering event, either by altering the validity, priority and cadence of the pointings inserted into the scheduler queue (see \aref{sec:ranking}) or by customising the commands issued by the pilot when each pointing is selected (see \aref{sec:pilot}).

The term \emph{strategy} is used deliberately to differentiate from the actions carried out by the on-site pilot, which are better called \emph{tactics}. Properly defined, strategy considers long-term aims and objectives (consider generals directing a war far from the front lines, or the coach of a sports team), whereas tactics are the on-the-ground implementation details used to move towards those objectives (determined by the captain in the trenches or on the pitch). The sentinel decides the observing strategy for a particular event, using the structures and functions within GOTO-alert. These decisions are then communicated to the pilot through the observation database, which decides what to do independently using the scheduler and the local conditions. Only then are commands sent to the hardware daemons to put the plan into action: like the soldiers on the battlefield or players on the pitch, theirs is not to reason why but to carry out their orders as issued.

The importance of such a distinction is that the strategies and objectives decided by the sentinel can only ever be aspirational, for the best-case scenario. It can decide a follow-up plan for an event, but if the location is not visible from La Palma, or it is currently raining, then the pilot will be unable to implement it. The sentinel is designed to be aspirational and not consider smaller details such as these in order to make it independent of physical hardware. Ultimately GOTO is envisioned to occupy multiple sites across the world, as described in \aref{sec:goto_expansion}, but the intention is that they will all still be taking orders from a single central sentinel and database (see \aref{sec:gtecs_future}).

\end{colsection}

\subsection{Determining observation strategies for events}
\label{sec:event_strategy}
\begin{colsection}

\begin{figure}[t]
    \begin{center}
        \includegraphics[width=\linewidth]{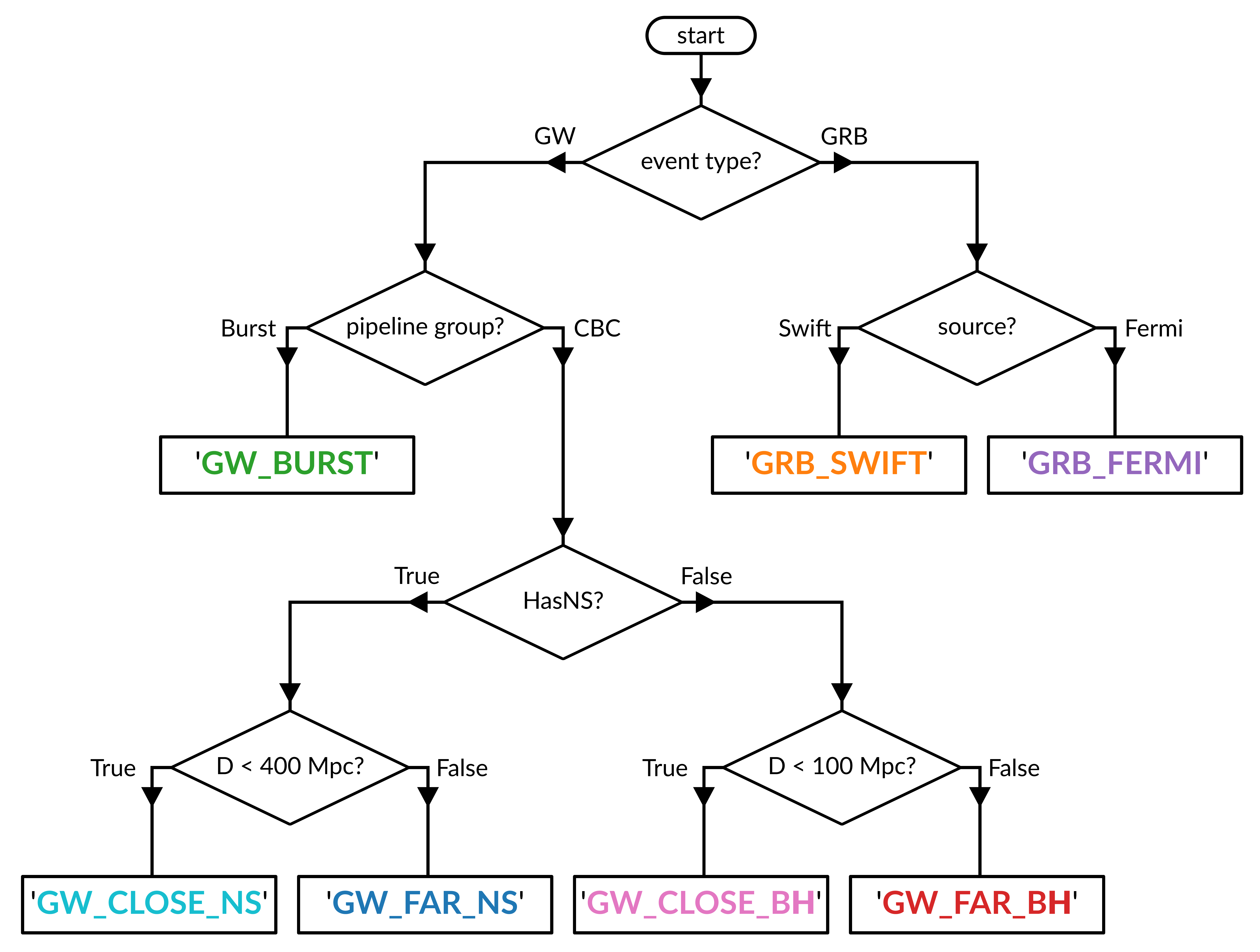}
    \end{center}
    \caption[Decision tree for determining event strategy]{
        Decision tree for determining event strategy.
    }\label{fig:strategy_flowchart}
\end{figure}

In order to decide the observation strategy for a given event, the GOTO-alert event handler uses the decision tree shown in \aref{fig:strategy_flowchart}. The codes at the end of the branches (\code{GW\_CLOSE\_NS}, \code{GRB\_FERMI}, etc) are the individual strategies, and they correspond to keys in the strategy dictionary defined within GOTO-alert as shown in \aref{tab:strategy_dict}. Each strategy corresponds to an integer rank as well as further keys relating to cadence, constraints and exposure sets which are keys in three additional dictionaries shown in \aref{tab:cadence_dict}, \aref{tab:constraints_dict} and \aref{tab:exposuresets_dict}.  Through this structure all the values required for inserting an event into the database are defined, and it is also simple to modify strategies or add in new ones. The reasoning behind each strategy is outlined below.

\clearpage

\begin{table}[!p]
    \begin{center}
        \begin{tabular}{lclll}
            Strategy & Rank & Cadence & Constraints & ExposureSets \\
            \midrule
            \code{GW\_CLOSE\_NS} &    2 & \code{NO\_DELAY}               & \code{LENIENT} & \code{3x60L} \\ %
            \code{GW\_FAR\_NS}   &   13 & \code{NO\_DELAY}               & \code{LENIENT} & \code{3x60L} \\ %
            \code{GW\_CLOSE\_BH} &   24 & \code{TWO\_NIGHTS}             & \code{LENIENT} & \code{3x60L} \\ %
            \code{GW\_FAR\_BH}   &  105 & \code{TWO\_NIGHTS}             & \code{LENIENT} & \code{3x60L} \\ %
            \code{GW\_BURST}     &   52 & \code{NO\_DELAY}               & \code{LENIENT} & \code{3x60L} \\ %
            \code{GRB\_SWIFT}    &  207 & \code{TWO\_FIRST\_ONE\_SECOND} & \code{NORMAL}  & \code{3x60L} \\ %
            \code{GRB\_FERMI}    &  218 & \code{TWO\_FIRST\_ONE\_SECOND} & \code{NORMAL}  & \code{3x60L} \\ %
        \end{tabular}
    \end{center}
    \caption[Event strategy dictionary keys]{
        Event strategy dictionary keys. The ranks are used by the scheduler to sort pointings (see \aref{sec:rank}). The cadence values are matched to \aref{tab:cadence_dict}, constraints values to \aref{tab:constraints_dict} and ExposureSets to \aref{tab:exposuresets_dict}.
    }\label{tab:strategy_dict}
\end{table}

\begin{table}[!p]
    \begin{center}
        \begin{tabular}{lccc}
            Cadence & Max visits & Visit delay (hours) & Valid days \\
            \midrule
            \code{NO\_DELAY}               & 99 &     0 & 3 \\
            \code{TWO\_NIGHTS}             &  2 &    12 & 3 \\
            \code{TWO\_FIRST\_ONE\_SECOND} &  3 & 4, 12 & 3 \\
        \end{tabular}
    \end{center}
    \caption[Cadence strategy dictionary keys]{
        Cadence strategy dictionary keys, used to define pointings in the observation database (see \aref{sec:obsdb}).
    }\label{tab:cadence_dict}
\end{table}

\begin{table}[!p]
    \begin{center}
        \begin{tabular}{lcccc}
            Constraints & Min Alt & Max Sun Alt & Min Moon Separation & Max Moon Phase \\
            \midrule
            \code{NORMAL}  & \SI{30}{\degree} & \SI{-15}{\degree} & \SI{30}{\degree} & Bright \\
            \code{LENIENT} & \SI{30}{\degree} & \SI{-12}{\degree} & \SI{30}{\degree} & Bright \\
        \end{tabular}
    \end{center}
    \caption[Constraints strategy dictionary keys]{
        Constraints strategy dictionary keys, used to define the constraints applied by the scheduler to determine if a pointing is valid (see \aref{sec:constraints}).
    }\label{tab:constraints_dict}
\end{table}

\begin{table}[!p]
    \begin{center}
        \begin{tabular}{lcccc}
            ExposureSets & Set position & Number of exposures & Exposure time & Filter \\
            \midrule
            \code{3x60L}   & 1/1 & 3 & \SI{60}{\second} & \textit{L} \\ %
            \code{3x60RGB} & 1/3 & 1 & \SI{60}{\second} & \textit{R} \\ %
                           & 2/3 & 1 & \SI{60}{\second} & \textit{G} \\ %
                           & 3/3 & 1 & \SI{60}{\second} & \textit{B} \\ %
        \end{tabular}
    \end{center}
    \caption[ExposureSets strategy dictionary keys]{
        ExposureSets strategy dictionary keys, used by the pilot when adding exposure sets to the exposure queue (see \aref{sec:exq}).
    }\label{tab:exposuresets_dict}
\end{table}

\clearpage

\subsubsection{Event sources}

The first distinction to make is between gravitational-wave and gamma-ray burst events. GOTO's primary aim is searching for GW counterparts, with any GRB follow-up being a useful but decidedly lower-priority use of GOTO's time. Therefore all GW events have ranks considerably higher than GRB events and, due to their importance to the project, GW events also use more lenient observing constraints (defined in \aref{tab:constraints_dict}) than the normal ones used by GRB events and the all-sky survey.

\subsubsection{Gravitational-wave sources}

The highest priority for GOTO should always be GW events that are predicted to contain a neutron star component, as they are the ones that are expected to produce an electromagnetic counterpart that GOTO could detect (see \aref{sec:gw_sources}, and the definitions of ``HasNS'' and ``HasRemnant'' in \aref{sec:event_classes}).

Neutron star events (which can include neutron star-black hole binaries and MassGap binaries) have no delay between visits (see \aref{tab:cadence_dict}), meaning once a pointing is completed it will be immediately re-inserted into the queue (for each observation the effective rank will be increased by 10, as described in \aref{sec:rank}). For these events, once GOTO has observed all of the visible tiles once it will immediately start covering the skymap again, and by default this will continue until it reaches the stop time three days after the event (or until it reaches 99 observations of each tile, which is just inserted as a nominal maximum and is not expected to be reached within three nights).

Binary black hole (BBH) events, on the other hand, have a more limited \code{TWO\_NIGHTS} strategy, which only requires two observations of each tile with at least 12 hours between them. The 12 hour delay in practice means observing on two subsequent nights; as the GOTO site on La Palma can not see the northern celestial pole circumpolar targets do not need to be considered, and even in winter the longest nights are less than 12 hours.

\newpage

For both classes of events it is expected that GOTO will want to respond immediately and attempt to image the localisation area as quickly as possible. The sole current detection of an electromagnetic counterpart to a gravitational-wave event, AT~2017gfo associated with GW170817 (see \aref{sec:followup}), was first observed by Swope 10.9 hours after the event, and there remains a lot of uncertainty in how kilonova appear in their early stages \citep{GW_kilonova_early}. Therefore even an early non-detection by GOTO would provide valuable information to constrain the lightcurve.

\subsubsection{Gravitational-wave distances}

For both neutron star and black hole events a distinction is also made between ``close'' and ``far'' events based on their reported source distance: a close neutron star event is defined to be within \SI{400}{\mega\parsec} while a close binary black hole event is within \SI{100}{\mega\parsec}.

Swope observed AT~2017gfo at \textit{i}$=17.057\pm0.018$ mag \citep{GW170817_Swope}, which at a distance of \SI{40}{\mega\parsec} corresponds to an absolute magnitude of -16. Using the equation for apparent magnitude
\begin{equation}
    m-M = -5 +5\log_{10}(d),
    \label{eq:absolute_magnitude}
\end{equation}
AT~2017gfo would have peaked above 19th magnitude out to a distance of \SI{100}{\mega\parsec} and above 22nd magnitude out to a distance of \SI{400}{\mega\parsec}. Binary-black hole mergers are not expected to produce the same amounts of ejected matter that would produce a kilonova, but some predictions suggest there may be material ejected from disks around one or more of the black holes which might reach 22nd magnitude if closer than \SI{100}{\mega\parsec} \citep{BBH_EM}. As a first approximation, therefore, the 22nd magnitude limits were adopted for the close/far distinction. Note this is a arbitrary division, not particularly based on GOTO's capability but a more general division into sources that could have a counterpart discoverable by existing wide-field follow-up projects (see the limiting magnitudes in \aref{tab:rivals} in \aref{sec:goto_motivation}).

\newpage

The only difference in strategy between ``close'' and ``far'' events is that they are assigned different initial ranks when inserted into the database (shown in \aref{tab:strategy_dict}). The rest of the strategy values, including cadence and constraints, are identical, and therefore the division is completely academic except in the case where \emph{multiple} events exist in the observing queue at the same time. Should this happen, and tiles for both events are visible at the same time, then the ranking system provides a quick method to prioritise events for observations. Using the ranks given in \aref{tab:strategy_dict}, events from close neutron star mergers would be inserted at rank 2, while far mergers of the same type would be inserted at rank 13. As the rank of a pointing is increased by 10 every time it is observed (see \aref{sec:rank}), this would prioritise two passes of the ``close'' skymap before the ``far'' skymap. The first observation would be at rank 2, the second at rank 12, and by the third it would be in the queue at rank 22. This is lower than the initial rank of 13 for the ``far'' event, so the latter would then be higher in the queue. Any following observations would alternate between the two events: ``close'' observation 3 at rank 22, ``far'' observation 2 at rank 23, ``close'' observation 4 at rank 32 etc. The ranks for the other events are likewise carefully chosen: close BBH events would be inserted at rank 24 and therefore fall behind the first three passes of a close NS or two passes of a far NS.\@ Far BBH events are very unlikely to produce any visible optical counterparts, so although GOTO will still follow-up the alerts they are inserted at rank 105, well below several passes of any more promising GW events.

What is currently not considered by the strategy outlined above is a \textit{maximum} valid distance, beyond which GOTO would not respond to the alert. LVC detections have already reached out to the gigaparsec scale, and at those distances the chance of there being any optical counterpart visible from Earth is very low. At the time of writing GW events are rare enough that there is little reason for GOTO not to follow up every alert, but in the future if necessary it would be easy to limit which events GOTO follows up based on distance (or another parameter, such as the false-alarm rate).

\newpage

\subsubsection{Gravitational-wave burst alerts}

Gravitational-wave events from the LVC burst pipelines \citep{GW_burst} are hard to categorise, as there is very little information to base any observation strategy on (as noted in \aref{sec:event_classes}, alerts from unmodelled burst detections do not contain source classifications or predicted distances). As a compromise they use the same cadence strategy as neutron star events, but are inserted at rank 52, below any more promising GW events but above binary-black hole events. As described in \aref{sec:gw_sources}, to date no burst alerts, e.g.\ from supernovae, have been released by the LVC.\@ The only detections from the burst pipelines (that have been made public) have corresponded to compact binary coalescence events detected by the other pipelines.

\subsubsection{Gamma-ray burst alerts}

Gamma-ray burst alerts from \textit{Fermi} and \textit{Swift} are also processed by the sentinel and inserted into the observation database. As shown in \aref{tab:strategy_dict}, pointings from GRB events are inserted at ranks above 200, ensuring that GOTO will always prioritise gravitational-wave follow-up. GRB events use a different cadence strategy of \code{TWO\_FIRST\_ONE\_SECOND}, which, as detailed in \aref{tab:cadence_dict}, consists of three observations: two in the first night separated by at least 4 hours and another on the second night. This cadence was recently implemented to attempt to account for the fast-fading nature of GRB afterglows, and is an example of more complex cadences allowed by the G-TeCS scheduler.

For gamma-ray burst events the only distinction is made between events originating from \textit{Swift} and \textit{Fermi}. Typically \textit{Swift}'s \acro{bat} detections are much better localised than those from \textit{Fermi}'s \acro{gbm}, and so, in cases where a source is detected by both, the \textit{Swift} detection should be prioritised. Further division could be made based on the burst being classified by \textit{Fermi} as Long or Short, but this is not currently implemented.

\subsubsection{Exposure sets}

Each strategy currently defined in \aref{tab:strategy_dict} uses the same exposure set definition: \code{3x60L}. From \aref{tab:exposuresets_dict} this comprises of three sequential \SI{60}{\second} exposures in the \textit{L} filter, the same as the all-sky survey. A different set of exposures using the coloured filters instead is shown in \aref{tab:exposuresets_dict} as \code{3x60RGB}, this a possible example of what could be defined using the G-TeCS system but is not used as part of any current strategy.  %

\subsubsection{Subsiquent strategy alterations}

The strategies detailed in the above sections are designed to inform the default, automatic reaction of GOTO to any incoming alert. They are not alone intended to be a perfect reaction for every case, and later, human-guided input is to be expected. For example, the default strategy for a gravitational-wave alert from a close neutron star source is to observe the tiles inserted over and over until the three day limit has passed. In practice, it should be clear after the first few passes if there is any counterpart candidate, in which case a human could intervene and direct GOTO to go take observations of particular tiles with promising candidates. Likewise, if after the first pass of a distant binary black hole event skymap no candidates are detected, and nothing has been reported from other facilities, the decision could be made not to bother with the second pass the day after and just return to the all-sky survey.

Another possible modification to a follow-up campaign could be the inclusion of feedback from the GOTOphoto detection pipeline (described in \aref{sec:gotophoto}). In the same way as a human could take over observations described above, should the pipeline detect a promising source it could be allowed to trigger GOTO to re-observe that tile, either automatically or after human vetting. This would require significant development to the pipeline and sentinel which is not a current priority, and is given as an example of possible future work in \aref{sec:software_future}.

\newpage

\end{colsection}

\subsection{Inserting events into the observation database}
\label{sec:event_insert}
\begin{colsection}

Once the strategy has been determined for an event, based on the details described in \aref{sec:event_strategy}, then the sentinel needs to insert pointings into the observation database so they are visible to the scheduler (see \aref{sec:obsdb} and \aref{sec:scheduler}). This involves mapping the event skymap on to the all-sky grid, as well as accounting for any previous detections of the same event.

\subsubsection{Previous records}

Before inserting any new pointings, the sentinel event handler first checks for any existing records of the new event in the observation database \code{events} table. This is done to update event pointings as revised VOEvent alerts are received, or in order to process retraction events. As described in \aref{sec:event_classes} there are several types of LVC alerts: ``Preliminary'' alerts are released first, followed by ``Initial'' alerts when the detections are confirmed, updated notices are released as ``Update'' alerts, and ``Retraction'' alerts are issued if the detection is later retracted. At any of these stages the event skymap might be modified, shifting the area to observe, and the pointings in the observation database will therefore need to updated. This is most common for gravitational-wave events as the initial skymaps are typically created using the rapid BAYESTAR pipeline \citep{BAYESTAR}, while later updated skymaps are made using using the slower LALInference code \citep{LALInference}. If a previous entry for a new event is detected then all of the old pointings that are still pending in the queue are deleted, before the new ones are added as described below. Should the event be of type \code{GWRetractionEvent} then this is where the event handler exits, as once the previous pointings are deleted then there are no more to replace them.

\subsubsection{Mapping onto the all-sky grid}

Once any existing pointings have been removed, new entries in the database need to be created. All GOTO pointings from the sentinel are defined \textit{on-grid}, meaning that they need to be mapped onto the current GOTO-tile sky grid (see \aref{sec:gototile}). The database \code{grids} table contains the field of view and overlap parameters of the current grid as well as the algorithm used (see \aref{sec:algorithms}), allowing it to be reconstructed using GOTO-tile within the event handler. Once a GOTO-tile \code{SkyGrid} class has been created, then a corresponding \code{SkyMap} class is made based on the information in the event. If the event was from a gravitational-wave alert then it should have a URL to download the LVC-created skymap (shown in \aref{fig:voevent_xml}). If instead it just has a coordinate and error radius, i.e.\ it is a GRB alert, then a new Gaussian skymap is constructed as described in \aref{sec:grb_skymaps}. Once both grid and map are ready then the skymap is mapped onto the grid as described in \aref{sec:mapping_skymaps}, using the class method \code{SkyGrid.apply\_skymap(SkyMap)}. This returns a table of tiles and associated contained probabilities, which are used to create the database pointings. %

\subsubsection{Selecting tiles}

The tile probability table created by GOTO-tile contains entries for every one of the thousands of tiles in the all-sky grid, of which the vast majority will contain only a very small amount of the overall probability for any reasonably-well located skymap. Adding an entry to the database for every tile would therefore be unnecessary, and even harmful to the follow-up observations. The scheduling system is designed to complete a full pass of the visible tiles before going on to re-observe those already completed, a consequence of the effective rank increasing by 10 each time a tile is observed (see \aref{sec:rank}), and so adding excess tiles would delay subsequent observations. On the other hand, it is still important to add in enough tiles covering enough of the probability area to maximise the chance of detecting the source. Therefore there is a balance required between adding too many or too few tiles into the database.

There are multiple ways to chose which tiles to select. Initially GOTO-alert used a simple cut-off in terms of each tile's contained probability, selecting tiles that contain greater than, for example, $1\%$ of the total probability. This hard limit however quickly proved to be unsuitable, as large, spread-out skymaps might have few if any tiles which reach the limit, while for well-localised events adding tiles down to the $1\%$ level is redundant and would waste time observing them compared to revisiting higher probability tiles. An attempt to correct this was to modify the probability cut-off for each skymap, making it a function of the highest tile probability. For example, a well-localised event (such as GW170817, see \aref{fig:170817_gw}) might result in the highest-probability tile containing $40\%$ of the probability; with a $P=0.1P_\text{max}$ cut-off all tiles containing 4\% or above would therefore be added to the database. On the other, hand a spread-out skymap might have a highest tile containing only 2\% of the probability, so then all tiles with 0.2\% or above would be added. Ultimately, a hard probability cut-off proved to be unresponsive to the spread of the skymap, and determining the relative limit (e.g. 0.1 in the above example) was hard to balance between large and small maps. The preferred method to select tiles is instead based on the 90\% probability contour, this method was discussed previously in \aref{sec:mapping_skymaps}.

\subsubsection{Adding database entries}

Once the tiles to add have been selected then entries can be inserted into the appropriate tables in the observation database (see \aref{sec:obsdb}).

First a new entry in the \code{events} table is added for the event, containing the unique VOEvent IVORN, the event source (LVC, \textit{Fermi} etc), type (GW or GRB), and an event ID (for example S190425z for an LVC event, \textit{Fermi} and \textit{Swift} have their own trigger IDs). Then an entry in the \code{surveys} table is also created, in order to group together all of the pointings from this particular event. The database does allow multiple surveys per event; for example, there could be a quick initial survey in a wide-passband filter that prioritises possible host galaxies, followed by a slower survey using the colour filters and longer exposure times that focuses on covering the skymap. However, at the time of writing each event only has a single survey defined.

Finally, the individual tiles are inserted as entries in the \code{mpointings} table. The most important entries for the mpointings are determined by the event strategy as described in \aref{sec:event_strategy}: the rank (taken from \aref{tab:strategy_dict}), cadence parameters (from \aref{tab:cadence_dict}), the target constraint values (minimum altitude, moon phase etc; from \aref{tab:constraints_dict}), and the exposure settings (exposure time, filter and number in each set; from \aref{tab:exposuresets_dict}). Each mpointing is connected to an entry in the \code{grid\_tiles} table for this particular grid, and the tile probabilities are stored as corresponding weights in the \code{survey\_tiles} table. Once the mpointings are defined, the first pointings are also created and added to the \code{pointings} table with the status \code{pending}, to insure they are immediately valid in the queue (see \aref{sec:scheduler}).

Once all of the entries described above have been added to the observation database the event has been successfully handled. At this point the GOTO-alert event handler sends the final visibility report to Slack (see \aref{sec:event_slack}) and exits. In total the entire event handling process, from the alert being received to the pointings being added to the database, takes under 10 seconds. The majority of this time for a gravitational-wave event is downloading the often quite large skymap files from GraceDB, and actually processing the event only takes a few seconds. Once the event pointings are added to the database they should be ready to observe the next time the scheduler fetches the queue, and if they are valid the highest priority pointing will be sent to the pilot to immediately begin follow-up observations.

\end{colsection}

\section{Summary and Conclusions}
\label{sec:alerts_conclusion}

\begin{colsection}

In this chapter I described how the functions within the GOTO-alert process astronomical transient event alerts.

The GOTO sentinel alert listener receives gravitational-wave alerts from the LIGO-Virgo Collaboration in the form of GCN Circulars, which are formatted XML documents following the VOEvent schema. These VOEvents contain the key properties of the detection and a link to a skymap localisation file, which is then mapped onto the GOTO all-sky grid (see \aref{chap:tiling}).

One of the important features of the GOTO-alert event handler is the ability to automatically select the observation strategy for different alerts based on the contents of the VOEvent. Gravitational-wave detections predicted to come from near-by binary neutron star or neutron star-black hole binaries are the highest priority to follow up, followed by other classes of events. Gamma-ray burst events are also processed and added to the observations database by the G-TeCS sentinel (see \aref{chap:autonomous}), but always at a lower priority than the gravitational-wave alerts.

At the time of writing GOTO has been following-up gravitational-wave events for several months since the start of the third LIGO-Virgo observing run. The results of the observing run so far are detailed as part of the overall conclusions in \aref{chap:conclusion}.

\end{colsection}

\chapter{On-Site Commissioning}
\label{chap:commissioning}

\chaptoc{}

\section{Introduction}
\label{sec:commissioning_intro}

\begin{colsection}

In this chapter I describe commissioning the GOTO hardware and parallel software developments.
\begin{itemize}
    \item In \nref{sec:hardware_commissioning} I give an outline of the commissioning period, focusing on my own involvement with trips to La Palma and building additional hardware for the dome.
    \item In \nref{sec:software_commissioning} I describe how the control software was developed alongside the hardware, including creating nightly observing routines to take flat fields and focus the telescopes, and how challenges arising from hardware issues were overcome.
\end{itemize}
All work described in this chapter is my own unless otherwise indicated, and has not been published elsewhere. Commissioning the GOTO hardware on La Palma was carried out along with several members of the GOTO collaboration; in particular Vik Dhillon and Stu Littlefair from Sheffield; Danny Steeghs, Krzysztof Ulaczyk and Paul Chote from Warwick; Kendall Ackley from Monash; and several undergraduate and postgraduate students from the GOTO member institutions.

\end{colsection}

\section{Deploying the hardware}
\label{sec:hardware_commissioning}

\begin{colsection}

GOTO commissioning began with the installation of the prototype telescope on La Palma in the spring of 2017. Over 2017 and 2018 I spent a total of nine weeks on-site, helping deploy the hardware as well as commissioning and developing the G-TeCS software described in previous chapters.

\end{colsection}

\subsection{Deployment timeline}
\label{sec:timeline}
\begin{colsection}

GOTO was envisioned as a quick, simple and cheap project that could provide a large field of view to cover the early gravitational-wave skymaps produced by LIGO.\@ When I first interviewed to join the project in February 2015 it was anticipated that GOTO would be up and running imminently, perhaps before the end of that year. Ultimately that did not happen, as can be seen in the project timeline given in \aref{tab:timeline}, and GOTO did not see first light until June 2017. In hindsight the delay was irrelevant, as the first (and, at the time of writing, only) gravitational-wave event to have an electromagnetic counterpart occurred two months later in August 2017 --- and was only visible from the southern hemisphere \citep{GW170817,GW170817_followup}.

GOTO's deployment date was repeatedly set back for a variety of reasons, including planning permission being held up by local tax disputes and delays in manufacturing the mount and optics. The site was ready months before the telescope was, with the first dome being built in November 2016. My first visit to La Palma took place in March 2017, while the telescopes and mount were still in the factory. Vik Dhillon and I went out to the site to develop the dome control and conditions monitoring systems as described in \aref{sec:dome} and \aref{sec:conditions}. During this trip we also installed the additional dome hardware systems I had built, which are described in \aref{sec:arduino}.

\begin{figure}[p]
    \begin{center}
        \begin{tabular}{cl|@{\tls}l} %
            2015 & July      & Collaboration meeting in Warwick (29 Jul) \\
                 &           & Site planning application submitted \\
                 & September & Research collaboration agreement signed \\
                 &           & \textit{LIGO's first observing run (O1) begins} \\
                 &           & \textit{First observation of gravitational waves (GW150914)} \\
            \midrule
            2016 & January   & \textit{O1 ends} \\
                 & August    & Planning permission granted \\
                 & September & Site construction begins \\
                 & November  & First dome assembled \\
                 &           & \textit{LIGO's second observing run (O2) begins} \\
            \midrule
            2017 & March     & \textcolor{Blue}{Trip 1 (23--31 Mar) --- install dome systems} \\
                 & May       & Telescope hardware shipped \\
                 & June      & \textbf{GOTO first light (10 Jun)} \\
                 &           & Collaboration meeting in Warwick (19--20 Jun) \\
                 &           & \textcolor{Blue}{Trip 2 (22 Jun--7 Jul) --- install control software} \\
                 & July      & Inauguration ceremony (3 July) \\
                 &           & Dec axis encoder fails \\
                 &           & \textcolor{Blue}{Trip 3 (20--28 Jul) --- pilot commissioning} \\
                 &           & Robotic operations begin \\
                 & August    & UT3 mirrors sent back to manufactures \\
                 &           & \textit{Virgo joins O2} \\
                 &           & \textit{First gravitational-wave counterpart detected (GW170817)} \\
                 &           & \textit{O2 ends} \\
                 & November  & Drive motors upgraded, arm extensions installed \\
                 &           & \textcolor{Blue}{Trip 4 (9--16 Nov) --- on-site monitoring} \\
                 & December  & Second dome assembled \\
            \midrule
            2018 & January   & \textcolor{Blue}{Trip 5 (14 Jan--5 Feb) --- on-site monitoring} \\
                 & April     & Collaboration meeting in Warwick (11--13 Apr)\\
                 & May       & On-site monitoring program ends \\
                 & June      & Refurbished mirrors installed into UT4, old mirrors sent back \\
                 & July      & \textcolor{Blue}{Trip 6 (5--13 Jul) --- software development} \\
                 &           & New UT mounting brackets installed \\
                 & December  & \textit{LIGO-Virgo Engineering Run 13 (14--18 Dec)} \\
            \midrule
            2019 & February  & Refurbished mirrors reinstalled into UT3 \\
                 &           & Current 4-UT all-sky survey begins \\
                 & April     & \textit{LIGO-Virgo's third observing run (O3) begins} \\
        \end{tabular}
    \end{center}
    \caption[Timeline of the GOTO project]{
        A timeline of the GOTO project from when I joined up until the time of writing, including the six trips I made to La Palma during commissioning (in \textcolorbf{Blue}{blue}) and concurrent developments in the field of gravitational waves (in \textit{italics}).
    }\label{tab:timeline}
\end{figure}

\clearpage

\begin{figure}[t]
    \begin{center}
        \includegraphics[width=\linewidth]{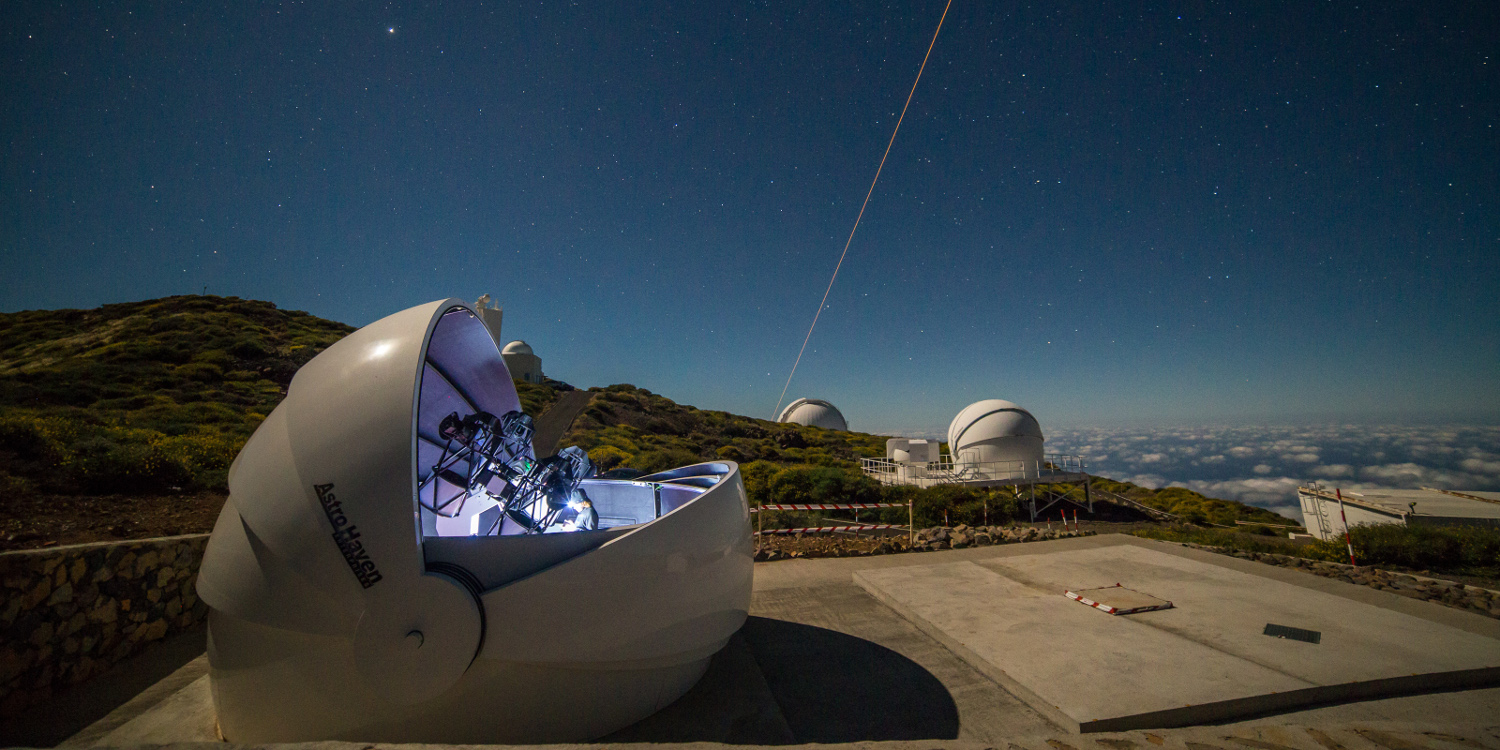}
    \end{center}
    \caption[Working in the GOTO dome prior to the inauguration in June 2017]{
        Working in the GOTO dome prior to the inauguration in June 2017.
        Photo taken looking west towards W1m and the WHT (with the orange CANARY laser visible). The second dome was installed on empty platform to the right later in the year.
    }\label{fig:inauguration}
\end{figure}

Ultimately, the mount and first four unit telescopes were shipped to La Palma in late May 2017, and GOTO officially saw first light on the 10th of June 2017. I went out to the site a few weeks later, in order to install the G-TeCS software (an image of the site at the time is shown in \aref{fig:inauguration}). By the time of the inauguration ceremony on the 3rd of July the hardware control system was in place and working well, and I was able to demonstrate the telescope that evening to the assembled dignitaries.

I returned to the site less than two weeks later with Stu Littlefair, in order to do further work on the control software. We commissioned the pilot and developed the observing routines described in \aref{sec:software_commissioning}, and oversaw the telescope's first fully-autonomous night on the 27th of July.

Unfortunately, in the months after the inauguration problems began to surface with the hardware. The first problem was the failure of the declination motor encoder shortly after the inauguration (prior to my second visit). We were able to operate GOTO in a limited RA survey mode (described in \aref{sec:challenges}), however this greatly limited the capability of the telescope. There were also other problems with the mounting brackets that hold the unit telescopes to the boom arm becoming loose, as well as the boom arms being short enough that the unit telescopes could hit the mount pier. These issues meant that for the first few months of commissioning someone always needed to be present in the dome to stop the mount moving if it was in danger of damaging itself. Once the second LIGO observing run (O2) finished at the end of August there was less of a reason to be observing in this limited mode, so GOTO was shut down during the autumn of 2017 until hardware upgrades could be installed at the start of November.

At the same time, problems with the optical performance of the unit telescopes had become apparent, which were blamed on the mirror quality and issues with collimation. A program of sending each set of mirrors back to the manufacturer one at a time was decided on, allowing GOTO to continue operating with the remaining three unit telescopes. The worst performing telescope, UT3, had its mirrors taken out and returned in August 2017. Once the telescope was reactivated in November, the remaining three unit telescopes were aligned to form a single 3$\times$1 footprint, shown in \aref{fig:3ut_footprint}. Counterweights were placed in the empty UT3 tube to allow the mount to maintain balance.

GOTO operated in this mode for over a year. The gap between LIGO runs gave time to fully test the control software, as well as develop the GOTOphoto image pipeline (see \aref{sec:gotophoto}). The first set of mirrors were returned to the site in June 2018 and were placed into UT4, which was the second-worst performing telescope. The old UT4 mirrors were then sent back to the manufacturer, and GOTO continued to operate with three unit telescopes until February 2019. At this point, based on the imminent start of the third LIGO-Virgo observing run (O3), it was decided to leave the UT1 and UT2 mirrors in place, and operate from then on in the 4-UT configuration. The resulting 2$\times$2 footprint is shown \aref{fig:4ut_footprint}. Note the unit telescopes are arranged with overlapping fields of view, to counteract for the poor image quality off-axis. With future optical improvements this overlap could be reduced, increasing the overall field of view.

\begin{figure}[p]
    \begin{center}
        \includegraphics[width=0.7\linewidth]{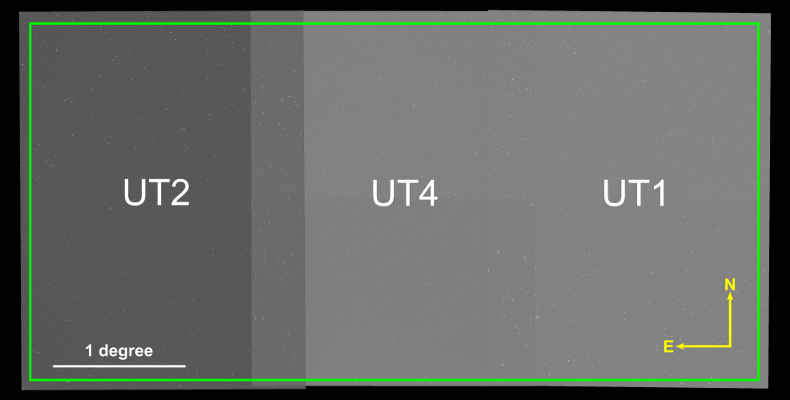}
    \end{center}
    \caption[The previous 3-UT GOTO footprint]{
        The 3-UT GOTO footprint, used from August 2017 to February 2019.
        The initial \SI{5.5}{\degree} $\times$ \SI{2.6}{\degree} tile area used by GOTO-tile (see \aref{chap:tiling}) is shown in \textcolorbf{Green}{green}. Note that a reasonable amount of space is left around the edge of the tile.
    }\label{fig:3ut_footprint}
\end{figure}

\begin{figure}[p]
    \begin{center}
        \includegraphics[width=0.55\linewidth]{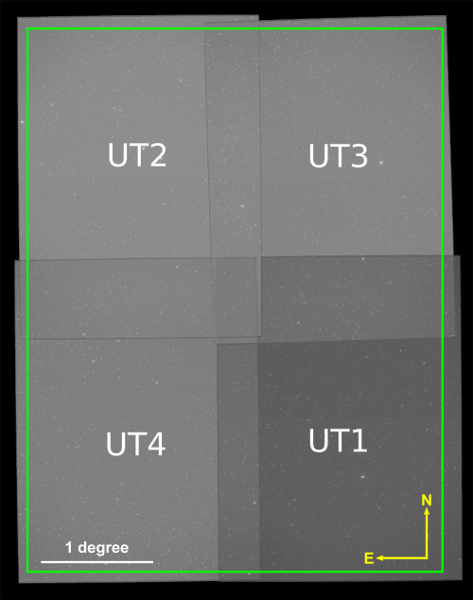}
    \end{center}
    \caption[The current 4-UT GOTO footprint]{
        The current 4-UT GOTO footprint, in use from February 2019 onwards.
        The revised \SI{3.7}{\degree} $\times$ \SI{4.9}{\degree} tile area is shown in \textcolorbf{Green}{green}. The future four unit telescopes are expected to be arranged in two more columns on the left and right as shown in \aref{fig:fov}, creating an approximately \SI{7.8}{\degree}--wide footprint.
    }\label{fig:4ut_footprint}
\end{figure}

\newpage

Another problem found during commissioning was excessive scattered light entering the system, in particular light from the Moon entering the corrector lens (see the optical design in \aref{fig:ota}). This was solved by adding covers around the telescope tubes, which prevented light from entering the corrector but made the system more susceptible to wind-shake (the reason that the tubes were open in the first place). Ultimately the covers were found to provide enough benefits, including protecting the mirrors from dust, that it has been decided that future unit telescopes will have closed tubes.

\begin{figure}[t]
    \begin{center}
        \includegraphics[width=0.4\linewidth]{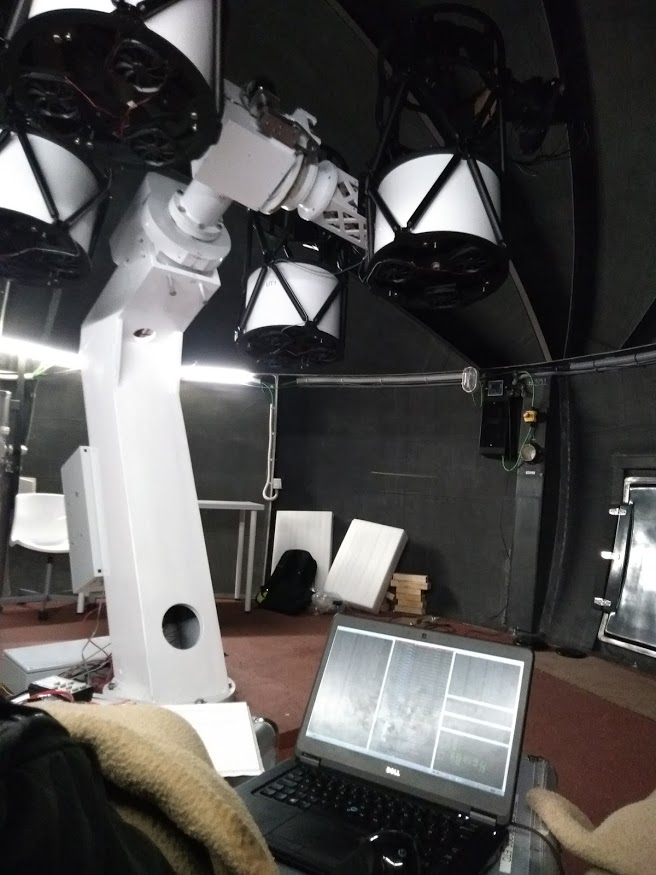}
        \includegraphics[width=0.4\linewidth]{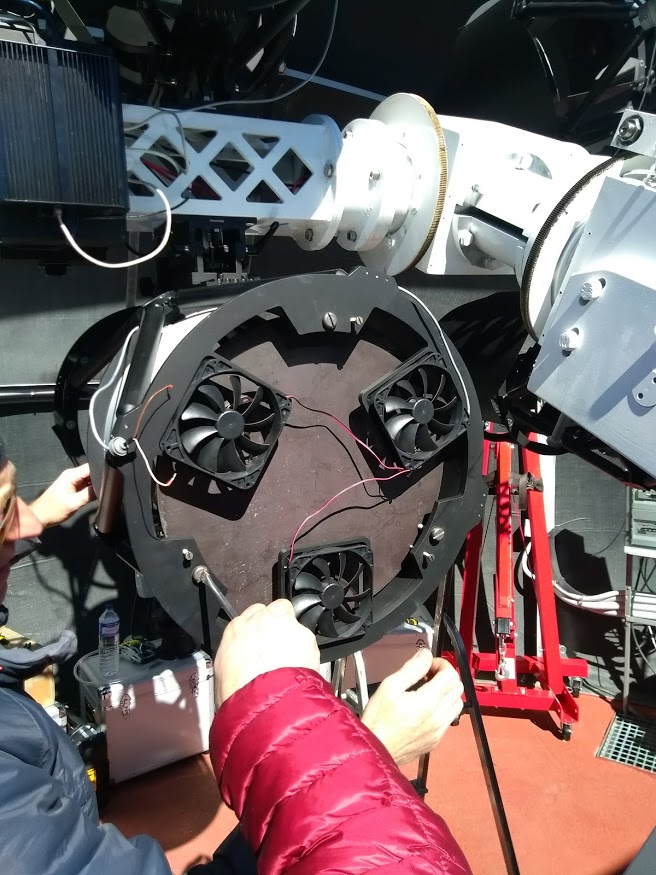}
    \end{center}
    \caption[Photos from GOTO commissioning]{
        Photos from GOTO commissioning: monitoring the telescope from inside the dome in November 2017 on the left, and fitting the mirror counterweights in UT3 in January 2018 on the right.
    }\label{fig:commissioning}
\end{figure}

The second commissioning period ran from the telescope being reactivated in November 2017 through to May 2018. During this time the telescope was typically running robotically each night, however there was always a member of the collaboration on site monitoring it in case of problems. Over that period the monitor moved from physically sitting in the dome (shown in \aref{fig:commissioning}), to sitting in the relative comfort of the neighbouring SuperWASP server room, until finally in the last few months being able to monitor from the observatory residencia or one of the other large telescopes on site.

I visited La Palma twice during this period, with the first trip in November 2017 covering the first week of the monitoring period immediately following the telescope being reactivated. Other volunteers monitored the system on-site until Christmas, then commissioning halted over the holiday period before I returned to the site in January for a three week visit. Kendall Ackley from Monash was also on site during the first week, and the first and third weeks overlapped with a team from Sheffield including Vik Dhillon and Stu Littlefair. During this period we replaced the counterweights (also shown in \aref{fig:commissioning}), rebalanced the mount and realigned the unit telescopes, and I continued the software work with a major update to the observation database. During the second week I monitored the telescope alone from SuperWASP, and continued developing the pilot so it was able to run automatically with no human supervision. In the third week I was due to remain on site and continue to monitor the telescope, however a severe snowstorm stopped all observing.

Due to the cold weather and ice build up GOTO was unable to open throughout all of February 2018. It was during this period that issues arose from the weight of the ice on the dome shutters, described in \aref{sec:challenges}. On-site monitoring resumed in the spring, once the snow had melted, and monitors continued to be on site for several more months, in between hardware upgrade trips lead by the Warwick team. Eventually in May the software was deemed robust enough to allow GOTO to run unsupervised. The pilot output is still regularly monitored remotely, especially from Australia by the Monash team, who have the benefit of a more convenient timezone.

By the time the 4-UT system was recommissioned, in February 2019, the G-TeCS pilot and hardware control systems had been fully tested and were operating reliably. By then my focus had shifted to the alert follow-up systems detailed in \aref{chap:alerts}, in advance of the start of O3 in April 2019. Since then GOTO has been reliably running and responding to gravitational-wave alerts, as detailed in \aref{sec:conclusion}.

\end{colsection}

\subsection{Additional dome systems}
\label{sec:arduino}
\begin{colsection}

GOTO uses a clamshell dome manufactured by Astrohaven, the same company that made the pt5m dome \citep{pt5m}. Based on experience with p5tm, there were several hardware systems which we decided to add to the GOTO dome. In fact, the entire pt5m dome control unit was replaced by a custom one designed and manufactured in Durham, but we wanted to avoid taking such a drastic step. Several limitations of the stock Astrohaven dome are outlined below.
\begin{itemize}
    \item First, there was no easily-accessible emergency stop button to cut power to the dome in an emergency (e.g.\ something gets caught in the motors). This is a serious concern for pt5m, as when the dome is open the shutters completely cover the access hatch, making it dangerous for anyone to be passing through the hatch when the dome is moving. Therefore, one of the additions to pt5m was an emergency stop button within arms reach of the hatch entrance. As the GOTO dome is taller the hatch is mostly uncovered when the dome is open, but installing an emergency stop button was still a priority for safety reasons.
    \item The dome does not come with a siren to sound when it opens or closes. This is an important safety feature when operating a robotic observatory, as the dome will be operated entirely through software and it is important to warn anyone on site several seconds before it is about to move. When members of the GOTO team are on-site the robotic systems can be disabled entirely by going into engineering mode (see \aref{sec:mode}), however it is still important to make sure that there is no chance of the dome moving without prior warning. In addition, the GOTO site on La Palma is publicly accessible, and it is not unknown for tourists or hikers, or other astronomers, to be around the dome when it is unsupervised.
    \item By default, the dome \acro{plc} only provides limited information about the status of the dome shutters. As described in \aref{sec:dome}, the PLC only returns a single status byte in response to a query. This is not enough to distinguish whether the dome is fully or only partially open, and if one side is moving the status of the other side is unknown. Adding our own sensors would allow the complete status of the dome to be determined. The dome comes with two in-built magnetic sensors on each side, which should detect when the shutters are either fully closed or fully open. However these have been known to be unreliable and tricky to align. In some cases the switches failed to trigger when the shutter reached its open limit, leading to the dome continuing to drive the belts and the shutter embedding itself into the ground. Therefore having secondary, independent sensors was a priority to ensure this did not happen.
    \item The dome does not include a sensor on the hatch door to detect if it is open, and there is no way to close the hatch remotely in case of bad weather. As GOTO will be operating without anyone on site, the hatch should normally remain closed at all times, and by adding a sensor to the hatch this could be confirmed and an alert issued if the hatch is detected as open (see \aref{sec:conditions_flags}).
    \item One final proposed addition was a quick-close button, which acts as a simple and direct way to communicate with the dome daemon. The motivation is a practical one: in the case that the weather turns bad and the telescope is exposed, assuming the automatic monitoring systems fail and we can not connect to close it remotely, it is a lot easier and quicker to direct someone on site to access the dome and press the prominent ``close'' button than direct them to log on to the in-dome computer, open a terminal and type \code{dome~close}. This was also true during the commissioning phase, when the software was still being tested and someone has to be on-site all night. One requirement was that this button could not be easily confused with the emergency stop button (i.e.\ it should not be coloured red), as instead of stopping the dome this button will prompt it to move.
\end{itemize}

\begin{figure}[t]
    \begin{center}
        \includegraphics[width=0.9\linewidth]{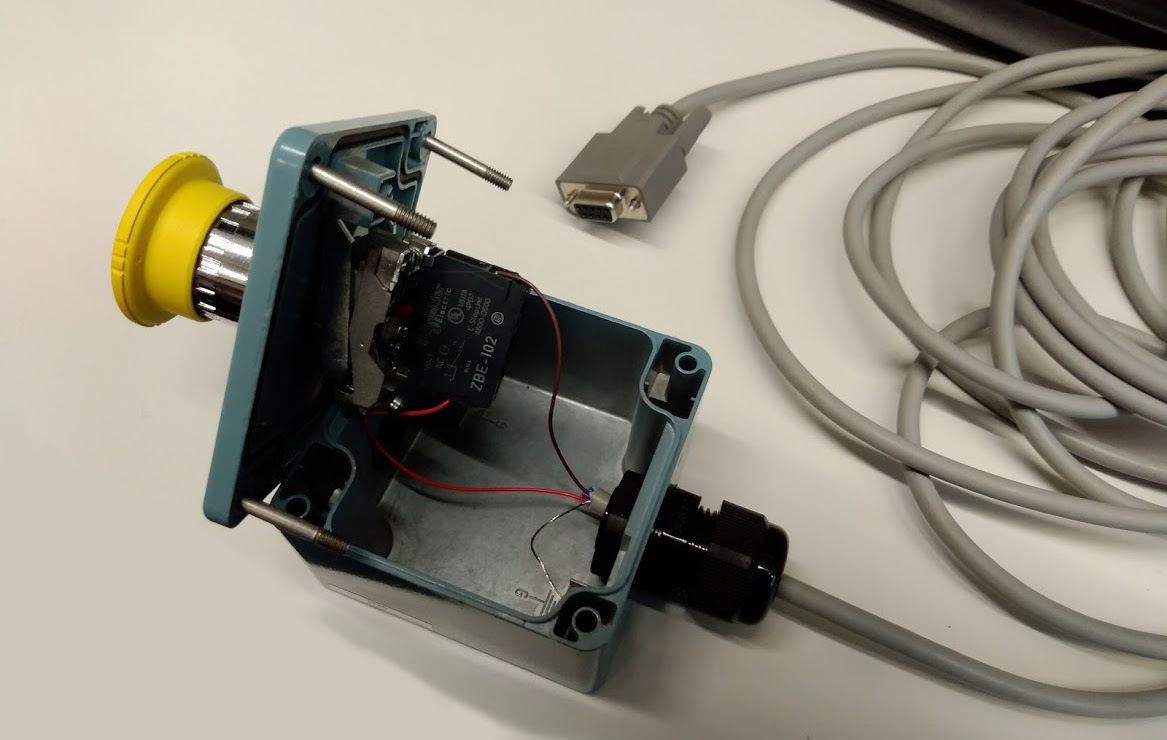}
    \end{center}
    \caption[Building the quick-close button]{
        Building the quick-close button in the lab. The transmit (\textcolorbf{Brown}{brown}) and receive (\textcolorbf{Red}{red}) wires from the serial cable are attached to the button NC pins.
    }\label{fig:quickclose_button}
\end{figure}

Creating a quick-close button involved attaching a ``normally closed'' (NC) push button in series between the transmit and receive wires of an RS 232 serial cable, as shown in \aref{fig:quickclose_button}. By doing this a simple feedback loop can be set up within the dome daemon, by sending a test signal out through the serial connection and listening for it to be returned to the same port. If after three tries the signal does not return then the button is assumed to have been pressed, and the dome daemon triggers a lockdown (see \aref{sec:dome}). By using a locking push button the loop will remain broken, and the dome closed, until the button is released. The bright yellow button was was labelled and attached to the wall of the dome near the computer rack (shown in \aref{fig:arduino_button_dome}).

Adding an emergency stop button to the dome was fairly simple. There was enough slack on the PLC power cable to install a prominent red button on the wall of the dome, as shown in \aref{fig:estop_plc}. When the button is pressed the power to the PLC and the dome motors is cut, which stops the dome moving.

\begin{figure}[p]
    \begin{center}
        \includegraphics[width=0.8\linewidth]{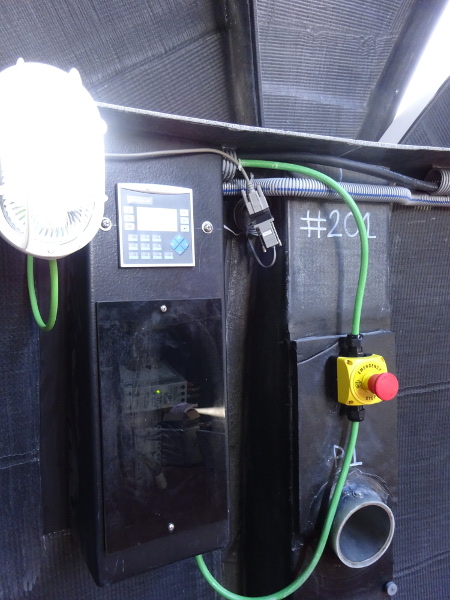}
    \end{center}
    \caption[The dome PLC and emergency stop button]{
        The dome \acro{plc} and emergency stop button. The green cable comes directly from the dome power supply at the top, passes through the button unit and into the PLC at the bottom.
    }\label{fig:estop_plc}
\end{figure}

\newpage

Adding a siren and additional sensors to the dome required an additional system to power, monitor and (for the siren) activate them. This was done using a small Arduino Uno microcontroller\footnote{\url{https://www.arduino.cc}} running a simple HTML server, which reports the status of the switches and can be queried in order to activate the siren. The circuit design for this system is shown in \aref{fig:arduino_circuit}. In order to power the siren from the Arduino a bipolar MOSFET (metal-oxide-semiconductor field-effect transistor)\acroadd{mosfet} was used to connect to one of the board input/output pins, with a large enough resistor to prevent the voltage from destroying the board. The Arduino and siren were mounted within a weatherproof case, with output connectors for the dome switches as well as for power and an ethernet connection. Photos of the box during construction are shown in \aref{fig:arduino_wip}, and its installation in the dome is shown in \aref{fig:arduino_installed} and \aref{fig:arduino_button_dome}.

\begin{figure}[t]
    \begin{center}
        \includegraphics[width=0.75\linewidth]{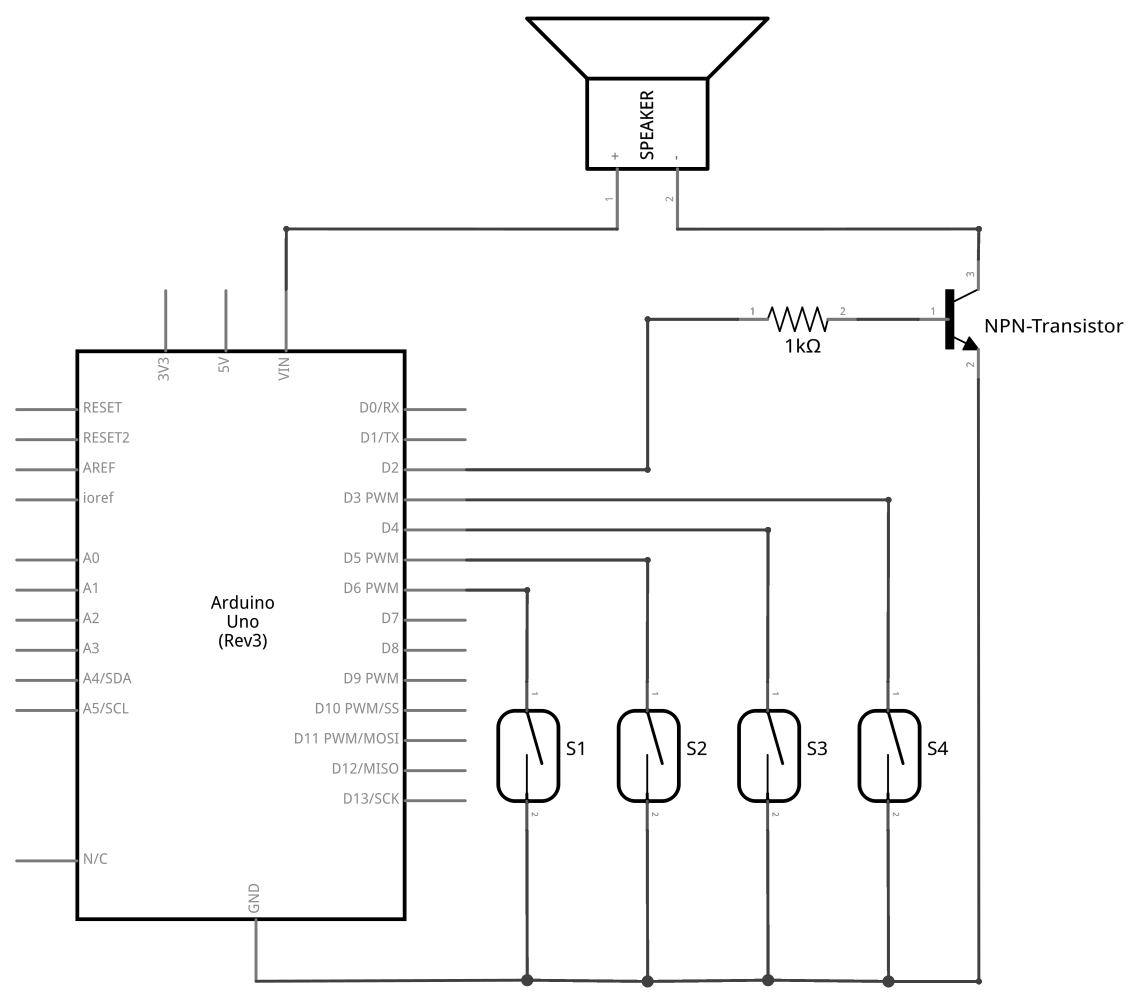}
    \end{center}
    \caption[Circuit design for the Arduino box]{
        Circuit design for the Arduino box.
    }\label{fig:arduino_circuit}
\end{figure}

\newpage

\begin{figure}[p]
    \begin{center}
        \includegraphics[width=\linewidth]{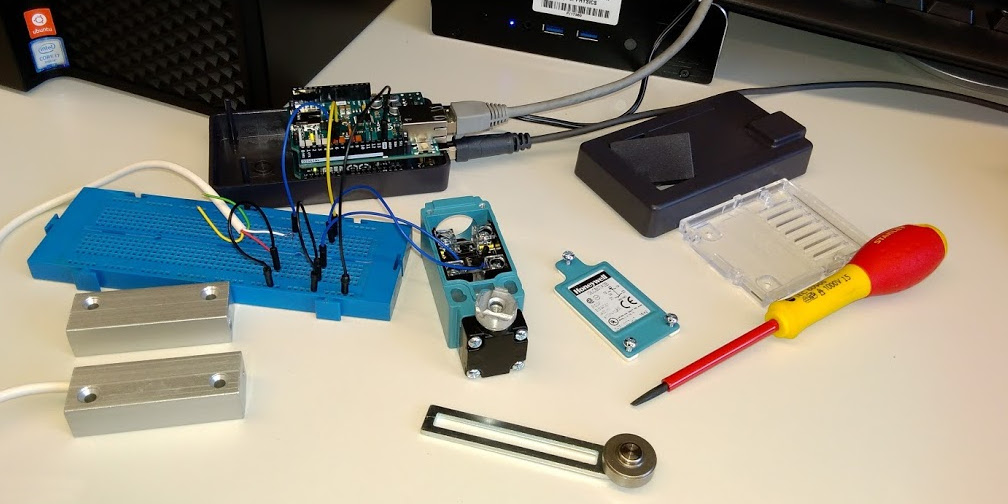}
        \includegraphics[width=\linewidth]{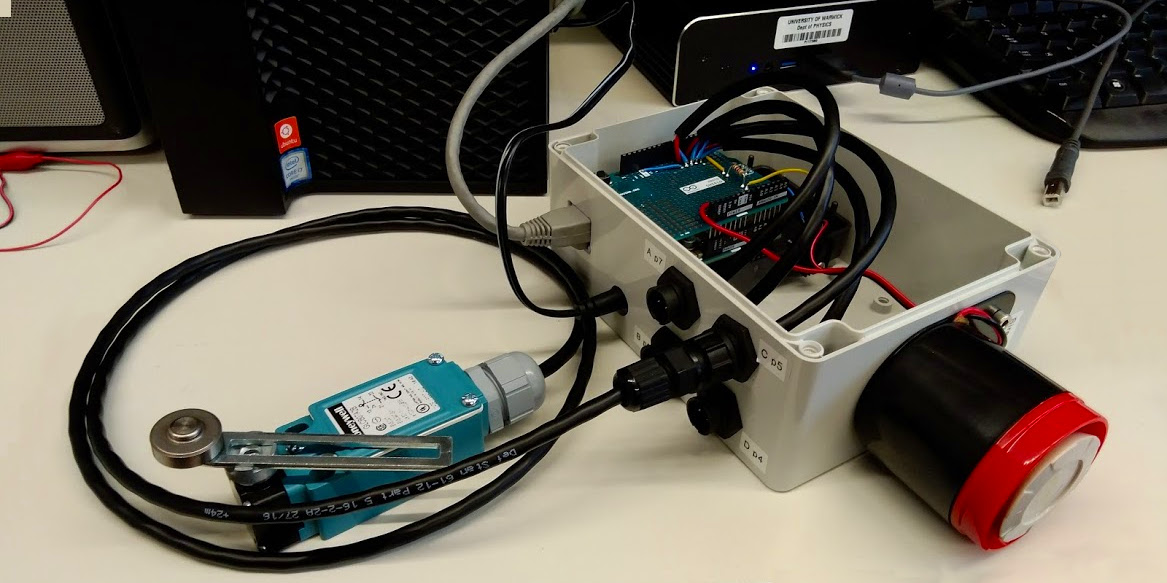}
    \end{center}
    \caption[Building the dome Arduino box]{
        Building the dome Arduino box in the lab in Sheffield. The top image shows the circuit design being tested, with the Arduino (circuit board in the back) and two of the dome switches: a magnetic proximity switch (grey blocks on the left) and a Honeywell limit switch (cyan unit in the centre, with the cover and arm detached). The lower image shows the completed weatherproof box with the siren, power and ethernet cables and one of the Honeywell switches attached.
    }\label{fig:arduino_wip}
\end{figure}

\newpage

\begin{figure}[p]
    \begin{center}
        \includegraphics[width=0.75\linewidth]{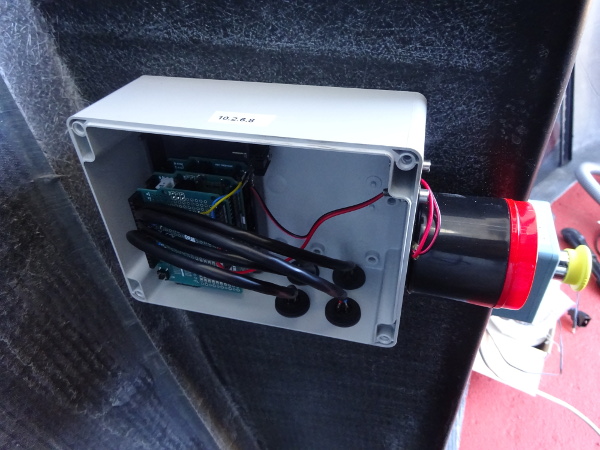}
    \end{center}
    \caption[The Arduino box installed in the GOTO dome]{
        The Arduino box installed in the GOTO dome during my first trip to La Palma in March 2017. This photo was taken before the cover and cables were attached.
    }\label{fig:arduino_installed}
\end{figure}

\begin{figure}[p]
    \begin{center}
        \includegraphics[width=0.75\linewidth]{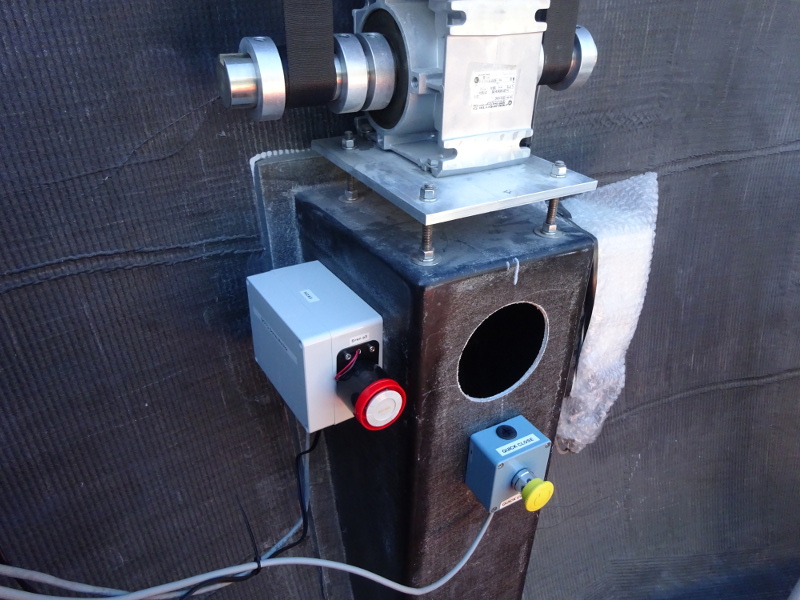}
    \end{center}
    \caption[The Arduino box and quick-close button in the GOTO dome]{
        The Arduino box and yellow quick-close button in the GOTO dome, attached to the southern pillar under the dome drive. The cables run to the computer rack which is just off to the left of the photo.
    }\label{fig:arduino_button_dome}
\end{figure}

\clearpage

Four additional sensors were added to the dome, each connected to a port on the Arduino through the connectors on the bottom of the weatherproof box. Two Honeywell limit switches were attached to the rim of the dome wall, set to be triggered when the dome was fully open; additional magnetic proximity switches were added to the two inner-most shutters, which trigger when the dome is fully closed; and a magnetic proximity switch was added to the dome hatch. Each of these are shown in \aref{fig:dome_switches}.

\begin{figure}[p]
    \begin{center}
        \includegraphics[width=0.8\linewidth]{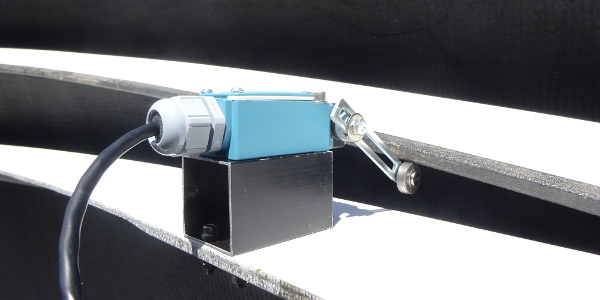}
        \includegraphics[width=0.8\linewidth]{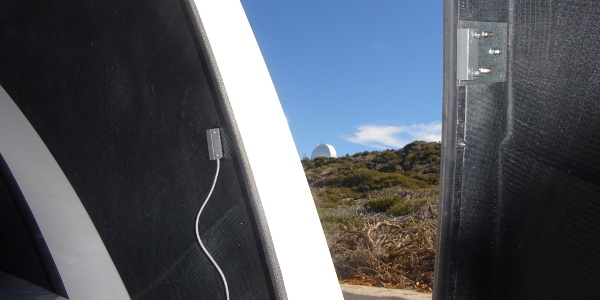}
        \includegraphics[width=0.8\linewidth]{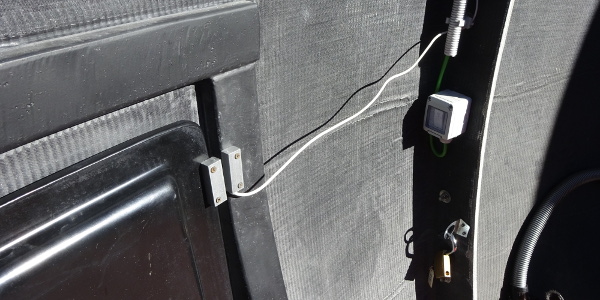}
    \end{center}
    \caption[Additional sensors added to the dome]{
        Additional sensors added to the dome. The top photo shows one of the two Honeywell limit switches added to the rim of the dome wall to detect when the dome is fully open, the middle photo shows magnetic switch added to the inner-most shutters to detect when the dome is fully closed, and the bottom photo shows the magnetic switch added to the dome hatch to detect when it is open.
    }\label{fig:dome_switches}
\end{figure}

Using a combination of these new switches and the PLC output it is possible for the dome daemon to build a complete picture of the status of the dome, as described in \aref{sec:dome}. Having a sensor on the hatch also allowed it to be added as a conditions flag, as described in \aref{sec:conditions_flags}, meaning the pilot will stop and report if the hatch is open when in robotic mode. As of yet, the hatch flag or in-dome buttons have not been needed in an emergency, however they are important as an insurance policy just in case. It is anticipated that the same work will need to be done in the second GOTO dome when it is commissioned. Comments have also been passed on to the dome manufacturer to suggest that they could include some of the features described in this section in their own stock hardware.

One further addition to the GOTO dome should be mentioned: the ``heartbeat'' monitor designed and installed by Paul Chote at Warwick. As described in \aref{sec:dome}, in the event that the dome daemon crashes, or the control NUC itself fails for whatever reason, the dome would be left vulnerable --- especially if it is already open. As a backup, Paul created his own Arduino system that connects to the dome PLC and sends it commands to close in the event that it does not receive a regular signal from the dome daemon. This system was installed into the GOTO dome in April 2018, along with the other Warwick domes on the site, and was one of the final stages required before GOTO could safely leave the commissioning phase and not require an on-site monitor. The two Arduino systems may be merged when the second GOTO dome is commissioned.

\end{colsection}

\section{Developing the software}
\label{sec:software_commissioning}

\begin{colsection}

While the GOTO hardware was being commissioned the G-TeCS control software was also being developed. There were several important parts of the software that could not reasonably be developed without access to the actual telescope, for example the observing routines for taking flat fields and autofocusing.

This section focuses on the software side of commissioning, and does not include pure hardware issues such as with the mount drive, mirrors or UT brackets outlined in \aref{sec:timeline}. These were dealt with by the core GOTO hardware team at Warwick, and although I spent many hours on La Palma balancing the mount, adjusting mirror positions and aligning unit telescopes, it had limited impact on the control software development.

\end{colsection}

\subsection{Taking flat fields}
\label{sec:flats}
\begin{colsection}

GOTO uses the twilight sky for taking flat fields. Some care has to be taken to design a reasonable flat-field routine, as taking sky flats is not simple for a wide-field instrument such as GOTO \citep{flats3, flats2}. As described in \aref{sec:night_marshal}, the night marshal runs the \code{take\_flats.py} observing script at twilight twice a day, once in the evening and again in the morning. In the evening the script begins after the dome is opened, when the Sun has set below \SI{0}{\degree} altitude, and in the morning the routine is run in reverse, starting after observations have finished and running until the Sun rises above \SI{0}{\degree}.

First, the telescope needs to slew to a chosen position. Based on the analysis of the twilight sky gradient in \citet{flats}, GOTO slews to the ``anti-sun'' position, which is at an azimuth of \SI{180}{\degree} opposite the position of the Sun and at an altitude of \SI{75}{\degree}. This should be the position where the sky gradient is minimised and therefore the field is flattest. In the pt5m version of the script, the telescope would slew to one of a predefined set of empty sky regions, however with GOTO's large field of view there are no large enough regions devoid of bright stars (to counter this, the mount moves slightly between images so that median stacking the frames will remove any stars).

Once the telescope is in position, glance exposures (see \aref{sec:cam}) are taken until the sky brightness has reached an appropriate level. In the evening, exposures start at $E_0=\SI{3}{\second}$ soon after sunset, and the first images will almost always be saturated. Images are taken until the mean count level has fallen below the target level of 25,000 counts per pixel. In the morning, exposures start at $E_0=\SI{40}{\second}$ while the sky is still dark, and exposures are taken until the mean count level is above the same target level.

Once the sky has reached the target level of brightness, exposures are taken at increasing exposure times in the evening, or decreasing in the morning. The exposure time sequence is determined using the method of \citet{flats3}, which defines the delay between exposures iteratively from $t_0=0$ using
\begin{equation}
    t_{i+1} = \frac{\ln{(a^{t_i+\Delta t} + a^{E_i} -1)}}{\ln{a}},
    \label{eq:sky}
\end{equation}
where $\Delta t$ is the time between exposures (including readout time and any offset slew time) and $a$ is a scaling factor which depends on the twilight duration $\tau$ in minutes as
\begin{equation}
    a = 10^{\pm 0.125/\tau}.
    \label{eq:sky2}
\end{equation}
The twilight duration can be calculated easily using Astropy, and $a$ is taken as less than 1 in the evening (the delay between exposures decreases) or greater than 1 in the morning (the delay increases). Note that $t$ is the time delay \emph{between} exposures, the actual exposure time of each exposure is given by
\begin{equation}
    E_{i+1} = t_{i+1} - (t_i + \Delta t).
    \label{eq:sky3}
\end{equation}

Using this method a sequence of exposure times is determined iteratively either until a target number of flat fields have been taken (by default 5 in each filter) or the exposure times pass a given limit (greater than \SI{60}{\second} in the evening, less than \SI{1}{\second} in the morning). Between each exposure the telescope is stepped 10~arcminutes in both RA and declination, enough to ensure that any objects in the field do not fall on the same CCD pixels. This means any stars in the field can be removed by median combining the individual flat field images.

Every time the script is run, flat fields are taken in each of the Baader \textit{LRGB} filters used by GOTO (see \aref{sec:filters}). In the evening flats start in the \textit{B} filter (as the sky progressively reddens as the Sun sets), progresses through \textit{G} and \textit{R}, and finishes on \textit{L} (as the \textit{L} filter has the widest bandpass). In the morning the sequence is reversed. Once the first set of flats is taken in the starting filter (\textit{B} in the evening, \textit{L} in the morning) a new starting exposure time ($E_0$) is calculated based on the relative difference in the filter bandpasses (see \aref{sec:filters}).

This method allows a reasonable set of flat fields to be taken in each filter most nights. The GOTOphoto pipeline (\aref{sec:gotophoto}) creates new master flat frames each month (the same is true of bias and dark frames); this means that taking new flats each night is important but not critical. If, for example, the Moon is too close to the anti-Sun point then flats can be skipped without causing any disruption. The routine also assumes a clear night and does not account for the presence of clouds in the field, however any poor-quality images are rejected by the pipeline when creating the master frames, and by taking new flats twice each night there should always be enough to create a valid master frame each month.

\end{colsection}

\subsection{Focusing the telescopes}
\label{sec:autofocus}
\begin{colsection}

The GOTO unit telescopes are designed to keep a stable focus through the night, and use carbon-fibre trusses to minimise any changes due to temperature fluctuations. Based on images taken during commissioning this is generally true, and the pilot only has to refocus the telescope once at the start of each night. Once the flats routine has finished the night marshal within in the pilot runs the \code{autofocus.py} observing script (see \aref{sec:night_marshal}). To save time, all of the unit telescopes are focused at the same time, although completely independently.

The autofocus routine is based on the V-curve method of \citet{autofocus}, which measures the focus using the \acro{hfd}. The HFD is defined as the diameter of a circle centred on a star in which half of the total flux lies inside the circle and half is outside. As this parameter is based only on the total spread of the flux and not on the maximum peak it is not disrupted due to seeing effects. Importantly, the HFD should vary linearly with focus position, forming a V-shaped curve with a fixed gradient either side of the best focus (unlike the FWHM, which forms a U-shaped, non-linear curve). The gradient of this V-curve (shown in \aref{fig:autofocus}) is a function of the telescope hardware: changes in seeing move the curve up and down while changes in temperature will move the curve side-to-side, but the shape should remain the same. Therefore, if the V-curve has been defined for the given telescope and you can find which part of the curve you are on, you can use the known gradient and intercept to move directly to the minimum point, which will give the best focus.

\begin{figure}[t]
    \begin{center}
        \includegraphics[width=0.8\linewidth]{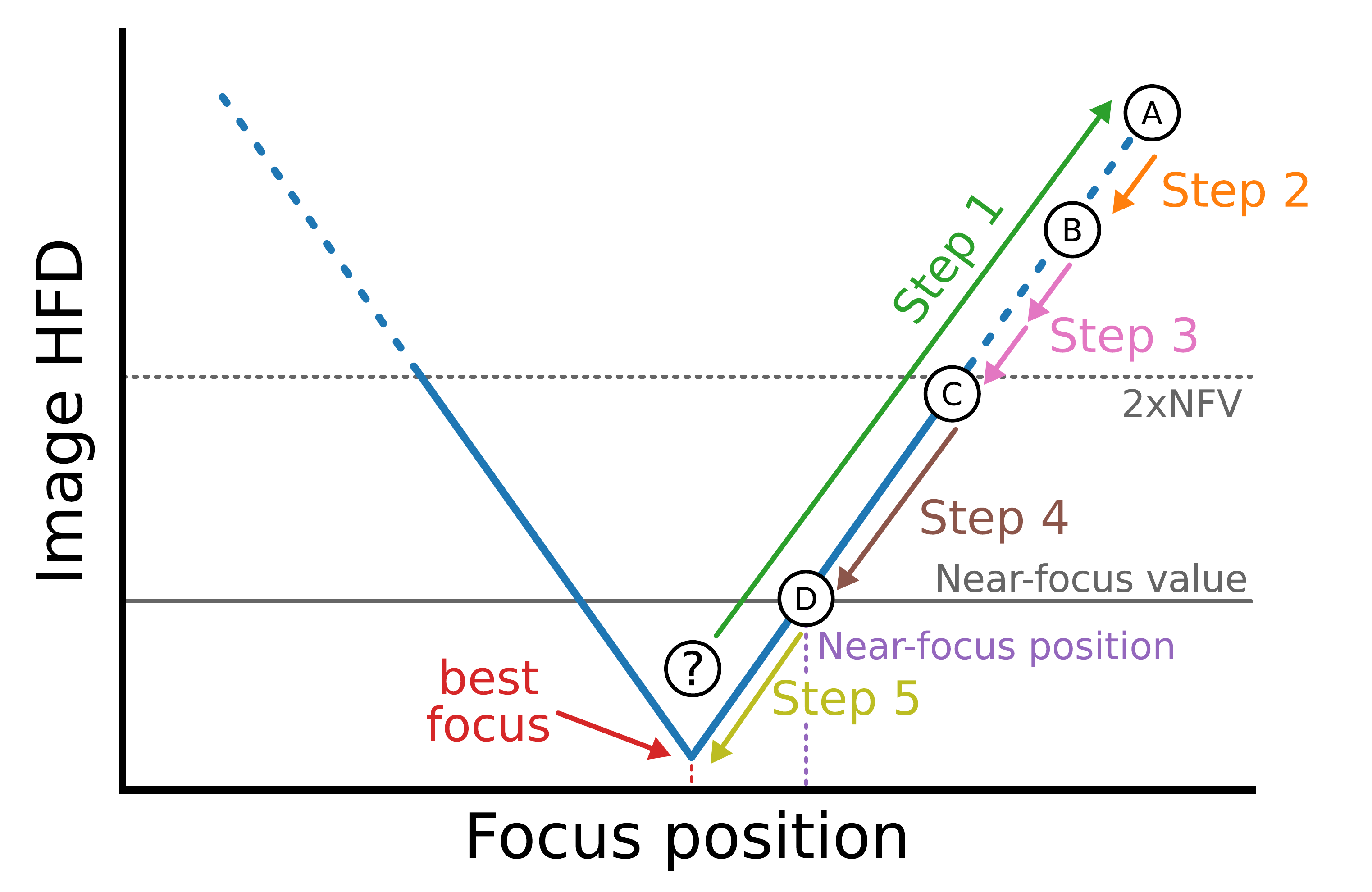}
    \end{center}
    \caption[Steps to find the best focus position using the HFD V-curve]{
        Steps to find the best focus position using the HFD V-curve.
    }\label{fig:autofocus}
\end{figure}

As GOTO is a wide-field instrument no particular focus star needs to be picked, and instead focusing takes place at zenith assuming that there are always enough stars to focus on within the field. Images are windowed to just a central 2000$\times$2000 pixel area, to avoid any distortions around the edge of the field. Sources are extracted and the half flux diameter measured using the SEP Python package (Source Extraction and Photometry, \pkg{sep}\footnote{\url{https://sep.readthedocs.io}}) which implements the Source Extractor algorithms in Python \citep{SE}. This is done for all objects in the frame with a signal of more than $5\sigma$ above the background sky, excluding any sources that do not match a Gaussian shape i.e.\ are not point sources. This typically results in several hundred sources in each unit telescope, and the mean of all of the points is taken as the representative HFD for that image.

At the start of the focus routine an initial image is taken and starting HFD values are measured. The starting focus position is also recorded at this point, so that if the script fails for any reason the initial focus can be restored. It is assumed that the initial value should be fairly close to the ideal focus position, but it is not known which side of the ideal position it is (i.e.\ if it is on the positive or negative gradient side of the V-curve). This starting point is shown by the \textbf{?} marker in \aref{fig:autofocus}.

The first step of the routine is to move the focuser by a large positive quantity, to point \textbf{A} in \aref{fig:autofocus}, and measure the HFD.\@ This is done to make sure that the image is completely de-focused, and we are on a known side of the best focus position. At this point the V-curve might not even be linear, but as long as the measured HFDs have increased compared to the starting value then we can proceed.

The second step is a small step back in the opposite direction, towards the best focus position (point \textbf{B}). The HFD is measured again, and it should now be smaller than when measured at point \textbf{A} but still larger than the starting value. If this is not true then the script returns an error, as it is not possible to determine if we are on the correct side of the V-curve.

The next stage is to continue taking small steps in the same (negative) direction, until the measured HFD is less than double the \emph{near-focus value} (NFV) \acroadd{nfv}. The NFV is chosen for each telescope to be a HFD value in pixels that is approximately equal to the expected best-focus value, based on previous measurements (for GOTO the NFV is 7 pixels). Once at this point (point \textbf{C}) we should be well within the linear portion of the V-curve. At this stage the exact HFD values are important, so three consecutive images are taken at this focus position and the smallest of the HFD values is taken as the first point on the V-curve. The HFD values between images will change due to external factors, such as seeing or windshake, but these will only ever make the HFD worse than the ``true'' value, never better. Therefore taking the minimum reduces the effect of these external factors on the measured HFD values.

Once the HFD value has been well measured at point \textbf{C} then the near-focus position (point \textbf{D}), the position that should produce a HFD equal to the near-focus value, can be found with
\begin{equation}
    F_\text{NF} = F + \frac{\text{NFV} - D(F)}{m_\text{R}}
    \label{eq:nearfocus}
\end{equation}
where $F$ is the current focus position, $D(F)$ is the current HFD and $m_\text{R}$ is the known negative gradient of the right-hand side of the V-curve --- this is just applying the equation of a straight line between two points.

Once the near-focus position has been found the focuser is moved to that position (point \textbf{D}) and the HFD is measured three times again. Now that we have a known $F_\text{NF}$ and $D(F_\text{NF})$ on the right-hand side of the V-curve the best focus position ($F_\text{BF}$) is given by the meeting point of the two lines shown in \aref{fig:autofocus}, which can be calculated using
\begin{equation}
    \begin{split}
                c_1 & = D(F_\text{NF}) - m_\text{R} F_\text{NF}, \\
                c_2 & = m_\text{L}(\frac{c_1}{m_\text{R}} - \delta), \\
        F_\text{BF} & = \frac{c_2 - c_1}{m_\text{R} - m_\text{L}},
    \end{split}
    \label{eq:bestfocus}
\end{equation}
where $m_\text{L}$ and $m_\text{R}$ are the gradients of the lines and $\delta$ is the difference between their intercepts. The focuser is then moved to the best focus position, the HFD values are recorded and the script is complete. This method has proven reliable to focus the GOTO telescopes on a nightly basis, although the optical aberrations produce badly-focused regions in the corners of the frames (described in \aref{sec:timeline}).

\aref{fig:focus_time} shows the focus values (half-flux diameter) measured from every image taken over a single night of observing. The HFD values vary between 3--5 pixels, and although there are some fluctuations there is no clear focus drift over the course of the night. \aref{fig:focus_temp} shows the same values plotted instead as a function of temperature, again no trends are visible. This is just a single sample from one night, and over a longer term there will be shifts in the best focus position. However, refocusing just once in the evening appears to produce a stable-enough position to last through the night.

It has been suggested that future GOTO unit telescopes might use an enclosed, prime-focus tube instead of the current Newtonian design with carbon fibre trusses (see \aref{sec:optics}). Solid metal tubes are much more sensitive to temperature variations and therefore need to continually be refocused during the night. To account for this a refocus step could be added into the exposure queue daemon (see \aref{sec:exq}) at the same stage that the filters are changed.

\begin{figure}[p]
    \begin{center}
        \includegraphics[width=0.95\linewidth]{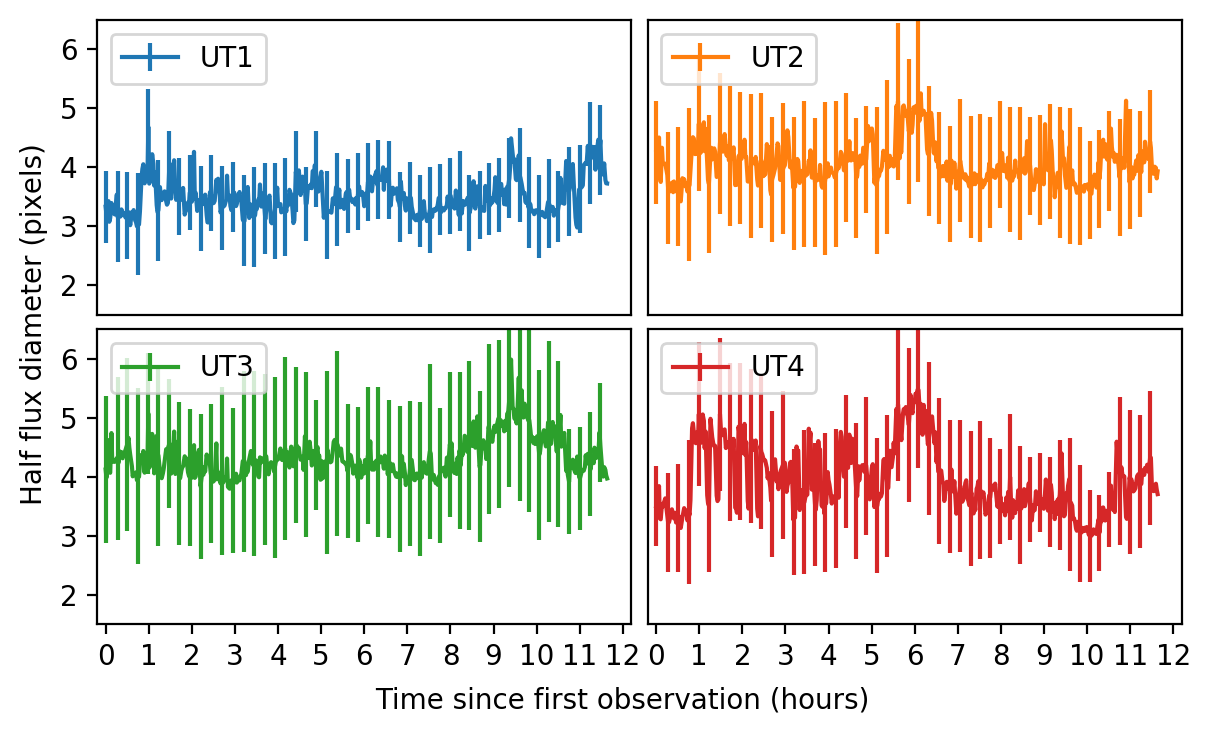}
    \end{center}
    \caption[Measured focus values over a night of observations]{
        Measured focus values (mean half flux diameter) over a night of observations. For clarity error bars are only plotted on every 10th point.
    }\label{fig:focus_time}
\end{figure}

\begin{figure}[p]
    \begin{center}
    \includegraphics[width=0.95\linewidth]{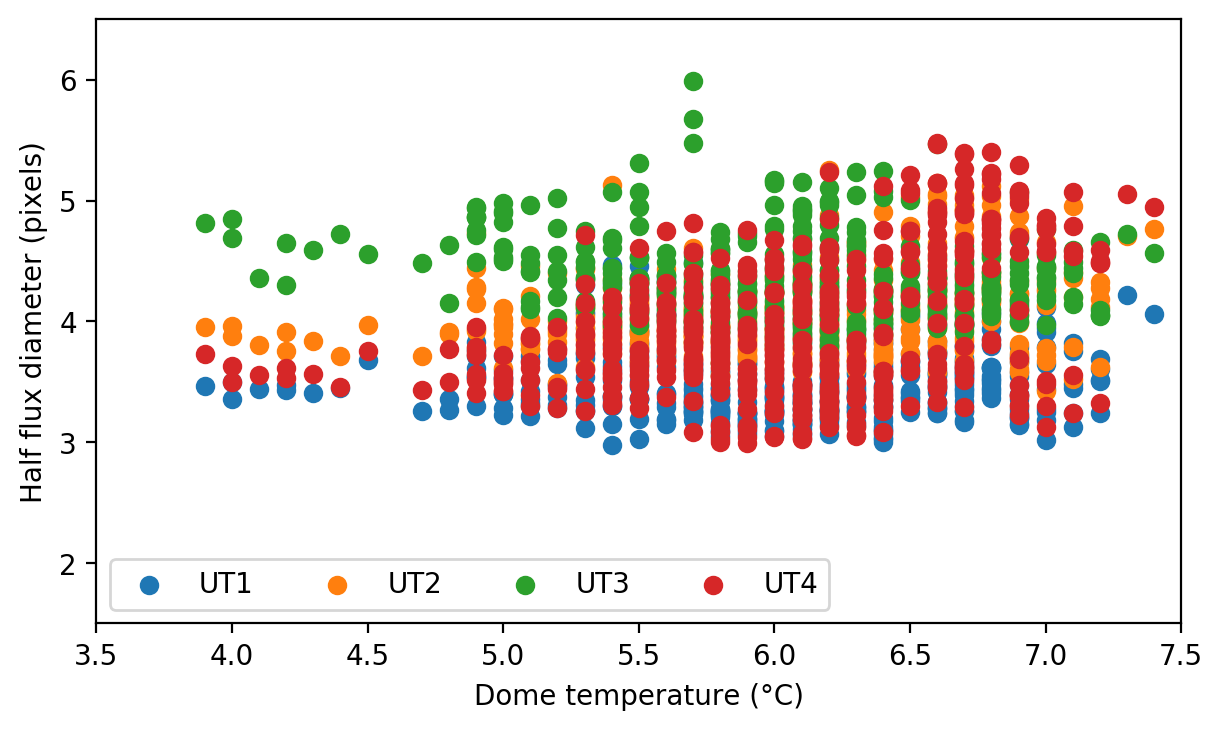}
    \end{center}
    \caption[Measured focus values against temperature]{
        Measured focus values against temperature inside the dome.
    }\label{fig:focus_temp}
\end{figure}

\clearpage

\end{colsection}

\subsection{Mount pointing and stability}
\label{sec:pointxp}
\begin{colsection}

As described in \aref{sec:mount}, using the SiTech-provided mount software (SiTechEXE) required a Windows computer for it to run on, as well as additional development effort to allow the rest of the software to interact with it. However once this was implemented it allowed us to use the various utilities built into SiTechEXE, including the pointing modelling software PointXP.\@ Using this software meant it was not necessary to create our own pointing model within the mount daemon, as once a model is created with PointXP any commands sent to SiTechEXE have the model applied before slewing.

In order to create a pointing model using PointXP the camera output from one of the telescopes needs to be connected to the Windows NUC running the software, and the rest of the G-TeCS software must be disabled to ensure PointXP has full control. The software creates a grid of equally spaced pointings at a range of altitudes and azimuths, as shown on the sky chart in \aref{fig:pointing_model}, then takes CCD images at each position, extracts the position of sources in each frame, and calculates the pointing model transformations to best convert requested coordinates to mount axis positions.

\begin{figure}[t]
    \begin{center}
        \includegraphics[width=\linewidth]{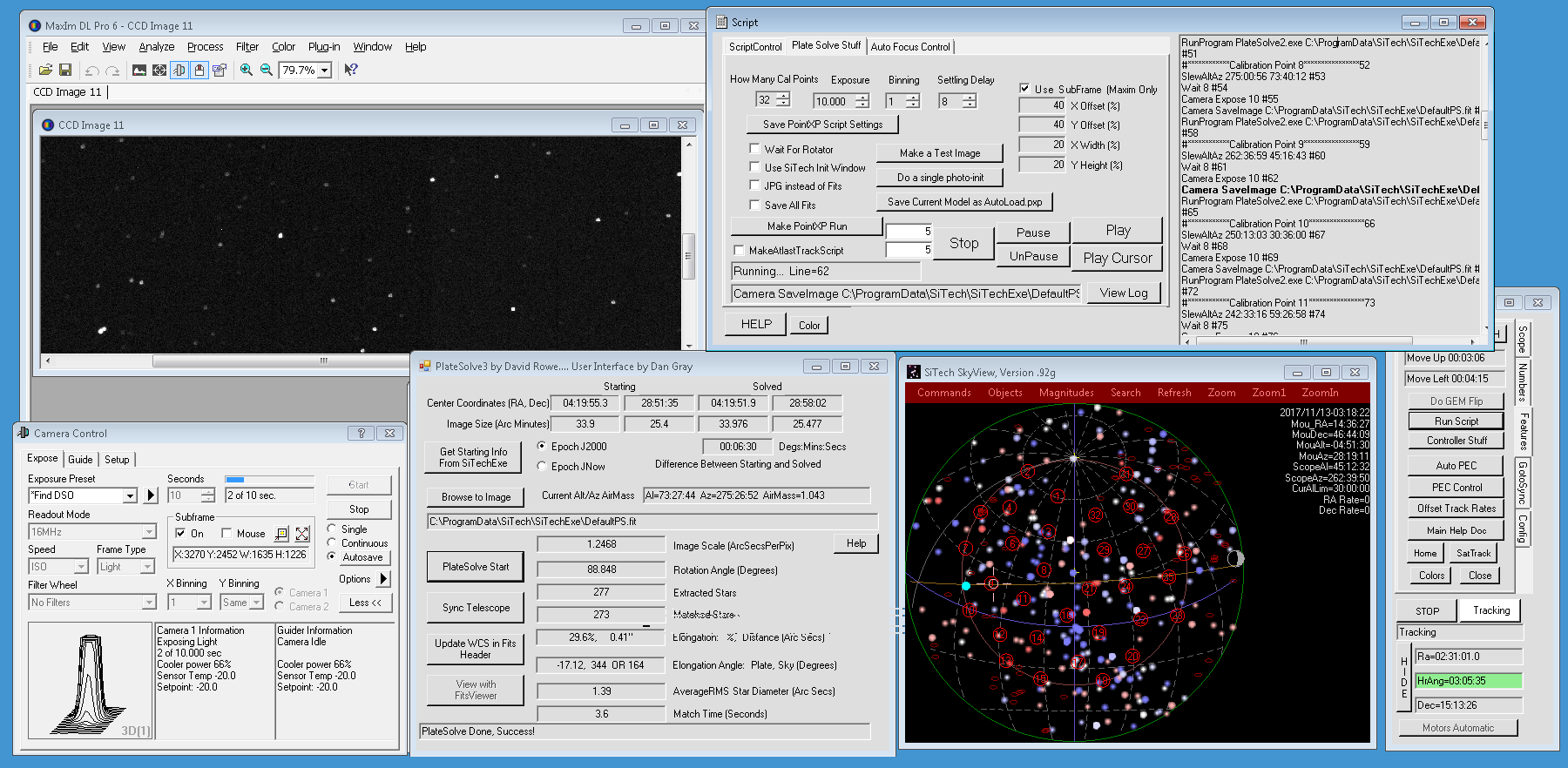}
    \end{center}
    \caption[Creating a pointing model using PointXP]{
        Creating a pointing model using PointXP.\@
    }\label{fig:pointing_model}
\end{figure}

One of the complications when creating a pointing model for GOTO is that the model can only be based on the output of a single unit telescope (as PointXP only expects a single camera input). When aligned into the 3-UT configuration, as shown in \aref{fig:3ut_footprint}, it was simple to create the model using the central telescope (UT4), but in the 4-UT configuration (\aref{fig:4ut_footprint}) this is not an option. The chosen camera needs to physically be disconnected from the interface NUC and connected to the Windows mount NUC PointXP is running on, which prevents creating a pointing model unless there is someone on-site. One suggestion has been to add a small guide telescope in the centre of the array which could be permanently connected to the Windows NUC PointXP is on, this could then be used to base the pointing model on.

\begin{figure}[t]
    \begin{center}
        \includegraphics[width=\linewidth]{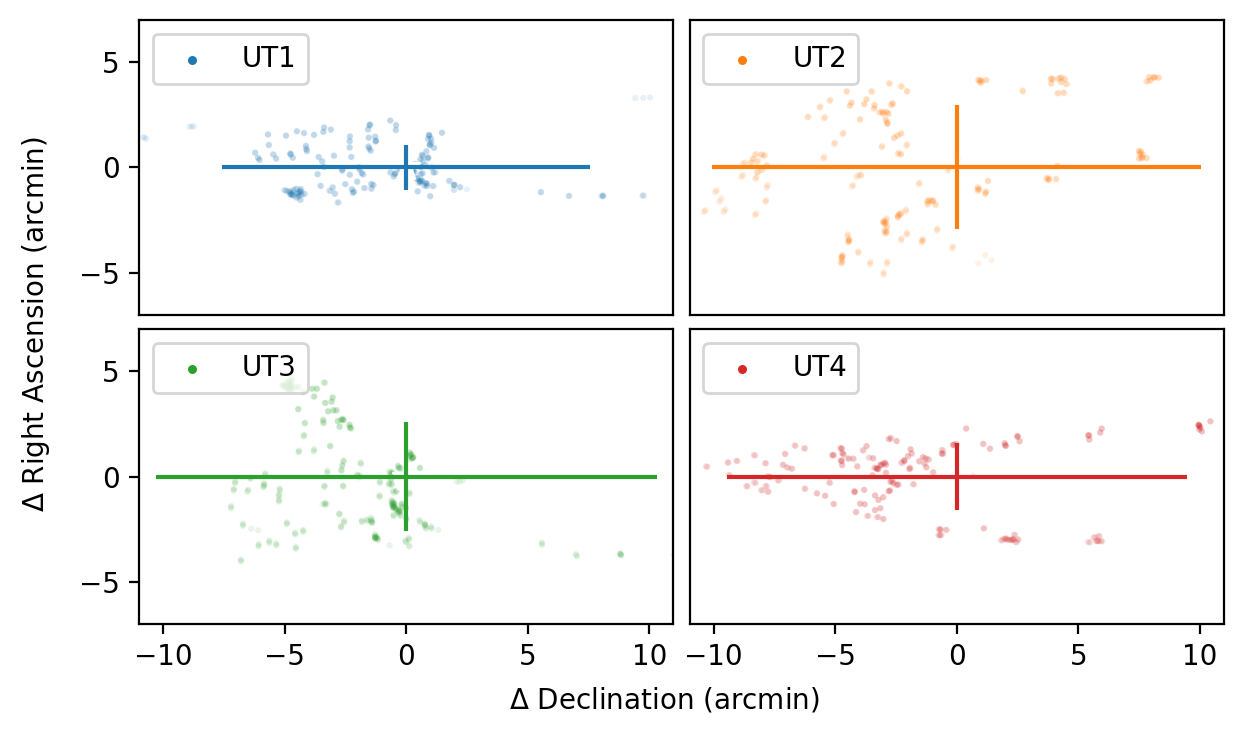}
    \end{center}
    \caption[Pointing errors over a single night]{
        Pointing errors (the difference between the target and actual image positions) taken from a single night of observations, 489 exposures in total. The error bars for each UT show the standard deviation of points in each axis.
    }\label{fig:pointing}
\end{figure}

\begin{figure}[t]
    \begin{center}
        \includegraphics[width=\linewidth]{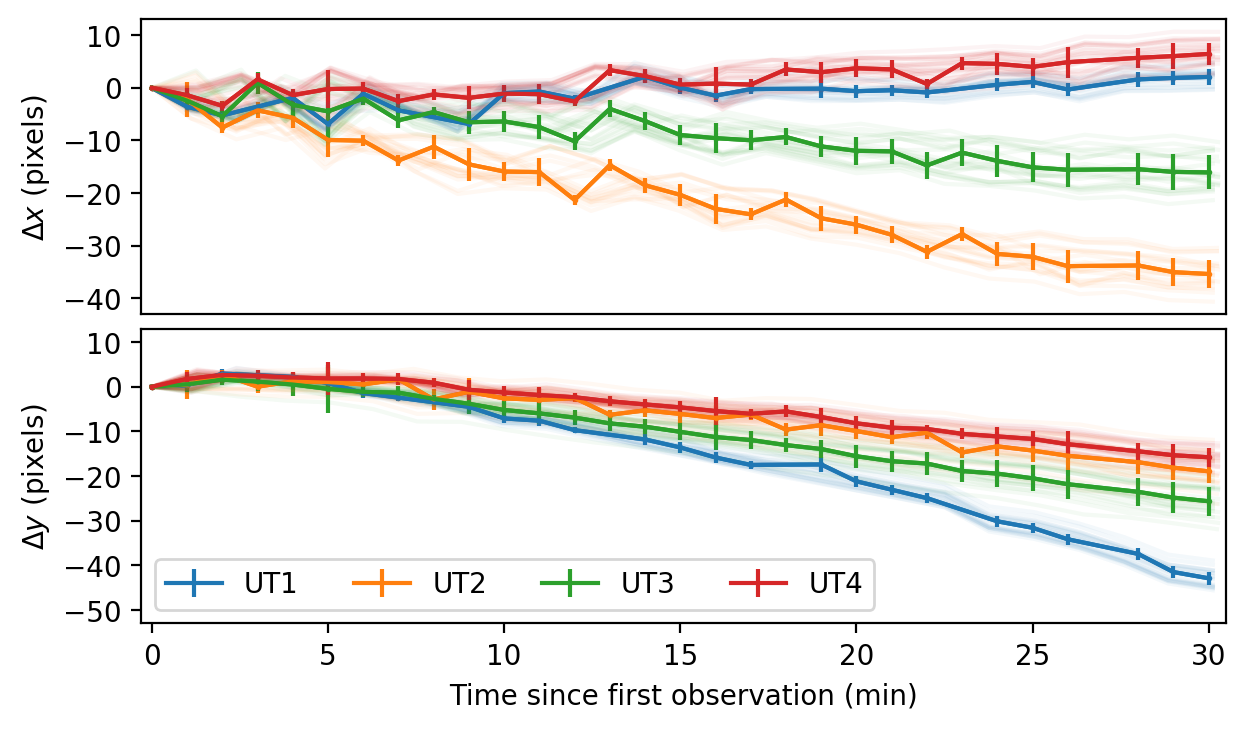}
    \end{center}
    \caption[Position drift over a 30 minute observation]{
        Position drift of sources in a series of images taken over 30 minutes. The darker solid lines show the average of all the sources detected in each UT.\
    }\label{fig:tracking}
\end{figure}

\aref{fig:pointing} shows the pointing error for each unit telescope from all images taken in a single night of normal observations after creating a new pointing model with PointXP, at various different altitudes and azimuths. The error is found as the difference between the target position reported by the mount and the actual image centre found by the GOTOphoto pipeline. As each UT has a unique offset each subplot is centred on the mean offset for that UT.\ The pointing errors vary between 8--\SI{10}{\arcmin} in right ascension and 1--\SI{3}{\arcmin} in declination, with clear variations between the unit telescopes. The pattern of points for each unit telescope also appears to be unconnected. This illustrates one of the major complications due to the GOTO mount design: each unit telescope will flex and shift slightly on their own mounts attaching them to the boom arm, in addition to the pointing of the overall mount. The original brackets mounting the unit telescopes to the boom arm were much worse and occasionally allowed unit telescopes to drift several degrees from their original position. These were replaced in July 2018, and since then the models created using PointXP have been perfectly good for use with GOTO.\@ In practice even being a few arcminutes off the desired pointing is minor compared to GOTO's large field of view, and can easily be accounted for when the images are processed by the GOTOphoto pipeline.

GOTO does not include an autoguiding system, although the proposed guide scope mentioned above could be used as one. In the absence of this the telescope must be able to track accurately. The initial mount drives suffered from tracking problems, and were very sensitive to any imbalances in the weight distribution. The motors were replaced in November 2017, and since then have been much more reliable, although it is still important to ensure that the mount is balanced. \aref{fig:tracking} shows the drift of sources taken from a series of exposures of the same target over 30 minutes, revealing a maximum drift of approximately 80 pixels/hour, or \SI{1.65}{\arcmin} (using the plate scale of \SI[per-mode=symbol]{1.24}{\arcsec\per\pixel}). Again each unit telescope has a slightly different drift in different directions (for example UT1 is very stable in the $x$ direction (right ascension) but has the biggest drift in $y$ (declination)). In practice GOTO usually switches targets every few minutes, so the long-term tracking performance over several hours is not a major concern. GOTO also typically only takes \SI{60}{\second} exposures, and no significant trailing is seen in these images.

\end{colsection}

\subsection{Other commissioning challenges}
\label{sec:challenges}
\begin{colsection}

In this section I outline a few of the changes that had to be made to the software based on experience with the hardware. This is not an exhaustive list, but gives some examples of the challenges that are typical when commissioning a facility such as GOTO.\@

\subsubsection{Filter wheel serial numbers}

One of the hardware issues that was identified early on concerned identifying the filter wheels when they were connected to the interface NUCs. The usual way to connect to specific hardware units through the FLI-API code is to search the connected USB devices for their unique FLI serial numbers (for example the serial numbers of the GOTO cameras given in \aref{tab:cameras}). However, the initial set of CFW9--5 filter wheels delivered to us by FLI did not have serial numbers defined in their firmware. Two filter wheels are connected to each interface NUC, and this problem made it impossible to tell between them or send a command to a particular filter wheel.

A solution was eventually found using the pyudev Python package (\pkg{pyudev}\footnote{\url{https://pyudev.readthedocs.io}}), which uses the Linux udev device manager to identify devices using the \code{/dev/} name, and therefore create a pseudo-serial number based on which port each device is connected to. Using this method it is possible to tell filter wheels apart as long as which physical USB port each is plugged into is known. This is not an ideal solution, but as long as the USB cables remain connected it is not an issue even if the NUCs are rebooted.

\subsubsection{Downloading images from the cameras}

One of the more complicated parts of the camera control software is reading images from the FLI cameras. Once an exposure has finished, photo-electrons from the CCD are read out and stored as counts in a memory buffer on the camera, where they can be downloaded by USB.\@ A last-minute change led to the GOTO cameras using new, larger detectors than originally designed for the cameras, which led to there not being enough space in the camera buffer to store a full-frame image. This was discovered when corrupted images such as the one shown in \aref{fig:cam_readout} were being produced; the regions in the lower third of the image are just duplicates of the data in the upper third, meaning the original data in this section was lost.

The cameras return a \code{DataReady} status once the exposure has finished and the data is reading out. However, when installing G-TeCS on site it was clear that the camera daemon was not able to reliably start downloading the data from the cameras quickly enough to clear space in the internal buffer before it starts being overwritten. The solution was to add an internal image queue within the camera class, which would immediately begin downloading from the cameras as soon as they reported the exposure was finished. This means the camera data is stored within the memory on the interface NUCs, and then the camera daemon queries the interface (see \aref{sec:fli}) to download the image across the network and write it to disk.

\begin{figure}[t]
    \begin{center}
        \includegraphics[width=0.7\linewidth]{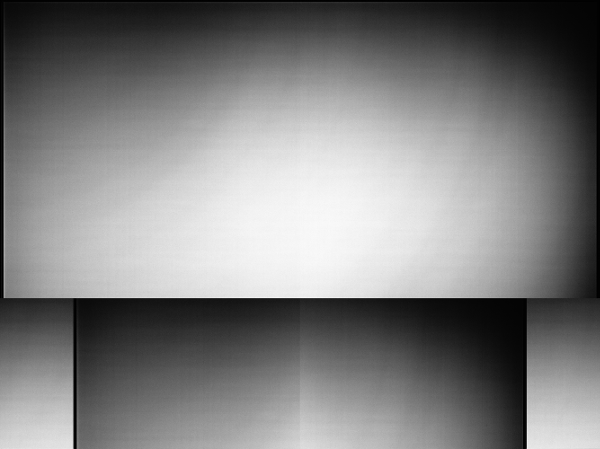}
    \end{center}
    \caption[A corrupted image which was not read out fast enough]{
        An example of a corrupted image from one of the FLI MicroLine cameras which was not read out fast enough.
    }\label{fig:cam_readout}
\end{figure}

This solution did run into a few problems with a feature of the Python programming language called the \acro{gil}, which prevents multiple threads accessing the same Python object at once (the exact workings of the GIL are outside of the scope of this thesis). In short, this prevented reading out the cameras to the internal queue in parallel, which added an extra delay. Luckily, the FLI-API package was not written in standard Python (technically CPython) but in Cython, which interfaces between the FLI SDK written in C and the rest of the control system written in Python. Cython contains a GIL to maintain compatibility with CPython, however it is not required and can be disabled. Doing this allowed the cameras to be read out in parallel as intended.

\newpage

\subsubsection{Declination axis encoder failure}

Just a few weeks after the inauguration the mount declination axis encoder failed, preventing any automated slewing in the declination axis (the telescope could still be moved manually with the hand-pad). Of the two axes this was by far the better one to fail, had the RA axis failed instead then the telescope would not have been able to track and so no observations could have been taken. Instead, the telescope could at least still take on-sky images, and commissioning the optics and the camera control software was able to continue by manually moving the mount.

However, the SiTech control software was not able to cope with the disabled declination axis, which meant that the telescope could not be operated in robotic mode. When sent a command to slew to a position, the mount would move to the correct RA but would never reach the target (as it could not move in declination) and therefore would not start tracking. Even sending commands to move only in RA (i.e.\ keeping the same declination position) did not work. The mount would reach the correct position but the declination encoder would never register reaching the target, so the slew was never registered as `complete' and the mount would not start tracking.

A workaround was therefore coded into the mount daemon: a separate thread which monitored the RA position and stopped the mount moving when the target RA coordinates were reached, regardless of the declination coordinates. This forced the SiTech slew command to reset, meaning it could start tracking. When this modification was in place GOTO was able to observe `normally', and was able to carry out a survey in a limited declination band of the sky. Had an important gravitational-wave alert come through during this period, the person on-site would have had to move the telescope to the correct declination and then manually carry out observations. Thankfully this was not needed, and, as described in \aref{sec:timeline}, once O2 ended GOTO was shut down until the motors could be replaced.

\subsubsection{Dome movement}

The Astrohaven clamshell dome is driven by internal belts attached to the dome shutters. It is important when moving the dome not to put undue stress on these belts, as should one of them dislodge or snap there is nothing preventing the dome from falling open. As described in \aref{sec:dome}, the dome motors are deliberately moved in short bursts rather than continually when opening. This prevents the shutters being pulled down too fast, which can cause the upper shutter to fall and put excess stress on the belts.

As mentioned in \aref{sec:arduino}, the dome has also occasionally opened past its limits when the in-built switches fail to trigger, meaning the motors drive the shutters into the ground. The extra limit switches we installed provide a backup in order to cut the motors when they are triggered, and also give a method to detect when the shutters overshoot and let the dome daemon move the shutters back up.

One of the more serious hardware problems occurred a few days after I left La Palma in February 2018. During freezing conditions, ice had built up on the upper dome shutter, and eventually was heavy enough to partially pull the shutter open past its limits, exposing the telescope to the elements (see \aref{fig:ice_internal}). I was able to remotely move the dome and drag the shutter closed, but a large amount of ice fell into the dome. Luckily, Vik Dhillon and Stu Littlefair were still at the observatory, along with Tom Marsh, from Warwick, and my replacement GOTO monitor, Tom Watts from Armagh. As shown in \aref{fig:ice_external}, they were able to get up the mountain to clear ice from inside and outside the dome, as well as place a tarpaulin over the telescope.

Once informed about the event, Astrohaven manufactured brackets to fit inside the dome to prevent the upper shutter from being forced open in this way. The dome falling open was accompanied by a sharp drop in the internal temperature, which was the motivation to add the \code{internal} and \code{ice} flags into the conditions monitor (see \aref{sec:conditions}), to alert us should similar conditions occur in the future.

\begin{figure}[p]
    \begin{center}
        \includegraphics[width=0.45\linewidth]{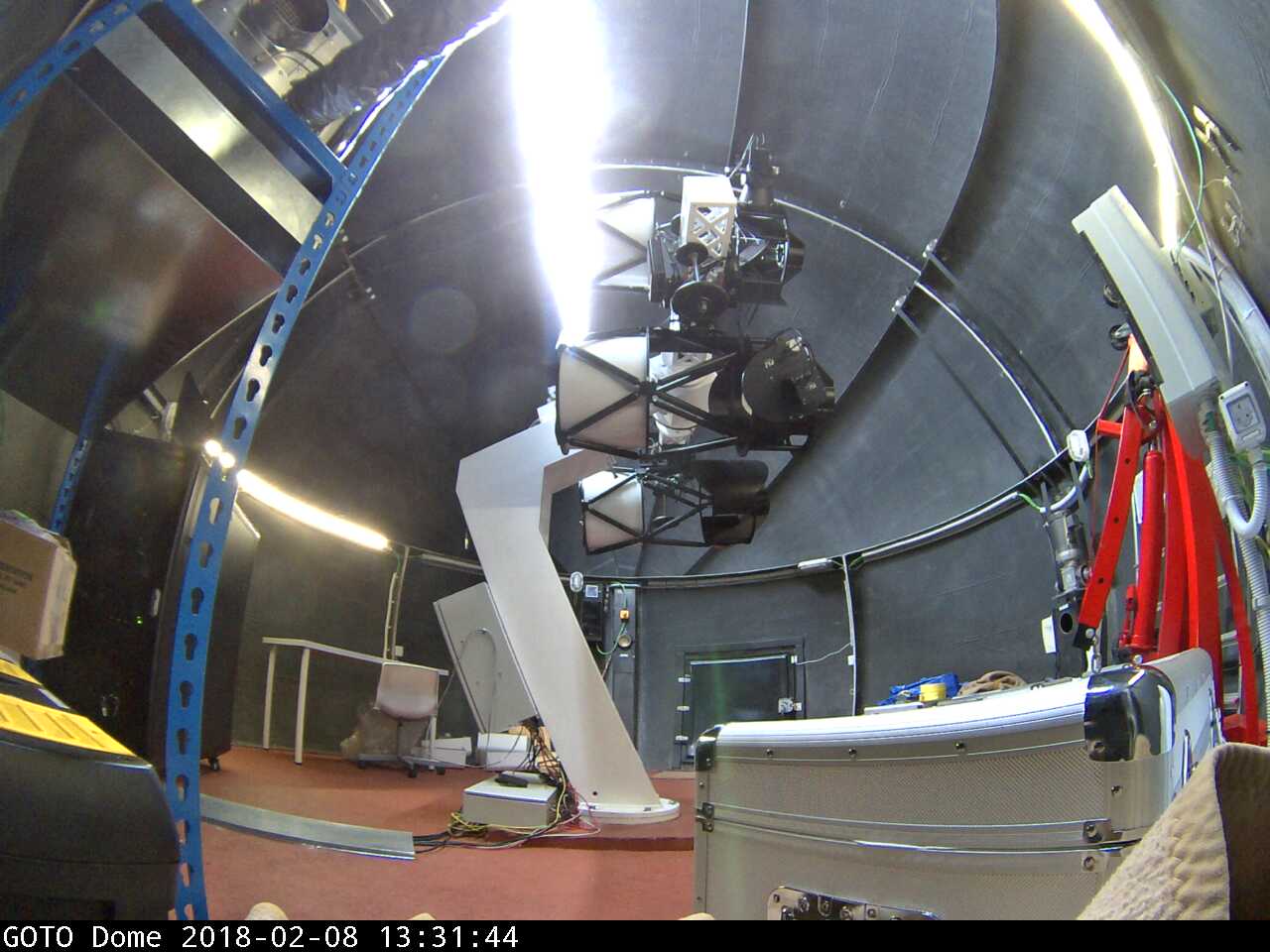}
        \includegraphics[width=0.45\linewidth]{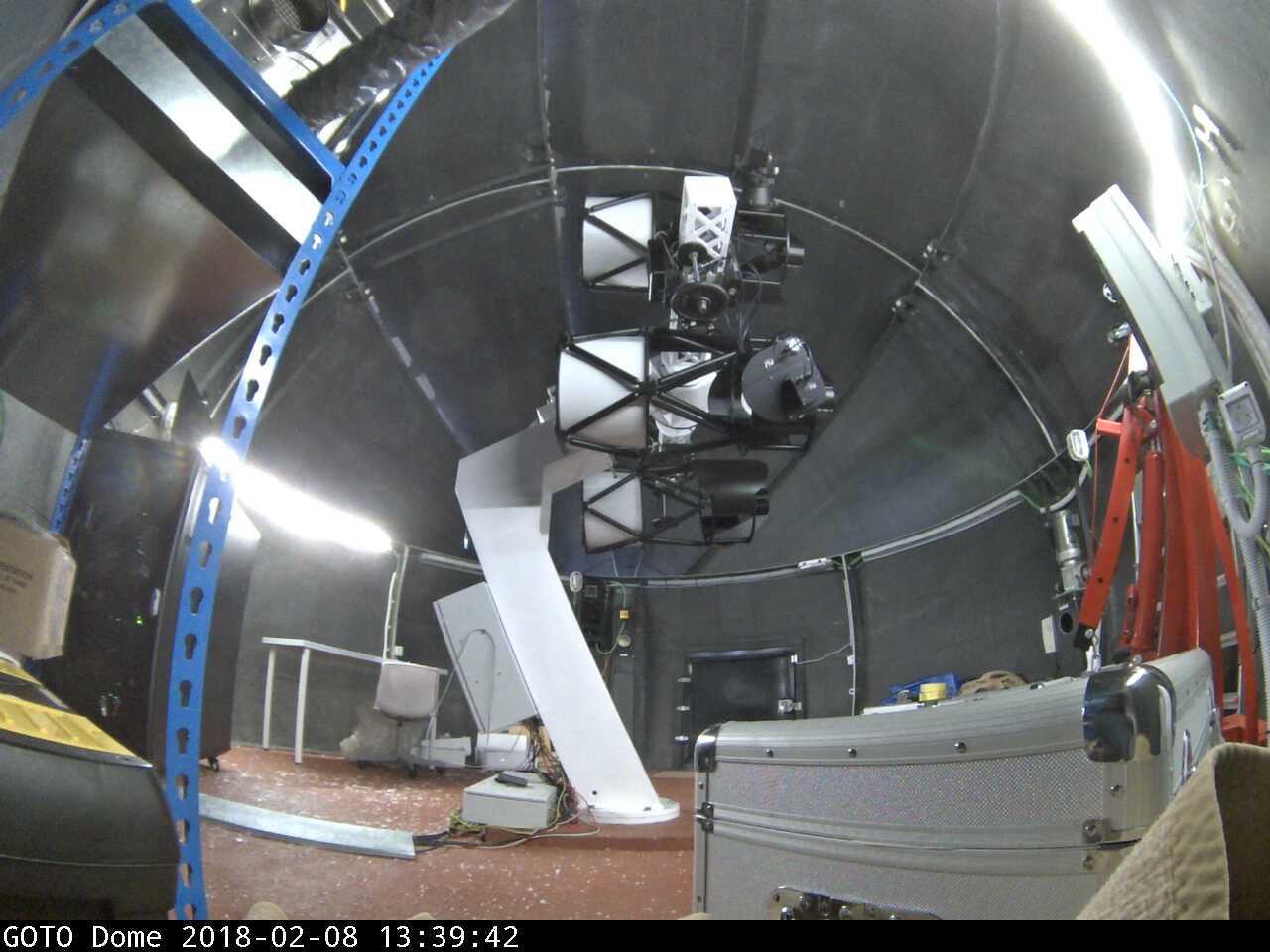}
    \end{center}
    \caption[Internal webcam images showing the dome open during a snowstorm]{
        Internal webcam images showing the dome open during the 2018 snowstorm. The image on the left was taken when opening was discovered, with the upper shutter (which normally closes on the south side, to the left of the image) having been open by the weight of ice built up on the north side. The image on the right was taken after closing the shutter remotely. Moving the dome caused a large amount of ice to dislodge and fall into the dome, thankfully missing the mirrors and camera hardware.
    }\label{fig:ice_internal}
\end{figure}

\begin{figure}[p]
    \begin{center}
    \includegraphics[width=0.88\linewidth]{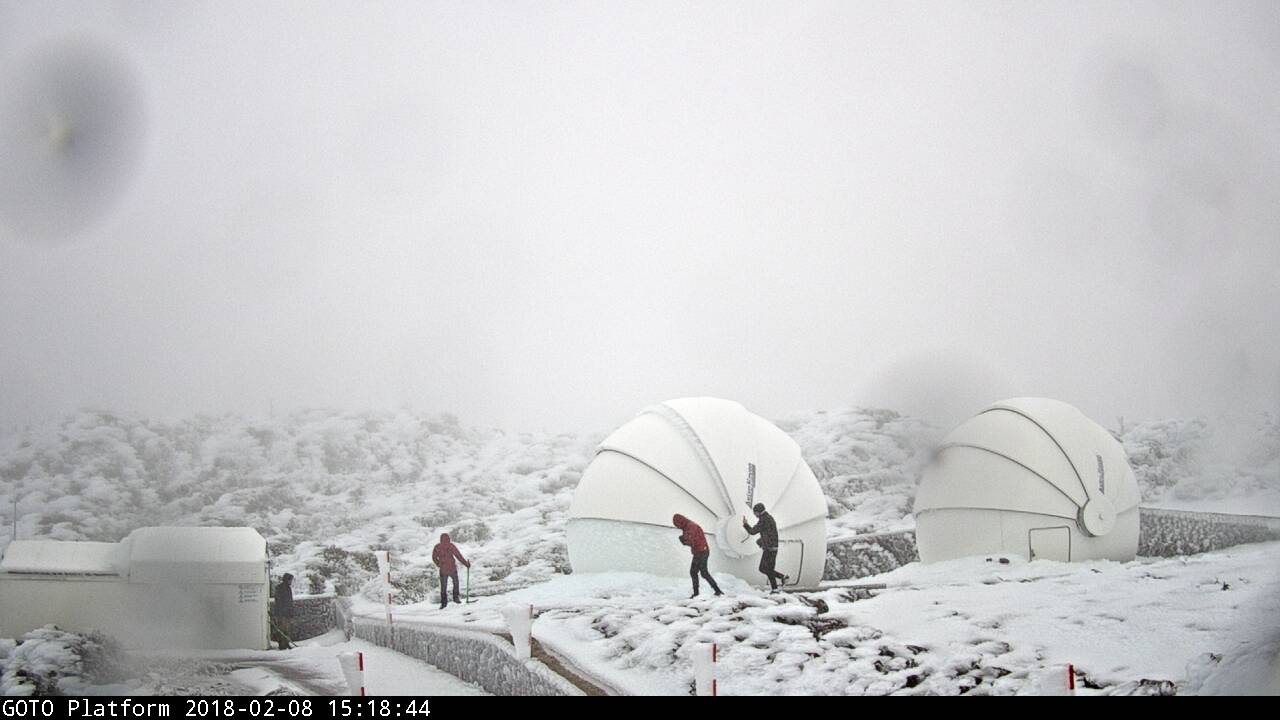}
    \end{center}
    \caption[External webcam image showing the ice rescue team]{
        External webcam image showing the ice rescue team. Note the build up of ice visible on the northern side of the upper shutter of the (empty) left-hand dome. A similar build up caused the upper shutter on the right-hand dome containing GOTO to be pulled open.
    }\label{fig:ice_external}
\end{figure}

\clearpage

\subsubsection{Errors processing GW alerts}

Ensuring the stability of the GOTO-alert event handling system described in \aref{chap:alerts} was a high priority when commissioning the GOTO software in the run-up to the third LIGO-Virgo observing run (O3). The LVC began producing test GCN notices in December 2018 prior to the beginning of the run, which included mock skymaps similar to the ones used for the simulations in \aref{sec:scheduler_sims} and \aref{sec:gw_sims}. These events allowed a full system test of the GOTO-alert code, from the VOEvent being received to the pointings being added to the observation database.

Since the start of O3 in April 2019 until the end of August there were 32 gravitational-wave alerts, of which seven were ultimately retracted, and the G-TeCS sentinel (see \aref{sec:sentinel}) received each event and processed it using the GOTO-alert event handler code. A full run-down of the response to each event is given in \aref{sec:gw_results}. A few complications did arise during O3 which required changes to the software, mostly due to problems at the LVC's end. These are outlined below.

\begin{itemize}
    \item Initially each notice sent out by the LVC had to be approved manually by a LIGO-Virgo member, which lead to some delays to follow-up observations being triggered. The initial alert for S190421ar was delayed until several hours after the event, even though the skymap had already been uploaded by the LVC to the GraceDB service. A possible addition to the G-TeCS sentinel was proposed, a separate thread that could query GraceDB to check for new skymaps. If it detected on the skymap could quickly be processed to start GOTO observing, even before the ``official'' notice was sent out. However, since the first few events the delay between the gravitational-wave detection and the notice being sent out has been much shorter, and this modification was never implemented.
    \item An updated skymap for the S190426c event was uploaded to GraceDB by the LVC with the wrong permissions, causing the sentinel to raise an error when it could not download it from the URL in the notice. Unfortunately the event handler (see \aref{sec:event_handler}) had already deleted the existing pointings from the observation database before crashing, preventing GOTO from observing the previous tiles. After this event the order of functions in the event handler was changed, so that existing pointings were only removed once the new skymap had been downloaded and processed.
    \item The second skymap for the S190521g event was initially uploaded in an uncommon HEALPix format (see \aref{sec:healpix}) used internally in the LVC, which could not be read by GOTO-tile. Due to the above changes made following the S190426c event this did not interrupt GOTO observations, and the LVC have since clarified their filetypes.
    \item Finally, the updated skymap for the S190814bv event was incorrectly uploaded by the LVC to GraceDB with the same filename (\code{bayestar.fits.gz}) as the initial skymap (typically updates are called \code{bayestar1.fits.gz}, etc). By default, the Astropy FITS download function used within the sentinel caches each file, and if asked to download a file from the same URL will instead use the cached version. This lead to GOTO continuing to observe the large initial skymap instead of focusing on the smaller region given in the updated map. Again, the LVC have said that this will be prevented in the future, but just in case the sentinel was patched to disable the caching feature.
\end{itemize}

\end{colsection}

\section{Summary and Conclusions}
\label{sec:commissioning_conclusion}

\begin{colsection}

In this chapter I described work carried out during the GOTO commissioning period on La Palma.

The GOTO prototype suffered several delays before finally being deployed in the summer of 2017. After that a series of hardware issues and failures lead to several elements being replaced, in particular two of the sets of mirrors. The final full prototype with four unit telescopes started reliable operations in February 2019, in time for the start of the third LIGO-Virgo observing run.

Amongst the hardware problems I installed, commissioned and developed the control software as described in the previous chapters. The primary G-TeCS hardware control systems (\aref{chap:gtecs}) were primarily developed before and during the delay in deployment, in particular I built and integrated several hardware units in the dome to ensure the safety of the telescope and any operators on site. The rest of the commissioning period was focused on developing the autonomous systems (\aref{chap:autonomous}), until in May 2018 the telescope was trusted to operate entirely robotically without full-time supervision. The G-TeCS software has proven itself to be reliable, and should provide a framework to build upon as GOTO expands.

\end{colsection}

\chapter{A Multi-Telescope Observatory}
\label{chap:multiscope}

\chaptoc{}

\section{Introduction}
\label{sec:multiscope_intro}

\begin{colsection}

In this chapter I describe the potential future expansion of the GOTO project, with additional telescopes at the current site on La Palma and a future second site in Australia.
\begin{itemize}
    \item In \nref{sec:multi_tel} I give an outline of the additional work required to create a multi-site scheduling system, and how the existing simulation code can be modified to approximate the required functionality.
    \item In \nref{sec:gw_sims} I describe simulations showing the benefits of additional telescopes observing gravitational-wave alerts in order to locate the counterpart source.
    \item In \nref{sec:survey_sims} I describe further simulations detailing the effects of additional telescopes on the all-sky survey.
\end{itemize}
All work described in this chapter is my own unless otherwise indicated, and has not been published elsewhere.

\end{colsection}

\section{Scheduling for multiple telescopes}
\label{sec:multi_tel}

\begin{colsection}

As described in \aref{sec:goto_expansion}, the ultimate aim of the GOTO project is to have multiple nodes around the world. Specifically, the plan calls for two full GOTO-8 systems on La Palma and another two at a second site in Australia (either at Siding Spring Observatory in New South Wales or Mt Kent Observatory in Queensland). It is anticipated that the G-TeCS scheduling system described in \aref{sec:observing} will be extended to cover all these telescopes, so that they each query a single observation database and a master scheduler decides what target each telescope should be observing at a given time. This will require a large amount of work to modify both the database structure and the scheduling functions and, as this is not currently implemented into the existing scheduler, several workarounds are needed in order to create realistic multi-telescope simulations.

\end{colsection}

\subsection{Multiple observing telescopes}
\label{sec:multi_tel_scheduling}
\begin{colsection}

One of the current restrictions in the scheduling functions (as described in \aref{chap:scheduling}) is that they only ever expect a single pointing in the observation database to be marked as \code{running} at any one time. It is explicitly coded into the scheduler that detecting multiple running pointings should raise a critical error, as certain bugs early in development could lead to this undesired state to occur. Obviously once the system is to be expanded to multiple telescopes this restriction will have to be lifted, but for running simulations a simplification was required to work around it.

It is currently planned that each telescope will have its own pilot and hardware daemons completely independent of each other, with the only point of overlap being the shared scheduler (and, for each site, the conditions daemon). This makes the master scheduler even more complicated, as each pilot will be querying it completely out-of-sync. If telescope 1 has just finished observing and makes a scheduler check, the scheduler will need to know what telescope 2 is observing, so as not to return the same pointing to telescope 1 (although in some cases having both telescopes observe the same target might be desired, adding yet another level of complexity). But should both telescopes finish observing at the same time then the scheduler will need some way to decide which telescope is assigned which target, perhaps based on the slew time to each target from the telescope's current position.

As none of the above has yet been implemented into the existing code, a simplified system was required in order to simulate multiple telescopes. The existing fake pilot code (described in \aref{sec:goto_sims}) already contains calls to the real scheduling functions, which return the highest priority pointing at given time. The first simplification was to make the function instead return the top $N$ highest pointings, where $N$ is the number of currently observing telescopes. In lieu of any better algorithm to decide which telescope observes which target, the code simply gives the highest priority pointing to telescope 1, the second highest to telescope 2, and so on. Should there only be one valid pointing returned then only telescope 1 will observe, while telescope 2 will remain ``parked'' until it is needed (in reality the second telescope would default to observing the all-sky survey until it also has something to do).

The second simplification was to ensure the telescopes always stay in sync when observing. This was achievable for the simulations described in this chapter because every pointing uses the same exposure set (three \SI{60}{\second} exposures), and therefore they take the same amount of time to observe. However, in reality each telescope would take a different amount of time to slew to its target, and so they would quickly get out of sync. Slew time is included in the fake pilot code for each telescope to acquire its new target, and so in order to remain synchronised with multiple telescopes the simulations simply wait the required amount of time for the telescope with the furthest distance to slew. This ensures both telescopes start and finish their observations at the same time, although it does mean a small amount of observing time is ``wasted'' while one telescope is waiting for the other to be in position.

\newpage

\end{colsection}

\subsection{Multiple observing sites}
\label{sec:multi_site_scheduling}
\begin{colsection}

The modifications to the scheduler described in the previous section provide a good approximation of the response of an arbitrary number of telescopes observing at one site. However, expanding the code further to simulate observations from multiple sites adds further complexity.

The scheduler functions (see \aref{sec:ranking}) need to know which site observations are being made from in order to correctly sort pointings. The visibility constraints (see \aref{sec:constraints}) check if each target is above the local horizon, as well as the local Sun altitude, and the tiebreak parameter (see \aref{sec:breaking_ties}) takes into account the airmass of each target. These are simple parameters to calculate if you are only observing from one site, but once there are telescopes at multiple sites querying the scheduler at the same time then the responses will need to take the position of each into account.

This could lead to problems when returning the highest priority pointings. For example, with two telescopes observing from different sites the scheduler could return the highest priority pointing visible from each. If they are different then each telescope can then observe the best target for its site; however if both telescopes were observing at the same time, and the visible portions of the sky from both sites overlapped, then it is very possible that the same pointing would be the highest priority from both sites. Assuming they should not both observe the same target at the same time, the scheduler would need to choose which telescope to assign that pointing to and then recalculate a different target for the other telescope. What would be better is to use the same method described in \aref{sec:multi_tel_scheduling}, and have the scheduler always return the top $X$ pointings, where $X = N_\text{site1} + N_\text{site2} + \cdots$ is the total number of telescopes across the globe. In reality targets would need to be assigned to telescopes based on their airmass at each site, or the slew time from the current target, but as in \aref{sec:multi_tel_scheduling} for the simulations they can just be assigned to each telescope in order.

\newpage

\begin{figure}[t]
    \begin{center}
        \includegraphics[width=\linewidth]{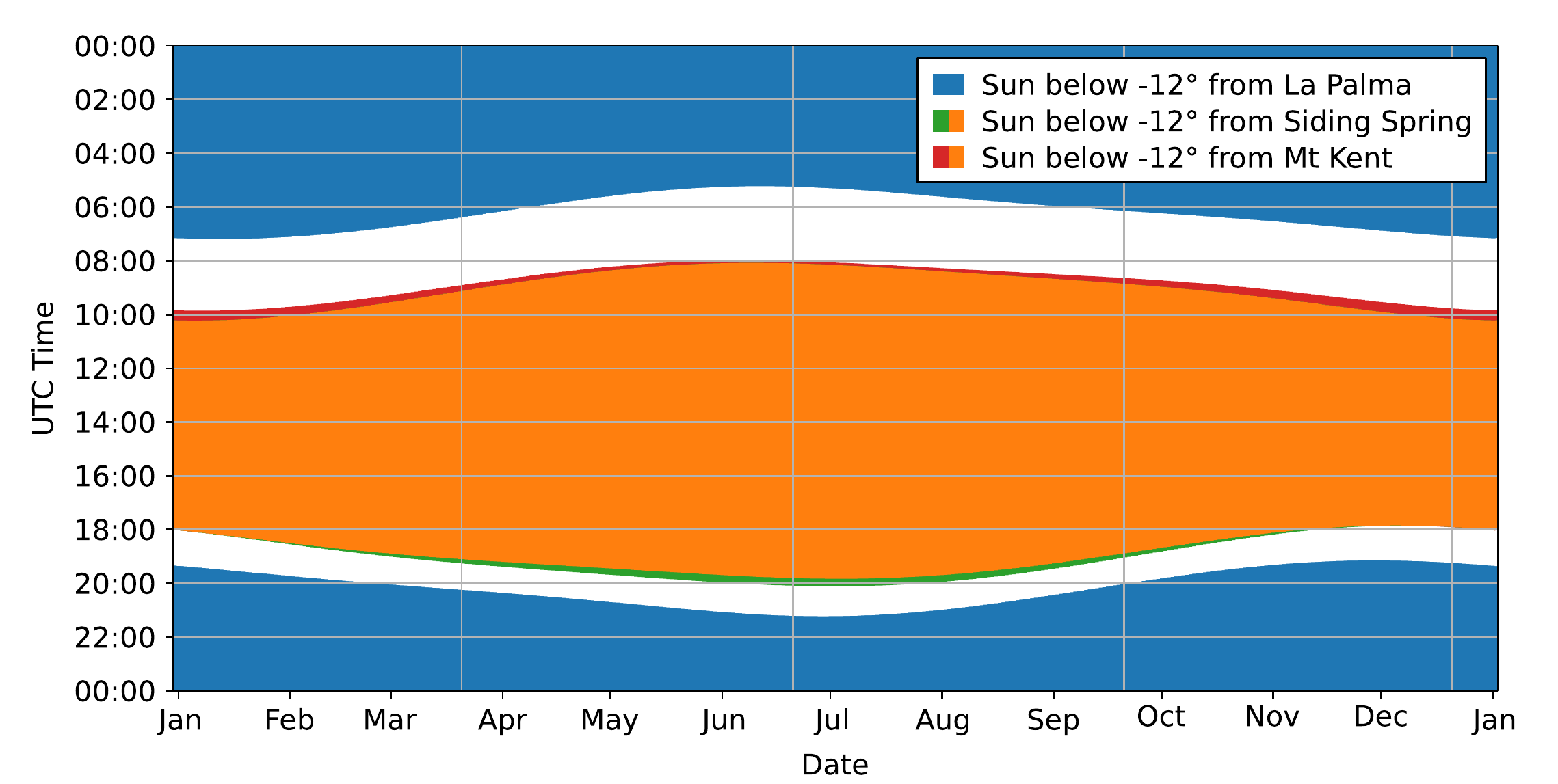}
    \end{center}
    \caption[Night times throughout the year for GOTO sites]{
        Night times throughout the year for GOTO sites. Night here is defined as when the Sun is \SI{12}{\degree} below the local horizon.
    }\label{fig:nights}
\end{figure}

However, saying that the scheduler needs to find the top $X$ pointings, where $X$ is the total number of telescopes at all sites, is not strictly true --- it actually only needs to return enough pointings to satisfy the telescopes at the sites that are currently observing. In other words, if there are two sites but one is shut down, due to weather or because it is daytime there, the scheduler only needs to consider the single site. Conveniently, for simulating the proposed GOTO network this is always true: by defining night as when the Sun is below \SI{-12}{\degree} altitude, the periods of darkness between La Palma and either of the two proposed Australian sites never overlap. This is shown in \aref{fig:nights}, where there is a constant ``buffer zone'' between night ending at one site and beginning at the other. This case only applies for a very limited number of combinations of sites. As shown in \aref{fig:site_nights} there is a tear-drop-shaped area on the Earth's surface which contain the locations where the local night will never overlap with night on La Palma, comprising only of eastern Australia, New Zealand and Melanesia. For a Sun altitude limit of \SI{-12}{\degree} this area contains just $6.6 \%$ of the Earth's surface.

\begin{figure}[p]
    \begin{center}
        \includegraphics[width=\linewidth]{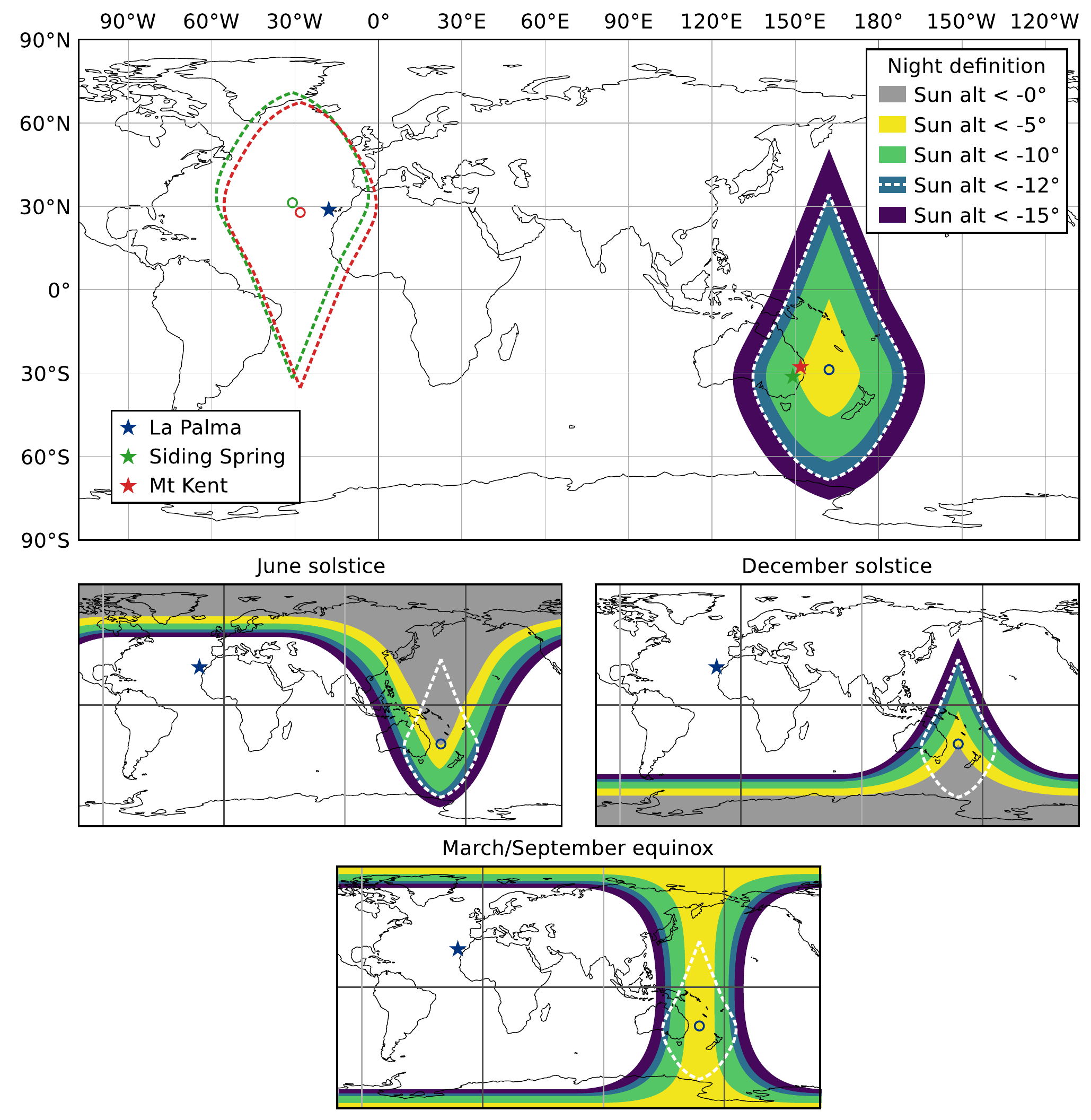}
    \end{center}
    \caption[Locations on the Earth with non-overlapping night times]{
        Finding locations on the Earth with non-overlapping night times. In the upper plot the filled areas show the locations on the Earth where local night never overlaps with night on La Palma, at any time of the year. The different colours denote different night time definitions, with the \SI{-12}{\degree} definition also being surrounded by a white dashed line. The location of the GOTO sites considered (\textcolor{NavyBlue}{La Palma}, \textcolor{Green}{Siding Spring} and \textcolor{Red}{Mt Kent}) are marked with stars and their antipodes are marked with hollow circles. The equivalent areas for Siding Spring and Mt Kent are also shown by the coloured dashed lines surrounding their antipodes, for the \SI{-12}{\degree} night definition only. The lower plots show how the region varies over the course of a year, from the solstices via the equinox (the plot is identical for the two equinoxes and therefore is only shown once).
    }\label{fig:site_nights}
\end{figure}

The fortuitous location of the proposed Australian sites means implementing telescopes at multiple sites into the simulation code was fairly simple. Each simulation would consider either only La Palma, or La Palma and one of the Australian sites, as these are the only anticipated scenarios for GOTO.\@ In the two-site cases only the telescopes at one of the sites would ever be observing at any one time, and so the simulation only requires the multi-telescope implementation as described in \aref{sec:multi_tel_scheduling}.

Should simulations be desired for other sites on the Earth that are not within the small area with non-overlapping nights shown in \aref{fig:site_nights}, for example other potential GOTO-South sites in South Africa or Chile, then the simulation code would need to be modified to take this into account. This has not yet been done as it was not required for the scenarios described here. In principle, the fact the sites overlap could be ignored, and the simulations can be run for each site as a stand-alone observatory and then combined afterwards with the results from other, stand-alone simulations. This, however, removes the benefit of the sites acting together and using a common observation database, and would lead to multiple observations of the same targets from each site.

\end{colsection}

\subsection{Simulating different survey grids}
\label{sec:multi_grid_scheduling}
\begin{colsection}

One fundamental feature of the existing G-TeCS code is that observations are carried out on a fixed all-sky grid, as defined in \aref{chap:tiling}. When considering multiple telescopes this is both useful in some ways and limiting in others. Having a fixed grid that is common to all telescopes is vital for the GOTO image subtraction pipeline GOTOphoto, as it requires observations of the same part of the sky to create reference frames for difference imaging (see \aref{sec:gotophoto}). This is why a common grid is anticipated to form the base of the global system. By sharing the same tiles each telescope can contribute to the same all-sky survey grid, as well as efficiently coordinate mapping out a gravitational-wave skymap.

\newpage

However, sharing the grid requires all of the telescopes to have essentially the same field of view. There is some leeway in the exact field of view of each telescope array; the grid tiles are defined to leave a slight overlap around the edge (see \aref{fig:4ut_footprint}). But if the field of view of the telescope array is much larger than the tile size then the pointings will be too close together and therefore inefficient. Even worse, if the field of view of the telescope array is much smaller than the defined tile size it would lead to gaps in the sky coverage.

For the proposed GOTO system with near-identical GOTO-8 units around the world this is not an issue, but it should be recognised as a limitation of not just the simulations but the whole G-TeCS control system. One potential case where this may be an issue is when commissioning GOTO-South. If it spends time as a GOTO-4 system similar to La Palma before getting the second set of unit telescopes to bring it up to a full set of eight, then it will be observing concurrently with one or two GOTO-8 systems on La Palma. This is a likely enough situation that it was considered in the gravitational-wave simulations as described in \aref{sec:gw_sims}, using the workaround of two independent simulations mentioned previously. How this scenario would be dealt with within a real implementation of G-TeCS is a problem that needs development in the future, should it prove to be necessary.

\end{colsection}

\section{Gravitational-wave follow-up simulations}
\label{sec:gw_sims}

\begin{colsection}

As the primary mission of the GOTO project is to follow up gravitational-wave detections, it is important to consider what benefit additional telescopes will bring to the project. In order to do this, simulations were run on the LIGO First Two Years mock skymaps \citep{First2Years}, a small selection of which were previously used for the scheduler simulations described in \aref{sec:scheduler_sims}. The full sample contained 1105 events, each based on simulating a binary neutron star coalescence at a particular sky position and distance. Each event had two skymaps generated: the first using the rapid BAYESTAR pipeline \citep{BAYESTAR}, which is typically available minutes after the event, and the second using the LALInference code \citep{LALInference}, which can take hours or days to complete. For these simulations, therefore, only the BAYESTAR skymaps were considered in order to focus on GOTO's initial follow-up, although an extension to the simulations could include the effects of the second updated skymap being processed and added to the database some hours after the event.

\end{colsection}

\subsection{Event visibility}
\label{sec:gw_visability}
\begin{colsection}

The simulations were designed to begin at the time the event was detected, and then simulate the next 24 hours of observations. This guaranteed one night's worth of observing at each site, although split into two halves if the event occurred during the night. The time each event occurred was taken from the simulated skymaps, and does not account for the delay between the event being detected and the alert being issued and processed by the G-TeCS sentinel. Events were uniformly distributed in time of occurrence during the day, and they all occurred over a two month period spanning either side of the 2010 September equinox as shown in \aref{fig:f2y_times}. It is not clear why this range of dates was selected, although surrounding one of the equinoxes might have been an attempt to reduce bias towards observers from either hemisphere. However, the events are not entirely equally distributed either side of the equinox (03:09 UTC on 2010--09--23): the first event occurs 33 days before the equinox and the last 27 days after. Overall 64\% of events occurred before the September equinox and 36\% after, which leads to a slight bias in visibility towards southern telescopes as they experience longer nights before the equinox (in the southern winter).

\begin{figure}[t]
    \begin{center}
        \includegraphics[width=\linewidth]{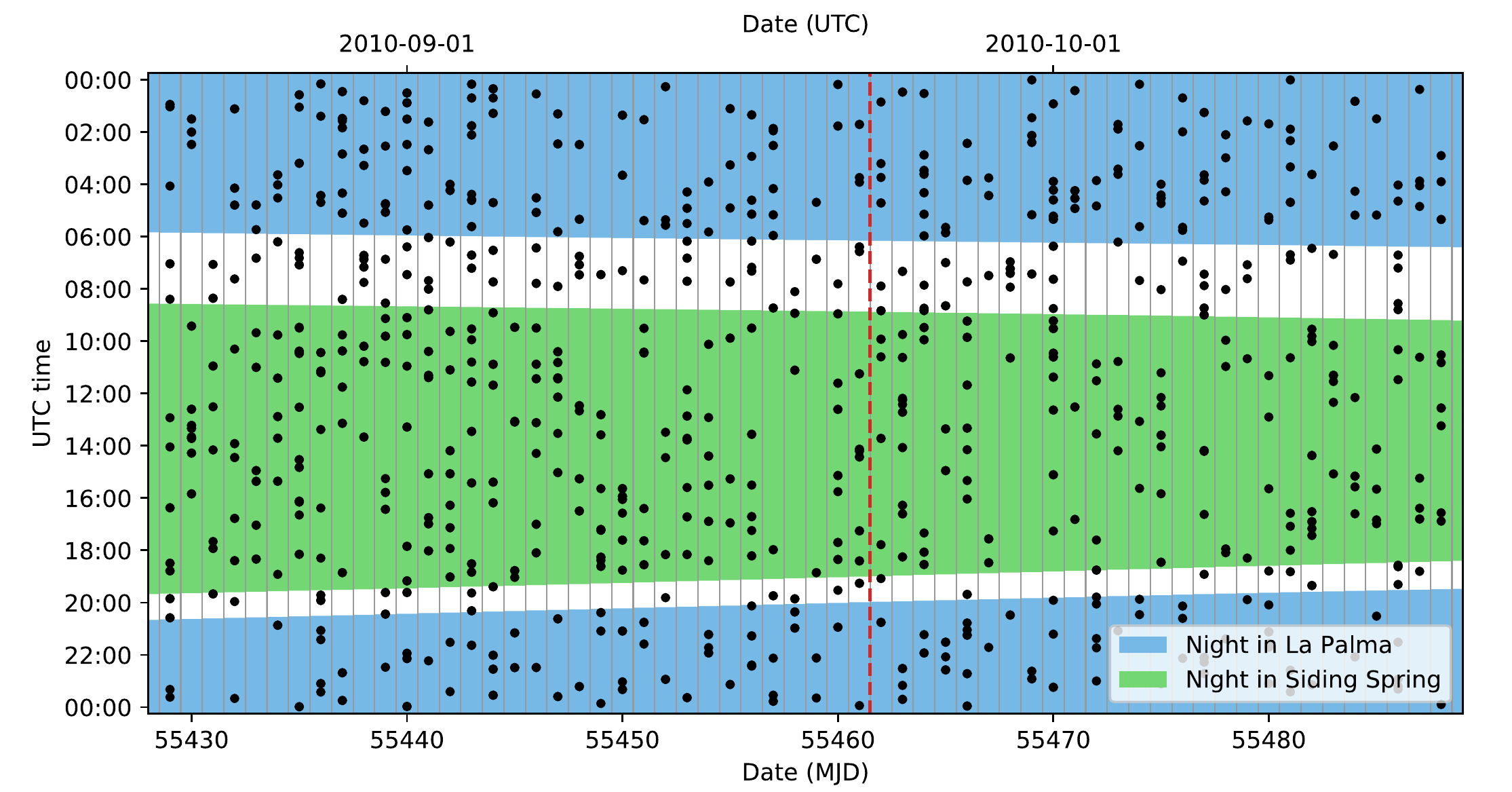}
    \end{center}
    \caption[Date and time distribution of events in the First Two Years sample]{
        Date and time distribution of the ``First Two Years'' events. The night periods are shown in \textcolorbf{NavyBlue}{blue} for La Palma and \textcolorbf{Green}{green} for Siding Spring, and the date of the equinox is marked by the \textcolorbf{Red}{red} dashed line.
    }\label{fig:f2y_times}
\end{figure}

Events were uniformly distributed across the sky, and are uniform in distance cubed \citep{First2Years}. Although each source included distance information this was not taken into account in the simulations, aside from determining the event strategy to use (all were well within the \SI{400}{\mega\parsec} definition for close neutron star events defined for GOTO-alert in \aref{sec:event_strategy}). Future simulations could use the distance to the event to estimate a light curve based on the observed kilonova for GW170817 \citep{GW170817_followup} and use it to predict how long each event would be visible for GOTO for (the GW170817 transient AT~2017gfo faded below GOTO's 20 mag limit after 2.5 days).

\newpage

The first stage of simulating observations of each event was to determine if the source location was visible from the chosen site(s) in the 24 hours after the event occurred. In the cases where this was not true there was no point running the full simulation, as it the source would never be observed. Each event was classified into one of four categories: %

\begin{itemize}
    \item \textbf{Not visible --- too close to the Sun}. These events had sources that were too close to the Sun to observe within the 24 hour period after the event, regardless of the site considered. This was defined as the source location being within \SI{42}{\degree} of the Sun (\SI{-12}{\degree} from the definition of night time plus \SI{30}{\degree} from the altitude limit). It is a fixed fraction of the sky: 5280 sq deg, or 13\% of the celestial sphere\footnote{The area of a circle with radius $r$ on the surface of a sphere with radius $R$ is $2\pi R^2(1-\cos(r))$. The radius of the celestial sphere $R=\SI{360}{\degree}/2\pi \approx \SI{57.3}{\degree}$.}.

    \item \textbf{Not visible --- below declination limit}. The sources for these events fell within the region of the sky that is never visible due to the limited declination range visible from a given site. For example, using the GOTO \SI{30}{\degree} altitude limit a telescope on La Palma (latitude \SI{28}{\degree} N) can see a band of sky between \SI{+88}{\degree} and \SI{-32}{\degree} declination. Sources outside of this region (that are not already excluded due to being too close to the Sun) would therefore never be observable from the site, but could be observed from other locations. At the equator this band covers 87\% of the sky over the course of a year, at latitudes of $\pm \SI{30}{\degree}$ 75\% of the sky is visible, falling to just 25\% at the poles\footnote{The area of a segment on a sphere between angles $\theta$ and $\phi$ is $2 \pi R^2 (\cos(\theta)-\cos(\phi))$.}.

    \item \textbf{Not visible --- daytime}. These event sources are within the visible declination range, but are not observable from a given site during the 24 hour period after the event as they are only above the horizon during the day. Unlike the fraction of the sky within the circular \SI{42}{\degree} region around the Sun, these positions could still be observable from other sites at different latitudes.

    \item \textbf{Visible}. The source for this event falls outside of either of the above three areas, and therefore is nominally above the \SI{30}{\degree} altitude limit at some point during night time within 24 hours after the event. The portion of the sky that is visible in one night from a given site depends on the latitude of the site and the time of year.
\end{itemize}

\begin{table}[t]
    \begin{center}
        \begin{tabular}{c|ccc} %
            \multirow{3}{*}{Night} & \multicolumn{3}{c}{Site} \\
                      & La Palma             & Siding Spring  & Mt Kent \\
                      & (\SI{28}{\degree} N) &  (\SI{31}{\degree} S) &  (\SI{27}{\degree} S) \\
                      \midrule
                      \\
            March     & \textcolorbf{Green}{57.1\% visible}
                      & \textcolorbf{Green}{56.3\% visible}
                      & \textcolorbf{Green}{59.6\% visible}
                      \\
            equinox   & {\scriptsize(\textcolorbf{Orange}{12.9\%} $\cdot$
                                     \textcolorbf{NavyBlue}{23.4\%} $\cdot$
                                     \textcolorbf{Blue}{6.7\%})}
                      & {\scriptsize(\textcolorbf{Orange}{12.9\%} $\cdot$
                                     \textcolorbf{NavyBlue}{24.6\%} $\cdot$
                                     \textcolorbf{Blue}{6.2\%})}
                      & {\scriptsize(\textcolorbf{Orange}{12.9\%} $\cdot$
                                     \textcolorbf{NavyBlue}{22.5\%} $\cdot$
                                     \textcolorbf{Blue}{5.0\%})}
                      \\[0.5cm]
            June      & \textcolorbf{Green}{50.4\% visible}
                      & \textcolorbf{Green}{61.9\% visible}
                      & \textcolorbf{Green}{62.7\% visible}
                      \\
            solstice  & {\scriptsize(\textcolorbf{Orange}{12.9\%} $\cdot$
                                     \textcolorbf{NavyBlue}{24.2\%} $\cdot$
                                     \textcolorbf{Blue}{12.4\%})}
                      & {\scriptsize(\textcolorbf{Orange}{12.9\%} $\cdot$
                                     \textcolorbf{NavyBlue}{20.9\%} $\cdot$
                                     \textcolorbf{Blue}{4.3\%})}
                      & {\scriptsize(\textcolorbf{Orange}{12.9\%} $\cdot$
                                     \textcolorbf{NavyBlue}{19.0\%} $\cdot$
                                     \textcolorbf{Blue}{5.4\%})}
                      \\[0.5cm]
            September & \textcolorbf{Green}{57.0\% visible}
                      & \textcolorbf{Green}{56.4\% visible}
                      & \textcolorbf{Green}{57.7\% visible}
                      \\
            equinox   & {\scriptsize(\textcolorbf{Orange}{12.9\%} $\cdot$
                                     \textcolorbf{NavyBlue}{23.4\%} $\cdot$
                                     \textcolorbf{Blue}{6.7\%})}
                      & {\scriptsize(\textcolorbf{Orange}{12.9\%} $\cdot$
                                     \textcolorbf{NavyBlue}{24.5\%} $\cdot$
                                     \textcolorbf{Blue}{6.2\%})}
                      & {\scriptsize(\textcolorbf{Orange}{12.9\%} $\cdot$
                                     \textcolorbf{NavyBlue}{22.4\%} $\cdot$
                                     \textcolorbf{Blue}{7.0\%})}
                      \\[0.5cm]
            December  & \textcolorbf{Green}{62.5\% visible}
                      & \textcolorbf{Green}{48.9\% visible}
                      & \textcolorbf{Green}{51.0\% visible}
                      \\
            solstice  & {\scriptsize(\textcolorbf{Orange}{12.9\%} $\cdot$
                                     \textcolorbf{NavyBlue}{19.7\%} $\cdot$
                                     \textcolorbf{Blue}{4.8\%})}
                      & {\scriptsize(\textcolorbf{Orange}{12.9\%} $\cdot$
                                     \textcolorbf{NavyBlue}{25.8\%} $\cdot$
                                     \textcolorbf{Blue}{12.4\%})}
                      & {\scriptsize(\textcolorbf{Orange}{12.9\%} $\cdot$
                                     \textcolorbf{NavyBlue}{23.2\%} $\cdot$
                                     \textcolorbf{Blue}{12.9\%})}
                      \\
        \end{tabular}
    \end{center}
    \caption[Sky visibility over a year]{
        Sky visibility over a year from the three different GOTO sites. The upper value in \textcolorbf{Green}{green} shows the fraction of the sky that is visible during the night. The lower values break down the remaining fraction of the sky into the three non-visible categories: too close to the Sun in \textcolorbf{Orange}{orange}, below the declination limit in \textcolorbf{NavyBlue}{light blue} and only visible during the day in \textcolorbf{Blue}{dark blue}.
    }\label{tab:visibility}
\end{table}

The region of the sky visible during the night for a given site changes over the course of the year. \aref{tab:visibility} shows the fractions of the sky in each of the four categories above at the solstices and equinoxes. \aref{fig:visibility} plots the regions on the celestial sphere, in order to better visualise how they change depending on observing site and time of year.

\begin{figure}[p]
    \begin{center}
        \includegraphics[width=\linewidth]{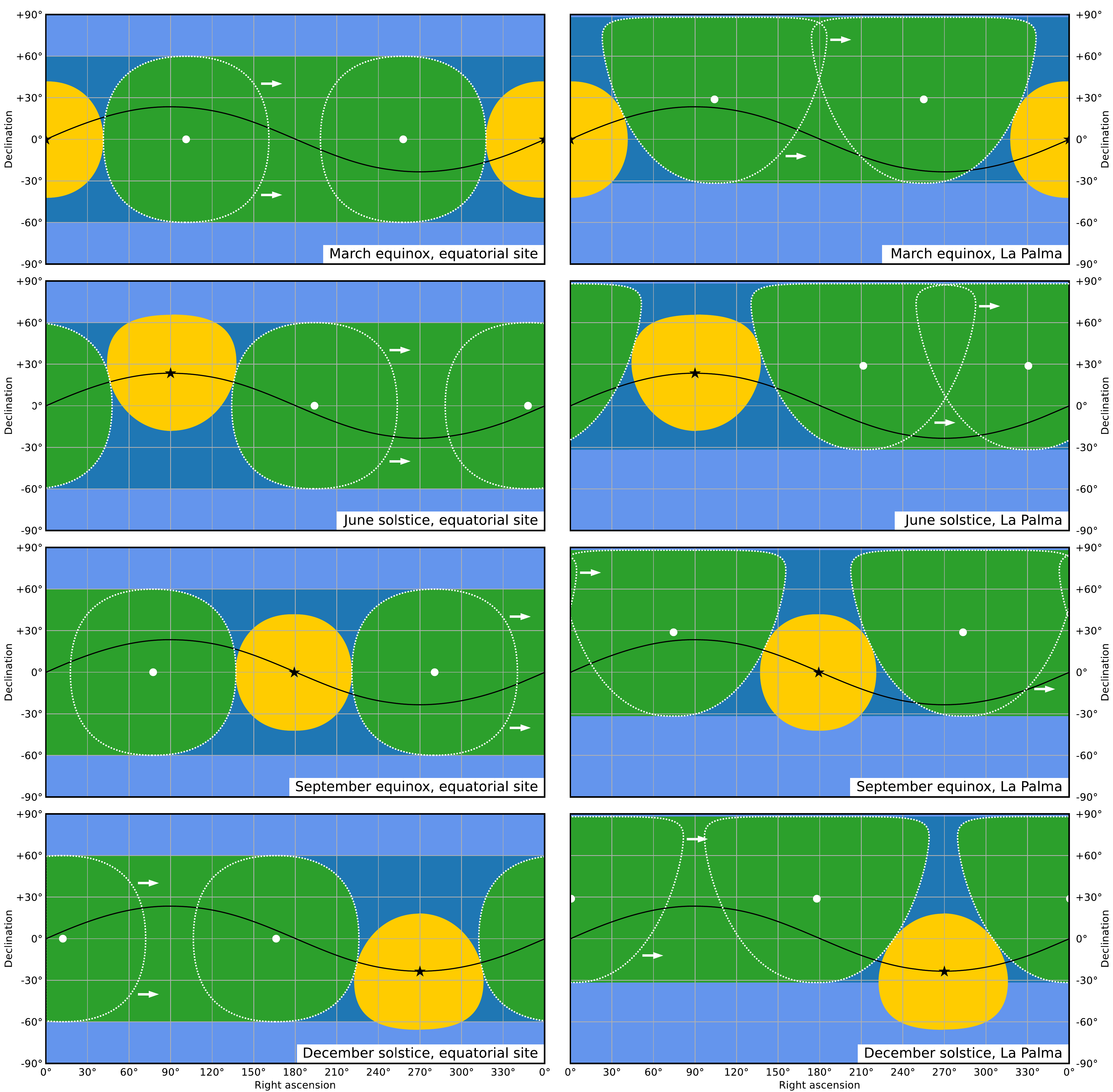}
    \end{center}
    \caption[Plotting sky visibility regions over a year]{
        Changing sky visibility regions over a year, plotted on the celestial sphere.
        Visibility is shown at the equinoxes and the solstices for two different sites: the left column shows visibility for an observer located on the Earth's equator, the right column shows visibility from La Palma. The \textcolorbf{Green}{green} regions are visible during the night. The area above the \SI{30}{\degree} altitude limit at sunset is shown by a white dashed line, with the local zenith marked with a white point, and the arrows show how the visible region moves in RA during the night until sunrise. The position of the Sun on the ecliptic is shown by the black star, the \textcolorbf{Orange}{orange} region is within \SI{42}{\degree} of the Sun and is therefore not visible from anywhere on Earth. The \textcolorbf{NavyBlue}{light blue} regions are permanently out of the visible declination range of the site, and the regions in \textcolorbf{Blue}{dark blue} would only be visible on that day when the Sun is above the horizon (but are visible from other sites).
    }\label{fig:visibility}
\end{figure}

\clearpage

\end{colsection}

\subsection{Selecting event tiles}
\label{sec:gw_selecting}
\begin{colsection}

Even if the source of a gravitational-wave event is visible within 24 hours from a given site, or combination of sites, there is one further criterion that would prevent the source being observed --- whether or not the source is located within any of the tile pointings added to the database. The issue of determining which tiles to add to the database is detailed in \aref{sec:selecting_tiles}, but is ultimately a matter of probability: if a telescope covers the 90\% confidence region for every gravitational-wave event then it would be expected to observe 90\% of the sources.

GOTO-alert uses the mean contour level method to select tiles, as described in \aref{sec:event_insert}. For simulations described in this chapter a mean contour selection value of $0.9$ was used for the GOTO-4 grid and $0.95$ for the GOTO-8 grid. Using these values, 92\% of GW events had sources within at least one of the selected tiles for the GOTO-4 grid, and 95\% for the GOTO-8 grid. Two events where the source location lay outside the selected tiles are shown in \aref{fig:poor_selection}. In the following simulation results the tile selection was only considered after the visibility restrictions in the previous section had been applied, as the visibility would be true for any telescopes at the relevant sites while the tiles are specific to GOTO.\@ In other words, if the source location was visible but was not within the tiles selected by GOTO, this is only GOTO's problem, and other telescopes might still have observed it.

In order to find the optimal selection levels, further simulations could be run using the same sample of skymaps but altering the selection level. As discussed in \aref{sec:event_insert}, there is a trade-off between adding too few tiles and missing the source, and adding too many and increasing the time to cover them all. Additional telescopes in the GOTO network would lead to the skymap being covered faster, which could make adding less probable tiles worthwhile.

\begin{figure}[p]
    \begin{center}
        \includegraphics[width=\linewidth]{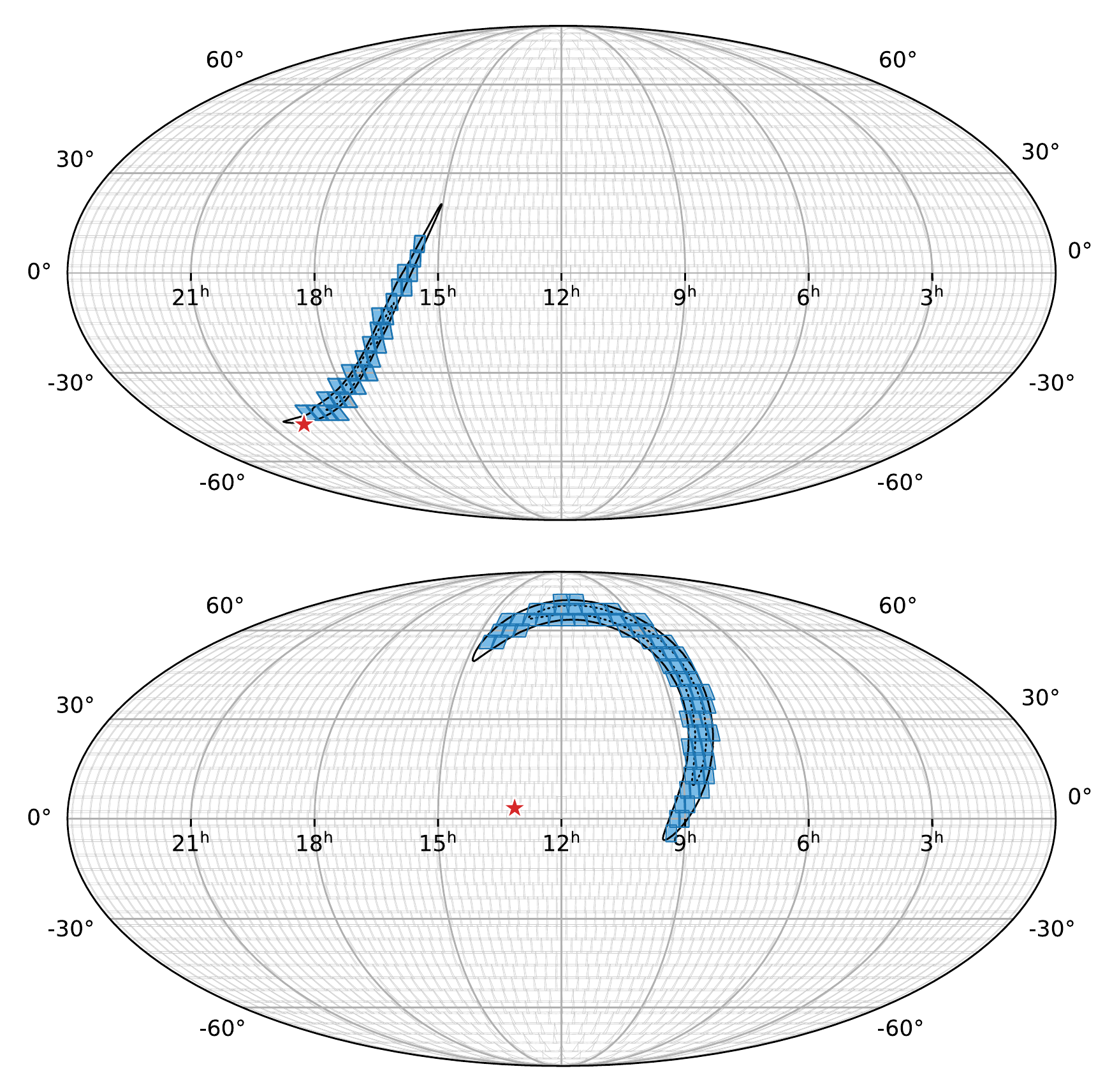}
    \end{center}
    \caption[Examples of mock GW event sources falling outside of the selected tiles]{
        Two examples of mock GW event sources (marked by the \textcolorbf{Red}{red} star) falling outside the selected tiles (the tiles highlighted in \textcolorbf{NavyBlue}{blue}). In the upper case (trigger ID 13630) the source fell just outside of the selected tiles, while in the lower case (trigger ID 930001) the source was on completely the other side of the sky.
    }\label{fig:poor_selection}
\end{figure}

\clearpage

\end{colsection}

\subsection{Multi-telescope simulation results}
\label{sec:gw_sim_results}
\begin{colsection}

In order to simulate the response of different GOTO systems, each of the 1105 First Two Years skymaps \citep{First2Years} were simulated using a script \code{sim\_skymaps.py}. The object of the simulations was to find how quickly the event source would be observed. For events that fell into one of the exceptions described previously, either the source was not visible within 24 hours or the source tile was not selected to be added to the database, the simulation was aborted early, and the result recorded. The remaining events were classified as ``observable'', and for these the full fake pilot simulation was run for up to 24 hours after the time the event occurred. The fake pilot knew which tiles the event source fell within, and once any of those tiles were recorded as being observed the simulation ended. The time of the observation and the alt/az position the tile was observed at were recorded. Any events which were simulated for the full 24 hours without the source being observed were counted as failures, and were classed as ``not observed''.

Simulations were carried out for a variety of possible GOTO systems. Each simulation was assigned a code based on how many telescopes of each type were located at each site. Two possible GOTO ``models'' were considered: the GOTO-4 prototype with four unit telescopes and the intended GOTO-8 design with eight (see \aref{sec:goto_design}). In the following section the code \textbf{1N4} refers to one GOTO-4 mount on La Palma (the current system at the time of writing), \textbf{2N8+1S4} is two GOTO-8 telescopes on La Palma and one GOTO-4 in Siding Spring, \textbf{2N8+1K4} would be the same but the southern telescope is at Mt Kent.

The results of the simulations for six key scenarios are given in the following plots: \aref{fig:gw_sim_1n4}, \aref{fig:gw_sim_1n8} and \aref{fig:gw_sim_2n8} show results for the evolving site on La Palma, while \aref{fig:gw_sim_2n8+1s4}, \aref{fig:gw_sim_2n8+2s8} and \aref{fig:gw_sim_2n8+2k8} shows the effect of adding three different southern facilities. A summary of the key results from all of the simulations that were carried out is given in \aref{tab:gw_sim_results}.

\newpage

\begin{figure}[p]
    \begin{center}
        \begin{minipage}[t]{0.15\linewidth}\vspace{0.6cm}
            \includegraphics[trim={.5cm 0 .5cm 0},clip,width=\linewidth]{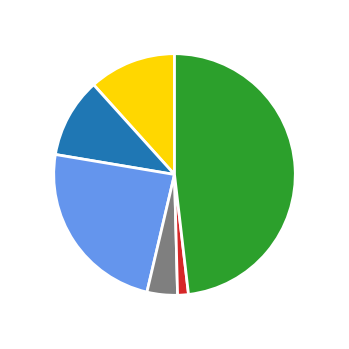}
        \end{minipage}
        \begin{minipage}[t]{0.45\linewidth}\vspace{0pt}
            \begin{tabular}{lrr}
                \multicolumn{3}{c}{\textbf{Simulation results}} \\
                \midrule
                \textcolor{Green}{Observed} & 532 & 48.1\% \\
                \textcolor{Red}{Not observed} & 16 & 1.4\% \\
                \textcolor{darkgray}{Not selected} & 45 & 4.1\% \\
                \textcolor{NavyBlue}{Never above dec limit} & 265 & 24.0\% \\
                \textcolor{Blue}{Not visible at night} & 118 & 10.7\% \\
                \textcolor{Orange}{Too close to Sun} & 129 & 11.7\% \\
                \midrule
                Visible events & 593 &  53.7\% \\
            \end{tabular}
        \end{minipage}
        \begin{minipage}[t]{0.37\linewidth}\vspace{0pt}
            \begin{tabular}{lr}
                \multicolumn{2}{c}{\textbf{System: 1N4}} \\
                \midrule
                Observing efficiency & 89.7\% \\
                \midrule
                Mean delay after     & \multirow{2}{*}{9.96 h} \\
                event time           & \\
                Mean delay after     & \multirow{2}{*}{1.58 h} \\
                becoming visible     & \\
                \midrule
                Mean airmass         & 1.64 \\
            \end{tabular}
        \end{minipage}
    \end{center}
    \caption[GW simulation results: 1N4 system]{
        Simulation results for a 1N4 system. The pie chart and the table on the left shows which category each of the 1105 events fell into. ``Visible events'' include only the top three categories, and the ``observing efficiency'' is the fraction of these events which were subsequently observed. The table on the right gives the mean delay and airmass of the source observation, for events where the source was observed.
    }\label{fig:gw_sim_1n4}
\end{figure}

\begin{figure}[p]
    \begin{center}
        \begin{minipage}[t]{0.15\linewidth}\vspace{0.6cm}
            \includegraphics[trim={.5cm 0 .5cm 0},clip,width=\linewidth]{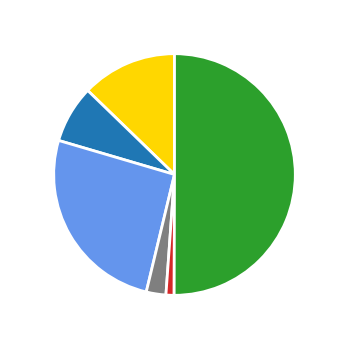}
        \end{minipage}
        \begin{minipage}[t]{0.45\linewidth}\vspace{0pt}
            \begin{tabular}{lrr}
                \multicolumn{3}{c}{\textbf{Simulation results}} \\
                \midrule
                \textcolor{Green}{Observed} & 553 & 50.0\% \\
                \textcolor{Red}{Not observed} & 12 & 1.1\% \\
                \textcolor{darkgray}{Not selected} & 29 & 2.6\% \\
                \textcolor{NavyBlue}{Never above dec limit} & 285 & 25.8\% \\
                \textcolor{Blue}{Not visible at night} & 85 & 7.7\% \\
                \textcolor{Orange}{Too close to Sun} & 141 & 12.8\% \\
                \midrule
                Visible events & 594 &  53.8\% \\
            \end{tabular}
        \end{minipage}
        \begin{minipage}[t]{0.37\linewidth}\vspace{0pt}
            \begin{tabular}{lr}
                \multicolumn{2}{c}{\textbf{System: 1N8}} \\
                \midrule
                Observing efficiency & 93.1\% \\
                \midrule
                Mean delay after     & \multirow{2}{*}{10.06 h} \\
                event time           & \\
                Mean delay after     & \multirow{2}{*}{1.60 h} \\
                becoming visible     & \\
                \midrule
                Mean airmass         & 1.66 \\
                & \\
            \end{tabular}
        \end{minipage}
    \end{center}
    \caption[GW simulation results: 1N8 system]{
        Simulation results for a 1N8 system. Note the distribution of events changes due to the different grid used, and the biggest gain in events observed is from the decreased number of events with sources not included in the selected tiles.
    }\label{fig:gw_sim_1n8}
\end{figure}

\begin{figure}[p]
    \begin{center}
        \begin{minipage}[t]{0.15\linewidth}\vspace{0.6cm}
            \includegraphics[trim={.5cm 0 .5cm 0},clip,width=\linewidth]{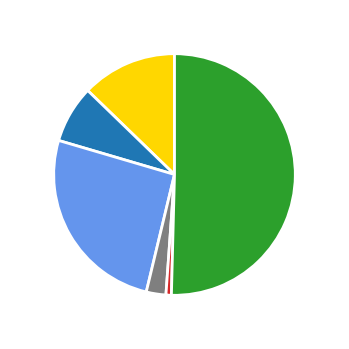}
        \end{minipage}
        \begin{minipage}[t]{0.45\linewidth}\vspace{0pt}
            \begin{tabular}{lrr}
                \multicolumn{3}{c}{\textbf{Simulation results}} \\
                \midrule
                \textcolor{Green}{Observed} & 557 & 50.4\% \\
                \textcolor{Red}{Not observed} & 8 & 0.7\% \\
                \textcolor{darkgray}{Not selected} & 29 & 2.6\% \\
                \textcolor{NavyBlue}{Never above dec limit} & 285 & 25.8\% \\
                \textcolor{Blue}{Not visible at night} & 85 & 7.7\% \\
                \textcolor{Orange}{Too close to Sun} & 141 & 12.8\% \\
                \midrule
                Visible events & 594 &  53.8\% \\
            \end{tabular}
        \end{minipage}
        \begin{minipage}[t]{0.37\linewidth}\vspace{0pt}
            \begin{tabular}{lr}
                \multicolumn{2}{c}{\textbf{System: 2N8}} \\
                \midrule
                Observing efficiency & 93.8\% \\
                \midrule
                Mean delay after     & \multirow{2}{*}{9.89 h} \\
                event time           & \\
                Mean delay after     & \multirow{2}{*}{1.53 h} \\
                becoming visible     & \\
                \midrule
                Mean airmass         & 1.67 \\
                & \\
            \end{tabular}
        \end{minipage}
    \end{center}
    \caption[GW simulation results: 2N8 system]{
        Simulation results for a 2N8 system. The improvements over the 1N8 system are a small gain in observing efficiency and a decrease in the mean delay time.
    }\label{fig:gw_sim_2n8}
\end{figure}

\newpage

\begin{figure}[p]
    \begin{center}
        \begin{minipage}[t]{0.15\linewidth}\vspace{0.6cm}
            \includegraphics[trim={.5cm 0 .5cm 0},clip,width=\linewidth]{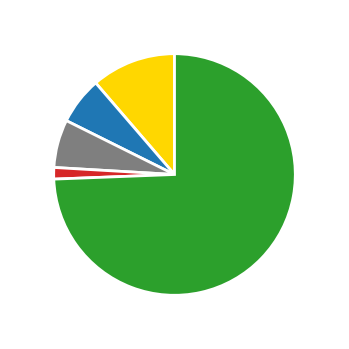}
        \end{minipage}
        \begin{minipage}[t]{0.45\linewidth}\vspace{0pt}
            \begin{tabular}{lrr}
                \multicolumn{3}{c}{\textbf{Simulation results}} \\
                \midrule
                \textcolor{Green}{Observed} & 822 & 74.4\% \\
                \textcolor{Red}{Not observed} & 17 & 1.5\% \\
                \textcolor{darkgray}{Not selected} & 71 & 6.4\% \\
                \textcolor{NavyBlue}{Never above dec limit} & 0 & 0.0\% \\
                \textcolor{Blue}{Not visible at night} & 70 & 6.3\% \\
                \textcolor{Orange}{Too close to Sun} & 125 & 11.3\% \\
                \midrule
                Visible events & 910 &  82.4\% \\
            \end{tabular}
        \end{minipage}
        \begin{minipage}[t]{0.37\linewidth}\vspace{0pt}
            \begin{tabular}{lr}
                \multicolumn{2}{c}{\textbf{System: 2N8+1S4}} \\
                \midrule
                Observing efficiency & 90.3\% \\
                \midrule
                Mean delay after     & \multirow{2}{*}{8.16 h} \\
                event time           & \\
                Mean delay after     & \multirow{2}{*}{1.66 h} \\
                becoming visible     & \\
                \midrule
                Mean airmass         & 1.63 \\
                & \\
            \end{tabular}
        \end{minipage}
    \end{center}
    \caption[GW simulation results: 2N8+1S4 system]{
        Simulation results for a 2N8+1S4 system. As these sites use different grids they were simulated independently and the results combined.
    }\label{fig:gw_sim_2n8+1s4}
\end{figure}

\begin{figure}[p]
    \begin{center}
        \begin{minipage}[t]{0.15\linewidth}\vspace{0.6cm}
            \includegraphics[trim={.5cm 0 .5cm 0},clip,width=\linewidth]{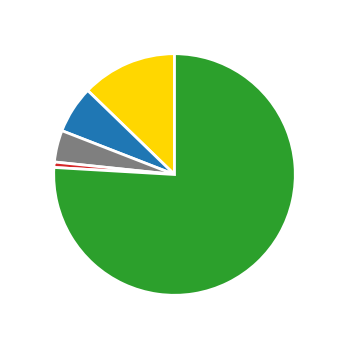}
        \end{minipage}
        \begin{minipage}[t]{0.45\linewidth}\vspace{0pt}
            \begin{tabular}{lrr}
                \multicolumn{3}{c}{\textbf{Simulation results}} \\
                \midrule
                \textcolor{Green}{Observed} & 839 & 75.9\% \\
                \textcolor{Red}{Not observed} & 8 & 0.7\% \\
                \textcolor{darkgray}{Not selected} & 47 & 4.3\% \\
                \textcolor{NavyBlue}{Never above dec limit} & 0 & 0.0\% \\
                \textcolor{Blue}{Not visible at night} & 70 & 6.3\% \\
                \textcolor{Orange}{Too close to Sun} & 141 & 12.8\% \\
                \midrule
                Visible events & 894 &  80.9\% \\
            \end{tabular}
        \end{minipage}
        \begin{minipage}[t]{0.37\linewidth}\vspace{0pt}
            \begin{tabular}{lr}
                \multicolumn{2}{c}{\textbf{System: 2N8+2S8}} \\
                \midrule
                Observing efficiency & 93.8\% \\
                \midrule
                Mean delay after     & \multirow{2}{*}{7.69 h} \\
                event time           & \\
                Mean delay after     & \multirow{2}{*}{1.57 h} \\
                becoming visible     & \\
                \midrule
                Mean airmass         & 1.64 \\
                & \\
            \end{tabular}
        \end{minipage}
    \end{center}
    \caption[GW simulation results: 2N8+2S8 system]{
        Simulation results for a 2N8+2S8 system. The obvious improvement over the northern hemisphere-only 2N8 system (\aref{fig:gw_sim_2n8}) is the removal of the declination-limited events, meaning more event sources are visible. The observing efficiency remains the same, but there is a notable improvement in the post-event delay times.
    }\label{fig:gw_sim_2n8+2s8}
\end{figure}

\begin{figure}[p]
    \begin{center}
        \begin{minipage}[t]{0.15\linewidth}\vspace{0.6cm}
            \includegraphics[trim={.5cm 0 .5cm 0},clip,width=\linewidth]{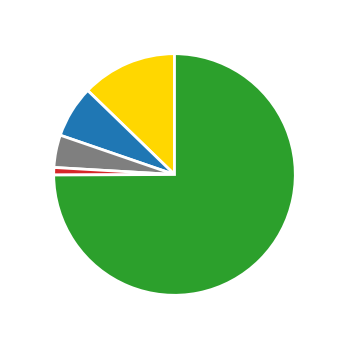}
        \end{minipage}
        \begin{minipage}[t]{0.45\linewidth}\vspace{0pt}
            \begin{tabular}{lrr}
                \multicolumn{3}{c}{\textbf{Simulation results}} \\
                \midrule
                \textcolor{Green}{Observed} & 828 & 74.9\% \\
                \textcolor{Red}{Not observed} & 11 & 1.0\% \\
                \textcolor{darkgray}{Not selected} & 48 & 4.3\% \\
                \textcolor{NavyBlue}{Never above dec limit} & 0 & 0.0\% \\
                \textcolor{Blue}{Not visible at night} & 77 & 7.0\% \\
                \textcolor{Orange}{Too close to Sun} & 141 & 12.8\% \\
                \midrule
                Visible events & 887 &  80.3\% \\
            \end{tabular}
        \end{minipage}
        \begin{minipage}[t]{0.37\linewidth}\vspace{0pt}
            \begin{tabular}{lr}
                \multicolumn{2}{c}{\textbf{System: 2N8+2K8}} \\
                \midrule
                Observing efficiency & 93.3\% \\
                \midrule
                Mean delay after     & \multirow{2}{*}{7.69 h} \\
                event time           & \\
                Mean delay after     & \multirow{2}{*}{1.55 h} \\
                becoming visible     & \\
                \midrule
                Mean airmass         & 1.64 \\
                & \\
            \end{tabular}
        \end{minipage}
    \end{center}
    \caption[GW simulation results: 2N8+2K8 system]{
        Simulation results for a 2N8+2K8 system. Comparing to \aref{fig:gw_sim_2n8+2s8} it makes very little difference to the results if the southern site is at Siding Spring or Mt Kent, when compared to the huge gain from having either available instead of just La Palma (\aref{fig:gw_sim_2n8}).
    }\label{fig:gw_sim_2n8+2k8}
\end{figure}

\begin{table}[p]
    \begin{center}
        \begin{tabular}{c|cccc|c|cc} %
            \multirow{2}{*}{System} &
            \multicolumn{4}{c|}{Source observed within \ldots} &
            {\small Observing} &
            \multicolumn{2}{c}{Mean delay after \ldots} \\
                & 1h & 6h & 12h & 24h & efficiency & {\small the event} & {\small becoming visible} \\
            \midrule
                 1N4 &  5.9\% & 16.5\% & 26.2\% & 48.1\% & 89.7\% &  9.96 h & 1.58 h \\
                 1N8 &  6.8\% & 16.9\% & 27.1\% & 50.0\% & 93.1\% & 10.06 h & 1.60 h \\
                 2N8 &  7.2\% & 17.3\% & 27.9\% & 50.4\% & 93.8\% &  9.89 h & 1.53 h \\
            &&&&&&&\\
                 1S4 &  8.8\% & 18.8\% & 27.9\% & 47.1\% & 87.7\% &  9.39 h & 1.64 h \\
                 1S8 & 10.3\% & 20.6\% & 31.1\% & 49.2\% & 91.0\% &  8.81 h & 1.52 h \\
                 2S8 & 11.1\% & 21.2\% & 31.8\% & 49.9\% & 92.1\% &  8.67 h & 1.47 h \\
            &&&&&&&\\
                 2K8 & 10.8\% & 20.9\% & 31.3\% & 50.9\% & 92.3\% &  9.02 h & 1.46 h \\
            &&&&&&&\\
             1N4+1S4 & 14.7\% & 34.9\% & 48.9\% & 71.5\% & 89.6\% &  7.99 h & 1.64 h \\
             1N8+1S8 & 17.1\% & 36.8\% & 52.5\% & 74.8\% & 92.5\% &  7.82 h & 1.63 h \\
             2N8+1S8 & 17.6\% & 37.2\% & 52.9\% & 75.2\% & 93.0\% &  7.75 h & 1.59 h \\
             2N8+2S8 & 18.4\% & 37.7\% & 53.8\% & 75.9\% & 93.8\% &  7.69 h & 1.57 h \\
            &&&&&&&\\
            2N8+2K8  & 18.2\% & 38.0\% & 52.8\% & 74.9\% & 93.3\% &  7.69 h & 1.55 h \\
            &&&&&&&\\
            2N8+1S4* & 16.0\% & 35.7\% & 50.3\% & 74.4\% & 90.3\% &  8.16 h & 1.66 h \\
            2N8+1S8* & 17.6\% & 37.2\% & 53.0\% & 75.2\% & 93.0\% &  7.76 h & 1.60 h \\
            2N8+2S8* & 18.4\% & 37.7\% & 53.7\% & 75.9\% & 93.8\% &  7.70 h & 1.58 h \\
        \end{tabular}
    \end{center}
    \caption[GW simulation results summary table]{
        Summary of simulation results. The fraction of events where the source was observed is given for different time delays after the event, along with the overall observing efficiency after 24 hours. The mean delay between the event and the source being observed is given, along with the mean time it took to observe the source tile after it became visible. Systems marked with an asterisk (*) were not simulated together, but were instead combined from the individual simulations for each site.
    }\label{tab:gw_sim_results}
\end{table}

\clearpage

\end{colsection}

\subsection{Analysis of simulation results}
\label{sec:gw_sim_analysis}
\begin{colsection}

The results of the gravitational-wave follow-up simulations support two conclusions: the addition of the southern site provides a huge benefit to the number of sources that can be observed, while adding further telescopes at a single site provides a much more modest benefit. \aref{fig:gw_sim_results} summarises the simulated post-event delay times for the different possible stages of GOTO deployment.

\begin{figure}[t]
    \begin{center}
        \includegraphics[width=\linewidth]{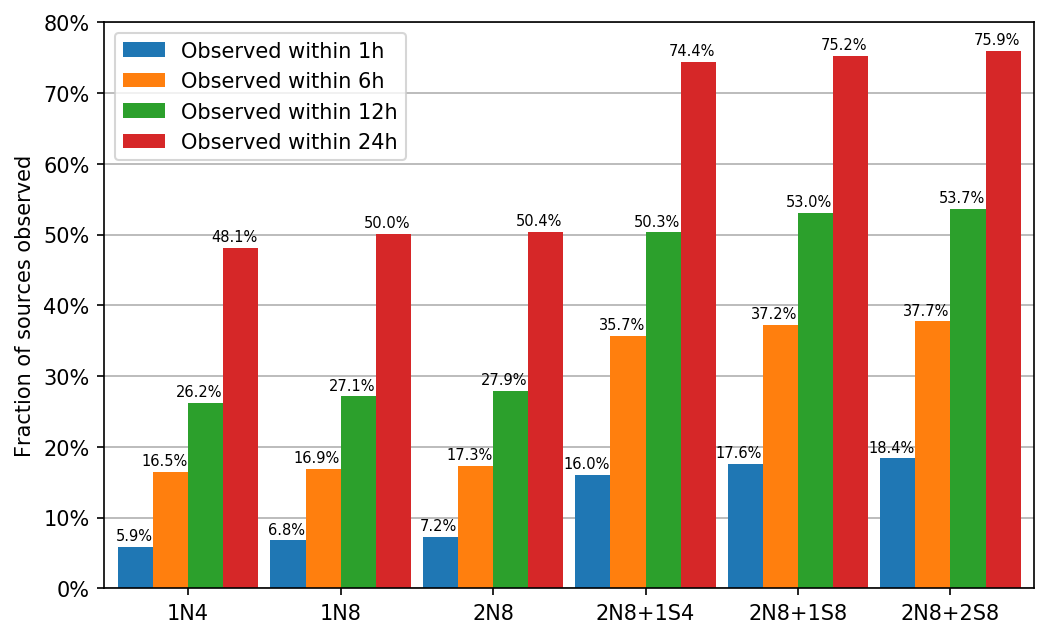}
    \end{center}
    \caption[Simulation delay time for different GOTO systems]{
        Post-event delay in observing gravitational-wave event sources for six possible deployment stages of the GOTO system.
    }\label{fig:gw_sim_results}
\end{figure}

The reason for the first conclusion is obvious: adding a site in the southern hemisphere opens up a large number of sources that are physically incapable of being observed from La Palma. The second comes about essentially because the efficiency of a single GOTO system is already very high. \aref{fig:gw_sim_1n8} shows that the 1N8 system already observes 93.1\% of all sources visible from La Palma, and the majority of those not observed were outside of the selected tiles (29 events) compared to just not being observed (12 events). The addition of the second GOTO-8 system, as shown in \aref{fig:gw_sim_2n8}, moves just 4 events from ``not observed'' to ``observed''; adding a second telescope can not make any changes to any of the other categories. The mean delay time does decrease, but again only by a small amount.

The above conclusions are perhaps most visible by comparing the results in \aref{tab:gw_sim_results} for the 2N8 system to the 1N8+1S8 system, where there is a clear increase in the number of sources observed within 24 hours (from 50.4\% to 74.8\%). Therefore, based on these metrics alone, it would be far better to prioritise deploying a second mount in Australia before adding another on La Palma. There are numerous practical reasons why this is not the priority of the collaboration, and \aref{sec:survey_sims} illustrates that multiple telescopes at a single site are much more important to the all-sky survey cadence (which in practice would benefit the counterpart search by producing more recent reference images).

Regarding the choice of southern site, there is very little difference between results for Siding Spring and for Mt Kent. Comparing the 2S8 and 2K8 simulation results in \aref{tab:gw_sim_results} shows that Mt Kent has a small advantage in terms of the number of events observed, but Siding Spring has a lower mean delay time. Overall, there is no real difference between the two sites, and in most cases the ``S'' simulations are considered as representative of either site.

Another factor to emerge from these simulations is the difference between two independent systems in each hemisphere verses one combined system that uses a common database. This emerged as important as it was desired to simulate the 2N8+1S4 system, as a plausible future stage of GOTO's deployment. As considered in \aref{sec:multi_grid_scheduling}, the simulations require all telescopes to be observing using the same grid, as otherwise it is impossible at this time to share the common tiles between them. However, it was possible to consider the two cases, 2N8 and 1S4, separately as independent simulations, and then combine the results. For the event counts the logic is fairly straightforward: if an event is observed by \emph{either} site (or both) it counts as being observed. In cases where the event source was observed by both sites independently, only the earlier observation is considered. Using this method the results shown in \aref{fig:gw_sim_2n8+1s4} were derived. The same method could also be used in situations where the two sites could be simulated together; for example, comparing the 2N8+2S8 simulation to the combined results of the 2N8 and 2S8 simulations, given as 2N8+2S8* in \aref{tab:gw_sim_results}. The same events fell into the same categories as shown in \aref{fig:gw_sim_2n8+2s8}, and the only difference is a small increase in the delay time when the sites are not simulated together. For large skymaps with tiles within the shared area of the sky visible from both sites (roughly $\pm$\SI{30}{\degree} declination) one site can complete observations of the region even if it can not see the source, meaning that once the other telescope opens and starts observing, a large area of the skymap that does not contain the source has already been excluded. This is only the case when both sites are observing using a shared database, as the second site needs to know what the first site has already observed.

\end{colsection}

\section{All-sky survey simulations}
\label{sec:survey_sims}

\begin{colsection}

As described in \aref{sec:goto_motivation}, carrying out the all-sky survey is just as critical to the GOTO project as the gravitational-wave follow-up operations, as up-to-date reference images will always be required to detect any counterpart sources. Therefore, in parallel to the gravitational-wave simulations, further simulations were carried out in order to quantify what benefit additional telescopes and sites will have on carrying out the all-sky survey. It was also an opportunity to consider different sky survey methods before implementing them in the real scheduling system. Unlike the gravitational-wave simulations, it is also possible to compare the simulated results for a single GOTO-4 system on La Palma to the actual observations the live GOTO system has taken since it began observing the current survey in February 2019.

\end{colsection}

\subsection{Simulating sky survey observations}
\label{sec:survey_sim_methods}
\begin{colsection}

Simulating the all-sky survey is more straightforward than the gravitational-wave simulations, as there is no added complication of processing the LVC skymap or checking the visibility of the source coordinates. Instead, all that is needed is to fill the observation database with the sky survey pointings (see \aref{sec:obsdb}) and run the fake pilot (see \aref{sec:goto_sims}).

The only drawback to this method is the time taken to perform the simulations. Unlike the gravitational-wave simulations, the sky survey simulation can not be finished early if the source is not visible, or stopped once the source has been observed. Instead, the fake pilot needs to simulate the full 24 hours of observations, for however many days the simulation is run for. The same simplifications detailed previously still apply, so each loop still skips approximately 4 minutes of simulation time until the observation of each tile has been completed. A full simulation of a year of observations including both sites (therefore observing for approximately 20 hours each day) requires 1.1 million steps, and with each simulation loop taking approximately 2 seconds of CPU time (the scheduler check takes the majority of this time), the full simulation takes approximately 60 hours. This compares to at most 16 hours for the multi-site gravitational-wave simulations.

Due to the full sky-survey simulations requiring a large time investment, only a few of them could be carried out in the time available. A simplified version of the simulation code was therefore developed, which could produce the same results much faster. This `lite' script did away with the scheduler and database code, and instead at each step just finds the highest altitude tiles that have been observed the fewest times. This is a major simplification of the scheduling functions described in \aref{chap:scheduling}, but the results are effectively the same, and can be obtained 15--20 times faster. Therefore, a majority of the simulations discussed in this section use this much faster `lite' script. The other benefit of this method was making it much easier to modify the scheduling function to test different surveying methods, as discussed in \aref{sec:survey_sim_meridian}, without needing to rewrite the actual G-TeCS scheduler.

\end{colsection}

\subsection{Multi-telescope simulation results}
\label{sec:survey_sim_results}
\begin{colsection}

Sky-survey simulations were carried out for different combinations of GOTO telescopes and sites, similar to the gravitational-wave event simulations detailed in \aref{sec:gw_sims}. Simulations were run for 365 days starting semi-arbitrarily on the 21st of February 2019, which was the date that the current ongoing GOTO all-sky survey started on La Palma.

Fewer simulations were carried out when compared to the gravitational-wave simulations. This is partially as they take longer to run, but there were also fewer possible cases to simulate. It is not possible to combine the results of telescopes observing the sky survey on different grids, as the results depend explicitly on the grid used. Therefore, unlike the gravitational-wave simulations, it was not possible to combine a GOTO-8 telescope in the north and a GOTO-4 telescope in the south.

The results of the sky-survey simulations are given in \aref{tab:survey_sim_results}. \aref{fig:survey_sim_1n4} shows the final tile-coverage map for the 1N4 system, in which each tile in the GOTO-4 grid is coloured by the number of times it was observed. \aref{fig:survey_sim_2n8+2s8} shows the same information for the final 2N8+2S8 system.

\begin{figure}[p]
    \begin{center}
        \includegraphics[height=190pt]{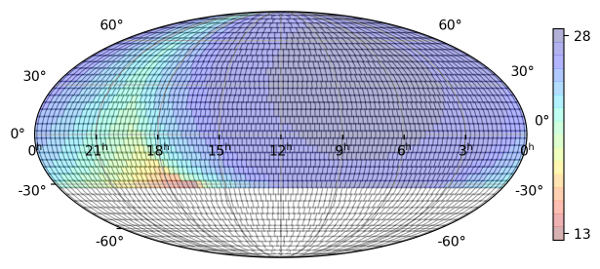}
    \end{center}
    \caption[All-sky survey simulation results: 1N4 system]{
        All-sky survey simulation coverage map for a 1N4 system, as currently deployed on La Palma. Tiles are coloured by the number of times they were observed over the 365 simulated nights. Tiles in white are those not visible from the northern site.
    }\label{fig:survey_sim_1n4}
\end{figure}

\begin{figure}[p]
    \begin{center}
        \includegraphics[height=190pt]{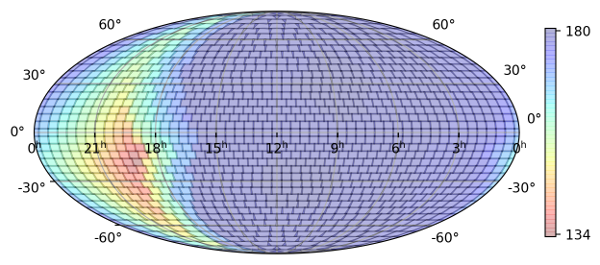}
    \end{center}
    \caption[All-sky survey simulation results: 2N8+2S8 system]{
        All-sky survey simulation coverage map for a 2N8+2S8 system, the ultimate design goal of the GOTO collaboration. Note the colour scale has changed from \aref{fig:survey_sim_1n4}, the grid has changed to the GOTO-8 tiles and the region which was previously not visible from just the north has been filled in.
    }\label{fig:survey_sim_2n8+2s8}
\end{figure}

\begin{table}[t]
    \begin{center}
        \begin{tabular}{c|cc|c|c|c} %
            \multirow{2}{*}{System} &
            \multicolumn{2}{c|}{Fraction of sky observed} &
            No.\ of times &
            Mean cadence &
            {\small Mean airmass}
            \\
            &
            each night &
            over 1y &
            tiles observed &
            (days) &
            observed
            \\
            \midrule
            1N4* & 4.3\%--6.5\% & 76.5\% & 26 (13--28) & $10.0\pm1.8$ & $1.6\pm0.4$ \\
            &&&&&\\
            1N4 & 4.3\%--6.4\% & 76.5\% & 26 (13--28) & $10.1\pm1.8$ & $1.6\pm0.3$ \\
            1N8 & 9.5\%--14.0\% & 74.2\% & 58 (34--62) & $4.6\pm0.7$ & $1.6\pm0.4$ \\
            2N8 & 19.0\%--28.1\% & 74.2\% & 117 (68--123) & $2.3\pm0.4$ & $1.6\pm0.4$ \\
            &&&&&\\
            1N4+1S4 & 10.5\%--11.0\% & 99.9\% & 39 (30--42) & $7.3\pm0.7$ & $1.5\pm0.4$ \\
            1N8+1S8 & 23.2\%--24.1\% & 99.9\% & 87 (67--91) & $3.4\pm0.3$ & $1.5\pm0.4$ \\
            2N8+1S8 & 33.0\%--37.3\% & 99.9\% & 130 (98--138) & $2.3\pm0.2$ & $1.6\pm0.4$ \\
            2N8+2S8 & 46.3\%--48.1\% & 99.9\% & 173 (134--180) & $1.7\pm0.1$ & $1.5\pm0.4$ \\
            &&&&&\\
            2N8+2K8 & 46.8\%--48.3\% & 99.8\% & 174 (134--181) & $1.7\pm0.1$ & $1.6\pm0.4$ \\
        \end{tabular}
    \end{center}
    \caption[All-sky survey simulation results summary table]{
        Summary of all-sky survey simulation results. The first 1N4 simulation, marked with an asterisk (*), was the only one carried out using the full scheduler and database system; all the other simulations used the `lite' script. The fraction of the sky observed each night is given as a range over the course of a year, as well as the total fraction of the sky observed over the whole year. The number of times each tile was observed is given as an average over all tiles observed and, in parenthesis, the minimum and maximum. The mean cadence between observations of each tile is also given, along with the mean airmass of each observation, over the whole year.
    }\label{tab:survey_sim_results}
\end{table}

\end{colsection}

\subsection{Analysis of simulation results}
\label{sec:survey_sim_analysis}
\begin{colsection}

The results of the all-sky survey simulations show, as expected, that the greatest benefit to the survey cadence comes from increasing the number of telescopes at each site. \aref{fig:survey_sim_results} plots the change in mean cadence and fraction of the sky observed each night for the planned stages of GOTO deployment.

\begin{figure}[t]
    \begin{center}
        \includegraphics[width=\linewidth]{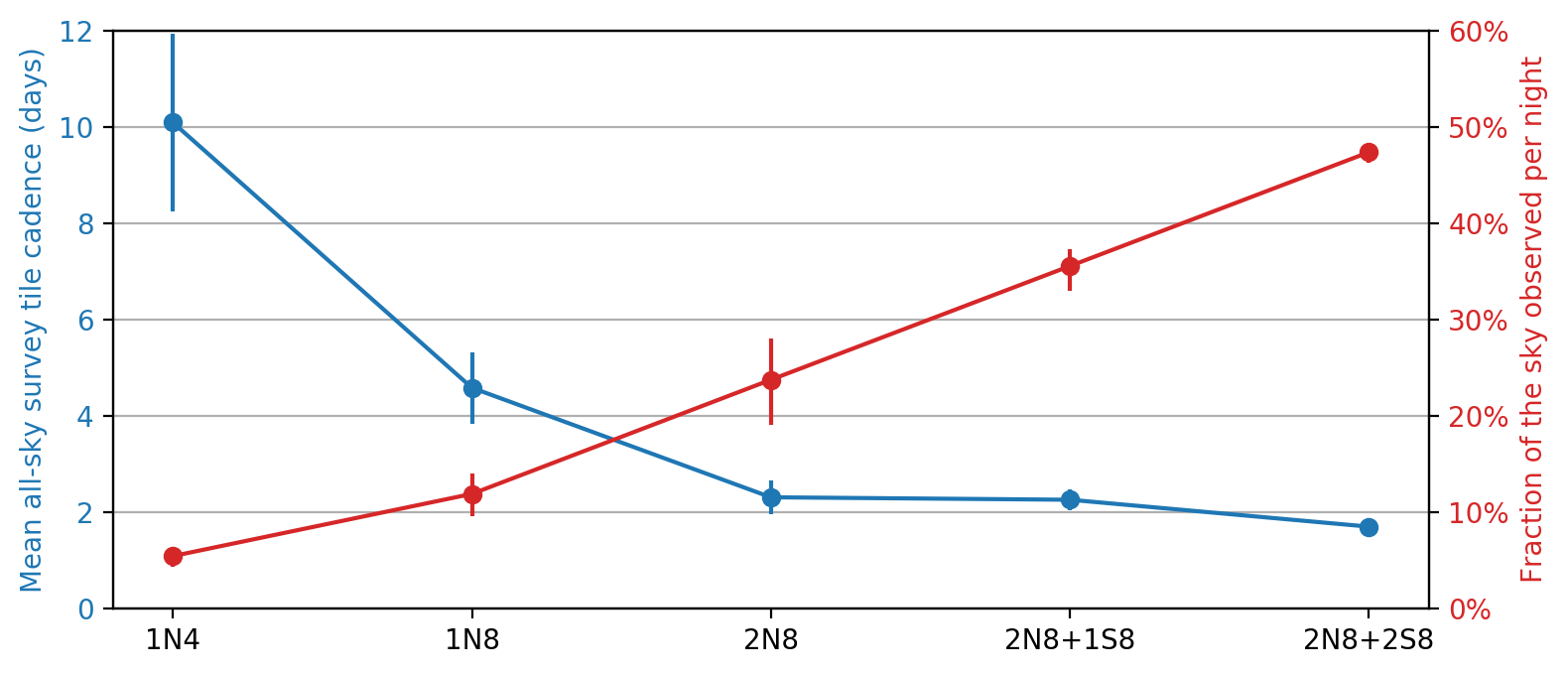}
    \end{center}
    \caption[Tile cadence and nightly sky observation for different GOTO systems]{
        Mean tile cadence (\textcolorbf{NavyBlue}{blue}) and fraction of the sky observed each night (\textcolorbf{Red}{red}) for five deployment stages of the GOTO system. Error bars on the cadence show the standard deviation for all of the tiles observed across the sky, while the error bars on the observed fraction show the minimum and maximum nightly observed fraction arising from differing night lengths throughout the year.
    }\label{fig:survey_sim_results}
\end{figure}

The improvement in tile observation cadence roughly follows the expected trend that doubling the instantaneous field of view would double the number of observations carried out in one night, and therefore halve the time between observations of a given tile. With the current GOTO-4 system (1N4) the simulations predict approximately 10 days between tile observations, reducing to approximately 5 days with the upgrade to the full GOTO-8 system (1N8) and then halving again to 2.5 days with the addition of the second GOTO-8 telescope on La Palma (2N8). Adding a single GOTO-8 telescope in Australia (2N8+1S8) leaves the tile cadence effectively unchanged. Adding a site in Australia increases the total amount of the sky that is visible over the year from roughly 75\% to almost 100\%, an increase of one third, and adding one more GOTO-8 telescope in the south corresponds to a one third increase in observing capability. Therefore the overall efficiency remains the same as in the 2N8 case. Adding the second telescope in Australia correspondingly decreases the cadence further; as this increases the overall instantaneous field of view by one third, the cadence is reduced by a third, from 2.5 to 1.66 days.

\newpage

As also shown in \aref{fig:survey_sim_results}, the increase in the fraction of the sky observed each night as more telescopes are added is fairly linear. The variation over the course of the year, shown by the error bars, comes from seasonal variation in the length of the night; the variation increases as more telescopes are added in the north but is then reduced to effectively zero with equal numbers of telescopes in both hemispheres.

Together, the sky-survey simulations confirm that having more telescopes decreases the survey cadence, as would be expected. Having a fast all-sky survey is critical for rapidly detecting candidate sources to gravitational-wave events, so it is necessary to consider the results of both sets of simulations together. Counter to the conclusions from \aref{sec:gw_sims}, the simulations in this section suggest it would not necessarily be best to prioritise adding a telescope in the south compared to adding a second telescope in the north. While going from the 1N8 case to the 2N8 case makes very little difference to the number of gravitational-wave events that can be observed, it would halve the survey cadence from 4.6 days to 2.3, thereby making it far easier to identify candidates for the events which are visible. The decision of what order to deploy the GOTO telescopes therefore will come down to more practical considerations. Having any telescopes in the southern hemisphere will increase the number of possible gravitational-wave sources that could be observed, but without a high-cadence sky survey identifying the counterpart will be much more difficult.

Overall, the two sets of simulations together suggest that the proposed full GOTO network, the ``2N8+2S8'' system, should expect to observe the position of over 75\% of gravitational-wave sources within 24 hours, and over 50\% within 12 hours. On average there should be a reference image taken of the same position within the past 1.7 days, which will greatly help in narrowing down potential candidates. When fully deployed, GOTO would therefore be a powerful system for rapidly finding counterparts to gravitational-wave detections.

\newpage

\end{colsection}

\subsection{Comparison of simulations to real observations}
\label{sec:survey_sim_150}
\begin{colsection}

The results of the 1N4 simulation can be compared to the real observations carried out by the existing telescope on La Palma, in order to confirm how good a model it is of the real system. The current phase of the GOTO project began on the night of the 21st of February 2019 (see \aref{sec:timeline}); this was the first night of fully robotic observations with the set of four unit telescopes, and marks the start of the ongoing all-sky survey. The first 5 months of observations span 150 days up to the night ending on the 21st of July, and this provides the benchmark to compare with simulations of the same period.

\begin{figure}[p]
    \begin{center}
        \includegraphics[width=\linewidth]{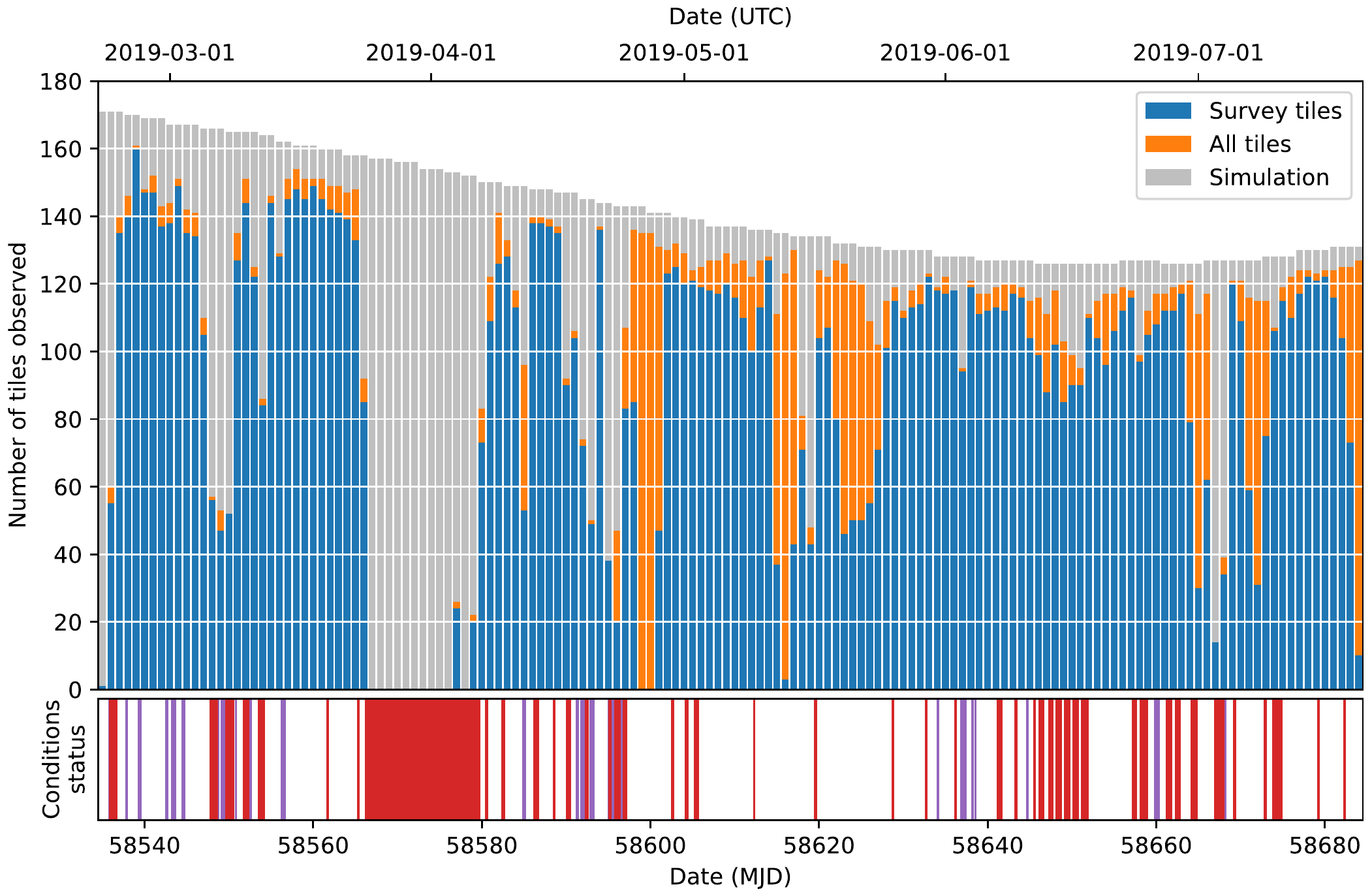}
    \end{center}
    \caption[Observations carried out in the first 150 days of the all-sky survey]{
        Observations carried out in the first 150 days of the current all-sky survey. The number of observations each night is shown in the upper plot, with the number of all-sky survey tiles observed shown in \textcolorbf{NavyBlue}{blue}, and any extra observations (of gravitational-wave events, GRB triggers or manually-inserted pointings) shown in \textcolorbf{Orange}{orange}. The background \textcolorbf{Gray}{grey} bars show the number of tiles observed on the same nights by the 1N4 survey simulation. The lower ``barcode'' plot shows the periods when the conditions flags were recorded as bad, either due to weather (\textcolorbf{Red}{red}) or hardware errors (\textcolorbf{Purple}{purple}).
    }\label{fig:150}
\end{figure}

\aref{fig:150} shows the number of tiles observed each night by the real GOTO on La Palma and the corresponding 1N4 simulation, restricted to the first 150 days. Over the 5-month period, 16,146 on-grid observations were carried out by the telescope on La Palma, of which 85\% were survey pointings, and over the same period the simulation produced 21,300 observations. The simulated observations can be considered the idealised case, and the real observations carried out differ from simulation in three ways.

First, the real system on La Palma is affected by bad conditions which prevent observations from being taken, shown in \aref{fig:150} by the red and purple bars below the main plot. There was one particularly bad period in late March and early April when the dome could not open for over a week. The simulations do not currently include the effects of bad weather, although the code exists to simulate periods of bad conditions, and future simulations could include the real weather conditions over the same period. There were also other reasons for observations to be stopped on some nights, for example switching to manual mode to carry out calibration tests or on-site work.

Secondly, the real system had to deal with multiple distractions from observing the all-sky survey. The orange bars in \aref{fig:150} show non-survey observations, which take up a significant amount of time (15\% of all observations). Some nights are almost entirely orange, corresponding to LVC gravitational-wave triggers with particularly large skymaps visible from La Palma (see \aref{sec:gw_results}). These include S190425z and S190426c in late April, multiple events during May and S190720a just before the end of the period in mid-July. Other orange patches represent observations of smaller gravitational-wave skymaps, gamma-ray burst triggers, or other manually-inserted targets; these were also not considered in the sky-survey simulations.

Finally, there is still a regular offset in \aref{fig:150} between the number of real observations taken in nights with clear conditions and the number predicted by the simulations. This discrepancy is due to the values used within the simulation for camera readout and slew time not matching up precisely with the actual times; future simulations will need to be calibrated more accurately against real data.

\aref{fig:survey_real_150} shows the sky coverage map for the real observations in the first 150 days, while \aref{fig:survey_sim_1n4_150} shows the same for the 1N4 simulation. The real sky coverage is very similar in extent to the simulations: the real observations cover 2135 of the 2913 GOTO-4 grid tiles at least once while the simulation covers 2187. The reason for the small discrepancy is that initially the real system used an altitude limit of \SI{35}{\degree} for the all-sky survey, which was lowered to \SI{30}{\degree} in May. This change is visible in the bottom row of tiles in \aref{fig:survey_real_150}. The simulations all assume a constant \SI{30}{\degree} limit.

The major difference in the coverage between the real and simulated results is in the number of times each tile was observed. \aref{fig:survey_real_150} shows the most a real tile was observed was nine times, while \aref{fig:survey_sim_1n4_150} shows that the simulated results included up to 13 observations of a single tile. The mean tile cadence of the real observations is $14\pm4$, compared to $10\pm2$ from the 1N4 simulation. It is clear that future simulations need to take into account the time lost to weather and other non-survey observations in order to accurately predict the output of the real system. Overall though, aside from the constant offset visible in \aref{fig:150}, the simulations do seem to provide a reasonable approximation of what GOTO could observe in this idealised case.

\begin{figure}[p]
    \begin{center}
        \includegraphics[height=190pt]{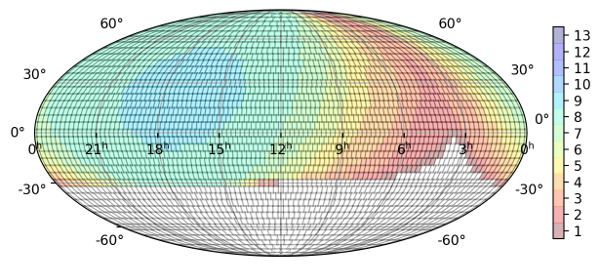}
    \end{center}
    \caption[Real survey observations over 150 days]{
        Real all-sky survey map of observations over the first 150 days, from 21st February to 21st July 2019. Tiles are coloured by the number of times they were observed during this period (the most a single tile was observed was 9 times), and white tiles were never observed.
    }\label{fig:survey_real_150}
\end{figure}

\begin{figure}[p]
    \begin{center}
        \includegraphics[height=190pt]{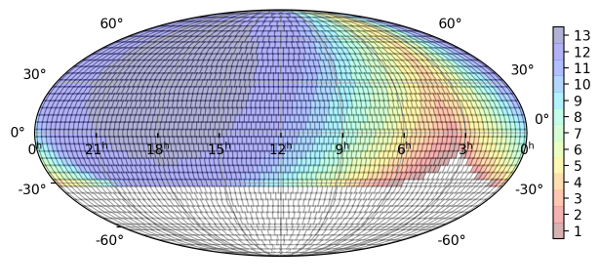}
    \end{center}
    \caption[1N4 survey simulation observations over 150 days]{
        Simulated 1N4 all-sky survey map of observations over the first 150 days, using the same scale as \aref{fig:survey_real_150} above. Compare to \aref{fig:survey_sim_1n4} for the coverage over the entire 1-year simulation.
    }\label{fig:survey_sim_1n4_150}
\end{figure}

\clearpage
\newpage

\end{colsection}

\subsection{An alternative, meridian-limited sky survey method}
\label{sec:survey_sim_meridian}
\begin{colsection}

One of the problems with the current scheduling system for the all-sky survey is that it leads to observations being carried out at high airmasses. Due to how the scheduler ranks tiles (see \aref{sec:ranking}), when all of the visible survey tiles have been observed the same number of times the scheduler will then choose between them based on the airmass tiebreak parameter (unlike tiles linked to skymaps, all survey tiles have equal weights, so the tiebreaking algorithm developed in \aref{sec:scheduler_tiebreaker} is simplified). This results in the scheduler always selecting tiles as soon as they rise if they have been observed fewer times than any others currently visible, and this means survey tiles are often observed at low altitudes and therefore high airmasses --- leading to poor data quality.

One possible method to fix this problem, and improve the data quality, is to implement stricter limits when observing survey tiles. This should not be based on altitude or airmass, because that would exclude tiles close to the site declination limits (such as near the north celestial pole from La Palma) which could never rise above the limit. Instead, observations should be limited based on distance from the observer's meridian, which in practice limits the target's hour angle. Limiting observations by hour angle defines a strip surrounding the observer's meridian within which survey tiles are valid and outside of which they are not.

In order to see the consequences of this method several simulations were carried out using the 1N4 system, but modified to limit the hour angle of each target. The results of the simulations are given in \aref{tab:survey_sim_meridian}, for different hour angle limits and the unlimited case for comparison. \aref{fig:survey_sim_airmass_365} shows the change in distribution of airmasses between the existing unlimited method and when restricting observations to a \SI{20}{\degree} wide strip ($\pm\SI{10}{\degree}$) around the observer's meridian. \aref{fig:survey_sim_airmass_normal} shows the mean airmass each tile is observed in the first month of a survey using the existing method, while \aref{fig:survey_sim_airmass_meridian} shows the same thing but for a simulation using the hour angle limit.

By restricting observations to be closer to the observer's meridian the mean airmass of the observations is decreased, as expected. This is shown by the mean airmasses in \aref{tab:survey_sim_meridian} but is even clearer in the distributions shown in \aref{fig:survey_sim_airmass_365}. The optimal value of the hour angle limit will depend on several factors, including the number of telescopes being used (as the instantaneous field of view increases, the tiles within the meridian strip will be observed faster, and so the hour angle limit should be increased).

A side effect of this method is that, as the width of the meridian strip is decreased and the effective visible sky is reduced, each tile within the strip is observed more often, and therefore the mean cadence decreases. However, this also restricts the overage area and leads to fewer unique tiles being observed, as shown in \aref{fig:survey_sim_airmass_meridian}. From \aref{tab:survey_sim_meridian} the fraction of sky observed within a single month reduces from almost 60\% using the unlimited method to below 40\% with the strictest hour angle limit. This would have a knock-on effect on the effectiveness of the sky survey, as although lower-airmass observations are desirable, so are more recent observations of the tiles for difference imaging. In practice it might be necessary to have two concurrent surveys, one optimised for minimum airmass and the other optimised for cadence.

\begin{table}[t]
    \begin{center}
        \begin{tabular}{c|cc|c|c|c} %
            Survey &
            \multicolumn{2}{c|}{Fraction of sky observed} &
            Mean cadence &
            Mean observed
            \\
            method &
            1st month &
            whole year &
            (days) &
            airmass
            \\
            \midrule
            Meridian  \SI{\pm5}{\degree} & 39.4\% & 76.5\% &  $6.2\pm0.5$ & $1.2\pm0.2$ \\
            Meridian \SI{\pm10}{\degree} & 41.3\% & 76.5\% &  $6.6\pm0.7$ & $1.2\pm0.3$ \\
            Meridian \SI{\pm30}{\degree} & 48.5\% & 76.5\% &  $7.9\pm0.8$ & $1.3\pm0.3$ \\
            Meridian \SI{\pm45}{\degree} & 53.3\% & 76.5\% &  $8.8\pm0.9$ & $1.3\pm0.3$ \\
            No limit                     & 57.0\% & 76.5\% & $10.1\pm1.8$ & $1.6\pm0.3$ \\
        \end{tabular}
    \end{center}
    \caption[Comparison of survey simulations using a meridian limit]{
        Comparison of 1N4 survey simulations using different meridian limits.
    }\label{tab:survey_sim_meridian}
\end{table}

\begin{figure}[p]
    \begin{center}
        \begin{minipage}[t]{0.49\linewidth}\vspace{10pt}
            \includegraphics[height=140pt]{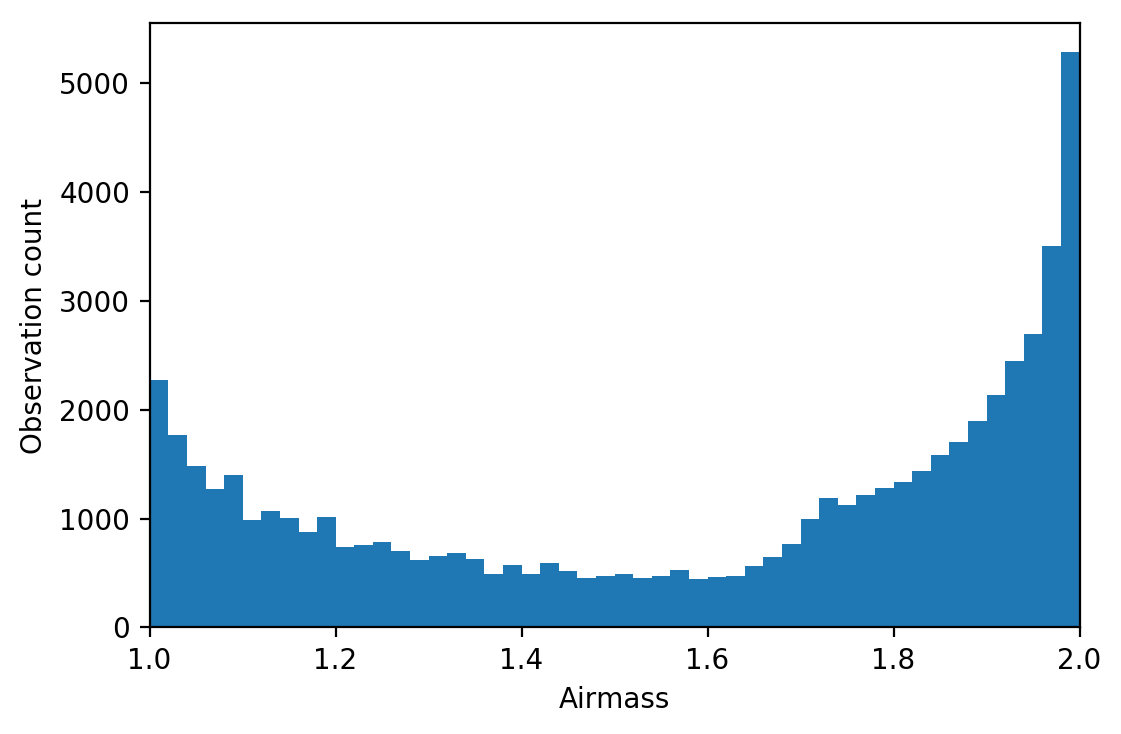}
        \end{minipage}
        \begin{minipage}[t]{0.49\linewidth}\vspace{10pt}
            \includegraphics[height=140pt]{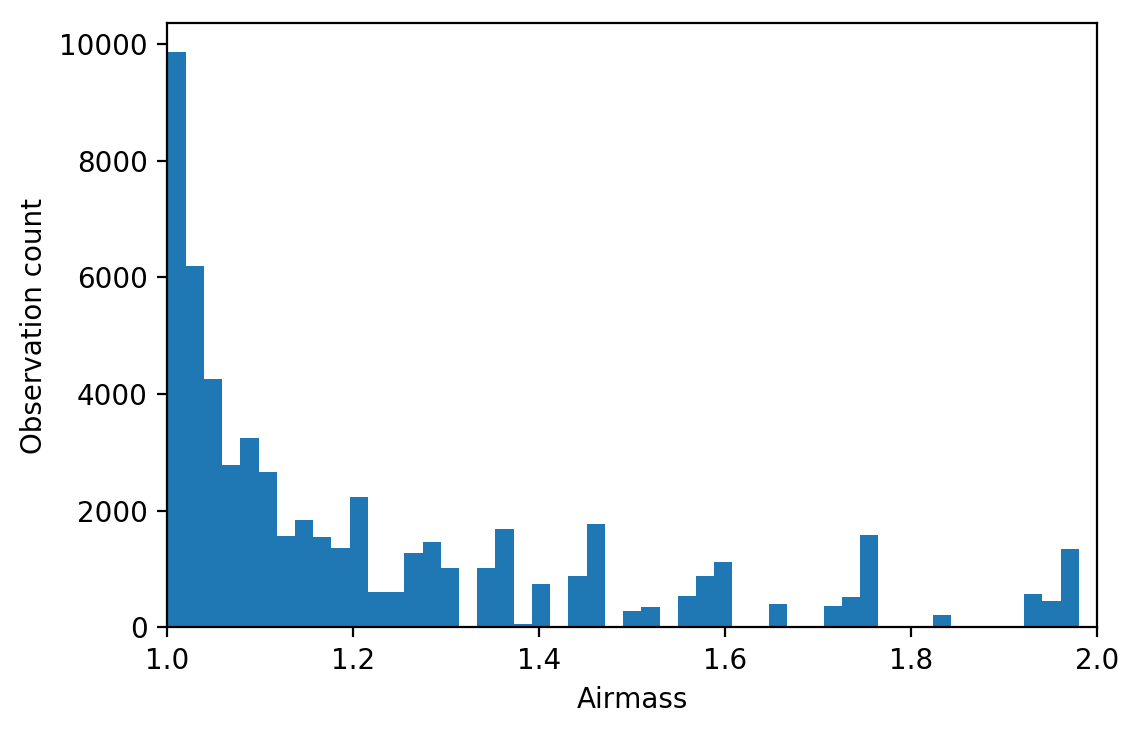}
        \end{minipage}
    \end{center}
    \caption[Airmass distribution over a year of observations]{
        Airmass distribution over a year of observations with the 1N4 system, for the normal unlimited case (left) and limited to the observer's meridian $\pm\SI{10}{\degree}$ (right).
    }\label{fig:survey_sim_airmass_365}
\end{figure}

\begin{figure}[p]
    \begin{center}
        \includegraphics[width=0.7\linewidth]{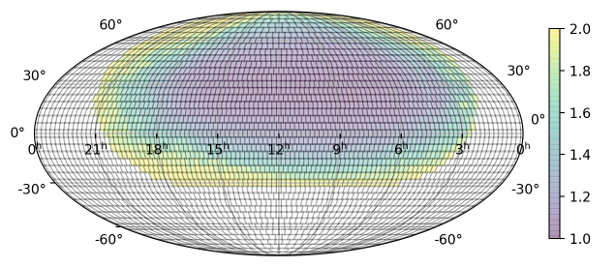}
    \end{center}
    \caption[Mean observation airmasses for the 1N4 survey simulation]{
        Mean observation airmasses for the first month of the 1N4 survey simulation, with no hour angle limit.
    }\label{fig:survey_sim_airmass_normal}
\end{figure}

\begin{figure}[p]
    \begin{center}
        \includegraphics[width=0.7\linewidth]{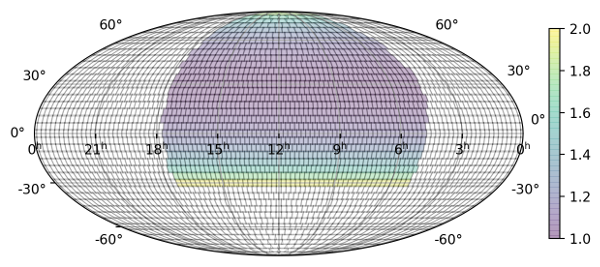}
    \end{center}
    \caption[Mean observation airmasses using the meridian scanning method]{
        Mean observation airmasses for the first month of a survey using the meridian scanning method, restricting observations to tiles within $\pm\SI{10}{\degree}$ of the observer's meridian. Using the hour angle limit it possible to optimise the airmass of observations, but at the cost of sky coverage.
    }\label{fig:survey_sim_airmass_meridian}
\end{figure}

\end{colsection}

\section{Summary and Conclusions}
\label{sec:multiscope_conclusion}

\begin{colsection}

In this chapter I exampled different possible expansion options for the GOTO project, and their effect on the core science.

The intention of the GOTO project has always been to include two complimentary sites in the northern and southern hemispheres, La Palma was he natural choice for the north and a second site in Australia would provide an ideal counterpart. The intention is to have four telescopes in total, two at each site, and all four will operate together as a single global observatory. The core G-TeCS control software outlined in \aref{chap:gtecs} and \aref{chap:autonomous} can be fairly easily duplicated for each system, but in order to achieve optimal coordination between the sites the G-TeCS scheduling system detailed in \aref{chap:scheduling} will need to be expanded to schedule all the active telescopes at once. This is given as a major area of future work in \aref{chap:conclusion}.

In order to examine different possible GOTO configurations, and to make the case for the full deployment described above, I carried out two major series of simulations. The first focused on the combined system's ability to follow-up gravitational-wave alerts, and determined that expanding to the southern hemisphere provides a large improvement to the ability to observe counterparts --- purely by allowing more of the sky to be surveyed. The second set of simulations made the case for hosting two telescopes at each site, as only then will the fast cadences required to reject candidates be achieved. Together the full proposed GOTO system, four 40 square degree field of view mounts located across the two sites, will be a world-leading facility, able to observe the entire visible sky every 1--2 days and provide the best chance to locate and identify optical counterparts to future gravitational-wave detections.

\end{colsection}

\chapter{Conclusions and Future Work}
\label{chap:conclusion}

\chaptoc{}

\section{Summary and Conclusions}
\label{sec:conclusion}

\begin{colsection}

In this thesis I have described my work as part of the GOTO project, primarily working on the control software in order to create a fully-autonomous robotic telescope. After several years of development and commissioning the prototype GOTO telescope is fully operational, and observing from La Palma most nights with no human interaction.

\end{colsection}

\subsection{Telescope control}
\label{sec:control_results}
\begin{colsection}

The core of my work has been the \acro{gtecs}, a Python software package that controls every aspect of the telescope. The hardware control daemons interface with the dome, mount and cameras (see \aref{chap:gtecs}) while the ``pilot'' master control program and its associated systems allow the telescope to function with no human involvement (see \aref{chap:autonomous}). GOTO has now been operating successfully for years with the pilot in full control. The conditions monitoring systems have proven robust enough to trust the dome to close in bad weather, and when the occasional unexpected hardware issues do occur the pilot recovery systems can fix the problem and resume observing in the majority of cases, often before a human even has time to log in. Of course, commissioning was not entirely without incident, as described in \aref{chap:commissioning}. However all of the software challenges were overcome, and the majority of the delays to GOTO were due to hardware faults which were out of my purview.

Each set of exposures taken with the G-TeCS camera daemon are assigned an incremental run number. From the initial installation in the summer of 2017 up until September 2019, GOTO had taken over 185,000 such exposure sets, and produced many tens of terabytes of data. The current all-sky survey that began in February 2019 has almost completely covered the northern sky, and at the time of writing the GOTO photometry database contains approximately 642 million sources from almost 500,000 individual frames taken since the start of the survey.

\newpage

\end{colsection}

\subsection{Scheduling and alert follow-up}
\label{sec:gw_results}
\begin{colsection}

As outlined in \aref{sec:control_requirements}, GOTO needed an observation scheduling system that could deal with both the survey and gravitational-wave follow-up modes. The scheduler used by G-TeCS is a just-in-time system (see \aref{chap:scheduling}), where the highest priority target is recalculated every time the scheduler is called. This makes it very reactive to transient alerts, which was a requirement of the project.

As previously detailed in \aref{sec:gw_detections}, in the first 5 months of the third LIGO-Virgo observing run (O3) 32 gravitational-wave events were detected, all predicted to come from compact binary mergers. These events are listed in \aref{tab:gw_log}. Of the 32 alerts 7 were ultimately retracted by the LVC, leaving 25 real events, and only three of these (S190425z, S190426c and S190814bv) are thought to have originated from sources that could produce visible optical counterparts (binary neutron stars or neutron star-black hole binaries). However, events are currently infrequent enough for GOTO to react to all of them, even if there was a very low chance of there being a visible counterpart.

The G-TeCS sentinel received and reacted to every one of these alerts. In a few cases the event handler initially failed to process the VOEvent or the skymap, as described in \aref{sec:challenges}; this was usually due to a problem on the LVC end, and, after each time, changes were made to the GOTO-alert code to work around the problem should it happen again. For every alert the sentinel received the VOEvent packet and passed it to the event handler code as described in \aref{chap:alerts}, which added pointings to the observation database which could then be observed by the pilot (see \aref{chap:autonomous}). Details of GOTO's reaction to every event are given in \aref{tab:obs_log}. Observations were taken for 25 of the 32 events; of the remaining seven events four were received during the day on La Palma and were then retracted before sunset, meaning the pointings were deleted from the database, and the last three events had no part of the skymap visible from La Palma.

\newpage

\makeatletter
\setlength{\@fptop}{0pt}
\makeatother

\begin{table}[t]
    \begin{footnotesize}
    \begin{center}
        \begin{tabular}{l|ccrrl} %
                                       & GW signal             & Source                      & \multicolumn{1}{c}{Dist.} & \multicolumn{1}{c}{90\% area} & \multicolumn{1}{c}{False Alarm} \\
            \multicolumn{1}{c|}{Event} & detection time        & Classification              & \multicolumn{1}{c}{(Mpc)}    & \multicolumn{1}{c}{(sq deg)}     & \multicolumn{1}{c}{Rate} \\
            \midrule
            \textcolor{Red}{S190405ar} & 2019--04--05 16:01:30 & Terrestrial                                                                    &  268 & 2677 & 1 per 0.00015 yrs      \\ %
                            S190408an  & 2019--04--08 18:18:02 & \textcolorbf{BrickRed}{BBH}                                                    & 1473 &  386 & 1 per \SI{1.1e+10} yrs \\ %
                            S190412m   & 2019--04--12 05:30:44 & \textcolorbf{BrickRed}{BBH}                                                    &  812 &  157 & 1 per \SI{1.9e+19} yrs \\ %
                            S190421ar  & 2019--04--21 21:38:56 & \textcolorbf{BrickRed}{BBH}                                                    & 1628 & 1443 & 1 per 2.1 yrs          \\ %
                            S190425z   & 2019--04--25 08:18:05 & \textcolorbf{Cerulean}{BNS}                                                    &  156 & 7461 & 1 per 69882 yrs        \\ %
                            S190426c   & 2019--04--26 15:21:55 & \textcolorbf{Cerulean}{BNS}/\textcolorbf{Purple}{NSBH}/\textcolorbf{Green}{MG} &  377 & 1131 & 1 per 1.6 yrs          \\ %
                            S190503bf  & 2019--05--03 18:54:04 & \textcolorbf{BrickRed}{BBH}                                                    &  421 &  448 & 1 per 19.4 yrs         \\ %
                            S190510g   & 2019--05--10 02:59:39 & \textcolorbf{Cerulean}{BNS}                                                    &  227 & 1166 & 1 per 3.6 yrs          \\ %
                            S190512at  & 2019--05--12 18:07:14 & \textcolorbf{BrickRed}{BBH}                                                    & 1388 &  252 & 1 per 16.7 yrs         \\ %
                            S190513bm  & 2019--05--13 20:54:28 & \textcolorbf{BrickRed}{BBH}                                                    & 1987 &  691 & 1 per 84922 yrs        \\ %
                            S190517h   & 2019--05--17 05:51:01 & \textcolorbf{BrickRed}{BBH}                                                    & 2950 &  939 & 1 per 13.4 yrs         \\ %
            \textcolor{Red}{S190518bb} & 2019--05--18 19:19:19 & \textcolorbf{Cerulean}{BNS}                                                    &   28 &  136 & 1 per 3.2 yrs          \\ %
                            S190519bj  & 2019--05--19 15:35:44 & \textcolorbf{BrickRed}{BBH}                                                    & 3154 &  967 & 1 per 5.6 yrs          \\ %
                            S190521g   & 2019--05--21 03:02:29 & \textcolorbf{BrickRed}{BBH}                                                    & 3931 &  765 & 1 per 8.3 yrs          \\ %
                            S190521r   & 2019--05--21 07:43:59 & \textcolorbf{BrickRed}{BBH}                                                    & 1136 &  488 & 1 per 100 yrs          \\ %
            \textcolor{Red}{S190524q}  & 2019--05--24 04:52:06 & \textcolorbf{Cerulean}{BNS}                                                    &  192 & 5685 & 1 per 4.5 yrs          \\ %
                            S190602aq  & 2019--06--02 17:59:27 & \textcolorbf{BrickRed}{BBH}                                                    &  797 & 1172 & 1 per 16.7 yrs         \\ %
                            S190630ag  & 2019--06--30 18:52:05 & \textcolorbf{BrickRed}{BBH}                                                    &  926 & 1483 & 1 per 220922 yrs       \\ %
                            S190701ah  & 2019--07--01 20:33:06 & \textcolorbf{BrickRed}{BBH}                                                    & 1849 &   49 & 1 per 1.7 yrs          \\ %
                            S190706ai  & 2019--07--06 22:26:41 & \textcolorbf{BrickRed}{BBH}                                                    & 5263 &  825 & 1 per 16.7 yrs         \\ %
                            S190707q   & 2019--07--07 09:33:26 & \textcolorbf{BrickRed}{BBH}                                                    &  781 &  921 & 1 per 6023 yrs         \\ %
                            S190718y   & 2019--07--18 14:35:12 & Terrestrial                                                                    &  227 & 7246 & 1 per 0.9 yrs          \\ %
                            S190720a   & 2019--07--20 00:08:36 & \textcolorbf{BrickRed}{BBH}                                                    &  869 &  443 & 1 per 8.3 yrs          \\ %
                            S190727h   & 2019--07--27 06:03:33 & \textcolorbf{BrickRed}{BBH}                                                    & 2839 &  152 & 1 per 230 yrs          \\ %
                            S190728q   & 2019--07--28 06:45:10 & \textcolorbf{BrickRed}{BBH}                                                    &  874 &  105 & 1 per \SI{1.3e+15} yrs \\ %
            \textcolor{Red}{S190808ae} & 2019--08--08 22:21:21 & \textcolorbf{Cerulean}{BNS}                                                    &  208 & 5365 & 1 per 0.9 yrs          \\ %
                            S190814bv  & 2019--08--14 21:10:39 & \textcolorbf{Purple}{NSBH}                                                     &  267 &   24 & 1 per \SI{1.6e+25} yrs \\ %
            \textcolor{Red}{S190816i}  & 2019--08--16 13:04:31 & \textcolorbf{Purple}{NSBH}                                                     &  261 & 1467 & 1 per 2.2 yrs          \\ %
            \textcolor{Red}{S190822c}  & 2019--08--22 01:29:59 & \textcolorbf{Cerulean}{BNS}                                                    &   35 & 2767 & 1 per \SI{5.2e+09} yrs \\ %
                            S190828j   & 2019--08--28 06:34:05 & \textcolorbf{BrickRed}{BBH}                                                    & 1946 &  228 & 1 per \SI{3.7e+13} yrs \\ %
                            S190828l   & 2019--08--28 06:55:09 & \textcolorbf{BrickRed}{BBH}                                                    & 1528 &  358 & 1 per 685.0 yrs        \\ %
            \textcolor{Red}{S190829u}  & 2019--08--29 21:05:56 & \textcolorbf{Green}{MG}                                                        &  157 & 8972 & 1 per 6.2 yrs          \\ %
        \end{tabular}
    \end{center}
    \end{footnotesize}
    \caption[GW detections from O3 so far]{
        All 32 detections of gravitational-wave signals made during O3, up to the end of August 2019. Events in \textcolorbf{Red}{red} were ultimately retracted by the LVC.\@ All the given values are from the latest issued alert and final skymaps. The distance is the peak of the distribution included in the alert, and the area given is the area contained within the 90\% skymap contour level (see \aref{sec:healpix}).
    }\label{tab:gw_log}
\end{table}

\clearpage

\begin{table}[t]
    \begin{footnotesize}
    \begin{center}
        \begin{tabular}{l|ccrrrr} %
                                       & Time alert & Time of first & \multicolumn{1}{c}{Time}  &                &                  &                                    \\
            \multicolumn{1}{c|}{Event} & received   & observation   & \multicolumn{1}{c}{delay} & $N_\text{obs}$ & $N_\text{tiles}$ & \multicolumn{1}{c}{$P_\text{obs}$} \\
            \midrule
            \textcolor{Red}{S190405ar} & 2019--04--12 15:07:26 & \multicolumn{5}{l}{\textcolor{Thistle}{\textit{(Not observed --- Event retracted before becoming visible)}}} \\
                            S190408an  & 2019--04--08 19:02:50 & 2019--04--09 05:40:39 &              \SI{10.63}{\hour} &  17 &   9 & 22.5\% \\
                            S190412m   & 2019--04--12 06:31:39 & 2019--04--12 20:28:35 &              \SI{13.95}{\hour} &  36 &  18 & 96.1\% \\
                            S190421ar  & 2019--04--22 16:26:24 & 2019--04--23 21:54:59 &              \SI{29.48}{\hour} &  49 &   7 & 10.2\% \\
                            S190425z   & 2019--04--25 09:00:56 & 2019--04--25 20:38:22 &              \SI{11.62}{\hour} & 306 & 173 & 22.6\% \\
                            S190426c   & 2019--04--26 15:47:11 & 2019--04--26 20:38:45 &              \SI{ 4.86}{\hour} &  96 &  49 & 55.6\% \\
                            S190503bf  & 2019--05--03 19:30:15 & \multicolumn{5}{l}{\textcolor{Thistle}{\textit{(Not observed --- Skymap not visible from La Palma)}}} \\
                            S190510g   & 2019--05--10 04:21:59 & 2019--05--10 04:22:55 &  \textcolorbf{NavyBlue}{56\,s} &   7 &   7 &  0.2\% \\
                            S190512at  & 2019--05--12 18:59:01 & 2019--05--12 20:53:20 &              \SI{ 1.91}{\hour} & 201 &  19 & 89.1\% \\
                            S190513bm  & 2019--05--13 21:21:51 & 2019--05--13 21:26:19 & \textcolorbf{NavyBlue}{4\,min} &  38 &   7 & 30.2\% \\
                            S190517h   & 2019--05--17 06:26:48 & 2019--05--17 21:42:06 &              \SI{15.26}{\hour} &   9 &   7 & 15.4\% \\
            \textcolor{Red}{S190518bb} & 2019--05--18 19:25:49 & \multicolumn{5}{l}{\textcolor{Thistle}{\textit{(Not observed --- Event retracted before becoming visible)}}} \\
                            S190519bj  & 2019--05--19 17:01:40 & 2019--05--19 20:55:19 &              \SI{ 3.89}{\hour} & 139 &  42 & 78.7\% \\
                            S190521g   & 2019--05--21 03:08:49 & 2019--05--21 03:09:17 &  \textcolorbf{NavyBlue}{28\,s} &  58 &  24 & 44.5\% \\
                            S190521r   & 2019--05--21 07:50:27 & 2019--05--21 22:54:03 &              \SI{15.06}{\hour} &  90 &  45 & 94.0\% \\
            \textcolor{Red}{S190524q}  & 2019--05--24 04:58:40 & 2019--05--24 04:59:33 &  \textcolorbf{NavyBlue}{53\,s} &   2 &   2 & 14.2\% \\
                            S190602aq  & 2019--06--02 18:06:01 & \multicolumn{5}{l}{\textcolor{Thistle}{\textit{(Not observed --- Skymap not visible from La Palma)}}} \\
                            S190630ag  & 2019--06--30 18:55:47 & 2019--06--30 21:14:49 &              \SI{ 2.32}{\hour} & 149 &  75 & 62.9\% \\
                            S190701ah  & 2019--07--01 20:38:06 & \multicolumn{5}{l}{\textcolor{Thistle}{\textit{(Not observed --- Skymap not visible from La Palma)}}} \\
                            S190706ai  & 2019--07--06 22:44:31 & 2019--07--06 22:45:09 &  \textcolorbf{NavyBlue}{38\,s} &  70 &  35 & 27.0\% \\
                            S190707q   & 2019--07--07 10:13:24 & 2019--07--07 21:54:47 &              \SI{11.69}{\hour} & 116 &  58 & 41.0\% \\
                            S190718y   & 2019--07--18 15:03:13 & 2019--07--18 21:08:53 &              \SI{ 6.09}{\hour} & 135 &  15 & 61.5\% \\
                            S190720a   & 2019--07--20 00:11:26 & 2019--07--20 00:11:57 &  \textcolorbf{NavyBlue}{31\,s} & 175 &  87 & 83.9\% \\
                            S190727h   & 2019--07--27 06:12:02 & 2019--07--27 21:03:40 &              \SI{14.86}{\hour} &  94 &  47 & 42.4\% \\
                            S190728q   & 2019--07--28 06:59:32 & 2019--07--28 21:29:58 &              \SI{14.51}{\hour} &  36 &   9 & 90.5\% \\
            \textcolor{Red}{S190808ae} & 2019--08--08 22:28:00 & 2019--08--08 22:28:31 &  \textcolorbf{NavyBlue}{31\,s} &  75 &  31 & 17.3\% \\
                            S190814bv  & 2019--08--14 21:31:44 & 2019--08--14 22:59:27 &              \SI{ 1.46}{\hour} & 141 &  45 & 95.4\% \\
            \textcolor{Red}{S190816i}  & 2019--08--16 13:11:35 & \multicolumn{5}{l}{\textcolor{Thistle}{\textit{(Not observed --- Event retracted before becoming visible)}}} \\
            \textcolor{Red}{S190822c}  & 2019--08--22 01:37:00 & 2019--08--22 01:37:30 &  \textcolorbf{NavyBlue}{30\,s} &  17 &   8 &  2.9\% \\
                            S190828j   & 2019--08--28 06:50:14 & 2019--08--28 22:38:25 &              \SI{15.80}{\hour} &  54 &  27 &  9.3\% \\
                            S190828l   & 2019--08--28 07:17:46 & 2019--08--28 23:48:38 &              \SI{16.51}{\hour} &  56 &  28 &  2.0\% \\
            \textcolor{Red}{S190829u}  & 2019--08--29 21:17:14 & \multicolumn{5}{l}{\textcolor{Thistle}{\textit{(Not observed --- Event retracted before becoming visible)}}} \\
        \end{tabular}
    \end{center}
    \end{footnotesize}
    \caption[GOTO observation log for O3 events so far]{
        GOTO observation log for O3 events (from \aref{tab:gw_log}). Seven events were not observed by GOTO (in \textcolorbf{Thistle}{pink}); four were retracted before observations could begin and three had skymaps never visible from La Palma. The time delay is the delay between the sentinel receiving the alert and observations beginning (including event processing and slew time), events with a delay in \textcolorbf{NavyBlue}{blue} were received during night on La Palma and had tiles immediately visible. $N_\text{obs}$ is the total number of pointings observed by GOTO for each event, $N_\text{tiles}$ is the number of tiles observed within those pointings and $P_\text{pbs}$ is the total contained skymap probability within the observed tiles.
    }\label{tab:obs_log}
\end{table}

\clearpage

\makeatletter
\setlength{\@fptop}{0\p@ \@plus 1fil} %
\makeatother

\newpage

\begin{figure}[p]
    \begin{center}
        \includegraphics[width=0.95\linewidth]{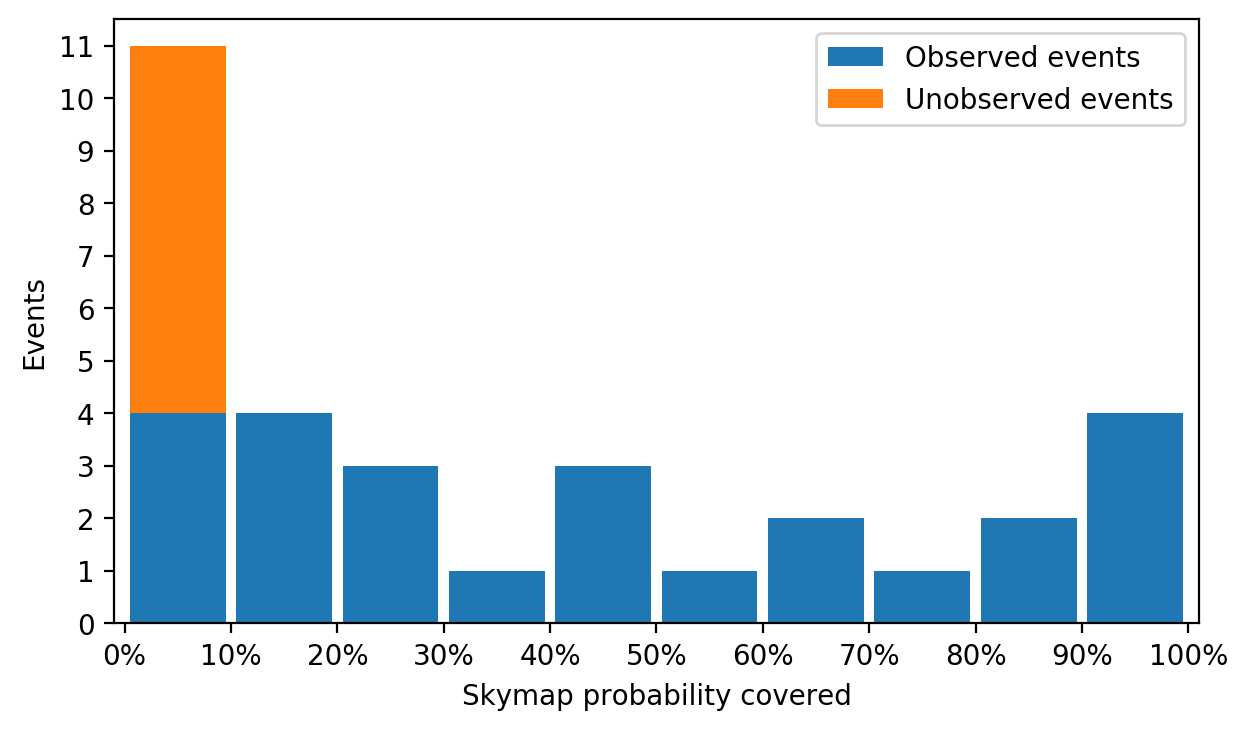}
    \end{center}
    \caption[Histogram of probabilities covered for O3 events]{
        Histogram of the skymap probability covered by GOTO for O3 events. \\
        7 of the 32 events were never observed (see \aref{tab:obs_log}).
    }\label{fig:events_prob}
\end{figure}

\begin{figure}[p]
    \begin{center}
    \includegraphics[width=0.95\linewidth]{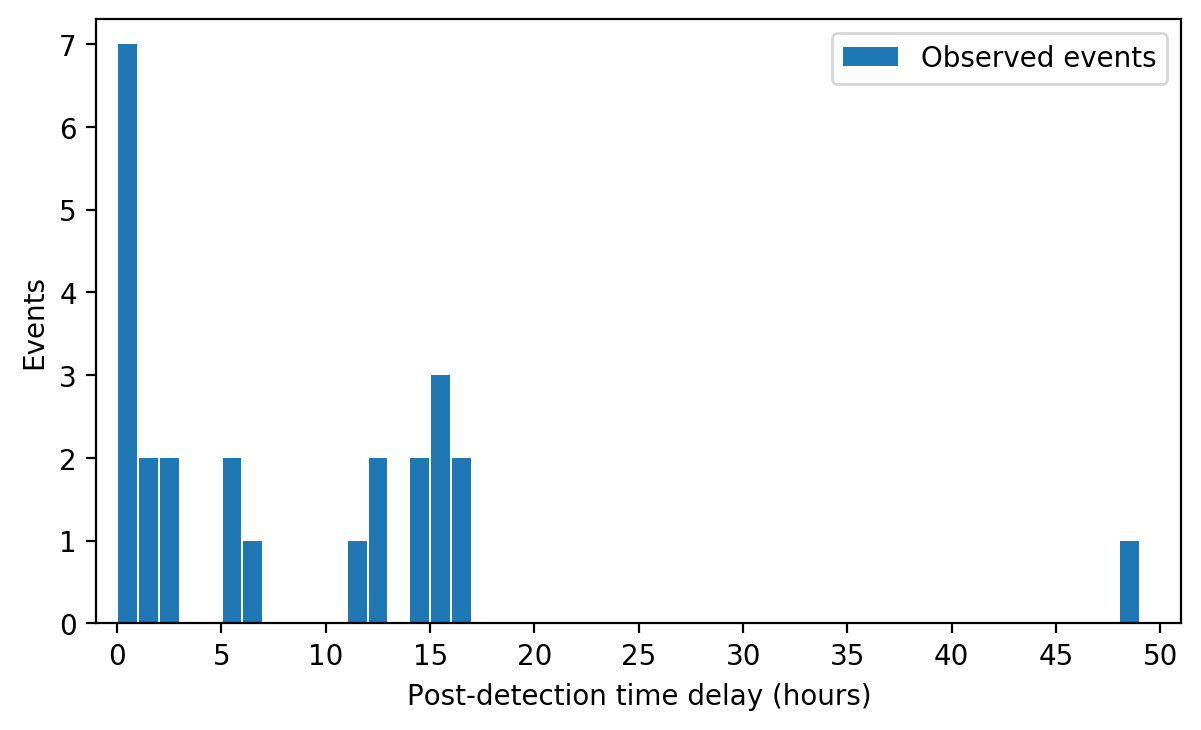}
    \end{center}
    \caption[Histogram of post-event delay for O3 events]{
        Histogram of the delay between the GW signal being detected by the LVC and GOTO commencing observations for O3 events. Events that GOTO did not observe are excluded. The outlier is S190421ar (due to a delay sending out the alert).
    }\label{fig:events_delay}
\end{figure}

\clearpage

\aref{fig:events_prob} shows the percentage of each event skymap covered by GOTO.\@ These value will depend on a variety of factors: for a particular event if only a small fraction of the skymap is covered that might be because of the position of the skymap in the sky, or a period of bad conditions forcing the dome to close. The coverage level is relatively even across all the events, once the seven events that GOTO did not observe are removed. \aref{fig:events_delay} shows the delay between the event being detected by the gravitational-wave detectors (from \aref{tab:gw_log}) and GOTO starting observations (from \aref{tab:obs_log}). For seven events GOTO was able to start observing less than an hour after the event was detected. This value includes factors such as the delay between the event being detected and the LVC releasing their alert, which was the cause for observations of S190421ar being delayed so long (see \aref{sec:challenges}). For this reason the time delay between the alert being received by the G-TeCS sentinel and the start of observations (given in \aref{tab:obs_log}) is a better indicator of the performance of the G-TeCS software.

Eight alerts were received while it was night time on La Palma, and the GOTO-alert event handling system allowed the pilot to immediately begin observations of the visible skymap. As shown in \aref{tab:obs_log}, in all but one of the eight cases the first exposure was started less than \SI{60}{\second} after the sentinel received the alert. The time delay varies between 28 and 56 seconds, primarily depending on how far the mount had to slew from its previous target. Of the remaining $\sim$\SI{25}{\second} delay, a significant amount is due to having to download the LVC skymaps, with the rest due to various small delays in the event handler, sentinel and pilot, such as the pilot needing to wait up to \SI{10}{\second} for the next scheduler check (see \aref{sec:checks}). Future optimisation could potentially reduce these delays further. The one exception was event S190513bm, which was immediately visible but observations were delayed by 4 minutes. At the time the alert was received the pilot was already observing a pointing from the S190512at event received the previous day; as both events were black hole binaries they were inserted at the same rank, and, as detailed in \aref{sec:toos}, equal-rank ToO pointings will not interrupt each other, so the new pointing had to wait until the previous one was completed. In all other cases the pilot was observing a lower-rank target, usually a survey tile, which was immediately aborted when the scheduler check returned the ToO gravitational-wave pointing.

\begin{table}[t]    \begin{center}
        \begin{tabular}{l|ccccc} %
                                       & Post-detection   & Probability &  Area covered & $5\sigma$ limiting & GOTO \\
            \multicolumn{1}{c|}{Event} & time delay       & covered     & (sq deg)      & magnitude          & GCN \\
            \midrule
                             S190425z  & \SI{12.3}{\hour} & 22.6\%      & 2857          & $g=20.1$           & 24224 \tablefootnote{~~\citet{S190425z_GOTO}} \\
                             S190426c  &  \SI{5.3}{\hour} & 55.6\%      &  841          & $g=19.9$           & 24291 \tablefootnote{~~\citet{S190426c_GOTO}} \\
                             S190814bv &  \SI{1.8}{\hour} & 95.4\%      &  811          & $g=18.9$           & 25337 \tablefootnote{~~\citet{S190814bv_GOTO}} \\
        \end{tabular}
    \end{center}
    \caption[GOTO follow-up results for three key O3 events]{
        GOTO follow-up results for three key O3 events.
    }\label{tab:events_3key}
\end{table}

The three events that were the most likely to have a potential electromagnetic counterpart (originating from either a binary neutron star or neutron star-black hole binary) were S190425z, S190426c and S190814bv. The GOTO response to each was reported in a public GCN Notice, and the key values are given in \aref{tab:events_3key}.

S190425z was the second detection of gravitational waves from a binary neutron star after GW170817 \citep{S190425z}. Unlike GW170817 however, the signal was only detected by a single detector, LIGO-Livingston (LIGO-Hanford was offline at the time, and while Virgo did detect a signal it was below the valid signal-to-noise threshold). This resulted in a very large initial skymap, shown in \aref{fig:190425_goto}, with a 90\% contour area of 10,183 square degrees. Many other projects aside from GOTO followed-up this event, and the \acro{ztf} efforts were described in \aref{sec:followup}. Unfortunately a large portion of the skymap was located behind the Sun at the time and was therefore unobservable (on the right of \aref{fig:190425_goto}). The alert was received at 09:00 UTC, a few hours after GOTO had closed in the morning, meaning observations from La Palma could not begin for just over 12 hours until sunset that evening. The final skymap reduced the search area to 7,416 square degrees but shifted the probability even further into the unobservable region of the sky, meaning in the end GOTO only covered 22.6\% of the final probability.

\begin{figure}[t]
    \begin{center}
        \includegraphics[width=0.9\linewidth]{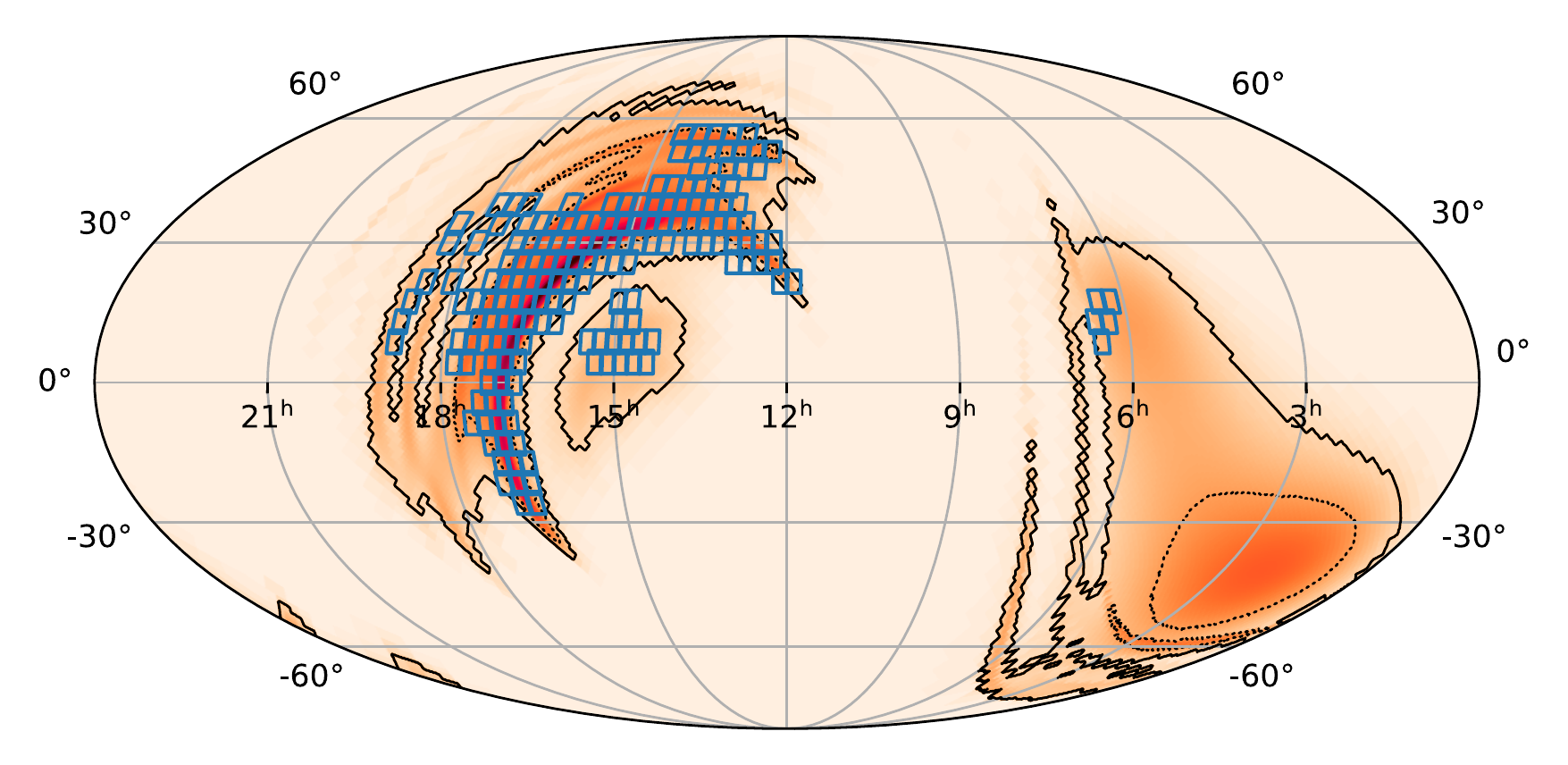}
    \end{center}
    \caption[Follow-up observations of S190425z with GOTO]{
        GOTO follow-up observations of GW event S190425z \citep{S190425z_GOTO}. The tiled observations are shown in \textcolorbf{NavyBlue}{blue} over the initial skymap. Compare to \aref{fig:ztf}, which shows ZTF's coverage of the same event.
        }\label{fig:190425_goto}
\end{figure}

\begin{figure}[t]
    \begin{center}
        \includegraphics[width=0.9\linewidth]{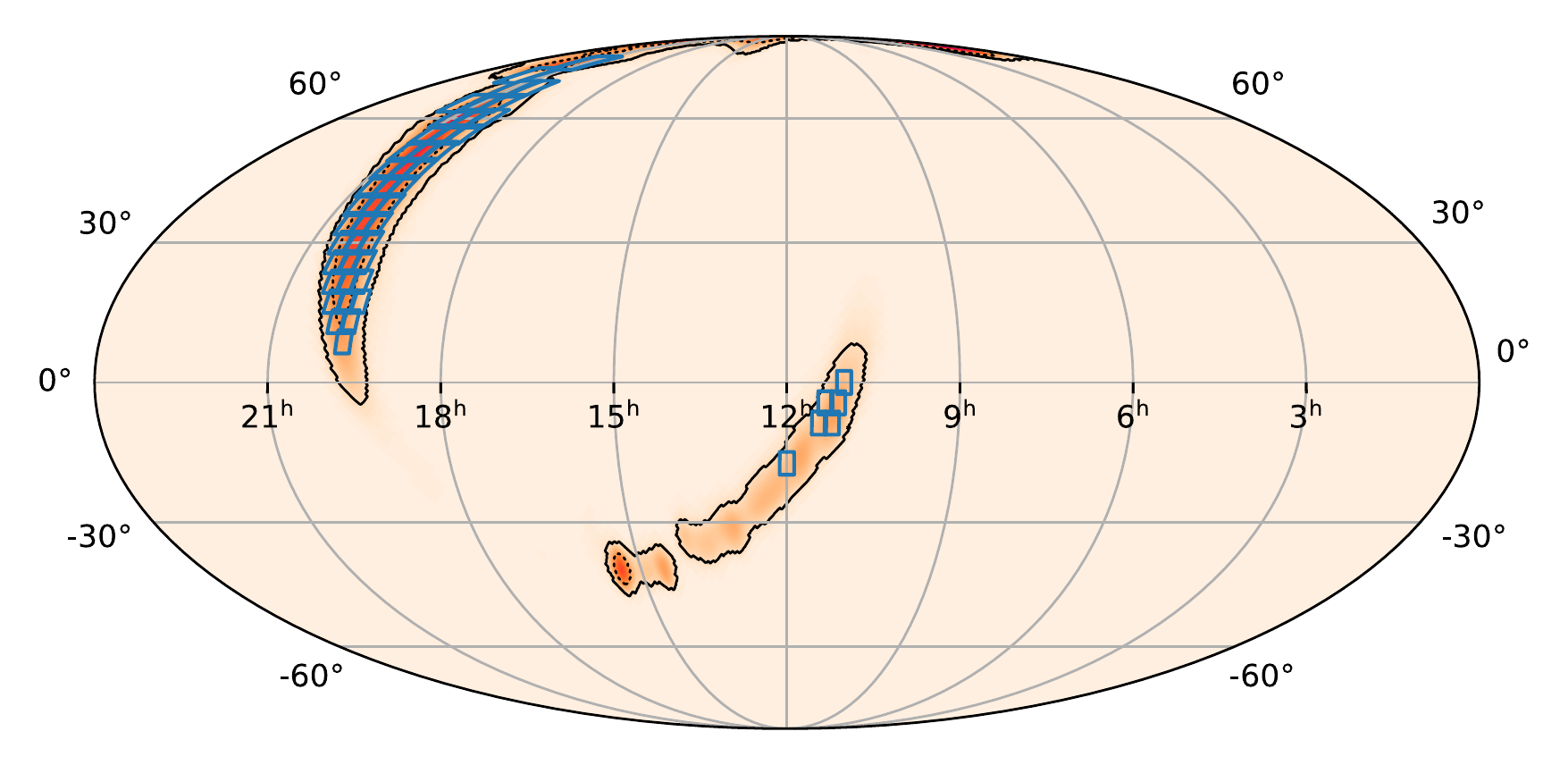}
    \end{center}
    \caption[Follow-up observations of S190426c with GOTO]{
        GOTO follow-up observations of GW event S190426c \citep{S190426c_GOTO}. The tiled observations are shown in \textcolorbf{NavyBlue}{blue} over the initial skymap.
        }\label{fig:190426_goto}
\end{figure}

190426c was detected just 31 hours after S190425z \citep{S190425z}. This meant that on the night of the 26th GOTO was completing both the first pass of the 190426c tiles and the second pass of the S190425z tiles which had been first observed the previous night. The initial 190426c skymap and the tiles observed are shown in \aref{fig:190426_goto}. This time event was detected by all three detectors; the initial skymap had an area of 1,932 square degrees, and changed very little in the final skymap. The skymaps from the two events also did not overlap. Following-up two events at once had been considered in the design of the G-TeCS scheduling system (described \aref{chap:scheduling}), and as planned GOTO alternated between the two as tiles were observed (therefore lowering the effective rank as described in \aref{sec:rank}).

The latest event with a high chance of an optical counterpart was S190814bv \citep{S190814bv}, the first confirmed detection of gravitational waves from a neutron star-black hole binary. As shown in \aref{tab:gw_log} several other binary neutron star and neutron star-black hole binary events have been detected but have since been retracted. The initial skymap only included the contribution from the LIGO-Livingston and Virgo detectors, it covered 772 square degrees and is shown in \aref{fig:190826_goto}. A few hours later a revised skymap including LIGO-Hanford data was released, which reduced the 90\% contour region to just 38 square degrees. Unfortunately due to an error in how the LVC uploaded the skymap this was not immediately processed by GOTO (see \aref{sec:challenges}). The lower plot of \aref{fig:190826_goto} shows the GOTO tiles over the final skymap, and although a portion was below the observable horizon from La Palma GOTO still covered 95.4\% of the probability in just 5 pointings. The Moon was full at the time, hence the worse limiting magnitude in \aref{tab:events_3key}.

No electromagnetic counterparts were found for any of the three events, either by GOTO or other projects. But the G-TeCS follow-up code has proven to be fast and reliable, and GOTO will continue to follow-up LIGO-Virgo alerts. Based on the time from previous alerts, if, or when, another GW170817-like event is detected GOTO could potentially be observing the counterpart within seconds of the alert being received.

\newpage

\makeatletter
\setlength{\@fptop}{0pt}
\makeatother

\begin{figure}[p]
    \begin{center}
        \includegraphics[width=0.9\linewidth]{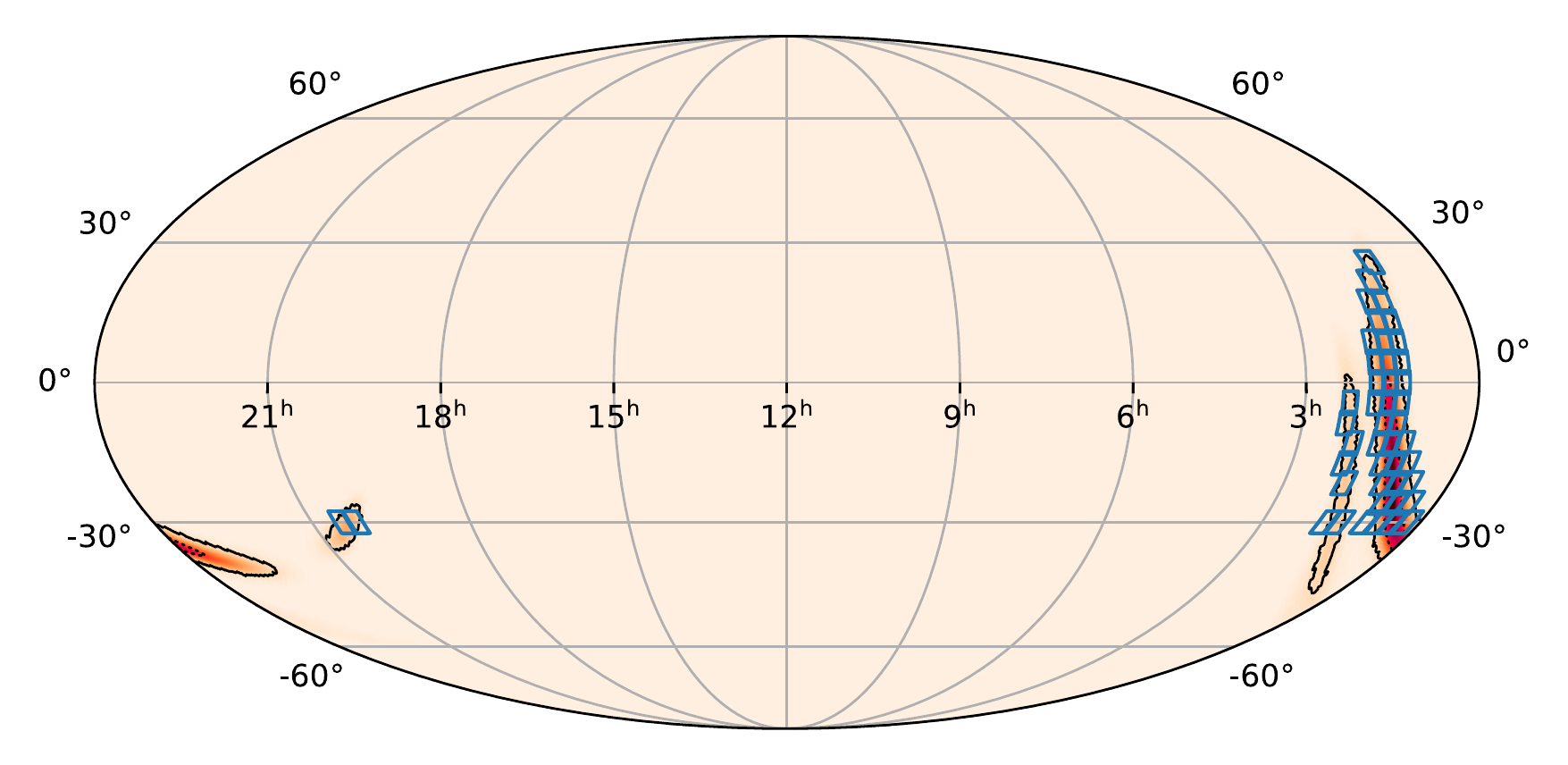}
        \includegraphics[width=0.8\linewidth]{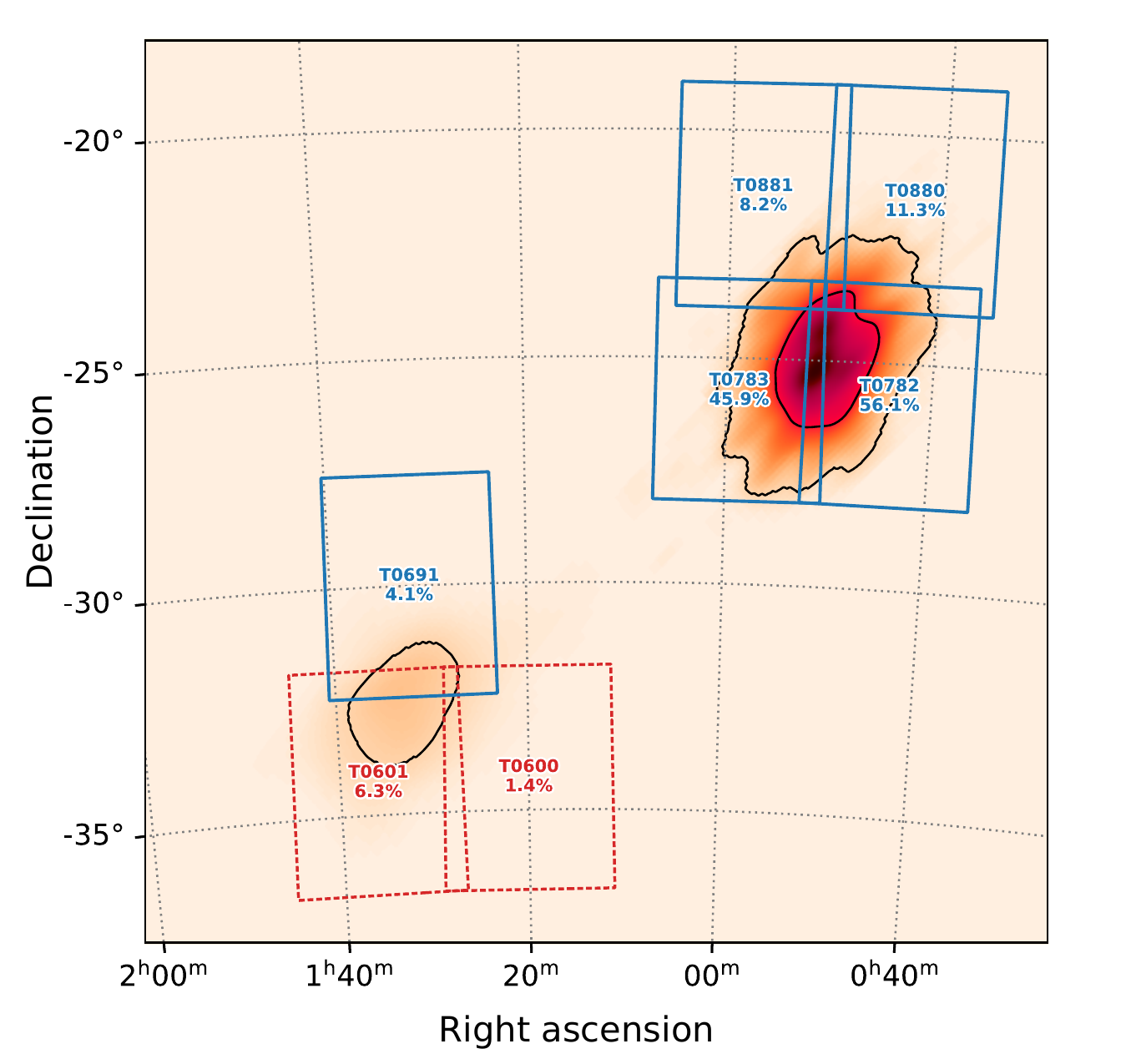}
    \end{center}
    \caption[Follow-up observations of S190814bv with GOTO]{
        GOTO follow-up observations of GW event S190814bv \citep{S190814bv_GOTO}. In the upper plot the tiled observations are shown in \textcolorbf{NavyBlue}{blue} over the initial skymap. The final skymap for this event is shown in the lower plot, with the seven GOTO tiles needed to cover the 90\% probability region. The two tiles in \textcolorbf{Red}{red} were below the local horizon and so were not observed.
        }\label{fig:190826_goto}
\end{figure}

\clearpage

\makeatletter
\setlength{\@fptop}{0\p@ \@plus 1fil} %
\makeatother

\newpage

\end{colsection}

\section{Future work}
\label{sec:future}

\begin{colsection}

This thesis only details the beginning of the GOTO project, and the work described will need to be continued and built upon as the project expands in the future.

\end{colsection}

\subsection{The global control system}
\label{sec:gtecs_future}
\begin{colsection}

Stage 1 of the GOTO project, the first mount and four unit telescopes, is currently observing from La Palma. The obvious direction of future work will be adapting and expanding G-TeCS in order to match the expansion of GOTO, as described in \aref{sec:goto_expansion}.

\subsubsection{Stage 2}

Adding the second set of four unit telescopes to the existing mount on La Palma should require nothing more than a few configuration changes to handle the new interface daemons. On the scheduling side, the observation database will need to be reset with a new all-sky grid, based on the GOTO-8 field of view instead of the existing GOTO-4 tiles (see \aref{fig:fov}). As each tile will cover a larger area, the tile selection algorithm described in \aref{sec:mapping_skymaps} might need to be adjusted, and some of the observing strategy detailed in \aref{chap:alerts} could be revisited. Otherwise, no major changes are anticipated to be required, and the pilot should be able to resume observations immediately.

\subsubsection{Stage 3}

The addition of the second mount in the second dome on La Palma will require more control system development. With two telescopes of the same design it should be simple to copy the hardware control daemons, and some systems could be shared between the two domes (for example, there is no need to have two conditions daemons both monitoring the same weather masts). A proposed system diagram is shown in \aref{fig:flow2}. But, as described in \aref{sec:multi_tel_scheduling}, the great benefit of having two telescopes is having them share a common scheduling system. This could be as simple as the system adopted for the multi-telescope simulations in \aref{chap:multiscope}, marking one telescope as the primary that always observes the highest-priority pointing and having the other always observe the second-highest. But in reality the telescopes will never be perfectly in sync, and there are more benefits to be gained from a more advanced scheduling system.

\begin{figure}[t]
    \begin{center}
        \includegraphics[width=\linewidth]{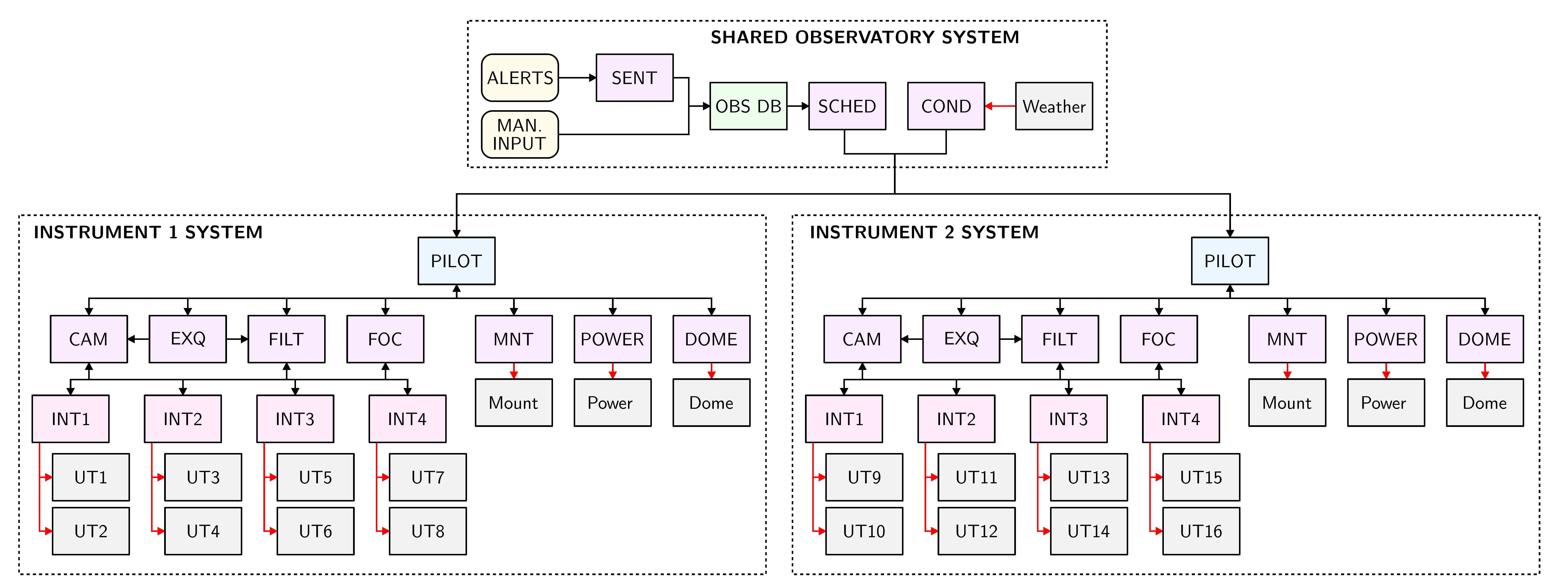}
    \end{center}
    \caption[Future G-TeCS system architecture for two telescopes]{
        Proposed future G-TeCS system architecture for controlling two telescopes at the same site (``Stage 3''). Note that the two pilots share the same scheduler and conditions daemons but are otherwise independent.
    }\label{fig:flow2}
\end{figure}

Just as the current scheduler described in \aref{chap:scheduling} has to decide what to observe based on various constraints, the next-generation scheduler will need to optimise which target is being observed by each telescope. Although several of the constraints will be the same for both mounts (e.g.\ the Moon phase or Sun altitude) it is possible that the two telescopes could have different artificial horizons and therefore altitude limits. The scheduler tie-breaking system will need to be revised, to account for distributing pointings to either telescope. One possible parameter to add to decide which telescope to send a pointing to would be the distance each mount would need to slew from the current target, and the scheduler could also account for the time left on the current observations (e.g.\ telescope 1 might be ready to observe while telescope 2 has 10 seconds left on the current target; if the difference in slew times to the new pointing is greater than \SI{10}{\second} then it would be better to wait for telescope 2 to finish).

The move from one 8-UT mount to two 8-UT mounts should mean the same all-sky grid can be used, as long as both have the same field of view (see \aref{sec:multi_grid_scheduling} for the problems inherent in observing with different grids). However, having multiple mounts opens up opportunities for more advanced observing strategies. They could both observe different parts of the sky to cover the survey or gravitational-wave skymaps more quickly, or they could observe the same tiles to achieve a greater depth when the images are stacked. The simulations in \aref{chap:multiscope} assumed coverage was the priority, but the future scheduling system should be designed to allow concurrent observations of the same tiles if required. The presence of the coloured filters adds even more possibilities. It should be possible to have the two telescopes observe the same target simultaneously but using different filters, to get immediate colour information on any sources. It might also be desirable to accept the impact on survey cadence and have each telescope carry out an independent survey in different filters, or perhaps have one taking rapid \SI{60}{\second} exposures while the other surveys the sky more slowly using the meridian scanning method detailed in \aref{sec:survey_sim_meridian}. The possibilities are endless, and although the ultimate decision will be made depending on the science requirements of the GOTO collaboration, ideally the future G-TeCS scheduling system should be able to handle whatever strategy is desired.

\subsubsection{Stage 4}

The final form of the GOTO project is intended to include multiple telescopes at different sites across the globe. This is unlikely to happen before the second mount is built on La Palma, so by the time an Australian site is added the advanced systems described under Stage 3 above should already be in place, and ideally the next-generation scheduler should be able to delegate observations to multiple telescopes wherever they are in the world. There are several existing projects that operate in this manner which GOTO can emulate, such as the \acro{lco} network \citep{LCO_scheduling}.

\newpage

\end{colsection}

\subsection{Continued development}
\label{sec:software_future}
\begin{colsection}

The work described in \aref{sec:gtecs_future} will be important to carry out as GOTO expands, but the timescales on which it will be needed will be dictated by the hardware status of the project. There is still plenty of software development work to do that is less dependent on the GOTO funding situation.

\subsubsection{Pipeline integration}

One particular area of importance is better integration between the control system and the analysis pipeline and candidate marshal described in \aref{sec:gotophoto}. To achieve a fully autonomous system, the pipeline should be able to reschedule observations independently, for example if an image is affected by clouds. More excitingly, a future transient detection algorithm might be permitted to automatically schedule follow-up observations of promising candidates.

\subsubsection{Unifying scheduler targets}

The scheduler system described in \aref{chap:scheduling} works well for both the all-sky survey and gravitational-wave follow-up events, as discussed in \aref{sec:conclusion}. However, one aspect that could be improved is the integration of both roles. For example, observing a particular tile as part of a gravitational-wave follow-up survey should be counted within the observation database as an observation of that tile for the all-sky survey as well, assuming they use the same filter. The GOTOphoto difference imaging pipeline already looks for reference images for difference imaging in all prior observations of that tile, regardless of what purpose it was taken for. In other words, observing tiles as part of a gravitational-wave skymap should also count towards the all-sky survey cadence: the same tiles are being observed, just in a different order.

\newpage

Another proposed addition in the same vein is linking tile observations between events. It is not uncommon for the same gamma-ray burst event to be detected by both \textit{Fermi} and \textit{Swift}; as described in \aref{sec:event_strategy}, both alerts are processed by the GOTO-alert event handler, and the only difference is that \textit{Swift} events are inserted into the observation database at a higher rank. This is intentional, as \textit{Swift} events typically are very well-localised, easily within a single GOTO tile, while the large skymaps for \textit{Fermi} events cover many tiles (see \aref{sec:grb_skymaps}). Because of this, if the sentinel detects an alert from both facilities within a few minutes, and the two skymaps overlap, then covering the large \textit{Fermi} skymap should be unnecessary, as the event source should be given by the much better localised \textit{Swift} position.

Independent detections of gamma-ray bursts are a common-enough example to test this behaviour, but where it could be very useful to GOTO is for coincident GRB and gravitational-wave events. The GW170817 event was notable as being also detected by \textit{Fermi} as GRB 170817A \citep{GW170817_Fermi}, and the LVC is investigating putting out automated alerts for future coincident GW-GRB events using the RAVEN pipeline \citep{RAVEN, LVC_userguide}. Were the G-TeCS sentinel able to achieve a similar result, simply prioritising GW skymap tiles that overlap with the GRB skymap, it could reduce the delay before observing the all-important tile that contains the counterpart kilonova.

\subsubsection{Further simulations}

The test code written to simulate GOTO observations was a vital tool for optimising the G-TeCS scheduler (see \aref{sec:scheduler_sims}), and in \aref{chap:multiscope} it was used to model the benefits of GOTO's plans for future expansion. As described in \aref{sec:scheduler_sim_future}, it would be good to revisit the scheduler simulations with the benefit of subsequent code development and more-realistic simulation parameters based on the live GOTO system. Other scheduler simulations have also been proposed, for example to find optimal tile-selection limits for gravitational-wave skymaps (see \aref{sec:selecting_tiles}). A majority of the future work proposed in this chapter will also require further simulations, and making the simulation code as realistic as possible is a priority.

\subsubsection{Code generalisation and availability}

Another potential future project that is being considered is the generalisation of some or all of the G-TeCS code, removing the GOTO-specific parts and making it usable by other projects. For example, the GOTO-tile code for creating survey grids and mapping skymaps to them described in \aref{chap:tiling} is not at all GOTO-specific, and would only require a small amount to work to rewrite into a separate Astropy-compatible Python package (probably along with a new name). On a wider scale, the G-TeCS control system could be adapted for other telescopes. A parallel version is already being used by the other Warwick telescopes on La Palma, and using G-TeCS is also being considered for other robotic telescope projects, such as the SAMNET solar telescope network \citep{SAMNET}. Currently all GOTO code is private, restricted only to GOTO collaboration members. However, if my code was reconfigured to be usable by other projects I would hope to make it publicly available and open-source.

\bigskip

Overall the GOTO Telescope Control System is still under active development, and this will continue as the GOTO project evolves. Based on the initial results from O3 the system has been working well and fulfilling its requirements, and it is therefore most likely only a matter of time until GOTO observes its first gravitational-wave counterpart.

\end{colsection}

\pagestyle{plain}
\emergencystretch=3em
\printbibliography[heading=bibintoc]{}
\clearpage

\end{document}

